\documentclass{dune}
\pdfoutput=1

\usepackage[pdftex,bookmarks,hidelinks]{hyperref}

\graphicspath{ {graphics/} }

\newif\ifdp
\newif\ifsp

\def\expshort{DUNE\xspace}
\def\dune{\expshort}
\def\explong{The Deep Underground Neutrino Experiment\xspace}

\def\thedocsubtitle{Deep Underground Neutrino Experiment (DUNE)} 
\def\tdrtitle{Technical Design Report}
\def\voltitleexec{Introduction to DUNE\xspace}
\def\volnumberexec{I}

\def\voltitlephysics{DUNE Physics\xspace}
\def\volnumberphysics{II}

\def\physchtools{Volume~\volnumberphysics{}, \voltitlephysics{}, Chapter~4\xspace}
\def\physchlbl{Volume~\volnumberphysics{}, \voltitlephysics{}, Chapter~5\xspace}
\def\physchndk{Volume~\volnumberphysics{}, \voltitlephysics{}, Chapter~6\xspace}

\def\voltitletc{DUNE Far Detector Technical Coordination\xspace}
\def\volnumbertc{III}

\def\voltitlesp{The DUNE Far Detector Single-Phase Technology\xspace}
\def\volnumbersp{IV}

\def\spchdaq{Volume~\volnumbersp{}, \voltitlesp{}, Chapter~7\xspace}

\def\voltitledp{The DUNE Far Detector Dual-Phase Technology\xspace}
\def\volnumberdp{V}

\newcommand{\refsec}[2]{Volume~\csname volnumber#1\endcsname \xspace Section~#2}
\newcommand{\refch}[2]{Volume~\csname volnumber#1\endcsname \xspace Chapter~#2}
\newcommand{\refinch}[2]{#2 in Volume~\csname volnumber#1\endcsname \xspace}

\newcommand{\numu}{\ensuremath{\nu_\mu}\xspace}
\newcommand{\nue}{\ensuremath{\nu_e}\xspace}

\newcommand{\anumu}{\ensuremath{\bar\nu_\mu}\xspace}
\newcommand{\anue}{\ensuremath{\bar\nu_e}\xspace}

\newcommand{\dm}[1]{\ensuremath{\Delta m^2_{#1}}\xspace} 

\newcommand{\sinst}[1]{\ensuremath{\sin^2\theta_{#1}}\xspace} 
\newcommand{\sinstt}[1]{\ensuremath{\sin^22\theta_{#1}}\xspace}  

\newcommand{\deltacp}{\ensuremath{\delta_{\rm CP}}\xspace}   
\newcommand{\mdeltacp}{\ensuremath{\delta_{\rm CP}}}   

\newcommand{\ptoknubar}{\ensuremath{p\rightarrow K^+ \overline{\nu}}\xspace}
\newcommand{\ptoepizero}{\ensuremath{p \rightarrow e^+ \pi^0}\xspace}

\def\argon40{${}^{40}$Ar}       
\def\Ar39{$^{39}$Ar}
\def\Cl40{$^{40}$Cl}
\def\K40{$^{40}$K}
\def\B8{$^{8}$B}

\def\fdfiducialmass{\SI{40}{\kt}\xspace}
\def\lartemp{\SI{88}\,K\xspace}
\def\larmass{\SI{17.5}{\kt}\xspace} 

\def\cryostatht{\SI{17.8}{\meter}\xspace} 
\def\cryostatlen{\SI{65.8}{\meter}\xspace} 
\def\cryostatwdth{\SI{18.9}{\meter}\xspace} 

\def\cryostathtinner{\SI{14.0}{\meter}\xspace} 
\def\cryostatleninner{\SI{62.0}{\meter}\xspace} 
\def\cryostatwdthinner{\SI{15.1}{\meter}\xspace} 

\def\nominalmodsize{\SI{10}{kt}\xspace} 
 
\def\spmaxfield{\SI{500}{\volt/\centi\meter}\xspace} 
\def\spactivelarmass{\SI{10}{\kt}\xspace} 
\def\spmaxdrift{\SI{3.5}{\m}\xspace}
\def\tpcheight{\SI{12.0}{\meter}\xspace} 
\def\sptpclen{\SI{58.2}{\meter}\xspace} 

\def\sptargetdriftvolt{$-\SI{180}{\kilo\volt}$\xspace} 
\def\sptargetdriftvoltpos{\SI{180}{\kilo\volt}\xspace} 
\def\coldbox{cold box\xspace} 

\def\dpactivelarmass{\SI{12.1}{\kt}\xspace} 
\def\dptpclen{\SI{60.0}{\meter}\xspace} 
\def\dptpcwdth{\SI{12.0}{\meter}\xspace} 
\def\dpswchpercrp{\num{36}\xspace} 
\def\dpnumswch{\num{2880}\xspace} 
\def\dptotcrp{\num{80}\xspace} 
\def\dpnumcrpch{\num{153600}\xspace} 
\def\dpnumpmtch{\num{720}\xspace} 
\def\dpstrippitch{\SI{3.1}{\milli\meter}\xspace} 
\def\dpnominaldriftfield{\SI{500}{\volt/\cm}\xspace} 
\def\dptargetdriftvoltpos{\SI{600}{\kV}\xspace} 

\def\surffnalbw{\SI{100}{\Gbps}\xspace}

\newcommand{\efield}{E field\xspace}

\newcommand{\threed}{3D\xspace}
\newcommand{\twod}{2D\xspace}
\newcommand{\fdth}{feedthrough\xspace} 
\newcommand{\phel}{photoelectron\xspace} 
\newcommand{\frfour}{FR-4\xspace} 

\newcommand{\lsim}{{\;\raise0.3ex\hbox{$<$\kern-0.75em\raise-1.1ex\hbox{$\sim$}}\;}}
\newcommand{\gsim}{{\;\raise0.3ex\hbox{$>$\kern-0.75em\raise-1.1ex\hbox{$\sim$}}\;}}
\newcommand{\beq}{\begin{equation}}
\newcommand{\eeq}{\end{equation}}
\newcommand{\bea}{\begin{eqnarray}}
\newcommand{\eea}{\end{eqnarray}}

\mathchardef\minus="002D

\newcommand{\sdlwavailable}{April 2022\xspace}
\newcommand{\cucbenocc}{October 2022\xspace}
\newcommand{\accesscuccountrm}{April  2023\xspace}
\newcommand{\accesstopfirstcryo}{January 2024\xspace}
\newcommand{\startfirsttpcinstall}{August 2024\xspace}
\newcommand{\accesstopsecondcryo}{January 2025\xspace}
\newcommand{\firsttpcinstallend}{May 2025\xspace}
\newcommand{\startsecondtpcinstall}{August 2025\xspace}
\newcommand{\secondtpcinstallend}{May 2026\xspace}

\newcommand{\rrt}[1]{}

\newcommand{\superk}{Super--Kamiokande\xspace} 
\newcommand{\microboone}{MicroBooNE\xspace} 
\newcommand{\minerva}{MINERvA\xspace} 
\newcommand{\nova}{NOvA\xspace} 

\newcommand{\larnd}{LAr ND\xspace}

\newcommand{\lartpc}{LArTPC\xspace}

\newcommand{\fnal}{Fermilab\xspace} 
\newcommand{\surf}{SURF\xspace}

\newcommand{\dual}{DP\xspace}

\newcommand{\single}{SP\xspace}

\newcommand{\lar}{LAr\xspace}
\newcommand{\lntwo}{LN$_2$\xspace}  

\DeclareSIUnit \s {\second}
\DeclareSIUnit \MB {\mega\byte}
\DeclareSIUnit \GB {\giga\byte}
\DeclareSIUnit \TB {\tera\byte}
\DeclareSIUnit \PB {\peta\byte}
\DeclareSIUnit \Mbps {\mega\bit/\s}
\DeclareSIUnit \Gbps {\giga\bit/\s}
\DeclareSIUnit \Tbps {\tera\bit/\s}
\DeclareSIUnit \Pbps {\peta\bit/\s}
\DeclareSIUnit \kton {\kilo\tonne} 
\DeclareSIUnit \kt {\kilo\tonne}
\DeclareSIUnit \Mt {\mega\tonne}
\DeclareSIUnit \eV {\electronvolt}
\DeclareSIUnit \keV {\kilo\electronvolt}
\DeclareSIUnit \MeV {\mega\electronvolt}
\DeclareSIUnit \GeV {\giga\electronvolt}
\DeclareSIUnit \m {\meter}
\DeclareSIUnit \cm {\centi\meter}
\DeclareSIUnit \in {\inchcommand}
\DeclareSIUnit \km {\kilo\meter}
\DeclareSIUnit \kV {\kilo\volt}
\DeclareSIUnit \kW {\kilo\watt}
\DeclareSIUnit \MW {\mega\watt}
\DeclareSIUnit \MHz {\mega\hertz}
\DeclareSIUnit \mrad {\milli\radian}
\DeclareSIUnit \year {year}
\DeclareSIUnit \POT {POT}
\DeclareSIUnit \sig {$\sigma$}
\DeclareSIUnit\parsec{pc}
\DeclareSIUnit\lightyear{ly}
\DeclareSIUnit\foot{ft}
\DeclareSIUnit\ft{ft}
\DeclareSIUnit \ppb{ppb}
\DeclareSIUnit \ppt{ppt}
\DeclareSIUnit \samples{S}

\newcommand{\ktyr}{\si{\kt\year}\xspace}
\newcommand{\ktMWyr}{\si{\kt\MW\year}\xspace}

\newcommand{\beamreprate}{\SI[round-mode=places,round-precision=2]{0.833333333333}{\hertz}\xspace}

\usepackage[toc]{glossaries}
\makeglossaries

\newcommand{\dshort}[1]{\glsentrytext{#1}}  
\newcommand{\dshorts}[1]{\glsentryshortpl{#1}}  

\newcommand{\dfirst}[1]{\glsfirst{#1}\glsunset{#1}}
\newcommand{\dfirsts}[1]{\glsfirstplural{#1}\glsunset{#1}}

\newcommand{\dword}[1]{\gls{#1}}
\newcommand{\dwords}[1]{\glspl{#1}}
\newcommand{\Dword}[1]{\Gls{#1}}
\newcommand{\Dwords}[1]{\Glspl{#1}}

\newcommand{\newduneword}[3]{
    \newglossaryentry{#1}{
        text={#2},
        long={#2},
        name={\glsentrylong{#1}},
        first={\glsentryname{#1}},
        firstplural={\glsentrylong{#1}\glspluralsuffix},
        description={#3},
        sort={#2}
    }
}

\newcommand{\newduneabbrev}[4]{
  \newglossaryentry{#1}{
    text={#2},
    long={#3},
    shortplural={{#2}\glspluralsuffix},
    longplural={{#3}\glspluralsuffix{}},
    name={\glsentrylong{#1}{} (\glsentrytext{#1}{})},
    first={#3 (#2)},
    firstplural={#3\glspluralsuffix{} (\glsentrytext{#1}\glspluralsuffix{})},
    description={#4},
    sort={#2}
  }
}

\newcommand{\newduneabbrevs}[5]{
  \newglossaryentry{#1}{
    text={#2},
    long={#3},
    plural={#4},
    shortplural={{#2}\glspluralsuffix},
    longplural={#4},
    name={\glsentrylong{#1}{} (\glsentrytext{#1}{})},
    first={#3 (#2)},
    firstplural={#4 (\glsentrytext{#1}\glspluralsuffix{})},
    description={#5},
    sort={#2}    
  }
}

\newduneword{dword}{DUNE Word}{A term in the DUNE lexicon}

\newduneword{nasa}{NASA}{U.S. National Aereonautics and Space Administration}

\newduneabbrev{nd}{ND}{near detector}{Refers to the detector(s) 
 installed close to the neutrino source at \fnal }

\newduneabbrev{fd}{FD}{far detector}{The \SI{70}{kt} total (\fdfiducialmass fiducial) mass \gls{lartpc} DUNE detector, composed of four \larmass total (\nominalmodsize fiducial) mass modules,  
  to be installed at the far site at \surf in
  Lead, SD, USA}

\newduneabbrev{sp}{SP}{single-phase}{Distinguishes one of the DUNE far detector technologies by the fact that it operates using argon in its liquid phase only}

\newduneabbrev{dp}{DP}{dual-phase}{Distinguishes one of the DUNE far detector technologies by the fact that it operates using argon 
 in both gas and liquid phases}

\newduneabbrev{pds}{PD system}{photon detection system}{The detector 
  subsystem sensitive to light produced in the \lar }

\newduneabbrev{hvs}{HVS}{high voltage system}{The detector 
  subsystem that provides the \gls{tpc} drift field}

\newduneabbrev{tpc}{TPC}{time projection chamber}{A type of particle detector that uses an \efield together with a sensitive volume of gas or liquid, e.g., \gls{lar}, to perform a \threed reconstruction of a particle trajectory or interaction. The activity is recorded by digitizing the waveforms of current
  induced on the anode as the distribution of ionization charge passes by
  or is collected on the electrode} 

\newduneabbrev{lartpc}{LArTPC}{liquid argon time-projection chamber}{A \gls{tpc} filled with liquid argon; 
the basis for the \gls{dune} \gls{fd} modules} 

\newduneabbrevs{apa}{APA}{anode plane assembly}{anode plane assemblies}{A unit of the \single
  detector module containing the elements sensitive to ionization in the \lar. 
  It contains two faces each of three planes of wires, and interfaces to the cold
  electronics and photon detection system} 

\newduneabbrev{awg}{AWG}{American wire gauge} {U.S. standard set of non-ferrous wire conductor sizes}

\newduneabbrev{ufer}{Ufer}{concrete encased electrode} {U.S. National Electrical Code grounding method refered to as Concrete Encased Electrode}

\newduneabbrev{cro}{CRO}{charge readout}{The system for detecting
  ionization charge distributions in a \dual detector module}

\newduneabbrev{lro}{LRO}{light readout}{The system for detecting
  scintillation photons in a \dual  detector module}

\newduneabbrev{shv}{SHV}{safe high voltage}{Type of bayonet mount
connector used on coaxial cables that has additional insulation 
compared to standard BNC and MHV connectors that makes it safer
for handling \gls{hv} by preventing accidental contact with the
live wire connector in an unmated connector or plug}

\newduneabbrev{fe}{FE}{front-end}{The front-end refers a point that is
  ``upstream'' of the data flow for a particular subsystem. 
  For example the \gls{sp} front-end electronics is where the cold electronics
  meet the sense wires of the TPC and the front-end \gls{daq} is where the
  \gls{daq} meets the output of the electronics}

\newduneabbrev{daqrou}{DAQ RU}{DAQ readout unit}{The first element in the data flow of the \gls{daq}}

\newduneabbrev{cots}{COTS}{commercial off-the-shelf}{Items, typically hardware such as 
computers, that may be purchased whole, without any custom design or fabrication and 
thus at normal consumer prices and availability}

\newduneabbrev{i2c}{I2C}{Inter-Integrated Circuit}{I$^2$C or I2C is a synchronous, 
multi-master, multi-slave, packet switched, single-ended, serial computer bus widely used 
for attaching lower-speed peripheral ICs to processors and microcontrollers in short-distance, 
intra-board communication} 

\newduneabbrev{spi}{SPI}{Serial Peripheral Interface}{The Serial Peripheral Interface is a 
synchronous serial communication interface specification used for short distance 
communication, primarily in embedded systems}

\newduneabbrev{miso}{MISO}{master in slave out}{The Master In Slave Out is a logic
signal on the \gls{spi} bus on which the data from the slave are transmitted once
a request from the master is received} 

\newduneabbrev{mosi}{MOSI}{master out slave in}{The Master Out Slave In is a logic
signal on the \gls{spi} bus on which the data from the master is transmitted} 

\newduneabbrev{uart}{UART}{Universal Asynchrous Receiver/Transmitter}{A universal 
asynchronous receiver-transmitter is a computer hardware device for asynchronous 
serial communication in which the data format and transmission speeds are configurable}

\newduneword{cr}{CR}{Capacitance-Resistance} 

\newduneword{dc}{DC}{direct coupling} 

\newduneword{ac}{AC}{capacitive coupling}  

\newduneabbrev{pll}{PLL}{Phase-Locked Loop}{A control system that generates an
output signal whose phase is related to the phase of an input signal}  

\newduneword{fifo}{FIFO}{First-In-First-Out} 

\newduneword{tsmc}{TSMC}{Taiwan Semiconductor Manufacturing Company}

\newduneword{saci}{SACI}{\gls{slac} \gls{asic} Control Interface}

\newduneword{om3}{OM3}{Type of multi-mode fiber optic cable, typically capable of \SI{10}{Gbps} data transmission at lengths up to \SI{300}{m}}

\newduneword{om4}{OM4}{Type of multi-mode fiber optic cable, typically capable of \SI{10}{Gbps} data transmission at lengths up to \SI{550}{m}}

\newduneword{qfp}{QFP}{Quad Flat Package} 

\newduneabbrev{ams}{AMS}{analog and mixed signal}{Verilog-AMS is a derivative of the Verilog hardware description language that includes analog and mixed-signal extensions (AMS) in order to define the behavior of analog and mixed-signal systems}

\newduneabbrev{hepa}{HEPA}{High Efficiency Particulate Air}{The High Efficiency Particulate Air filters are a type of air filter that remove \num{99.97}\% of particles that have a size greater han or equal to \SI{0.3}{$\mu$m}}  

\newduneabbrev{uvm}{UVM}{universal verification methodology}{The Universal Verification Methodology is a standardized methodology for verifying integrated circuit designs}   

\newduneword{lhc}{LHC}{Large Hadron Collider}

\newduneabbrev{lsb}{LSB}{least significant bit}{The bit with the lowest numerical value in a binary number}

\newduneabbrev{ldo}{LDO}{low-dropout regulator}{A low-dropout or LDO regulator is a \gls{dc} linear voltage regulator that can regulate the output voltage even when the supply voltage is very close to the output voltage}

\newduneabbrev{adc}{ADC}{analog-to-digital converter}{A sampling of a voltage
  resulting in a discrete integer count corresponding in some way to
  the input}

\newduneabbrev{inl}{INL}{integral non-linearity}{A commonly used measure of performance in \glspl{adc}. It is the deviation between the ideal input threshold value and the measured threshold level of a certain output code}

\newduneabbrev{dnl}{DNL}{differential non-linearity}{A commonly used measure of performance in \glspl{adc}. The DNL error is defined as the difference between an actual step width and the ideal value of one \gls{lsb}}

\newduneword{pnp}{PNP}{Type of bipolar junction transistor consistning of a
layer of N-doped semiconductor sandwiched between two layers of P-doped material}

\newduneword{spice}{SPICE}{SPICE
(``Simulation Program with Integrated Circuit Emphasis'') is a general-purpose, 
open-source analog electronic circuit simulator. It is a program used in integrated 
circuit and board-level design to check the integrity of circuit designs and to 
predict circuit behavior}

\newduneabbrev{daq}{DAQ}{data acquisition}{The data acquisition system
  accepts data from the detector \gls{fe} electronics, buffers
  the data, performs a \gls{trigdecision}, builds events from the selected
  data and delivers the result to the offline \gls{diskbuffer}}

\newduneword{detmodule}{detector module}{The entire DUNE far detector is
  segmented into four modules, each with a nominal \SI{10}{\kton}
  fiducial mass}

\newduneword{detunit}{detector unit}{A 
portion of a \gls{detmodule} may be further partitioned into a number of similar parts.   For example the \gls{sp} \gls{tpc} 
is made up of \gls{apa}  units (and other elements)}

\newduneword{diskbuffer}{secondary DAQ buffer}{A secondary
  \dshort{daq} buffer holds a small subset of the full rate as
  selected by a \gls{trigcommand}. 
  This buffer also marks the interface with the DUNE Offline}

\newduneabbrev{om}{OM}{online monitoring}{Processes that run inside
  the \gls{daq} on data ``in flight,'' specifically before landing on the
  offline disk buffer, and that provide feedback on the operation of
  the \gls{daq} itself and the general health of the data it is marshalling}

\newduneabbrev{dqm}{DQM}{data quality monitoring}{Analysis of the raw
  data to monitor the integrity of the data and the performance of the
  detectors and their electronics. This type of monitoring may be
  performed in real time, within the \gls{daq} system, or in later
  stages of processing, using disk files as input}

\newduneword{dumpbuffer}{DAQ dump buffer}{This \gls{daq} buffer
  accepts a high-rate data stream, in aggregate, from an associated
  portion of a \gls{detmodule} sufficient to collect all data likely relevant to
  a potential \gls{snb}}

\newduneabbrev{etl}{ETL}{external trigger logic}{Trigger processing
  that consumes \gls{detmodule} level \gls{trignote} information
  and other global sources of trigger input and emits
  \gls{trigcommand} information back to the \glspl{mtl}}
\newduneabbrev{daqeti}{ETI}{external trigger interface}{Interface between \glspl{mtl} and external source and sinks of relevant trigger information}

\newduneword{trignote}{trigger notification}{Information provided by
  \gls{mtl} to \gls{etl} about \gls{trigdecision} 
  processing}

\newduneword{trigprimitive}{trigger primitive}{Information derived by
  the \gls{daq} \gls{fe} hardware that describes a region of space (e.g.,
  one or several neighboring channels) and time (e.g., a contiguous set
  of \gls{adc} sample ticks) associated with some activity}

\newduneword{externtrigger}{external trigger candidate}{Information
  provided to the \gls{mtl} about events external to a
  \gls{detmodule} so that it may be considered in forming
  \glspl{trigcommand}}

\newduneabbrev{daqoob}{OOB dispatcher}{out-of-band trigger command
  dispatcher}{This component is responsible for dispatching a \gls{snb} dump
  command to all \glspl{daqfer} in the \gls{detmodule}}

\newduneabbrev{mtl}{MTL}{module trigger logic}{Trigger processing
  that consumes \gls{detunit} level \gls{trigcommand} information
  and emits \glspl{trigcommand}. 
  It provides the \gls{etl} with \glspl{trignote} and receives back any
  \glspl{externtrigger}}

\newduneword{octant}{octant}{Any of the eight parts into which 4$\pi$
  is divided by three mutually perpendicular axes. 
  In particular in referencing the value for the mixing angle
  $\theta_{23}$}

\newduneword{trigcandidate}{trigger candidate}{Summary information derived
  from the full data stream and representing a contribution toward
  forming a \gls{trigdecision}}

\newduneword{trigcommand}{trigger command}{Information derived from
  one or more \glspl{trigcandidate}  that directs elements of the
  \gls{detmodule} to read out a portion of the data stream}

\newduneabbrev{tcm}{TCM}{trigger command message}{A message flowing
  down the trigger hierarchy from global to local context.  Also see \gls{tpm}}

\newduneabbrev{mlt}{MLT}{module level trigger}{The \gls{daq} component responsible for producing a \gls{trigdecision} that will be used to command the readout of a detector module}

\newduneword{trigdecision}{trigger decision}{The process by which
  \glspl{trigcandidate} are converted into \glspl{trigcommand}}

\newduneabbrev{tpm}{TPM}{trigger primitive message}{A message flowing
  up the trigger hierarchy from local to global context.  Also see \gls{tcm}}

\newduneabbrev{ipc}{IPC}{inter-process communication}{A system for software elements to exchange information between threads, local processes or across a data network.  An IPC system is typically specified in terms of protocols  composed of message types and their associated data schema}

\newduneword{daqdispre}{discovery and presence}{As used in the context of the \gls{ipc}, a system that provides mechanisms for a node on a communication network to learn of the existence of peers and their identity (discovery) as well as determine if they are currently operational or have become unresponsive (presence)}

\newduneabbrev{pubsub}{PUB/SUB}{publish-subscribe communication pattern}{An \gls{ipc} communication pattern where one element, the publisher, sends data to all connected elements, the subscribers.  Each subscriber may connect to multiple publishers.  A variant is PUB/SUB with topics where a subscriber may register an identifier, the topic, to limit the information received to just an associated subset}

\newduneabbrev{eb}{EB}{event builder}{A software agent that executes \glspl{trigcommand}  for one  \gls{detmodule} by reading out the requested data}

\newduneabbrev{daqdfo}{DFO}{data flow orchestrator}{The process by which trigger commands are executed in parallel and asynchronous manner by the back-end output subsystem of the \gls{daq}}

\newduneabbrev{daqubi}{UBI}{upstream DAQ buffer interface}{The process which provides read-only access to data residing in the upstream \gls{daq} buffers to processes on the network}

\newduneabbrev{cob}{COB}{cluster on board}{An ATCA motherboard housing four RCEs}

\newduneabbrev{rce}{RCE}{reconfigurable computing element}{Data processor located outside of the cryostat on a \gls{cob} that contains \gls{fpga}, RAM and \gls{ssd} resources, responsible for buffering data, producing trigger primitives, responding to triggered requests for data and synching \gls{snb} dumps}

\newduneabbrev{bow}{BOW}{Bump On Wire}{A working name for the front-end readout computing elements used in the nominal \gls{daq} design to interface the \dual  crates to the \gls{daq} front-end computers}

\newduneabbrev{atca}{ATCA}{Advanced Telecommunications Computing
  Architecture}{An advanced computer architecture specification developed for the telecommunications, military, and aerospace industries that incorporates the latest trends in  high-speed interconnect technologies, next-generation processors, and improved reliability, availability and serviceability} 

\newduneabbrev{utca}{$\mu$TCA}{Micro Telecommunications Computing Architecture}{The computer architecture specification followed by the crates that house charge and light readout electronics in the \gls{dpmod}} 

\newduneabbrev{udp}{UDP}{user datagram protocol}{A simple,
  connectionless Internet protocol that supports data integrity
  checksums, requires no handshaking, and does not guarantee packet delivery}

\newduneabbrev{amc}{AMC}{advanced mezzanine card}{Holds digitizing
  electronics and lives in \gls{utca} crates}

\newduneabbrev{rf}{RF}{radio frequency}{Electromagnetic emissions
  that are within the (radio) frequency band of sensitivity of the detector
  electronics}

\newduneabbrev{fpga}{FPGA}{field programmable gate array}{An
integrated circuit technology that allows the hardware to be reconfigured to
execute different algorithms after its manufacture and deployment}

\newduneabbrev{fmc}{FMC}{FPGA mezzanine card}{Boards holding \glspl{fpga} and other integrated circuitry that attach to a motherboard}

\newduneabbrev{felix}{FELIX}{Front-End Link eXchange}{A
  high-throughput interface between \gls{fe} and trigger electronics
  and the standard PCIe computer bus}

\newduneword{daqpart}{DAQ partition}{A cohesive and
 coherent collection of \gls{daq} hardware and software working together to trigger and read out some portion of one detector module; it consists of an integral number of \glspl{daqfrag}. 
 Multiple \gls{daq} partitions may operate simultaneously, but each instance operates independently}

\newduneabbrev{fec}{DAQ FEC}{DAQ front-end computer}{The portion of one
  \gls{daqpart} that hosts the \gls{daqdr}, \gls{daqbuf} and
  \gls{daqds}.  It hosts the \gls{daqfer} and corresponding portion of the \gls{daqbuf}}

\newduneword{daqfrag}{DAQ front-end fragment}{The portion of one
  \gls{daqpart} relating to a single \gls{fec} and corresponding to an
  integral number of \glspl{detunit}.  See also \gls{datafrag}}

\newduneword{datafrag}{data fragment}{A block of data read out from a single \gls{daqfrag} that
span a contiguous period of time as requested by a \gls{trigcommand}}

\newduneabbrev{daqfer}{FER}{DAQ front-end readout}{The portion of a
  \gls{daqfrag} that accepts data from the detector electronics and
  provides it to the \gls{fec}}

\newduneabbrev{daqdr}{DDR}{DAQ data receiver}{The portion of the
  \gls{daqfrag} that accepts data from the \gls{daqfer}, emits
  trigger candidates produced from the input trigger primitives, and
  forwards the full data stream to the \gls{daqbuf}}

\newduneword{daqbuf}{DAQ primary buffer}{The portion
  of the \gls{daqfrag} that accepts full data stream from the
  corresponding \gls{detunit} and retains it sufficiently long for it
  to be available to produce a \gls{datafrag}}

\newduneword{daqds}{data selector}{The portion of the \gls{daqfrag}
  that accepts \glspl{trigcommand} and returns the corresponding
  \gls{datafrag}.  Not to be confused with \gls{daqdsn}}

\newduneword{daqdsn}{data selection}{The process of forming a trigger decision for selecting a subset of detector data for output by the \gls{daq} from the content of the detector data itself.  Not to be confused with \gls{daqds}}

\newduneabbrev{daqros}{DAQ RO}{DAQ readout subsystem}{The subsystem of the \gls{daq} for accepting and buffering data input from detector electronics}

\newduneabbrev{daqdss}{DAQ DS}{DAQ data selection subsystem}{The subsystem of the \gls{daq} responsible for forming a trigger decision based on a portion of the input data stream.  The majority subset of the \gls{daqtrs}}

\newduneabbrev{daqtrs}{DAQ TS}{DAQ trigger subsystem}{The subsystem of the \gls{daq} responsible for forming a trigger decision}

\newduneabbrev{daqbes}{DAQ BE}{DAQ back-end subsystem}{The portion of the \gls{daq} that is generally toward its output end.  It is responsible for accepting and executing trigger commands and marshaling the data they address to output storage buffers}

\newduneabbrev{daqtss}{DAQ TSS}{DAQ timing and synchronization subsystem}{The portion of the \gls{daq} that provides for timing and synchronization to various components}

\newduneabbrev{femb}{FEMB}{front-end mother board}{Refers a unit of
  the \gls{sp} \gls{ce} that contains the \gls{fe} amplifier
  and \gls{adc} \glspl{asic} covering 128 channels}

\newduneword{asic}{ASIC}{application-specific integrated circuit}

\newduneword{lv}{LV}{low voltage}

\newduneabbrev{iceberg}{ICEBERG}{ICEBERG R\&D cryostat and electronics}{Integrated Cryostat and Electronics Built for Experimental Research Goals:
a new double-walled cryostat built and installed at \gls{fnal} 
for liquid argon detector R\&D and for testing of DUNE detector components}

\newduneword{coldadc}{ColdADC}{A newly developed 16-channels \gls{asic} providing analog to digital conversion}

\newduneword{coldata}{COLDATA}{A 64-channel control and communications \gls{asic}}

\newduneword{cryo}{CRYO}{Integrated ASIC including \gls{fe} circuitry providing signal amplification and pulse shaping, analog to digital conversion, and control and communication functionalities for 64 channels}

\newduneword{larasic}{LArASIC}{A 16-channel \gls{fe} \gls{asic} that provides signal amplification and pulse shaping}

\newduneword{cmos}{CMOS}{Complementary metal-oxide-semiconductor}

\newduneabbrev{enc}{ENC}{equivalent noise charge}{The equivalent noise charge is the input charge that corresponds to a 
$\gls{snr}=1$}

\newduneword{sar}{SAR}{successive approximation register}

\newduneword{protodune}{ProtoDUNE}{Either of the two DUNE prototype detectors constructed at \gls{cern}. 
  One prototype implements \gls{sp} technology and the other \gls{dp}}
  
\newduneword{protodune2}{ProtoDUNE-2}{The second run of a \gls{protodune} detector}

\newduneword{pdsp}{ProtoDUNE-SP}{The \gls{sp} \gls{protodune} detector at \gls{cern}}

\newduneword{pddp}{ProtoDUNE-DP}{The \gls{dp} \gls{protodune} detector at \gls{cern}}

\newduneword{wa105}{WA105 DP demonstrator}{The \SI[product-units=power]{3x1x1}{m} WA105 \gls{dp} prototype detector at \gls{cern}}

\newduneword{rawevent}{DAQ event block}{The unit of data output by the
  \gls{daq}.  
  It contains trigger and detector data spanning a unique, contiguous
  time period and a subset of the detector channels}

\newduneabbrev{ssd}{SSD}{solid-state disk}{Any storage device that
  may provide sufficient write throughput to receive, both collectively and
  distributed, the sustained full rate of data from a \gls{detmodule}
  for many seconds}
\newduneabbrev{nvme}{NVMe}{Non-volatile memory express}{A specification for an interface to storage media attached via PCIe}

\newduneabbrev{hlt}{HLT}{high-level trigger}{This is actually a filter applied to data that has been triggered and aggregated in order to further reduce or characterize it}

\newduneabbrev{pid}{PID}{particle ID}{Particle identification}

\newduneword{readout window}{readout window}{A fixed, atomic and
  continuous period of time over which data from a \gls{detmodule}, in
  whole or in part, is recorded. 
  This period may differ based on the trigger that initiated the
  readout}

\newduneabbrev{zs}{ZS}{zero-suppression}{Used to delete some portion of a
  data stream that does not significantly deviate from zero or
  intrinsic noise levels. 
  It may be applied at different granularity from per-channel to per
  \gls{detunit}}

\newduneabbrev{rc}{RC}{run control}{The system for configuring,
  starting and terminating the \gls{daq}}

\newduneword{r-c}{RC}{resistive-capacitive (circuit)}

\newduneabbrev{daqccm}{CCM}{DAQ control, configuration and monitoring subsystem}{A system for controlling, configuring and monitoring other systems in particular those that make up the \gls{daq} where the CCM encompasses \gls{rc}}

\newduneword{daqrun}{DAQ run}{A period of time over which relevant data taking conditions and \gls{daq} configuration are asserted to be unchanged. 
  Multiple \gls{daq} runs may occur simultaneously when multiple \glspl{daqpart} are active. 
  This term should not be confused with DUNE experiment or beam ``runs'' that typically span many \gls{daq} runs}
\newduneword{daqrunnum}{DAQ run number}{A monotonically increasing count that uniquely and globally identifies a \gls{daqrun}}

\newduneabbrev{snb}{SNB}{supernova neutrino burst}{A prompt 
  increase in the flux of low-energy neutrinos emitted in the first few seconds of a core-collapse supernova.  It can also refer to a trigger command type that may be due to this phenomenon,
  or detector conditions that mimic its interaction signature}

\newduneabbrev{snble}{SNB/LE}{supernova neutrino burst and low
  energy}{Supernova neutrino burst and low-energy physics program}

\newduneabbrev{snews}{SNEWS}{SuperNova Early Warning System}{A global
  supernova neutrino burst trigger formed by a coincidence of \gls{snb} 
  triggers collected from participating experiments}

\newduneabbrev{pps}{1PPS signal}{one-pulse-per-second signal}{An
  electrical signal with a fast rise time and that arrives in real
  time with a precise period of one second}

\newduneabbrev{sls}{SLS}{spill location system}{A system residing at
  the DUNE far detector site that provides information, possibly
  predictive, indicating periods of time when neutrinos are being
  produced by the \fnal Main Injector beam spills}

\newduneabbrev{wib}{WIB}{warm interface board}{Digital electronics
  situated just outside the \gls{sp} cryostat that receives digital data
  from the \glspl{femb} over cold copper connections and sends it to the \gls{rce}
  \gls{fe} readout hardware}

\newduneabbrev{gps}{GPS}{Global Positioning System}{A satellite-based system that provides a highly accurate \gls{pps} that may be used to synchronize clocks and determine location}

\newduneabbrev{ntp}{NTP}{Network Time Protocol}{A networking protocol that allows synchronizing of clocks to within a few \si{\milli\second} of a time standard on a local network and within a few tens of \si{\milli\second} over the Internet} 

\newduneabbrev{ptproto}{PTP}{Precision Time Protocol}{A networking protocol that allows synchronizing of clocks to within a few \si{\micro\second} of a time standard on a local network} 

\newduneabbrev{irig}{IRIG}{inter-range instrumentation group}{A standards body that defined a time-code standard for transferring timing information}

\newduneabbrev{nic}{NIC}{network interface controller}{Hardware for controlling the interface to a communication network.  Typically, one that obeys the Ethernet protocol}

\newduneabbrev{wiec}{WIEC}{warm interface electronics crate}{Crates mounted on the signal flanges that contain the \glspl{wib}}

\newduneabbrev{ptc}{PTC}{power and timing card}{Cards that provide further processing and distribution of the signals entering and exiting the \gls{sp} cryostat}

\newduneabbrev{ptb}{PTB}{power and timing backplane}{Backplane used to connect the \gls{wib}s and the \gls{ptc}s on the \gls{wiec}. Also connects the \gls{ce} flange on the cryostat penetration}

\newduneabbrev{sipm}{SiPM}{silicon photomultiplier}{A solid-state
  avalanche photodiode sensitive to single \phel signals}

\newduneabbrev{cisc}{CISC}{cryogenic instrumentation and slow controls}{Includes equipment to monitor all detector  components and  \gls{lar} quality and behavior, and provides a control system for many of the detector components}

\newduneword{fte}{FTE}{full-time equivalent. A unit of labor
  for the project. One year of work from one person}

\newduneword{art}{art}{A software framework implementing an
  event-based execution paradigm} 

\newduneabbrev{sam}{SAM}{sequential
  access via metadata}{A data-handling system to store and retrieve
  files and associated metadata, including a complete record of the
  processing that has used the files}

\newduneword{artdaq}{artdaq}{A data acquisition toolkit for data transfer, aggregation and processing}

\newduneword{beamline}{beamline}{A sequence of control and monitoring devices used for the formation of a directed collection of particles}
\newduneabbrev{cdr}{CDR}{conceptual design report}{A formal project
  document 
   that describes the experiment
  at a conceptual level}

\newduneabbrev{cf}{CF}{conventional facilities}{Pertaining to
  construction and operation of buildings and conventional infrastructure, and for \gls{lbnf-dune}, CF includes the excavation caverns}

\newduneabbrev{cp}{CP}{charge parity}{Product of charge and parity
  transformations}

\newduneabbrev{cpt}{CPT}{charge, parity, and time reversal symmetry}{product of charge, parity
  and time-reversal transformations}

\newduneabbrev{cpv}{CPV}{charge-parity symmetry violation}{Lack of
  symmetry in a system before and after charge and parity
  transformations are applied. 
  For CP symmetry to hold,  a particle turns into its
 corresponding antiparticle under a charge transformation, and a parity
transformation inverts its space coordinates, i.e., 
produces the mirror image}

\newduneword{doe}{DOE}{U.S. Department of Energy}

\newduneabbrev{fra}{FRA}{Fermi Research Alliance}{A joint partnership of the University of Chicago and the Universities Research Association (URA) that manages and operates Fermilab on behalf of the \gls{doe}}

\newduneabbrev{dune}{DUNE}{Deep Underground Neutrino Experiment}{A leading-edge, international experiment for neutrino science and proton decay studies}

\newduneabbrev{esh}{ES\&H}{environment, safety and health}{A discipline and specialty that studies and implements practical aspects of environmental protection and safety at work} 

\newduneabbrev{ppe}{PPE}{personnel protective equipment}{Equipment worn to minimize exposure to hazards that cause serious workplace injuries and illnesses}

\newduneabbrev{odh}{ODH}{oxygen deficiency hazard}{a hazard that occurs when inert gases such as nitrogen, helium, or argon displace room air and thus reduce the percentage of oxygen below the level required for human life}

\newduneabbrev{feshm}{FESHM}{Fermilab Environment, Safety and Health Manual}{The document that contains Fermilab's policies and procedures designed to manage environment, safety, and health in all its programs}

\newduneabbrev{fscf}{FSCF}{far site conventional facilities}{The
  \gls{cf} at the DUNE far detector site, \surf}
  
\newduneabbrev{nscf}{NSCF}{near site conventional facilities}{The
  \gls{cf} at the DUNE near detector site, \fnal}

\newduneabbrevs{gut}{GUT}{grand unified theory}{grand unified theories}{A class of theories that unifies the electro-weak and strong forces}

\newduneabbrev{lar}{LAr}{liquid argon}{Argon in its liquid phase; it is a cryogenic liquid with a boiling point of $\SI{-90}{^\circ{C}}$ (\SI{87}{K}) and density of \SI{1.4}{g/ml}}

\newduneabbrev{lbl}{LBL}{long-baseline}{Refers to the distance between the 
  neutrino source  and the \gls{fd}.  It can also refer to the distance between the near and far detectors. 
  The ``long'' designation is an approximate and relative distinction. For DUNE, this distance  (between \gls{fnal} and \gls{surf}) is approximately \SI{1300}{km}}

\newduneabbrev{lbnf}{LBNF}{Long-Baseline Neutrino Facility}{The
  organizational entity responsible for developing the neutrino beam, the cryostats
  and cryogenics systems, and the conventional facilities for DUNE}
  
\newduneabbrev{lbnf-dune}{LBNF/DUNE}{LBNF and DUNE project}{The overall global project, including \gls{lbnf} and \gls{dune}}

\newduneabbrev{lbnc}{LBNC}{Long-Baseline Neutrino Committee}{The committee, composed of internationally prominent scientists with relevant expertise, charged by the \gls{fnal} director to review the scientific, technical, and managerial progress, plans and decisions associated with \gls{dune}}

\newduneabbrev{ncg}{NCG}{Neutrino Cost Group}{A group of internationally prominent scientists with relevant experience that is charged by the \gls{fnal} director to review the cost, schedule, and associated risks for the \gls{dune} experiment}

\newduneabbrev{mh}{MH}{mass hierarchy}{Describes the separation
  between the mass squared differences related to the solar and
  atmospheric neutrino problems}

\newduneabbrev{mi}{MI}{Fermilab Main Injector}{An accelerator at
  \fnal that provides a beam of high-energy protons that upon
  striking a target produce secondaries that decay to provide the
  neutrinos directed toward the DUNE far detector}

\newduneabbrev{pot}{POT}{protons on target}{Typically used as a unit
  of normalization for the number of protons striking the neutrino
  production target}

\newduneabbrev{qa}{QA}{quality assurance}{The set of actions taken to provide confidence that quality requirements are fulfilled, and to detect and correct poor results}

\newduneabbrev{qc}{QC}{quality control}{An aggregate of activities (such as design analysis and inspection for defects) performed to ensure adequate quality in manufactured products}

\newduneabbrev{sm}{SM}{standard model}{Refers to a theory describing
  the interaction of elementary particles}

\newduneabbrev{tdr}{TDR}{technical design report}{A formal project
  document 
  that describes the experiment at a technical level}

\newduneabbrev{prelimdr}{PDR}{preliminary design report}{A formal project
  document 
  that describes the experiment at a preliminary design level}

\newduneabbrev{tp}{IDR}{interim design report}{An intermediate
milestone on the path to a full \gls{tdr}} 

\newduneabbrev{ckm}{CKM matrix}{Cabibbo-Kobayashi-Maskawa
  matrix}{Refers to the matrix describing the mixing between mass and
  weak eigenstates of quarks}

\newduneabbrev{cl}{CL}{confidence level}{Refers to a probability
  used to determine the value of a random variable given its
  distribution}

\newduneabbrev{pmns}{PMNS}{Pontecorvo-Maki-Nakagawa-Sakata}{A type of matrix that describes the mixing between mass and weak eigenstates of
  the neutrino}

\newduneabbrevs{cpa}{CPA}{cathode plane assembly}{cathode plane assemblies}{The component of the \single detector module that provides the drift HV cathode}

\newduneabbrev{fc}{FC}{field cage}{The component of a \gls{lartpc} that contains and shapes the applied \efield}

\newduneword{cpafc}{CPA/FC}{A pair of \gls{cpa} panels and the top and bottom \gls{fc} portions that attach to the pair; an intermediate assembly for installation into the \gls{spmod} }

\newduneabbrev{topfc}{top FC}{top field cage}{The horizontal portions of the \gls{sp} \gls{fc}   on the top of the \gls{tpc}}

\newduneabbrev{botfc}{bottom FC}{bottom field cage}{The horizontal portions of the \gls{sp} \gls{fc} on the bottom of the \gls{tpc}}

\newduneabbrev{ewfc}{endwall FC}{endwall field cage}{The vertical portions of the \gls{sp} \gls{fc} near the wall}

\newduneabbrev{gp}{GP}{ground plane}{An electrode held electrically neutral relative to Earth ground voltage; it is mounted on the \gls{fc} in a \gls{spmod} to protect the cryostat wall}

  \newduneword{gg}{ground grid}{An electrode held electrically neutral relative to Earth ground voltage; it is installed between the cathode and the \glspl{pd} in a \gls{dpmod} to protect the \glspl{pmt}, maintaining high transparency to light}

\newduneabbrev{alara}{ALARA}{as low as reasonably
  achievable}{Typically used with regard management of radiation
  exposure but may be used more generally. It means making every
  reasonable effort to maintain e.g., exposures, to as far below the
  limits as practical, consistent with the purpose for that the
  activity is undertaken}

\newduneabbrev{ecal}{ECAL}{electromagnetic calorimeter}{A detector
  component that measures energy deposition of traversing particles (in the near detector conceptual design)}

\newduneabbrev{hv}{HV}{high voltage}{Generally describes a voltage
  applied to drive the motion of free electrons through some media, e.g., LAr}

\newduneword{spmod}{SP module}{single-phase DUNE \gls{fd} module}
\newduneword{dpmod}{DP module}{dual-phase DUNE \gls{fd} module}

\newduneabbrev{tcoord}{TC}{technical coordinator}{A member of the \gls{dune} management team responsible for organizing the technical aspects of the project effort; is head of \gls{tc}}

\newduneabbrev{rcoord}{RC}{resource coordinator}{A member of the \gls{dune} management team responsible for coordinating the financial resources of the project effort}

\newduneword{tc}{technical coordination}{The DUNE organization responsible for overall integration 
of the detector elements and successful execution of the detector
construction project; areas of responsibility include 
general project oversight, systems engineering, \gls{qa} 
and safety}

\newduneabbrev{exb}{EB}{executive board}{The highest level DUNE
  decision-making body for the collaboration}

\newduneabbrev{tb}{TB}{technical board}{The DUNE organization responsible for
  evaluating technical decisions}

\newduneabbrev{rrb}{RRB}{Resources Review Board}{A part of \gls{dune}'s international project governance structure, composed of representatives of all funding agencies that sponsor the project, and of  \gls{fnal} management, established to provide coordination among funding partners and oversight of \gls{dune}}

\newduneabbrev{inc}{INC}{International Neutrino Council}{A highest-level international advisory body to the U.S. \gls{doe} and the  \gls{fnal} directorate on matters related to the  \gls{lbnf} and the  \gls{pip2} projects. This council is composed of representatives from the international funding agencies and  \gls{cern} that make major contributions the infrastructure}

\newduneabbrev{cc}{CC}{charged current}{Refers to an interaction
  between elementary particles where a charged weak force carrier
  ($W^+$ or $W^-$) is exchanged}

\newduneabbrev{dis}{DIS}{deep inelastic scattering}{Refers to the 
  interaction of an elementary charged particle with a nucleus in an
  energy range where the interaction can be modeled as taking place with
  individual nucleons}

\newduneabbrev{fsi}{FSI}{final-state interactions}{Refers to
  interactions between elementary or composite particles subsequent to
  the initial, fundamental particle interaction, such as may occur as
  the products exit a nucleus}

\newduneword{geant4}{Geant4}{A
  software toolkit for the simulation of the passage of particles
  through matter using \gls{mc} methods}

\newduneabbrev{genie}{GENIE}{Generates Events for Neutrino Interaction
  Experiments}{Software providing an object-oriented neutrino
  interaction simulation resulting in kinematics of the products of
  the interaction}

\newduneabbrev{mc}{MC}{Monte Carlo}{Refers to a method of numerical
  integration that entails the statistical sampling of the integrand
  function. 
  Forms the basis for some types of detector and physics simulations}

\newduneabbrev{qe}{QE}{quasi-elastic}{Refers to interaction between
  elementary particles and a nucleus in an energy range where the
  interaction can be modeled as occurring between constituent quarks
  of one nucleon and resulting in no bulk recoil of the resulting
  nucleus}

\newduneabbrev{mou}{MoU}{memorandum of understanding}{A document
  summarizing an agreement between two or more parties}

\newduneabbrev{pip2}{PIP-II}{Proton Improvement Plan II}{A \gls{fnal} project for
  improving the protons on target delivered delivered by the \gls{lbnf} neutrino production beam. 
  This is version two of this plan and it is planned to be followed by a PIP-III}
  
\newduneabbrev{sdsta}{SDSTA}{South Dakota Science and Technology
  Authority}{The legal entity that manages \gls{surf}, in Lead, S.D}
  
\newduneabbrev{sdsd}{SDSD}{Fermilab South Dakota Services Division}{A Fermilab division responsible providing host laboratory functions at SURF in South Dakota}

\newduneabbrev{firus}{FIRUS}{Facility Information Reporting Utility System}
 {The safety system at \surf}

\newduneabbrev{bsi}{BSI}{building and site infrastructure}
 {The work package for outfitting of the \gls{lbnf} underground infrastructure}

\newduneabbrev{wbs}{WBS}{work breakdown structure}{An organizational
  project management tool by which the tasks to be performed are
  partitioned in a hierarchical manner}

\newduneabbrev{br}{BR}{branching ratio}{A fractional probability for a
  decay of a composite particle to occur into some specified set or
  sets of products}
\newduneword{bsm}{BSM}{beyond the standard model}

\newduneabbrev{dm}{DM}{dark matter}{The term given to the unknown
  matter or force that explains measurements of galaxy motion 
  that are otherwise inconsistent with the amount of mass associated
  with the observed amount of photon production}
  
  \newduneabbrev{bdm}{BDM}{boosted dark matter}{A new model that describes a relativistic dark matter particle boosted by the annihilation of heavier dark matter participles in the galactic center or the sun}

\newduneabbrev{cern}{CERN}{European Organization for Nuclear
Research}{The leading particle physics laboratory in Europe and home to the ProtoDUNEs. (In French, the Organisation Europ\'{e}enne pour la Recherche Nucl\'{e}aire, derived from Conseil Europ\'{e}en pour la Recherche Nucl\'{e}aire}

\newduneabbrev{dsnb}{DSNB}{diffuse supernova neutrino background}{The
  term describing the pervasive, constant flux of neutrinos due to all
  past supernova neutrino bursts}

\newduneabbrev{espp}{ESPP}{European Strategy for Particle Physics}{The
cornerstone of Europe's
decision-making process for the long-term future of the
field. Mandated by the \gls{cern} Council, it is formed through a broad
consultation of the grass-roots particle physics community, it
actively solicits the opinions of physicists from around the world,
and it is developed in close coordination with similar processes in
the USA and Japan in order to ensure coordination between regions and
optimal use of resources globally}

\newduneabbrev{gar}{GAr}{gaseous argon}{argon in its gas phase}
\newduneabbrev{gartpc}{GArTPC}{gaseous argon time-projection chamber}{A \gls{tpc} filled with gaseous argon; a possible technology choice for the \gls{nd}}

\newduneabbrev{globes}{GLoBES}{General Long-Baseline Experiment
  Simulator}{A software package for simulating energy spectra of
  neutrino flux, interactions, and energy spectra measured after application of some
  model of a detector response)}

\newduneabbrev{snowglobes}{SNOwGLoBES}{SuperNova
Observatories with GLoBES} {From the official description~\cite{snowglobes}: 
SNOwGLoBES is public software for computing interaction rates and distributions of observed quantities for \gls{snb} neutrinos in common detector materials} 

\newduneword{l/e}{L/E}{length-to-energy ratio}
\newduneword{lri}{LRI}{long-range interactions}

\newduneabbrev{nc}{NC}{neutral current}{Refers to an interaction
  between elementary particles where a neutrally charged weak force carrier
  ($Z^0$) is exchanged}

\newduneabbrev{nh}{NH}{normal hierarchy}{Refers to the neutrino mass
  eigenstate ordering whereby the sign of the mass squared difference
  associated with the atmospheric neutrino problem is positive}

\newduneabbrev{ih}{IH}{inverted hierarchy}{Refers to the neutrino mass
  eigenstate ordering whereby the sign of the mass squared difference
  associated with the atmospheric neutrino problem is negative}

\newduneabbrev{no}{NO}{normal ordering}{Refers to the neutrino mass
  eigenstate ordering whereby the sign of the mass squared difference
  associated with the atmospheric neutrino problem is positive}

\newduneabbrev{io}{IO}{inverted ordering}{Refers to the neutrino mass
  eigenstate ordering whereby the sign of the mass squared difference
  associated with the atmospheric neutrino problem is negative}

\newduneabbrev{msw}{MSW}{Mikheyev-Smirnov-Wolfenstein effect}{Explains
  the oscillatory behavior of neutrinos produced inside the sun as
  they traverse the solar matter}

\newduneabbrev{nsi}{NSI}{nonstandard interaction}{A general class of
  theory of elementary particles other than the Standard Model}

\newduneabbrev{pfive}{P5}{Particle Physics Project Prioritization
Panel}{The Particle Physics Project Prioritization Panel (P5) was a
subpanel of the High Energy Physics Advisory Panel (HEPAP). It completed
its Report, a ten-year strategic plan for high energy physics in the
U.S., in 2014. This report included a recommendation that ``host a world-leading neutrino
program that will have an optimized set of short- and long-baseline neutrino oscillation experiments, and its long-term focus
is a reformulated venture referred to here as the Long Baseline
Neutrino Facility (LBNF)''}

\newduneabbrev{sme}{SME}{standard-model extension}{an effective field theory that contains the \gls{sm}, general relativity, and all possible operators that break Lorentz symmetry (Wikipedia)}

\newduneabbrev{susy}{SUSY}{supersymmetry}{Theoretical symmetry between a fermion and a boson}

\newduneabbrev{wimp}{WIMP}{weakly-interacting massive particle}{A
  hypothesized particle that may be a component of dark matter}

\newduneabbrev{ce}{CE}{cold electronics}{Analog and digital readout electronics that operate at cryogenic temperatures}

\newduneabbrev{crp}{CRP}{charge-readout plane}{In the \gls{dp} technology, a  collection of
  electrodes in a planar arrangement placed at a particular voltage
  relative to some applied \efield such that drifting electrons
  may be collected and their number and time may be measured}

\newduneabbrev{dram}{DRAM}{dynamic random access memory}{A computer memory technology}

\newduneabbrev{fnal}{Fermilab}
{Fermi National Accelerator Laboratory}{U.S. national laboratory in Batavia, IL. It is the laboratory that hosts \gls{dune} and serves as its near site}

\newduneabbrev{bnl}{BNL}{Brookhaven National Laboratory}{US national laboratory in Upton, NY}

\newduneabbrev{slac}{SLAC}{SLAC National Accelerator Laboratory}{US national laboratory in Menlo Park, CA}

\newduneabbrev{lbnl}{LBNL}{Lawrence Berkeley National Laboratory}{US national laboratory in Berkeley, CA}

\newduneabbrev{anl}{ANL}{Argonne National Laboratory}{US national laboratory in Lemont, IL}

\newduneabbrev{lanl}{LANL}{Los Alamos National Laboratory}{US national laboratory in Los Alamos, NM}

\newduneabbrev{fs}{FS}{full stream}{Relates to a data stream that has not undergone selection, compression or other form of reduction}

\newduneabbrev{lem}{LEM}{large electron multiplier}{A micro-pattern detector suitable for use in ultra-pure argon vapor; LEMs consist of copper-clad PCB boards with sub-millimeter-size holes through which electrons undergo amplification}

\newduneabbrev{lng}{LNG}{liquefied natural gas}{Pertaining to natural gas in its liquid phase}

\newduneabbrev{mip}{MIP}{minimum ionizing particle}{Refers to a
  particle traversing some medium such that the particle's mean energy loss is  
  near the minimum}

\newduneabbrev{pd}{PD}{photon detector}{The detector
  elements involved in measurement of the number and arrival times of
  optical photons produced in a detector module} 

\newduneabbrev{pmt}{PMT}{photomultiplier tube}{A device that makes use
  of the photoelectric effect to produce an electrical signal from the
  arrival of optical photons}

\newduneabbrev{ppm}{ppm}{parts per million}{A concentration equal to one part in $10^{-6}$}
\newduneabbrev{ppb}{ppb}{parts per billion}{A concentration equal to one part in $10^{-9}$}
\newduneabbrev{ppt}{ppt}{parts per trillion}{A concentration equal to one part in $10^{-12}$}

\newduneword{rio}{RIO}{reconfigurable input output}

\newduneabbrev{s/n}{S/N}{signal-to-noise}{signal-to-noise ratio}
\newduneword{snr}{\mbox{S/N}}{signal-to-noise ratio}

\newduneword{ssp}{SSP}{SiPM signal processor}

\newduneabbrev{sbn}{SBN}{Short-Baseline Neutrino}{A \gls{fnal} program consisting of three collaborations, \gls{microboone}, \gls{sbnd}, and \gls{icarus}, to perform sensitive searches for $\nue$ appearance and $\numu$ disappearance in the Booster Neutrino Beam}

\newduneword{stt}{STT}{straw tube tracker}

\newduneword{wire board}{wire board}{At the head end of the APA in the \single TPC, stacks of electronics boards referred to as ``wire boards'' are arrayed to anchor the wires.  They also provide the connection between the wires and the cold electronics} 

\newduneabbrev{wls}{WLS}{wavelength-shifting}{A material or process by
  which incident photons are absorbed by a material and photons are
  emitted at a different, typically longer, wavelength}
  
\newduneabbrev{tpb}{TPB}{tetra-phenyl butadiene}{A 
\gls{wls} material}

\newduneabbrev{ptp}{PTP}{p-terphenyl}{A 
\gls{wls} material}

\newduneabbrev{sft}{SFT}{signal feedthrough}{A cryostat penetration allowing for the passage of cables or other extended parts}
\newduneabbrev{sftchimney}{SFT chimney}{signal feedthrough chimney}{In the \dual technology, a volume above the cryostat penetration used for a signal feedthrough}

\newduneabbrev{catiroc}{CATIROC}{charge and time integrated readout chip}{A complete read-out chip manufactured in AustriaMicroSystem designed to read arrays of 16 photomultipliers}

\newduneabbrev{wr}{WR}{White Rabbit}{A component of the timing system that forwards clock signal and time-of-day reference data to the master timing unit}

\newduneabbrev{mch}{MCH}{MicroTCA Carrier Hub}{An network switching device}

\newduneabbrev{wrmch}{WR-MCH}{White Rabbit \gls{utca} Carrier Hub}{A card mounted in \gls{utca} crate that recieves time syncronization information and trigger data packets over \gls{wr} network and disributes them to the \gls{amc} over \gls{utca} backplane} 

\newduneabbrev{wrtsn}{WR-TSN}{White Rabbit TimeStamping Node}{A unit on the \gls{wr} network that timestamps the trigger signals and sends out trigger data packets to \gls{wrmch}}

\newduneword{cmp}{CMP}{configuration management plan}
\newduneword{qap}{QAP}{quality assurance plan} 
\newduneword{ieshp}{IESHP}{integrated environmental, safety and health plan}
\newduneword{dmp}{DMP}{data management plan} 
\newduneword{qam}{QAM}{quality assurance manager} 

\newduneabbrev{dss}{DSS}{detector support system}{The system used to support a \gls{sp} \gls{detmodule} within its cryostat}

\newduneabbrev{ddss}{DDSS}{DUNE detector safety system}{The system used to manage key aspects of detector safety}

\newduneabbrev{lc}{LC}{logistics center}{A facility where \gls{lbnf} and \gls{dune} components will be received and transhipped to \gls{surf}}

\newduneabbrev{tco}{TCO}{temporary construction opening}{An opening in the side of a cryostat through which detector elements are brought into the cryostat; utilized during construction and installation}

\newduneabbrev{surf}{SURF}{Sanford Underground Research Facility}{The laboratory in South Dakota where the \gls{lbnf} \gls{fscf} will be constructed and the \gls{dune} \gls{fd} will be installed and operated}

\newduneabbrev{sit}{SIT}{surface installation team}{An organizational unit responsible for logistics and integration in South Dakota}

\newduneabbrev{uit}{UIT}{underground installation team}{An organizational unit responsible for installation in the underground area at the \gls{surf} site}

\newduneabbrev{cmgc}{CMGC}{construction manager/general contractor}{The organizational unit responsible for management of the construction of conventional facilities at the underground area at the \surf site}

\newduneword{cdrev}{conceptual design review}{A project management device by which a conceptual design is reviewed} 
\newduneword{pdr}{preliminary design review}{A project management device by which an early design is reviewed} 
\newduneword{fdr}{final design review}{A project management device by which a final design is reviewed}
\newduneword{prr}{production readiness review}{A project management device by which the production readiness is reviewed}
\newduneword{irr}{installation readiness review}{A project management device by which the plan for installation is reviewed}
\newduneword{orr}{operational readiness review}{A project management device by which the operational readiness is reviewed}
\newduneword{ppr}{production progress review}{A project management device by which the progress of production is reviewed} 
\newduneabbrev{edms}{EDMS}{engineering document management system}{A computerized document management system developed and supported at \gls{cern} in which some DUNE documents, drawings and engineering models are managed}
\newduneabbrev{ecr}{ECR}{engineering change request}{The first step in the change control process in which a proposed change is described}
\newduneabbrev{docdb}{DocDB}{Document DataBase}{A computerized document management system developed and supported at \gls{fnal} in which virtually all LBNF and most DUNE documents are managed}

\newduneword{wrgm}{WR grandmaster}{White Rabbit grandmaster}

\newduneabbrev{larsoft}{LArSoft}{Liquid Argon Software}{A shared base of physics software across \lartpc experiments}
\newduneword{nova}{NOvA}{The \nova off-axis neutrino oscillation experiment at \gls{fnal}}
\newduneword{minerva}{MINERvA}{The \minerva neutrino cross sections experiment at  \gls{fnal}}
\newduneword{microboone}{MicroBooNE}{The \lartpc-based \microboone neutrino oscillation experiment at  \gls{fnal}}
\newduneword{sbnd}{SBND}{The Short-Baseline Near Detector experiment at  \gls{fnal}}
\newduneabbrev{nexo}{nEXO}{Enriched Xenon Observatory}{Experiment at Lawrence Livermore National Laboratory (U.S. national lab in Livermore, CA)searching for new physics with neutrinoless double-beta decay}
\newduneword{argoneut}{ArgoNeuT}{The ArgoNeuT test-beam experiment and \gls{lartpc} prototype at  \gls{fnal}}
\newduneword{icarus}{ICARUS}{A neutrino experiment that was located at the Laboratori Nazionali del Gran Sasso (LNGS) in Italy, then refurbished at \gls{cern} for re-use in the same neutrino beam from \gls{fnal} used by the MiniBooNE, \gls{microboone} and \gls{sbnd} experiments. The ICARUS detector is being reassembled at \gls{fnal}}
\newduneword{atlas}{ATLAS}{One of two general-purpose detectors at the \gls{lhc}. It investigates a wide range of physics, from the search for the Higgs boson to extra dimensions and particles that could make up \gls{dm}}

\newduneword{lbne}{LBNE}{Long Baseline Neutrino Experiment (a terminated US project that was reformulated in 2014 under the auspices of the new \gls{dune} collaboration, an internationally coordinated and internationally funded program, with \gls{fnal} as host)}

\newduneabbrev{lbno}{LBNO}{Long Baseline Neutrino Observatory} {A terminated European project that, during its six-year duration, assessed the feasibility of a next-generation deep underground neutrino observatory in Europe)}

\newduneword{wirecell}{Wire-Cell}{A tomographic automated \threed neutrino event reconstruction method for \lartpc{}s}
\newduneabbrev{wct}{WCT}{Wire-Cell Toolkit}{A software toolkit with data flow processing components for \lartpc noise and signal simulation, noise filtering, signal processing, and tomographic \threed ionization activity imaging}
\newduneword{ftslite}{F-FTS-lite}{Light-weight version of the \fnal File Transfer system used for rapid data transfers out of the online systems}
\newduneabbrev{fts}{FTS}{File Transfer System}{A file transfer system developed at \fnal to catalog and move data to permanent storage}

\newduneword{35t}{35 ton prototype}{A prototype cryostat and \gls{sp} detector built at \fnal before the \gls{protodune} detectors}

\newduneabbrev{mcr}{MCR}{main communications room}{Space at the \gls{fd} site for cyber infrastructure}

\newduneabbrev{cuc}{CUC}{central utility cavern}{The utility cavern at the 4850L of \gls{surf} located between the two detector caverns. It contains utilities such as central cryogenics and other systems, and the underground data center and control room}

\newduneabbrev{cfd}{CFD}{computational fluid dynamics}{High performance computer-assisted modeling of fluid dynamical systems}
\newduneword{vuv}{VUV}{vacuum ultra-violet}
\newduneword{tallbo}{TallBo}{A cylindrical cryostat at \gls{fnal} primarily used for developing scintillation light collection technologies for \gls{lartpc} detectors}

\newduneword{root}{ROOT}{A modular scientific software toolkit. It provides all the functionalities needed to deal with big data processing, statistical analysis, visualisation and storage. It is mainly written in C++ but integrated with other languages such as Python and R}

\newduneabbrev{eos}{EOS}{EOS}{The XRootD-based distributed file system developed by CERN}
\newduneabbrev{ehn1}{EHN1}{Experiment Hall North One}{Location at CERN of the ProtoDUNE experiments}
\newduneword{led}{LED}{Light-emitting diode}
\newduneabbrev{rtd}{RTD}{resistance temperature detector}{A temperature sensor consisting of a material with an accurate and reproducible resistance/temperature relationship}
\newduneword{swc}{SWC}{Software \& Computing}
\newduneabbrev{las}{LAS}{LEM-anode Sandwich}{In the \dual technology, a \gls{lem} and its corresponding anode are mounted together in a module called a LEM-anode sandwich}

\newduneword{roi}{ROI}{region of interest}
\newduneabbrev{hpc}{HPC}{high-performance computing}{high-performance computing facilities; generally computing facilities emphasizing parallel computing with aggregate power of more than a teraflop}

\newduneword{comfund}{common fund}{The shared resources of the collaboration}
\newduneabbrev{ims}{IMS}{integrated master schedule}{A project management device consisting of linked tasks and milestones}

\newduneword{hvdb}{HVDB}{HV divider board}
\newduneword{sas}{SAS}{Another term for the materials airlock; a pass-through chamber used to ensure safe transfer of materials into a clean room, avoiding contamination in both directions}

\newduneabbrev{fea}{FEA}{finite element analysis}{Simulation of a physical phenomenon using the numerical technique called Finite Element Method (FEM), a numerical method for solving problems of engineering and mathematical physics}

\newduneword{fss}{FSS}{field shaping strips}
\newduneword{lvds}{LVDS}{low-voltage differential signaling}

\newduneword{esd}{ESD}{electrostatic discharge}

\newduneabbrev{rp}{RP}{resistive panel}{Resistive panels form the constant potential surfaces for a \gls{spmod} \gls{cpa}; they are composed of a thin layer of carbon-impregnated Kapton and laminated to both sides of a \frfour sheet}

\newduneword{uhmwpe}{UHMWPE}{ultra-high molecular weight polyethylene}

\newduneword{cts}{CTS}{Cryogenic Test System}
\newduneword{plc}{PLC}{programmable logic controller}

\newduneword{mppc}{MPPC}{\SI{6}{mm}$\times$\SI{6}{mm} Multi-Pixel Photon Counters produced by Hamamatsu\texttrademark{} Photonics K.K}

\newduneabbrev{sfp}{SFP}{small form-factor pluggable}{a particular standard for optical transceivers}

\newduneabbrev{minipod}{MiniPOD}{miniature parallel optical device}{a family of types of multi-channel optical transceivers}

\newduneword{ccc}{CCC}{configuration change command}
\newduneword{act}{ACT}{activation time stamp}
\newduneword{lcm}{LCM}{light calibration module}
\newduneword{lpm}{LPM}{light pulser module}
\newduneword{dac}{DAC}{digital-to-analog converter}
\newduneword{arapuca}{ARAPUCA}{A \gls{pds} design that consists of a light trap that captures wavelength-shifted photons inside boxes with highly reflective internal surfaces until they are eventually detected by \gls{sipm} detectors or are lost}
\newduneword{sarapu}{S-ARAPUCA}{Standard \gls{arapuca} design with different \gls{wls} coatings on both faces of the dichroic filter window(s) of the cell}
\newduneword{xarapu}{X-ARAPUCA}{Extended \gls{arapuca} design with \gls{wls} coating on only the external face of the dichroic filter window(s) but with a \gls{wls} doped plate inside the cell}
\newduneword{feb}{FEB}{front-end board}

\newduneabbrev{lsnd}{LSND}{Liquid Scintilator Neutrino Detector}{A scintillation detector and associated experiment located at Los Alamos National Laboratory}

\newduneabbrev{cvn}{CVN}{convolutional visual network}{An algorithm for identifying neutrino interactions based on their topology and without the need for detailed reconstruction algorithms}

\newduneword{pandora}{Pandora}{The Pandora multi-algorithm approach to pattern recognition} 

\newduneabbrev{pma}{PMA}{Projection Matching Algorithm}{A reconstruction algorithm that combines \twod reconstructed objects to form a \threed representation}
\newduneabbrev{bdt}{BDT}{boosted decision tree}{A method of multivariate analysis}
\newduneabbrev{cnn}{CNN}{convolutional neural network}{A deep learning technique most commonly applied to analyzing visual imagery}
\newduneword{pdg}{PDG}{Particle Data Group}

\newduneword{pci}{PCI}{Peripheral Component Interconnect}

\newduneword{labview}{LabVIEW}{Laboratory Virtual Instrument Engineering Workbench is a system-design platform and development environment for a visual programming language from National Instruments}

\newduneword{pcb}{PCB}{printed circuit board}

\newduneword{crio}{cRIO}{Compact Reconfigurable Input Output}

\newduneword{dcs}{DCS}{Distributed Communications System}

\newduneword{opc-ua}{OPC-UA}{OPC  Unified Architecture is a machine to machine communication protocol for industrial automation developed by the OPC Foundation. OPC stands for Object Linking and Embedding for Process Control}

\newduneword{cabangle}{Cabibbo angle}{A quark mixing parameter that governs the coupling of up quarks to strange quarks}
\newduneword{valor}{VALOR}{A neutrino oscillation fitting framework that is used by \gls{t2k}; the name stands for VALencia-Oxford-Rutherford, the original three institutions that developed it}
\newduneword{cafana}{CAFAna}{Common Analysis File Analysis}
\newduneabbrev{pca}{PCA}{principal component analysis}{A statistical procedure that uses an orthogonal transformation to convert a set of observations of possibly correlated variables into a set of values of linearly uncorrelated variables called principal components (Wikipedia)}
\newduneword{numi}{NuMI}{a set of facilities at \fnal, collectively called ``Neutrinos at the Main Injector.''  The NuMI neutrino beamline target system converts an intense proton beam into a focused neutrino beam}
\newduneword{gibuu}{GiBUU}{Giessen Boltzmann-Uehling-Uhlenback Project; a unified theory and transport framework in the MeV and GeV energy regimes for elementary reactions on nuclei }
\newduneabbrev{rpa}{RPA}{random phase approximation} {an approximation method commonly used for describing the dynamic linear electronic response of electron systems (Wikipedia)}
\newduneword{t2k}{T2K}{T2K (Tokai to Kamioka) is a long-baseline neutrino experiment in Japan studying neutrino oscillations}
\newduneword{mptdet}{MPT detector}{multipurpose tracking detector}

\newduneword{lariat}{LArIAT}{The repurposed ArgoNeuT \gls{lartpc}, modified for use in a charged particle beam, dedicated to the calibration and precise characterization of the output response of these detectors}

\newduneword{captain}{CAPTAIN}{Experimental program sited at \gls{lanl} that is designed to make measurements of scientific importance to \gls{lbl} neutrino physics and physics topics that will be explored by large underground detectors}

\newduneword{dayabay}{Daya Bay}{a neutrino-oscillation experiment in Daya Bay, China, designed to measure the mixing angle $\Theta_{13}$  using antineutrinos produced by the reactors of the Daya Bay and Ling Ao nuclear power plants}

\newduneword{nuwro}{NuWro}{neutrino interaction generator}

\newduneabbrev{neut}{NEUT}{neutrino interaction generator}{A neutrino interaction simulation program library for the studies of atmospheric accelerator neutrinos}

\newduneword{minos}{MINOS}{A long-baseline neutrino experiment, with a near detector at \gls{fnal} and a far detector in the Soudan mine in Minnesota, designed to observe the phenomena of neutrino oscillations (ended data runs in 2012)}

\newduneabbrev{efig}{EFIG}{Experimental Facilities Interface Group}{The body responsible for the required high-level coordination between the \gls{lbnf} and \gls{dune} projects}
\newduneword{ashriver}{Ash River}{The Ash River, Minnesota, USA \gls{nova} experiment far site, used as an assembly test site for \gls{dune}} 

\newduneword{ipd}{project integration director}{Responsible for integration and installation of \gls{lbnf} and \gls{dune} deliverables in South Dakota. Manages the \gls{integoff}}

\newduneabbrev{jpo}{JPO}{Joint Project Office}{The framework through which team members from the LBNF project office, \gls{integoff}, and DUNE \gls{tc} work together to provide coherence in project support functions across the global enterprise. 
Its functions include global project configuration and integration, installation planning and coordination, scheduling, safety assurance, technical review planning and oversight, development of partner agreements, and financial reporting}

\newduneword{ifbeam}{IFbeam}{Database that stores beamline information 
indexed by timestamp}

\newduneabbrev{marley}{MARLEY}{Model of Argon Reaction Low Energy
Yields}{Developed at UC Davis, MARLEY is the first realistic model of
neutrino electron interactions on argon for enegies less than \SI{50}{MeV}. This includes the energy range important for \gls{snb}
neutrinos and also solar 8--boron neutrinos}

\newduneabbrev{es}{ES}{elastic scattering}{Events in which a neutrino
elastically scatters off of another particle}

\newduneabbrev{cno}{CNO}{carbon nitrogen oxygen}{The CNO cycle (for carbon-nitrogen-oxygen) is one of the two known sets of fusion reactions by which stars convert
hydrogen to helium, the other being the proton-proton chain reaction
(pp-chain reaction). In the CNO cycle, four protons fuse, using
carbon, nitrogen, and oxygen isotopes as catalysts, to produce one
alpha particle, two positrons and two electron neutrinos}

\newduneabbrev{sdwf}{SDWF}{South Dakota Warehouse Facility}{Warehousing operations in South Dakota responsible for receiving LBNF and DUNE goods and coordinating shipments to the Ross shaft at \gls{surf}}

\newduneabbrev{wms}{WMS}{warehouse management system}{Commercial software package used to track shipments and interface to freight forwarders. This includes a database for shipping}

\newduneabbrev{dcdb}{DCDB}{DUNE construction database}{Database used by DUNE to track the history and testing of all parts of each \gls{detmodule}}

\newduneabbrev{aup}{AUP}{acceptance for use and possession}{Required for beneficial occupancy of the underground areas at SURF for LBNF and DUNE}

\newduneabbrev{bms}{BMS}{building management system}{Part of the safety system at \gls{surf} that includes the fire and life safety system}
\newduneabbrev{fls}{FLS}{fire and life safety system}{Part of the safety system at \gls{surf}}

\newduneabbrev{sno}{SNO}{Sudbury Neutrino Observatory}{The Sudbury
Neutrino Observatory was a detector built 6800 feet under ground, in
INCO's Creighton mine near Sudbury, Ontario, Canada. SNO was a
heavy-water Cherenkov detector designed to detect neutrinos produced
by fusion reactions in the sun}

\newduneword{sk}{Super-Kamiokande}{Experiment sited in the Kamioka-mine, Hida-city, Gifu, Japan that uses a large water Cherenkov detector to study neutrino properties through the observation of solar neutrinos, atmospheric neutrinos and man-made neutrinos}

\newduneabbrev{id}{ID}{inner diameter}{Inner diameter of a tube}

\newduneabbrev{od}{OD}{outer diameter}{Outer diameter of a tube}

\newduneabbrev{rms}{RMS}{root mean square}{The square root of the arithmetic mean of the squares of a set of values, used as a measure of the typical magnitude of a set of numbers, regardless of their sign}

\newduneabbrev{orc}{ORC}{operational readiness clearance}{Final safety approval prior to the start of operation}

\newduneabbrev{gsc}{GSC group}{global safety coordination group}{DUNE group that evaluates applicable codes and standards, including international code equivalency, for the design, assembly, and installation of the \gls{fd}}

\newduneabbrev{ha}{HA}{hazard analysis}{A first step in a process to assess risk; the result of hazard analysis is the identification of the hazards present for a task or process}
\newduneword{har}{HAR}{hazard analysis report}

\newduneabbrev{tap}{TAP}{trip action plan}{A document required for any trip by a worker to the underground area at \gls{surf}, per that site's access control program; 
it describes the work to be accomplished during the trip} 

\newduneword{em}{EM}{emergency management}
\newduneword{ert}{ERT}{emergency response team}

\newduneabbrev{ndk}{NDK}{nucleon decay}{The hypothetical, baryon number violating decay of a proton or a bound neutron into lighter particles}

\newduneabbrev{emi}{EMI}{electromagnetic interference}{Disturbance generated by an external source that affects an electrical circuit by electromagnetic induction, electrostatic coupling, or conduction}

\newduneabbrev{pe}{PE}{photoelectron}{An electron ejected from the surface of a material by the photoelectric effect}

\newduneabbrev{spe}{SPE}{single photoelectron}{A single photoelectron}

\newduneabbrev{fwhm}{FWHM}{full width at half maximum}{Width of a distribution measured between those points at which the distribution is equal to half of its maximum amplitude}

\newduneabbrev{gdml}{GDML}{geometry description markup language}{An application-indepedent, geometry-description format based on XML}

\newduneabbrev{xml}{XML}{extensible markup language}{A markup language that defines a set of rules for encoding documents in a format that is both human-readable and machine-readable}

\newduneabbrev{crt}{CRT}{cosmic ray tagger}{Detector external to the TPC designed to tag TPC-traversing cosmic ray particles}

\newduneabbrev{sn}{SN}{supernova}{Event that occurs upon the death of certain types of stars}

\newduneabbrev{wg}{WG}{working group}{A group of persons working together to achieve specified goals}

\newduneabbrev{ctsf}{CTSF}{coating, testing and storage facility}{A facility where the the \dual photon detectors will be coated, tested, and stored}

\newduneword{rucio}{Rucio}{Data management system originally developed
by \gls{atlas} but now open-source and shared across HEP}
\newduneabbrev{doma}{DOMA}{data organization, management, and
access}{data organization, management, and access efforts through the
HEP Software Foundation}

\newduneabbrev{hsf}{HSC}{HEP Software Foundation Collaboration}{A foundation that facilitates cooperation and common efforts in high energy physics software and computing internationally}

\newduneabbrev{wlcg}{WLCG}{Worldwide LHC Computing Grid}{Worldwide LHC
Computing Grid}
\newduneabbrev{osg}{OSG}{Open Science Grid}{Open Science Grid}
\newduneabbrev{sci}{SCI}{Scientific Computing Infrastructure}{Proposed
extension of the infrastructure component of \gls{wlcg} to other
experiments}
\newduneabbrev{csc}{CSC}{computing and software consortium}{DUNE
computing and software consortium}

\newduneword{dirac}{DIRAC}{Computing workflow management designed for
LHCb and now used by many HEP experiments}

\newduneword{frp}{FRP}{fiber-reinforced plastic}
\newduneabbrev{hdpe}{HDPE}{high-density polyethylene}{High-density polyethylene plastic}
\newduneword{hvps}{HVPS}{\gls{hv} power supply}
\newduneword{aisi}{AISI}{American Iron and Steel Institute}
\newduneword{ific}{IFIC}{Instituto de Fisica Corpuscular (in Valencia, Spain)}
\newduneabbrev{rsds}{RSDS}{radioactive source deployment system}{Proposed calibration system based on the deployment of
radioactive sources inside the \gls{dune} cryostat}
\newduneword{2p2h}{2p2h}{two particle, two hole}
\newduneabbrev{duneprism}{DUNE-PRISM}{\gls{dune} Precision Reaction-Independent Spectrum Measurement}{a mobile near detector that can perform measurements over a range of angles off-axis from the neutrino beam direction in order to sample many different neutrino energy distributions}
\newduneword{arcube}{ArgonCube}{The name of the core part of the \gls{dune} \gls{nd}, a \gls{lartpc}}

\newduneabbrev{citf}{CITF}{cryogenic instrumentation test facility}{A facility at \fnal with small ($<\,\SI{1}{ton}$) to intermediate ($\sim\,\SI{1}{ton}$) volumes of instrumented, purified TPC-grade \lar, used for testing devices intended for use in \gls{dune}}

\newduneabbrev{3dst}{3DST}{3D scintillator tracker}{The core part of the \threed projection scintillator tracker spectrometer in the near detector conceptual design}
\newduneabbrev{3dsts}{3DST-S}{3D scintillator tracker spectrometer}{The \threed projection scintillator tracker spectrometer  in the near detector conceptual design}
\newduneabbrev{mpd}{MPD}{multi-purpose detector}{A component of the near detector conceptual design; it is a magnetized system consisting of a \gls{hpgtpc} and a surrounding \gls{ecal}}
\newduneabbrev{hpg}{HPG}{high-pressure gas}{gas at high pressure to be used in a \gls{hpgtpc}} 
\newduneabbrev{hpgtpc}{HPgTPC}{high-pressure gaseous argon TPC}{A \gls{tpc} filled with gaseous argon; a possible component of the \gls{dune} \gls{nd}}

\newduneword{src}{SRC}{short-range correlated nucleon-nucleon interactions}
\newduneword{larpix}{LArPix}{ \gls{asic} pixelated charge readout for a \gls{tpc} }
\newduneword{arclt}{ArCLight}{a light detector \gls{arcube} effort}
\newduneword{fhc}{FHC}{forward horn current ($\numu$ mode)}
\newduneword{rhc}{RHC}{reverse horn current ($\overline{\nu}_{\mu}$ mode)}
\newduneword{mwpc}{MWPC}{multi-wire proportional chamber}
\newduneword{na61}{NA61}{CERN hadron production experiment}
\newduneword{pdnd}{ProtoDUNE-ND}{a prototype \gls{dune} \gls{nd}}
\newduneword{ccqe}{CCQE}{charged current quasielastic interaction} 
\newduneabbrev{roc}{ROC}{readout chamber}{readout chamber for gaseous argon \gls{tpc}}
\newduneabbrev{iroc}{IROC}{inner readout chamber}{inner (radial) readout chamber for gaseous argon \gls{tpc}}
\newduneabbrev{oroc}{OROC}{outer readout chamber}{outer (radial) readout chamber for gaseous argon \gls{tpc}}

\newduneword{lux}{LUX}{Large Underground Xenon (LUX) dark matter detector at \gls{surf} }

\newduneword{mjdemo}{Majorana Demonstrator}{Experiment sited at \gls{surf} that  seeks to determine whether neutrinos are their own antiparticles}

\newduneword{lz}{LZ}{Experiment sited at \gls{surf} that  seeks to detect faint interactions between galactic dark matter and regular matter}

\newduneword{mu2e}{Mu2e}{An experiment sited at \gls{fnal} that searches for charged-lepton flavor violation and seeks to discover physics beyond the \gls{sm}}

\newduneword{pdsp2}{ProtoDUNE-SP-2}{A second test run in the singe-phase
ProtoDUNE test stand at CERN, acting as a validation of the final
single-phase detector design}

\newduneword{osha}{OSHA}{Occupational Safety and Health Administration (USA Department of Labor) formed by the Occupational Safety and Health Act of 1970}
\newduneabbrev{pns}{PNS}{pulsed neutron source}{Calibration system based
on neutron capture gamma showers spread out in the whole detector}

\newduneabbrev{fv}{FV}{fiducial volume}{The detector volume within the \gls{tpc} 
that is selected for physics analysis through cuts on reconstructed event position}

\newduneword{p6}{P6}{framework used to plan and status the resource-loaded schedule of activities associated with the USA contributions to \gls{lbnf} and \gls{dune} }
\newduneabbrev{evms}{EVMS}{earned value management system}{Earned Value Management is a systematic approach to the integration and measurement of cost, schedule, and technical (scope) accomplishments on a project or task. It provides both the government and contractors the ability to examine detailed schedule information, critical program and technical milestones, and cost data (text from the US DOE); the EVMS is a system that implements this approach}

\newduneword{core}{CORE}{CORE contributions are in either monetary units or labor hours. They can be technical components for the facility or experiment and the effort of the staff needed to produce, install, and test them;  major facilities for the experiment; or other products and services relevant for the completion of the facility or experiment} 

\newduneabbrev{ahj}{AHJ}{Authority Having Jurisdiction}{An organization, office, or individual responsible for enforcing the requirements of a code or standard, or for approving equipment, materials, an installation, or a procedure (OSHA)}
\newduneword{cte}{CTE}{coefficient of thermal expansion}

\newduneabbrev{opc}{OPC}{open platform communications}{Open platform communications is a series of standards and specifications for industrial telecommunication} 
\newduneword{scada}{SCADA}{supervisory control and data acquisition}
\newduneword{ln}{LN$_2$}{liquid nitrogen}
\newduneabbrev{lapd}{LAPD}{Liquid Argon Purity Demonstrator}{Cryostat at Fermilab for long-term studies requiring a large volume of argon}

\newduneabbrev{pab}{PAB}{Proton Assembly Building}{Home of several \gls{lar} facilities at Fermilab}
\newduneword{hep}{HEP}{high energy physics}
\newduneword{sc}{SC}{scientific computing}  
\newduneword{cms}{CMS}{Compact Muon Solenoid experiment at CERN}
\newduneword{alice}{ALICE}{A Large Ion Collider Experiment, at CERN}
\newduneword{gpib}{GPIB}{general purpose interface bus}

\newduneabbrev{pfparticle}{PFParticle}{particle flow particle}{Each of the individual reconstructed particles in the hierarchy (or particle flow) describing the reconstructed event interaction}

\newduneabbrev{mcparticle}{MCParticle}{Monte Carlo Particle}{Individual true simulated particle}
\newduneword{au}{AU}{astronomical unit}
\newduneword{nufit}{NuFIT 4.0}{The NuFIT 4.0 global fit to neutrino oscillation data}

\newduneabbrev{sgft}{SGFT}{term}{add def (DP install)}
\newduneword{uhv}{UHV}{ultra high vacuum}
\newduneword{lps}{LPS}{laser positioning system}

\newduneword{unicamp}{UNICAMP}{University of Campinas, Sao Paulo, Brazil}
 
\newduneabbrev{fbk}{FBK}{Fondazione Bruno Kessler}{FBK is a research non-profit entity in Trento, Italy that partners in the development of technology with applications in various fields including High Energy Physics}

\newduneword{fft}{FFT}{fast Fourier transform}
\newduneabbrev{enob}{ENOB}{effective number of bits}{The effective number of bits is a measure of the dynamic range of an \gls{adc} and its associated circuitry. The resolution of an \gls{adc} is specified by the number of bits used to represent the analog value, in principle giving 2N signal levels for an N-bit signal. However, all real \gls{adc} circuits introduce noise and distortion. ENOB specifies the resolution of an ideal \gls{adc} circuit that would have the same resolution as the circuit under consideration}
\newduneabbrev{sndr}{SNDR}{signal to noise and distortion ratio}{Also known as SINAD. Ratio of the \gls{rms} signal amplitude to the mean value of the root-sum-square of all other spectral components, including harmonics, but excluding \gls{dc} levels. It is a good indication of the overall dynamic performance of an \gls{adc} because it includes all components which make up noise and distortion}
\newduneabbrev{sfdr}{SFDR}{spurious free dynamic range}{Spurious free dynamic range is the ratio of the \gls{rms} value of the signal to the \gls{rms} value of the worst spurious signal regardless of where it falls in the frequency spectrum. The worst spur may or may not be a harmonic of the original signal}
\newduneabbrev{thd}{THD}{total harmonic distortion}{Total harmonic distortion is the ratio of the \gls{rms} value of the fundamental signal to the mean value of the root-sum-square of its harmonics} 
\newduneword{tvs}{TVS}{transient voltage suppression}

\newduneword{riskprob}{risk probabilities}{The risk probability, after taking into account the planned mitigation activities, is ranked as 
L (low $<\,$\SI{10}{\%}), 
M (medium \SIrange{10}{25}{\%}), or 
H (high $>\,$\SI{25}{\%}). 
The cost and schedule impacts are ranked as 
L (cost increase $<\,$\SI{5}{\%}, schedule delay $<\,$\num{2} months), 
M (\SIrange{5}{25}{\%} and 2--6 months, respectively) and 
H ($>\,$\SI{20}{\%} and $>\,$2 months, respectively)}

\newduneabbrev{lbls}{LBLS}{laser beam location system}
{Auxiliary calibration system providing an independent location measurement of the ionization laser beams direction}

\newduneabbrev{lsst}{LSST}{Large Synoptic Survey Telescope}{8.4 m telescope with 3.2G-pixel camera that will start taking data in 2023}
\newduneabbrev{ska}{SKA}{Square Kilometer Array}{International radio telescope array planned to start data-taking in 2027}
\newduneabbrev{hyperk}{HyperK}{Hyper Kamiokande}{260 kt water Cerenkov neutrino detector to begin construction at Kamiokande in 2020}
\newduneword{lhcb}{LHCb}{LHC experiment dedicated to forward physics}
\newduneword{belleii}{Belle II}{B-factory experiment now running at KEK}

 \newduneabbrev{ldm}{LDM}{light-mass dark matter}{Refers to dark matter particles with mass values much lower than the electroweak scale, specifically below the 1~GeV level}
 
\newduneabbrev{bnv}{BNV}{baryon-number violating}{Describing an interaction where \gls{baryonnumber} is not conserved}

\newduneword{bugey}{Bugey}{Neutrino experiment that operated at the Bugey nuclear power plant in France}

\newduneword{minosplus}{MINOS$+$}{The successor to the \gls{minos} experiment, utilizing the same detectors and beam line, but operating at higher beam energy tune than \gls{minos}, parasitic with \gls{nova}}

\newduneword{baryonnumber}{baryon number}{A quantity expressing the total number of baryons in a system minus the number of antibaryons}

\newduneword{np04}{NP04}{CERN North Area hadron beamline used for the \gls{sp} test beam run}

\newduneword{ua1}{UA1}{UA1 (Underground Area 1) was a particle detector at \gls{cern}'s  Super Proton Synchrotron (SPS). It ran from 1981 until 1990, when the SPS was used as a proton-antiproton collider, searching for traces of W and Z particles in collisions. (CERN) The UA1 dipole magnet was reused in the NOMAD experiment and currently provides the magnetic field for the \gls{t2k} ND280 detector}

\newduneword{ssc}{SSC}{The Superconducting Super Collider was to be a huge underground ring complex beneath the area near Waxahachie, Texas, USA, that would have been the world’s most energetic particle accelerator. It was begun in 1990, but canceled by the U.S. Congress in 1993 (scientificamerican.com Oct 2013)}

\newduneword{daphne}{DAPHNE}{Detector electronics for Acquiring PHotons from NEutrinos is a custom-developed warm front-end waveform digitizing electronics module derived from the readout system developed at Fermilab for the Mu2e experiment}
 
\newduneword{nersc}{NERSC}{National Energy Research Computing Facility at \gls{lbnl}}

  \newduneword{integoff}{integration office}{The office that incorporates the onsite team responsible for coordinating integration and installation activities at SURF}

\newduneabbrev{sma}{SMA}{SubMiniature version A}{Connector interface for coaxial cables
with a screw-type coupling mechanism}

\newduneword{kloe}{KLOE}{KLOE is a $e^+ e^-$ collider detector spectrometer operated at DAFNE, the $\phi$-meson factory at Frascati, Rome. In DUNE it will consist of a \SI{26}{cm} Pb+scintillating fiber ECAL surrounding a cylindrical open detector region that is  \SI{4.00}{m} in diameter and \SI{4.30}{m} long. The ECAL and detector region are embedded in a \SI{0.6}{T} magnetic field created by a \SI{4.86}{m} diameter superconducting coil and a \SI{475}{tonne} iron yoke}

\newduneword{ro}{review office}{An office within the \gls{integoff} that organizes reviews }

\newduneabbrev{doecd}{CD}{critical decision}{The U.S. DOE's Order 413.3B outlines a series of staged project approvals, each of which is referred to as a critical decision (CD)}

\newduneabbrev{lbnfspac}{LBNF SPAC}{LBNF Strategic Project Advisory Committee}{A committee charged by the host laboratory director to provide expert, independent advice on significant issues and strategies related to LBNF project organization, management, and risks}

\newduneabbrev{sand}{SAND}{System for on-Axis Neutrino Detection}{The beam monitor component of the near detector that remains on-axis at all times and serves as a dedicated neutrino spectrum monitor}

\newduneword{4850l}{4850L}{The depth in feet (1480 m) of the top of the cryostats underground at SURF; used more generally to refer to the DUNE underground area. Called the ``4850 level'' or ``4850L''}

\hypersetup{
    pdftitle={\expshort TDR \thedocsubtitle},
    pdfauthor={\expshort Collaboration},
    final=true,
    colorlinks=false,
    linktocpage=true,
    linkbordercolor=blue,
    citebordercolor=green,
    urlbordercolor=magenta,
    filecolor=black,
    pdfpagemode=UseOutlines,
    pdfborderstyle={/S/U},  
}

\renewcommand\thedoctitle{\voltitleexec} 
\newcommand\thevolumenumber{\volnumberexec}

\begin{document}

\pagestyle{titlepage}
\includepdf[pages={-}]{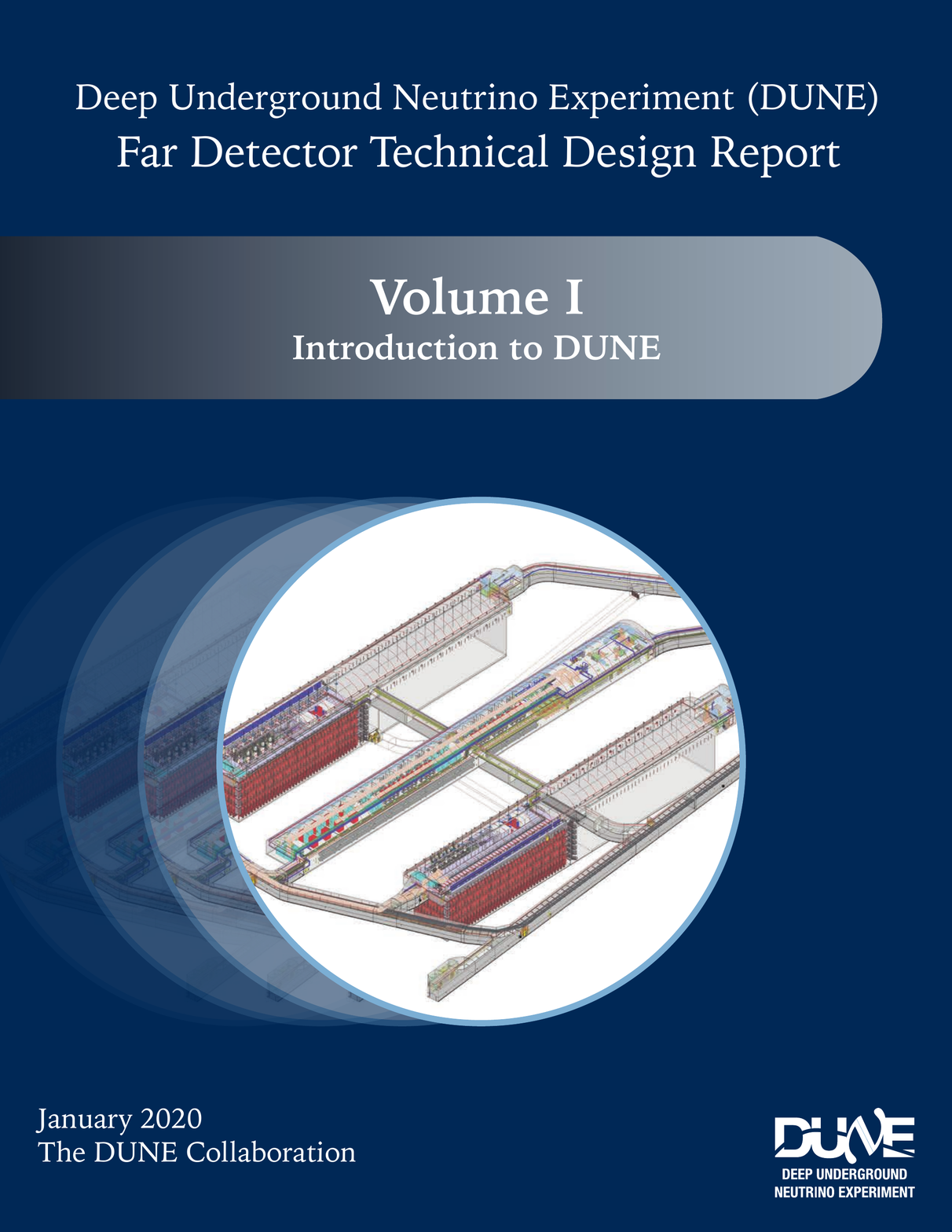}
\cleardoublepage

\cleardoublepage
\vspace*{16cm} 
  {\small  This document was prepared by the DUNE collaboration using the resources of the Fermi National Accelerator Laboratory (Fermilab), a U.S. Department of Energy, Office of Science, HEP User Facility. Fermilab is managed by Fermi Research Alliance, LLC (FRA), acting under Contract No. DE-AC02-07CH11359.
  
The DUNE collaboration also acknowledges the international, national, and regional funding agencies supporting the institutions who have contributed to completing this Technical Design Report.  
  }
\includepdf[pages={-}]{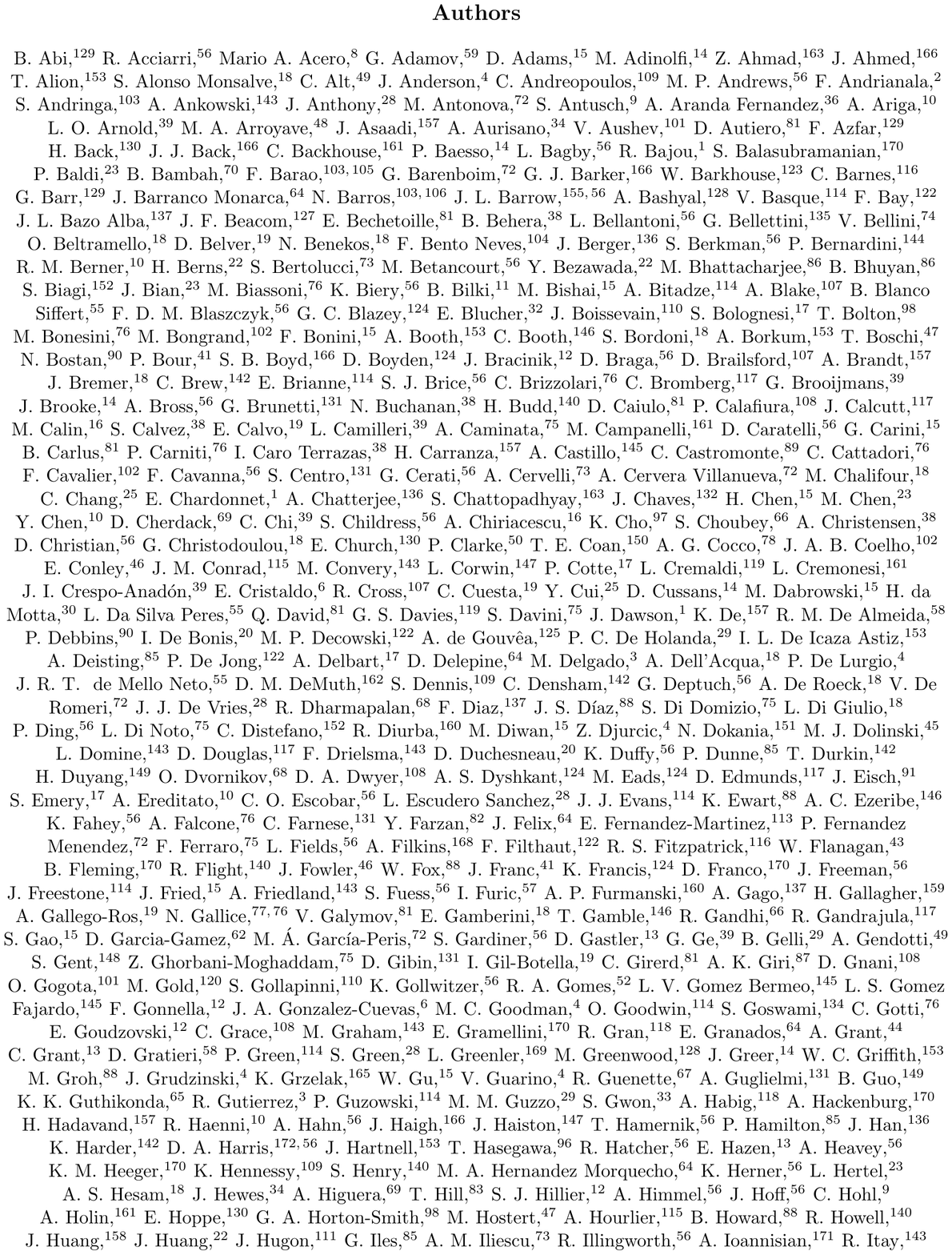}

\renewcommand{\familydefault}{\sfdefault}
\renewcommand{\thepage}{\roman{page}}
\setcounter{page}{0}

\pagestyle{plain}

\textsf{\tableofcontents}

\textsf{\listoffigures}

\textsf{\listoftables}
  \vspace{4mm}
  \addcontentsline{toc}{chapter}{A Roadmap of the DUNE Technical Design Report}

\iffinal\else
\textsf{\listoftodos}
\clearpage
\fi

\renewcommand{\thepage}{\arabic{page}}
\setcounter{page}{1}

\pagestyle{fancy}

\renewcommand{\chaptermark}[1]{%
\markboth{Chapter \thechapter:\ #1}{}}
\fancyhead{}
\fancyhead[RO,LE]{\textsf{\footnotesize \thechapter--\thepage}}
\fancyhead[LO,RE]{\textsf{\footnotesize \leftmark}}

\fancyfoot{}
\fancyfoot[RO]{\textsf{\footnotesize The DUNE Technical Design Report}}
\fancyfoot[LO]{\textsf{\footnotesize \thedoctitle}}
\fancypagestyle{plain}{}

\renewcommand{\headrule}{\vspace{-4mm}\color[gray]{0.5}{\rule{\headwidth}{0.5pt}}}



\cleardoublepage
\chapter*{A Roadmap of the DUNE Technical Design Report}

The \dword{dune} \dword{fd} \dword{tdr} describes the proposed physics program,  
detector designs, and management structures and procedures at the technical design stage.  

The TDR is composed of five volumes, as follows:

\begin{itemize}
\item Volume~\volnumberexec{} (\voltitleexec{}) provides an overview of all of DUNE for science policy professionals.

\item Volume~\volnumberphysics{} (\voltitlephysics{}) describes the DUNE physics program.

\item Volume~\volnumbertc{} (\voltitletc{}) outlines DUNE management structures, methodologies, procedures, requirements, and risks. 

\item Volume~\volnumbersp{} (\voltitlesp{}) and Volume~\volnumberdp{} (\voltitledp{}) describe the two \dword{fd} \dword{lartpc} technologies.

\end{itemize}

The text includes terms that hyperlink to definitions in a volume-specific glossary. These terms  appear underlined in some online browsers, if enabled in the browser's settings.

\cleardoublepage

\cleardoublepage

\chapter{Executive Summary}
\label{ch:exec-overall}

\section{Overview}
\label{sec:exec-overall-1}

\subsection{The DUNE Science Program}
\label{sec:exec-overall-sci}

The preponderance of matter over antimatter in the early universe, the dynamics of the \dwords{snb} that produced the heavy elements necessary for life, and whether protons eventually decay --- these mysteries at the forefront of particle physics and astrophysics are key to understanding the early evolution of our universe, its current state, and its eventual fate. The \dword{dune} is an international world-class experiment dedicated to addressing these questions.

Experiments carried out over the past half century have revealed that neutrinos are found in three states, or \textit{flavors}, and can transform from one flavor into another. These results indicate that each neutrino flavor state is a mixture of three different nonzero mass states, and to date offer the most compelling evidence for physics beyond the standard model. In a single experiment, \dword{dune} will enable a broad exploration of the three-flavor model of neutrino physics with unprecedented detail. Chief among its potential discoveries is that of matter-antimatter asymmetries (through the mechanism of \dword{cpv}) in neutrino flavor mixing --- a step toward
unraveling the mystery of matter generation in the early universe. Independently, determination of the unknown neutrino mass ordering and precise measurement of neutrino mixing parameters by \dword{dune} may reveal new fundamental symmetries of nature.

Neutrinos emitted in the first few seconds of a core-collapse supernova carry with them the potential for great insight into the evolution of the universe. \dword{dune}'s capability to collect and analyze this high-statistics neutrino signal from a supernova within the Milky Way would provide a rare opportunity to peer inside a newly formed neutron star and potentially witness the birth of a black hole.

\Dwords{gut}, which attempt to describe the unification of the known forces, predict rates for proton decay that cover a range directly accessible with the next generation of large underground detectors such as the \dword{dune} far detector. The experiment's sensitivity to key proton decay channels will offer unique opportunities for the ground-breaking discovery of this phenomenon.

\subsection{The DUNE Detectors and Supporting Facilities}
\label{sec:exec-overall-det-sppt}

To achieve its goals, the international \dword{dune} experiment, hosted by the U.S. Department of Energy's \dword{fnal} in Illinois, comprises three central components: (1) a new, high-intensity neutrino source generated from a megawatt-class proton accelerator at \dword{fnal}, (2)  a massive \dword{fd} situated \SI{1.5}{\km} underground at the \dword{surf} in South Dakota, and (3) a composite \dword{nd} installed just downstream of the neutrino source. 
Figure~\ref{fig:lbnf} illustrates the layout of these components. The far detector, the subject of this \dword{tdr}, will be a modular \dword{lartpc} 
with a fiducial (sensitive) mass of
\fdfiducialmass{}\footnote{For comparison, this is nearly twice the mass of the Statue of Liberty and nearly four times that of the Eiffel Tower.} 
(\SI{40}{\giga\gram}) 
of \dword{lar}, a cryogenic liquid that must be kept at \lartemp{} (\SI{-185}{\degree}C). This detector will be able to  uniquely reconstruct neutrino interactions with image-like precision and unprecedented resolution~\cite{Adams:2013qkq}. 

\begin{dunefigure}[ 	
Configuration of the LBNF beam and the DUNE detectors]{fig:lbnf}{ 	
Cartoon illustrating the configuration of the LBNF beamline at Fermilab, in Illinois, and the DUNE detectors in Illinois and South Dakota, separated by \SI{1300}{km}.}
\includegraphics[width=0.9\textwidth]{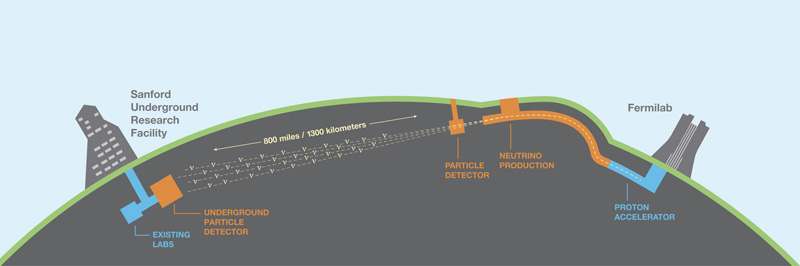}
\end{dunefigure}

The \dword{lbnf} project, also hosted by \dword{fnal}, provides the beamline and the civil construction, called \dword{cf}, for the \dword{dune} experiment.  
The organization and management of \dword{lbnf} is separate from that of the experiment; its design and construction are organized as a U.S. \dword{doe}/\dword{fnal} project incorporating international partners. 

The \dword{lbnf} 
beamline at \dword{fnal} will deliver the world's most intense neutrino beam to the near and far detectors in an on-axis configuration. 
The upgrade to the  \dword{pip2}~\cite{pip2-2013}, a leading-edge, superconducting, linear proton accelerator under construction at \dword{fnal}, will deliver between \num{1.0} and \SI{1.2}{MW} of proton beam power from the \dword{fnal} Main Injector to \dword{lbnf}, which will aim and focus the beam, whereupon the protons,  in a wide energy band of \SIrange{60}{120}{\GeV}, will collide with a high-power production target, creating a secondary beam from which the intense neutrino flux will emerge, traveling in the direction of the \dword{dune} detectors (Figure~\ref{fig:beamline}).   A further planned upgrade 
of the accelerator complex could provide up to \SI{2.4}{\MW} of beam power by 2030, 
potentially extending the \dword{dune} science reach. The upgrade will also increase the reliability of the \dword{fnal} accelerator complex and provide the flexibility to produce customized  beams tailored to specific scientific needs.

\begin{dunefigure}[Neutrino beamline and DUNE near detector hall in Illinois
]{fig:beamline}{Neutrino beamline and \dshort{dune} near detector hall at \dshort{fnal} in Illinois}
\includegraphics[width=0.9\textwidth]{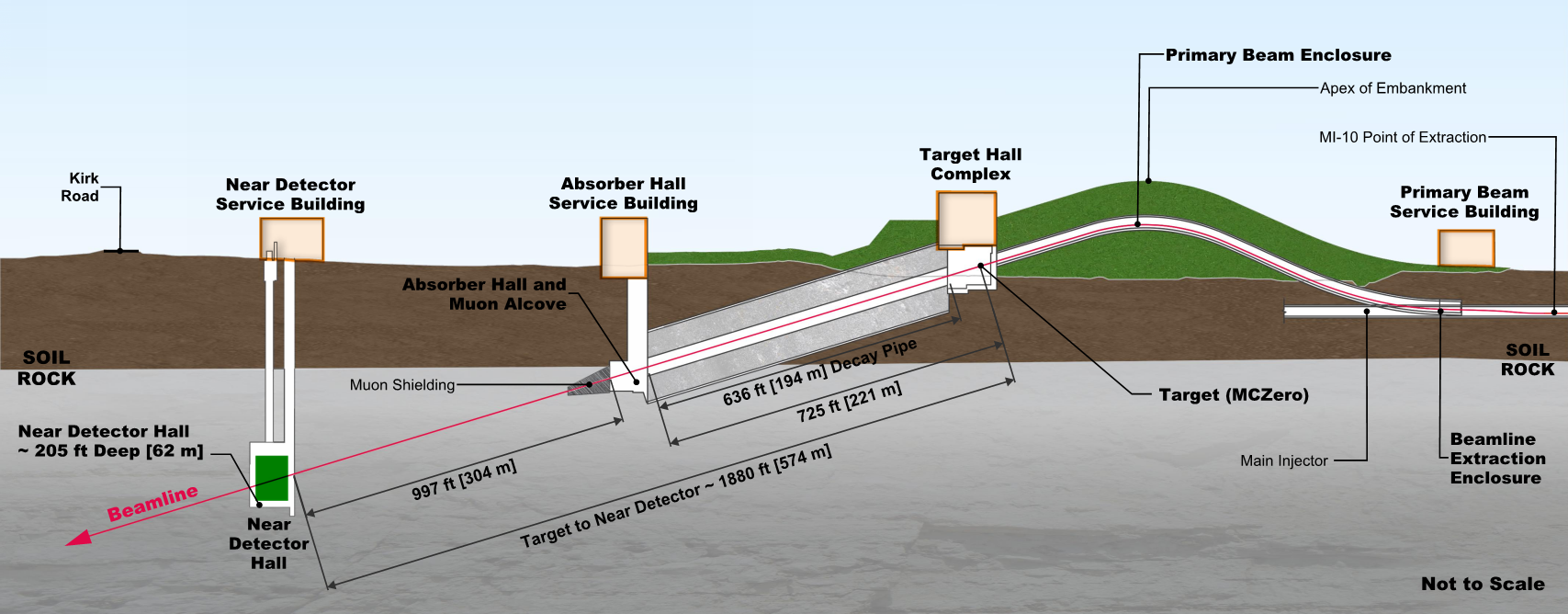}
\end{dunefigure}

The intense, wide-band neutrino beam, the massive \dword{lartpc} detector at the far site, and the composite (i.e., multi-purpose) 
\dword{nd} will provide a rich ancillary science program for the \dword{dune} experiment, beyond its primary goals,   
including accelerator-based neutrino flavor-transition measurements with sensitivity to physics beyond the standard model, measurements of tau neutrino appearance, measurements of neutrino oscillation phenomena using atmospheric neutrinos, and  a rich neutrino interaction physics program using the \dword{dune} \dword{nd}, including a wide range of measurements of neutrino cross sections, studies of nuclear effects, and searches for dark matter. 
Further advances in \dword{lartpc} 
technology during  \dword{fd} construction may open up possibilities to observe very low-energy phenomena such as solar neutrinos or even the diffuse supernova neutrino flux -- measurements that require a sensitivity that is presently beyond our reach.

\subsection{The DUNE Collaboration}

The \dword{dune} collaboration is a global organization with more than \num{1000} scientists and engineers from \num{31} countries (Figure~\ref{fig:map2}). It represents the combination of several worldwide efforts that developed independent paths toward a next-generation long-baseline neutrino experiment over the last decade. \dword{dune} was formed in April 2015, combining the strengths of the \dword{lbne} project in the U.S. and the \dword{lbno} project in Europe, adding many new international partners in the process. \dword{dune} thus represents the convergence of a substantial fraction of the worldwide neutrino-physics community around the opportunity provided by the large investment planned by the U.S. \dword{doe} and \dword{fnal} to support a significant expansion of the underground infrastructure at \dword{surf} in South Dakota and to create a megawatt neutrino-beam facility at \dword{fnal}. 

\begin{dunefigure}[DUNE collaboration global map]{fig:map2}{The international \dshort{dune}
collaboration. Countries with \dshort{dune} membership are in light brown.}
\includegraphics[width=0.9\textwidth]{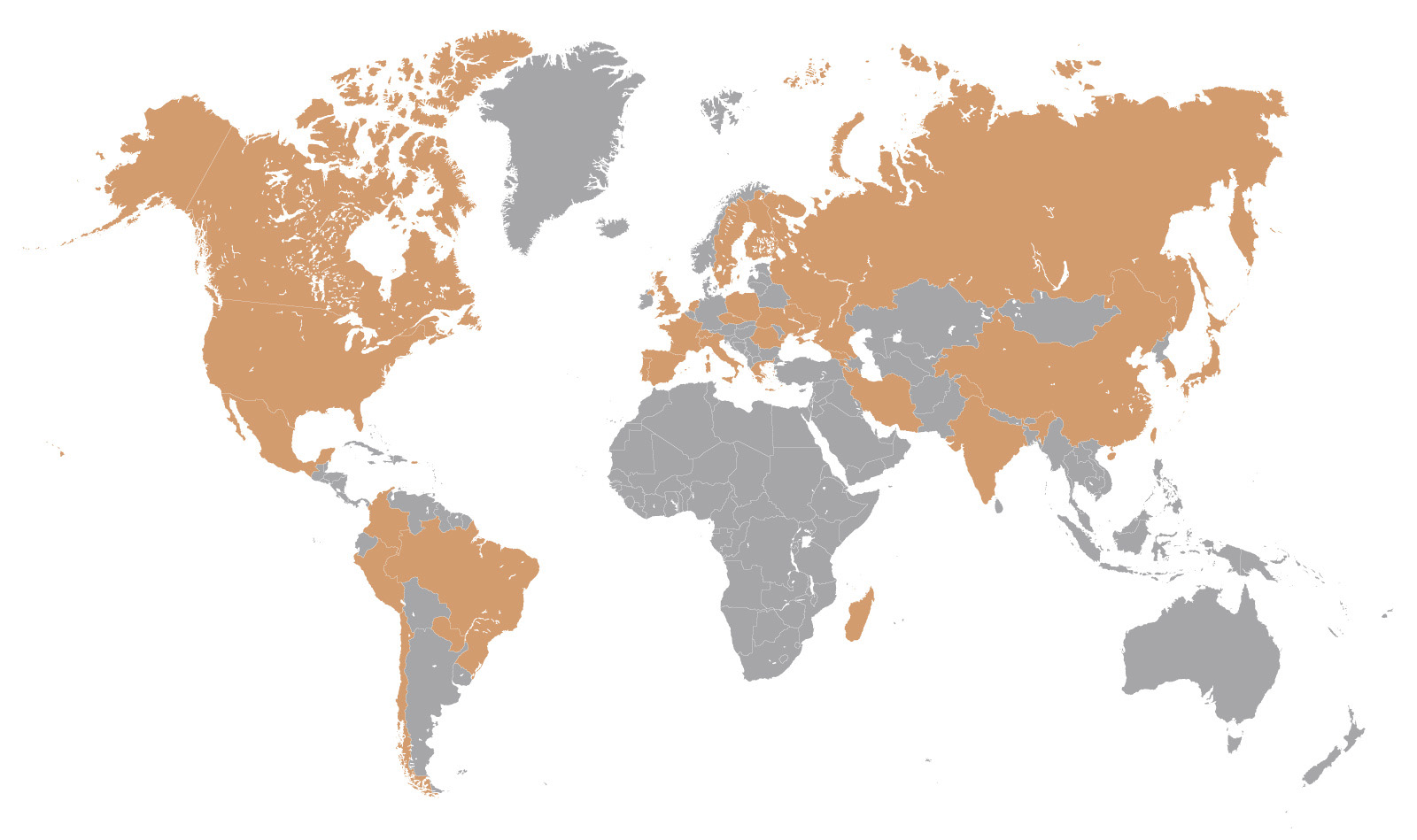}  
\end{dunefigure} 

\subsection{Strategy for the DUNE Far Detector Design}
\label{exec:strategy-fd-des}

\dword{dune} and \dword{lbnf} have developed the strategy presented in this \dword{tdr} to meet the requirements laid 
out in the report of the U.S. Particle Physics Project Prioritization Panel (P5) in 2014. The strategy also takes into account the recommendations of the \dword{espp} adopted by the \dword{cern} Council in 2013, which classified the long-baseline neutrino program as one of the four scientific objectives requiring significant resources, sizable collaborations, and sustained commitment.

The P5 report~\cite{p5report} set the goal of 
determining leptonic \dword{cpv} with a precision of three standard deviations (\num{3}$\,\sigma$) or greater (i.e., a precision of at least \SI{0.03}{\%} ), over more than \SI{75}{\%}   of the range of possible values  (0 through 2$\pi$) of the unknown \dshort{cp}-violating phase \deltacp. It is impossible to provide coverage of \SI{100}{\%}  since \dword{cpv} effects vanish as the value of \deltacp approaches $0$ or $\pi$. 
 Based partly on this goal, the report stated that ``the 
minimum requirements to proceed are the identified capability to reach an exposure 
of \SI{120}{\ktMWyr}\footnote{\SI{1}{\ktMWyr} is the amount of exposure a detector of target mass \SI{1}{kt} of \dword{lar} would get in one year using a \SI{1}{MW} proton beam to generate the neutrinos. Exposure increases linearly as each of the factors increases.}
 by the 2035 time frame, the far detector situated underground 
with cavern space for expansion to at least \fdfiducialmass \dword{lar} fiducial volume, and \SI{1.2}{MW} beam power upgradable to multi-megawatt power.
The experiment should have the demonstrated 
capability to search for \dfirsts{snb} and for proton decay, providing a significant 
improvement in discovery sensitivity over current searches for proton decay.'' 

Here we briefly address how the \dword{lbnf-dune} strategy and designs will satisfiy each of these requirements and deliver a world-leading neutrino program. 
The following chapters and the other volumes of this \dword{tdr} elaborate on these concepts, providing a full picture of this ambitious enterprise. 

\textit{Reach at least \SI{120}{\ktMWyr} exposure by the 2035 timeframe and \SI{1.2}{\MW} beam power: } To reach the necessary precision on its measurements, \dword{dune} will need to collect a few thousand neutrino interactions over a period of about ten years. The number of interactions is the product of (1) the intensity of the neutrino beam, (2) the probability that a neutrino will oscillate (approximately \num{0.02}), (3) the interaction cross section, and (4) the detector mass.  Currently, the highest-power proton beam that a beam target can safely withstand is between \num{1} and \SI{2}{\MW}, which caps the achievable neutrino beam intensity. This points to a required a detector mass in the tens-of-kilotons range. 
The \dword{dune} \dword{fd}  cryostats will hold \larmass of \dword{lar} each, for a total of nearly \SI{70}{kt}, of which at least \fdfiducialmass will be encompassed by the four detector modules as fiducial mass. Higher usable fractions of the total volume are under study.

Moreover, the \dword{dune} concept
    builds on the notion that a highly-performant detector technology  
    with excellent neutrino energy reconstruction
    and background rejection capabilities can
    optimize sensitivity and cost with an on-axis exposure to
    an intense, wide-band, conventional (magnetic horn-focused) beam.
   The current generation of long-baseline neutrino experiments
	have benefited from narrow-band beam characteristics 
	associated with off-axis detector deployment, which offers 
	a low background rate in both electron neutrino (\nue) appearance 
	and muon neutrino (\numu) disappearance channels. 
	However, this advantage comes at a cost of flux and 
	spectral information relative to an on-axis detector 
    configuration~\cite{Adams:2013qkq,Agarwalla:2014tca}.

\textit{Situated underground:}
Given the rate of cosmic rays at the surface (\SI{165}{kHz}) and the neutrino beam parameters, the ratio of neutrino events to cosmic rays would be less than one to a million and the discovery potential for \dword{dune}'s oscillation physics goals would be vanishingly small.  Roughly \SI{1500}{m} underground at the \dword{surf} site, this ratio becomes slightly higher than 1, raising the discovery potential to a very achievable level.  Supernova neutrinos have energies on the order of \num{100} times lower than beam neutrinos, and despite the fact that they arrive in a few-second burst, would be nearly impossible to identify on the surface. A meaningful search for nucleon decay is not possible at the surface. All three of the experiment's primary goals require significant overburden for the \dword{fd}, which the \dword{surf} site provides. 

\textit{Use of liquid argon (\lar):}
This requirement implies the use of the \dword{lartpc} technology, which enables finer resolution for kiloton-scale particle
detectors than earlier technologies do. The enhanced resolution leads to greater efficiency in distinguishing signal events from background, which in turn leads to a reduction in the necessary size of the detector and potentially broadens the physics program.
    It is especially important for the long-baseline program with a    
    wide-band neutrino beam.
    Additionally, the choice of \dword{lartpc} technology provides 
    valuable complementarity to other 
    existing and planned detectors pursuing many
    of the same goals.  As an example,
    the sensitivity of \dword{dune} to the \nue component of supernova 
    neutrino flux, prevalent in the neutronization phase of the 
    explosion, provides distinct information relative to that 
    provided by water or organic scintillator-based detectors in 
    which electron antineutrino (\anue) interactions dominate. 

\textit{Sensitivity to \dword{cpv}:}
The physics that \dword{dune} will pursue demands measurements at the few-percent level. With just a \dword{fd}, the neutrino fluxes would be known only to about \SI{10}{\%} and interaction rates to at best \SI{20}{\%}.  To adequately reduce these uncertainties, is necessary to measure the neutrinos at a near location, e.g., \SI{500}{m} from the neutrino source and  at a far location, e.g., \SI{1300}{km} away, using the same target nucleus in both detectors, and to extract the physics measurements from differences between the two.  The \dword{nd} can be smaller than the \dword{fd}, but it must be multi-functional since the differences in the measurements are not due solely to oscillations.  The detector rate is the product of the neutrino flux, the detector response, and the interaction cross section, the first two of which will differ between the \dword{nd} and \dword{fd} due to other factors as well, e.g.,  event rate and geometry. The \dword{nd} must be able to measure the factors that go into the detector rate separately.  

The optimal \dword{fd} distance (baseline) to determine the \dword{mh}, observe \dword{cpv}, and observe \deltacp is between \num{1000} and \SI{2000}{km}; at shorter baselines the optimal neutrino energy is lower, the second oscillation maximum is too low in energy to be visible, and \dword{cp} sensitivity is reduced by ambiguities from the unknown mass ordering. At longer baselines \dword{cp} sensitivity is harmed by matter effects that increase with baseline. The \SI{1300}{km} baseline offered by locating the \dword{fd} at \dword{surf} is optimized for the neutrino oscillation physics goals of the \dword{dune} program.

 The scientific basis for \dword{dune}'s foundational experimental
design choices has been examined and validated through extensive
review, undertaken at all stages of \dword{dune} development.
Recent experimental and theoretical developments have only
strengthened the scientific case for \dword{dune} and its
basic configuration.  The technical underpinnings for
these choices have also been strengthened over time through a worldwide
program of R\&D and engineering development, as described in a suite
of \dword{lbnf-dune} project documents including this \dword{tdr}, as
well as 
through independent experiments and development
activities.

\section{The Long-Baseline Neutrino Facility (LBNF)} 
\label{sec:exec:lbnf}

As mentioned above, the \dword{lbnf} project will provide the beamline and the \dfirst{cf} for both detectors of the \dword{dune} experiment. At the far site, \dword{surf} in South Dakota, \dword{lbnf} will construct a facility to house and provide infrastructure for the \dword{dune}  \nominalmodsize fiducial mass \dword{fd} modules; in particular \dword{lbnf} is responsible for:

\begin{itemize}

\item the excavation of three underground caverns at \dword{surf}, north and south detector caverns and a \dword{cuc} for the detector's ancillary systems; this requires the removal of \SI{800}{kt} of rock\footnote{This is roughly the mass of the Golden Gate Bridge including its anchorage and approaches.};

 \item free-standing, steel-supported cryostats to contain each \dword{detmodule} in a bath of \larmass of \dword{lar};   
 \item the required cryogenics systems for rapidly deploying the first two modules;

\item surface, shaft, and underground infrastructure at  \dword{surf} to support installation, commissioning, and operation of the detector; and

\item the \dword{lar} required to fill the first two cryostats.
\end{itemize}

\dword{dune} intends to install the third and fourth \dword{fd} modules as rapidly as funding will 
allow. When finished, the north and south caverns will each house two modules and the \dword{cuc} will house cryogenics and data acquisition facilities for all four modules.

Figure~\ref{fig:caverns} shows the cavern layout for the \dword{fd} in the \dword{surf} underground area, also referred to as the 4850 (foot) level or \dword{4850l}.

\begin{dunefigure}[ 	
Underground caverns for DUNE in South Dakota]{fig:caverns}{Underground caverns for the \dword{dune}{} \dword{fd}{} and cryogenics systems at \dword{surf} in South Dakota. The drawing shows the cryostats (red) for the first two \dword{fd} modules  in place at the \dword{4850l}. The Ross Shaft, the vertical shaft that will provide access to the \dword{dune} underground area, appears on the right. 
Each cryostat is \cryostatlen long (\SI{216}{ft}, approximately the length of two and a half tennis courts), \cryostatwdth wide, and \cryostatht tall (about three times as tall as an adult giraffe).
The two detector caverns are each 
\SI{144.5}{\meter} long, \SI{19.8}{\meter} wide,  and \SI{28.0}{\meter} high, providing some room around the cryostats.}
\includegraphics[width=0.7\textwidth]{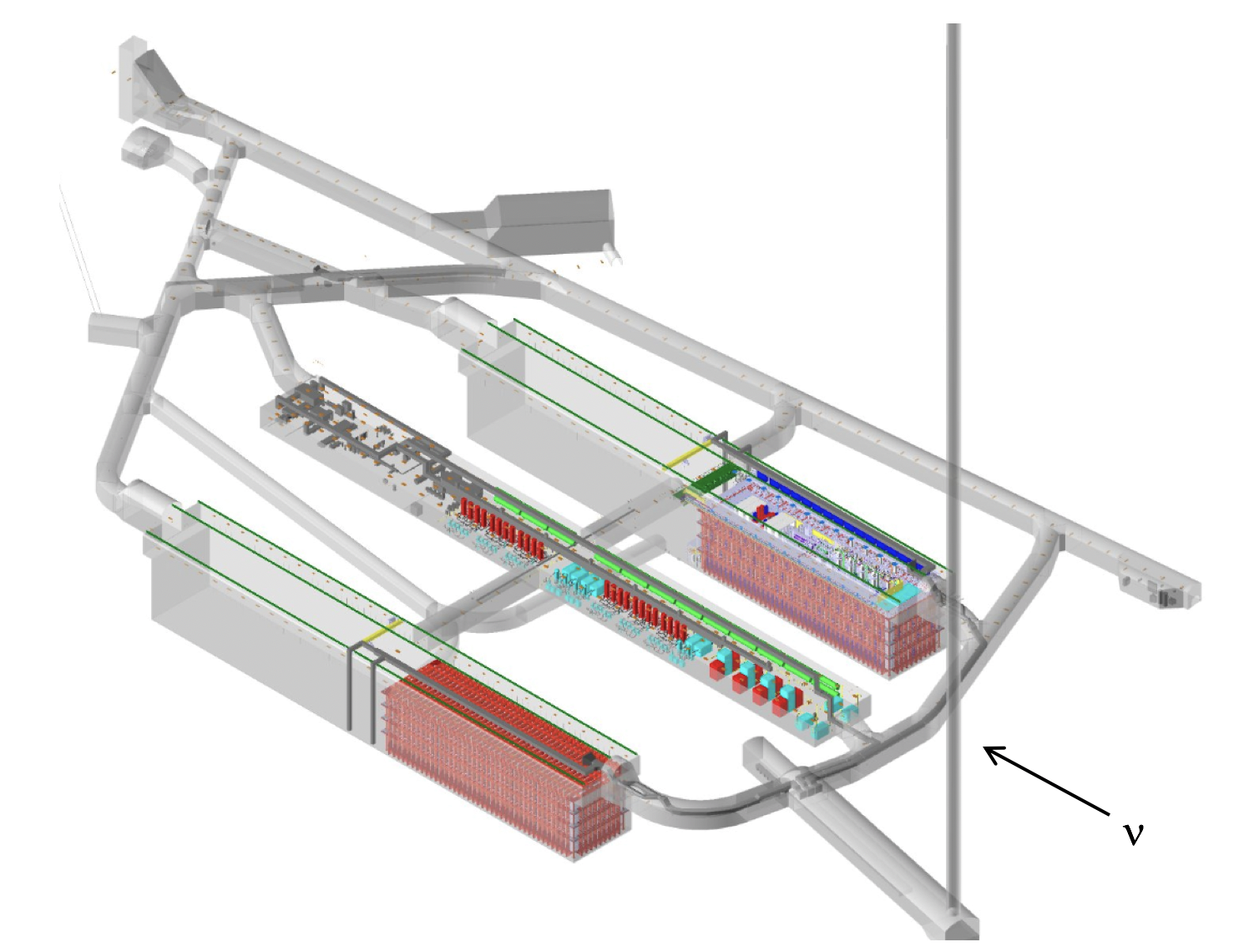}
\end{dunefigure}

\section{The DUNE Detectors}

\subsection{Far Detector}
\label{ch:dune-det-tech-ov-fd}

The \dword{dune} 
\dword{fd} will consist of four \dword{lartpc} \dwords{detmodule}, each with a \dword{lar} mass in the sensitive region of the cryostat (fiducial mass) of at least \nominalmodsize, installed approximately \SI{1.5}{km} underground. excellent tracking and calorimetry performance, making it an ideal 
choice. Each \dword{lartpc} fits inside a cryostat of internal dimensions
\cryostatwdthinner (w) $\times$ \cryostathtinner (h) $\times$ \cryostatleninner~(l) containing a total \dword{lar} mass of about \larmass{}.
 The design of the four identically sized modules is sufficiently flexible for staging construction and evolving the \dword{lartpc} technology.

\dword{dune} is planning for and currently developing two \dword{lartpc} technologies: \dword{sp} in which all the detector elements inside the cryostat are immersed in liquid; and \dword{dp}, in which some components operate in a layer of gaseous argon above the liquid.

\begin{itemize}
\item In the \dword{sp} technology, ionization charges drift horizontally in the \dword{lar} under the influence of an electric field (\efield) towards 
a vertical anode, where they are read out. 
This design requires very low-noise electronics to achieve readout with a good \dword{s/n} ratio because no signal amplification occurs inside the cryostat. 
This technology was pioneered in the \dword{icarus} project, and after several decades of worldwide R\&D, is now mature. It is the technology used for \dword{fnal}'s currently operating \dword{microboone} detector, as well as the \dword{sbnd} detector, which is under construction. Figure~\ref{fig:LArTPC1ch1} shows the operating principle of an \dword{sp} \dword{lartpc}.

\item The \dword{dp} technology was pioneered at a large scale by the \dword{wa105} collaboration at \dword{cern}. It is less mature than the \dword{sp} technology, and whereas it presents some challenges, it offers several advantages.  Here, ionization charges drift vertically upward in \dword{lar} and are transferred into a layer of argon gas above the liquid. Devices called \dwords{lem} amplify the signal charges in the gas phase before they reach a horizontal anode. The gain achieved in the gas reduces the stringent requirements on the electronics noise and the overall design increases the possible drift length, which, in turn, requires a correspondingly higher voltage. 
Figure~\ref{fig:DPprinciplech1} shows the operating principle of a \dword{dp} \dword{lartpc}. 
\end{itemize}
 In both technologies, the drift volumes are surrounded by a \dword{fc} that defines the active detector volume and ensures uniformity of the \efield within that volume.
 
\begin{dunefigure}[The single-phase (SP) LArTPC operating principle]{fig:LArTPC1ch1}
{The general operating principle of the \dword{sp} \dword{lartpc}. Negatively charged ionization electrons from the neutrino interaction drift horizontally opposite to the \efield in the \dword{lar} and are collected on the  
 anode, which is made up of the U, V and X sense wires.  The right-hand side represents the time projections in two dimensions as the event occurs. Light ($\gamma$) detectors (not shown) will provide the $t_0$ of the interaction.}
\includegraphics[width=0.75\textwidth]{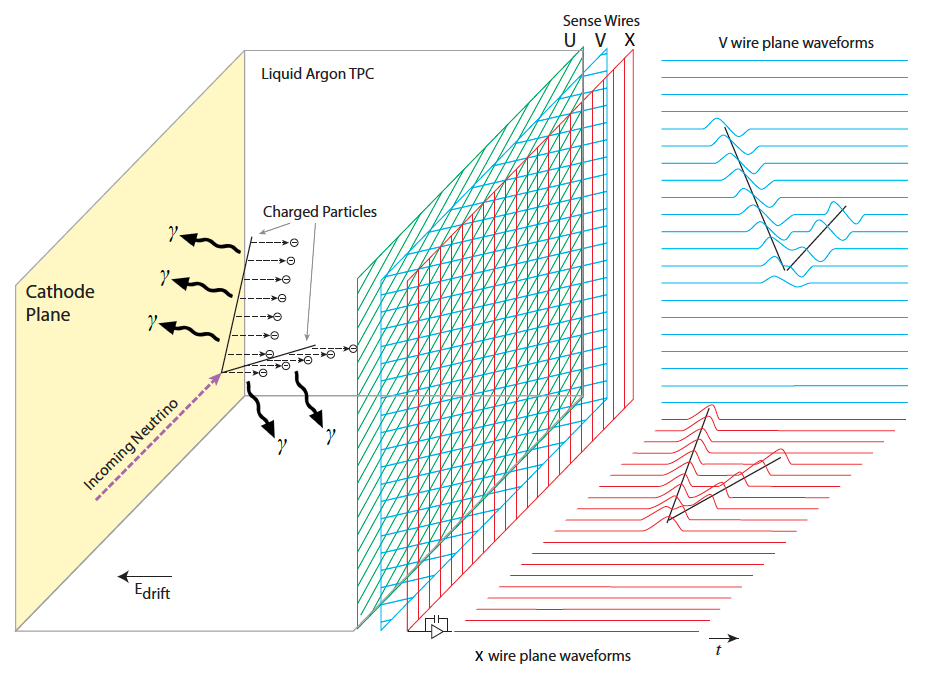} 
\end{dunefigure}

\begin{dunefigure}[The dual-phase (DP) LArTPC operating principle]{fig:DPprinciplech1}{The general operating principle of the \dword{dp} \dword{lartpc}. The ionization charges drift vertically upward in \dword{lar} and are transferred into a layer of argon gas above the liquid where they are amplified before collection on the anode. The light detectors (PMTs) sit under the cathode.}
\includegraphics[width=0.5\textwidth]{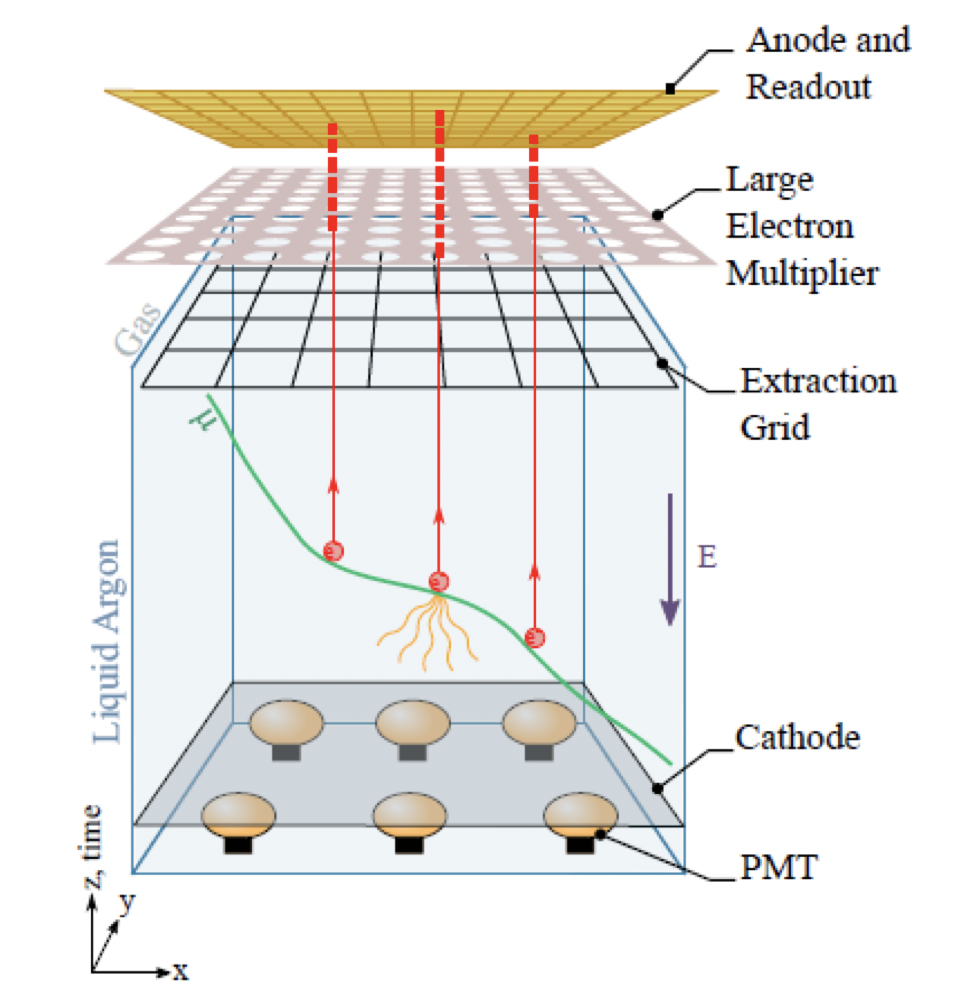}
\end{dunefigure}

Argon is an excellent scintillator at a wavelength of \SI{126.8}{\nano\meter} (UV), a property that both detector designs exploit. This fast scintillation light (photons), once shifted into the visible spectrum, is collected by \dwords{pd} in both designs. The light collection provides an initial start time ($t_{0}$) for every event recorded by the \dword{tpc}, indicating when the ionization electrons begin to drift. Comparing the time at which the ionization signal reaches the anode relative to this start time allows reconstruction of the event topology in the drift coordinate (i.e., horizontal and transverse to the beam for \dword{sp} and vertical for \dword{dp}); the precision of the measured $t_{0}$, therefore, directly corresponds to the precision of the spatial reconstruction in this direction. 

Two key factors affect the performance of the \dword{dune} \dwords{lartpc}: \dword{lar} purity and noise on the readout electronics.  First, the \dword{lar} purity must be 
quite high to minimize charge and light attenuation over the longest drift lengths in the \dword{detmodule}.    
The \dword{sp} and \dword{dp} designs have slightly different purity requirements (expressed in minimum electron lifetimes of \SI{3}{ms} versus \SI{5}{ms}) due to the different maximum drift lengths.
Second, the electronic readout of the \dword{lartpc} requires very low noise levels for the signal from the drifting electrons to be clearly discerned over the baseline of the electronics.  This requires using low-noise cryogenic electronics, especially in the case of the \dword{sp} design.

The \dword{dune} collaboration is committed to deploying both technologies. The full \dword{dune} \dword{fd} requires four modules. In this \dword{tdr}, we describe plans for the first three modules: two \dwords{spmod}, one of which will be the first module installed, and one \dword{dpmod}.  
The actual sequence of \dword{detmodule} installation will depend on results from the prototype detectors, described below, and on available resources. Plans for the fourth \dword{detmodule}, which may use a more advanced design, remain to be determined. 

The plans for the \dword{sp} and \dword{dp} modules are described briefly in the following sections, more fully in Chapters~\ref{ch:exec-sp} and~\ref{ch:exec-dp}, and finally in great detail in Volumes~\volnumbersp{} and~\volnumberdp{} of this \dword{tdr}.

\subsubsection{A Single-Phase Far Detector Module}
\label{sec:fdsp-exec-splar}

The operating principle of an \dword{sp} \dword{lartpc} (Figure~\ref{fig:LArTPC1ch1}) has been demonstrated by  \dword{icarus}~\cite{Icarus-T600}, \dword{microboone}~\cite{microboone}, \dword{argoneut}~\cite{Anderson:2012vc}, \dword{lariat}~\cite{Cavanna:2014iqa}, and \dword{pdsp}~\cite{Abi:2017aow}. Charged particles passing through the \dword{tpc} ionize the argon, and the ionization electrons drift in an \efield{} to the anode planes. Figure~\ref{fig:DUNESchematic1ch1} shows the configuration of a \dword{dune} \dword{spmod}. Each of the four drift volumes of \dword{lar} is subjected to a strong \efield{} of \spmaxfield, corresponding to a cathode \dword{hv} of \sptargetdriftvoltpos. The maximum drift length is \spmaxdrift.

\begin{dunefigure}[A \nominalmodsize DUNE far detector SP module]{fig:DUNESchematic1ch1}
{A \nominalmodsize \dword{dune} \dword{fd} \dword{spmod}, showing the alternating \sptpclen{} long (into the page), \tpcheight{} high anode (A) and cathode (C) planes, as well as the \dfirst{fc} that surrounds the drift regions between the anode and cathode planes. On the right-hand cathode plane, the foremost portion of the \dword{fc} is shown in its undeployed (folded) state.}
\includegraphics[width=0.65\textwidth]{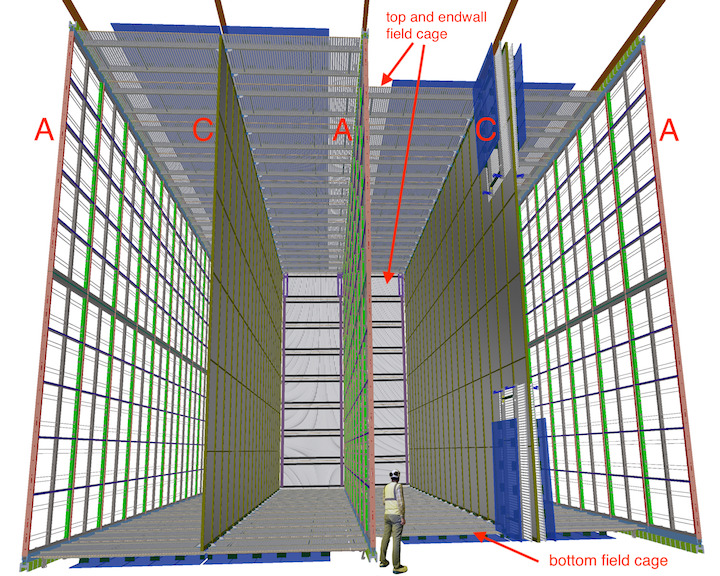}
\end{dunefigure}

An \dword{spmod} is instrumented with three module-length (\sptpclen) anode planes constructed from \SI{6}{m} high by \SI{2.3}{m} wide \dwords{apa}, stacked two \dword{apa}s high and 25 wide, for 50 \dword{apa}s per plane, and 150 total. Each \dword{apa} 
consists of an alumninum frame with three layers of active wires, strung at angles chosen to reduce ambiguities in event reconstruction, that form a grid on each side of the \dword{apa}. The relative voltage between the layers is chosen to ensure transparency to the drifting electrons of the first two layers ($U$ and $V$). These layers produce bipolar induction signals as the electrons pass through them. The final layer ($X$) collects the drifting electrons, resulting in a unipolar signal. The pattern of ionization collected on the grid of anode wires provides the reconstruction in the remaining two coordinates perpendicular to the drift direction (Figure~\ref{fig:LArTPC1ch1}).

Novel \dword{sipm} based \dfirsts{pd} called \dwords{arapuca}\footnote{An ``arapuca'' is a South American bird trap; the name is used here as an analogy to the way the devices trap photons. These devices have been developed by DUNE collaborators in Brazil.}
 are placed in the inactive space between the innermost wire planes of the \dword{apa}s, installed through slots in the \dword{apa} frame. 
Each \dword{apa} holds ten \dword{pd} modules, for a total of \num{1500} per \dword{spmod}.  Of these, \num{500} are mounted in the \dword{apa}s of the central anode plane and collect light from both directions, 
and \num{500} each are mounted in the outer \dword{apa} frames and collect light from only the inner-facing direction. 

\FloatBarrier
\subsubsection{A Dual-Phase Far Detector Module}
\label{sec:fddp-exec-splar}

The \dword{dp} operating principle, illustrated in Figure~\ref{fig:DPprinciplech1}, is very similar to that of the \dword{sp}. 
 Charged particles that traverse the active volume of the \dword{lartpc} ionize the medium while also producing scintillation light.  The ionization electrons drift, in this case vertically, along an \efield toward a segmented anode where they deposit their charge. Any scintillation light that is produced is measured in  \dwords{pd} that view the interior of the volume from below. 
 
 In this design, shown in Figure~\ref{fig:DPdet1ch1}, ionization electrons drift upward toward an extraction grid just below the liquid-vapor interface. 
After reaching the grid, an \efield stronger than the \dpnominaldriftfield{} drift field extracts the electrons from the liquid up into the gas phase. Once in the gas, the electrons encounter micro-pattern gas detectors, called \dwords{lem}, with high-field regions
in which they are amplified. 
The amplified charge is then collected and recorded on a \twod anode
consisting of two sets of 
gold-plated copper strips that provide the $x$ and $y$ coordinates (and thus two views) of an event. 
An array of \dwords{pmt} coated with a wavelength-shifting material sits below the cathode to record the time ($t_{0}$) and pulse characteristics of the incident light.

\begin{dunefigure}[A \nominalmodsize DUNE far detector DP module]{fig:DPdet1ch1}
  {Schematic of a \nominalmodsize \dword{dune} \dword{fd} \dword{dp} \dword{detmodule} with cathode, \dwords{pmt}, \dword{fc}, and anode plane with \dwords{sftchimney}. The drift direction is vertical in the case of a DP module. The scale is indicated by the figures of two people standing in front of the model.}
  \includegraphics[width=0.9\textwidth]{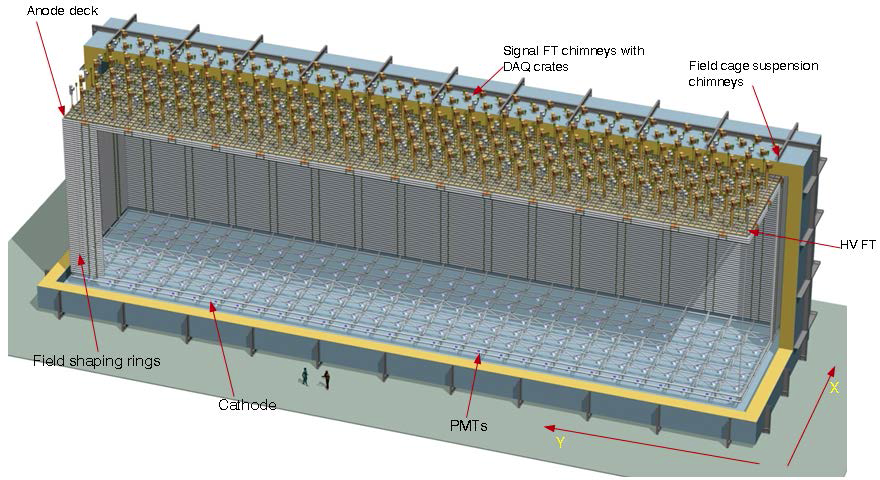}
\end{dunefigure}

The extraction grid, \dword{lem}, and anode are assembled into three-layered sandwiches with precisely defined inter-stage distances and inter-alignment,  which are then connected horizontally into \num{9}~m$^2$ modular detection units called \dwords{crp}.

The precision tracking and calorimetry offered by both the \dword{sp} and \dword{dp} technologies provide excellent capabilities for identifying interactions of interest while mitigating sources of background.  Whereas the \dword{sp} design has multiple drift volumes, the \dword{dpmod} design allows a single, fully homogeneous \dword{lar} volume with a much longer drift length.

\FloatBarrier

\subsection{ProtoDUNEs: Far Detector Prototypes}
\label{sec:exec:overall:pdune}

The \dword{dune} collaboration has constructed and operated 
two large prototype detectors, \dword{pdsp}, and \dword{pddp}, at \dword{cern}.  
 Each is approximately one-twentieth the size of the planned \dword{fd} modules but uses components identical in size to those of the full-scale module. \dword{pdsp} has the same \spmaxdrift maximum drift length as the full \dword{spmod}. \dword{pddp} has a \SI{6}{m} maximum drift length, half that planned for the \dword{dpmod}.  Figure~\ref{fig:protodunes_northarea} shows the two cryostats, \dword{pdsp} in the foreground and \dword{pddp} at an angle in the rear. Figure~\ref{fig:protodunes_interior} shows one of the two drift volumes of \dword{pdsp} on the left and the single drift volume of \dword{pddp} on the right.

\begin{dunefigure}[ProtoDUNE cryostats at the CERN Neutrino Platform]
{fig:protodunes_northarea}
{\dword{pdsp}{} and \dword{pddp}{} cryostats in the \dword{cern}{} Neutrino Platform in \dword{cern}'s North Area.}
\includegraphics[width=0.9\linewidth]{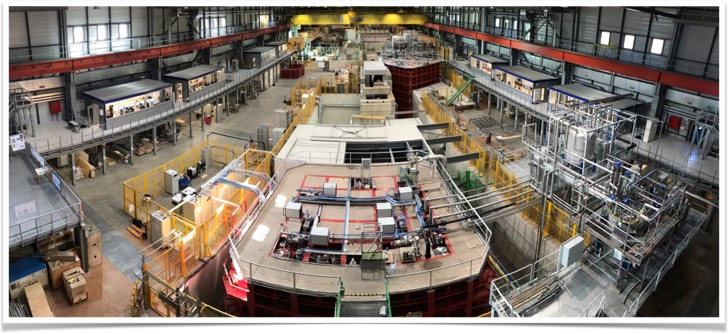}
\end{dunefigure}

\begin{dunefigure}[Interior views of the ProtoDUNEs]
{fig:protodunes_interior}
{Left: View of one of the two drift volumes in \dword{pdsp}; the \dword{apa} is on the left, the \dword{cpa} is on the right, and two of the four the \dword{fc} surfaces bounding the drift volume are at the center and bottom of the image.  Right:  the single \dword{pddp} drift volume (still incomplete when the image was taken), looking up; the \dwords{crp} (orange) are at the top. Three sides of the surrounding \dword{fc} are shown, but the cathode is not visible.}
\includegraphics[width=0.46\linewidth]{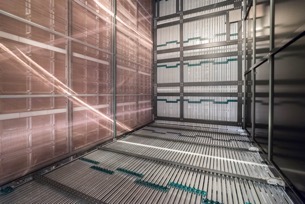}\hspace{0.05\linewidth}
\includegraphics[width=0.44\linewidth]{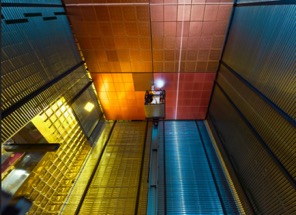}
\end{dunefigure}

This massive prototyping program was undertaken with both engineering and scientific goals in mind, namely: 
 
\begin{enumerate}
\item production of components: stress-test the production and \dword{qa} processes of detector components and mitigate the associated risks for the \dword{fd};
\item validation of installation procedures: test the interfaces between the detector elements
and mitigate the associated risks for the \dword{fd};
\item operation of the detector with cosmic rays: validate the detector designs and performance; and 
\item collection of test beam data: measure the physics response of the detector.
\end{enumerate}

Construction of the \dword{pdsp} detector was finished in July 2018 and filled with \dword{lar} the following month. It collected hadron beam and cosmic ray data during the fall of 2018 and continues to collect cosmic ray data.  Construction of the \dword{pddp} detector was complete in June of 2019, and the detector started operations in September 2019.

The data taken with \dword{pdsp} demonstrate the detector's excellent performance and have already provided valuable information on the design, calibration, and simulation of the \dword{dune} \dword{fd}. In all, \SI{99.7}{\%}  of the 15360 \dword{tpc} electronics channels are responsive in the \dword{lar}. The \dword{enc} amounts to $\approx 550$ $e^{-}$ on the collection wires and $\approx 650$ $e^{-}$ on the induction wires, roughly half of the allowed maximum. An average \dword{s/n} of 38 for the collection plane is measured using cosmic-ray muons, while for the two induction planes, the \dword{s/n} is 14 (U) and 17 (V), easily exceeding the requirement of 4 
for the \dword{dune} \dword{fd}. 

When an interaction occurs, we convert the charge deposited along the track to the energy loss ($dE/dx$) using stopping cosmic ray muons. Calibration constants have been derived with this method, which are applied to the energy deposits measured for the beam particles -- muons, pions, protons, and positrons. 
Figure~\ref{fig:pdtpcpd} (left) shows the calibrated $dE/dx$ values as a function of the track residual range for protons in the \SI{1}{GeV/$c$} beam, in good agreement with expectations. 

\begin{dunefigure}
  [Calibrated $dE/dx$ vs residual range and PD response in ProtoDUNE-SP]
  {fig:pdtpcpd}
  {Left: Calibrated $dE/dx$ (energy loss over distance) versus residual range measured by a \dword{tpc} for 1 GeV/c   
  stopping protons. Right: Response in \dword{pdsp} of an \dword{arapuca} \dword{pd} module in \dword{apa}{}3 as a function of incident electron kinetic energy.} 
  \includegraphics[width=0.47\linewidth]{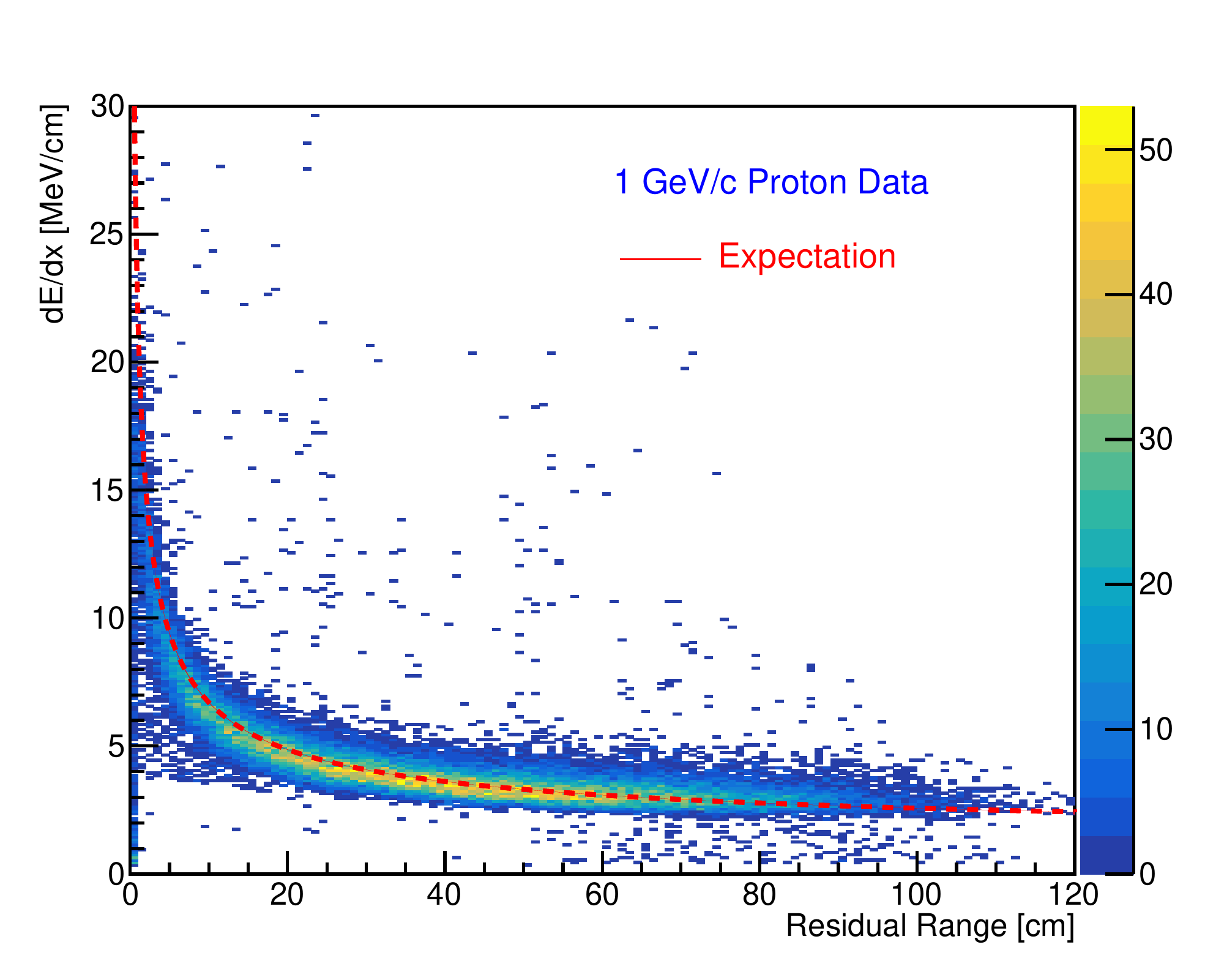}
  \includegraphics[width=0.5\linewidth]{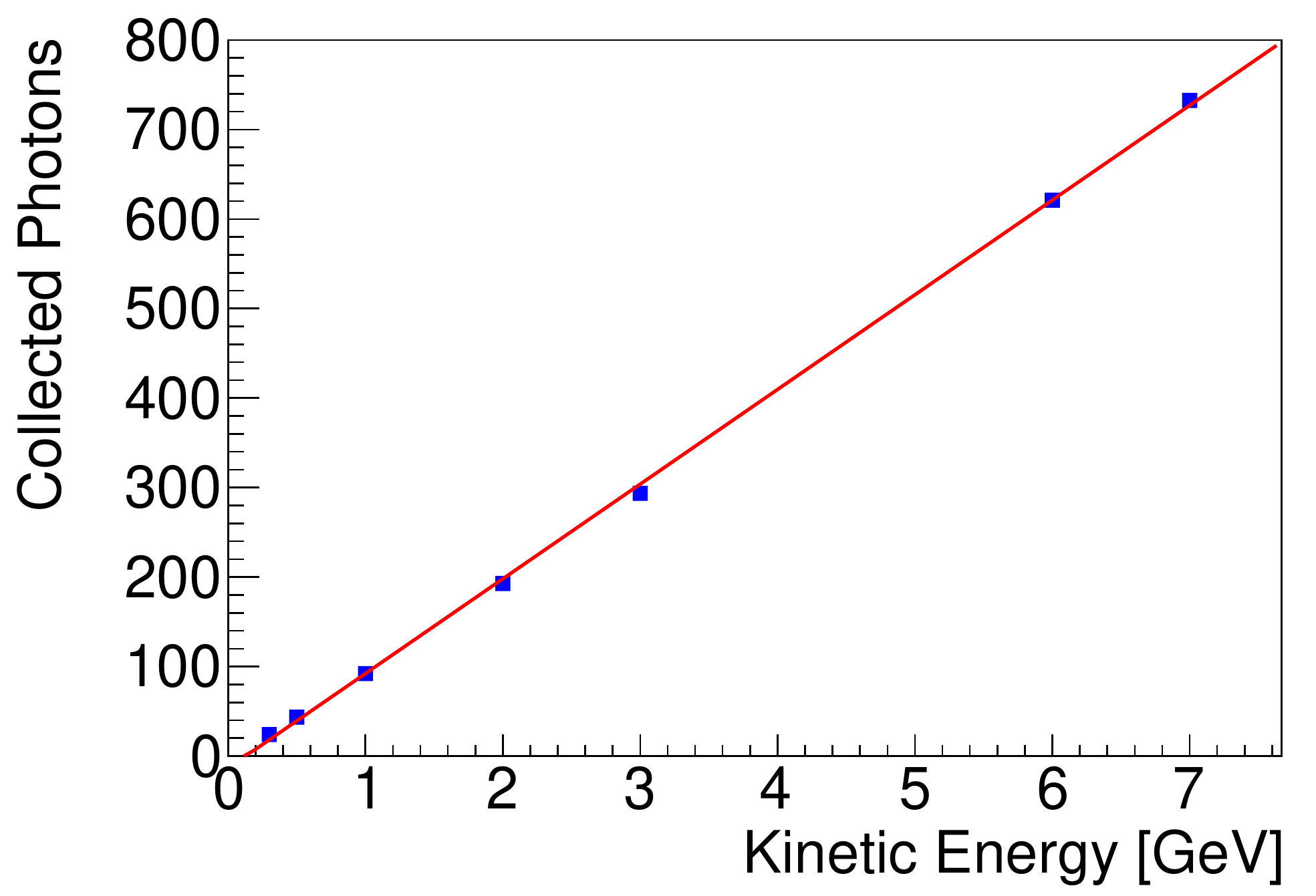}
\end{dunefigure}
  
  The \dword{pdsp} beam run provides a unique set of high-quality data for detector performance characterization, physics studies, and calibration, and will 
allow us to perform hadron-argon cross section measurements, which are relevant for future \dword{dune} neutrino oscillation analyses.
Data collected during the beam run will also be used to characterize the \dword{pds} response to light signals. Other useful data sets include
beam data with triggers determined by the beam instrumentation; cosmic ray data  from random triggers or from those in coincidence with the \dword{crt} modules; and calibration data, with triggers enabling programmed light pulses. 
The 
response and gain for each of the 256 readout channels of the \dword{pds} have been determined from calibration data, and 
the initial analysis results indicate very good performance and stability for this system. 

Figure~\ref{fig:pdtpcpd} (right) shows the response of an 
\dword{arapuca} \dword{pd} module (not corrected for geometry and detection efficiency) as a function of incident electron kinetic energy measured in \dword{pdsp}. 
 This preliminary analysis demonstrates the achieved energy linearity for beam electrons contained in the detector.  
In addition to verifying the \dword{pds} response and calibration, \dword{pdsp} shows excellent correlation between \dword{tpc} timing and the \dword{pds} timing. The latter will enable the further optimized physics reach of \dword{dune}. 

\subsection{Near Detector}
\label{sec:nd-verview}

Although not the subject of this \dword{tdr}, an understanding of \dword{dune}'s capabilities would be impossible without 
some description of the  \dword{nd}'s crucial contribution  to the experiment.
The \dword{nd} will serve as the experiment's control,
 constraining systematic errors and measuring the initial unoscillated \numu and \nue energy spectra (and that of the corresponding antineutrinos). 
Comparing the measured neutrino energy spectra near the beam source, before any oscillation takes place, and again at the far site allows us to disentangle the different energy-dependent effects that modulate the beam spectrum and to reduce the systematic uncertainties to the level required for discovering \dword{cpv}. Its other key role in this arena is to measure neutrino-argon interactions with high precision using both gaseous and liquid argon, which will further reduce the systematic uncertainties associated with modeling these interactions. 

The \dword{nd} will have a physics program of its own, as well, independent of the \dword{fd}.  This program will include measuring neutrino interactions to explore the two pillars of the standard model: electroweak physics and quantum chromodynamics. It will also explore physics beyond the standard model, searching for non-standard interactions, sterile neutrinos,  dark photons, and  other exotic particles.

The \dword{nd} will be located \SI{574}{m} downstream from the neutrino beam source and will include three primary detector components, illustrated in Figure~\ref{fig:neardetectors}: 

\begin{itemize}
\item a \dword{lartpc} called \dword{arcube}; 
\item a \dword{hpgtpc} surrounded by an \dword{ecal} in a \SI{0.5}{T} magnetic field, together called the \dword{mpd}; and 
\item an on-axis beam monitor called \dword{sand}.
\end{itemize}
These components serve important individual and overlapping functions in the mission of the \dword{nd}.  The first two can move off-axis relative to the beam, providing access to different neutrino energy spectra. The movement off-axis, called \dword{duneprism}, provides a crucial extra degree of freedom for the \dword{nd} measurements and is an integral part of the \dword{dune} \dword{nd} concept. 

\begin{dunefigure}[DUNE near detector (ND) components]
{fig:neardetectors}
{\dshort{dune} \dshort{nd}. The axis of the beam is shown as it enters from the right. Neutrinos first encounter
the \dword{lartpc} (right), the \dshort{mpd} (center), and then the on-axis beam monitor (left).}
\includegraphics[width=0.9\textwidth]{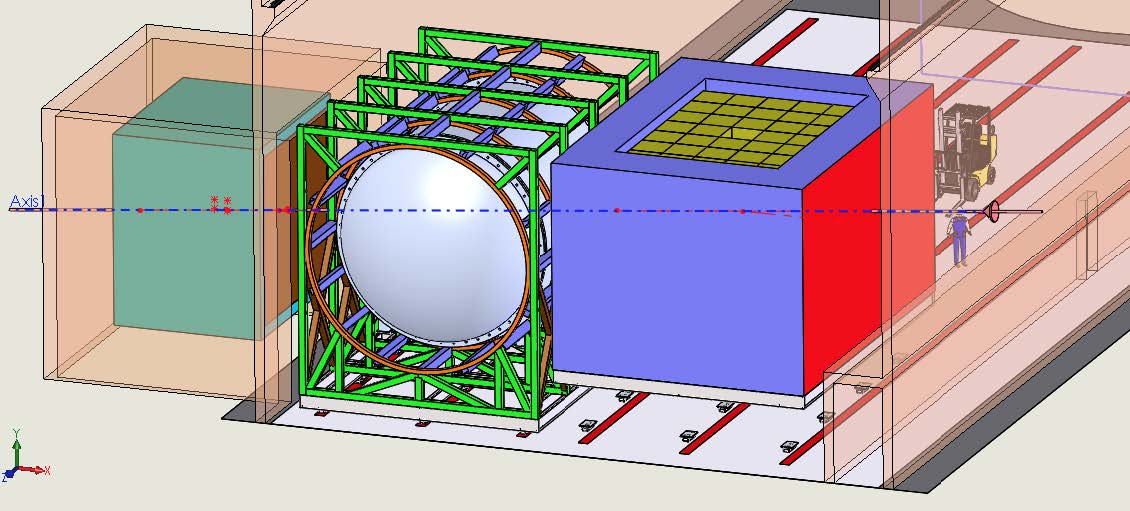}
\end{dunefigure}

The \dword{arcube} detector contains the same target nucleus and shares some aspects of form and functionality with the \dword{fd}. 
This similarity in target nucleus and, to some extent, technology, reduces sensitivity to nuclear effects and detector-driven systematic uncertainties in extracting the oscillation signal at the  \dword{fd}. 
\dword{arcube} is large enough to provide high statistics (\num{1e8} $\numu$  charged current events/year on-axis), and its volume is sufficiently large to provide good hadron containment.  The tracking and energy resolution, combined with the \dword{lar} mass, will allow measurement of the neutrino beam using several techniques. 

\begin{dunefigure}[DUNE ND hall with component detectors]
{fig:NDHallconfigs}
{\dword{dune} \dword{nd} hall shown with component detectors all in the on-axis configuration (left) and with the \dword{lartpc} and \dword{mpd} in an off-axis configuration (right). The 
beam monitor (\dshort{sand}) is shown in position on the beam axis in both figures. The beam is shown entering the hall at the bottom traveling from right to left.}
\includegraphics[width=0.49\textwidth]{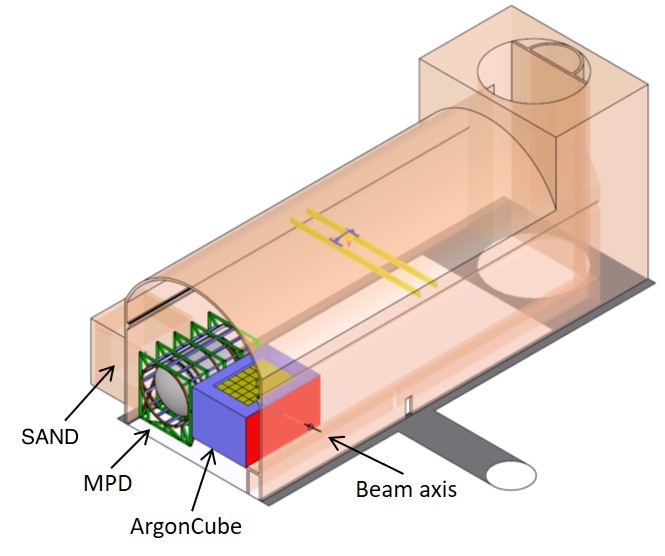}
\includegraphics[width=0.49\textwidth]{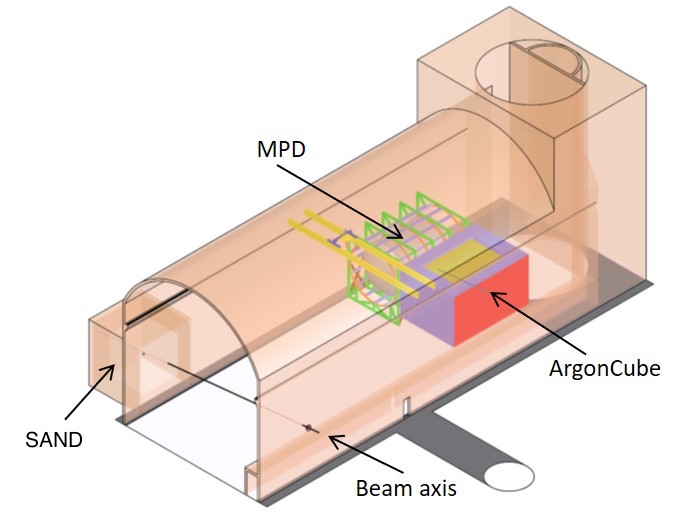}
\end{dunefigure}

A \dword{lartpc} energy acceptance falls off for muons with a measured momentum higher than $\sim\,$\SI{0.7}{GeV/c} because the muons will not be contained in the \dword{lartpc} volume.  Since muon momentum is critical to determining the incoming neutrino's energy, a magnetic spectrometer is needed downstream of the \dword{lartpc} to measure the momentum and charge of the muons.   
In the \dword{dune} \dword{nd} concept, the \dword{mpd} will make these measurements. The \dword{hpgtpc} provides a lower density medium with excellent tracking resolution for muons from the \dword{lartpc}.

The \dword{lartpc} and \dword{mpd} can be moved sideways up to \SI{33}{m} to take data in positions off the beam axis (\dword{duneprism}). As the detectors move off-axis, the incident neutrino flux spectrum changes, with the mean energy dropping and the spectrum becoming more monochromatic.  
The \dword{dune} concept is based on reconstructing the energy-dependent neutrino spectrum and comparing measurements at the far and near sites. The ability to take measurements at the near site in off-axis locations will allow us to 
disentangle otherwise degenerate effects due to systematic biases of the energy reconstruction.

The final component of the \dword{dune} \dword{nd} suite is the beam monitor that remains on-axis at all times and
serves as a dedicated  neutrino spectrum monitor. 
It can also provide an excellent on-axis neutrino flux determination that can be used as an important point of comparison and a systematic crosscheck for the flux as determined by \dword{arcube}. 

Chapter~\ref{ch:exsum-nd} of this \dword{tdr} volume presents a more complete introduction to the \dword{nd} and further details of the system can be found in the appendices. The \dword{dune} collaboration is now in the process of finalizing studies for the \dword{nd} Conceptual Design Report.

\section{DUNE Project Organization and Responsibilities} 
\label{es:ch1:intl-org-resp}

\dword{dune} is the first large-scale science project to be built in the U.S. 
conceived as a fully international collaboration with majority international participation. As such, \dword{dune} requires a new organizational and governance model that takes into account the international nature of the project and its relationship
to \dword{lbnf}.
The model used by \dword{cern} to manage constructing and operating the \dword{lhc} and its experiments served as a starting point for the 
management structure of both  \dword{dune} and \dword{lbnf}, 
and our model continues to evolve as the \dword{dune} project moves forward in concert with \dword{lbnf} to build this experiment and the supporting facilities.  The \dword{dune} project is 
organized by the \dword{dune} collaboration (Section~\ref{sec:exec:collab:org}) with appropriate oversight from all its  international stakeholders. In contrast, 
\dword{lbnf} (Section~\ref{sec:exec:lbnf}) 
is organized as a \dword{doe}-\dword{fnal} project incorporating international partners.

A set  of  organizational structures  has been established  to
coordinate  the  participating  funding agencies,
overseeing the \dword{lbnf} and \dword{dune} projects,
and coordinating and communicating between the 
two. These structures and the relationships among them are shown 
in Figure~\ref{fig:org}. They include 
the following committees\footnote{The \dword{lbnfspac} advises the \dword{fnal} director on \dword{lbnf} only, and thus is not described here. \dword{pip2} is introduced in Section~\ref{sec:exec-overall-det-sppt}.}:
\begin{itemize}
\item International Neutrino Council 

The \dword{inc} is part of the international project governance structure for the  \dword{lbnf} and the  \dword{pip2} projects. The \dword{inc} comprises representatives from the international funding agencies and  \dword{cern} that make major contributions to the infrastructure. 
The \dword{inc} acts as the highest-level international advisory body to the U.S.  \dword{doe} and the  \dword{fnal} directorate on anything related to the program, including coordination among the international partners. The associate director for HEP in the \dword{doe} Office of Science chairs the \dword{inc}, and the \dword{inc}{} includes the  \dword{fnal} director as a member. The council meets once a year and provides pertinent advice on the \dword{lbnf} and \dword{pip2}  projects.

\item Resources Review Board (RRB)

A \dword{rrb} is part of \dword{dune}'s international project governance structure, 
established to coordinate among funding partners and oversee the \dword{dune} project. It includes representatives from all funding agencies that sponsor the project and  from \dword{fnal} management. The  \dword{rrb} provides focused monitoring of the \dword{dune} collaboration and also receives updates on the progress of \dword{lbnf},  \dword{pip2}, and the \dword{sbn} program. The  \dword{rrb} receives periodic reports from both the \dword{lbnc} and \dword{ncg}, described here.  
A representative from the \dword{fnal} directorate chairs the \dword{rrb} and calls regular meetings to monitor progress on the \dword{dune} project.

\item Long-Baseline Neutrino Committee (LBNC)

The \dword{fnal} director has charged the \dword{lbnc} to review the scientific, technical, and managerial progress, as well as plans and decisions associated with the \dword{dune} project. 
The  \dword{lbnc}, comprising internationally prominent scientists with relevant expertise, 
provides regular external scientific peer review of the project. It also provides regular reports and candid assessments to the \dword{fnal} director, which are also made available to the \dword{rrb}, \dword{lbnf}, and \dword{dune} collaboration leadership, as well as the funding agencies that support these international projects. The  \dword{lbnc} reviews the \dwords{tdr} for \dword{dune} and, if acceptable, recommends endorsing the  \dwords{tdr} to the \dword{fnal} director and the \dword{rrb}. Upon request by the \dword{fnal} director, the  \dword{lbnc} may task other \dword{dune} and \dword{lbnf} groups with providing more detailed reports and evaluations of specific systems. The chair of the  \dword{lbnc} participates as a delegate to both the \dword{fnal}-managed \dword{rrb} and the \dword{doe}-managed \dword{inc}. At meetings of the \dword{rrb} and \dword{inc}, the \dword{lbnc} chair reports on  \dword{lbnc} deliberations to the international delegates. The chair of the  \dword{lbnc} is an ex-officio member of  \dword{fnal}'s Physics Advisory Committee.

\item Neutrino Cost Group (NCG)

The \dword{fnal} director has charged the \dword{ncg} to review the cost, schedule, and associated risks of the \dword{dune} project and to provide regular reports to the \dword{fnal} director and the \dword{rrb}. This group comprises internationally prominent scientists with relevant experience. 
The  \dword{ncg} reviews the  \dwords{tdr} for \dword{dune} and provides a recommendation to the \dword{fnal} directorate and the \dword{rrb} on endorsing the  \dwords{tdr}. The chair of the  \dword{ncg} participates as a delegate to both the 
\dword{rrb} and 
\dword{inc}. At meetings of the \dword{rrb} and \dword{inc}, the \dword{ncg} chair 
reports on  \dword{ncg} deliberations to the international delegates.

\item Experiment-Facility Interface Group (EFIG)

Coordination between the \dword{dune} and \dword{lbnf} projects must be close and continuous to ensure the success of the combined enterprise. The \dword{efig} (green box in Figure~\ref{fig:org}) oversees coordination between them, especially during design and construction, but will continue during experiment operations. 
This group examines interfaces between the detectors and their corresponding conventional facilities, between individual detector systems and the \dword{lbnf} infrastructure, and between design and operation of the \dword{lbnf} neutrino beamline,  which may have issues that affect both \dword{lbnf} and \dword{dune}.  

\end{itemize}

\begin{dunefigure}[Structure for oversight of the DUNE and LBNF projects]	
{fig:org}{Top-level organization structure for oversight of the \dword{dune}{} and \dword{lbnf}{} projects, and flowdown.}
\includegraphics[width=0.8\textwidth]{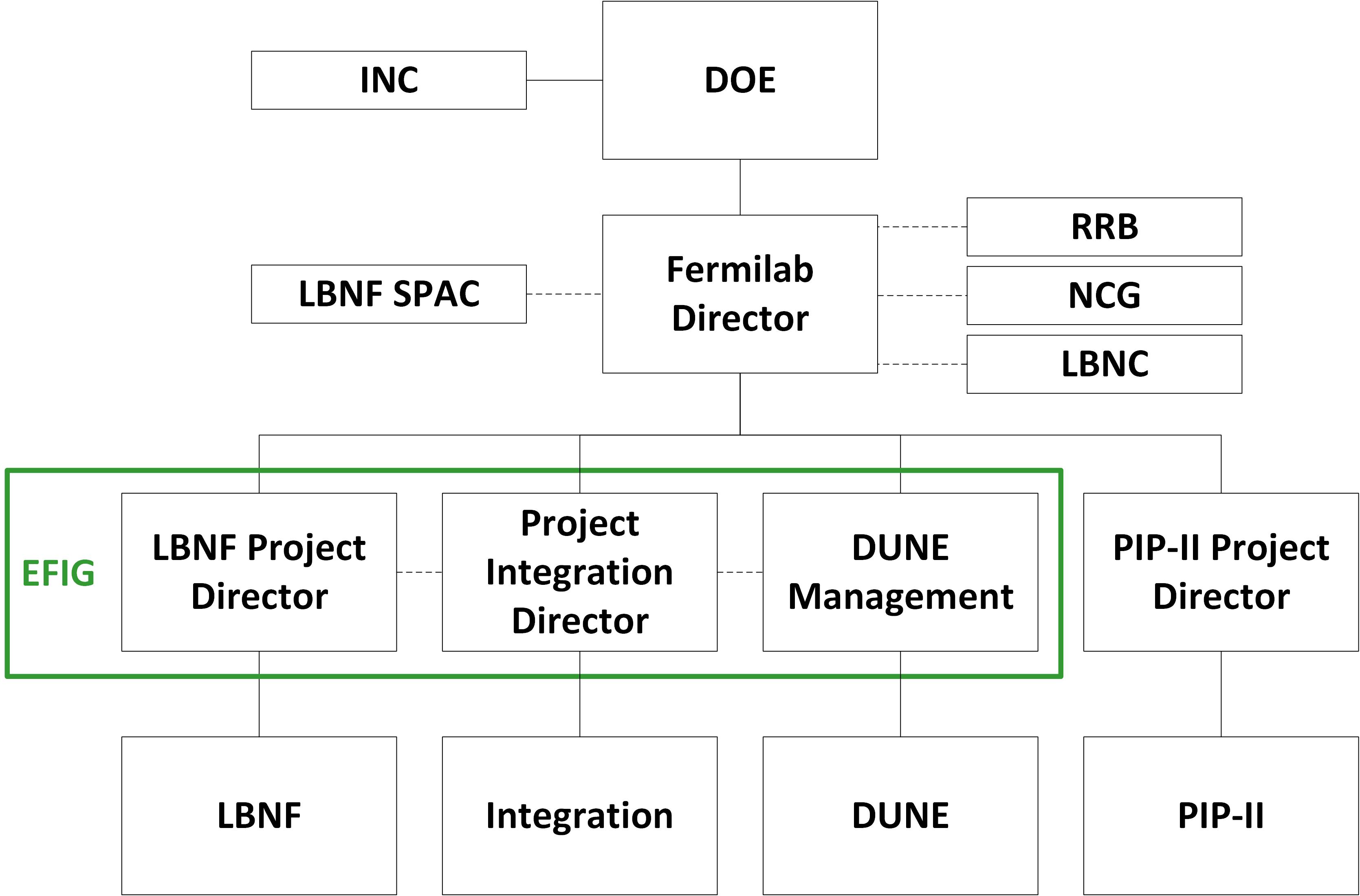}  
\end{dunefigure}

\section{DUNE Collaboration Organization and Management}
\label{sec:exec:collab:org}

The \dword{dune} collaboration organizes and manages \dword{dune} in its entirety.  Stakeholders include all collaborating institutions, 
the funding agencies participating in \dword{dune}, and \dword{fnal} as the host laboratory.  All collaborating institutions have a representative on the \dword{dune} institutional board (IB), which is responsible for establishing the governance rules of the collaboration and regulating governance-related issues. The collaboration is responsible for the design, construction, installation, commissioning, and operation of the detectors and prototypes used to pursue the scientific program. The \dword{dune} \dword{exb}, described below, is the primary management body of the collaboration and approves all significant strategic and technical decisions.

The top-level \dword{dune} collaboration management team consists of two elected co-spokespersons, a \dword{tcoord}, and a \dword{rcoord}. The \dword{tcoord} and \dword{rcoord} are selected jointly by the co-spokespersons and the \dword{fnal} director. The management team is responsible for the day-to-day management of the collaboration and for developing the overall collaboration strategy, which is presented for approval to the \dword{exb}. The \dword{exb} 
comprises the leaders of the main collaboration activities and currently includes 
the top-level management team, institutional board chair, physics coordinator, beam interface coordinator, computing coordinator, \dword{nd} coordinator, and leaders of the \dword{fd} consortia, described below. It is responsible for  ensuring that all stakeholders in the collaboration have a voice in making decisions (see Figure~\ref{fig:eb}). 
Once the \dword{dune} \dword{fd}  \dword{tdr} is accepted, consortium leaders and coordinators of other major collaboration activities will become elected positions.

\begin{dunefigure}[DUNE executive board]	
{fig:eb}{\dword{dune}{} Executive Board.}
\includegraphics[width=1.3\textwidth, angle=90]{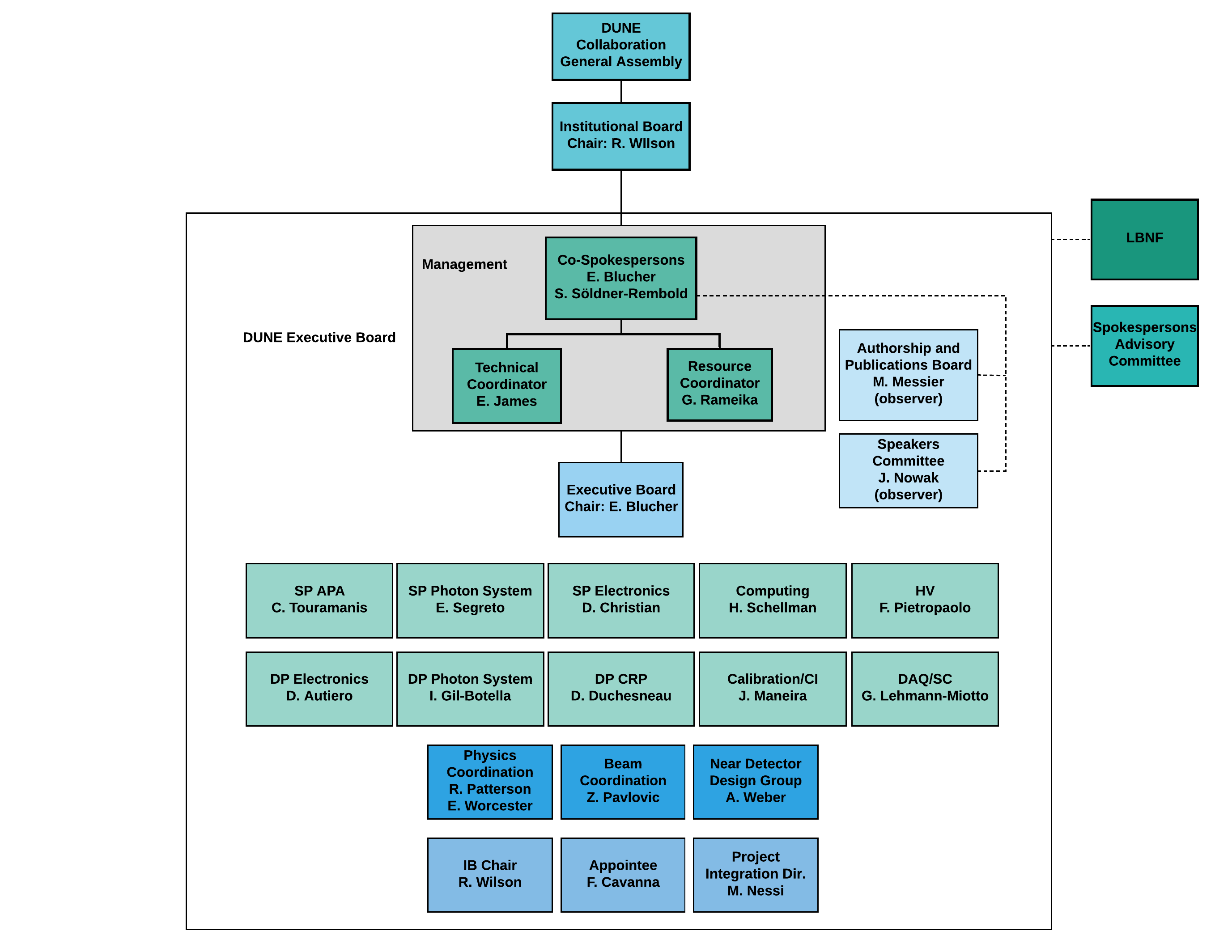}
\end{dunefigure}

To carry out design and construction work for the \dword{dune} \dword{fd}, \dword{dune} has  formed consortia of institutions, each of which is responsible for an individual detector subsystem. A similar structure will be formed for the \dword{nd} once the final detector concept is selected. The \dword{fd} currently includes eleven consortia, including three specific to \dword{sp}, three specific to \dword{dp}, and five common to both technologies:
\begin{itemize}
\item (\single) \dfirsts{apa}, 
\item (\single) \dfirst{tpc} electronics, 
\item (\single) \dfirst{pds}, 
\item (\dual) \dfirsts{crp}, 
\item (\dual) \dfirst{tpc} electronics, 
\item (\dual) \dfirst{pds}, 
\item (common) \dword{hvs}, 
\item (common) \dword{daq},  
\item (common) \dword{cisc}, 
\item (common) calibration,  and 
\item (common) computing.
\end{itemize} 
 Each consortium has an overall leader, a technical lead, and a consortium board with representatives from each participating institution. The consortia have full responsibility for their subsystems and for developing a \dword{wbs}, and are expected to understand and document all interfaces with other systems, prepare final technical designs, and draft their own sections of the \dword{tdr}. Following approval of the  \dword{tdr}, they will be responsible for constructing their respective detector subsystems. 

Chapter~\ref{ch:exec-tc} of this volume 
introduces the \dword{dune} management and organization specifically as it relates to the \dword{fd}; and Volume~\volnumbertc{}, \voltitletc{}, of the \dword{tdr} provides more detail.

\section{Milestones for the First Two Far Detector Modules} 

The plan for construction and commissioning of the first two \dword{fd} modules includes a set of key milestones and dates 
that will be finalized once the international project baseline is established.  Table~\ref{tab:DUNE_schedule_milestones} shows some key dates and milestones (colored rows) and indicates how the detector consortia will add subsystem-specific milestones based on these dates (no background color).
 
\begin{dunetable}
[\dshort{dune} schedule milestones]
{p{0.75\textwidth}p{0.18\textwidth}}
{tab:DUNE_schedule_milestones}
{\dword{dune} schedule milestones for first two far detector modules. Key DUNE dates and milestones, defined for planning purposes in this TDR, are shown in orange.  Dates will be finalized following establishment of the international project baseline.}
Milestone & Date   \\ \toprowrule
Final design reviews  & 2020 \\ \colhline
Start of APA production & August 2020 \\ \colhline
Start photosensor procurement & July 2021 \\ \colhline
Start TPC electronics procurement  & December 2021 \\ \colhline
Production readiness reviews  &  2022    \\ \colhline
\rowcolor{dunepeach} South Dakota Logistics Warehouse available& \sdlwavailable      \\ \colhline
Start of ASIC/FEMB production   & May 2022   \\ \colhline
Start of DAQ server procurement &September 2022    \\ \colhline
\rowcolor{dunepeach} Beneficial occupancy of cavern 1 and \dshort{cuc}& \cucbenocc      \\ \colhline
Finish assembly of initial PD modules (80)      &March 2023    \\ \colhline
\rowcolor{dunepeach} \dshort{cuc} \dshort{daq} room available& \accesscuccountrm      \\ \colhline
Start of DAQ installation&      May 2023   \\ \colhline
Start of FC production for \dshort{detmodule} \#1       &September 2023   \\ \colhline
Start of CPA production for \dshort{detmodule} \#1&     December 2023   \\ \colhline
\rowcolor{dunepeach} Top of \dshort{detmodule} \#1 cryostat accessible& \accesstopfirstcryo      \\ \colhline
Start TPC electronics installation on top of \dshort{detmodule} \#1     & April 2024   \\ \colhline
Start FEMB installation on APAs for \dshort{detmodule} \#1 &    August 2024    \\ \colhline
\rowcolor{dunepeach}Start of \dshort{detmodule} \#1 \dshort{tpc} installation& \startfirsttpcinstall      \\ \colhline
\rowcolor{dunepeach} Top of \dshort{detmodule} \#2 cryostat accessible& \accesstopsecondcryo      \\ \colhline 
Complete FEMB installation on APAs for \dshort{detmodule} \#1   &March 2025    \\ \colhline
End DAQ installation    &May 2025    \\ \colhline
\rowcolor{dunepeach} End of \dshort{detmodule} \#1 \dshort{tpc} installation& \firsttpcinstallend      \\ \colhline 
\rowcolor{dunepeach}Start of \dshort{detmodule} \#2 \dshort{tpc} installation& \startsecondtpcinstall      \\ \colhline
End of FC production for \dshort{detmodule} \#1 &January 2026     \\ \colhline
End of APA production for \dshort{detmodule} \#1        &April 2026    \\ \colhline
\rowcolor{dunepeach} End \dshort{detmodule} \#2 \dshort{tpc} installation& \secondtpcinstallend      \\  \colhline
\rowcolor{dunepeach}Start detector module \#1 operations & July 2026 \\
\end{dunetable}

The schedule for the design and construction of \dword{lbnf} and \dword{dune} has two critical parallel paths: one for the far site (\dword{surf}) 
and another for the 
near site (\dword{fnal}). 
The schedule for initial work is driven by the design and construction of the conventional facilities (\dword{cf}).

During the initial phase of the project, the far site \dword{cf} has been given priority. 
Early far site preparation is complete as of fall 2019, following completion of the  
rehabilitation work on the Ross Shaft that leads to the \dword{dune} underground area, and excavation can now begin.  As each detector 
 cavern is excavated and sufficient utilities are installed, the cryostat construction and cryogenics system installation begins, followed by detector installation, filling with \dword{lar}, and commissioning. 
The first \dword{detmodule} is scheduled to be operational in 2026.

U.S. \dword{doe} project management requires approval at \dword{doecd} milestones before allowing the \dword{lbnf}/\dword{dune} project to move on to the next step. 
\dword{doecd}-1R was granted in 2015, and \dword{doecd}-3A for \dword{lbnf} far site construction was granted in 2016. 
In 2020, \dword{dune} and \dword{lbnf} will seek \dword{doecd}-2/3b and 
 \dword{doecd}-2/3 for the near site. 
The project will conclude with \dword{doecd}-4 approval to start operations.

\cleardoublepage

\chapter{DUNE Physics}
\label{ch:exec-phys}

\textit{This chapter provides a brief introduction to \dshort{dune} physics.  The text below closely follows the presentation in the introductory chapters of Volume~\volnumberphysics{}, 
where many more details may be found.}

\textit{Presented here in summary form are   
(1) the scientific goals and opportunities, 
(2) the methodologies we have 
employed to evaluate the capabilities of \dshort{dune} to realize 
the science, and
(3) the corresponding results for selected program elements.}  


\section{Goals of the DUNE Science Program}
\label{sec:exec-phys-key-goals}

The primary goals and ancillary science program elements listed in the previous 
chapter represent discovery opportunities at the forefront of particle physics and 
astrophysics.  The \dword{dune} has been designed to capitalize on these opportunities with 
a unique set of experimental conditions and capabilities.  In this section we 
elaborate on elements of the science program that motivate the operating principles   
%
of \dword{dune} 
that were presented in the previous chapter. 

The focus of the presentation here is on physics opportunities offered 
by the \dword{dune} \dword{fd}.
The \dword{dune} \dword{nd} will have its own physics program, only a small portion 
of which is covered here.  
The full \dword{nd} physics program will be described in the \dword{nd} \dword{cdr}, 
which is in progress as of this writing.

\subsection{Neutrino Oscillations: Masses, Mixing Angles and CP Violation}

Neutrino oscillations imply nonzero neutrino masses and flavor-mixing 
in the leptonic \dword{cc} interactions.  The nonzero neutrino mass is among the most important discoveries in fundamental 
particle physics of the twenty-first century. Understanding the mechanism 
behind nonzero neutrino masses is among the unresolved mysteries that 
drive particle physics today; they remain one of the few unambiguous 
facts that point to the existence of new particles and interactions, 
beyond those that make up the remarkable standard model of 
particle physics.

Almost all neutrino data can be understood within the three-flavor paradigm 
with massive neutrinos,
the simplest extension of the standard model capable of reconciling 
theory with observations.  
It consists of introducing distinct, nonzero, masses for at least two 
neutrinos, while maintaining the remainder of the standard model. Hence, neutrinos 
interact only via the standard model \dword{cc} and \dword{nc}  weak 
interactions. The neutrino mass eigenstates -- defined as $\nu_1,\nu_2, \nu_3$ 
with masses, $m_1, m_2, m_3$, respectively -- are distinct from the neutrino 
\dword{cc} interaction eigenstates, also referred to as the flavor 
eigenstates -- $\nu_e$, $\nu_{\mu}$, $\nu_{\tau}$, labeled according 
to the respective charged-lepton $e,\mu,\tau$ to which they couple in 
the \dword{cc} weak interaction. The flavor eigenstates can be expressed 
as linear combinations of the mass eigenstates: 
the coefficients of the respective linear combinations define a 
unitary $3\times 3$ mixing matrix, the \dword{pmns}
 matrix, as follows:
\begin{equation}
\left(\begin{array}{c} \nu_e \\ \nu_{\mu} \\ \nu_{\tau} \end{array}\right) = \left(\begin{array}{ccc} U_{e1} & U_{e2} & U_{e3} \\  U_{\mu1} & U_{\mu2} & U_{\mu3}  \\  U_{\tau1} & U_{\tau2} &  U_{\tau3}  \end{array}\right) \left(\begin{array}{c} \nu_1 \\ \nu_2 \\ \nu_3 \end{array}\right).
\end{equation}

Nonzero values for at least some of the off-diagonal elements, coupled with nonzero 
differences in the masses of $\nu_1$, $\nu_2$ and $\nu_3$, lead to the phenomenon of 
neutrino oscillations, in which a neutrino -- 
produced in a  
flavor eigenstate -- acquires an oscillating probability 
of interacting as a different flavor (with an oscillation frequency proportional to 
the differences of the squares of the neutrino masses, $\Delta m^2_{ij}\equiv m_i^2-m_j^2$). 

The \dword{pmns} matrix is the leptonic-equivalent of the \dword{ckm} 
that describes the \dword{cc} interactions of quark mass 
eigenstates. If the neutrinos are Dirac fermions, the \dword{pmns} matrix, 
like the \dword{ckm}, can be unambiguously parameterized with three mixing 
angles and one complex phase.\footnote{Additional nontrivial phases are present 
if neutrinos are Majorana fermions, but these do not affect oscillations at 
an observable level.} By convention~\cite{Tanabashi:2018oca}, the mixing angles 
are denoted $\theta_{12}$, $\theta_{13}$, and $\theta_{23}$, defined as
\begin{eqnarray}
\sin^2\theta_{12} &\equiv& \frac{|U_{e2}|^2}{1-|U_{e3}|^2}, \\
\sin^2\theta_{23} &\equiv& \frac{|U_{\mu3}|^2}{1-|U_{e3}|^2}, \\
\sin^2\theta_{13} &\equiv& |U_{e3}|^2,
\end{eqnarray} 
and one phase $\deltacp$, which in the conventions of~\cite{Tanabashi:2018oca}, is given by
\begin{equation}
\deltacp \equiv -{\rm arg}(U_{e3}).
\end{equation}
For values of $\deltacp\neq 0,\pi$, and assuming none of the $U_{\alpha i}$ 
vanish ($\alpha=e,\mu,\tau$, $i=1,2,3$), the neutrino mixing matrix is complex 
and \dword{cp}-invariance is violated in the lepton sector. This, in turn, manifests 
itself as different oscillation probabilities, in vacuum, for neutrinos 
and antineutrinos: $P(\nu_{\alpha}\to\nu_{\beta})\neq 
P(\bar{\nu}_{\alpha}\to\bar{\nu}_{\beta})$, $\alpha,\beta=e,\mu,\tau$, $\alpha\neq\beta$.

The central aim of the worldwide program of 
neutrino experiments past, present and planned, is to explore 
the phenomenology of neutrino oscillations in the context of the 
three-flavor paradigm, and, critically, 
to challenge its validity with measurements at progressively 
finer levels of precision.  The world's neutrino data significantly constrain 
all of the oscillation parameters in the three-flavor paradigm, but 
with precision that varies considerably from one parameter to the next.

Critical questions remain open. The neutrino mass ordering -- whether $\nu_3$ is 
the heaviest (``normal'' ordering) or the lightest (``inverted'' ordering) -- is unknown. 
Current data prefer the normal ordering, but the inverted one still 
provides a decent fit to the data. The angle $\theta_{23}$ is known to be 
close to the maximal-mixing value of $\pi/4$, but assuming it is not exactly 
so, the octant 
(whether $\sin^2\theta_{23}<0.5$ [$\theta_{23}<\pi/4$] or 
$\sin^2\theta_{23}>0.5$ [$\theta_{23}>\pi/4$]) is also unknown. 
The value of \deltacp is only poorly constrained. 
While positive values of $\sin\deltacp$ are disfavored, all $\deltacp$ values between $\pi$ and $2\pi$, including the \dword{cp}-conserving values $\deltacp=0,\pi$, are consistent with the world's neutrino data.\footnote{It should be noted that recent results from the \dword{t2k} experiment~\cite{Abe:2019vii} show only marginal consistency with \dword{cp}-conserving values of $\deltacp$.}
That the best fit to the world's data 
favors large \dword{cpv} is intriguing, providing further impetus 
for experimental input to resolve this particular question.
It is central to the \dword{dune} mission that all of the questions 
posed here can be addressed 
by neutrino oscillation experiments.

Conventional horn-focused beams, where either \numu or \anumu 
is the dominant species (depending on horn current polarity), provide 
access to these questions for experiments at long baselines as in the case 
of \dword{dune} and the \dword{lbnf}.  By virtue 
of the near-maximal value of $\theta_{23}$, oscillations are mainly in the  
mode $\nu_\mu \rightarrow \nu_\tau$.  For realizable baselines, 
this channel is best studied by 
measuring the $\nu_\mu$ disappearance probability as a function of 
neutrino energy rather than through direct observation of $\nu_\tau$ 
appearance.  This is because oscillation maxima occur at energies 
below the threshold for $\tau$-lepton production in $\nu_\tau$ 
\dword{cc} interactions in the detector.
On the other hand, the sub-dominant $\nu_\mu \rightarrow \nu_e$ channel is 
amenable to detailed study through the energy dependence of the
$\nu_e$ and $\bar\nu_e$ appearance probabilities, 
which is directly sensitive (in a rather complex way) to 
multiple \dword{pmns} matrix parameters, as described below.  

Specifically, the oscillation probability of \numu $\rightarrow$ \nue 
through matter in a constant density approximation is,  
to first order~\cite{Nunokawa:2007qh}:
\begin{eqnarray}
P(\nu_\mu \rightarrow \nu_e) & \simeq & \sin^2 \theta_{23} \sin^2 2 \theta_{13} 
\frac{ \sin^2(\Delta_{31} - aL)}{(\Delta_{31}-aL)^2} \Delta_{31}^2 \nonumber \\
& & + \sin 2 \theta_{23} \sin 2 \theta_{13} \sin 2 \theta_{12} \frac{ \sin(\Delta_{31} - aL)}{(\Delta_{31}-aL)} \Delta_{31} \frac{\sin(aL)}{(aL)} \Delta_{21} \cos (\Delta_{31} + \mdeltacp) \nonumber\\
& & + \cos^2 \theta_{23} \sin^2 2 \theta_{12} \frac {\sin^2(aL)}{(aL)^2} \Delta_{21}^2, 
\label{eqn:appprob}
\end{eqnarray}
where $\Delta_{ij} = \Delta m^2_{ij} L/4E_\nu$, $a = G_FN_e/\sqrt{2}$, $G_F$ is the Fermi constant, $N_e$ is the number density of electrons in the Earth, $L$ is the baseline in km, and $E_\nu$ is the neutrino energy in GeV. 
In the equation above, both \deltacp and $a$ 
switch signs in going from the
$\nu_\mu \to \nu_e$ to the $\bar{\nu}_\mu \to \bar{\nu}_e$ channel; i.e.,
a neutrino-antineutrino asymmetry is introduced both by \dword{cpv} (\deltacp)
and the matter effect ($a$). As is evident from Equation~\ref{eqn:appprob}, 
the matter effect introduces a sensitivity to the sign of $\Delta_{31}$, 
which specifies the neutrino mass ordering.
The origin of the matter effect asymmetry 
is simply the presence of electrons and absence of positrons in the Earth.  
In the few-GeV energy range, the asymmetry from the matter effect increases 
with baseline as the neutrinos
pass through more matter; therefore an experiment with a longer baseline will be
more sensitive to the neutrino mass ordering. For baselines longer than 
$\sim$\SI{1200}\km, the degeneracy between the asymmetries from matter
and \dword{cpv} effects can be resolved~\cite{Bass:2013vcg}.

The electron neutrino appearance probability, $P(\nu_\mu \rightarrow \nu_e)$, 
is plotted in 
Figure~\ref{fig:oscprob} at a baseline of \SI{1300}\km{} as a function of neutrino 
energy for several values of \deltacp. As this figure illustrates, the value 
of \deltacp affects both the amplitude and phase of
the oscillation. The difference in probability amplitude
for different values of \deltacp is larger at higher oscillation nodes, which 
correspond to energies less than 1.5~GeV. Therefore, a broadband experiment, 
capable of measuring not only the rate of \nue appearance but of mapping out the 
spectrum of observed oscillations down to energies of at least 500~MeV, is desirable. 

\begin{dunefigure}[Appearance probabilities for \nue and \anue at \SI{1300}{\km}]{fig:oscprob}{The appearance probability at a baseline of \SI{1300}\km{},
  as a function of neutrino energy, for \deltacp = $-\pi/2$ (blue), 
  0 (red), and $\pi/2$ (green), for neutrinos (left) and antineutrinos
  (right), for normal ordering. The black line indicates the oscillation
  probability if $\theta_{13}$ were equal to zero. Note that the \dword{dune} \dword{fd} will be built at a baseline of \SI{1300}{\km}.}
\includegraphics[width=0.45\linewidth]{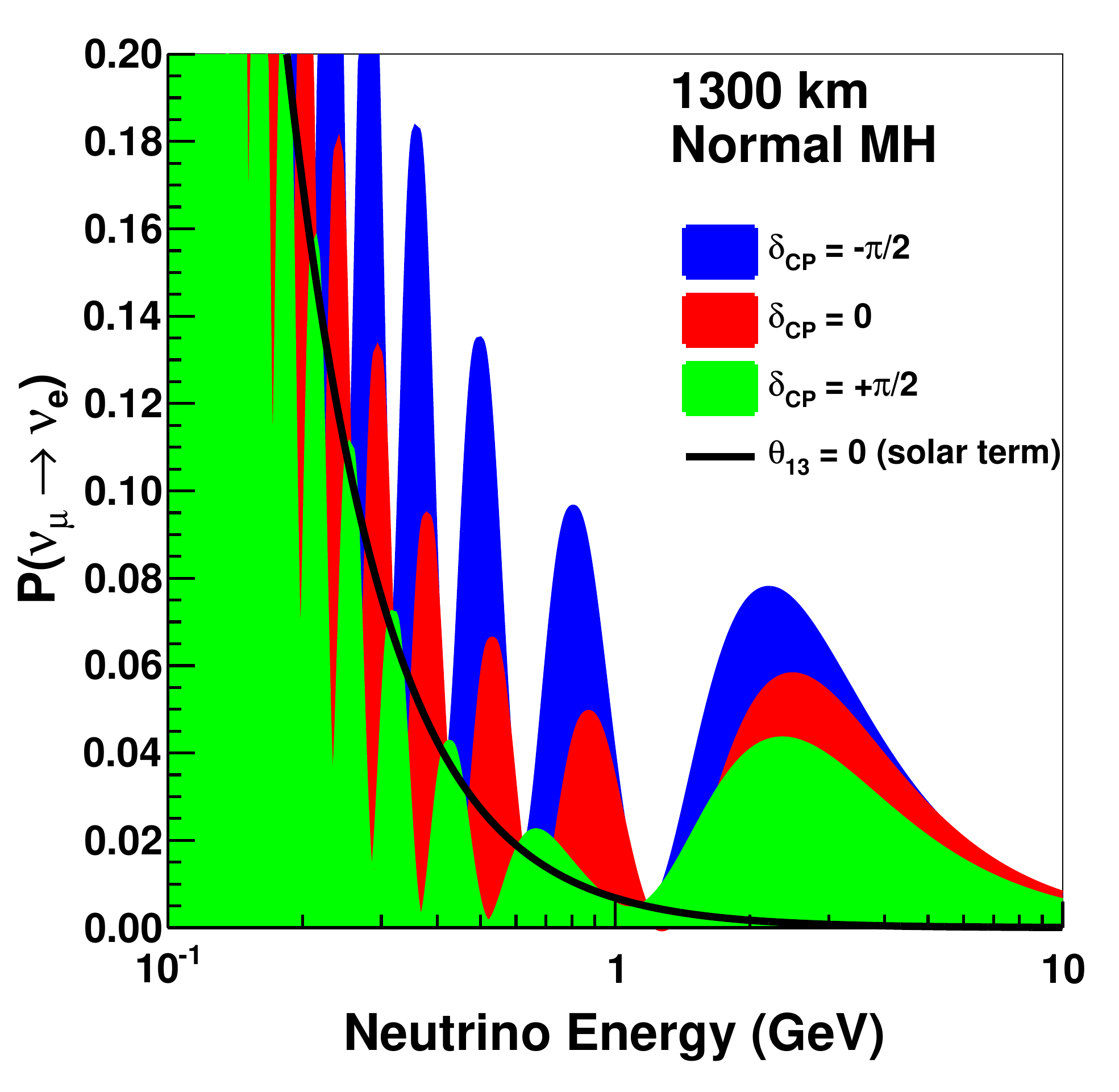}
\includegraphics[width=0.45\linewidth]{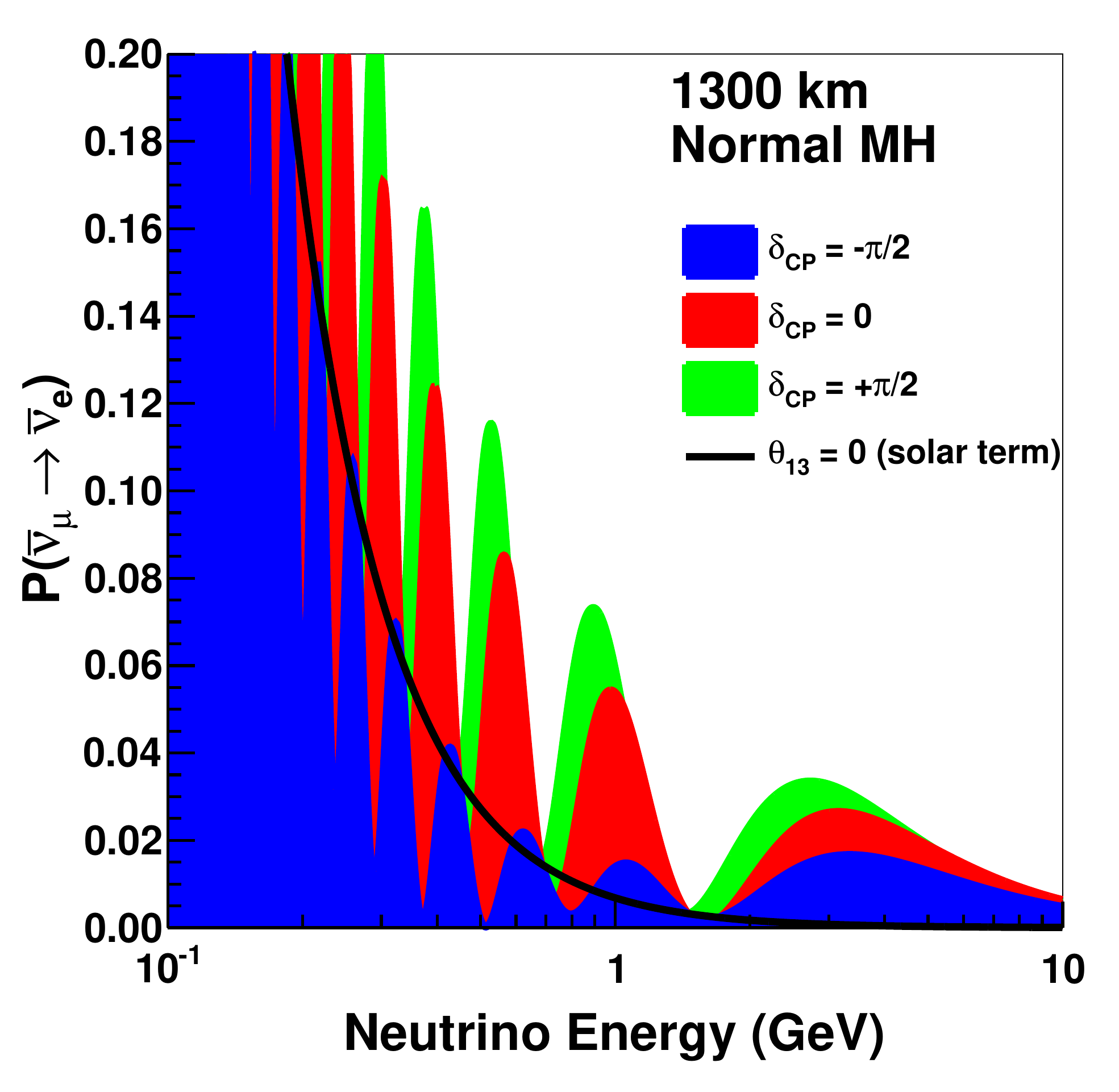}
\end{dunefigure}

\dword{dune} is designed to address the questions articulated above, 
to over-constrain the three-flavor paradigm, 
and to reveal what may potentially lie beyond.  
Even if consistency is found, the precision measurements 
obtained by \dword{dune} will have profound implications. As just one example, 
the discovery of \dword{cpv} in neutrino oscillations would provide 
strong circumstantial evidence for the leptogenesis mechanism as 
the origin of the baryon asymmetry of the universe.  

Going further, the patterns defined by the fermion masses and 
mixing parameters have been the subject of intense theoretical 
activity for the last several decades.  \Dwords{gut} 
posit that quarks and leptons are different manifestations of the same 
fundamental entities, and thus their masses and mixing parameters 
are related. Different models make different predictions but, 
in order to compare different possibilities, it is important that 
lepton mixing parameters be known as precisely as quark mixing parameters.
To enable equal-footing comparisons between quark and lepton mixing 
it is required 
that the mixing angles be determined at the few percent level 
while $\deltacp$ should be measured at the 10\% level or better.
Measurements with precision at these levels are expected from \dword{dune} 
for the mixing angles $\theta_{23}$ and $\theta_{13}$, 
and the \dword{cp} phase $\deltacp$.   
These measurements will thus open a new era of flavor physics, 
with the potential to offer insight on deep questions on which the 
\dword{sm} 
 is essentially silent.

\subsection{Baryon Number Violation}

Are protons 
stable? Few questions within elementary 
particle physics can be posed as simply and at the same time 
have implications as immediate.  The 
apparent stability of protons suggests that baryon number 
is conserved in nature, although no known symmetry 
requires it to be so.  Indeed, baryon number conservation is 
implicit in the formulation of the \dword{sm} Lagrangian, and 
thus observation of \dword{bnv} processes such 
as nucleon decay or neutron-antineutron oscillation 
would be evidence for physics beyond the \dword{sm}.
On the other hand, continued non-observation of \dword{bnv} processes will 
demand an answer to what new symmetry is at play that forbids 
them.
 
Especially compelling is that the observation of \dword{bnv} processes 
could be the harbinger for \dwords{gut}, in which strong, weak and 
electromagnetic forces are unified.  Numerous \dword{gut} models 
have been proposed, each with distinct features.  Yet, \dword{bnv} processes 
are expected on general grounds, and it is a feature of many models 
that nucleon decay channels can proceed at experimentally 
accessible rates.  This is illustrated for several key nucleon 
decay channels relevant for \dword{dune} in 
Figure~\ref{fig:theoryexplimitsummary}, along with existing 
experimental limits.

\begin{dunefigure}[Summary of nucleon decay experimental limits and model predictions]{fig:theoryexplimitsummary}{Summary of nucleon decay experimental lifetime limits from past or currently running experiments for several modes, and the model predictions for the lifetimes in the two modes \ptoepizero and \ptoknubar.  The limits shown are 90\% \dword{cl} lower limits on the partial lifetimes, $\tau/B$, where $\tau$ is the total mean life and $B$ is the branching fraction. Updated from~\cite{Babu:2013jba}.}
\includegraphics[width=0.9\textwidth]{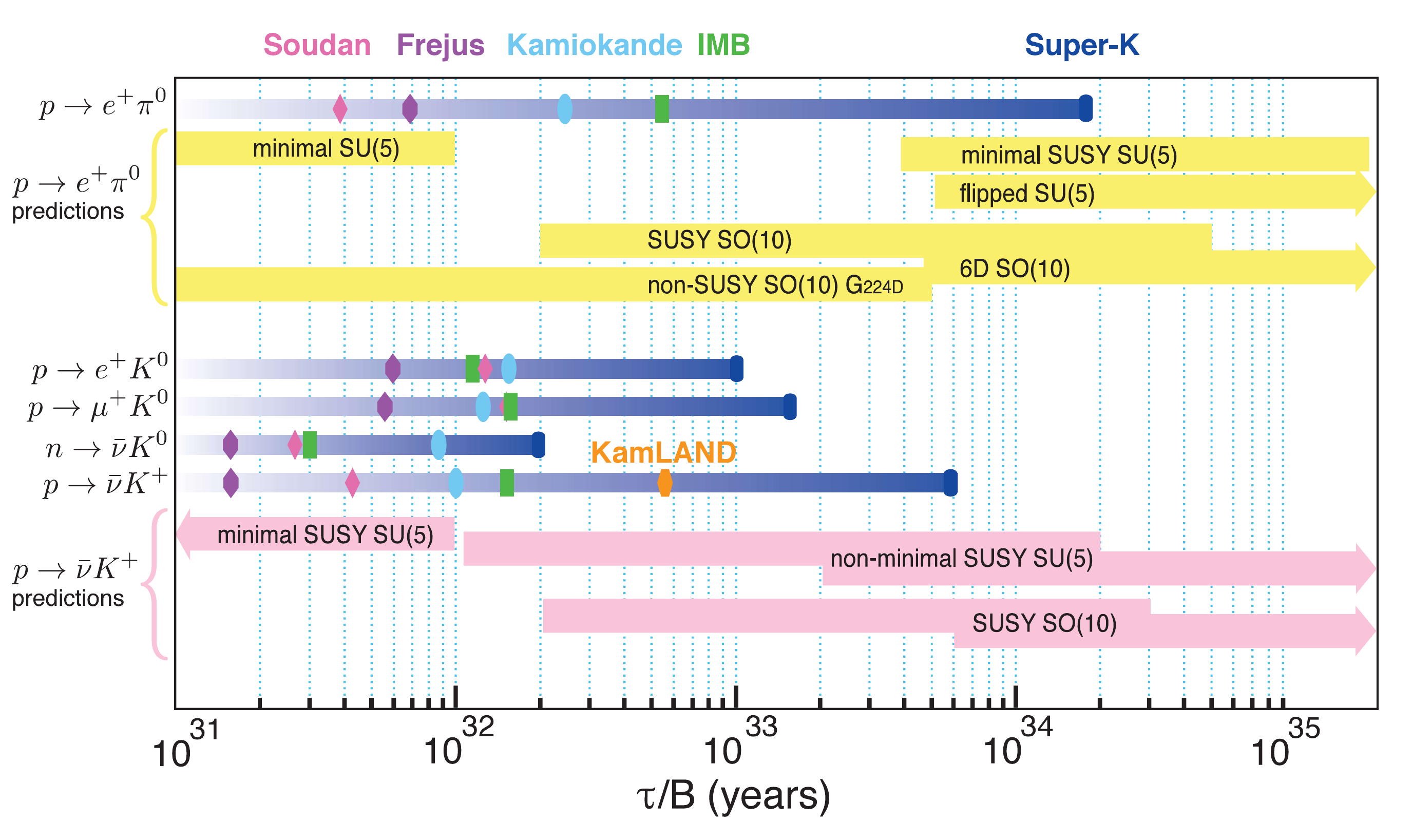}
\end{dunefigure}

Given the scale of energy deposition in the few hundred \si{\MeV} 
to few \si{\GeV} range, 
a detector optimized for neutrino oscillation physics at long 
baselines is naturally well suited for sensitive searches for 
nucleon decay and neutron-antineutron oscillations.
Thanks to the excellent imaging, calorimetric and particle 
identification capabilities of the \dword{lartpc}, 
backgrounds can in principle be reduced below the single-event 
level for key nucleon decay channels at exposures where other 
detector technologies are no longer background-free.  On the 
other hand, a challenge presented by an argon-based 
detector is the impact of \dword{fsi} 
on 
nucleon decay event reconstruction, which is 
expected to be more severe than for detectors based on water 
or liquid scintillator, for example.  
On balance, however, should nucleon decays occur at rates 
not far beyond current best limits, a handful of candidate 
events could be observed by \dword{dune} in a given decay mode.  
%
For the case of zero observed candidates, \dword{dune} has the  
potential to set partial lifetime limits for favorable channels at the 
\SI{e34}{year} level, as indicated in Section~\ref{sec:es:phys:pdecay}.

\subsection{Supernova Neutrino Bursts}

The burst of neutrinos~\cite{Bionta:1987qt,Hirata:1987hu} 
from the celebrated core-collapse supernova 
1987A in the Large Magellanic Cloud, about
\SI{50}{kiloparsecs} (kpc) from Earth, heralded the era of extragalactic neutrino
astronomy.  The few dozen recorded $\bar{\nu}_e$ events
have confirmed the basic physical
picture of core collapse and yielded constraints on a wide range of new
physics~\cite{Schramm:1990pf, Vissani:2014doa}.  

Core-collapse supernovae within a few hundred kpc of Earth --
within our own galaxy and nearby -- are quite rare on a human
timescale.  They are expected once every few decades in the Milky Way
(within about 20~kpc), and with a similar rate in Andromeda, about
700~kpc away.  These core collapses should be common enough to have
a reasonable chance of occurring during the few-decade long lifetime
of a typical large-scale neutrino detector. 

It is important that at least one module of the \dword{dune} \dword{fd} be online at all times to
observe this unpredictable and spectacular event, if and when it occurs.   
The rarity of these events makes it all the more critical for the community to
be prepared to capture every last bit of information from them.

The information in a \dword{snb} available in principle
to be gathered by experimentalists is the \textit{flavor, energy and
  time structure} of a several-tens-of-second-long, all-flavor,
few-tens-of-MeV neutrino burst~\cite{Mirizzi:2015eza, Horiuchi:2017sku}.  Imprinted on
the neutrino spectrum as a function of time is information about the
progenitor, the collapse, the explosion, and the remnant, as well as
information about neutrino parameters and potentially exotic new
physics.  Neutrino energies and flavor content of the burst can be
measured only imperfectly, due to the intrinsic nature of the weak
interactions of neutrinos with matter, as well as due to imperfect
detection resolution in any real detector.  For example, \dword{snb} 
 energies are below \dword{cc} threshold for $\nu_\mu$,
$\nu_\tau$, $\bar{\nu}_\mu$, and $\bar{\nu}_{\tau}$ (collectively
$\nu_x$), which represent two-thirds of the flux; so these flavors are
accessible only via \dword{nc}  interactions, which tend to have
low cross sections and indistinct detector signatures. 
These issues make a
comprehensive unfolding of neutrino flavor, time and energy structure
from the observed interactions a challenging problem.

The core-collapse neutrino signal starts with a short, sharp
\emph{neutronization} (or \emph{break-out}) burst primarily composed of
$\nu_e$. These neutrinos are messengers of the shock front breaking through the neutrinosphere (the surface of neutrino trapping): when this happens, iron is disintegrated, the neutrino scattering cross section drops, and the lepton number trapped just below the original neutrinosphere is suddenly released. This quick and intense burst is followed by an
\emph{accretion phase} lasting some hundreds of milliseconds, depending on the progenitor star mass, as matter falls onto the collapsed core and the shock is stalled at the distance of perhaps $\sim 200$ km. The gravitational binding energy of the accreting material is powering the neutrino luminosity during this stage. The later
\emph{cooling phase} over $\sim$10~seconds represents the main part of
the signal, over which the proto-neutron star sheds its trapped energy.  

The flavor content and spectra of the neutrinos emitted from the neutrinosphere change
throughout these phases, and the supernova's evolution can
be followed with the neutrino signal. 
Some fairly generic features of these emitted neutrino fluxes are
illustrated in Figure~\ref{params}.

\begin{dunefigure}[Expected time-dependent neutrino burst 
characteristics for a core-collapse 
supernova]{params}{Expected
  time-dependent signal for a specific flux model for an
  electron-capture supernova~\cite{Huedepohl:2009wh} at 10~kpc.  No oscillations are assumed; the effect of neutrino flavor transitions under different mass ordering assumptions can be dramatic and is described in Volume~\volnumberphysics{} Chapter~7. 
  Note that $\nu_x$  refers to $\nu_\mu$,
$\nu_\tau$, $\bar{\nu}_\mu$, and $\bar{\nu}_{\tau}$ collectively. The
  top plot shows the luminosity as a function of time ($\nu_x$ is the sum of all, the second plot
  shows average neutrino energy, and the third plot shows the $\alpha$
  (pinching) parameter.  The vertical dashed line at 0.02 seconds indicates
  the time of core bounce, and the vertical lines indicate different
  eras in the supernova evolution.  The leftmost time interval
  indicates the infall period.  The next interval, from core bounce to
  50~ms, is the neutronization burst era, in which the flux is
  composed primarily of $\nu_e$.  The next period, from 50 to 200~ms,
  is the accretion period. The final era, from 0.2 to 9~seconds, is
  the proto-neutron-star cooling period.  The general features are
  qualitatively similar for most core-collapse supernovae.}
\includegraphics[width=0.9\textwidth]{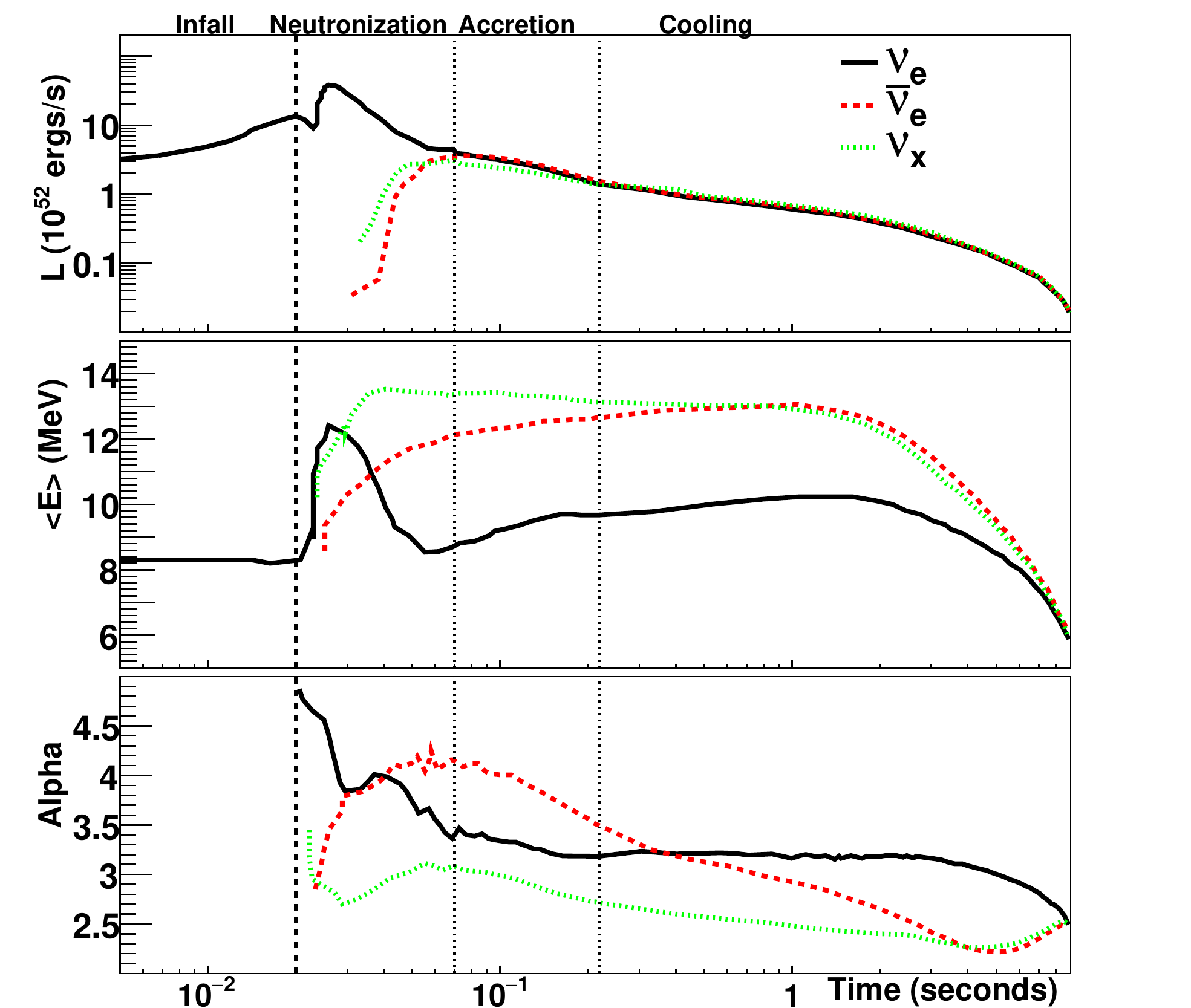}
\end{dunefigure}

In the world's current supernova neutrino flavor sensitivity
portfolio~\cite{Scholberg:2012id, Mirizzi:2015eza}, the sensitivity 
is primarily to electron antineutrino flavor, via inverse beta decay.  
There is only minor sensitivity to the $\nu_e$
component of the flux, which carries with it particularly interesting 
information content of the burst (e.g., neutronization burst neutrinos
are created primarily as $\nu_e$).  While there is some $\nu_e$
sensitivity in existing and other planned detectors via elastic 
scattering on electrons and via subdominant channels on nuclei, 
statistics are relatively small,
and it can be difficult to disentangle the flavor content.  It is 
in this respect that an experiment with an argon target such as \dword{dune} 
will be especially valuable, since the dominant process in 
this case is $\nu_e$ \dword{cc} scattering.

\subsection{Additional Beyond-Standard-Model Physics Signatures}

The capabilities that enable access to the physics program described 
in the previous sections open a myriad of opportunities to search for 
evidence of physics beyond the standard model.  Below we list the 
identified opportunities that we have investigated.  
Projected sensitivities are shown later in this chapter for only 
a few of these opportunities; we refer the reader to  
Volume~\volnumberphysics{}, \voltitlephysics{}, for a more complete demonstration of 
the potential impact of \dword{dune}'s searches for \dword{bsm} phenomena,  
which can justifiably be considered as prominent ancillary 
elements of the \dword{dune} science program.  At the same time, it is 
important to note that new physics may appear in ways that have 
not yet been considered: history has repeatedly shown that nature 
can reward new experimental approaches and sensitive detectors 
with the appearance of entirely unanticipated phenomena.

Opportunities in \dword{bsm} physics that have been considered 
as elements of the \dword{dune} science program include:

\textit{Search for active-sterile neutrino mixing:} \dword{dune} is sensitive over a broad range of potential sterile neutrino mass splittings by looking for disappearance of \dword{cc} and \dword{nc}  interactions over the long distance separating the \dword{nd} and \dword{fd}, as well as over the short baseline of the \dword{nd}. With a longer baseline, a more intense beam, and a high-resolution large-mass \dword{fd}, compared to previous experiments, \dword{dune} provides a unique opportunity to improve significantly on the sensitivities of the existing probes, and greatly enhance the ability to map the extended parameter space if a sterile neutrino is discovered.

\textit{Searches for non-unitarity of the \dword{pmns} matrix:} Deviation from 
unitarity of the $3 \times 3$ \dword{pmns} matrix due to extra heavy 
neutrino states may be observable. 
Parameters characterizing the degree 
of non-unitarity can
become sizable as the masses of the new states decrease.

\textit{Searches for \dwords{nsi}:} 
\Dwords{nsi} affecting neutrino propagation through the Earth can significantly modify the data to be collected by \dword{dune} as long as the new physics parameters are large enough.  If the \dword{dune} data are consistent with standard oscillations for three massive neutrinos, interaction effects of order 0.1 $G_{F}$ can be ruled out at \dword{dune}.

\textit{Searches for violation of Lorentz symmetry or \dword{cpt}:}  \dword{dune} can improve the present limits on Lorentz and \dword{cpt} violation in the neutrino sector by several orders of magnitude, contributing an important experimental test of these fundamental assumptions underlying quantum field theory.

\textit{Studies of neutrino trident production:}  
Interactions of neutrinos with the Coulomb field of a 
nucleus can lead to final states with a lepton-pair 
accompanying the lepton from the neutrino interaction 
vertex. 
With a predicted annual rate of over 100 dimuon 
neutrino trident interactions at the \dword{nd}, \dword{dune} will be 
able to measure deviations from the \dword{sm} rates and test 
the presence of new gauge symmetries.

\textit{Search for \dword{ldm}:}  The lack of evidence for \dword{wimp}  \dword{dm}  
candidates from direct detection and \dword{lhc} experiments has resulted in a reconsideration of the \dword{wimp} paradigm, and has revitalized the effort to search for \dword{ldm}  candidates 
of around a GeV or below in mass. High-flux neutrino beam experiments, such as \dword{dune}, have been shown to provide coverage of \dword{dm}+mediator parameter space that cannot be covered by either direct detection or collider experiments. \dword{dm} particles can be detected in the \dword{nd} through \dword{nc}-like interactions either with electrons or nucleons in the detector material and enable \dword{dune}'s search for \dword{ldm} to be competitive and complementary to other experiments.

\textit{Search for \dword{bdm}:} Using its large \dword{fd}, \dword{dune} will be able to search for  \dword{bdm}. In these models there are several  \dword{dm} particles with different masses and
properties concerning their interactions with \dword{sm} particles. \dword{dune} will search for such particles 
as generally produced anywhere in the cosmos or specifically through annihilation 
in the core of the sun, allowing competitive results for both production scenarios

\textit{Search for heavy neutral leptons (HNLs):} HNLs in the context of the $\nu$MSM model with 
masses less than 2 GeV can be produced in the beam-dump of the proton beam to generate  the 
\dword{dune} neutrino beam.  The \dword{nd} data can be used to search for HNL decays, and competitive
results with other present and proposed facilities can be obtained.

\section{Summary of Assumptions and Methods Employed}
\label{sec:exec-phys-assm-meth}

Scientific capabilities are determined assuming that \dword{dune}
is configured according to the general parameters described 
in Section~\ref{exec:strategy-fd-des}. 
Further assumptions regarding the neutrino beam and detector 
systems, and their deployment are stated here in
Sections~\ref{sec:exec-phys-assm-meth-beamdetector} and
\ref{sec:exec-phys-assm-meth-deployment}. 

Determination of experimental sensitivities relies on the
modeling of the underlying physics and background processes,
as well as the detector response, including calibration and
event reconstruction performance and the utilization of data
analysis techniques and tools.
Section~\ref{sec:exec-phys-assm-meth-simreco} gives a brief discussion of the strategies employed. 

\subsection{Beam and Detector}
\label{sec:exec-phys-assm-meth-beamdetector}

Physics sensitivities are based on 
the optimized design of a \SI{1.2}{MW} neutrino beam and
corresponding protons-on-target per year assumed to
be $1.1 \times 10^{21}$ POT.  These numbers assume a combined
uptime and efficiency of the \dword{fnal} accelerator complex and the
\dword{lbnf} beamline of 56\%.\footnote{This projection, from which one  
year of \dword{lbnf} beam operations (live time) can be expressed as \SI{1.7e7}{seconds}, 
is based on extensive 
experience with intense neutrino beams at \dword{fnal}, and in particular 
the \dword{numi} beam line, which incorporates elements like those in the  
proposed \dword{lbnf} beamline design and faces similar operating conditions.} 

For the neutrino oscillation physics program, it is assumed that
equal exposures (time-integrated beam power times fiducial mass) are obtained with both horn current polarities,
and therefore with the corresponding mix of primarily \numu
and \anumu data samples.

It is assumed that the \dword{dune} \dword{fd} will include some
combination of the different \nominalmodsize fiducial volume
implementations -- \dword{sp} or \dword{dp} -- of the \dword{lartpc} concept
for which technical designs have been developed.                                                     
For much of the science program it is expected that the
capabilities of the two proposed \dword{fd} module 
implementations will be comparable.  As a result of the
current state of reconstruction and analysis software development
(see Section~\ref{sec:exec-phys-assm-meth-simreco}), the
physics sensitivity studies reported in this \dword{tdr} are based on
the \dword{sp} \dword{lartpc} implementation,
documented in full in Volume~\volnumbersp{}. 

It is also assumed that validation of the \dword{dune} \dword{fd} 
designs will come from data and operational experience acquired 
with the large-scale \dword{protodune} detectors staged at the \dword{cern}, 
including single-particle studies of data obtained 
in test-beam running.  

The \dword{nd} for \dword{dune} has been under active development,
and a \dword{cdr} is in preparation.
Correspondingly, the descriptions 
in this \dword{tdr}
are consistent with this level of development.  

\subsection{Deployment Scenario}
\label{sec:exec-phys-assm-meth-deployment}

Where presented as a function of calendar year,
sensitivities are calculated with the following
assumed deployment plan, which is based on a
technically limited schedule:
\begin{itemize}
    \item Start of beam run: Two \dword{fd} module 
    volumes for total fiducial mass of \SI{20}{kt}, \SI{1.2}{MW} beam;
    \item After one year: Add one \dword{fd} module  volume for total fiducial mass of \SI{30}{kt};
    \item After three years: Add one \dword{fd} module  volume for total fiducial mass of \fdfiducialmass{}; and
    \item After six years: Upgrade to \SI{2.4}{MW} beam.
\end{itemize}

\subsection{Simulation, Reconstruction, and Data Analysis Tools}
\label{sec:exec-phys-assm-meth-simreco}

The development of algorithms and software infrastructure needed
to carry out physics sensitivity studies has been an active 
effort within \dword{dune} and the associated scientific community.  
Significant progress has been made: event reconstruction 
codes can be run on fully simulated neutrino interaction events 
in \dword{dune} \dword{fd} modules; the \dword{dune} computing infrastructure 
allows high-statistics production runs; and end-user interfaces 
are functioning.  Robust end-to-end analyses not previously
possible have now been performed and are being 
reported in this document.

For some aspects -- for example, beamline modeling
and GeV-scale neutrino interaction simulations --
well-developed and validated (with data) software packages have
been available throughout much of \dword{dune}'s design phase.
For others, corresponding tools did not exist and needed to be
either developed from scratch or adapted with substantial
modifications from other experimental programs.  Concurrent
with these development efforts, interim descriptions such
as parametric detector response modeling, necessarily simple
but based on reasonable extrapolation from experience and
dedicated studies, were employed to assess physics capabilities.
Even for the case of the better-developed tools -- again, neutrino 
interaction modeling is a good example -- significant incremental
improvements have been made as data from neutrino experiments
and other sources have become available and as theoretical
understandings have advanced.

As a result of the rapid pace of development as well as 
practical considerations including human 
resource availability, different levels
of rigor have been applied in the evaluation of physics 
capabilities for different elements of the program.  
The strategy adopted for
this \dword{tdr} has been to hold the primary elements of the program
to the highest standard of rigor, involving direct analysis
of fully simulated data, using actual event reconstruction
codes and analysis tools that could be applied to real data
from \dword{dune} \dword{fd} modules.  For other elements of the
program, sensitivities use realistic beam and
physics simulations, but employ parametric detector
response models in place of full reconstruction.

The implementation of this strategy comes with caveats
and clarifications that are discussed in the corresponding
chapters of Volume~\volnumberphysics{}.   
We mention some of these here.
\begin{itemize}
\item In the case of the long-baseline oscillation physics
      program, this approach requires a combination of the 
      full end-to-end analysis of simulated \dword{fd} data
      with the concurrent analysis of simulated data from
      \dword{nd} systems to capture in a realistic way 
      the level of 
      control over systematic errors.  Given the 
      current state of development of the \dword{dune} \dword{nd} design and 
      corresponding analysis tools,  it has been necessary to 
      employ parametric detector response modeling for \dword{nd} components. 

\item In the case of the nucleon decay searches, 
      reconstruction and analysis tools dedicated to 
      addressing the particular challenges presented are 
      not as well developed as in the case of the 
      beam-based oscillation physics program. Effort is 
      ongoing to improve the performance of these tools. 

\item The \dword{snb}  program relies on 
reconstruction of event signatures from \dword{lartpc} signals 
generated by low-energy (MeV-scale) particles (electrons 
and de-excitation gammas).  Full simulation and reconstruction 
is used for some studies, such as for those demonstrating 
the supernova pointing capability of \dword{dune}.  
For other studies, a modified strategy is employed in order 
to efficiently explore model space:  reconstruction metrics 
(resolution smearing matrices, for example) are derived 
from analysis of fully simulated and reconstructed low-energy 
particles and events in the \dword{fd}, and are applied to 
understanding the mean detector response over a range of signal predictions.

\item For scientific program elements where
      analysis of fully reconstructed simulated data has 
      not yet been performed, the parametric response models 
      used for the analyses presented here have
      been well characterized with dedicated studies
      and incorporation of results from other experiments.
      The demonstration of sensitivities for the long-baseline
      oscillation physics program (with full reconstruction) 
      that are comparable to those
      previously obtained based on parametric response
      validates this approach.
\end{itemize}

\section{Selected Results from Sensitivity Studies}
\label{sec:exec-phys-sensitiv-results}

In this section, selected sensitivity projections from the 
central elements of the \dword{dune} science program are presented.  
This selection is intended to convey just the headlines from 
what is an extensive and diverse program of frontier science.

\subsection{\dshort{cpv} in the Neutrino 
Sector and Precise Oscillation Parameter Measurements}
\label{sec:es:phys:cpv}

The key strength of the \dword{dune} design concept is its ability to 
robustly measure the oscillation patterns of \numu and \anumu 
over a range of energies spanning the first and second 
oscillation maxima. 

This is accomplished by a coordinated analysis of the 
reconstructed \numu, \anumu, \nue, and \anue energy spectra 
in near and far detectors, 
incorporating data collected with forward (neutrino-dominated) 
and reverse (antineutrino-dominated) horn current polarities.  

The statistical power of \dword{dune} relative to the current 
generation of long-baseline oscillation experiments 
is a result of many factors including  
(1) on-axis operations, (2) the \dword{lbnf} beam power, 
(3) long-baseline and correspondingly high energy 
oscillation maxima and strong separation of 
normal and inverted neutrino mass ordering scenarios, 
(4) detector mass, and (5) event 
reconstruction and selection capabilities. 
Tables~\ref{tab:execsumm-apprates} and 
\ref{tab:execsumm-disrates} 
give the expected event 
yields for the appearance (\nue and \anue) 
and disappearance (\numu and \anumu) channels, respectively, 
after seven years of operation.
For these estimates, $\mdeltacp = 0$ is assumed, and
values for other parameters are taken from the 
\dword{nufit}~\cite{Esteban:2018azc,nufitweb} global fit 
to world neutrino data.  (See also
\cite{deSalas:2017kay} and \cite{Capozzi:2017yic} for other recent global fits.) 
The \dword{dune} \nue and \anue event yields represent order-of-magnitude increases relative to those in the current \dword{nova}~\cite{Acero:2019ksn} and \dword{t2k}~\cite{Abe:2018wpn,Abe:2019vii} data samples, while the 
corresponding increases are even larger for the  
\numu and \anumu channels thanks to \dword{dune}'s on-axis exposure to the \dword{lbnf} beam.
%
%
Figures~\ref{fig:appspectra} and~\ref{fig:disspectra} show the corresponding distributions in reconstructed neutrino energy.
\begin{dunetable}
[\nue and \anue appearance yields]
{lrr}
{tab:execsumm-apprates}
{\nue and \anue appearance yields: Integrated yield of selected $\nu_e$ \dword{cc}-like events between 0.5 and 8.0~GeV assuming \num{3.5}-year (staged) exposures in the neutrino-beam and antineutrino-beam modes.  The signal yields are shown for both normal mass ordering (NO) and inverted mass ordering (IO), and all the background yields assume normal mass ordering.  All the yields assume $\mdeltacp = 0$, and \dword{nufit}~\cite{Esteban:2018azc,nufitweb} 
values for other parameters.}
& \multicolumn{2}{c}{Expected Events (3.5 years staged per mode)} \\ \toprowrule
 & $\nu$ mode & $\bar{\nu}$ mode  \\
 \colhline 
 \nue signal NO (IO) & 1092 (497) & 76 (36) \\
 \anue signal NO (IO) & 18 (31)   & 224 (470) \\
  \colhline
 Total signal NO (IO) & 1110 (528) & 300 (506) \\
  \colhline 
 Beam $\nu_{e}+\bar{\nu}_{e}$ \dword{cc} background & 190 & 117 \\
 \dword{nc} background & 81  & 38\\
 $\nu_{\tau}+\bar{\nu}_{\tau}$ \dword{cc} background & 32 & 20 \\
 $\nu_{\mu}+\bar{\nu}_{\mu}$ \dword{cc} background & 14 & 5 \\
  \colhline
 Total background & 317 & 180\\
 
\end{dunetable}

\begin{dunetable}
[\numu and \anumu disappearance yields]
{lrr}
{tab:execsumm-disrates}
{\numu and \anumu disappearance yields: Integrated yield of selected $\nu_{\mu}$ \dword{cc}-like events between 0.5 and 8.0~GeV assuming a \num{3.5}-year (staged) exposure in the neutrino-beam mode and antineutrino-beam mode.  The yields are shown for normal mass ordering and $\mdeltacp = 0$.}
& Expected Events (3.5 years staged)\\ \toprowrule
  $\nu$ mode & \\
 \colhline 
 \numu Signal & 6200 \\
 \colhline 
  \anumu \dword{cc} background & 389 \\
 \dword{nc} background & 200 \\
 $\nu_{\tau}+\bar{\nu}_{\tau}$ \dword{cc} background & 46 \\
 $\nu_e+\bar{\nu}_e$ \dword{cc} background & 8 \\
 \toprowrule
 $\bar{\nu}$ mode  & \\
\colhline 
 \anumu signal & 2303 \\
\colhline 
  \numu \dword{cc} background & 1129 \\
 \dword{nc} background & 101 \\
 $\nu_{\tau}+\bar{\nu}_{\tau}$ \dword{cc} background & 27 \\
 $\nu_e+\bar{\nu}_e$ \dword{cc} background & 2 \\
\end{dunetable}

\begin{dunefigure}[\nue and \anue appearance spectra]{fig:appspectra}
{\nue and \anue appearance spectra: Reconstructed energy distribution of selected \nue \dword{cc}-like events assuming 3.5 years (staged) running in the neutrino-beam mode (left) and antineutrino-beam mode (right), for a total of seven years (staged) exposure.  The plots assume normal mass ordering and include curves for $\mdeltacp = -\pi/2, 0$, and $\pi/2$.}
 \includegraphics[width=0.49\textwidth]{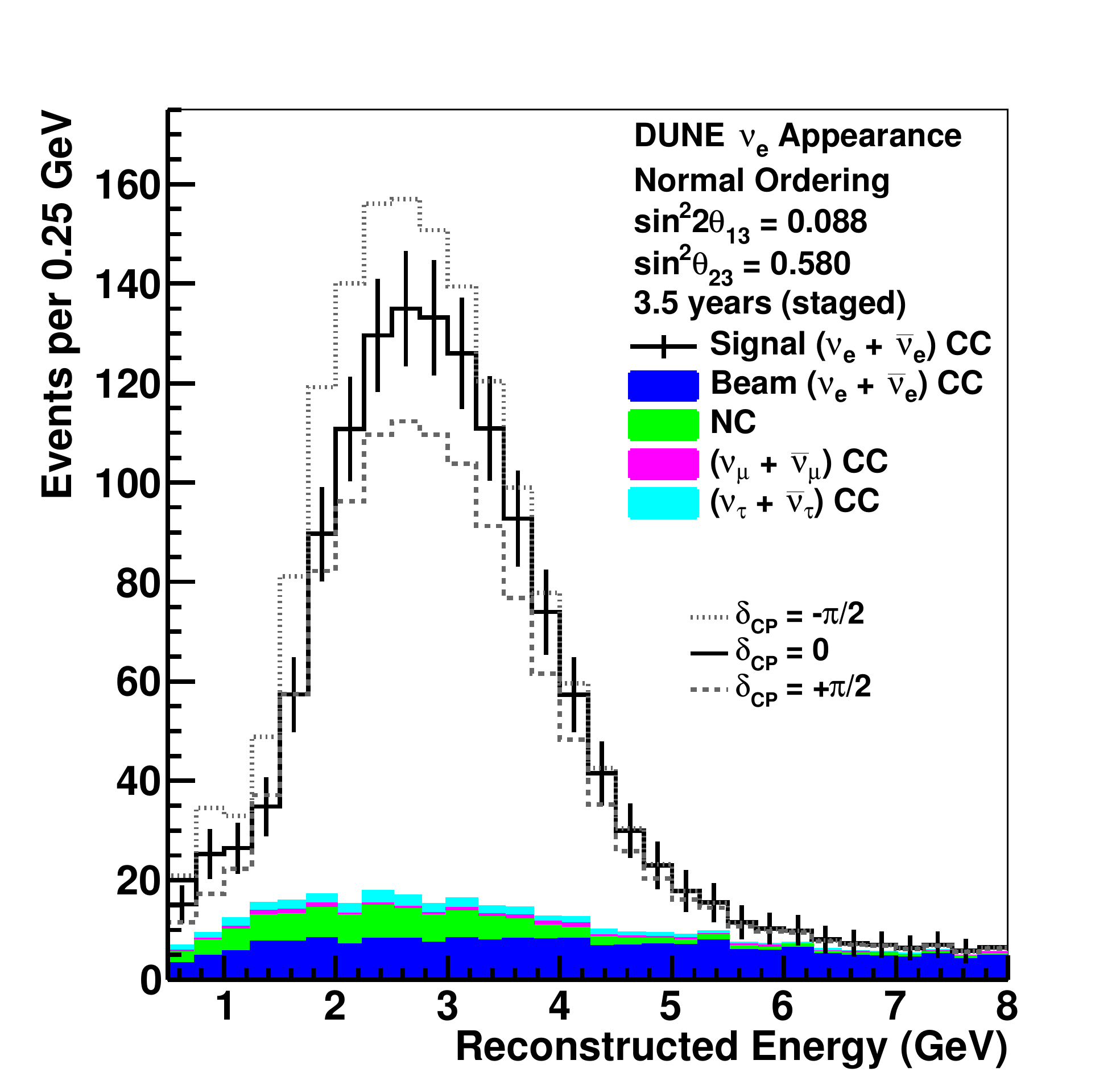}
 \includegraphics[width=0.49\textwidth]{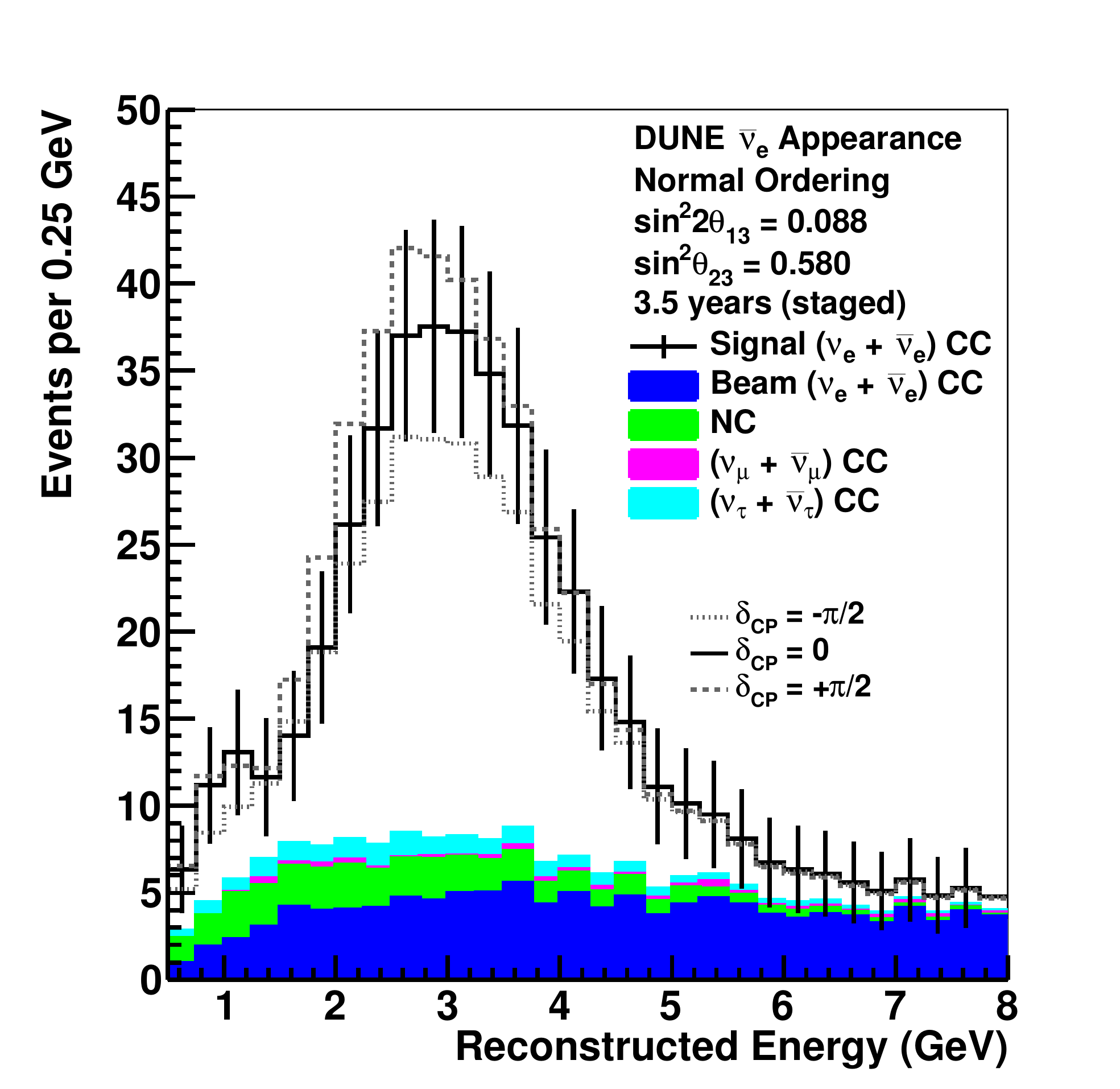}
\end{dunefigure}

\begin{dunefigure}[\numu and \anumu disappearance spectra]{fig:disspectra}
{\numu and \anumu disappearance spectra: Reconstructed energy distribution of selected $\nu_{\mu}$ \dword{cc}-like events assuming 3.5 years (staged) running in the neutrino-beam mode (left) and antineutrino-beam mode (right), for a total of seven years (staged) exposure. The plots assume normal mass ordering.}
\includegraphics[width=0.49\textwidth]{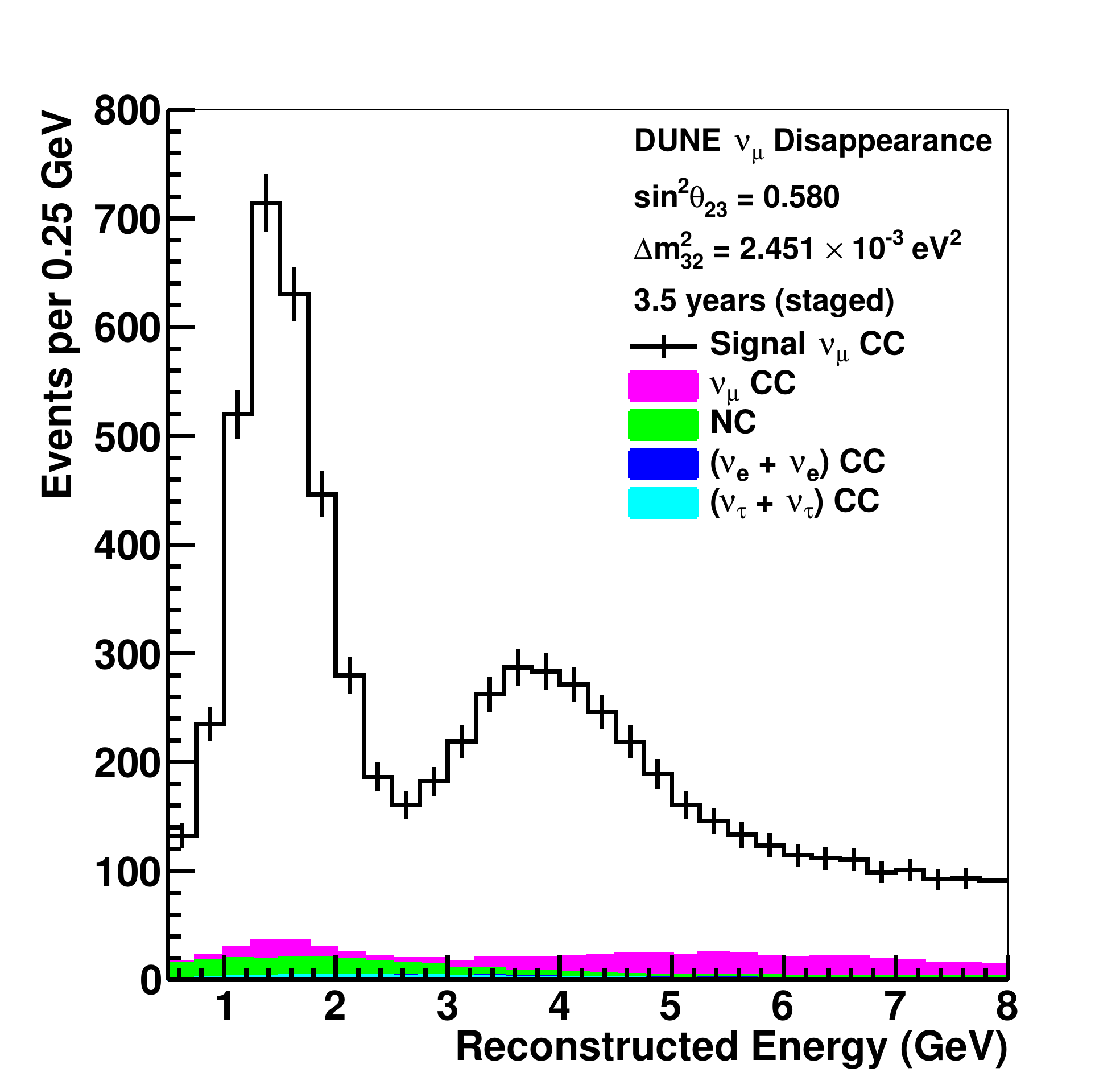}
\includegraphics[width=0.49\textwidth]{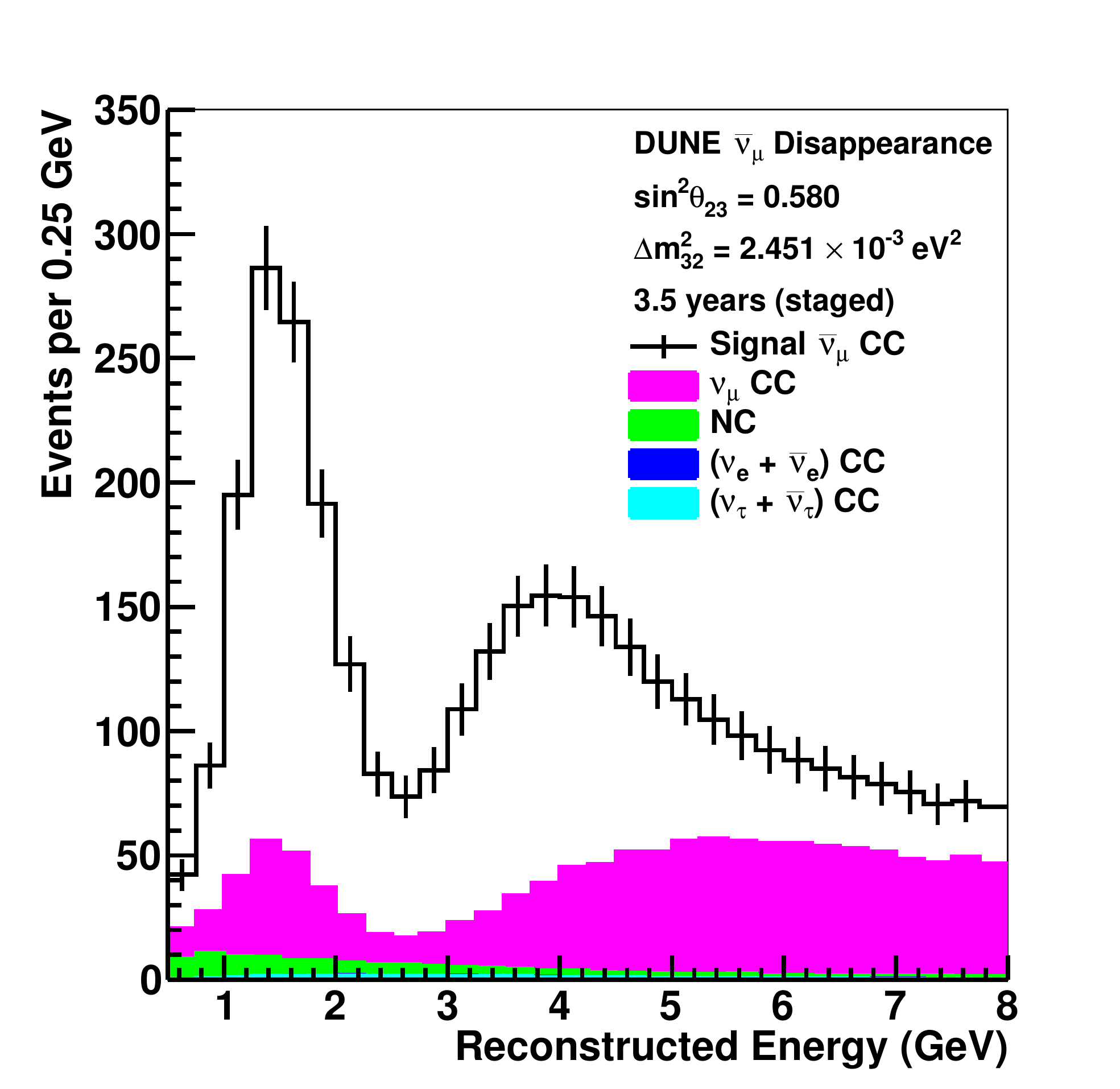}
\end{dunefigure}

Experimental sensitivities were evaluated 
based on the methodologies described in the preceding section, 
including incorporation of \dword{nd} simulations 
and uncertainties from all known sources of systematic error.  
Considerable attention and sophistication 
has been applied to the treatment of systematic errors and the crucial role of the \dword{nd}, 
both of which are documented in 
Volume~\volnumberphysics{}, \voltitlephysics{}.

A summary of representative sensitivity milestones for neutrino 
mass ordering and \dword{cpv} discovery, as well as precision on 
\deltacp and \sinstt{13} is given in 
Table~\ref{tab:milestones_execsumm-es}.  
The ultimate level of 
precision that can be obtained on oscillation parameters 
highlights the point that \dword{dune} will provide crucial input for  
flavor physics:  Patterns required by particular symmetries 
underlying fermion masses and mixing angles may appear.  The 
unitarity of the neutrino mixing matrix can be tested directly 
through comparisons of \sinstt{13} with the value obtained from 
reactor experiments.  In conjunction with \sinstt{13} and 
other parameters, the precise value of \deltacp can  
constrain models of leptogenesis that are leading 
candidates for explanation of the baryon asymmetry of the universe.

\begin{dunetable}[Projected DUNE oscillation physics milestones]
{lcc}
{tab:milestones_execsumm-es}
{Exposure in years, assuming true normal ordering and equal 
running in neutrino and antineutrino mode, required to reach 
selected physics milestones in the nominal analysis, using the 
NuFIT 4.0~\cite{Esteban:2018azc,nufitweb} best-fit values for the oscillation parameters. As 
discussed in \physchlbl, there are 
significant variations in sensitivity with the value of
\sinst{23}, so the exact values quoted here 
(using \sinst{23} = 0.580) are strongly dependent on that choice. 
The staging scenario presented in
Section~\ref{sec:exec-phys-assm-meth-deployment} is assumed. Exposures
are rounded to the nearest ``staged'' calendar year.
For reference, 30, 100, 200, 336, 624, and \SI{1104}{\ktMWyr} correspond to 1.2, 3.1, 5.2, 7, 10, and 15 staged years, respectively.
}
 Physics Milestone & Exposure (staged years) \\
 5$\sigma$ mass ordering & 1 \\
 \phantom{xxx}(\deltacp = -$\pi/2$) & \\ \colhline
 5$\sigma$ mass ordering & 2 \\
 \phantom{xxx}(100\% of \deltacp values) & \\ \colhline
 3$\sigma$ \dshort{cpv} & 3 \\
 \phantom{xxx}(\deltacp = -$\pi/2$) & \\ \colhline
 3$\sigma$ \dshort{cpv} & 5 \\
 \phantom{xxx}(50\% of \deltacp values) & \\ \colhline
 5$\sigma$ \dshort{cpv} & 7 \\
 \phantom{xxx}(\deltacp = $-\pi/2$) & \\ \colhline
 5$\sigma$ \dshort{cpv} & 10 \\
 \phantom{xxx}(50\% of \deltacp values) & \\ \colhline
 3$\sigma$ \dshort{cpv} & 13 \\
 \phantom{xxx}(75\% of \deltacp values) & \\ \colhline
 \deltacp resolution of 10 degrees & 8 \\
 \phantom{xxx}(\deltacp = 0) & \\ \colhline
 \deltacp resolution of 20 degrees & 12 \\
 \phantom{xxx}(\deltacp = -$\pi/2$) & \\ \colhline
 \sinstt{13} resolution of 0.004 & 15 \\ 
\end{dunetable}

The milestones presented in Table~\ref{tab:milestones_execsumm-es} 
form a coarse snapshot of the \dword{dune} program in oscillation physics, 
demonstrating the prospects for important results throughout 
the lifetime of the experiment.  More detail on the 
sensitivities for the individual program elements
is presented in the sections below.

\subsubsection{Discovery Potential for \dshort{cpv} and Neutrino 
Mass Ordering}

Figure~\ref{fig:cpv_staging_execsum} 
illustrates \dword{dune}'s ability to distinguish 
the value of the \dword{cp} phase \deltacp from \dword{cp}-conserving 
values (0 or $\pi$) as a function of time in calendar year.  
These projections incorporate a sophisticated treatment of systematic 
error, as described in detail in \physchlbl.  
Evidence ($>3\sigma$) for \dword{cpv} is obtained for 
favorable values (half of the phase space) of \deltacp after five 
years of running, leading to a $>5\sigma$ 
observation after ten years.

\begin{dunefigure}[Significance of the DUNE determination of 
CP violation]{fig:cpv_staging_execsum}
{Significance of the DUNE determination of CP-violation (i.e.: \deltacp 
$\neq 0$ or $\pi$) for the case when \deltacp=$-\pi/2$, and for 50\% and 
75\% of possible true \deltacp values, as a function of time in calendar 
years. True normal ordering is assumed. The width of the band shows the 
impact of applying an external constraint on \sinstt{13}.}
\includegraphics[width=0.75\linewidth]{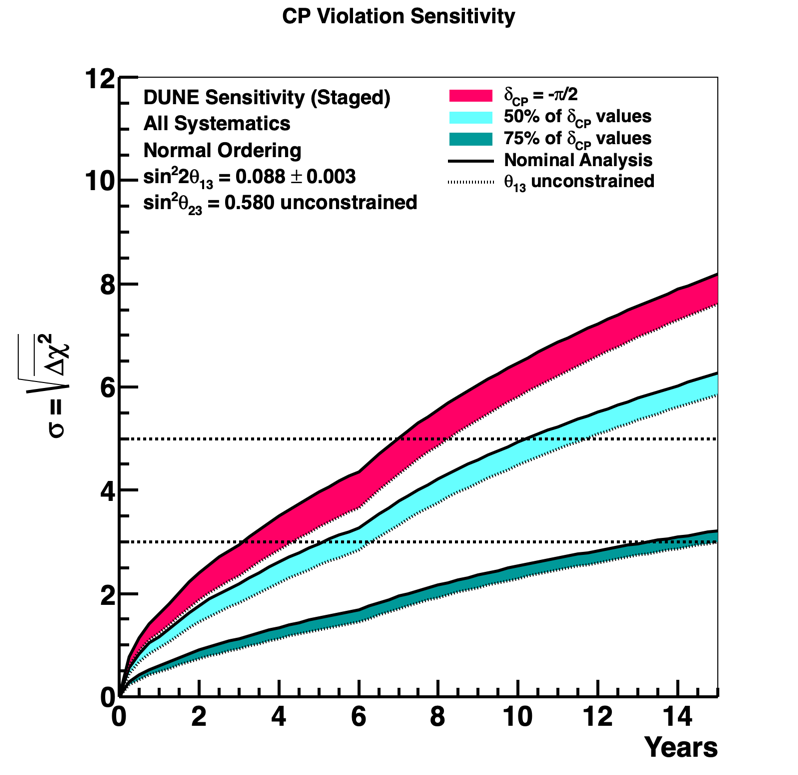}
\end{dunefigure}

Figure~\ref{fig:mh_staging} shows the significance
with which the neutrino mass ordering can be determined for 100\% of \deltacp values, and when $\deltacp=-\pi/2$, as a function of exposure in years. The width of the bands show the impact of applying an external constraint on \sinstt{13}. As \dword{dune} will be able to establish the neutrino mass ordering at the 5$\sigma$ level for 100\% of \deltacp values after 2-3 years, this plot extends only to seven years.

\begin{dunefigure}
[Significance of the DUNE neutrino mass ordering determination, as a function of time]
{fig:mh_staging}
{Significance of the DUNE determination of the neutrino mass ordering for the case when \deltacp=$-\pi/2$, and for 100\% of possible true \deltacp values, as a function of time in calendar years. True normal ordering is assumed. The width of the band shows the impact of applying an external constraint on \sinstt{13}.}
	\includegraphics[width=0.75\linewidth]{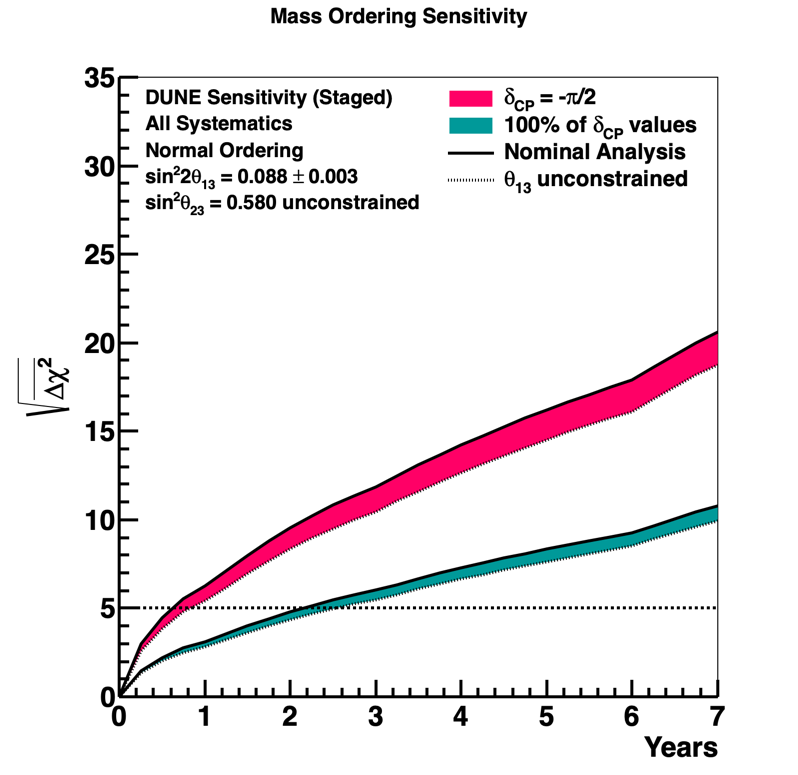}	
\end{dunefigure}

\subsubsection{Precision Measurement of Mass and Mixing Parameters}

In addition to the discovery potential for neutrino mass ordering and \dword{cpv}, 
\dword{dune} will improve the precision on key parameters that govern neutrino oscillations, including: \deltacp, $\sin^22\theta_{13}$, \dm{31}, $\sin^2\theta_{23}$, and the octant of $\theta_{23}$. 

Figure~\ref{fig:dcpresvdcp} shows the resolution, in degrees, of \dword{dune}'s measurement of \deltacp as a function of the true value of \deltacp. The resolution of this measurement is significantly better near \dword{cp}-conserving values of \deltacp, compared to maximally \dword{cp}-violating values. For fifteen years of exposure, resolutions between five and fifteen degrees are possible, depending on the true value of \deltacp. 

\begin{dunefigure}
[Resolution for the DUNE measurement of \deltacp as a function of its true value]
{fig:dcpresvdcp}
{Resolution in degrees for the DUNE measurement of \deltacp, as a function of the true value of \deltacp, for seven (blue), ten (orange), and fifteen (green) years of exposure. True normal ordering is assumed. The width of the band shows the impact of applying an external constraint on \sinstt{13}.}
		\includegraphics[width=0.75\linewidth]{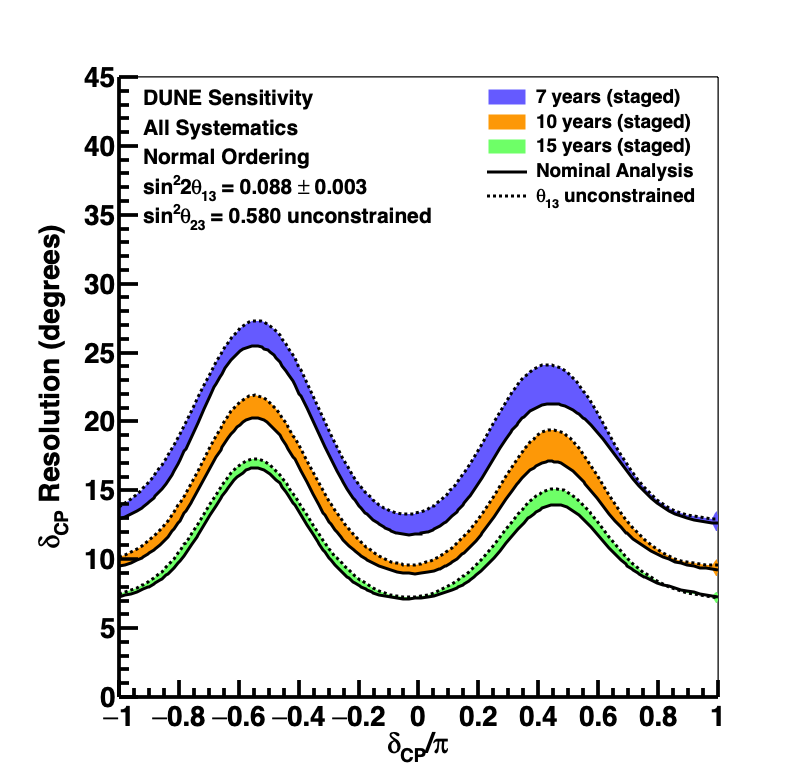}
\end{dunefigure}

Figures \ref{fig:appres_exp} and  \ref{fig:disres_exp} show the resolution of \dword{dune}'s measurements of \deltacp and \sinstt{13} and of \sinstt{23} and $\Delta m^{2}_{32}$, respectively, as a function of exposure in kt-MW-years. As seen in Figure~\ref{fig:dcpresvdcp}, the \deltacp resolution varies significantly with the true value of \deltacp, but for favorable values, resolutions near five degrees are possible for large exposure. The \dword{dune} measurement of \sinstt{13} approaches the precision of reactor experiments for high exposure, allowing a comparison between the two results, which is of interest as a test of the unitarity of the \dword{pmns} matrix. 

\begin{dunefigure}
[Resolution of DUNE measurements of \deltacp and \sinstt{13}, as a function of exposure]
{fig:appres_exp}
{Resolution of DUNE measurements of \deltacp (left) and \sinstt{13} (right), as a function of exposure in kt-MW-years. As seen in Figure~\ref{fig:dcpresvdcp}, the \deltacp resolution has a significant dependence on the true value of \deltacp, so curves for $\deltacp=-\pi/2$ (red) and $\deltacp=0$ (green) are shown. The width of the band shows the impact of applying an external constraint on \sinstt{13}. For the \sinstt{13} resolution, an external constraint does not make sense, so only the unconstrained curve is shown.  
For reference, 336, 624, and \SI{1104}{\ktMWyr} correspond to 7, 10, and 15 staged years, respectively.
}
		\includegraphics[width=0.475\linewidth]{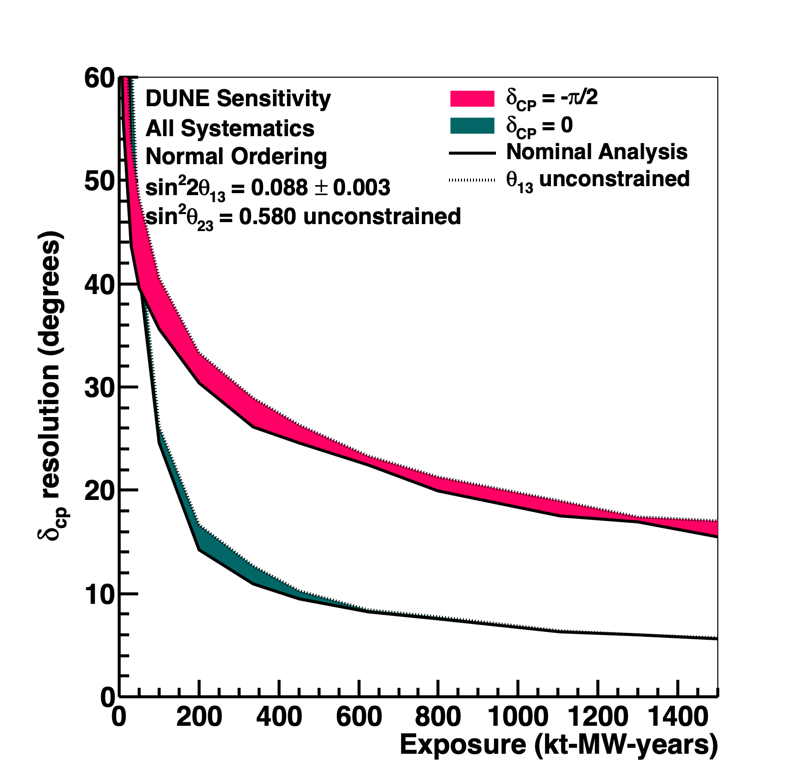} 
		\includegraphics[width=0.475\linewidth]{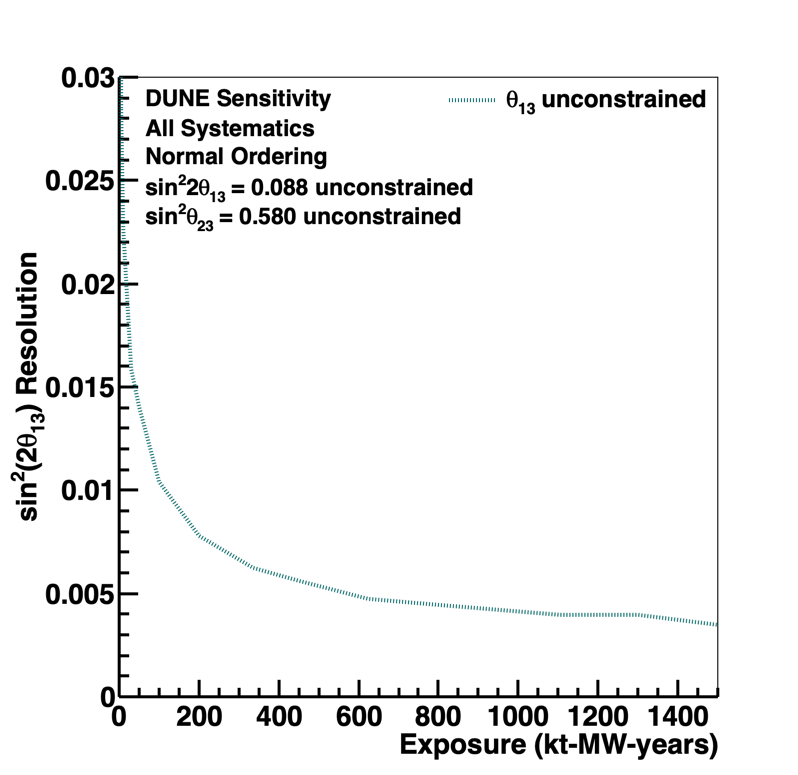} 
\end{dunefigure}

\begin{dunefigure}
[Resolution of DUNE measurements of \sinstt{23} and $\Delta m^{2}_{32}$, as a function of exposure]
{fig:disres_exp}
{Resolution of DUNE measurements of \sinstt{23} (left) and $\Delta m^{2}_{32}$ (right), as a function of exposure in kt-MW-years. The width of the band for the \sinstt{23} resolution shows the impact of applying an external constraint on \sinstt{13}. For the $\Delta m^{2}_{32}$ resolution, an external constraint does not have a significant impact, so only the unconstrained curve is shown.
For reference, 336, 624, and \SI{1104}{\ktMWyr} correspond to 7, 10, and 15 staged years, respectively.
}
		\includegraphics[width=0.475\linewidth]{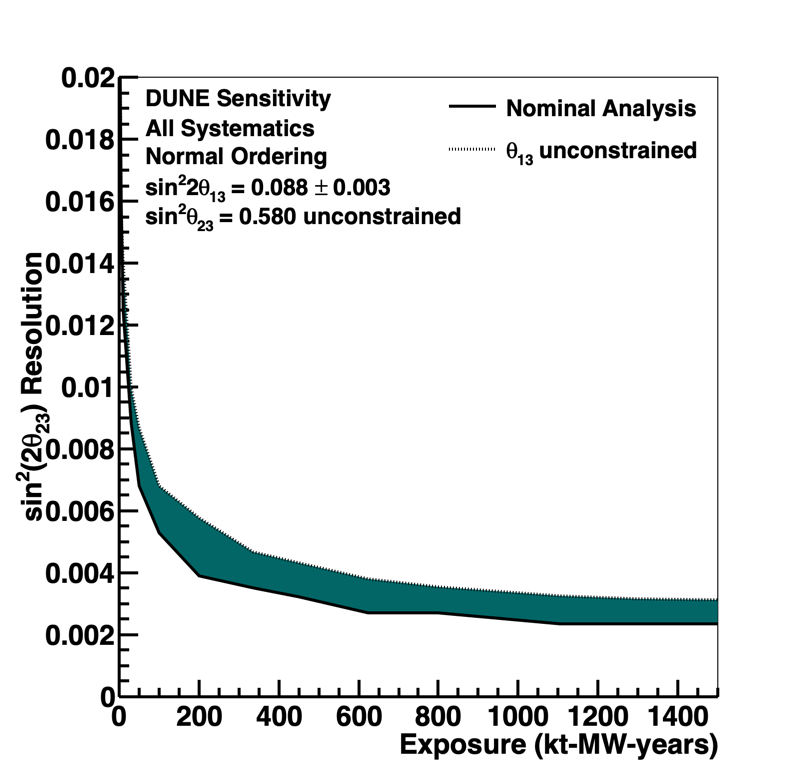} 
		\includegraphics[width=0.475\linewidth]{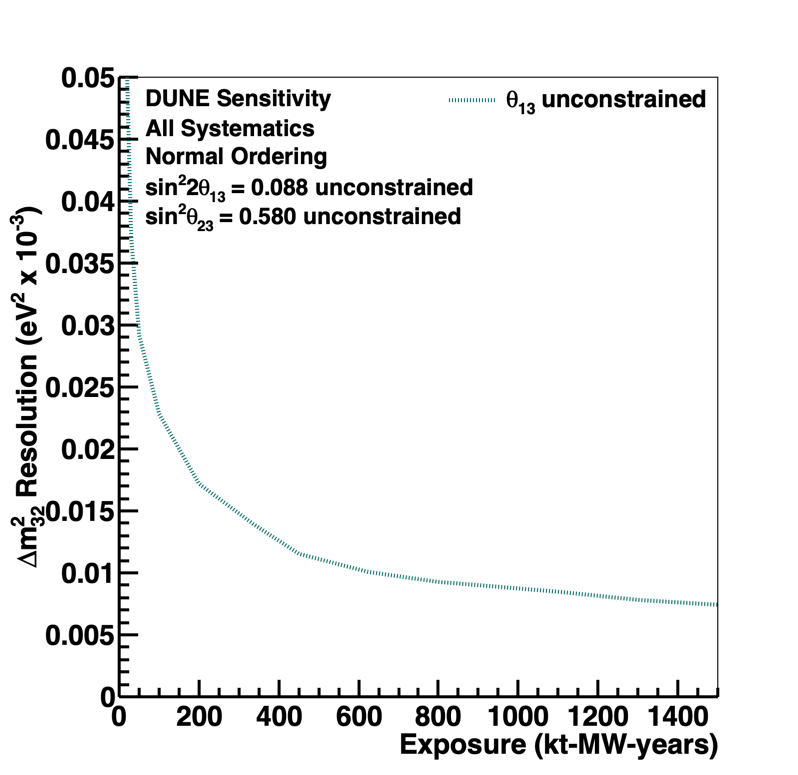} 
\end{dunefigure}

%
One of the principal strengths of \dword{dune} is its ability to simultaneously measure all oscillation parameters governing long-baseline neutrino oscillation, without a need for external constraints. As an example, Figure~\ref{fig:res_th13th23vdcp} shows the 90\% C.L.\ allowed regions for \sinstt{13} (left) and \sinst{23} (right) versus \deltacp for 7, 10 and 15 years of running, 
compared to the current measurements from world data.

\begin{dunefigure}
[Two-dimensional 90\% C.L. regions in \sinstt{13}/\sinst{23} vs.\ \deltacp]
{fig:res_th13th23vdcp}
{Left: Two-dimensional 90\% C.L. region in \sinstt{13} and \deltacp, for 7, 10, and 15 years of exposure, with equal running in neutrino and antineutrino mode. The 90\% C. L. region for the \dword{nufit} global fit is shown in yellow for comparison. The true values of the oscillation parameters are assumed to be the central values of the \dword{nufit} global fit and the oscillation parameters governing long-baseline oscillation are unconstrained.  Right: Corresponding region in \sinst{23} and \deltacp.  
In this case, \sinstt{13} is constrained by \dword{nufit}.
}
		\includegraphics[width=0.5\linewidth]{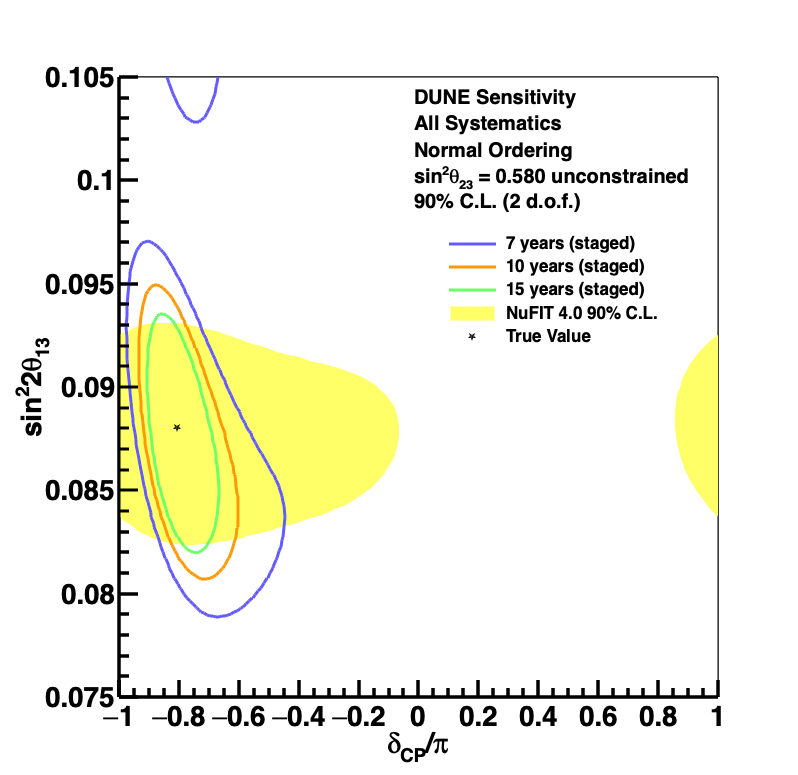}\hfill
		\includegraphics[width=0.5\linewidth]{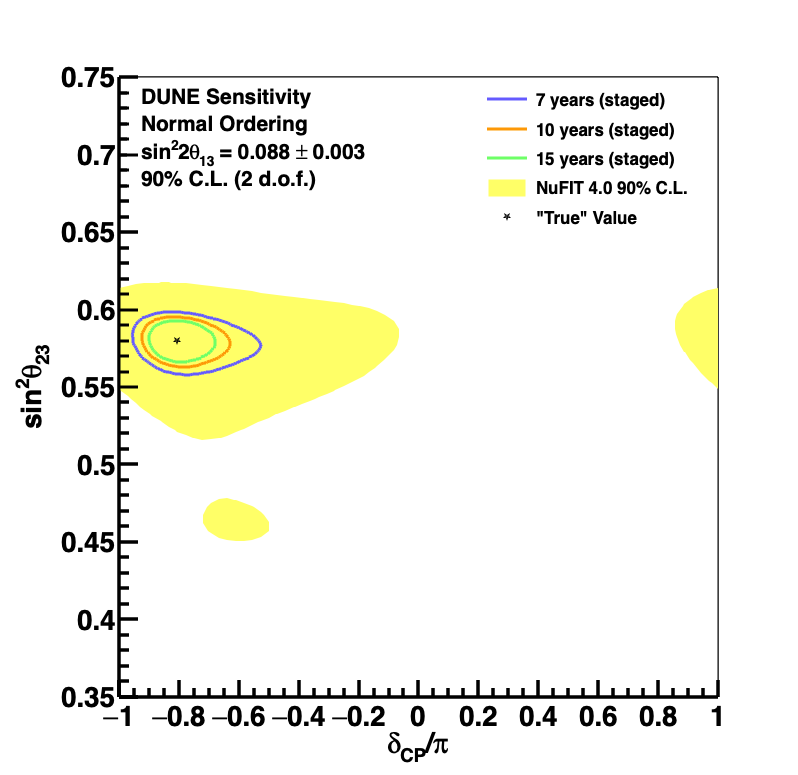}
\end{dunefigure}
%

The measurement of $\nu_\mu \rightarrow \nu_\mu$ oscillations is sensitive to $\sin ^2 2 \theta_{23}$, whereas the measurement of $\nu_\mu \rightarrow \nu_e$ oscillations is sensitive to $\sin^2 \theta_{23}$.  A combination of both $\nu_e$ appearance and $\nu_\mu$ disappearance measurements can probe both maximal mixing and
the $\theta_{23}$ octant.  
Figure~\ref{fig:lbloctant} shows the sensitivity to determining the octant as a function of the true value of $\sinst{23}$.

\begin{dunefigure}
[Sensitivity of determination of the $\theta_{23}$ octant as a function of \sinst{23}]
{fig:lbloctant}
{Sensitivity to determination of the $\theta_{23}$ octant as a function of the true value of \sinst{23}, for ten (orange) and fifteen (green) years of exposure. True normal ordering is assumed. The width of the transparent bands cover 68\% of fits in which random throws are used to simulate statistical variations and select true values of the oscillation and systematic uncertainty parameters, constrained by pre-fit uncertainties. The solid lines show the median sensitivity.}
		\includegraphics[width=0.5\linewidth]{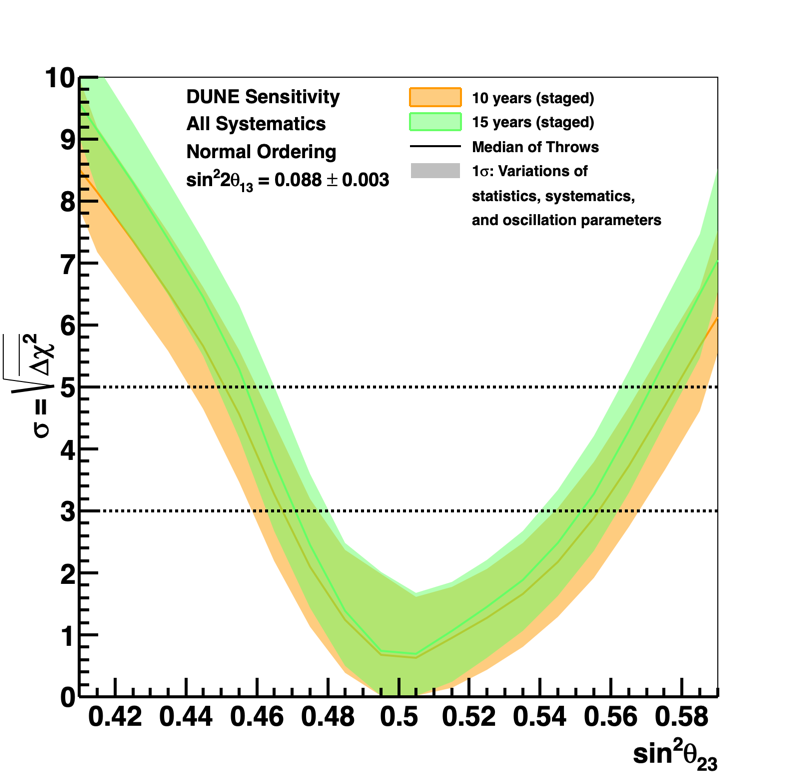}
\end{dunefigure}

\clearpage
\subsection{Proton Decay and Other 
Baryon-number Violating Processes}
\label{sec:es:phys:pdecay}

By virtue of its deep underground location and large fiducial 
mass, as well as its excellent event imaging, particle 
identification and 
calorimetric capabilities, the \dword{dune} \dword{fd} will be 
a powerful instrument 
to probe baryon-number violation.
\dword{dune} will be able to observe signatures of decays of protons and 
neutrons, as well as the phenomenon of neutron-antineutron mixing, 
at rates below the limits placed by the current generation of 
experiments.

Many nucleon decay modes are accessible to \dword{dune}.  
As a benchmark, a particularly compelling discovery channel 
is the decay of a proton to a positive kaon and a neutrino, 
\ptoknubar.  In this channel, the kaon and its decay products 
can be imaged, identified, and tested for kinematic consistency 
with the full decay chain, together with precision sufficient to 
reject backgrounds due to atmospheric muon and neutrino 
interactions. 
Preliminary analysis of single-particle beam and cosmic ray tracks 
in the \dword{pdsp} \dword{lartpc} is already demonstrating the particle 
identification capability of \dword{dune}, as illustrated in 
Figure~\ref{fig:pdsp_dedx_execsum}.  
The signature of the kaon track and its observable decay particles is 
sufficiently rich that a credible claim of evidence for 
proton decay could be made on the basis of just 
one or two sufficiently well-imaged events, for the case 
where background sources are expected to contribute much less 
than one event.

\begin{dunefigure}[Reconstructed $dE/dx$ of protons and muons in 
\dshort{pdsp}]{fig:pdsp_dedx_execsum}
{Energy loss of protons (left) and muons (right) in 1-GeV  
running with the \dword{pdsp} \dword{lartpc} at \dword{cern}, as a function of 
residual range.  The protons are beam particles identified from 
beamline instrumentation; the muons are reconstructed stopping 
cosmic rays collected concurrently.  
The red curves represent the mean of the 
corresponding expected signature.  Note the difference in 
the vertical scale of the two plots.  The kaon $dE/dx$ curve 
will lie between the two curves shown.}
\includegraphics[width=0.45\linewidth]{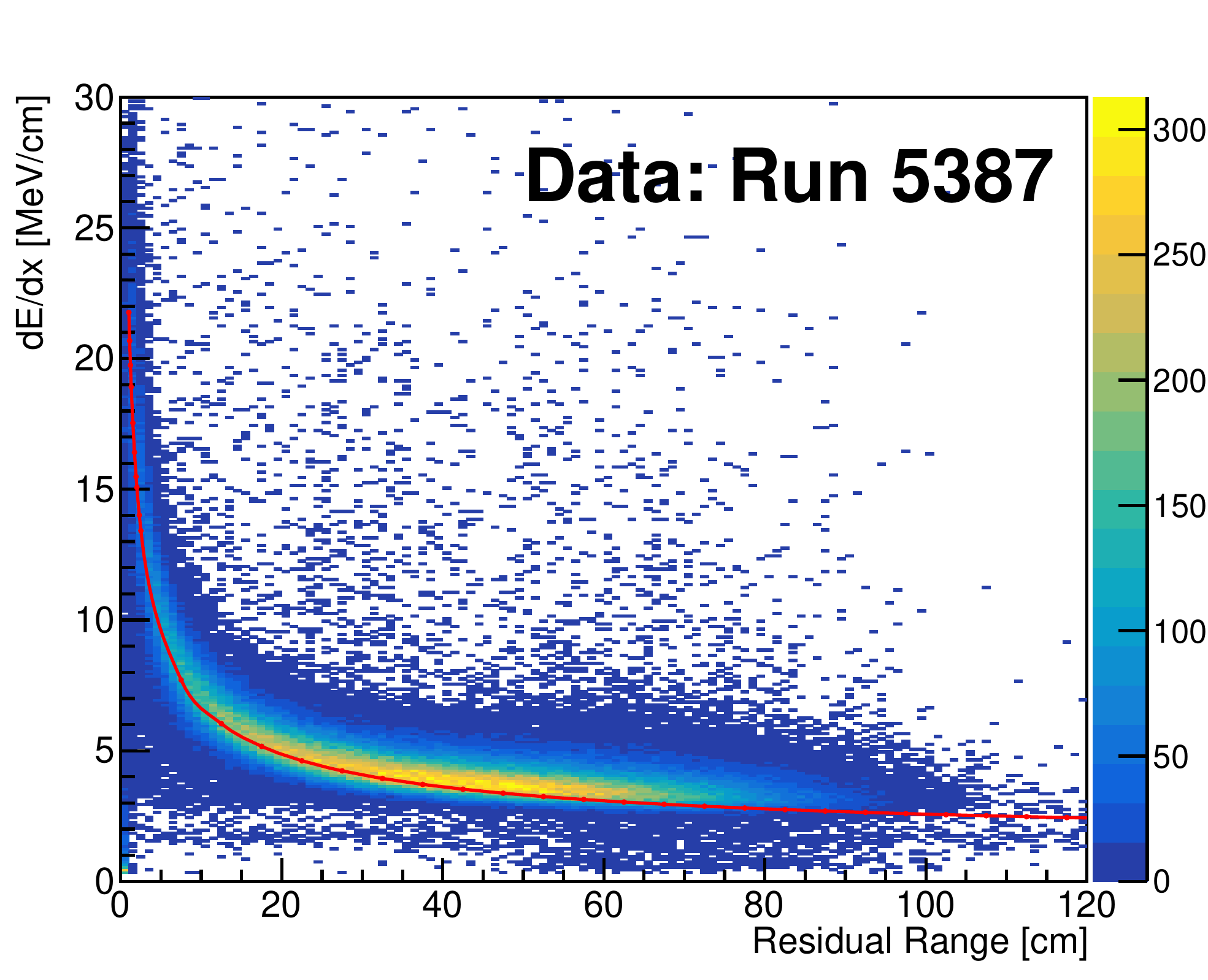}\hspace{0.05\linewidth}
\includegraphics[width=0.45\linewidth]{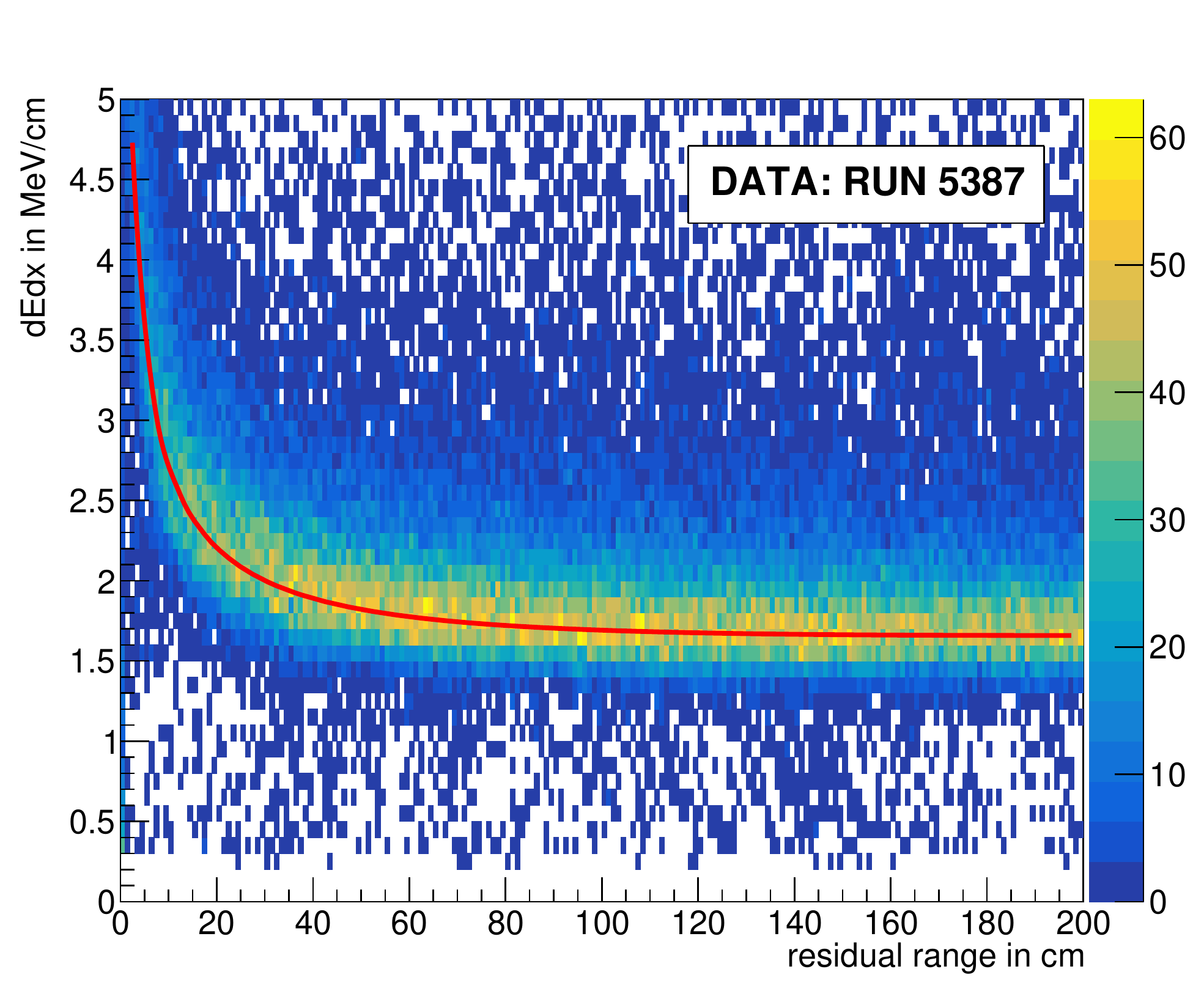}
\end{dunefigure}

Projecting from the current analysis of \ptoknubar in the \dword{dune} 
\dword{fd}, with a detection efficiency of \num{30}\% as 
described in \physchndk, the 
expected 90\% C.L. lower limit on lifetime divided by branching 
fraction is \SI{1.3e34}{years} for a  \SI{400}{\ktyr} 
exposure, assuming no candidate events are observed.  This 
is roughly twice the current limit of 
\SI{5.9e33}{years} from \superk~\cite{Abe:2014mwa}, 
based on an exposure of \SI{260}{\ktyr}.  Thus, should the rate 
for this decay be at the current \superk limit, five candidate 
events would be expected in \dword{dune} within ten years 
of running with four \dword{fd} modules.  Ongoing work is aimed 
at improving the efficiency in this and other channels.

\subsection{Galactic Supernovae via Measurements of Neutrino Bursts}
\label{sec:es:phys:galact}

As has been demonstrated with SN1987a, the observation 
of neutrinos~\cite{Bionta:1987qt,Hirata:1987hu} from a 
core-collapse supernova can reveal much about these  
phenomena that is not accessible in its  
electromagnetic signature.  Correspondingly, there is a 
wide range of predictions from supernova models for even 
very basic characteristics of the \dwords{snb}.  Typical  
models predict that a supernova explosion in the 
center of the Milky Way will result in several thousand 
detectable neutrino interactions in the \dword{dune} \dword{fd} 
occurring over an interval of up to a few tens of seconds.
The neutrino energy spectrum peaks around \SI{10}{\MeV}, 
with appreciable flux up to about \SI{30}{\MeV}.

\Dword{lar} based detectors are sensitive to the \nue 
component of the flux, while water Cherenkov and organic 
scintillator detectors are most sensitive to the \anue 
component.  Thus \dword{dune} is uniquely 
positioned to study the 
neutronization burst, in which \nue's are produced during the 
first few tens of milliseconds.  More generally,  
measurements of the (flavor-dependent) neutrino flux and energy 
spectrum as a function of time over the entirety of the burst 
can be sensitive to astrophysical properties of the supernova 
and its progenitor, and distortions relative to nominal 
expectations can serve as signatures for phenomena such 
as shock wave and turbulence effects, or even black hole 
formation.  

Below, we present the results of analyses 
of \dword{dune}'s capabilities for two elements of the \dword{snb} 
 program: (1) fits to the reconstructed 
neutrino energy spectrum and comparison to models in which 
none of the distortions listed above are present, and (2) 
neutrino flux direction determination for angular localization 
of the supernova position.

\subsubsection{Results from Fits to Pinched Thermal Neutrino Energy Spectrum}

The physics of neutrino decoupling and spectra formation is far from trivial, owing to the energy dependence of the cross sections and the roles played by both \dword{cc} and \dword{nc} reactions.
Detailed transport calculations using methods such as \dword{mc} or Boltzmann solvers have been employed. It has been observed that spectra coming out of such simulations can typically be parameterized at a given moment in time by the following ansatz (e.g.,~\cite{Minakata:2008nc,Tamborra:2012ac}):
\begin{equation}
        \label{eq:pinched}
        \phi(E_{\nu}) = \mathcal{N} 
        \left(\frac{E_{\nu}}{\langle E_{\nu} \rangle}\right)^{\alpha} \exp\left[-\left(\alpha + 1\right)\frac{E_{\nu}}{\langle E_{\nu} \rangle}\right] \ ,
\end{equation}
where $E_{\nu}$ is the neutrino energy, $\langle E_\nu \rangle$ is the
mean neutrino energy, $\alpha$ is a ``pinching parameter,'' and
$\mathcal{N}$ is a normalization constant that can be related to 
the total binding energy release of the supernova, denoted $\varepsilon$ 
in the discussion below.
Large $\alpha$ corresponds to a more ``pinched'' spectrum (suppressed
high-energy tail). This parameterization is referred to as a
``pinched-thermal'' form. The different $\nu_e$, $\overline{\nu}_e$ and
$\nu_x, \, x = \mu, \tau$ flavors are expected to have different
average energy and $\alpha$ parameters and to evolve differently in
time. 
The primary experimental task is to determine the true neutrino spectrum from the observed supernova event spectrum.  Given that the spectrum is well described by the functional form in Equation 2.7, this task is approximately equivalent to that of fitting the threepinched-thermal parameters  $(\alpha^0, \langle E_\nu \rangle^0, \varepsilon^0)$ that carry the spectral information.  

To evaluate \dword{dune}'s capabilities, we have developed a
forward fitting algorithm requiring a binned reconstructed 
neutrino energy spectrum expected for 
a supernova at a given distance generated with a ``true'' set of
these parameters. 
Figure~\ref{fig:example3params} shows an example of a resulting fit,
with the approximate parameters for several specific supernova models
superimposed to illustrate the potential for discrimination 
between them.

\begin{dunefigure}[Fit to three supernova neutrino pinched-thermal 
  spectrum parameters]{fig:example3params}{
    Estimated sensitivity regions for \dword{dune}'s determination of the three supernova spectra parameters at 90\% C.L. given the assumed true value indicated by the black star.  The simulated data were generated for a supernova at \SI{10}{kpc}
    with a neutrino interaction model appropriate for low energies, 
    realistic detector smearing, and a step efficiency function with a \SI{5}{\MeV}
    detected energy threshold. To indicate the expected range of possible true flux parameters, superimposed are parameters corresponding to the time-integrated flux for three different sets of models:
  Nakazato~\cite{Nakazato:2012qf}, Huedepohl black hole formation models, and Huedepohl
  cooling models~\cite{huedepohldb}.  For the Nakazato parameters (for which there is no
  explicit pinching, corresponding to $\alpha=2.3$), the parameters are
  taken directly from the reference; for the Huedepohl models, they are fit to a
  time-integrated flux.}
	\includegraphics[scale = 0.37]{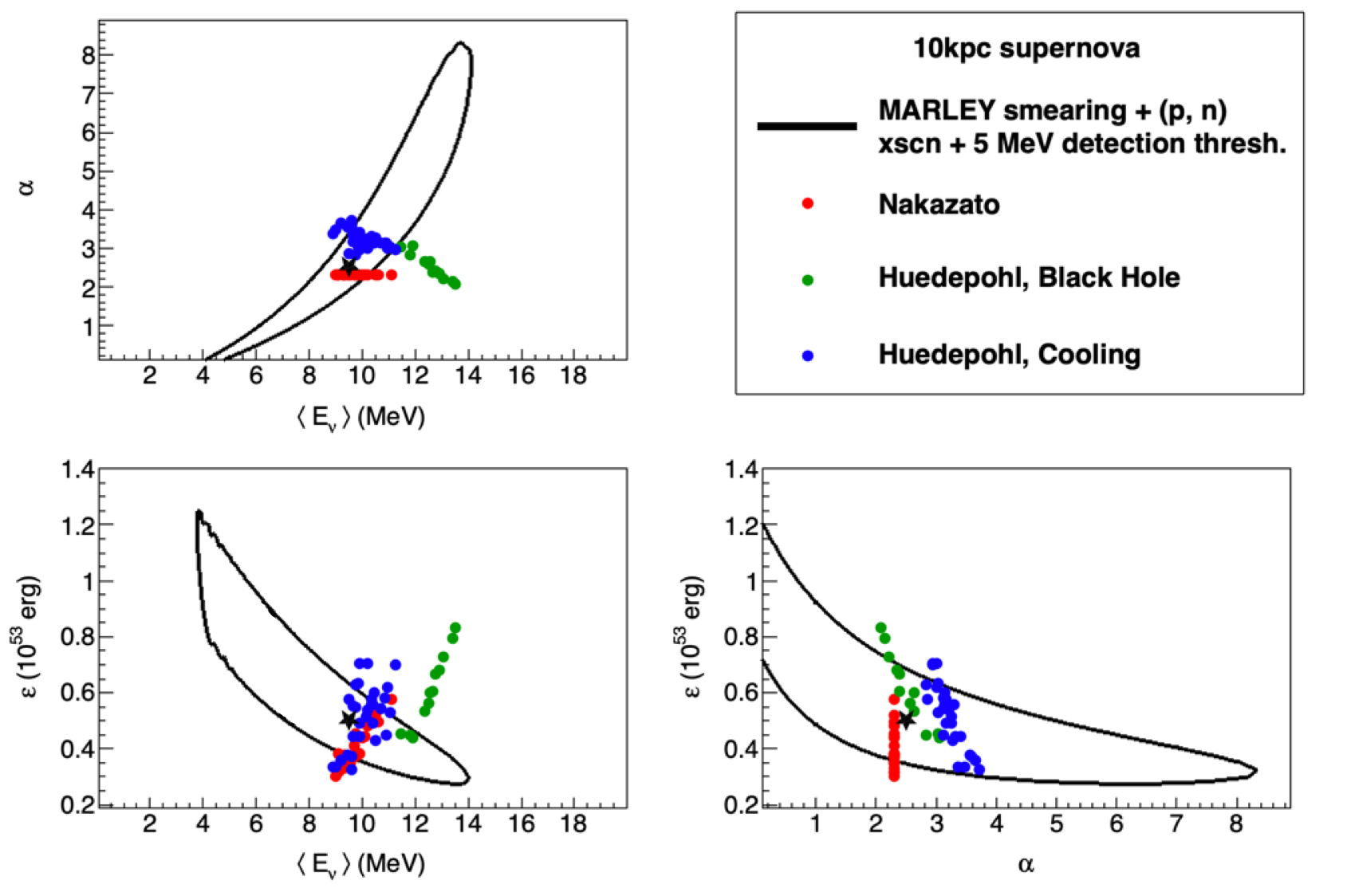}
  \end{dunefigure}

\subsubsection{Pointing Sensitivity of DUNE}

An illustration 
of another element of the \dword{dune} \dword{snb} program 
is given in Figure~\ref{fig:fullSN_execsum}, 
which indicates a pointing resolution of better than $5^\circ$ that 
can be obtained by analysis of both subdominant highly-directional $\nu$-$e$ elastic scattering 
events and dominant weakly-directional $\nu_e$ \dword{cc} events within a \dword{snb}, based 
on full reconstruction and analysis. The \dword{dune} results can be 
combined with corresponding measurements in other neutrino detectors to 
provide supernova localization from neutrinos alone in real time.
\begin{dunefigure}[Supernova direction determination from $\nu-e$ elastic scattering events]{fig:fullSN_execsum}{Left: Log
    likelihood values as a function of direction for a
    supernova sample with 260 $\nu$-$e$ elastic scattering (ES) events.  Right: Distribution of angular differences for
    directions to \SI{10}{kpc} supernova using a maximum likelihood
    method.}
  \includegraphics[width=0.44\textwidth]{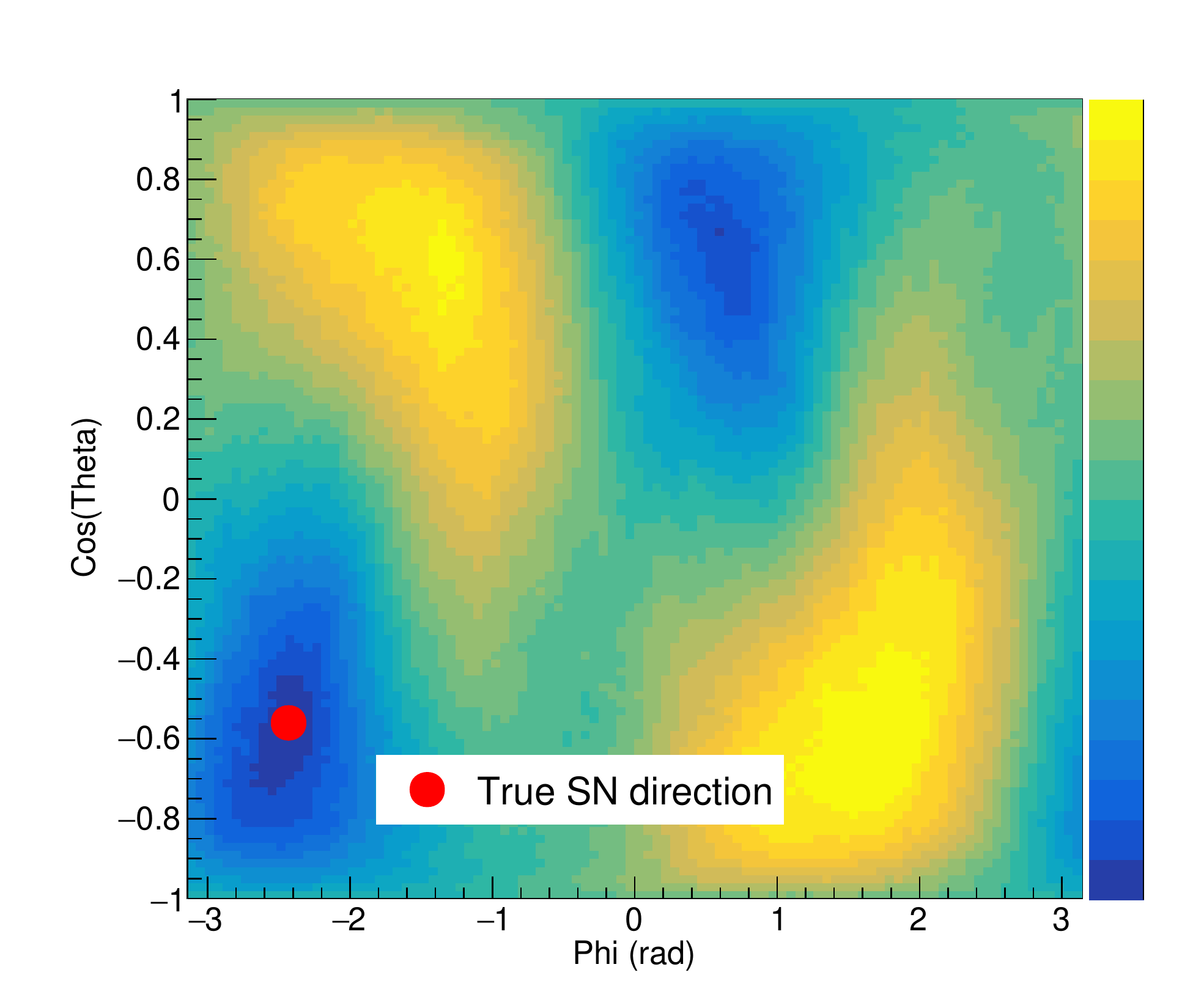}
  \includegraphics[width=0.48\textwidth]{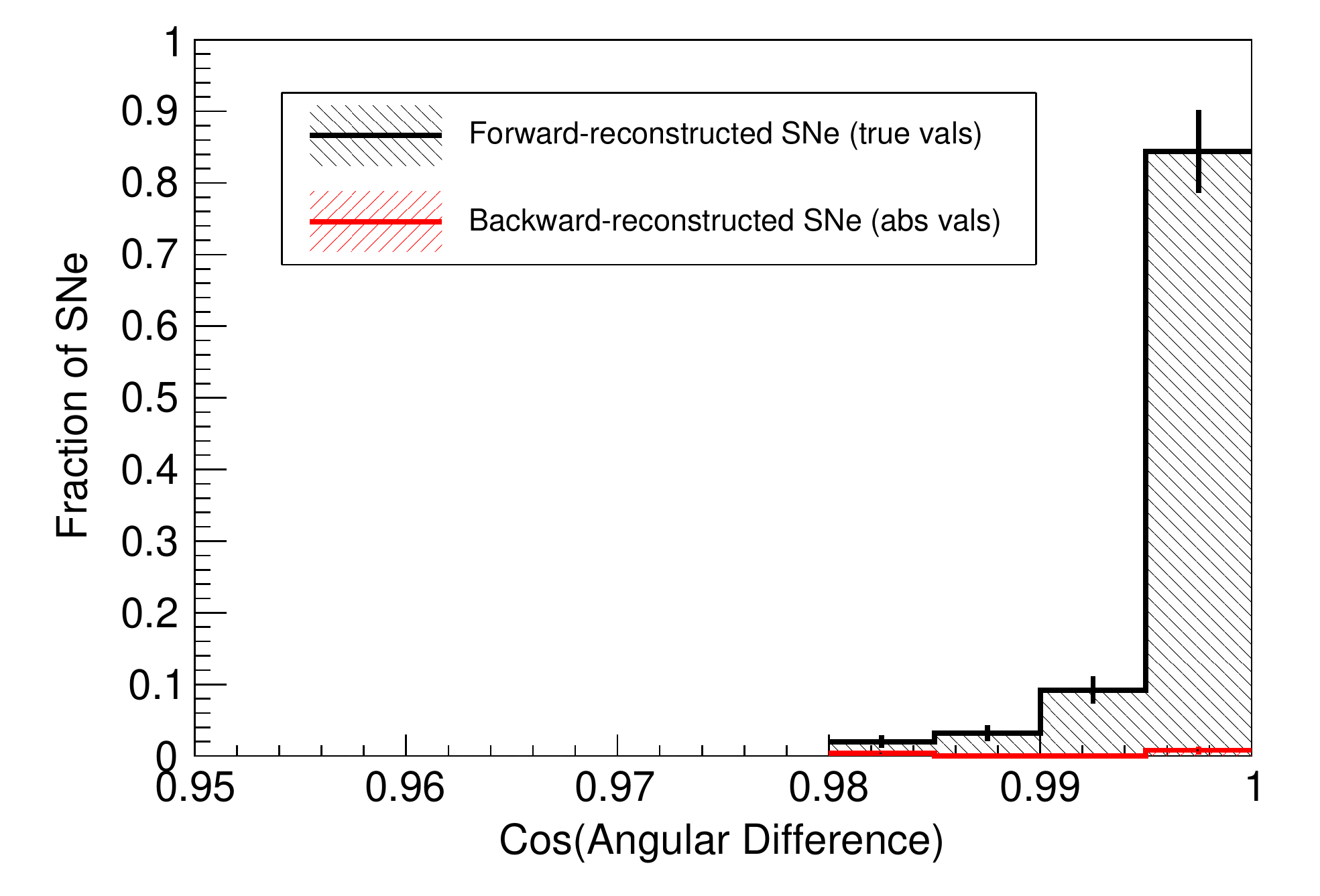}
\end{dunefigure}

\subsection{Searches for Beyond-Standard-Model Physics}

The unique combination of the high-intensity \dword{lbnf} neutrino beam with
\dword{dune}'s \dword{nd}  and massive \dword{lartpc} \dword{fd} modules at a 
\SI{1300}{\km} baseline enables a variety of probes of \dword{bsm} 
physics, either novel or with unprecedented sensitivity.

As examples of the potential impact of \dword{dune}, we present results from the 
analysis of simulated data sets for two \dword{bsm} scenarios, one with 
a sterile neutrino species participating in oscillations, 
and the other with anomalous ``neutrino trident'' events.
%
From the sterile neutrino analysis,  
the \dword{dune} sensitivities to the effective mixing angle $\theta_{\mu e}$
(which depends on new mixing angles $\theta_{14}$ and $\theta_{24}$), from the appearance and disappearance samples at the \dword{nd} and \dword{fd} are shown in Figure~\ref{fig:th_me2}. 

Considering a neutrino trident analysis in \dword{nd} data, 
existing constraints and projected sensitivity to parameters of a $Z^\prime$ boson resulting from the gauging of the difference between 
muon and tau lepton numbers, $L_\mu - L_\tau$, are presented 
in Figure~\ref{fig:LmuLtau2}.  This plot indicates that \dword{dune} 
can cover much of parameter space for which this model is 
able to explain the 
departure of the present observed muon $g-2$ central value 
from standard model expectations.

\begin{dunefigure}
[Sensitivity to effective mixing angle $\theta_{\mu e}$ from 
a DUNE sterile neutrino analysis]
{fig:th_me2}
{DUNE 90\% C.L. sensitivities to $\theta_{\mu e}$ from the appearance and disappearance samples at the \dword{nd} and \dword{fd} is shown along with a comparison with previous existing experiments and the sensitivity from the future \dword{sbn} program. Regions to the right of the DUNE contours are excluded.}
$\vcenter{\hbox{\includegraphics[width=0.70\columnwidth]{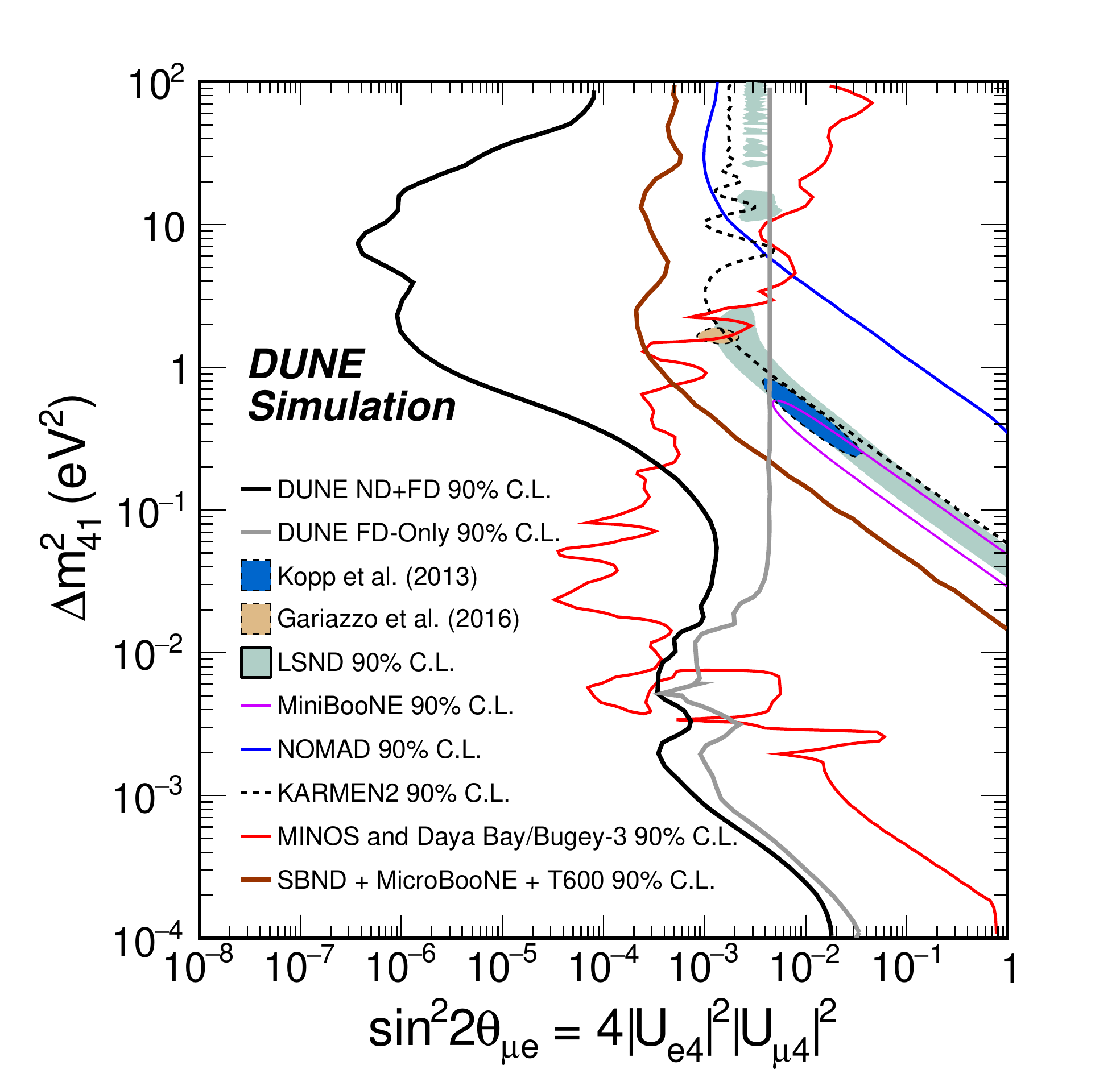}}}$
\end{dunefigure}

\begin{dunefigure}
[Projected sensitivity to BSM 
contributions to neutrino trident events]
{fig:LmuLtau2}
{Existing constraints and projected \dword{dune} sensitivity in the $Z^\prime$ parameter spaced associated with gauging $L_\mu - L_\tau$. Shown in green is the region where the $(g-2)_\mu$ anomaly can be explained at the $2\sigma$ level. The parameter regions already excluded by existing constraints are shaded in gray and correspond to 
a \dword{cms} search for $pp \to \mu^+\mu^- Z' \to \mu^+\mu^-\mu^+\mu^-$~\cite{Sirunyan:2018nnz} (``\dword{lhc}''), 
a BaBar search for $e^+e^- \to \mu^+\mu^- Z' \to \mu^+\mu^-\mu^+\mu^-$~\cite{TheBABAR:2016rlg} (``BaBar''), 
a previous measurement of the trident cross section~\cite{Mishra:1991bv,Altmannshofer:2014pba} (``CCFR''), 
a measurement of the scattering rate of solar neutrinos on electrons~\cite{Bellini:2011rx,Harnik:2012ni,Agostini:2017ixy} (``Borexino''), 
and bounds from Big Bang Nucleosynthesis~\cite{Ahlgren:2013wba,Kamada:2015era} (``BBN''). The \dword{dune} sensitivity shown by the solid blue line assumes a measurement of the trident cross section with $40\%$ precision.}
\includegraphics[width=0.75\textwidth]{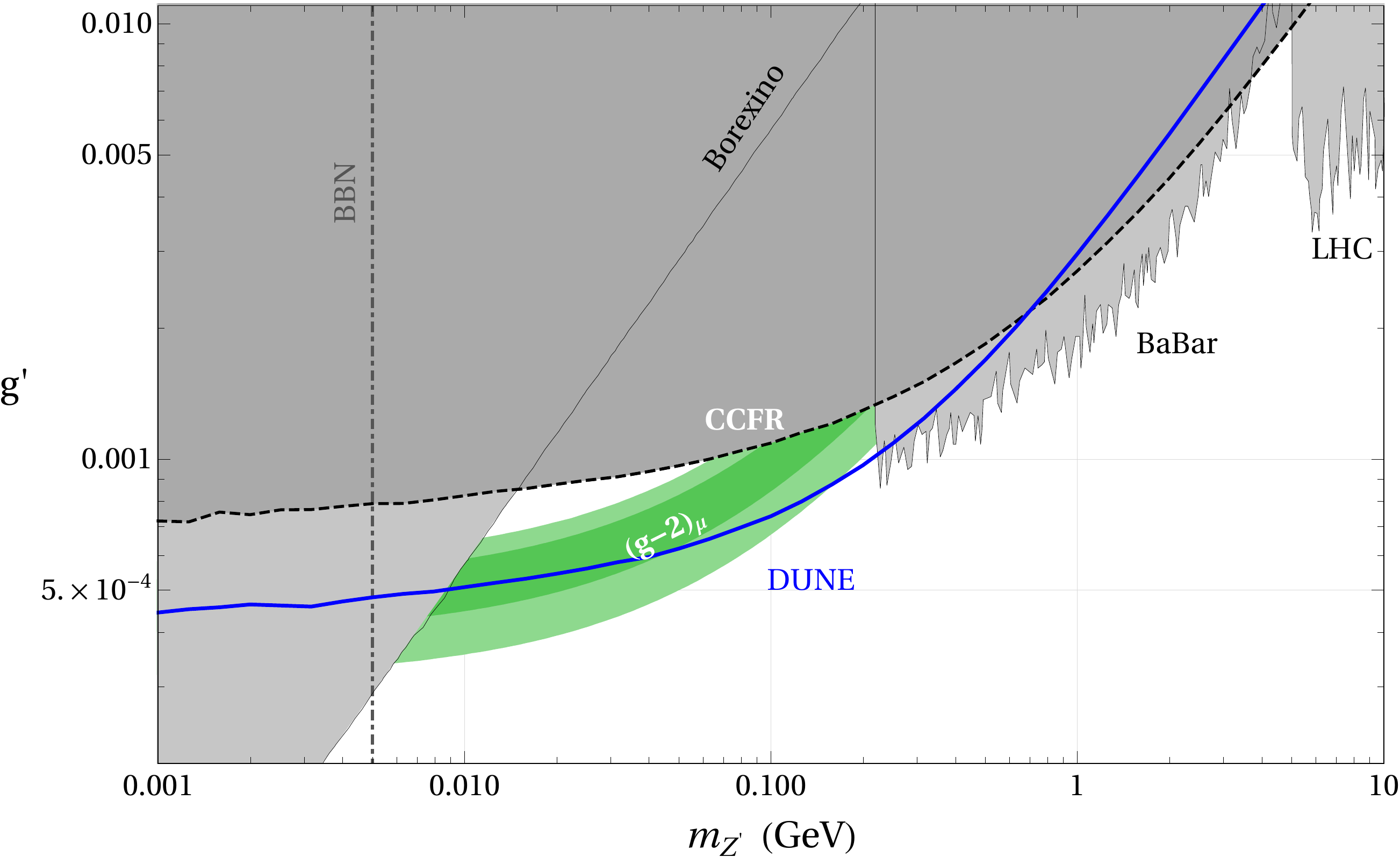}
\end{dunefigure}

\cleardoublepage

\chapter{Single-Phase Far Detector Technology}
\label{ch:exec-sp}

\textit{This chapter provides a brief introduction to the single-phase (SP) far detector technology.  The text below closely follows that found in the introductory chapter of Volume~\volnumbersp{}, \voltitlesp{}, where many more details may be found.}

\section{Overview}
\label{sec:exec-sp-over}

The overriding physics goals of \dword{dune} are to search for leptonic \dword{cpv} and for nucleon decay as a signature of a \dword{gut} underlying the \dword{sm}, as well as to observe neutrino bursts from supernovae. Central to achieving this physics program is constructing a detector that combines the many-kiloton fiducial mass necessary for rare event searches with sub-centimeter spatial resolution in its ability to image those events, allowing us to identify the signatures of the physics processes we seek among the many backgrounds. The \dword{sp} \dword{lartpc}~\cite{Rubbia:1977zz} allows us to achieve these dual goals, providing a way to read out with sub-centimeter granularity the patterns of ionization in $\SI{10}{kt}$ volumes of \dword{lar} resulting from the $O(\SI{1}{MeV})$ interactions of solar and supernova neutrinos up to the $O(\SI{1}{GeV})$ interactions of neutrinos from the \dword{lbnf} beam.

To search for leptonic \dword{cpv}, we must study \nue appearance in the \dword{lbnf} \numu beam. This requires the ability to separate electromagnetic activity induced by \dword{cc} \nue interactions from similar activity arising from photons, such as photons from $\pi^{0}$ decay. Two signatures allow this. First, photon showers are typically preceded by a gap prior to conversion, characterized by the \SI{18}{cm} conversion length in \dword{lar}. Second, the initial part of a photon shower, where an electron-positron pair is produced, has twice the $dE/dx$ of the initial part of an electron-induced shower. To search for nucleon decay, where the primary channel of interest is $p\rightarrow K^{+}\overline{\nu}$, we must identify kaon tracks as short as a few centimeters. It is also vital to accurately fiducialize these nucleon-decay events to suppress cosmic-muon-induced backgrounds, and here detecting argon-scintillation photons is important in determining the time of the event. Detecting a \dword{snb} poses different challenges: those of dealing with a high data rate and maintaining the high detector up-time required to ensure we do not miss one of these rare events. The signature of an \dword{snb} is a collection of MeV-energy electron tracks a few centimeters in length from \dword{cc} $\nu_{e}$ interactions, spread over the entire detector volume. To fully reconstruct an \dword{snb}, the entire detector must be read out, a data-rate of up to $\SI{2}{\tera\byte/\second}$, for \SIrange{30}{100}{s}, including a $\sim\!\SI{4}{s}$ pre-trigger window.

Figure~\ref{fig:LArTPC1ch1} in Section~\ref{ch:dune-det-tech-ov-fd} shows a schematic of the general operating principle of a \dword{sp} \dword{lartpc}. A large volume of \dword{lar} is subjected to a strong \efield of a few hundred volts per centimeter. Charged particles passing through the detector ionize the argon atoms, and the ionization electrons drift in the \efield to the anode wall (called an \dword{apa} array) on a timescale of milliseconds. 
The \dword{spmod} \dword{apa}s consist of layers of active wires strung at angles to each other to form a grid. The relative voltage between the layers is chosen to ensure that all but the final layer are transparent to the drifting electrons, and these first layers produce bipolar induction signals as the electrons pass through them. The final layer collects the drifting electrons, resulting in a monopolar signal.

\dword{lar} is also an excellent scintillator, emitting \dword{vuv} light at a wavelength of \SI{127}{\nano\meter}. This prompt scintillation light, which crosses the detector on a timescale of nanoseconds, is shifted into the visible and collected by \dwords{pd}. The \dword{pd}s can provide a $t_{0}$ determination for events, indicating when the ionization electrons began to drift. Relative to this $t_{0}$, the time at which the ionization electrons reach the anode allows reconstruction of the event topology along the drift direction, which is crucial to fiducialize nucleon-decay events and to apply drift corrections to the ionization charge.

The pattern of current observed on the grid of anode wires provides information for reconstruction in the two coordinates perpendicular to the drift direction. The wire pitch on the wire layers is chosen to optimize considerations of  spatial resolution, cost, and \dword{s/n} of the ionization measurement. \dword{s/n} is important because the measurement of the ionization collected is a direct measurement of the $dE/dx$ of the charged particles, which is what enables both calorimetry and particle identification (\dshort{pid}).

Figure~\ref{fig:DUNESchematic1ch1} in Section~\ref{sec:fdsp-exec-splar} shows a \nominalmodsize fiducial mass \dword{spmod} (\larmass total mass); the key parameters of a \dword{spmod} are listed in Table~\ref{tab:sp-key-parameters}. Inside a cryostat of outer dimensions \cryostatlen (L) by \cryostatwdth (W) by \cryostatht{} (H), shown in Figure~\ref{fig:Cryostat}, four \spmaxdrift drift volumes are created between five alternating \dword{apa} and \dword{cpa} arrays, each of dimensions \sptpclen (L) by \tpcheight (H).

\begin{dunetable}
[Key parameters for a \nominalmodsize  \dshort{fd} \dshort{spmod}]
{p{0.65\textwidth}p{0.25\textwidth}}
{tab:sp-key-parameters}
{Key parameters for a \nominalmodsize  \dshort{fd} \dshort{spmod}.}
Item & Quantity   \\ \toprowrule
TPC size & $\tpcheight{}\times\SI{14.0}{\meter}\times\sptpclen{}$ \\ \colhline
Nominal fiducial mass & \spactivelarmass \\ \colhline
\dshort{apa} size & $\SI{6}{\meter}\times\SI{2.3}{\meter}$ \\ \colhline
\dshort{cpa} size & $\SI{1.2}{\meter}\times\SI{4}{\meter}$ \\ \colhline
Number of \dshort{apa}s & 150 \\ \colhline
Number of \dshort{cpa}s & 300 \\ \colhline
Number of \dshort{xarapu} \dshort{pd} bars & 1500 \\ \colhline
\dshort{xarapu} \dshort{pd} bar size & $\SI{209}{\cm}\times\SI{12}{cm}\times\SI{2}{\cm}$ \\ \colhline
Design voltage & \sptargetdriftvolt \\ \colhline
Design drift field & \spmaxfield \\ \colhline
Drift length & \spmaxdrift \\ \colhline
Drift speed & $\SI{1.6}{\mm/\micro\second}$ \\
\end{dunetable}

\begin{dunefigure}[A far detector (\dshort{fd}) cryostat]{fig:Cryostat}{A  \cryostatlen (L) by \cryostatwdth (W) by \cryostatht{} (H) outer-dimension cryostat that houses a \nominalmodsize \dword{fd} module. A mezzanine (light blue) installed \SI{2.3}{m} above the cryostat supports both  detector and cryogenics instrumentation. At lower left, between the \lar recirculation pumps (green) installed on the cavern floor,  the figure of a person indicates the scale.}
\includegraphics[width=0.8\textwidth]{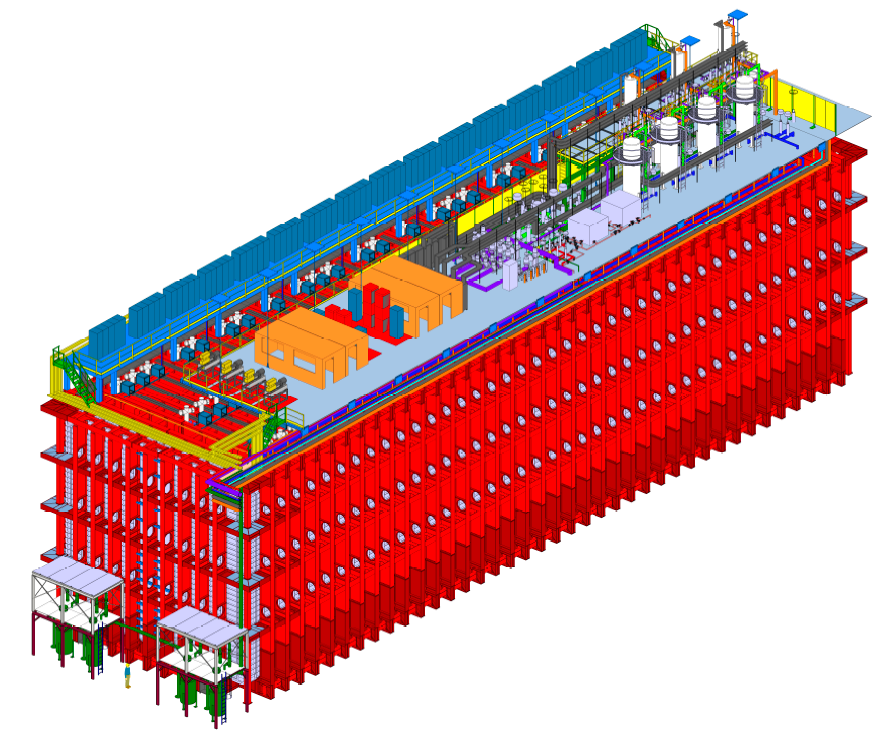}
\end{dunefigure}

The target purity from electronegative contaminants in the argon is $<\!100$ \dword{ppt} O$_{2}$ equivalent, enough to ensure a $>\!\SI{3}{\milli\second}$ ionization-electron lifetime at the nominal \SI{500}{\volt/\centi\meter} drift voltage. This target electron lifetime ensures \dword{s/n} of $>\!5$ for the induction planes and $>\!10$ for the collection planes, which are necessary to perform pattern recognition and two-track separation. 

Nitrogen contamination must be $<\!25$ \dword{ppm} to ensure we achieve our requirement of at least 0.5 \phel{}s per MeV detected for events in all parts of the detector, which in turn ensures 
that we can fiducialize nucleon decay events throughout the detector.

\section{Anode Planes}
\label{sec:exec-sp-apa}

The modular anode walls are each made up of 50 \dword{apa}s (25 along the module length and two high), each $\SI{6}{\meter}\times\SI{2.3}{\meter}$. Figure~\ref{fig:APA} shows a schematic and a photograph. As Figure~\ref{fig:APAStack} shows, the \dword{apa}s hang vertically. The \dword{apa}s are two-sided, with three active wire layers and an additional shielding layer, sometimes called a grid layer, wrapped around them. The wire spacing on the layers is $\sim\!\SI{5}{\mm}$. The collection layer is called the $X$ layer; the induction layer immediately next to that is called the $V$ layer; the next induction layer is the $U$ layer; and the shielding layer is the $G$ layer. The $X$ and $G$ layer wires run vertically when installed  (Figure~\ref{fig:APA} shows them horizontal); the $U$ and $V$ layer wires are at $\pm\,\SI{35.7}{\degree}$ to the vertical. The 
wire spacing on each plane defines the spatial resolution of the \dword{apa}; it is wide enough to keep readout costs low and \dword{s/n} high, but small enough to enable reconstruction of 
short tracks such as few-\si{\cm} kaon tracks from proton decay events.

\begin{dunefigure}[An anode plane assembly (APA)]{fig:APA}
{Top: a schematic of an \dword{apa}. The steel \dword{apa} frame is shown in black. The green and magenta lines indicate the directions of the induction wire layers. The blue lines indicate the directions of the induction and shielding (grid) wire layers. The blue boxes at the right end  are the \dword{ce}. Bottom: a \dword{pdsp} \dword{apa} in a wire-winding machine. The end on the right is the head end, onto which the \dword{ce} are mounted.}
\includegraphics[width=0.7\textwidth]{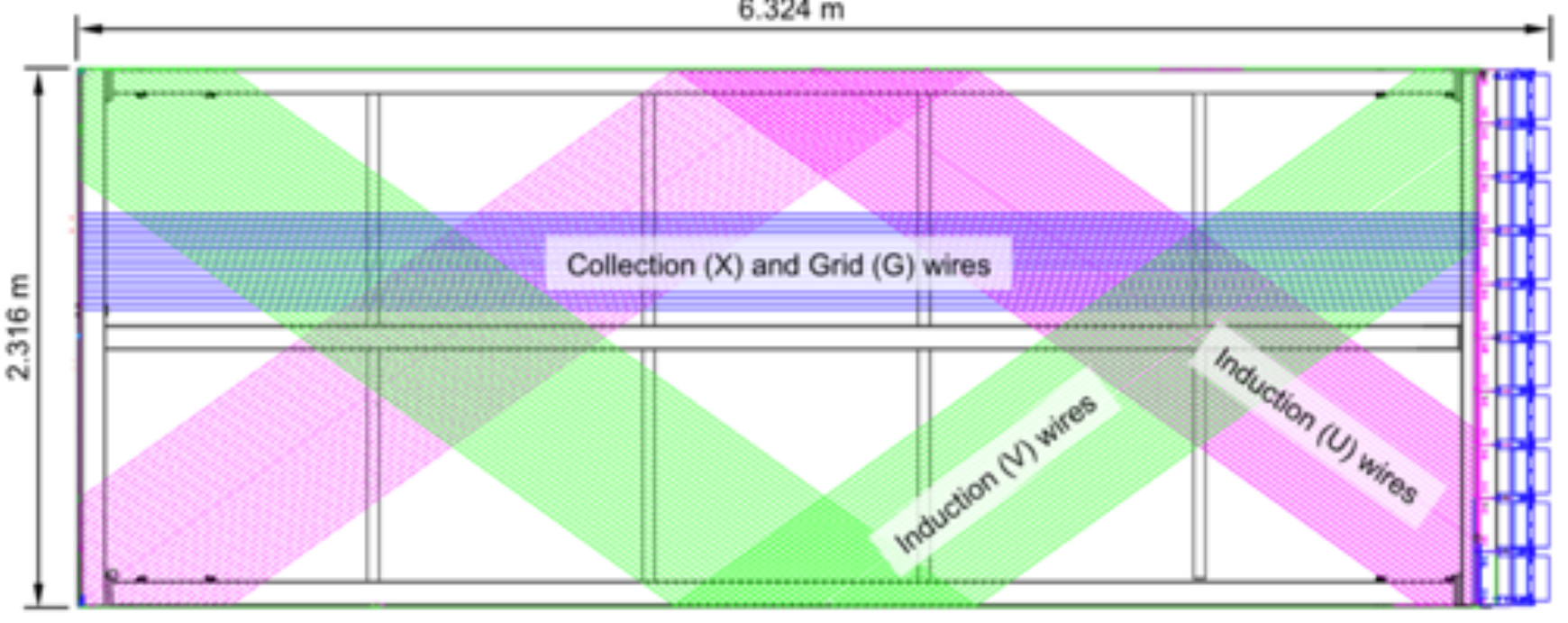}
\includegraphics[width=0.6\textwidth]{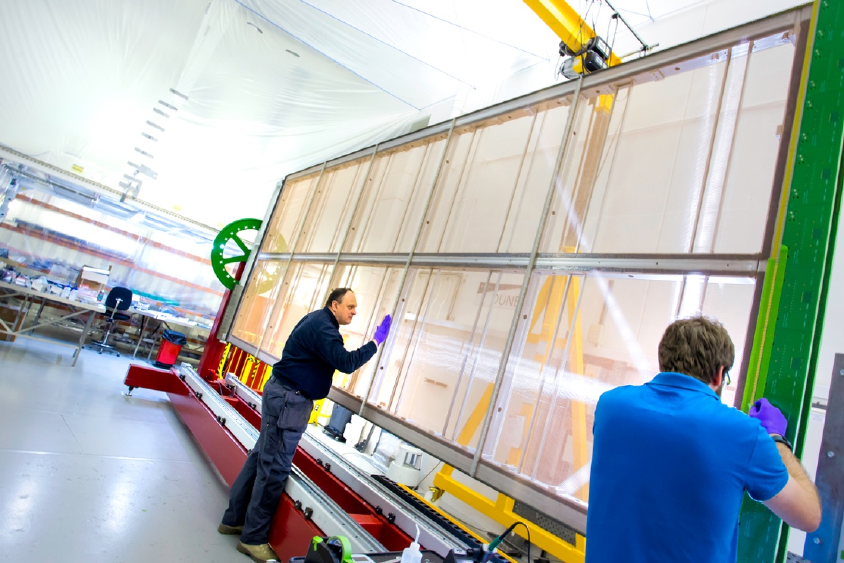}
\end{dunefigure}

\begin{dunefigure}[A stack of two APAs]{fig:APAStack}
{Left: two vertically linked \dword{apa}s form one unit of an \dword{apa} array. \dword{pd} bars are installed across the width of the \dword{apa}s. Right: a zoom into only the top and bottom ends of the \dword{apa} stack (notice the breaks in white). This shows the readout electronics and the center of the stack where the \dword{apa}s are connected.}
\includegraphics[width=\textwidth]{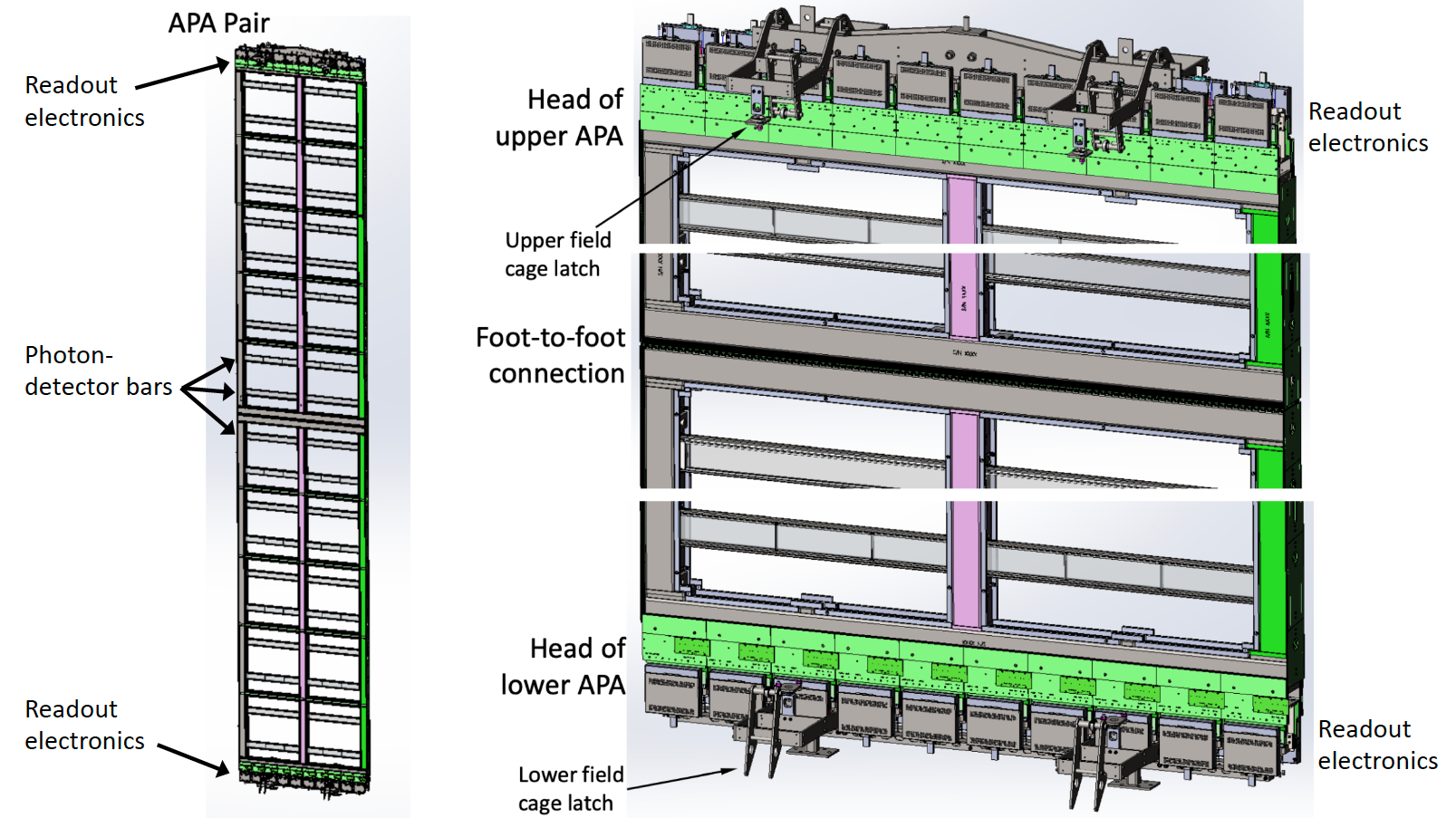}
\end{dunefigure}

\section{Cathode Planes and High Voltage}

Each of the module's two \dfirst{cpa} arrays is formed from 150 \dword{cpa}s (50 along the length, stacked three high), each of which is a $\SI{1.2}{\meter}\times\SI{4}{\meter}$ resistive panel. Each \dword{cpa} has its own independent \dword{hv} supply, providing a current of \SI{0.16}{\milli\ampere} at \sptargetdriftvolt{}. 
With the  \dword{apa} arrays held close to ground, this results in a uniform \spmaxfield \efield across the drift volume.  A typical \dword{mip} passing through the argon produces roughly 60k ionization electrons per centimeter that drift toward the anodes at approximately $\SI{1.6}{\mm/\micro\second}$. The time to cover the full drift distance is about $\SI{2.2}{\milli\second}$.

A \dword{fc} built from field-shaping aluminum profiles surrounds the drift volumes, keeping  the \efield uniform throughout the active \dword{tpc} volume to within 1\%.  The aluminum profiles are connected via a resistive divider chain; between each profile, two \SI{5}{\giga\ohm} resistors, arranged in parallel, provide  \SI{2.5}{\giga\ohm} resistance to create a nominal \SI{3}{\kilo\volt} drop. The \dword{fc} is illustrated in Figure~\ref{fig:FieldCage}.

\begin{dunefigure}[A section of the field cage (FC)]{fig:FieldCage}
{A section of the field cage, showing the extruded aluminum field-shaping profiles, with white polyethylene caps on the ends to prevent discharges.}
\includegraphics[width=0.5\textwidth]{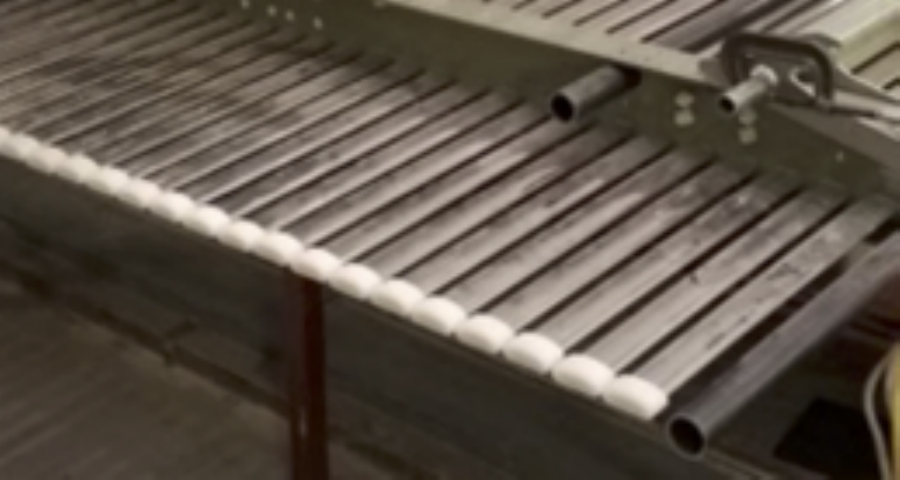}
\end{dunefigure}

\section{Electronics}
\label{sec:exec-sp-electronics}

\Dword{fe} readout electronics (in the \dword{lar}), called \dfirst{ce}, are attached to the top end of the top \dword{apa} and the bottom end of the bottom \dword{apa}. 
Benefitting from a reduction in thermal noise due to the low temperature, the \dword{ce}  
shape, amplify, and digitize the signals from the \dword{apa} induction and collection wires thanks to a series of three different types of \dwords{asic} through which all signals pass. 
Outside the cryostat, signals are passed to \dwords{wib} that put the signals onto \SI{10}{\giga\byte} optical fibers, ten per \dword{apa}, which will carry the signals to the upstream \dword{daq} system in the \dword{cuc}. Each \dword{detmodule} has an independent \dword{daq} system that allows it to run as an independent detector, thereby minimizing any chance of a complete \dword{fd} outage. Modules can, however, provide the others with an \dword{snb} trigger signal. The \dword{daq} system also provides the detector clock. 

To enable observation of low-energy particles, we plan to keep noise below $1000\,e^{-}$ per channel, which should be compared to the 20k -- 30k $e^{-}$ per channel collected from a \dword{mip} traveling parallel to the wire plane and perpendicular to the wire orientation. For large signals, we require a linear response up to 500k $e^{-}$, which ensures that fewer than 10\% of beam events experience saturation. This can be achieved using 12\, \dword{adc} bits. 
The \dword{ce} are designed with an \dword{fe} peaking time of \SI{1}{\micro\second}, which matches the time for the electrons to drift between wire planes on the \dword{apa}; this leads to a design sampling frequency of \SI{2}{\mega\hertz} to satisfy the Nyquist criterion.

\section{Photon Detection System}
\label{sec:fdsp-exec-pds}

In addition to the ionization, charged particles passing through the argon produce approximately 24,000 scintillation photons per \si{\mega\electronvolt}. The scintillation photons are fast, arriving at the \dfirsts{pd} nanoseconds after production. This scintillation light provides a $t_{0}$ for each event. Comparison of the arrival time of ionization at the anode with this $t_{0}$ enables reconstruction in the drift direction. 
The \dword{spmod} implementation enables $\sim\!\SI{1}{\mm}$ position resolution for \SI{10}{\mega\electronvolt} \dword{snb} events. The \dword{pd} $t_{0}$ is also vital in fiducializing nucleon decay events, which allows us to reject cosmic-muon-induced background events that will occur near the edges of the detector modules.

The photons are collected by devices called \dwords{xarapu}, which are mounted in the \dword{apa} frames between the sets of wire layers, as shown in Figure~\ref{fig:APAStack}. 
The \dwords{xarapu} 
consist of layers of dichroic filter and wavelength-shifter, illustrated in Figure~\ref{fig:ArapucaCell}, that shift the \dword{vuv} scintillation light into the visible range  trap  the visible photons, 
and transport them to \dword{sipm} devices. The signals from these \dwords{sipm} are sent along cables that pass through the hollow \dword{apa} frames, up to feedthroughs in the cryostat roof. The \dword{pd} and \dword{apa}-wire data-streams are merged at the \dword{daq}. 

\dword{pd} modules, shown in Figure~\ref{fig:PDModules}, are 
$\SI{209}{\cm}\times\SI{12}{\cm}\times\SI{2}{\cm}$ bars that each hold 24 \dwords{xarapu}. 
Ten \dword{pd}  modules  are mounted in each \dword{apa} between the wire layers. 

\begin{dunefigure}[An \dshort{xarapu} photon detector (PD) cell]{fig:ArapucaCell}
{Left: an \dword{xarapu} cell. Right: an exploded view of the \dword{xarapu} cell, where the blue sheet is the \dword{wls} plate and the yellow sheets are the dichroic filters.}
\includegraphics[width=\textwidth]{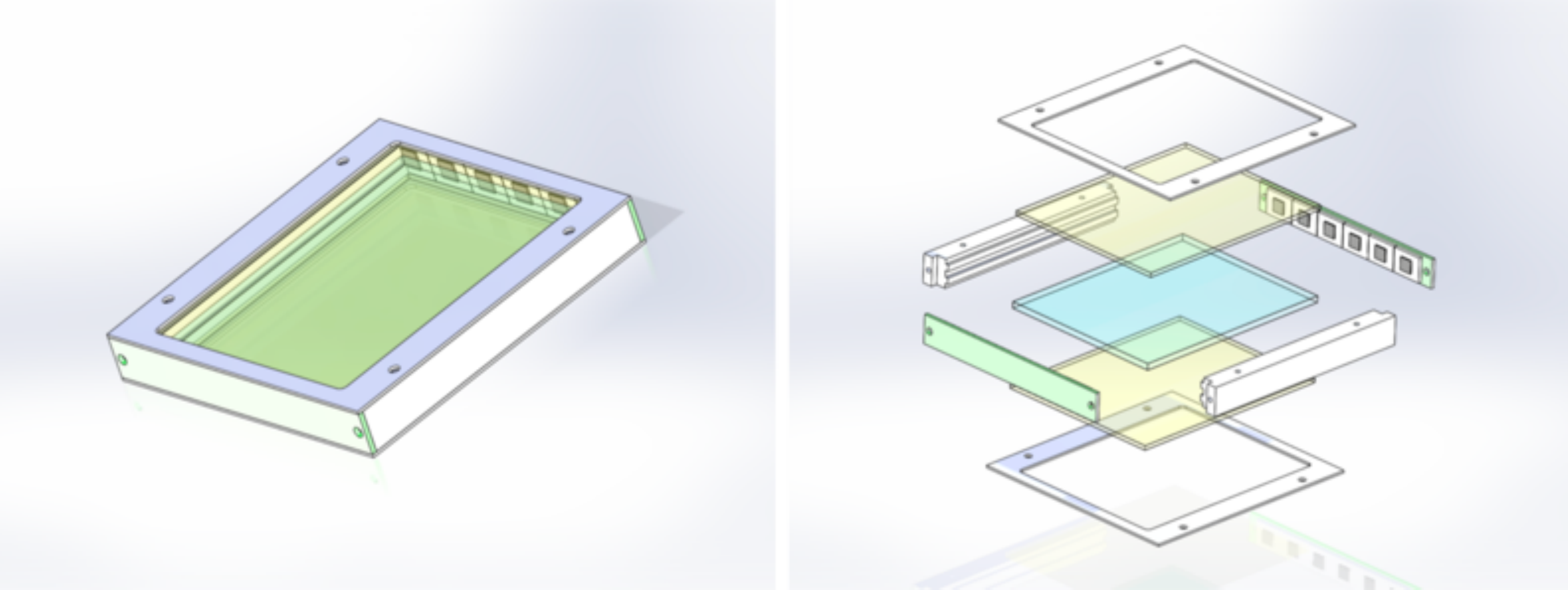}
\end{dunefigure}

\begin{dunefigure}[PD modules mounted in an APA]{fig:PDModules}
{Left: an \dword{xarapu} \dword{pd} module. The 48 \dwords{sipm} that detect the light from the 24 cells are along the long edges of the module. Right: \dword{xarapu} \dword{pd} modules mounted inside an \dword{apa}.}
\includegraphics[width=0.49\textwidth]{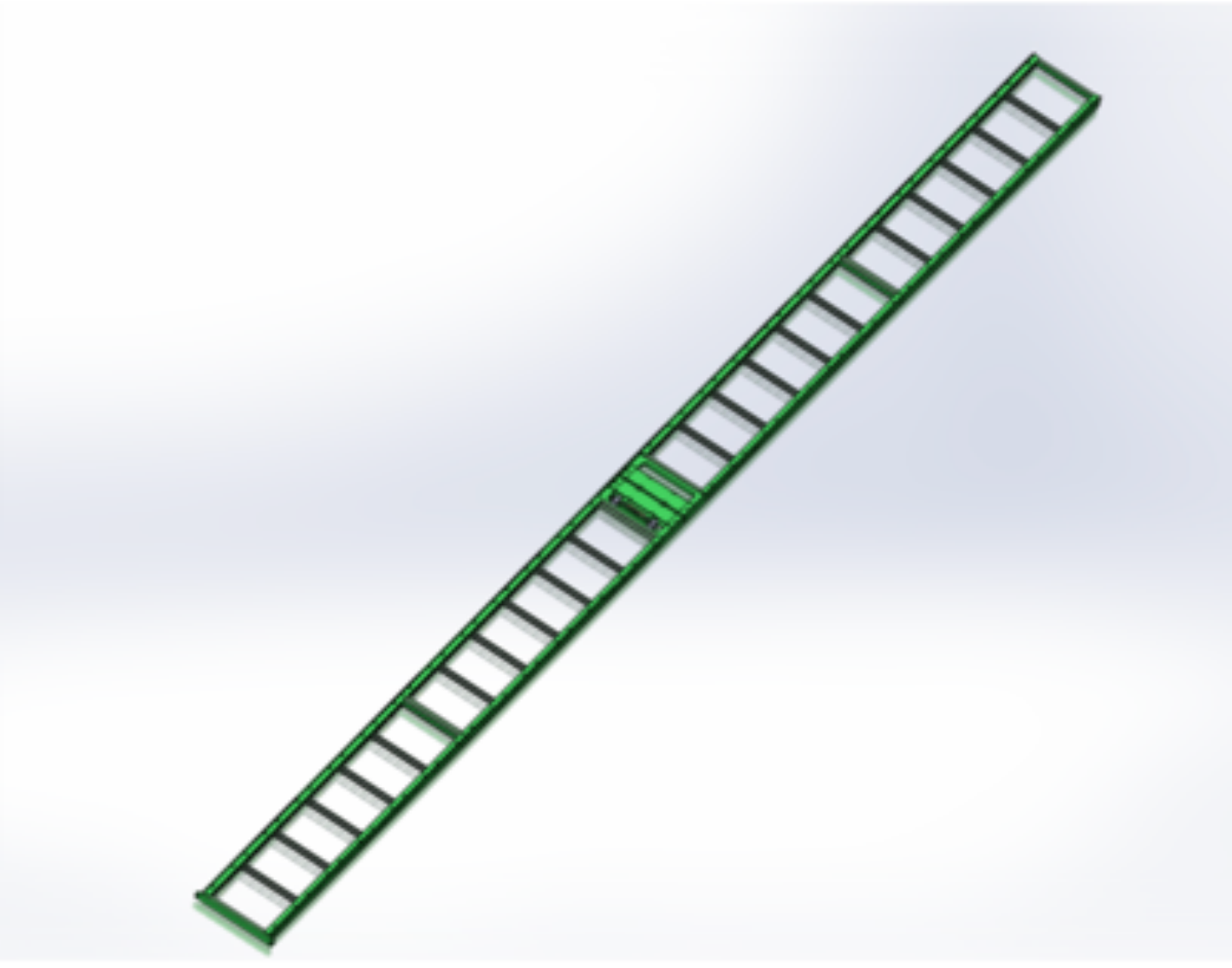}
\includegraphics[width=0.49\textwidth]{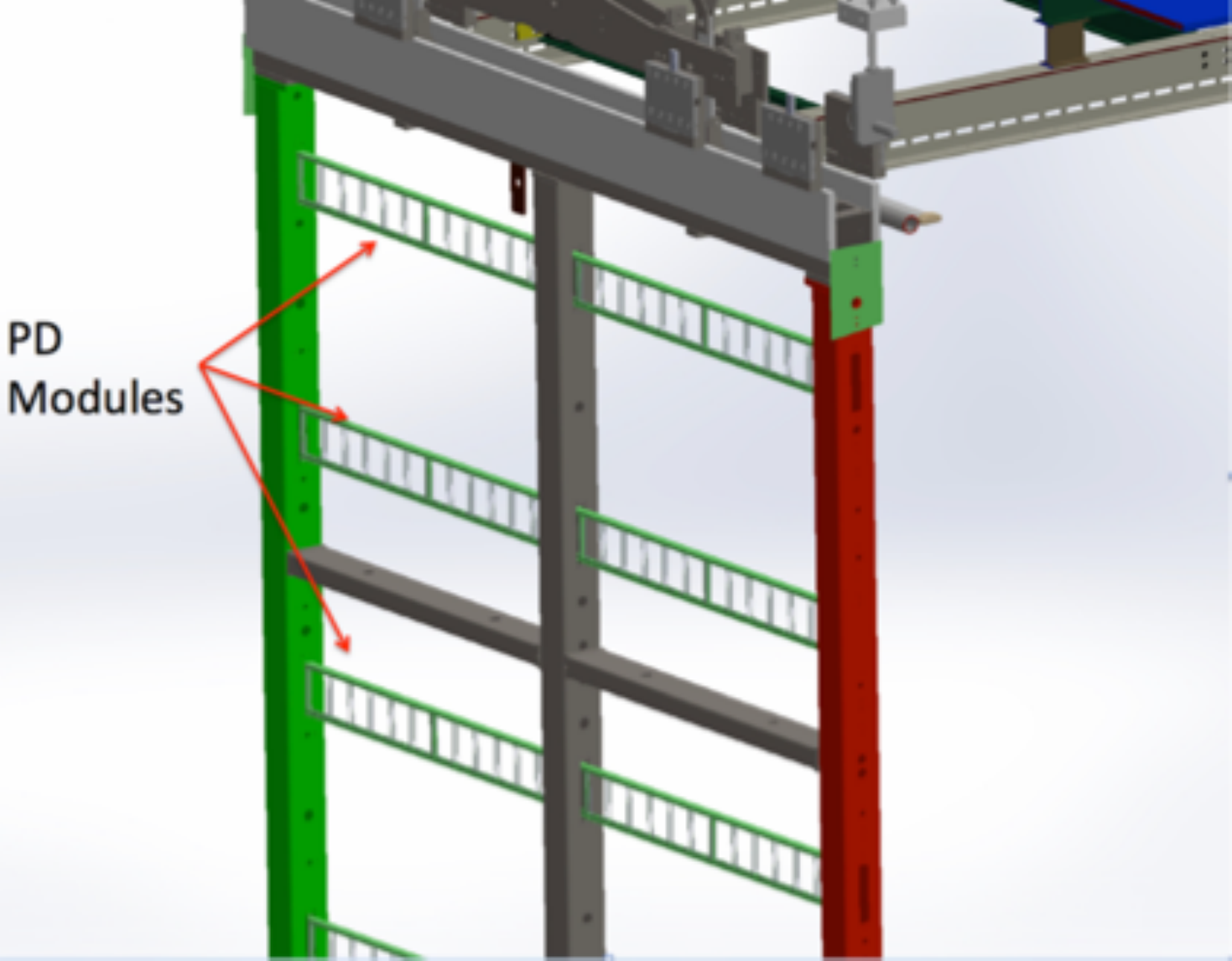}
\end{dunefigure}

The 48 \dword{sipm}s on each \dword{xarapu} supercell are ganged together, and the signals are collected by \dword{fe} electronics, mounted on the supercell. The design of the \dword{fe}  electronics is inspired by the system used for the \dword{mu2e} cosmic-ray tagger~\cite{bib:mu2e_tdr}, which uses commercial ultrasound \dwords{asic}. The \dword{fe} electronics define the \SI{1}{\micro\second} timing resolution of the \dword{pd} system.

\section{Calibration}
\label{sec:exec-sp-calibration}

The challenge of calibrating the \dword{dune} \dword{fd} is controlling the response of a huge cryogenic detector over a period of decades, a challenge amplified by the detector's location deep underground and therefore shielded from the cosmic muons that have typically been used as standard candles for 
\dwords{lartpc}.  The \dword{fd} calibration system  has been designed jointly for the \dword{sp} and \dword{dp} technologies, and uses the same strategies and systems for both.

To achieve our \si{\giga\electronvolt}-scale oscillation and nucleon decay physics goals, we must know our fiducial volume to 1 -- 2\% and have a similar understanding of the vertex position resolution; understand the \nue event rate to 2\%; and control our lepton energy scales to 1\% and hadron energy scales to 3\%. At the \si{\mega\electronvolt} scale, our physics requirements are driven by our goal of identifying and measuring the spectral structure of an \dword{snb}; here, we must achieve a 20 -- 30\% energy resolution, understand our event timing to the \SI{1}{\micro\second} level, and measure our trigger efficiency and levels of radiological background. 

The tools available to us for calibration include the \dword{lbnf} beam, atmospheric neutrinos, atmospheric muons, radiological backgrounds, and dedicated calibration devices installed in the detector. At the lowest energies, we have deployable neutron sources and intrinsic radioactive sources; in particular, the natural $^{39}$Ar component of the \dword{lar} with its \SI{565}{\kilo\electronvolt} end-point, given its pervasive nature across the detector, can be used to measure the spatial and temporal variations in electron lifetime. The possibility of deploying radioactive sources is also under study. 
In the \SIrange{10}{100}{\mega\electronvolt} energy range, we will use Michel electrons, photons from $\pi^{0}$ decay, stopping protons, and both stopping and through-going muons. We will also have built-in lasers, purity monitors, and thermometers, as well as the ability to inject charge into the readout electronics. Finally, data from the \dword{protodune} detectors  (Section~\ref{sec:exec:overall:pdune}) will be invaluable in understanding the response and \dword{pid} capabilities of the \dword{fd}.

Over time, the \dword{fd} calibration program will evolve as statistics from cosmic rays and the \dword{lbnf} beam amass and add to the information gained from the calibration hardware systems. These many calibration tools will work alongside the detector monitoring system, the \dword{cfd} models of the argon flow, and \dword{protodune} data to give us a detailed understanding of the \dword{fd} response across the \dword{dune} physics program.

\section{Data Acquisition}
\label{sec:exec-sp-daq}

The \dword{daq} systems for the \dword{sp} and \dword{dp} technologies have been designed jointly and are identical except for the architecture of the detector readout electronics.  
The output format of the generated data is common, and both are synchronized to the same global clock signals.  The \dword{daq} architecture is based on the \dword{felix} system designed at \dshort{cern} and used for the \dword{lhc} experiments. 

The \dword{daq} is divided between an upstream section, located underground in the \dword{cuc}, and a downstream \dword{daqbes} to be located above ground at the \dword{surf}. All trigger decisions are made 
upstream, and the data is buffered underground until the \dword{daqbes} indicates it is ready to receive data; this controls the rate of data flowing to the surface. An end-goal of the \dword{daq} is to achieve a data rate to tape of no more than \SI{30}{\peta\byte/\year}.

For the \dword{spmod}, the 150 \dword{apa}s are processed by 75 \dwords{daqrou}; each \dword{daqrou} contains one \dword{felix} board. The \dwords{pd} from the module will have a lower data rate because the \dword{pd} electronics, unlike the \dword{tpc} electronics, perform zero-suppression; therefore, the \dwords{pd} of a module will be processed by six to eight additional \dwords{daqrou}. The \dword{daq} can be partitioned: it will be possible to run multiple instances of the \dword{daq} simultaneously, so most of the detector can be taking physics data while other \dword{daq} instances are doing test runs for development or special runs such as calibration runs. 

Two basic triggers will be operating. Beam, cosmic, and nucleon decay events will be triggered using the localized high-energy trigger 
that will open a readout window of \SI{5.4}{\ms}, enough to read out the full \dword{tpc} drift around an event. For \dwords{snb}, we will use an extended low-energy trigger. This will look for coincident regions of low-energy deposits, below \SI{10}{\mega\electronvolt}, across an entire module and in a \SI{10}{\second} period. An extended high-energy trigger will open a readout window of \SI{100}{\second} to capture a full \dword{snb}. 

The \dword{daq} must also provide the system clock that keeps the detector components synchronized and 
timestamps all data. The timestamp derives from a \dword{gps} \dword{pps} fed into the \dword{daq} with \SI{1}{\micro\second} precision, adequate 
for beam and \dword{snb} events. To provide finer synchronization between detector components, a \SI{10}{\mega\hertz} reference clock drives the module's \SI{62.5}{\mega\hertz} master clock, which is fanned out to all detector components, providing an overall synchronization to a precision of \SI{1}{\nano\second}.

\section{Cryogenics Instrumentation and Slow Controls}
\label{sec:dp-execsum-sc}

\fixme{Anne rewrote this from text from the SP/DP volumes because the previous text was DP-centric. Somebody should check this.}

\dword{dune}'s \dword{cisc} system is responsible for recognizing and preventing fault conditions that could develop in a \dword{detmodule} over long periods of running. 
The \dword{sp} and \dwords{dpmod} will use a \dword{cisc} system that has been designed jointly.

Cryogenics instrumentation includes purity monitors,  various types of temperature monitors, and cameras with their associated light emitting systems. Also included are 
gas analyzers and \dword{lar} level monitors that are directly related to the external cryogenics system, which have substantial interfaces with \dword{lbnf}. 

Cryogenics instrumentation 
requires significant engineering, physics, and
simulation work, such as \efield simulations and cryogenics modeling
studies using \dfirst{cfd}. \efield simulations
identify desirable locations for instrumentation
devices in the cryostat, away from 
regions of high \efield, so that 
their presence does not induce large field distortions. 
\dword{cfd} simulations help identify 
expected temperature, impurity, and velocity flow distributions and guide the placement and distribution of instrumentation devices inside the cryostat.

The slow controls portion of \dword{cisc} consists of three main components: 
hardware, infrastructure, and software. The slow controls hardware and infrastructure comprises networking hardware, signal processing hardware, computing hardware, and associated rack infrastructure. The slow controls software provides, for every slow control quantity, the central slow controls processing architecture, databases, alarms, archiving, and control room displays.

\section{Installation}

A significant challenge for \dword{dune} is transporting all detector and infrastructure components down the \SI{1500}{\meter} Ross shaft and through drifts to a detector cavern. The 150 \dwords{apa}, each \SI{6.0}{m} high and \SI{2.3}{m} wide, and  weighing \SI{600}{kg} with $3500$ strung sense and shielding wires, must be taken down the shaft as special ``slung loads,'' presenting an extra challenge. 

Once the \dword{spmod} cryostat is ready, 
a \dword{tco} is left open at one end through which the detector components are installed. A cleanroom is built around the \dword{tco} to prevent any contamination entering the cryostat during installation. The \dword{dss} is then installed into the cryostat, ready to receive the \dword{tpc} components. 

In the cleanroom the \dword{apa}s are outfitted  with \dword{pd} units and passed through a series of qualification tests.
Two \dword{apa}s are linked into a vertical \SI{12}{m} high double unit and connected to readout electronics. 
They are tested in a \coldbox, then move into the cryostat to be installed at the proper location on the \dword{dss}, and have their cabling connected to \fdth{}s. 
The \dword{fc}, \dwords{cpa} and their \dword{hv} connections, elements of the \dword{cisc}, and detector calibration systems are installed in parallel with the \dword{apa}s. 

After twelve months of detector component installation, the \dword{tco} closes (the last installation steps occur in a confined space accessed through a narrow human-access port on top of the cryostat). 
Following leak checks, final electrical connection tests, and installation of the neutron calibration source, the process of filling the cryostat with \SI{17000000}{\kilo\gram} of \dword{lar} begins.

To help plan the installation phase, installation tests will be performed at the \dword{nova} \dword{fd} site in \dword{ashriver}, Minnesota, USA. These tests will allow us to develop our procedures, train installation workers, and develop our labor planning through time and motion studies. Throughout the project, safety, \dword{qa}, and \dword{qc} are written into all processes.

Safety of both personnel and  detector components is the paramount consideration throughout the installation process and beyond. Once the detectors are taking data, 
the \dword{ddss} will be monitoring for argon level drops, water leaks, and smoke. A detailed detector and cavern grounding scheme has been developed that not only guards against ground loops but also ensures that any power faults are safely shunted to the facility ground.

\cleardoublepage

\chapter{Dual-Phase Far Detector Technology}
\label{ch:exec-dp}

\textit{This chapter provides a brief introduction to the dual-phase far detector technology.  The text below closely follows that found in the introductory chapter of Volume~\volnumberdp{}, \voltitledp{}, where many more details may be found.}

\section{Overview}
\label{sec:dp-execsum-introduction}

\dword{dune}'s rich physics program, with discovery potential for \dword{cpv} in the neutrino sector and its capability to make significant observations of nucleon decay and astrophysical events is enabled by the exquisite resolution of the \dword{lartpc} detector technique, which the \dword{dp} design further augments relative to the \dword{sp} design. 
The operating principle of a \dword{dp} \dword{lartpc}, illustrated in Figure~\ref{fig:figure-label-DPprinciple}, is very similar to that of the \dword{sp} design (Figure~\ref{fig:LArTPC1ch1}).  Charged particles that traverse the active volume of the \dword{lartpc} ionize the medium while also producing scintillation light. 
In a \dword{dpmod} ionization charges drift vertically in \dword{lar} and are transferred
into a layer of gas above the liquid where they deposit their charge on a segmented anode. This design allows for a single, fully homogeneous \dword{lar} volume, offering a much longer drift length and reducing  the quantity of nonactive materials in the \dword{lar}.
While the longer drift length requires a higher voltage (up to \SI{600}{kV}) on the cathode, the \dword{dp} design improves the \dword{s/n} ratio in the charge readout, reducing the threshold for the smallest observable signals, while also achieving a finer readout granularity.  
Other advantages of the \dword{dp} design include accessible readout electronics and fewer detector components, reducing costs and simplifying installation logistics.
The precision tracking and calorimetry offered by the \dword{dp} technology provides excellent capabilities for identifying interactions of interest while mitigating sources of background.  

\begin{dunefigure}[Principle of the \dual readout]{fig:figure-label-DPprinciple}{Principle of the DP readout.}
\includegraphics[width=0.6\textwidth]{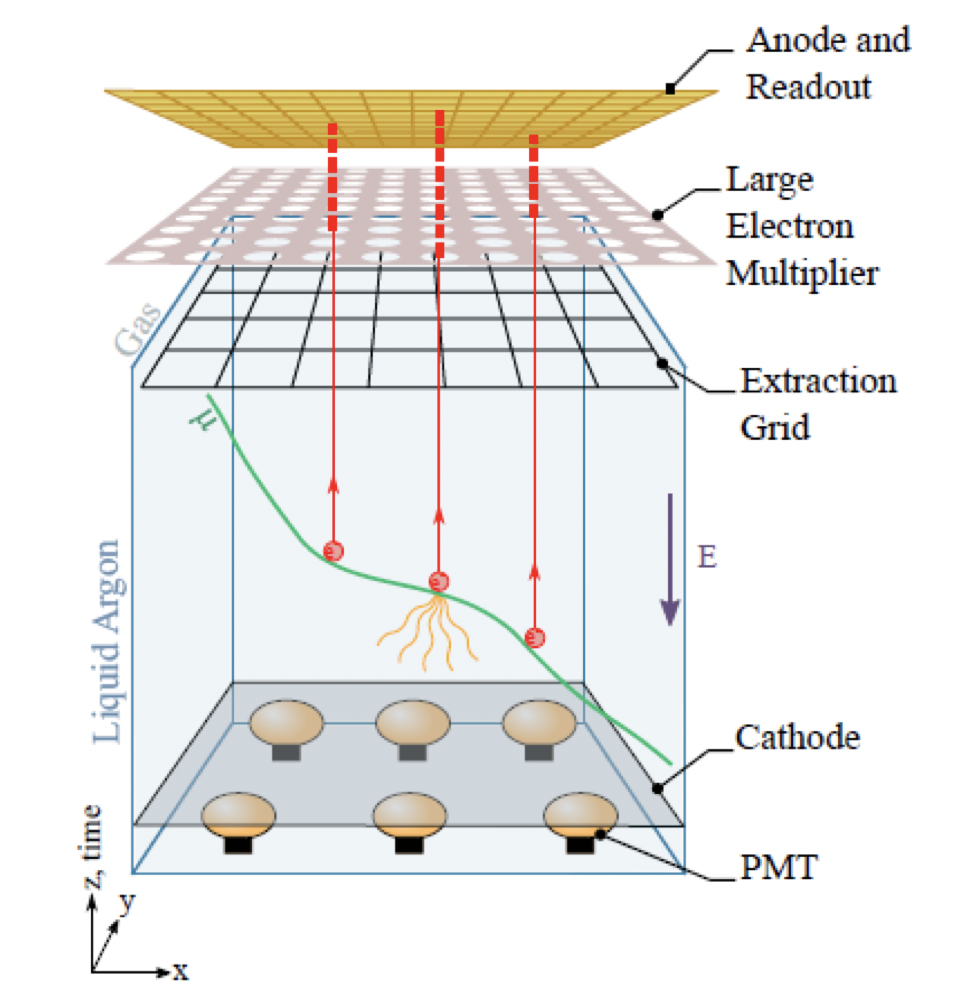}
\end{dunefigure}

The argon scintillation light, at a wavelength of  \SI{127}{nm}, is deep in the UV spectrum. It is recorded by an array of \dwords{pmt} located below the cathode at the bottom of the cryostat.  The \dwords{pmt}, coated with a \dword{wls} material, shift the light closer to the visible spectrum and record the time and pulse characteristics of the incident light.

Two of the key factors that affect the performance of a \dword{lartpc} are argon purity and noise.  The \dword{dp} and \dword{sp} designs have slightly different purity requirements to cope optimally with the different drift lengths. We express the purity level in terms of electron lifetime: a minimum of \SI{5}{ms} for \dword{dp} versus \SI{3}{ms} for \dword{sp}. In both cases the levels of electronegative contaminants in the \dword{lar} (e.g., oxygen and water) must remain 
at \dword{ppt} levels.  
To clearly discern the drifting electrons over the baseline of the electronics, the \dword{tpc}  electronic readout noise must be kept very low. This requires use of low-noise cryogenic electronics. 
Amplification of the electron signal in the gas phase mitigates the potential effect of both factors on the performance. 

\section{Features of the Dual-Phase Design}
\label{sec:dp-execsum-description}

A \dword{dpmod}, with a  \dpactivelarmass{} active mass \dword{lartpc} and dimensions (LWH) \dptpclen by \dptpcwdth by \tpcheight{}, includes all associated cryogenics, electronic readout, computing, and safety systems. The module is built as a single active volume, with the anode at the top in the gas volume, the cathode near the bottom, and an array of \dwords{pd} underneath the cathode. The active volume (see Figure~\ref{fig:DPdet1}) is surrounded by a \dword{fc}. 
The \dword{dp} design maximizes the active volume within the confines of the membrane cryostat while minimizing dead regions and the presence of dead materials in the drift region.  The detector elements are all modular to facilitate production and to allow for transport underground.

The key differentiating concept of the \dword{dp} design is the amplification of the ionization signal in an ``avalanche'' process.  Ionization electrons drift upward toward an extraction grid situated just below the liquid-vapor interface. After reaching the grid, an \efield stronger than the drift field extracts the electrons from the liquid upward into the ultra-pure argon gas. 
Once in the gas, the electrons encounter detectors, called \dwords{lem}, that have a micro-pattern of high-field regions in which the electrons are greatly amplified (via avalanches caused by Townsend multiplication). 
The amplified charge is collected on an anode. The use of avalanches to amplify the charges in the gas phase increases the \dword{s/n} ratio by at least
a factor of ten, with the goal of achieving a gain of about 20, which will significantly improve the
event reconstruction quality.

The modular extraction grids, \dwords{lem}, and anodes are assembled into three-layered sandwiches with precisely defined inter-stage distances and inter-alignment,  which are then connected together horizontally into modular detection units that are \num{9}~m$^2$. These composite detection units, called \dwords{crp}, are discussed in Section~\ref{sec:dp-execsum-crp}.
A \dword{crp} provides an adjustable charge gain and two independent, orthogonal readout views, each with a pitch of \dpstrippitch. It collects the charge projectively, with practically no dead region. Together, the time information  ($t_0$) from the \dword{lar} scintillation readout and the  \threed track imaging of the \dwords{crp} provide $dE/dx$ information.  

Slow-control \fdth{}s,  one per \dword{crp}, are used for level meter and temperature probe readout,  for pulsing calibration signals, and to apply \dword{hv} bias on the two sides of the \dwords{lem} and on the extraction grid. Calibration and \dword{cisc} systems for the \dword{sp} and \dword{dp} technologies have been designed jointly, and are discussed in Sections~\ref{sec:exec-sp-calibration} and~\ref{sec:dp-execsum-sc}, respectively.

Signals in each \dword{crp} unit are collected via three \dwords{sftchimney} on the roof of the cryostat that house the \dword{fe} cards with the (replaceable) cryogenic \dword{asic} amplifiers.  The only active electronics elements inside the cryostat are the \dword{pmt} bases.  Each \dword{sftchimney} is coupled to a \dword{utca} crate to provide signal digitization and \dword{daq}. These crates are connected  via optical fiber links to the \dword{daq} back-end. The total number of readout channels  per \nominalmodsize module is \dpnumcrpch.

Figure~\ref{fig:DPdet1} shows the \dword{dpmod}'s main components. The number of components and corresponding parameters for a \dpactivelarmass \dword{dpmod} are summarized in Table~\ref{tab:DP_numbers}.

\begin{dunefigure}[Diagram of a \dshort{dpmod}]{fig:DPdet1}
  {A \dword{dpmod} with cathode, \dwords{pmt}, \dword{fc}, and anode plane with \dwords{sftchimney}.}
  \includegraphics[width=0.9\textwidth]{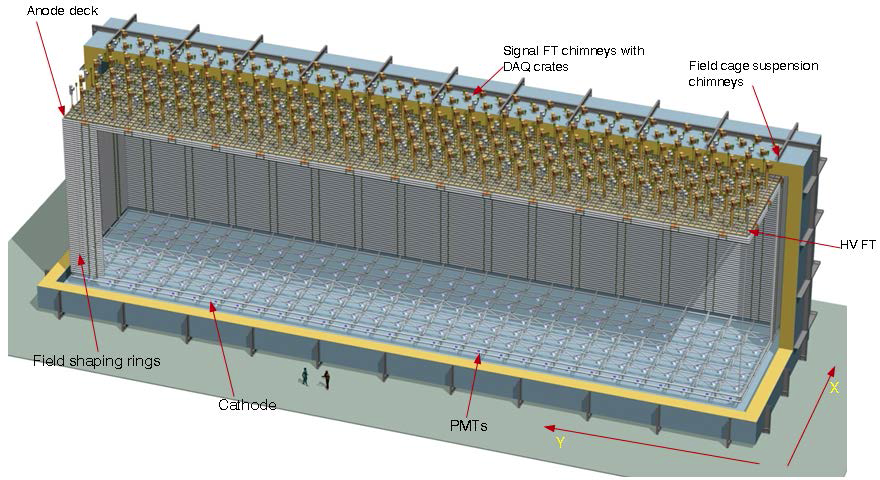}
\end{dunefigure}

\begin{dunetable}[\dshort{dpmod} component quantities and parameters]{ll}{tab:DP_numbers}{\dshort{dpmod} component quantities and parameters.}  Component & Value    \\ \toprowrule
Anode plane size & W = \dptpcwdth, L = \dptpclen \\ \colhline
\dshort{crp} unit size & W = \SI{3}{m}, L = \SI{3}{m}  \\ \colhline
\dshorts{crp} & \num{4}\,$\times$\,\num{20} = \dptotcrp \\ \colhline
\dshort{crp} channels & \dpnumcrpch  \\ \colhline 
\dshort{lem}-anode sandwiches per \dword{crp} unit & \dpswchpercrp \\ \colhline 
\dshort{lem}-anode sandwiches (total) & \dpnumswch \\ \colhline
\dshorts{sftchimney} per \dword{crp} unit & \num{3} \\ \colhline
\dshorts{sftchimney} & \num{240} \\ \colhline
Charge readout channels per \dword{sftchimney} & \num{640}  \\ \colhline
Charge readout channels (total) & \dpnumcrpch \\ \colhline
Suspension \fdth per \dword{crp} unit & \num{3}  \\ \colhline
Suspension \fdth{}s  (total)& \num{240}  \\ \colhline
Slow control \fdth{}s   (total)& \num{80} \\ \colhline
\dshort{hv} \fdth & \num{1}  \\ \colhline
Nominal drift \efield & \SI{0.5}{kV/cm}  \\ \colhline
Nominal/target \dshort{hv} for vertical drift & \dpnominaldriftfield{}/\dptargetdriftvoltpos \\ \colhline
\dshort{fc} voltage degrader resistive chains & \num{12} \\ \colhline
\dshort{fc} cathode modules & \num{15}  \\ \colhline
\dshort{fc} rings & \num{199}     \\ \colhline
\dshort{fc} modules (\SI{4}{m}$\times$\SI{12}{m}) & \num{36}  \\ \colhline
\dshorts{pmt}  & \dpnumpmtch (\num{1}/m$^2$) \\  \colhline
\dshort{pmt} channels & \dpnumpmtch  \\ 
\end{dunetable}

\section{Charge Readout Planes}
\label{sec:dp-execsum-crp}

The  collection, amplification, and readout components of the \dword{tpc} are combined into  layered modules called \dwords{crp}. The charge is collected in a finely segmented readout anode plane at the top of the gas volume and fed to the \dword{fe} readout electronics. The  \dword{crp}'s amplification components, the \dwords{lem}, are horizontally oriented \SI{1}{mm}-thick printed circuit boards (\dwords{pcb}) with electrodes on the top and bottom surfaces.  
The \dword{crp} structure also integrates the immersed extraction grid, which is an array of $x$- and $y$-oriented stainless steel wires, \SI{0.1}{mm} in diameter, with a \dpstrippitch pitch. Figure~\ref{fig:CRP_struct} shows the thicknesses and possible biasing voltages for the different \dword{crp} layers.

Each \dword{crp} is made up of several independent \num{0.5}\,$\times$\,\SI{0.5}{m$^2$} units, each of which is composed  of a \dword{lem}-anode ``sandwich.''  
The anode is a two-dimensional (\twod) \dword{pcb} with two sets of \SI{3.125}{mm}-pitch gold-plated copper strips that provide the $x$ and $y$ coordinates (and thus two views) of an event. Both the \dwords{lem} and anodes are produced in units of $50 \times 50\, $cm$^2$. 
The \dwords{crp} are embedded in a mechanically reinforced frame of \frfour and iron-nickel invar alloy. 

An extraction efficiency of \num{100}\,\% of the electrons from the liquid to the gas phase is achieved with an \efield of the order of \SI{2}{kV/cm} across the liquid-gas interface, applied between the  extraction grid immersed in the liquid and charge amplification devices situated above, in the argon gas. 

\begin{dunefigure}[Thicknesses and HV values for electron extraction from liquid to gaseous Ar]{fig:CRP_struct}
{Thicknesses and \dword{hv} values for electron extraction from liquid to gaseous argon, their  multiplication by \dwords{lem}, and their collection on the $x$ and $y$ readout anode plane. The \dword{hv} values are indicated for a drift field of \SI{0.5}{kV/cm} in \dword{lar}.}
\includegraphics[width=0.8\textwidth]{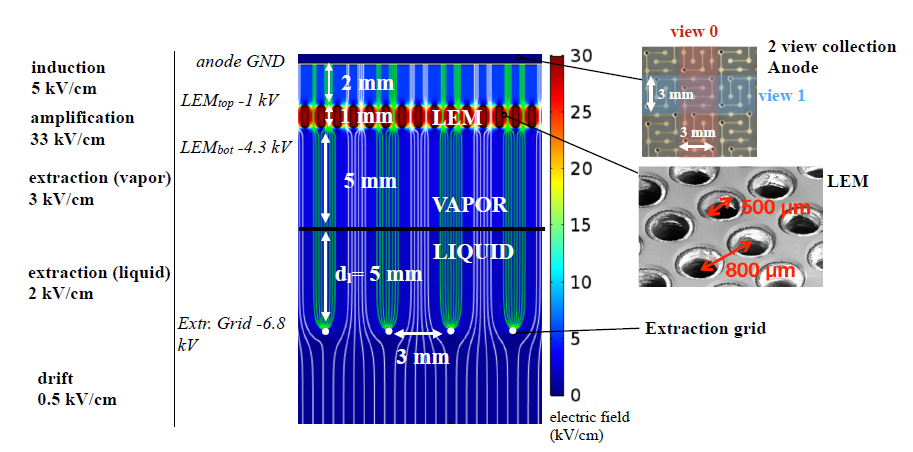}
\end{dunefigure}

The \dwords{lem} are drilled through with many tiny holes (these are the high-field regions), that collectively form a micro-pattern structure. When a \SI{3}{kV} potential difference is applied across the a \dword{lem}'s electrodes, the high \efield (\SI{30}{kV/cm}) produces avalanches (via Townsend multiplication) that amplify the ionization electrons.

Each \dword{crp} is independently suspended by three stainless-steel ropes linked to the top deck of the cryostat. This suspension system allows adjustment of the \dword{crp} height and level such that it remains parallel to the \dword{lar} surface and the extraction grid remains immersed.   

\section{Cathode, Field Cage, and HV System}
\label{sec:dp-execsum-cathode}

The drift field (nominal: E ${\simeq}$ \SI{0.5}{kV/cm}, minimum: E ${\simeq}$ \SI{0.25}{kV/cm}) inside the fully active \dword{lar} volume is produced by applying \dword{hv} to the cathode plane at the bottom of the cryostat and is kept uniform by the \dword{fc}, a stack of \num{199} equally spaced, field-shaping electrodes. 
These electrodes are set to linearly decreasing voltages starting from the cathode voltage at the bottom of the \dword{detmodule} to almost ground potential at the level of the \dword{crp}.

The cathode plane  is suspended from the \dword{fc} and hangs near the  bottom of the cryostat. It consists of 15 adjacent \SI{4x12}{m} 
modules to span the \dptpclen length of \dword{dpmod}. 

As shown in  Figure~\ref{fig:dp-execsum-dune-dp-cathode}, each cathode module is constructed of two \SI{12}{m} long trusses made from thin-walled stainless steel tubes with an  outer diameter of approximately \SI{50}{mm}.

\begin{dunefigure}[A \dshort{dp} cathode module]{fig:dp-execsum-dune-dp-cathode}
{Illustration of a \dword{dp} cathode module.  It is constructed using a pair of stainless steel trusses (blue) as the framework with an array of coated \dword{frp} rods. 
The lower-left inset shows the resistive interconnections and the lifting tab on the cathode truss structure. The upper-right inset is shows the resistive union.}
\includegraphics[width=0.9\textwidth]{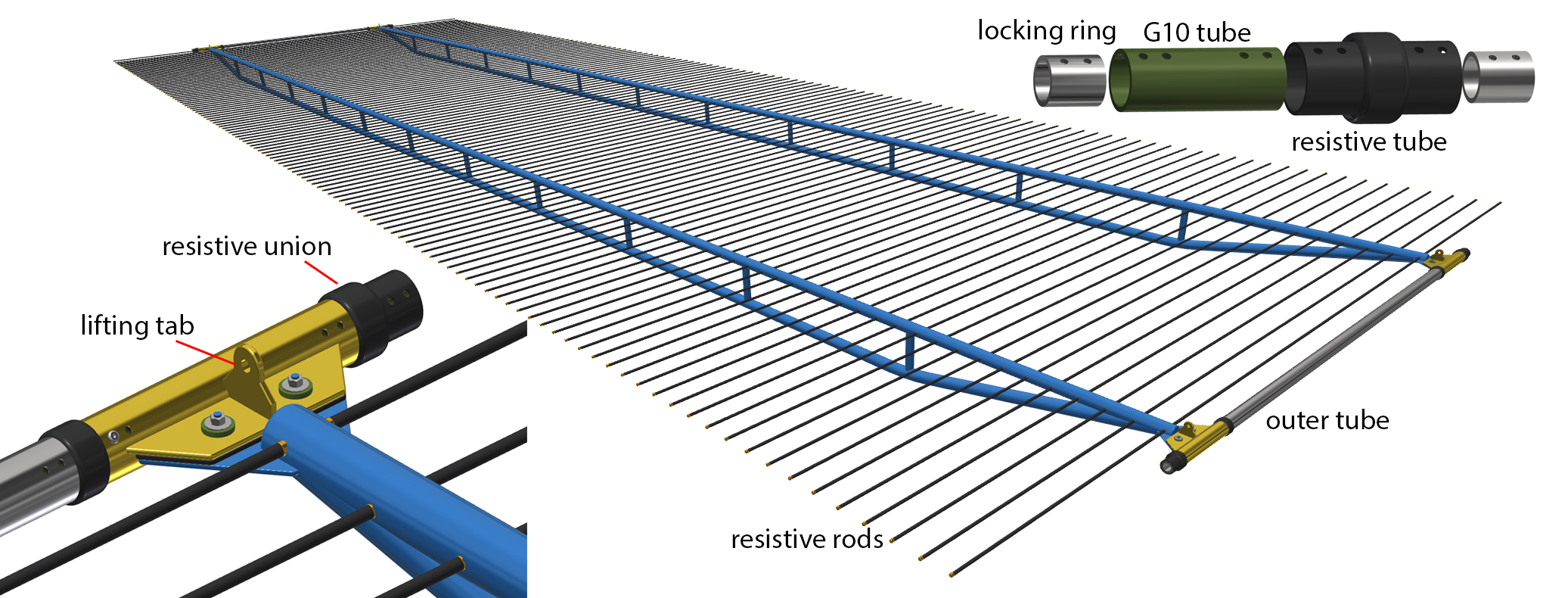}
\end{dunefigure}

A set of 80 \SI{3x3}{m} \dword{gg} modules, standing on the cryostat floor, are installed to protect the array of \dwords{pmt} against any electric discharge from the cathode.

The \dword{hv} for \dword{pddp} was designed for \SI{300}{kV}, corresponding to
a minimal requirement of \efield$\,\simeq\,$\SI{0.25}{kV/cm} for the \dword{dpmod}. We can test the equivalent
of a \SI{12.0}{m} drift in \dword{pddp} by further reducing the drift field, and in contradistinction, we expect to test the full \SI{600}{kV} in a second phase of \dword{pddp} after making some changes to the cathode and the \dword{hv} feedthrough. 
Fermilab is also considering building a test setup for \SI{600}{kV} operation and \SI{12.0}{m} drift. 

\section{Readout Electronics and Chimneys}
\label{sec:dp-execsum-electronics}

The electrical signals from the collected charges are passed to the outside of the cryostat via a set of dedicated \dwords{sftchimney}, tightly-fit pipes that penetrate the top layer of the cryostat insulation, and are therefore exposed to cryogenic temperatures at their lower ends and to room temperature above the cryostat. They are filled with nitrogen gas and closed at the top and  bottom by ultra-high-vacuum flanges (warm and cold).  

The cryogenic analog \dword{fe} electronics cards, mounted on \SI{2}{m} long blades that slide on lateral guides that are integrated into the mechanical structure of the \dword{sftchimney}, are installed at the bottom of the chimney and plugged into the top side of the cold flange. 
This arrangement allows access to and replacement of the cards from the outside. 
 The warm flange connects the analog differential signals to external digitization cards. In the other direction, it distributes the \dword{lv} and slow control signals to the \dword{fe} electronics.  The chimneys act also as Faraday cages, preventing the analog \dword{fe} electronics  from picking up possible noise from the digital electronics.   
 
The \dword{fe} cards are based on the analog cryogenic preamplifiers implemented in \dword{cmos} \dword{asic} circuits designed for high integration and large-scale affordable production. 
The \dword{asic} for the \dword{dpmod} circuits have been specially engineered to match the \dword{dpmod}'s signal dynamics. Inside the \dwords{sftchimney}, the cards are actively cooled to a temperature of approximately \SI{110}{K}.  The bottom sides of the cold flanges connect to  \dwords{crp} via flat \SI{0.5}{m} long cables intended to minimize the input capacitance to the preamplifiers. Each \dword{sftchimney} collects \num{640} readout channels. 

The digital electronics for the charge digitization system is installed on the cryostat roof. 
This makes it possible to use common design standards and benefit from commercially supported low-cost, high-speed networking technologies, such as \dword{utca}, which is used in the telecommunications industry.

Digitization cards in the \dword{amc} format read \num{64} channels per card. Each \dword{amc} card can digitize \num{64} channels at \SI{2.5}{MHz} and compress and transmit this continuous data stream, without zero-skipping, over a network link operating at \SI{10}{Gbit/s}. Lossless data compression is particularly effective thanks to the high \dword{s/n} ratio  of \dword{dp}, which limits noise contributions at the level of one \dword{adc} count. Each \dword{sftchimney} is coupled to a \dword{utca} crate that holds \num{10} \dword{amc} digitization cards and can therefore read  \num{640} channels. The \dword{amc} cards  transmit the data to the \dword{daq} back-end. A total of \num{240} \dword{utca} crates are required for reading the entire \dword{detmodule}.  

The light-readout digitization system uses \dword{utca} \dword{amc} card design derived from that of the charge readout system, but that implements a circuitry based on the \dword{catiroc} \dword{asic} to trigger the readout. 

The timing synchronization is based on the \dword{wr} standard. Specifically developed timing \dword{mch} connected to a \dword{wr} network ensures the distribution of clock, absolute timing, and trigger information on the backplane of the \dword{utca} crates. The \dword{wrmch} are connected via \SI{1}{Gbit/s} optical fibers to a system of \dword{wr} switches that interconnect the \dword{wr} network. This ensures that the digitization performed by the various \dword{amc} cards is completely aligned; it also refers to the absolute UTC time.

\section{Photon Detection System}
\label{sec:dp-execsum-pd}

The \dword{pds} is based on an array of \dwords{pmt} uniformly distributed below the cathode. 
The \dwords{pmt} have a \dword{tpb} coating on the photocathode's external glass surface that shifts the scintillation light from deep UV to visible light. The \dwords{pmt}  sit on the corrugated membrane cryostat floor, on 
mechanical supports that do not interfere with the membrane thermal contraction. 
Figure \ref{fig:dp-execsum-dppd_3_2} shows the \dword{pmt} with its support base attached to the bottom of the \dword{pddp} cryostat (Section~\ref{sec:exec:overall:pdune}).

\begin{dunefigure}[A Hamamatsu R5912-MOD20 \dshort{pmt} in \dshort{pddp}]{fig:dp-execsum-dppd_3_2}
{Picture of the cryogenic Hamamatsu R5912-MOD20 \dword{pmt} fixed on the membrane floor of \dword{pddp}. The optical fiber of the calibration system is also visible.}
\includegraphics[width=0.42\textwidth]{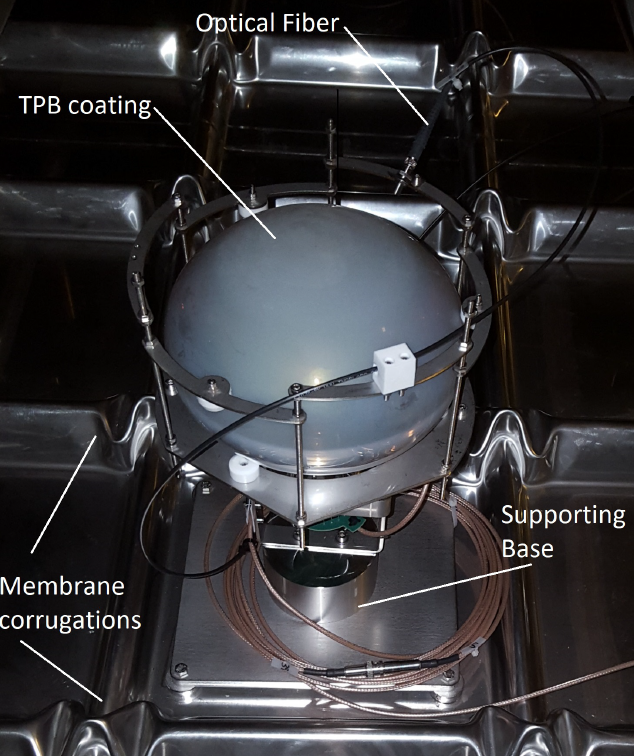}
\end{dunefigure}

In order to improve the light yield uniformity for signals generated in the top part of the drift volume a system of reflective panels with \dword{wls} coating is integrated on the  \dword{fc} walls.

\section{Data Acquisition}
\label{sec:dp-execsum-daq}

The \dword{daq} systems for both the \dword{sp} and \dword{dp} technologies have been designed jointly and are identical except for the architecture of the detector readout electronics.  The output format of the generated data is common, and both are synchronized to the same global clock signals. The shared \dword{daq} design is introduced in Section~\ref{sec:exec-sp-daq}.

The \dword{dp} readout architecture can be organized into \num{20} regions of interest (\dwords{roi}). 
Triggers are searched on the level-1 event builder machines, interconnecting multiple \dword{utca} crates, on a sliding windows of \SI{10}{s} contained in the event builder RAM.
Figure~\ref{fig:dp-execsum-daq-interface-scheme} illustrates the \dword{dp} readout architecture (bottom) and its interface to the \dword{daq} system.
  
\begin{dunefigure}[Interface of DP TPC electronics to DAQ]{fig:dp-execsum-daq-interface-scheme}
{Schematic illustration of the interface of \dword{dp} \dword{tpc} electronics to \dword{daq}.}
\includegraphics[width=0.95\textwidth]{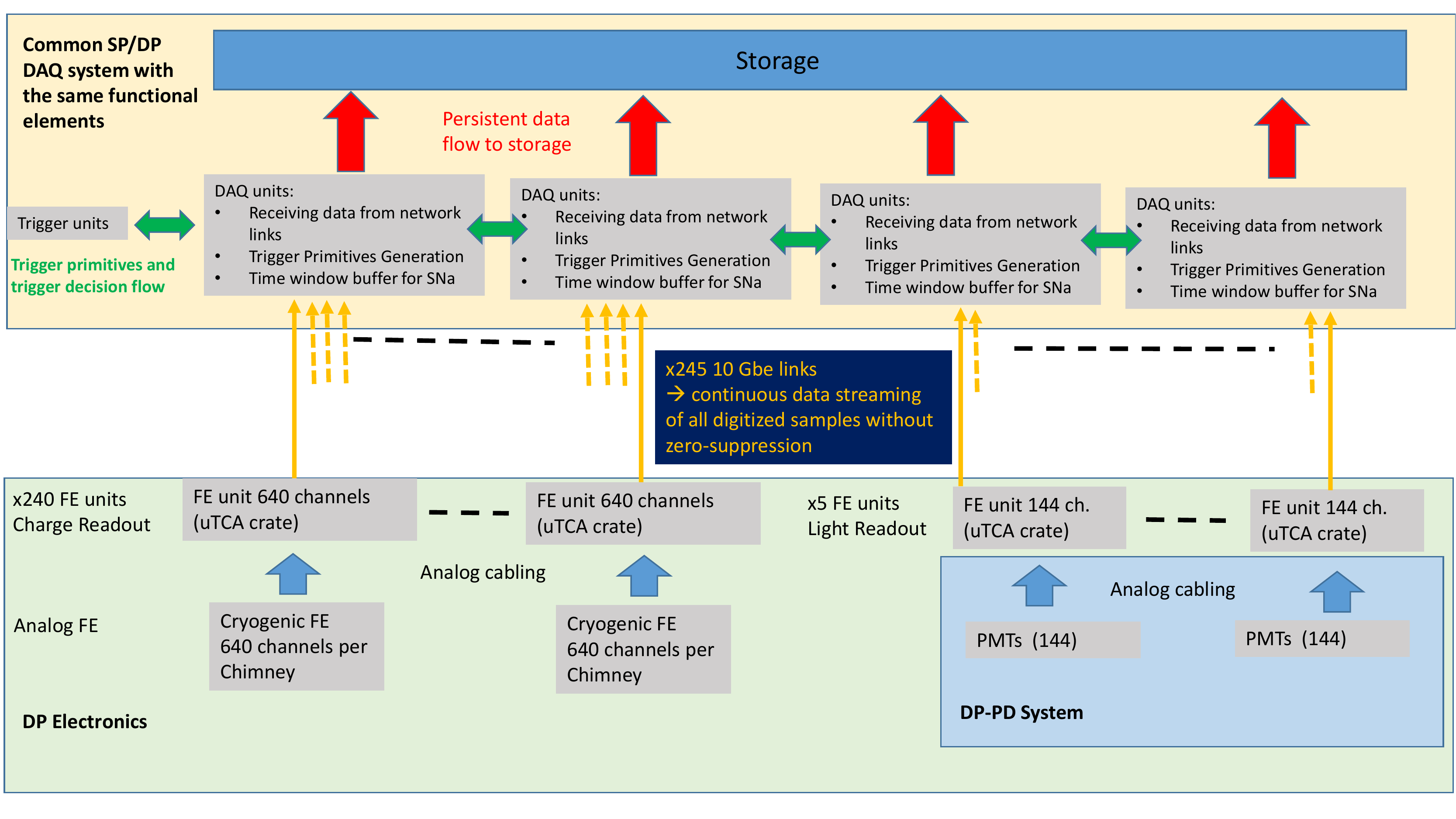}
\end{dunefigure}

\cleardoublepage

\chapter{The DUNE Near Detector} 
\label{ch:exsum-nd}

\textit{This chapter briefly introduces the DUNE near detector, emphasizing its role for the DUNE far detector physics program.  More details on the near detector may be found in  appendices of this TDR volume.  DUNE will issue a complete conceptual design report for the near detector in early 2020, with a technical design report to follow.}

\section{Overview of the DUNE Near Detector}
\label{sec:exsum-nd-overview}

\subsection{Motivation}
\label{sec:exsum-nd-BriefOverview-need}

The \dword{dune} experiment will measure oscillation probabilities for muon neutrinos or antineutrinos to either remain the same flavor or oscillate into their electron flavor counterparts as a function of the neutrino energy. This will allow the neutrino mass ordering to be definitively determined, as well as enable observation of leptonic \dword{cpv} for a significant range of $\delta_{\rm{CP}}$ values and precise measurement of 
neutrino mixing matrix parameters.

The \dword{nd} will serve as the experiment's control,
 constraining systematic errors and measuring the initial unoscillated \numu and \nue energy spectra (and that of the corresponding antineutrinos). 
 The energy spectra result from an energy-dependent convolution of flux, cross section, and detector response for each of the four 
 neutrino types (\nue, \numu, \anue, \anumu). 
  The \dword{nd} will make measurements that allow the three functions to be independently constrained and partially or fully deconvolved. The constraints will be used to improve the simulation program that is responsible for predicting the energy spectra at the \dword{fd} for particular choices of the oscillation parameters. This allows the actual oscillation parameters to be estimated from a fit to the \dword{fd} data.

The \dword{nd} will also have a physics program of its own, independent of the far detector.  This program will include measuring neutrino interactions to explore the two pillars of the standard model: electroweak physics and quantum chromodynamics. The \dword{nd} physics program will also explore physics beyond the standard model. This includes searches for non-standard interactions, sterile neutrinos,  dark photons, and  other exotic particles.

\subsection{Requirements}
\label{sec:exsum-nd-requirements}

The components of the  \dword{nd} must address their multiple missions in a complementary fashion. In this section, we list the key overarching requirements driving the \dword{nd} complex. Section~\ref{sec:appx-nd:requirements} in Appendix~\ref{ch:appx-nd} goes into more detail, discussing some thought experiments and case studies that illustrate how different parts of the complex work together. These case studies naturally suggest more detailed capabilities, performance statistics, and technical requirements; we are in the process of tabulating them. 

\begin{itemize}
    \item  \textit{Predict the neutrino spectrum at the \dword{fd}.} The \dword{nd} must predict the energy spectrum of \numu, \anumu, \nue and \anue at the \dword{fd}. The prediction must be provided as a function of the oscillation parameters, and systematic uncertainties must be small enough to achieve the required \dword{cp} coverage. This is the primary requirement of the \dword{dune} \dword{nd}.
    
    \item \textit{Measure interactions on argon.} The \dword{nd} must measure neutrino interactions on argon to reduce uncertainties due to nuclear modeling. The \dword{nd} must be able to determine the neutrino flavor and measure the full kinematic range of the interactions that will be seen at the \dword{fd}.
    
    \item \textit{Measure the neutrino energy.} The \dword{nd} must be able to reconstruct the neutrino energy in \dword{cc} events and control for any biases in energy scale or resolution, keeping them small enough to achieve the required \dword{cp} coverage. These measurements must also be transferable to the \dword{fd}. 
    
    \item \textit{Constrain the cross section model.} The \dword{nd} must measure neutrino cross sections in order to constrain the cross section model used in the oscillation analysis. In particular, cross section mismodeling that causes incorrect \dword{fd} predictions as a function of neutrino flavor and true or reconstructed energy must be constrained well enough to achieve the required \dword{cp} coverage. 
    
    \item \textit{Measure neutrino fluxes.} The \dword{nd} must measure neutrino fluxes as a function of flavor and neutrino energy. This allows neutrino cross sections to be measured and constrains the beam model and the extrapolation of neutrino energy spectra from the \dword{nd} to the \dword{fd}.
    
    \item \textit{Obtain data with different fluxes.} The \dword{nd} must measure neutrino interactions in different beam fluxes (especially ones with different mean energies) to disentangle flux and cross section, verify the beam model, and guard against systematic uncertainties on the neutrino energy reconstruction.
    
    \item \textit{Monitor the neutrino beam.} The \dword{nd} must monitor the neutrino beam energy spectrum with sufficient statistics to be sensitive to intentional or accidental changes in the beam that could affect the oscillation measurement. 
    
\end{itemize}

\subsection{Design}
\label{sec:exsum-nd-BriefOverview-design}

The \dword{dune} \dword{nd} is formed from three primary detector components and the capability of two of these components to move off the beam axis. The three detector components serve important individual and overlapping functions in the \dword{nd} mission.  Because these components have stand-alone features, the \dword{dune} \dword{nd} is often discussed as a suite or complex of detectors and capabilities.  The movement off-axis provides a valuable extra degree of freedom in the data. 
The power in the \dword{dune} \dword{nd} concept lies in the collective set of capabilities.  

Figure~\ref{fig:NDHallconfigs} in Chapter~\ref{ch:exec-overall} shows the  \dword{dune} \dword{nd} in the \dword{dune} \dword{nd} hall. 
Table~\ref{tab:NDsummch} provides a high-level overview of the three components of the \dword{dune} \dword{nd} along with the off-axis capability that is sometimes described as a fourth component.  

\begin{dunetable}[Components of the DUNE ND]
{p{.22\textwidth}p{.22\textwidth}p{.22\textwidth}p{.22\textwidth}}
{tab:NDsummch}{High-level breakdown of the three major detector components and the capability of movement for the DUNE ND, along with functions and primary physics goals.}
Component & Essential Characteristics & Primary function & Select physics aims \\ \toprowrule
LArTPC (\dshort{arcube}) & Mass  & Experimental control for the FD. & $\numu$($\overline{\nu}_{\mu}$) CC \\
          & Target nucleus Ar &  Unoscillated $E_\nu$ spectra measurements.   & $\nu$-e$^{-}$ scattering   \\
          &  Technology FD-like    &  Flux determination.  &  $\nue +$$\overline{\nu}_{e}$ CC  \\
          &  &  &  Interaction model \\ \colhline
Multipurpose detector (MPD) & Magnetic field & Experimental control for the LArTPCs. & $\numu$($\overline{\nu}_{\mu}$) CC \\
  &  Target nucleus Ar & Momentum-analyze $\mu$'s produced in LAr. & $\nue$ CC, $\overline{\nu}_{e}$ \\
  & Low density & Measure exclusive final states with low momentum threshold. & Interaction model \\  \colhline
DUNE-PRISM (capability) & \dshort{arcube}$+$\dshort{mpd} move off-axis & Change flux spectrum &  Deconvolve flux $\times$ cross section; \\ 
 & & & Energy response; \\
 & & & Provide FD-like energy spectrum at ND;\\ 
 & & & ID mismodeling. \\ \colhline
Beam Monitor (\dshort{sand}) & On-axis & Beam flux monitor &  On-axis flux stability \\ 
  & High-mass polystyrene target & Neutrons & Interaction model;  \\ 
& KLOE magnet &  & Atomic number (A) dependence; \\
    &  & & $\nu$-e$^{-}$ scattering. \\ 
\end{dunetable}

The core part of the \dword{dune} \dword{nd} is a \dword{lartpc} called \dword{arcube}.  
\dword{arcube} consists of an array of 35 modular \dwords{tpc} sharing a cryostat.  Figure~\ref{fig:es:ac_module} is a  drawing of a prototype of the modular \dwords{tpc}.  
This detector has the same target nucleus as the \dword{fd} and shares some aspects of form and functionality with it, where the differences are necessitated by the expected intensity of the beam at the \dword{nd}.  This similarity in target nucleus and technology reduces sensitivity to nuclear effects and detector-driven systematic errors in the extraction of the oscillation signal at the  \dword{fd}.  The \dword{lartpc} is large enough to provide high statistics 
($\num{e8}{\numu \text{-CC events/year}}$) and its volume  is sufficient to provide good hadron containment.  The tracking and energy resolution, combined with the mass of the \dword{lartpc}, will allow the flux in the beam to be measured using several techniques, including the well understood but rare process of $\numu$-e$^{-}$ scattering.

\begin{dunefigure}[ArgonCube 2$\times$2 demonstrator module]{fig:es:ac_module}
{Cutaway drawing of a \SI{0.67 x 0.67 x 1.81}{\metre} \dword{arcube} prototype module. For illustrative purposes, the drawing shows traditional field-shaping rings instead of a resistive field shell. The G10 walls will completely seal the module, isolating it from the neighboring modules and the outer \dword{lar} bath. The modules in this prototype system will not have individual pumps and filters.}
\includegraphics[width=0.8\textwidth]{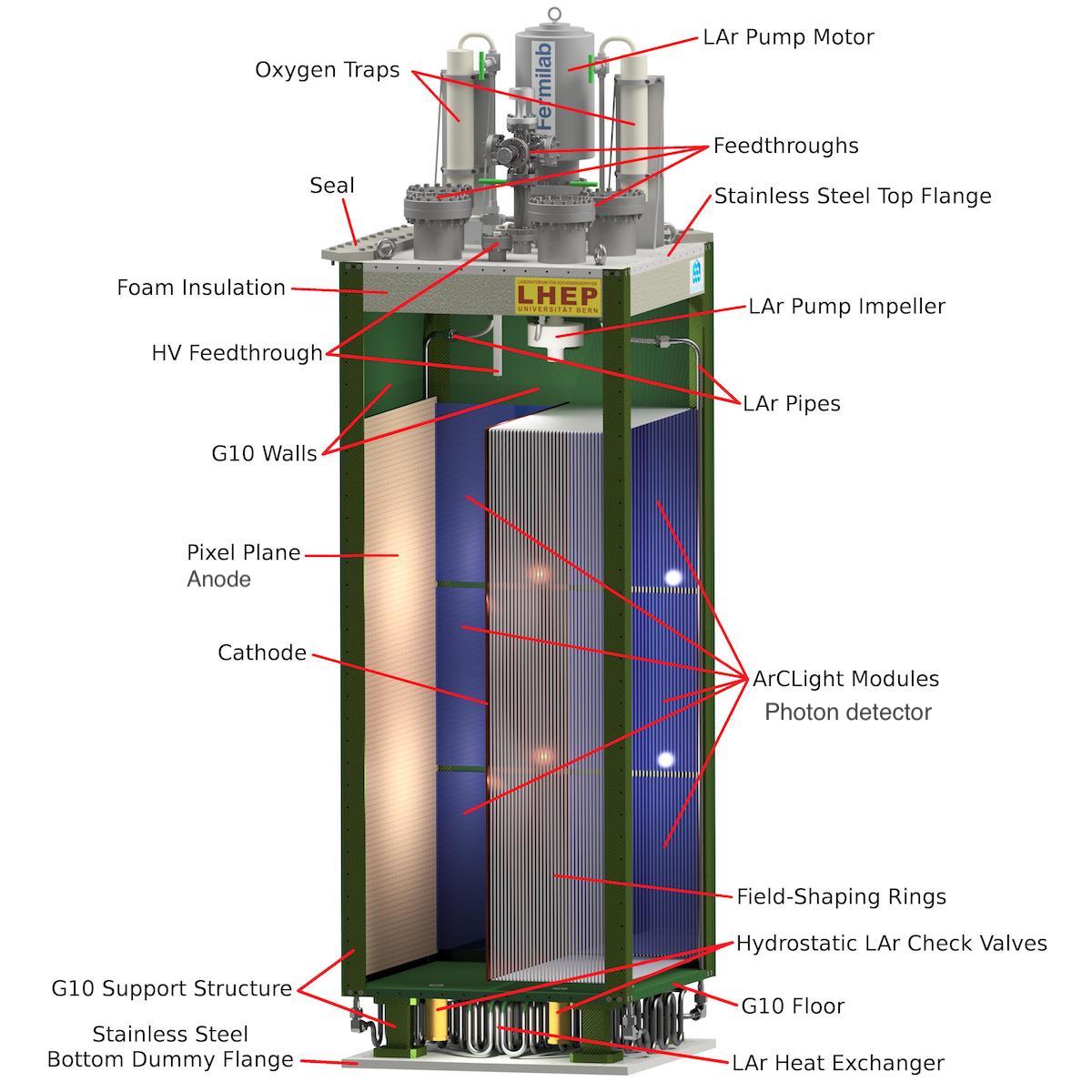}
\end{dunefigure}

The \dword{lartpc} acceptance falls off for muons with a measured momentum higher than \SI{0.7}{GeV/c} due to lack of containment.  Since the muon momentum is a critical component of the neutrino energy determination, a magnetic spectrometer is needed downstream of the \dword{lartpc} to measure the charge sign and momentum of these muons.  
The \dword{mpd} will accomplish this. It consists of a \dword{hpgtpc} surrounded by an \dword{ecal} in a \SI{0.5}{T} magnetic field (Figures~\ref{fig:es:ConceptDesign_NDECAL} and~\ref{fig:es:ConceptTile_NDECAL}). 

The \dword{hpgtpc} provides a lower-density medium with excellent tracking resolution for the muons from the \dword{lartpc}.  In addition, with this choice of technology for the tracker, neutrinos interacting on the argon in the gas \dword{tpc} constitute a sample of $\nu$-Ar events that can be studied with a very low charged-particle tracking threshold, excellent kinematic resolution, and systematic errors that differ from those of the liquid detector. 
The detector's high pressure will allow us to collect a sample of $\num{2e6}$ ${\numu \text{-CC events/year}}$ for these studies, events that will also  
be valuable for studying the charged particle activity near the interaction vertex 
since this detector can access lower-momentum protons than the \dword{lar} detector and 
provides better particle identification of charged pions.  The relative reduction in secondary interactions in these samples (compared to \dword{lar}) will help us to identify the particles produced in the primary interaction and to model secondary interactions in denser detectors, interactions that are known to be important \cite{Friedland:2018vry}.
In addition, using the \dword{ecal} we will be able to reconstruct 
many neutrons produced in neutrino interactions in the gaseous argon 
via time-of-flight.    
  

\begin{dunefigure}[\dshort{mpd} ECAL conceptual design]{fig:es:ConceptDesign_NDECAL}
{The conceptual design of the MPD system for the \dword{nd}. The TPC is shown in yellow inside the pressure vessel.  Outside the pressure vessel, the \dword{ecal} is shown in orange, and outside that are the magnet coils and cryostats.  The drawing illustrates the five-coil superconducting design.}
\includegraphics[width=0.8\textwidth]{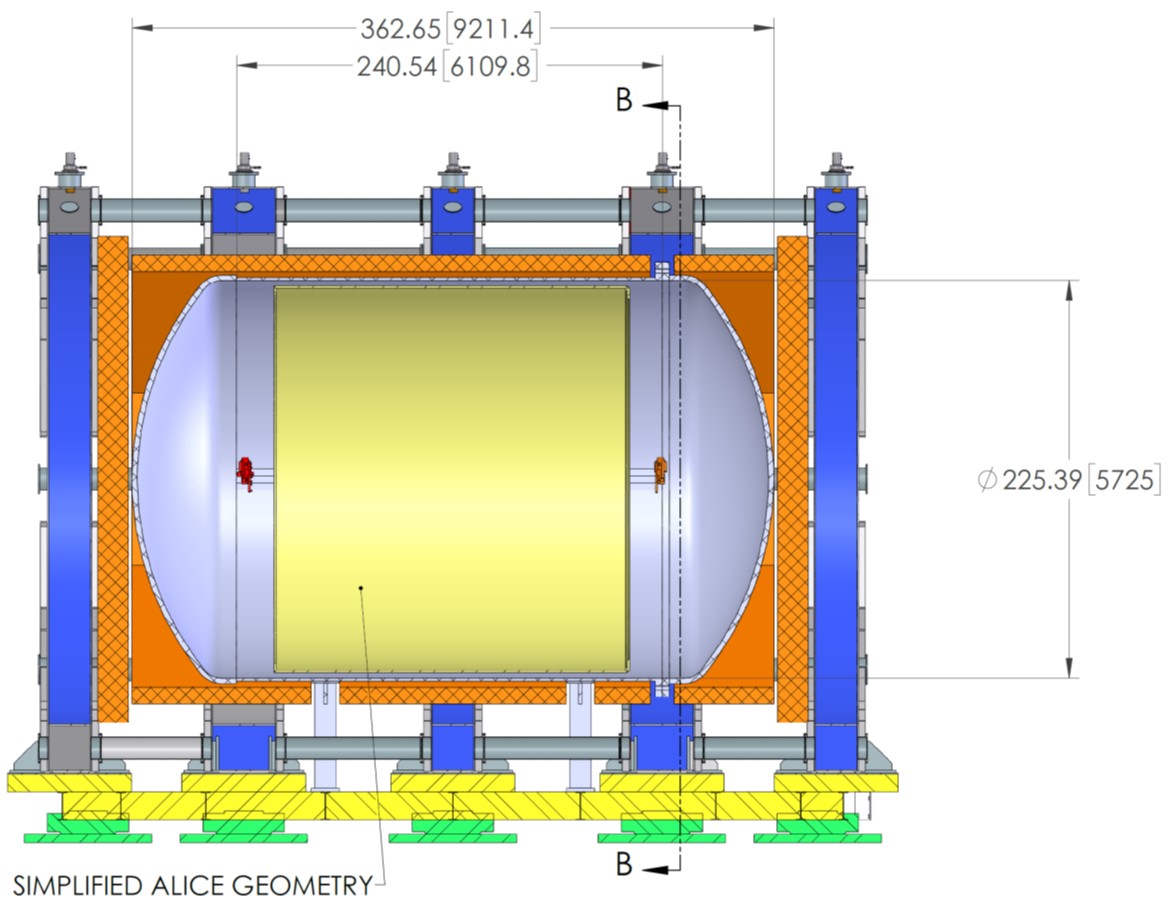}
\end{dunefigure}

\begin{dunefigure}[Conceptual layout of the \dshort{mpd} ECAL]{fig:es:ConceptTile_NDECAL}
{Conceptual layout of the calorimeter showing the absorber structure, scintillator tiles, \dwords{sipm}, and \dword{pcb}. The scintillating layers consist of a mix of tiles and cross-strips with embedded wavelength shifting fibers to achieve a comparable effective granularity.}
\includegraphics[width=0.8\textwidth]{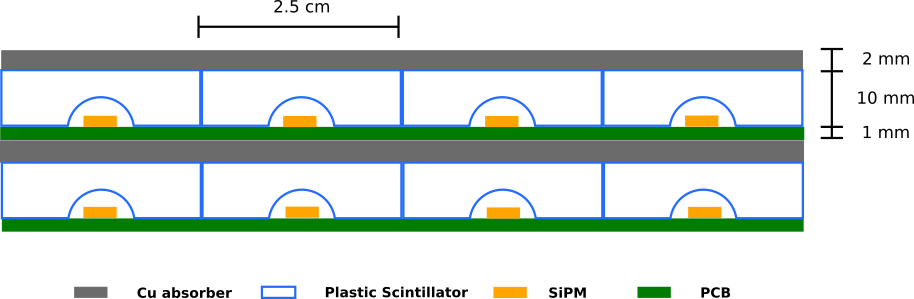}
\end{dunefigure}

The \dword{lartpc} and \dword{mpd} 
are able to move laterally to take data in positions off the beam axis.  This capability is referred to as \dword{duneprism}. As the detectors move off-axis, the incident neutrino flux spectrum changes:    the mean energy drops and the spectrum becomes more monochromatic.  Although the neutrino interaction rate drops, the intensity of the beam and the size of the \dword{lartpc} still combine to yield ample statistics. 
Figure~\ref{fig:es:offaxisfluxes} shows a sample of neutrino energy distributions taken at different off-axis angles.
Taking data at different off-axis angles allows the deconvolution of the neutrino flux and interaction cross section; it also allows mapping of the reconstructed versus true energy response of the detector.  This latter mapping is applicable at the \dword{fd} to the degree to which the near and far \dword{lar} detectors are similar.  Stated a different way, it is possible to use information from a linear combination of the different fluxes to create a data sample at the \dword{nd} with an effective neutrino energy distribution close to the oscillated spectrum at the \dword{fd}.  This data-driven technique will reduce systematic effects coming from differences in the energy spectra of the oscillated signal events in the \dword{fd} and the \dword{nd} samples used to constrain the interaction model. Finally, the off-axis degree of freedom may enable a sensitivity to some forms of mismodeling in the beam and/or interaction models. 

Figure~\ref{fig:es:duneprismfluxfits} shows linear combinations of off-axis fluxes giving \dword{fd} oscillated spectra for two sets of oscillation parameters. The procedure can model the \dword{fd} flux well for neutrino energies in the range of \SIrange{0.6}{3.6}{GeV}. 
The input spectra for the linear combinations, shown in Figure~\ref{fig:es:offaxisfluxes}, extend only slightly outside this range; they cannot be combined to model the flux in those extreme ranges while simultaneously fitting the central range well. 
The modeled range encompasses the range of data of interest for the oscillation program.

\begin{dunefigure}[Variation of neutrino energy spectrum as function of off-axis angle]{fig:es:offaxisfluxes}
{The variation in the neutrino energy spectrum shown as a function of detector off-axis position, assuming the nominal \dword{nd} location 574~m downstream from the production target.}
\includegraphics[width=0.8\textwidth]{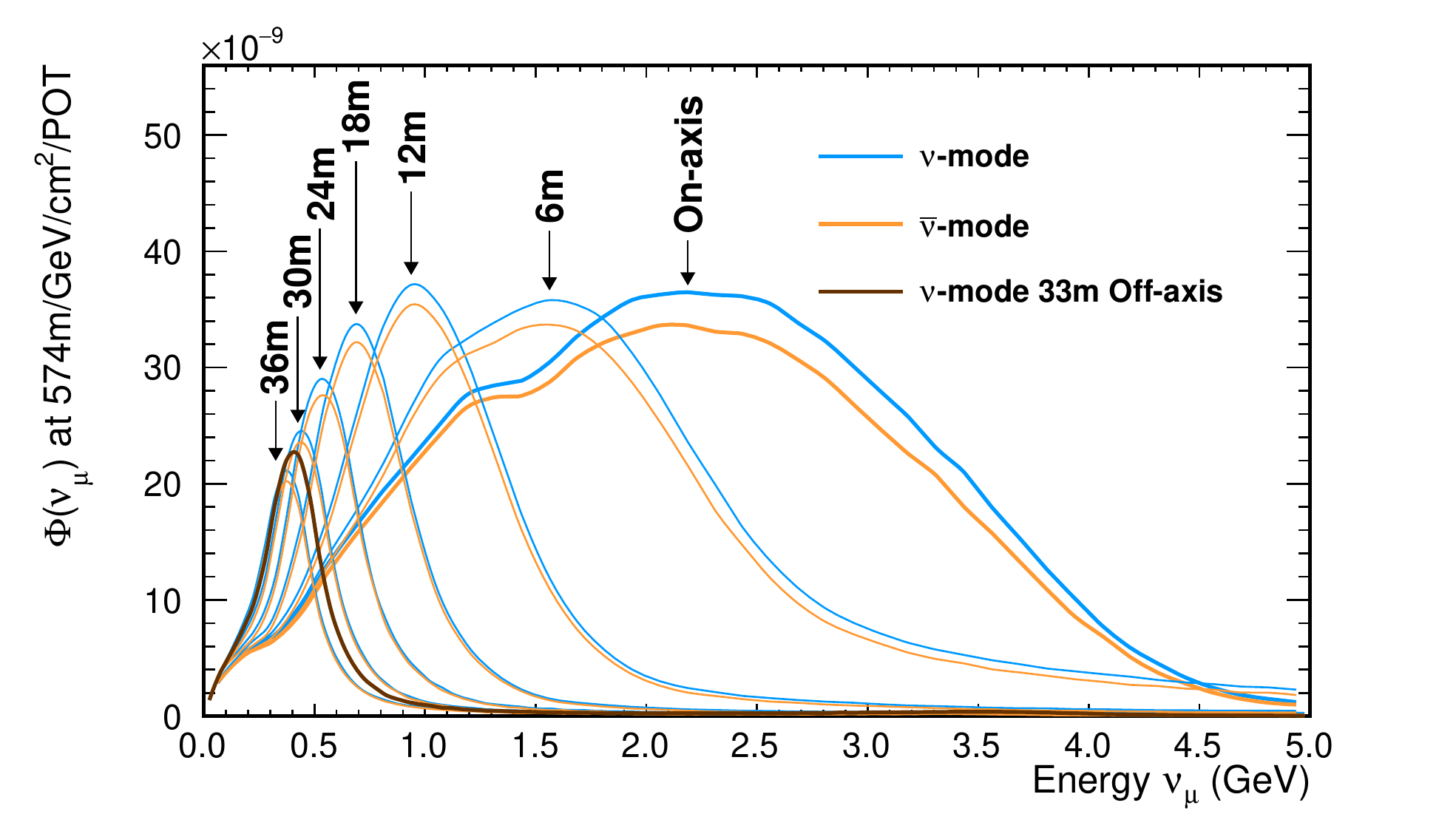}
\end{dunefigure}
\begin{dunefigure}[Linear combinations of off-axis fluxes giving FD oscillated spectra]{fig:es:duneprismfluxfits}
{Linear combinations of off-axis fluxes giving \dword{fd} oscillated spectra for a range of oscillation parameters. The  \dword{fd} oscillated flux is shown in black, the target flux is shown in green, and the linearly combined flux obtained with the nominal beam \dword{mc} is shown in red. Systematic effects due to 1$\,\sigma$ variations of the decay pipe radius (green), horn current (magenta), and horn cooling water layer thickness (teal) are also shown.}
	\includegraphics[width=0.85\textwidth]{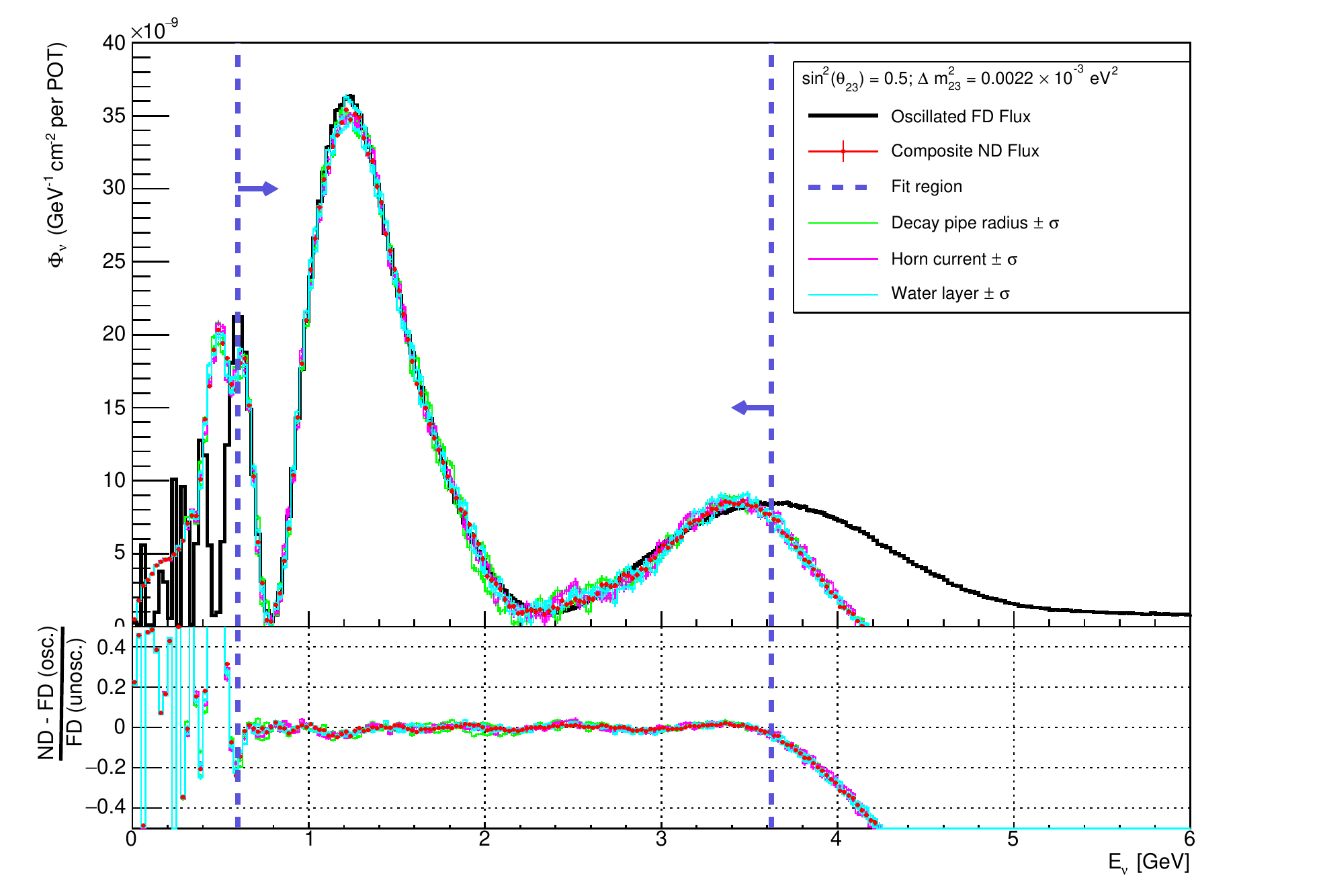}
	\includegraphics[width=0.85\textwidth]{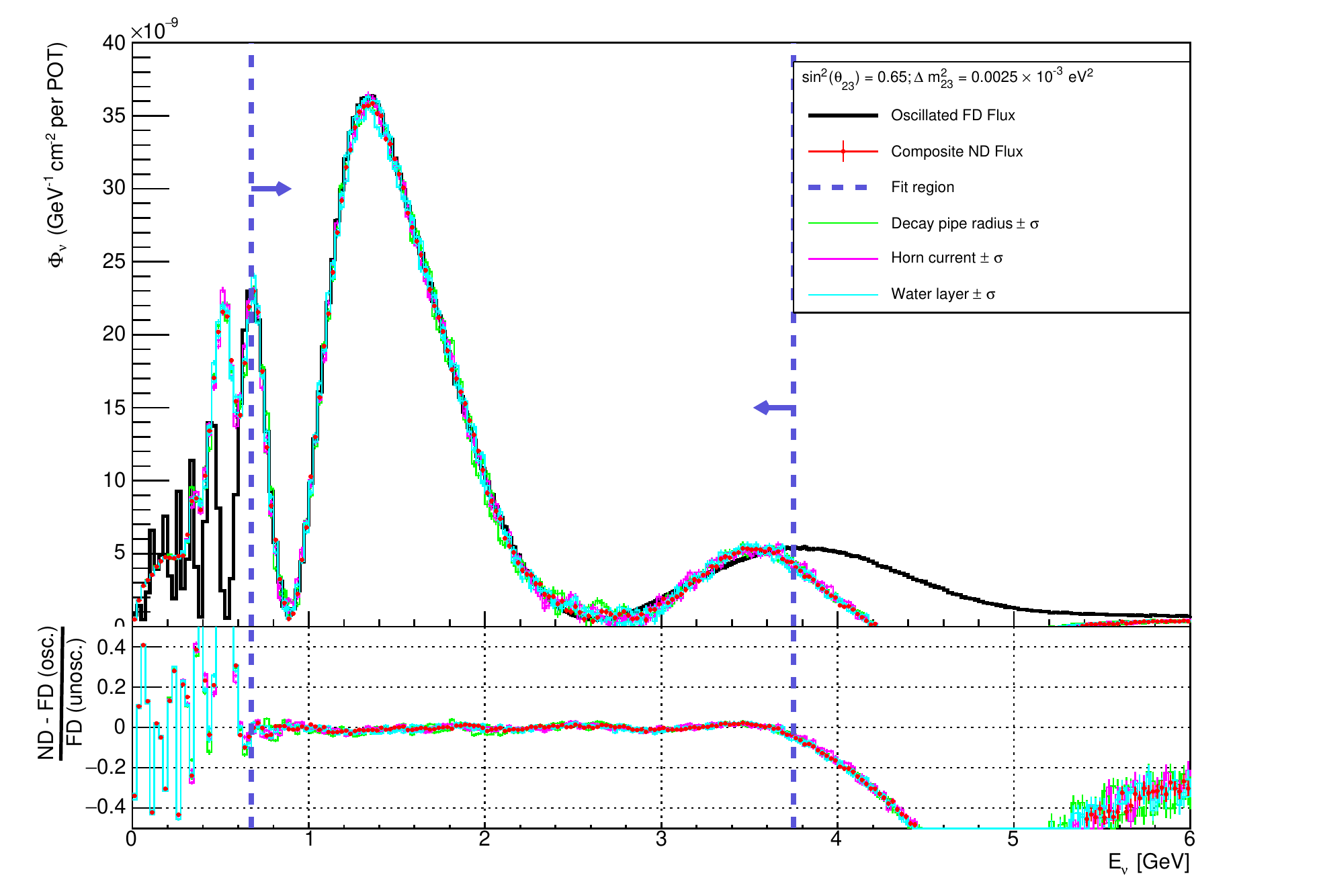}
\end{dunefigure}

The final component of the \dword{dune} \dword{nd} suite is the beam monitor, called the \dword{sand}.  The core part of it, the \dword{3dst}, is a plastic scintillator detector made of \SI{1}{\cubic\centi\meter} cubes read out along each of three orthogonal dimensions.  The design eliminates the typical planar-strip geometry common to scintillator detectors, leading to improved acceptance at large angles relative to the beam direction. It is mounted  
inside an envelope of high-resolution, normal pressure \dwords{tpc} and an \dword{ecal}, all 
of which are surrounded by a magnet, as illustrated in Figure~\ref{fig:es:sand-geometry}.  The reference design uses a repurposed magnet and \dword{ecal} from the \dword{kloe} experiment.

\begin{dunefigure}[The \dshort{sand} detector configuration]{fig:es:sand-geometry}
{The \dshort{sand} detector configuration with the \dword{3dst} inside the \dword{kloe} magnet. The drawing shows the \dword{3dst} in the center (white), \dwords{tpc} (magenta), \dword{ecal} (green), magnet coil (yellow), and the return yoke (gray).}
  \includegraphics[width=7.in]{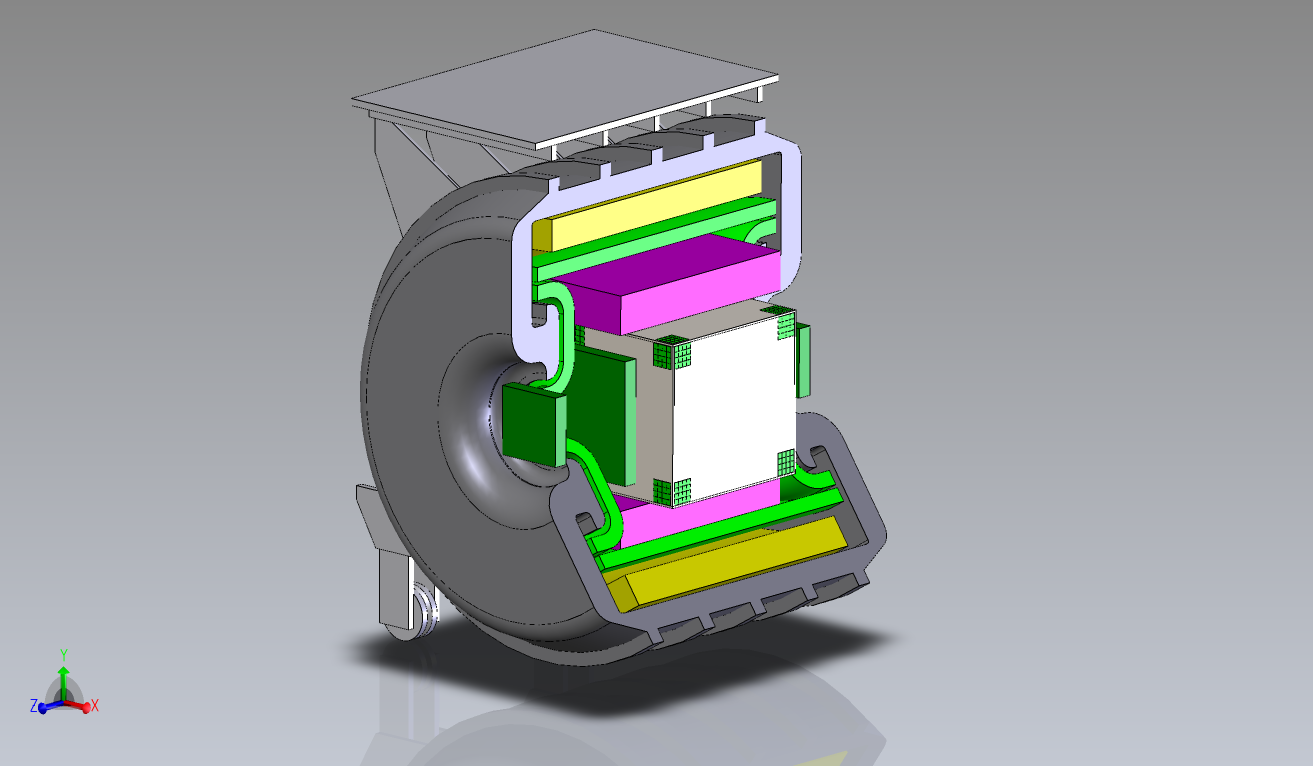}
\end{dunefigure}

\dword{sand} serves as a dedicated neutrino spectrum monitor that never moves off-axis. 
It also provides an excellent on-axis, neutrino flux determination using many of the methods discussed in Section~\ref{sec:appx-nd:fluxappendix}. 
The neutrino flux determined using this detector, with  
technologies, targets, and interaction systematic errors that are different from \dword{arcube}, is an important point of comparison and a systematic cross-check for the flux as determined by \dword{arcube}.

\dword{sand} provides very fast timing and can isolate small energy depositions from neutrons in three dimensions.  This provides the capability to  incorporate neutrons in the event reconstruction using energy determination via time-of-flight with a high efficiency. This capability should be useful for the low-$\nu$ flux determination\footnote{The low-$\nu$ technique involves measuring the flux for events with low energy transfer because the cross section is approximately constant with energy for this sample.  It provides a nice way to measure the shape of the spectrum.  This is discussed further in Section~\ref{sec:appx-nd:fluxappendix} of Appendix~\ref{ch:appx-nd}.} because it either allows events to be tagged with a significant neutron energy component or provides a way to include that energy in the calculation.  Including neutrons in detailed studies of neutrino interactions in \dword{sand}  using single transverse variables may prove useful in motivating improvements in the neutrino interaction model. Although the target for this device is carbon, not argon, basic insights into components of the interaction model may extend to argon.  For example, the multi-nucleon component of the interaction model will be used for argon although it was developed in response to observations made on plastic targets.

\section{Role of the ND in the DUNE Oscillation Program}
\label{sec:exsum-nd-role}

Neutrino oscillation experiments must accomplish three main tasks. First, they must identify the flavor of interacting neutrinos in \dword{cc} events or identify the events as \dword{nc} interactions. Second, they must measure the energy of the neutrinos because oscillations occur as a function of baseline length over neutrino energy, \dword{l/e}. Third, they must compare the observed event spectrum in the \dword{fd} to  predictions based on differing sets of oscillation parameters, subject to constraints from data observed in the \dword{nd}.  That comparison and how it varies with the oscillation parameters allows oscillation parameters to be measured.

The connection between the observations in the \dword{nd} and the \dword{fd} is made using a simulation that convolves models of the neutrino flux, neutrino interactions, nuclear effects, and detector response.
This gives rise to a host of complicating effects that 
muddy the simple picture. These complications come from two main sources. First, the identification efficiency is not \SI{100}{\%}, and there are
some background events (for example, \dword{nc} interactions with a $\pi^0$ present 
a background to \nue \dword{cc} interactions). Both the efficiency and background are imperfectly known. 
Because the background level tends to be similar in both the \dword{fd} and \dword{nd}, it helps if the \dword{nd} can characterize backgrounds better than the \dword{fd}. 

The second aspect that complicates the simple picture is that the \dword{fd} (and the similar \dword{nd}) must use a target material composed of heavy nuclei. 
The target nucleus affects 
neutrino interactions in ways that ultimately drive the design of the  \dword{nd} complex. In particular, in heavy nuclei, the nucleons interact with each other and exhibit Fermi motion, providing moving targets for neutrino interactions. 
The wavelength of an interaction 
depends on momentum transfer but is often long enough to simultaneously probe multiple nucleons.

Another complication is that neutrino-nucleus scattering models rely on neutrino-nucleus cross sections, but neutrino cross sections on \textit{free nucleons} are not generally well known in the kinematic range of interest to \dword{dune}.  
Since the \dword{nd} will enable high-statistics measurements on liquid and gaseous argon, rather than another nucleus, it will reduce nuclear model dependence. 
 A final complication comes about because neutrinos produce hadrons within the nucleus. After production the hadrons undergo \dword{fsi} and are thereby attenuated as they leave the target nucleus. Section~\ref{sec:appx-nd:exsum-nd-role} of Appendix~\ref{ch:appx-nd} discusses neutrino-nucleus scattering in more detail. 

Neutrons can be produced from the struck nucleus, as well as from follow-on interactions of the neutrino's reaction-products with other nuclei. The energy carried away by neutrons is difficult to detect and can bias the reconstructed neutrino energy. The \dword{sand}  and \dword{mpd} detectors have capabilities that allow neutron energy to be directly measured. The \dword{duneprism} program constrains the true-to-reconstructed energy relation and is thus also sensitive to energy carried by neutrons.

Heavy nuclei in the detector offer additional complications for particles that have left the struck nucleus, especially in the case where the detector is dense, e.g., in \dword{arcube}. Particles produced in a neutrino interaction may  reinteract inside the detector, creating electromagnetic and hadronic cascades. These cascades, particularly the hadronic ones, 
confuse the reconstruction program due to overlapping energy and event features. They also cause a degradation of the energy resolution and result in additional energy carried by neutrons that may go missing. Particle identification by $dE/dx$ is less effective for early showering particles, and  low-energy particle tracks in a dense detector may be too short to detect.  The \dword{hpgtpc} in the \dword{mpd} 
allows us to measure neutrino interactions 
on argon, but with significantly fewer secondary interactions and much lower-energy tracking thresholds.

Finally, setting aside complications due to heavy nuclei and dense detectors, we note that a significant fraction of the neutrino interactions in \dword{dune} will come from inelastic processes, not the simpler \dword{qe} scattering. 
This typically leads to a more complex morphology for events and greater challenges for the detector and the modeling.  The \dword{dune} \dword{nd} acts as a control for the \dword{fd} and is designed to be more capable than the \dword{fd} at measuring complicated inelastic events.

These complexities are incorporated imperfectly into the neutrino interaction model. The predicted signal in the \dword{nd} is a convolution of this interaction model with the beam model and the detector response model.  The critical role of  the  \dword{nd} is to supply the observations used to tune, or calibrate, this convolved model, thereby reducing the overall uncertainty in the expected signal at the \dword{fd}, which is used for extracting the oscillation parameters via comparison with the observed signal.  And with its high statistics and very capable subsystems, the \dword{nd} will produce data sets that will provide the raw material for 
improving the models beyond simple tuning.

\section{ND Hall and Construction}
\label{sec:exsum-nd-hall}

Figure~\ref{fig:es:NDhall} shows the current design of the underground hall required for the  \dword{nd} construction concept. The hall must house the detector components and enable the required off-axis movement. The layout shows the spaces required for the detector itself, and for safety and egress.  This  work is in progress.

\begin{dunefigure}[DUNE near detector hall and detectors, plan view]{fig:es:NDhall}
{\dword{dune}  \dword{nd} hall shown from above (top) and from the side transverse to the beam (bottom). The \dword{arcube}, \dword{mpd}, and \dword{sand}  are shown (in that order, bottom to top, in the upper figure) in position on the beam axis (black arrow) in both drawings. }
\includegraphics[width=0.8\textwidth]{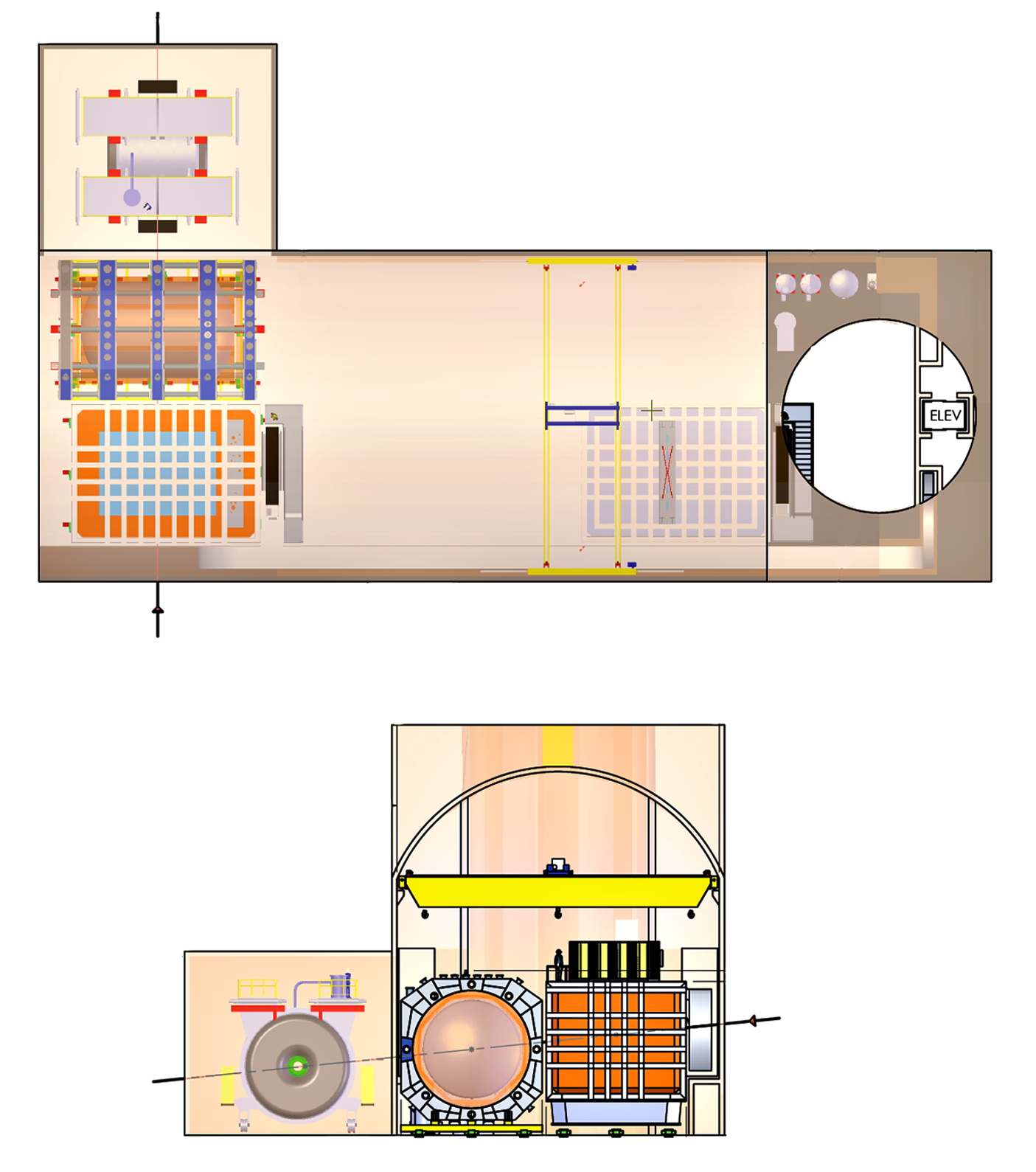}
\end{dunefigure}

The overall construction method means the \dword{cf} must 
provide a primary access shaft that is large enough for lowering the pressure vessel and the magnet coils. In the figure, \dword{arcube} is shown in its on-axis construction position, 
as is the \dword{mpd}. 
Because \dword{sand}  does not  move, it is placed in a dedicated alcove downstream of the other  detectors.

The \dword{duneprism} design requires that both the \dword{mpd} and \dword{arcube} be able to  move horizontally to a position off the beam axis. The direction of the motion is to one side of the beam, with a maximum displacement of approximately \SI{30.5}{m}. 
\cleardoublepage

\chapter{Computing in DUNE}
\label{ch:exec-comp}

\textit{This chapter briefly describes the DUNE computing model, which touches on all elements of the \dshort{nd} and \dshort{fd} and their physics programs.  More detailed aspects of the model may be found in Appendix~\ref{appx:es-comp}.  DUNE will produce a complete technical design report for computing over the next two years.}

\section{Overview}
\label{ch:exec-comp-es}


The \dword{dune} experiment will commission the first \nominalmodsize fiducial mass 
\dword{fd} module between 2024 and 2026 with a long data-taking run, and 
the remaining three modules between 2026 and 2036.  An active prototyping program is ongoing 
 at the \dword{cern}, where   
a short test beam run took place in 2018 with \dword{pdsp}, a \SI{700}{t}, 15,360 channel prototype \dword{lartpc} with \dword{sp} readout.  Tests of a \dword{dp} detector of similar size began in mid-2019.   The \dword{dune} experiment has already benefited greatly from these initial tests.  The collaboration has recently formed a formal \dword{csc}, with significant participation of European institutions and interest from groups in Asia, to develop common software and computing, and to formalize resource contributions.

The \dword{csc} resource model benefits from the existing \dword{osg}  and \dword{wlcg} infrastructure developed for the \dword{lhc} and broader \dword{hep} community.  \dword{dune} is already using global resources to simulate and analyze  \dword{pdsp} data.  Several European institutions are part of this resource pool, making significant contributions to the \dword{pdsp} and \dword{pddp} programs.  We expect this global computing consortium to grow and evolve as we begin gathering data from the full \dword{dune} detector in the 2020s.

The long-term \dword{dune} science program should produce volumes of raw data similar in scale to the data volumes that current \dword{lhc} Run-2 experiments have already successfully handled.  Baseline predictions for the \dword{dune} data, depending on actual detector performance and noise levels, are $\sim\,$\SI{30}{PB} of raw data per year.  These data, with simulations and derived analysis samples, 
will need to be made available to all collaborating institutions.  We anticipate that institutions around the world will both contribute and use storage and CPU resources for \dword{dune}.

The \dword{dune} computing strategy starts with considerable infrastructure in place for international computing collaboration, in part thanks to the \dword{lhc} program.  Additional large non-\dword{lhc} experiments,  such as \dword{lsst}, the \dword{belleii} B-factory experiment, and \dword{dune}  will begin operation over the next decade and must use and expand upon this model to encourage international cooperation.  The broader \dword{hep} community is organizing common efforts through the \dword{hsf}~\cite{Alves:2017she}.  The \dword{hsf} is an organization of interested parties using the extensive knowledge gained over the past two decades to anticipate the needs of experiments over the next two decades, and to develop a sustainable computing landscape for the \dword{hep} community.  The \dword{hsf} white papers and roadmaps emphasize common tools and infrastructure as the foundation of this landscape.

The \dword{dune} computing strategy heavily leverages the \dword{hsf} model of common tools and infrastructure, which encompass data movement and storage, job control and monitoring, accounting, and authentication.  
\dword{dune} recognizes that other large-scale experiments have similar needs and will encounter complementary issues, thus driving worldwide cooperation on common tools as the most cost-effective way to fulfill the scientific missions of the experiments.  \dword{dune} pilot programs already use this model.  Most recently in data management and storage, \dword{fnal}, \dword{cern}, Rutherford Appleton Laboratory, and other research and academic institutions in the 
UK are collaborating on adapting and using the \dword{rucio} data management systems~\cite{Barisits:2019fyl}  to serve as the core data management system for \dword{dune}.

This protoculture of international collaboration within \dword{dune} 
was demonstrated during the 2018 test beam run of the \dword{pdsp} detector, which  
produced raw data at rates of up to \SI{2}{GB/s}.  These data were transferred and stored in the archive facilities at \dword{cern} and \dword{fnal}, and replicated at sites in the UK and Czech Republic.  In a more recent commissioning test for the \dword{pddp} detector, similar data transmission rates 
were achieved to \dword{cern}, \dword{fnal}, and the CCIN2P3 computer center in Lyon, France.

In total, \SI{1.8}{PB} of raw data were produced during the ten-week test beam run, mimicking, within a factor of two, expected data rates and volumes from the initial running of the \dword{fd} complex.  The prototype run was used to examine and test the scalability of existing and proposed computing infrastructure and to establish operational experience within the institutions that have expressed interest in developing and constructing the \dword{dune} computing environment.  The planning for \dword{dune} is primarily based on the measurements and information gained from the \dword{protodune} experience.  These measurements are proofs-of-concept for many of the systems, and their behavior can be reliably extrapolated to the projected levels needed for the complete \dword{dune} experiment. 

The \dword{protodune} experience highlights the significant technical challenges that must be overcome by 2024 for \dword{dune}. Among the most significant will be  1) the design of \dword{dune}-specific systems able to integrate the large suite of ancillary data (e.g., configurations, calibrations, shower libraries) with the main \dword{tpc} data stream; 2) the potentially extreme size of some physics interactions, notably supernova bursts, that could produce enough data to overwhelm conventional processors; and 3) the continuing need for evolution of computing architectures and infrastructure over the next decade. These challenges 
are unique to \dword{dune} where the limiting factor will be human effort. 

In summary, the \dword{dune} computing strategy must be global, working with partners worldwide, and collaborative because many of the computational challenges we face are also 
faced by other, similar experiments.  We are extremely fortunate to have the \dword{protodune} experience and test data to exercise our computing infrastructure and develop algorithms for full \dword{dune} operations, although we know significant and interesting challenges lie ahead. 
 
\section{Computing Consortium}
\subsection{Overview}
\label{ch:exec-comp-ovr}

The mission of the \dword{dune} 
\dword{csc} is to acquire, process, and analyze both detector data and  simulations for the 
collaboration.  This mission must extend over all the primary physics drivers for the experiment and must do so both cost effectively and securely. The \dword{csc} provides the bridge between the online systems for \dword{daq} and monitoring and the physics groups who develop high-level algorithms and analysis techniques to perform measurements using the  \dword{dune} data and simulations. The \dword{csc} works with collaborating institutions to identify and provide computational and storage resources.  
It provides the software and computing infrastructure in the form of analysis frameworks, data catalogs, data transport systems, database infrastructure, code distribution mechanisms, production systems, and other support services essential for recording and analyzing the data and simulations. 

The \dword{csc} works with national agencies and major laboratories to negotiate use and allocation of computing resources.  This work includes support for near-term R\&D efforts 
such as \dword{protodune} runs, and extends to designing, developing, and deploying the \dword{dune} computing model and its requisite systems. 
This includes evaluating major software infrastructure systems to determine their usefulness 
in meeting the \dword{dune} physics requirements.  These evaluations should identify opportunities to adopt or adapt existing technologies, and to engage in collaborative ventures with \dword{hep} experiments outside of \dword{dune}. 

At first glance,  the \dword{dune} CPU and storage needs 
appear modest 
compared to the corresponding needs for the high-luminosity \dword{lhc} experiments.  
However, the  beam structure, event sizes, and analysis methodologies make \dword{dune} very unlike collider experiments 
in event processing needs and projected computational budgets. 
In particular, the large \dword{dune} event sizes (0.1-10 GB as opposed to 1-10 MB per detector readout) present a novel technical challenge when data processing and analysis are mapped onto  current and planned computing facilities. 
The advent of high-performance computing systems optimized for parallel processing of large data arrays is a great advantage for \dword{dune}. These architectures suit the uniform \dword{lartpc} raw data structure very well, in contrast to the more complex data structures and geometries present in conventional 
heterogeneous \dword{hep} data.

\dword{dune} will require significant effort to adapt to emerging  
global computing resources that 
  likely will be both more heterogeneous in computational capabilities (e.g., featuring CPU, GPU, and other advanced technologies) and more diverse in topological architectures and provisioning models.  The \dword{dune} \dword{csc} must 
be ready to fully exploit these global resources 
after 2026, allowing all collaborators to access the data and meet the scientific mission of the experiment.  

\subsection{Resources and Governance}
\label{sec:exec-comp-res}

The \dword{csc} was formed from an earlier ad hoc \dword{dune} computing and software group. 
The governance structure for the \dword{csc} is described in Ref.~\cite{bib:docdb12751}.  The consortium coordinates work across the collaboration, but funding comes from collaborating institutions, laboratories, and national funding agencies. 

The 
\dword{csc} has an elected consortium leader 
who is responsible for subsystem deliverables and represents the consortium in the overall \dword{dune} collaboration.
In addition, technical leads act as overall project managers for the \dword{csc}. The technical leads report to the overall consortium leader.
\dword{csc} has both a host laboratory technical lead to coordinate between the \dword{dune} project and \dword{fnal}, the host laboratory, 
and an international technical lead to coordinate with other entities.
At least one of the three leadership roles should be held by a scientist from outside the USA. 
\dword{csc} management currently 
appoints people to other roles, typically after a call for nominations.  A more formal structure for institutional contributions and commitments is under consideration. 

\begin{dunefigure}
[Organization chart for current computing and software consortium]
{fig:ch-exec-comp-org-es}
{Organization chart for current \dword{csc}.}
\includegraphics[height=4in]{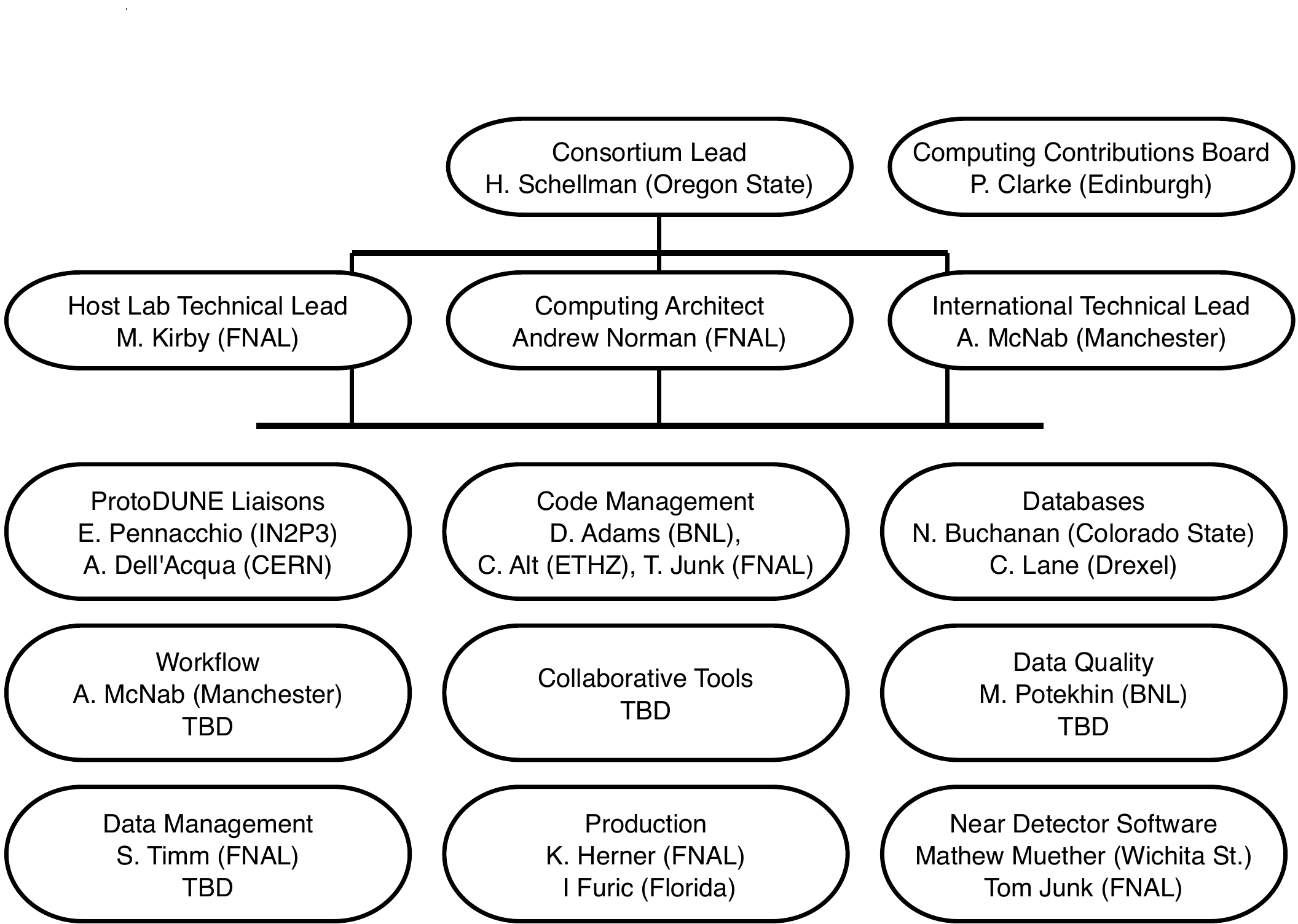}
\end{dunefigure}

\subsection{Scope of the Consortium}
\label{ch:exec-comp-gov-scope}

The \dword{csc} member institutions (Table~\ref{tab:exec-comp-consortium}) focus on the hardware and software infrastructure for offline computing.  Responsibility for developing algorithms resides 
with the physics groups, 
and online systems at experimental sites are governed by the \dword{daq} and \dword{cisc} consortia. 
The \dword{csc} defines interfaces, sets coding standards, and provides training. 
All groups coordinate closely to ensure that the full chain of data acquisition, processing, and analysis 
functions properly. Formal interfaces with the \dword{daq} and controls groups are described in~\cite{bib:docdb7123,bib:docdb7126}. 

The \dword{csc} operates at two levels: at the hardware level, where generic resources can be provided as in-kind contributions to the collaboration, and at the human level, where individuals and groups help develop common software infrastructure.  The technology for hardware contributions (e.g., grid CPU and storage) exists and was successfully used during the \dword{pdsp} data run and its associated 
simulation and reconstruction. Highlights of that effort are discussed below and in \physchtools{}. 

\begin{dunefigure}
[CPU wall-time from July 2018 to July 2019]
{fig:ch-exec-comp-cpupie-es}
{CPU wall-time from July 2018 to July 2019, the first peak shows  \dword{pdsp} reconstruction while the second is dominated by data analysis and \dword{pddp} simulation. A total of 31 million wall-hours were delivered with 24 M-hrs coming from \fnal.}
\includegraphics[height=3in]{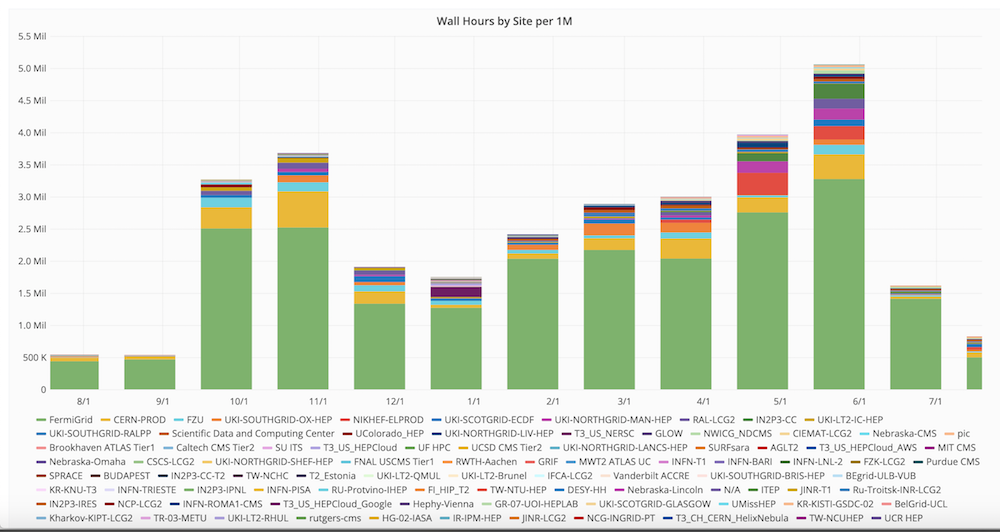}
\end{dunefigure}

\subsection{Hardware Resources} 
\label{sec:exec-comp-hi}

As illustrated in Figure~\ref{fig:ch-exec-comp-cpupie-es}, the \dword{dune} collaboration has already used substantial global resources through the \dword{wlcg} and \dword{osg} mechanisms. As the experiment evolves over the next five years, institutions and collaborating nations will be asked to formally pledge resources (both CPU and storage), and those resources will be accounted for and considered in-kind contributions to the collaboration.  A   computing resources board  will be set up to administer this process and serve as liaison to national resource providers. 

Several international partners are already contributing substantially to CPU resources, and we continue to  integrate additional large national facilities. Most CPU resources are opportunistic, but \dword{fnal} and \dword{cern} have committed several thousand processor cores and several PB of data storage. Additionally,  \dword{dune} is one of the first beneficiaries of the IRIS project (UK), which provides computing for astronomy and particle physics.  
We are working with \dword{osg} and \dword{wlcg} to integrate reporting mechanisms for CPU use, so accurate monitoring of hardware contributions will be in place for a second planned  \dword{protodune} run in 2021-2022 and the buildup to data taking in the mid 2020's. 

\begin{dunetable}
[Computing and software consortium institutions] 
{lll}{tab:exec-comp-consortium}{DUNE computing and software consortium members} 
Institution& Country \\ \colhline 
Centro Brasileiro de Pesquisas F\'isicas &	Brazil\\ \colhline
Universidade Estadual de Campinas &	Brazil\\ \colhline
York University & Canada\\ \colhline
European Organization for Nuclear Research &	CERN\\ \colhline
Institute of Physics of the Czech Academy of Sciences & Czech Republic\\ \colhline
Laboratoire d'Annecy-le-Vieux de Physique des Particules  &	France\\ \colhline
Tata Institute of Fundamental Research &	India\\ \colhline
Korean Institute for Science and Technology Information &	Korea\\ \colhline
Nikhef National Institute of Subatomic Physics &	Netherlands\\ \colhline
University of Bern &	Switzerland\\ \colhline
Centro de Investigaciones Energ\'eticas, Medeioambientales y Techn\'ologicas &	Spain\\ \colhline
University of Edinburgh &	UK\\ \colhline
The GridPP Collaboration &	UK\\ \colhline
University of Manchester &	UK\\ \colhline
Queen Mary University & UK\\ \colhline
STFC Rutherford Appleton Laboratory &	UK\\ \colhline
Argonne National Laboratory &	USA\\ \colhline
University of California, Berkeley &	USA\\ \colhline
Brookhaven National Lab &	USA\\ \colhline
Colorado State University &	USA\\ \colhline
University of Colorado Boulder &	USA\\ \colhline
Fermi National Accelerator Laboratory &	USA\\ \colhline
University of Florida& 	USA\\ \colhline
Lawrence Berkeley National Laboratory &	USA\\ \colhline
University of Minnesota &	USA\\ \colhline
Northern Illinois University &	USA\\ \colhline
University of Notre Dame &	USA\\ \colhline
Oregon State University&	USA\\ \colhline
SLAC National Accelerator Laboratory&	USA\\ \colhline
The University of Texas at Austin &	USA\\ \colhline
Valley City State University & USA\\ 

\end{dunetable}

\subsection{Personnel}
\label{sec:exec-comp-gov-personnel}

The advancement of a dedicated \dword{dune} computing team responsible for operations and development of new tools specific to the \dword{dune} experiment is ongoing. 
Figure~\ref{fig:ch-exec-comp-org-es} shows the current \dword{csc} organization, where 
only a few of the listed individuals are full-time on \dword{dune}.  Thus, we rely on common tools and techniques shared with other, smaller experiments at \dword{cern} and \dword{fnal}. In particular, \dword{dune} operates as one of several intensity frontier experiments at \dword{fnal} where a substantial amount of shared resources can be accessed, but with few personnel assigned specifically to \dword{dune}.  

The full \dword{dune} software and computing effort will be much larger than that for \dword{protodune}. 
The unique \dword{dune} data footprint and anticipated evolution in processor technologies will require a major undertaking to construct and operate the necessary computing infrastructure. 
This infrastructure must be in place  well before detector commissioning begins at \dword{surf}.


Two basic design tenets are cooperation with the broader community and reuse of tools. Collaboration scientists will develop much of the high-level algorithms, but a dedicated group of experts with advanced programming and project management skills are needed to build and operate the core software infrastructure for the experiment. 
Personnel resources are similar to those required for \dword{lhcb} and \dword{belleii}, which are collaborations of a similar size to \dword{dune}. 
Based on a comparison to those experiments, approximately 20 full-time-equivalent (\dword{fte}) workers should be dedicated to ensure primary \dword{dune} software and computing needs are met. 
This estimate of future personnel requirements follows from an assessment of the \dword{lhcb} organization structure.

Appendix~\ref{appx:comp-roles} describes computing personnel activities in detail.  In summary, we will need approximately 20 \dword{fte}, with 10 \dword{fte} for a software development team that will create and maintain the primary software needed to run \dword{dune} algorithms, and the distributed software infrastructure.  Some of these efforts will be completed jointly with other collaborations and \dword{hsf}/\dword{wlcg} projects, but in return, \dword{dune} must make substantive contributions to these common efforts. In addition to the software development effort,  \dword{dune} computing will require 
specific operational roles such as data manager, code librarian, and user support. 
Based on \dword{lhcb} experience, we have identified ten such roles 
each requiring an  \dword{fte} of 0.5 to 2.0.  These roles can be filled by experienced \dword{dune} collaborators or computing professionals, and their contributions to the experiment should be properly recognized as equivalent to efforts in construction or operation of the experiment. 

The \dword{csc} has instituted a series of workshops, which started with two on the topic of  ``Data Model and Infrastructure'' in the summer and fall of 2019, to set the scope of the subprojects to prepare for a formal computing \dword{tdr}. Table~\ref{tab:comp-milestones} gives a draft timeline for the computing project.

\begin{dunetable}[Milestones for DUNE computing development]{l l r}{tab:comp-milestones}{Milestones for \dword{dune} computing development.  Data volumes assume \SI{15}{PB/year} of compressed raw data starting in 2024.}
Year	&	Activity	&	Integrated data, PB	\\ \toprowrule%
2018	&  	&	10	\\ \colhline
	& 	\dshort{pdsp} beam run	&	\\ \colhline
2019	&		&	19	\\ \colhline%
	&	\dshort{pdsp} processing	&		\\ \colhline%
	&	\dshort{pddp} commissioning and data taking	&		\\ \colhline%
	&	Develop resource model	&		\\ \colhline%
	&	Develop high level task list	&		\\ \colhline%
2020	&		&	21	\\ \colhline%
	&	Continue \dshort{protodune} processing/operations	&		\\%
	&	Formalize international resource model	&		\\ \colhline%
	&	Build operations team	&		\\ \colhline%
	&	Evaluate data and computing models	&		\\ \colhline%
	&	Database design for hardware	&		\\ \colhline%
2021	&		&	25	\\ \colhline%
	&	Produce Computing TDR	&		\\ \colhline%
	&	Framework modifications for HPC 	&	\\ \colhline%
	&	Database design for conditions/configuration	&		\\ \colhline%
2022	&		&	39	\\ \colhline%
	&	\dshort{protodune} second beam run	&		\\ \colhline%
	&	Begin large scale purchases for FD commissioning	&		\\ \colhline%
2023	&		&	43	\\ \colhline%
	&	Reconstruct/analyze \dshort{protodune} results	&		\\ \colhline%
	&	Continue \dshort{protodune} processing/operations	&		\\ \colhline%
	&	Support FD commissioning	&		\\ \colhline%
	&	Conditions and configuration data fully integrated	&		\\ \colhline%
	&	Acquire storage for first year of data from one module	&		\\ \colhline%
2024	&		&	66	\\ \colhline%
	&	First real data from one FD module	&		\\ \colhline%
	&	Full operations team in place	&		\\ \colhline%
	&	Data analysis challenges	&		\\ \colhline%
2025	&		&	88	\\ \colhline%
	&	Complete provisioning of hardware/storage for first beam run	&		\\ \colhline%
2026	&		&	111	\\ \colhline%
	&	First beam run with two modules 	&	 	\\%
	\end{dunetable}

\subsection{Resource Contributions}
\label{sec:exec-comp-gov-rc}

The \dword{csc} resource board is developing a formal resource funding model. Currently, we would expect collaborating countries to contribute to computing physical resources and operational duties (e.g., shifts) distributed fairly and developed in consultation with the full \dword{dune} collaboration.  The core of the software development effort would mainly come from \dword{csc} members.  Contributions will be a mix of CPU resources, storage, and personnel, with the mix tailored to the resources and capabilities of each country and institution. To date, these contributions have been voluntary and opportunistic, but will evolve to a more formal model similar to the pledges in the \dword{lhc} experiments.

\section{Data Types and Data Volumes}
\label{sec:exec-comp-dtv}

Offline computing for \dword{dune} creates new and considerable challenges because of the experiment's large scale and 
diverse physics goals.  In particular, the advent of \dwords{lartpc} with an exquisite resolution and sensitivity, combined with the enormous physical volume of the \dword{dune} far detector and its large number of readout channels, presents challenges in acquiring, storing, reducing, and analyzing a 
prodigious amount of data --- orders of magnitude more data than in previous neutrino experiments. 

Neutrino experiments operate at low event rates, on the order of \SI{1}{Hz} even for near detectors. 
Despite these low event rates, 
\dword{dune} can generate enormous amounts of data from a single event. 
This leads to unique challenges in cataloguing, storing and reconstructing data, even though the total volume of data and CPU needs are significantly lower than in large collider experiments.  At a collider, each of the billions of triggered beam crossings is reasonably small and  effectively independent of the others. In contrast, a single \dword{dune} trigger readout can be many GB in size and, in the case of a supernova candidate, many TB. Maintaining the coherence of such large correlated 
volumes of data in a distributed computing environment presents a significant challenge. 

The rapidly changing computing landscape presents other challenges in this area, as 
the traditional \dword{hep} architecture of individual core processors running single-threaded applications 
is superseded by applications that efficiently use multiple processors and perhaps even require GPUs. At the same time, despite the rapid development of algorithms for \dword{lartpc} reconstruction, they are by no means mature. 
The \dword{pdsp} test at \dword{cern} in fall 2018 has contributed greatly to this development by providing a wealth of data that will inform the evolution of future \dword{dune} computing models.  

The \dword{csc},  \dword{sp} \dword{daq},  and host laboratory 
have agreed on a preliminary maximum data transfer rate from the \dword{fd} to \dword{fnal} of \surffnalbw{}, consistent with projected network bandwidths in the mid 2020s, and a limit of \SI{30}{PB/year} raw data stored to tape. 
Calibration for the \dword{fd} modules (\SIrange{10}{15}{PB/year/module}) and beam and cosmic ray interactions in the \dword{nd} will dominate uncompressed data volumes.  With a  factor of four for lossless compression, we anticipate a total compressed data volume of \SIrange{3}{5}{PB/year/module} for the full \dword{fd}; \dword{nd} rates are not yet established but are likely smaller.   

This section describes the make-up and rates of the data to be transfered. 

\subsection{Single-phase Technology Data Estimates}
\label{ch:exec-comp-dtv-sptde}

Each of the 150 \dword{spmod} \dwords{apa} (Section~\ref{sec:exec-sp-apa}) has 2,560 readout channels, each of which 
is sampled with 12 bit precision every \SI{500}{ns}. 
For a \dword{tpc} of this size, drift times in the \dword{lar} are approximately \SI{2.5}{ms} and the volume of raw data before compression is approximately \SI{6}{GB} 
per \SI{5.4}{ms} readout window.  With no triggering and no zero suppression or compression, the volume of raw data  for the four modules would be on the order of \SI{145}{exaB/year}. Table~\ref{tab:exec-comp-bigpicture-es} summarizes the relevant parameters for the \dword{sp} technology.  For our calculations of data volume, we assume lossless compression and partial, rather than full, readouts of regions of interest (\dwords{roi}) in the \dword{fd} modules.  We do not assume zero-suppression at the level of single channels. 

\begin{dunetable}[Useful quantities for computing \dshort{sp}
estimates]{lrr}{tab:exec-comp-bigpicture-es}
{Useful quantities for computing estimates for \dword{sp}
readout.}
Quantity&Value&Explanation\\
\toprowrule
{\bf Far Detector Beam:}\\ \colhline
Single APA readout &41.5 MB& Uncompressed 5.4 ms\\ \colhline
APAs per module& 150&\\
Full module readout &6.22  GB& Uncompressed 5.4 ms\\ \colhline
Beam rep. rate&\beamreprate&Untriggered\\ \colhline
CPU time/APA&100 sec&from MC/ProtoDUNE\\ \colhline
Memory footprint/APA&0.5-1GB&ProtoDUNE experience\\ \colhline
{\bf Supernova:}\\ \colhline
Single channel readout &300 MB& Uncompressed 100 s\\ \colhline
Four module readout& 450 TB& Uncompressed 100 s\\ \colhline
Trigger rate&1  per month&(assumption)\\
\end{dunetable}

\begin{dunefigure}[Expected physics-related activity
    rates in one FD module]{fig:daq-rates}{Expected physics-related activity
    rates in a single \nominalmodsize \dword{spmod}. Figure~from \spchdaq{}. \label{sec:comp:rates}
}
  \includegraphics[width=0.7\textwidth,clip,trim=6cm 6cm 10cm 2cm]{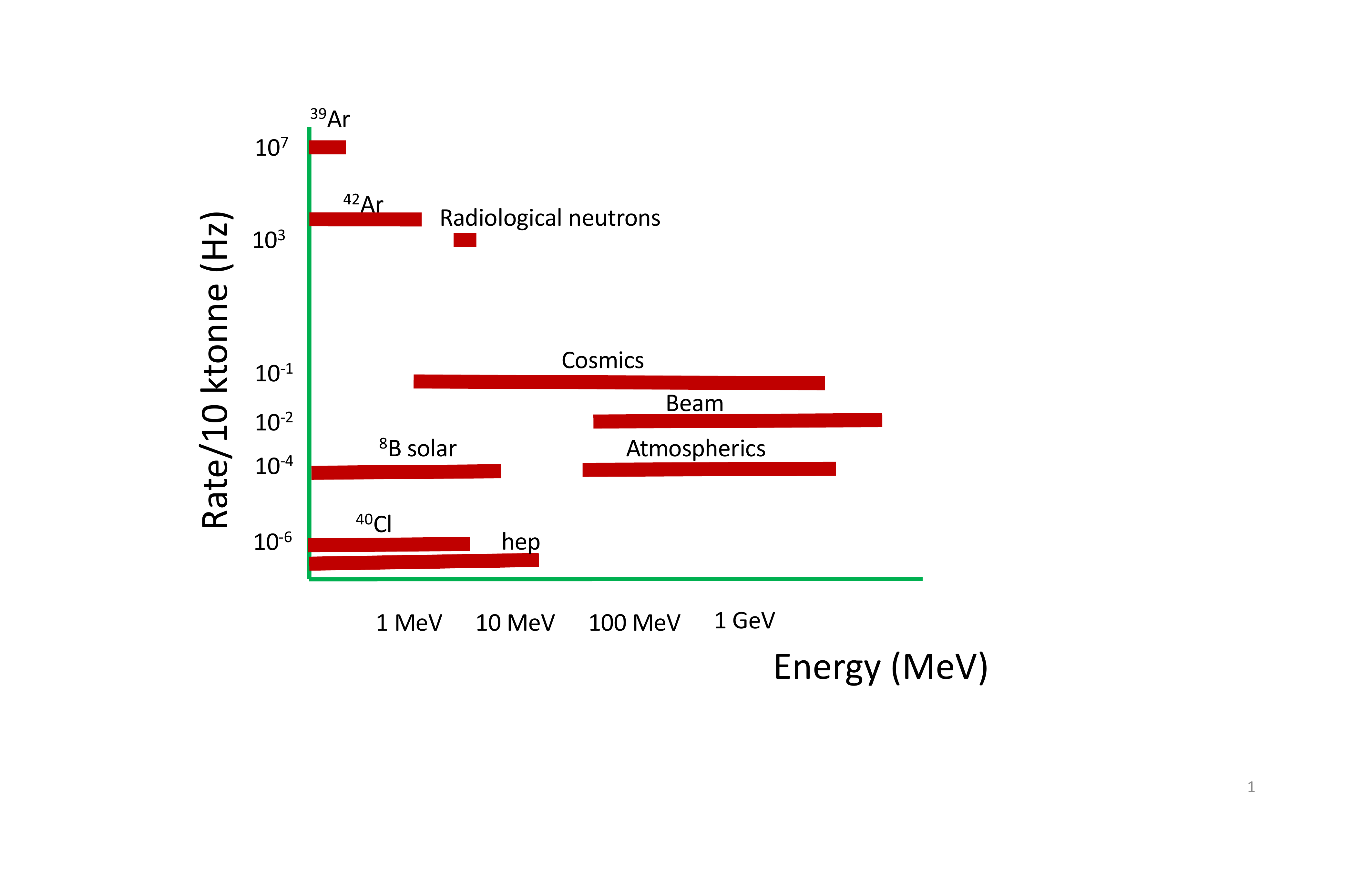}
\end{dunefigure}  

\begin{dunetable}
[Expected DAQ yearly data rates]
{p{0.3\textwidth}p{0.1\textwidth}p{0.5\textwidth}}
{tab:daq-data-rates-sp}
{Summary of expected data rates for a single \nominalmodsize \dword{spmod} (from Volume~\volnumbersp{}, \voltitlesp{}).  The rates assume no compression. \Ar39 decay candidates are 
temporarily stored for one to two months at a time. The same applies to fake \dword{snb} data. Improved readout algorithms will be developed and evaluated with the initial data and are expected to reduce the data volume by about a factor of ten while retaining efficiency.}
Source& Annual Data Volume & Assumptions \\ \toprowrule
Beam interactions & 27 TB & 10 MeV threshold in coincidence with beam
time, including cosmic coincidence; \SI{5.4}{\milli\second} readout \\ \colhline
$^{39}$Ar, cosmics and atmospheric neutrinos & 10 PB & \SI{5.4}{\milli\second} readout \\ \colhline
Radiological backgrounds & $<2$ PB & $<1$ per month fake rate for SNB
trigger\\\colhline
Cold electronics calibration & 200 TB & \\ \colhline
Radioactive source calibration & 100 TB & $<10$ Hz source rate; single
APA readout; \SI{5.4}{\milli\second} readout \\\colhline
Laser calibration & 200 TB & 10$^6$ total laser pulses; half the
TPC channels illuminated per pulse; lossy
compression (zero-suppression) on all channels\\\colhline
Random triggers & 60 TB & 45 per day\\
\end{dunetable}

\subsection{Dual-phase Technology Data Estimates}
\label{sec:exec-comp-dt-dptde}

A \dword{dpmod} will have 153,600 readout channels and a full drift time of \SI{7.5}{ms}. Given 20,000 samples in an \SI{8}{ms} readout, the uncompressed event size is \SI{4.2}{GB} (for one drift window).  Gas amplification leads to a high \dword{s/n} ratio, allowing lossless compression to be applied at the front end  with a compression factor of ten; this brings the event size per module to \SI{.42}{GB}.

An  \dword{fd} \dword{dpmod} can be treated as twenty smaller  detectors, each with a number of readout channels similar to  \dword{pddp} 
running in parallel,  and each defining a sub-module \dword{roi}. For beam or cosmic events, it is possible to record only the interesting \dwords{roi} with the compressed size of a single \dword{roi} at \SI{22}{MB}.

\subsection{Data Rates}
\label{sec:exec-comp-dt-dr}

Figure \ref{sec:comp:rates} illustrates the raw rates and energy ranges for relevant physical processes in a single \dword{dune} \dword{fd} module.

\subsubsection{Beam Coincident Rates}
\label{sec:exec-comp-dt-bcr}

 Requiring  coincidence with the \SI{10}{\micro\second} \dshort{lbnf} beam spill would reduce the effective live time from $\sim\,$\SI{1.2}{s} to a \SI{5.4}{ms} readout window (\SI{8}{ms} for \dword{dp}), leading to an uncompressed event size of approximately \SI{24}{GB} for four \dwords{spmod} (somewhat less for \dword{dp}) --- still too much to record permanently at full rate. 
Only a few thousand true beam interactions in the \dword{fd} modules are expected each year.  Compression and conservative triggering based on \dwords{pd} and ionization should reduce the data rate from beam interactions by several orders of magnitude without sacrificing efficiency.  Studies discussed in  \spchdaq{} indicate that high trigger efficiencies are achievable at an energy threshold of \SI{10}{MeV}, leading to event rates for beam-initiated  interactions of $\sim\,$6,400/year.
Table \ref{tab:daq-data-rates-sp}, adapted from Volume~\volnumbersp{} Chapter~7, 
 summarizes expected uncompressed rates from one \dword{spmod}. 

\subsubsection{Near Detector} 
\label{sec:exec-comp-dt-nd}

The \dword{nd} configuration is not yet fully defined,  but we do have substantial experience from \dword{t2k} and  \dword{microboone} at lower energies, and  \dword{minerva} and \dword{nova} at the  \dword{dune} beam energies on cosmic and beam interactions under similar conditions.  We expect that the \dword{dune} \dword{nd} will have $\sim\,$1 beam interaction per m$^3$ per beam pulse, and non-negligible rates of cosmic rays. Initial estimates indicate that zero-suppressed data rates will be of order \SI{10}{MB/s} with yearly data volumes less than a PB.  

\subsubsection{Processes not in Synchronization with the Beam Spill} 
\label{sec:exec-comp-dt-psbs}

Processes not associated with the beam spill 
include \dword{snb} physics, proton decay, neutron conversion, and atmospheric and solar neutrino interactions.  These processes generally have less energy, making triggering more difficult, and they are  asynchronous, thus requiring an internal or external trigger.  In particular, \dword{snb} signals will consist of a large number of low-energy interactions spread throughout the \dword{fd} volume over a period of 1-100 seconds. Buffering and storing 100 seconds of \dword{snb} data would require approximately 20,000 readout windows, or \SI{460}{TB} for a four-module  readout.  At a rate of one fake \dword{snb} event per month, this would generate  \SI{6}{PB} of uncompressed data per year.  Reconstructing and analyzing these data will require substantial evolution in our software frameworks, which were developed to process small (\SIrange{1}{100}{MB}) events on single processors. 
Updating the frameworks is a major upcoming task for the \dword{dune} computing R\&D. 

\subsubsection{Calibration}
\label{sec:exec-comp-cal}

The \dword{fd} modules will require continuous calibration and it is likely that these calibration samples 
will dominate the data volume. Cosmic-ray muons and atmospheric neutrino interactions will provide a substantial sample for energy and position calibrations.  Dedicated runs with radioactive sources and laser calibration will also generate substantial and extremely valuable samples. Table \ref{tab:daq-data-rates-sp} includes estimates for the 
\dword{spmod}. 

Electron lifetime in the detector volume can be monitored via $^{39}$Ar decays  at rates of $\sim\,$1/kg/sec. As discussed in the appendices for 
Volume~\volnumberphysics{}, \voltitlephysics{},  a single \SI{5}{ms} readout of the full detector would provide 50,000 decays for study.  A small number of such readouts per day would provide a global monitor of conditions at the 1\% level, but measurements sensitive on meter scales will require a factor of $10^4$ more data, and can become a significant fraction of the calibration data stream. In summary, $^{39}$Ar cosmic ray and atmospheric neutrino signals collected for calibration make up the bulk of the uncompressed \dword{sp} data volume at $\sim\,$\SI{10}{PB/year} per module and will dominate the rates from the \dword{fd} modules.  

\subsubsection{Zero Suppression}
\label{sec:exec-comp-cal-zs}

The data volumes discussed above are for non-zero-suppressed readout of the full \dword{fd}. A combination of local triggering, zero suppression, and  efficient lossless compression mechanisms can substantially reduce the final data volume. However, previous \dword{hep}experience  indicates that signal processing must be done carefully and 
is often done after data-taking, and when the data are well understood.  As a result, early running often generates large data volumes while algorithms are being tuned. 
Experience from  the \dword{sbn} and \dword{protodune} experiments will help us develop these data movement and processing algorithms, but they may be applied later in the processing chain for \dword{sp}.  No zero-suppression is planned for \dword{dp}.

\subsection{Simulation}
\label{sec:exec-comp-dt-sim}

The bulk of data collected with the \dword{fd} is likely to be background, with real beam interaction events in the \dword{fd} numbering in the thousands per year, not millions. Thus, the size of beam event simulation samples may be 
smaller than the unprocessed raw data considered above.  
Simulation of lower-energy events should reflect the fact that they are very rare; they could be simulated in sub-volumes of the whole detector. 
While simulation will be important to the experiment, it is not expected to dominate the data volume as it does in many experiments.  

Simulation inputs such as flux files, overlay samples, and shower libraries 
must be distributed to simulation jobs carefully.   Proper simulation requires that these inputs be distributed in unbiased parcels, which from a technical standpoint can be difficult to do efficiently in a widely distributed environment. This will require thoughtful design. 

\subsection{Analysis}
\label{sec:exec-comp-anal}

We anticipate that most analysis samples will be many times smaller than the raw data, however, since they are  
distinctive to particular analyses and even users,  producing and cataloguing them will require carefully designed tools and substantial oversight. 
\dword{dune} will need a mix of official samples, produced by physics groups and distributed through a common catalog and through common file transfer mechanisms, as well as small user samples on local disks. 

Final oscillation parameter scans with a large number of  
parameters can be quite CPU-intensive.  For example, the \dword{nova} collaboration's recent physics results required tens of millions of \dword{hpc} CPU hours at the \dword{nersc} 
facility at \dword{lbnl}. \dword{dune} collaborators used simpler models but the same techniques to generate some of the results presented in Volume~\volnumberphysics{}, \voltitlephysics{}. These large-scale analysis projects will require collaboration-wide coordination of resources and will benefit greatly from optimization for specific architectures.

\subsection{Data Storage and Retention Policies}
\label{sec:exec-comp-dsrp}

Some data samples, e.g., real neutrino and cosmic ray interactions in the \dword{fd}, most of the \dword{nd} data, and any real \dword{snb} events, will be extremely valuable and will require conservative and potentially redundant retention policies.    Calibration samples, and possibly fake \dword{snb} triggers, may be stored temporarily and discarded after processing.

\section{ProtoDUNE-SP Data}
\label{sec:exec-comp-proto-SP}

\subsection{Introduction}
\label{sec:exec-proto-intro}

\dword{pdsp} ran at \dword{cern} in the \dword{np04} beamline from September to November of 2018. Before that run, several data challenges at high rate validated the data transfer mechanisms. The run itself has already served as a  
substantial test of the global computing model, and studies of cosmic rays continue.

This section describes the \dword{protodune} data design and the lessons learned from our experience. 

\subsection{Data Challenges}

Starting in late 2017, a series of data challenges were performed with \dword{pdsp}.  Simulated data were passed through the full chain from the event builder machines to tape storage at \dword{cern} and \dword{fnal} at rates up to \SI{2}{GB/s}.  These studies allowed optimizing the network and storage elements well before the start of data taking in fall 2018. 
The full \dword{dune} \dword{fd}, in writing \SI{30}{PB/year}, will produce data at rates similar to  those in the 2018 data challenges. While accommodating the \dword{pdsp} data rates 
was not technically challenging, the integrated data volume from an experiment running 99\% of the time over several decades will be. 

\subsection{Commissioning and Physics Operations}

The first phase of operations involved commissioning the detector readout systems while the \dword{lar} reached full purity.  During this period data were taken with cosmic rays and beam. Once high \dword{lar} purity was reached, \dword{pdsp} collected physics data with beam through October and half of November 2018. 
After the beam run, we continued to take cosmic ray data under varying detector conditions, such as modified high voltage and purity, and new readout schemes. 

\subsection{Data Volumes}

\dword{pdsp} comprises a \dword{tpc} consisting of  six \dword{apa}s, their associated \dword{pd}s, and a \dword{crt}. 
In addition, the \dword{np04} beamline is instrumented with hodoscopes and Cherenkov counters to generate beam triggers. Random triggers  were generated at lower rates to collect unbiased cosmic ray information. During the test beam run, the \dword{tpc} readout dominated the data volume. The nominal readout window during the beam run was \SI{3}{ms} as a match to the drift time at full voltage (\SI{180}{kV}), which was maintained for most of the run.  
The \dword{tpc} alone produced \SI{138}{MB/event} without compression, not including headers. 
The uncompressed event size including all \dword{tpc} information, \dword{crt}, and \dword{pd} data was \SIrange{170}{180}{MB}. Data compression was implemented just before the October beam physics run, reducing this number to \SI{75}{MB}.  In all, 8.1 million beam events produced 
a total of \SI{572}{TB}.  An additional \SI{2.2}{PB} of commissioning data and cosmic ray data was also recorded. Table \ref{tab:exec-comp-pd-volumes} summarizes the data volumes recorded in \dword{pdsp} from October 2018 to October 2019.

\begin{dunetable}[Data volumes]{lrr}{tab:exec-comp-pd-volumes}{Data volumes  recorded by \dword{pdsp} as of October 2019.}
Type  & Events & Size\\ \rowtitlestyle
Raw Beam& 8.1 M& 520 TB \\ \colhline
Raw Cosmics&19.5 M&1,190 TB\\ \colhline
Commissioning&3.86 M& 388 TB\\ \colhline
Pre-commissioning&13.89 M&641 TB\\
\end{dunetable}

Events were written out in \SI{8}{GB} raw data files, each containing approximately 100 events. The beam was live for two \SI{4.5}{s} spills every \SI{32}{s} beam cycle, and data were taken at rates up to \SI{50}{Hz}, exceeding the typical \SI{25}{Hz}, leading to compressed \dword{dc} rates of \SIrange{400}{800}{MB/s} from the detector.  

\subsection{Reconstruction of ProtoDUNE-SP Data}
Thanks to substantial work by the \dword{35t}, \dword{microboone}, and the \dword{lartpc} community, high-quality algorithms were in place to reconstruct the \dword{tpc} data.  As a result, a first pass reconstruction of the \dword{pdsp} data with beam triggers was completed in December, 2018, less than a month after data taking ended.  Results from that reconstruction are presented in \physchtools 
with some highlights summarized here. 

\subsection{Data Preparation}

Before pattern recognition begins, data from the \dword{protodune} detector is
unpacked and copied to a standard format within the \dword{art} framework based on \dword{root} objects. 
This reformatted raw data includes the waveform for each channel, consisting of 6,000-15,000  12-bit, 0.5 $\mu$sec samples. 
The first step in reconstruction is data preparation, which entails the conversion of each \dword{adc} waveform into a calibrated charge waveform with
signals proportional to charge. Once the data are prepared, hit-level \dwords{roi} are identified, and data outside these regions are discarded, significantly reducing data size. The data preparation process is described more fully in~\cite{bib:docdb12349}. 

Figures~\ref{fig:ch-exec-comp-chtraw} and~\ref{fig:ch-exec-comp-chtroi} illustrate the transformation of \dword{tpc} data  during data preparation 
for one wire plane for \SI{3}{ms}.  A full \SI{5.4}{ms} readout of a single \nominalmodsize module would contain a factor of 750 times 
more information than this image.
\begin{dunefigure}
[Pedestal-subtracted data for a ProtoDUNE-SP  wire plane]
{fig:ch-exec-comp-chtraw}
{Example of pedestal-subtracted data for one \dword{pdsp}  wire plane.  The top pane shows the \dword{adc} values in a $V$ (induction) plane (Section~\ref{sec:exec-sp-apa}) with the $x$ axis as channel number and the $y$ axis as time slice. The bottom pane shows the bipolar pulses induced on one channel.}
\includegraphics[width=\textwidth,angle=0]{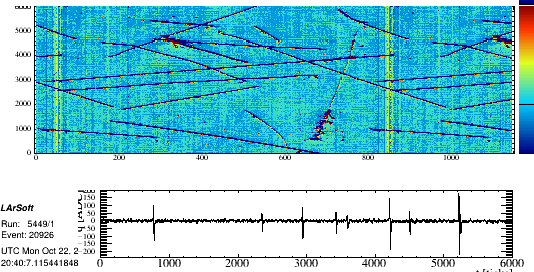}
\end{dunefigure}

\begin{dunefigure}
[Pedestal-subtracted data for a ProtoDUNE-SP  wire plane, after additional processing]
{fig:ch-exec-comp-chtroi}
{Pedestal-subtracted data for one \dword{pdsp} wire plane, as in Figure~\ref{fig:ch-exec-comp-chtraw}, after calibration, cleanup, deconvolution, and finding \dwords{roi}.}
 \includegraphics[width=\textwidth]{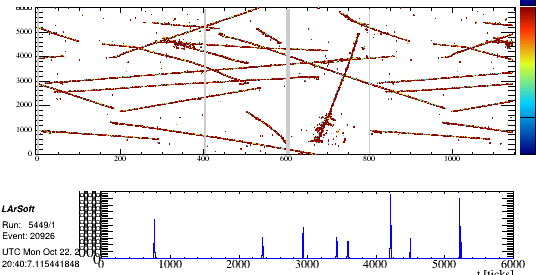}
\end{dunefigure}


\subsection{Computational Characteristics of Data Preparation and Deconvolution }

Decoding for \dword{pdsp} originally stored all six \dword{apa}s in memory. Each \SI{3}{ms} of \dword{apa} readout consists of more than \SI{15}{M} 16-bit values. Decompressing and converting this information to a floating point format causes substantial memory expansion. 
  Data with a \SI{7.5}{ms} window were also recorded. 
The input and output event sizes and reconstruction time scale were found to scale linearly with the readout window and with the number of \dword{apa}s processed. 

Processing each wire plane (three per \dword{apa}) independently reduces the memory footprint because electrical signals correlate between channels within an \dword{apa} wire plane but not between planes.
However,  although subdividing the detector into wire planes reduces the memory footprint for short (beam-related) readouts, this is  not a viable solution for the long readouts expected for \dword{snb} events. We are still seeking the best strategy for these much larger ($\times 10,000$) time windows. 

The \dword{daq} consortium is exploring methods for segmenting large events (such as \dwords{snb}) into  smaller \dword{roi}s in both time and space for efficient readout.  As long as those regions are on the scale of single interactions, the resulting data should fit into a reasonable memory budget at the expense of tracking and collating many distributed interactions. 

The operations performed in signal processing require few decisions but do include operations such as fast Fourier transforms and deconvolution.  These operations are well suited for GPU and parallel processing. We are actively exploring multi-threaded processing for all data preparation algorithms.

\subsection{Reconstruction Characteristics}

Once \dword{roi}s have been identified, several \threed  reconstruction packages are used. For the first reconstruction pass in November, the  \dword{pandora}~\cite{Acciarri:2017hat}, \dword{wirecell}~\cite{wirecell}, and \dword{pma}~\cite{ref:PMA}  frameworks were used. The results are described in Volume~\volnumberphysics{}, \voltitlephysics{}. 
Figure~\ref{fig:ch-exec-comp-tracking}, taken from that volume, illustrates the measured efficiency for the \dword{pandora} algorithm reconstructing a triggered beam particle as a function of momentum for the simulation and data for selected data taking runs. Figure~\ref{fig:ch-exec-comp-tracking} demonstrates that the efficiency is already high and reasonably well simulated.

Full reconstruction of these \dword{pdsp} interactions, with beam particles and approximately 60 cosmic rays per readout window, took  approximately \SI{600}{s/event} with \SI{200}{s} each for the signal processing and hit finding stages; the remaining time was divided among three different pattern recognition algorithms. Output event records were substantially smaller (\SI{22}{MB} compressed) but were still dominated by information for \dword{tpc} hits above threshold. 

All these algorithms are run on conventional 
Linux CPUs using \dword{osg}/\dword{wlcg} grid computing  infrastructure. Deep learning techniques based on image pattern recognition algorithms are also being developed. Many of these algorithms can be adapted to run on \dword{hpc} assets, although the optimal architecture for \threed reconstruction likely differs from that for hit finding.

\begin{dunefigure}
[Efficiency of reconstruction for the triggered test beam particle]
{fig:ch-exec-comp-tracking}
{The efficiency of reconstruction for the triggered test beam particle as a function of particle
momentum in data (red) and simulation (black). (Figure~4.30 from Volume~\volnumberphysics{}, \voltitlephysics{}.)}
\includegraphics[height=4in]{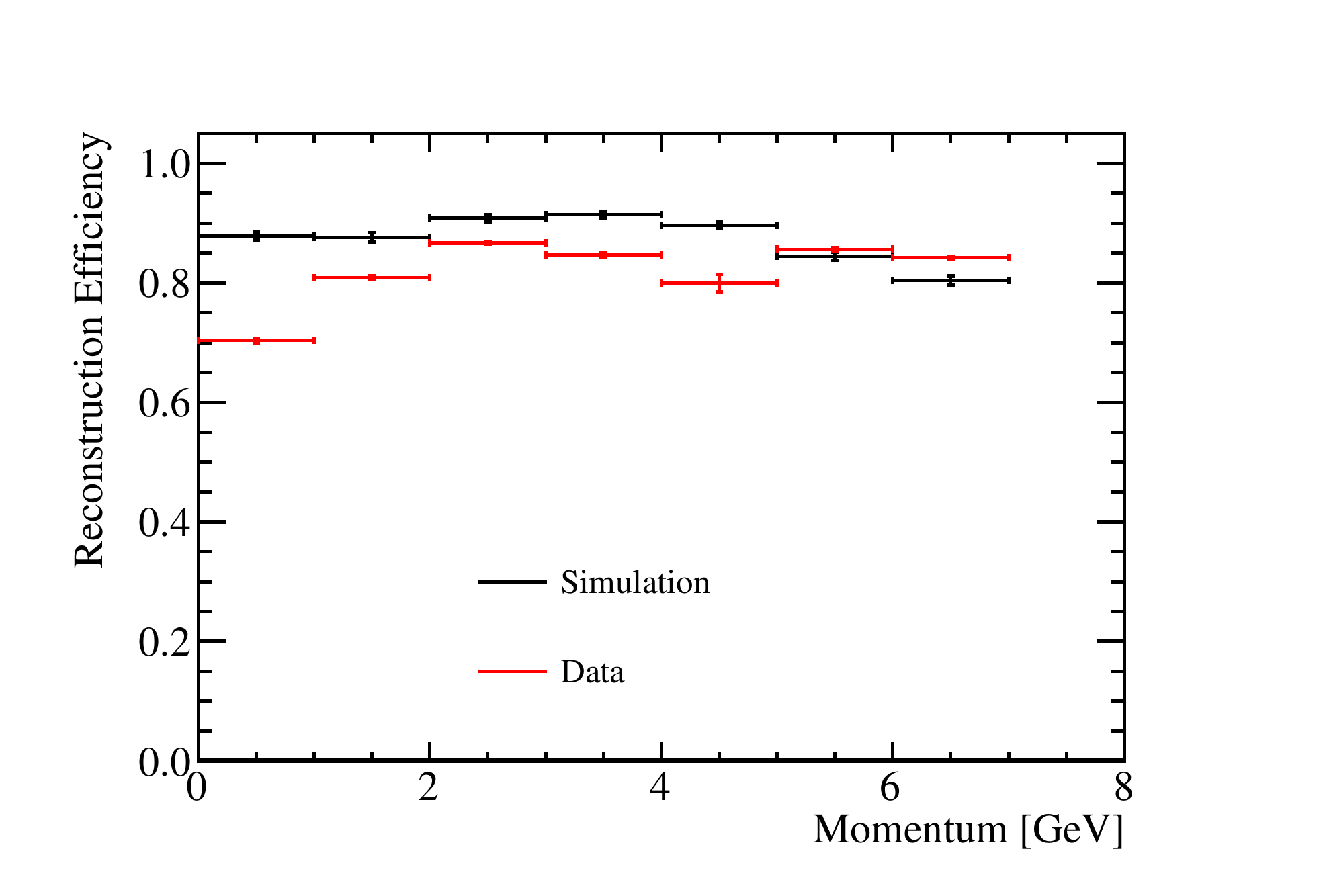}
\end{dunefigure}

\subsection{Lessons Learned}
The first \dword{pdsp} run has given us very valuable information for planning the full \dword{dune} computing model. 

\begin{itemize}
    \item Data and simulation challenges led to a reasonably mature and robust model for acquiring, storing, and cataloging the main data stream at design rates.
    \item The experiment integrated several existing grid sites and used substantial opportunistic resources.  This allowed initial processing of data within one month of the end of the run.
    \item Prototype infrastructure was in place for provisioning, authentication and authorization, data management, networking, file catalog, and workflow management. 
    \item Reconstruction algorithms were available, permitting immediate studies of detector performance and calibration. 
    \item Beam information was successfully integrated into processing through 
    an \dword{ifbeam} database.
    \item Auxiliary information from some systems, e.g., slow controls, was not fully integrated into processing. This led to a manual input of the running conditions by shift personnel, and offline incorporation of that information into the data catalog. This prompted a closer collaboration between the \dword{daq} and \dword{cisc} groups and the design of robust interfaces for configurations and conditions. 
\end{itemize}

Overall, the \dword{pdsp} data taking and processing was a success, despite too much reliance on manual intervention 
because  automated processes were not always in place. 
Considerable effort must go into integrating detector conditions, data migration, workflow systems, and \dword{hpc}s with multi-threaded and vectorized software.
 
\cleardoublepage

\chapter{DUNE Far Site Technical Coordination}
\label{ch:exec-tc}

\textit{This chapter provides a brief introduction to the DUNE far site technical coordination.  The text below closely follows that found in the introductory chapters of Volume~\volnumbertc{}, \voltitletc{}, where many more details may be found.}

\section{Overview}   

The \dword{dune} collaboration has  responsibility for the design 
and construction of the \dword{dune} \dword{fd}.  Groups of collaboration 
institutions, referred to as consortia, assume responsibility for 
the different detector subsystems.  The activities of the consortia are 
overseen and coordinated through the \dword{dune} \dword{tc} organization 
headed by the \dword{dune} \dword{tcoord}.  The \dword{tc} organization 
provides project support functions such as safety coordination, 
engineering integration, change control, document management, scheduling, 
risk management, and technical review planning.  \dword{dune} \dword{tc} 
manages internal, subsystem-to-subsystem interfaces, and is responsible 
for ensuring the proper integration of the different subsystems.

\dword{dune} \dword{tc} works closely with the support teams of its 
\dword{lbnf-dune} partners within the framework of a \dword{jpo} to 
ensure coherence in project support functions across the entire global 
enterprise.  To ensure consistency of the \dword{dune} \dword{esh} 
and \dword{qa} programs with those across \dword{lbnf-dune}, the 
\dword{lbnf-dune} \dword{esh} and \dword{qa} managers, who sit within 
the \dword{jpo}, are embedded within the \dword{dune} \dword{tc} 
organization.  

The \dword{lbnf-dune} \dword{integoff} under the 
direction of the \dword{ipd} incorporates the on-site team responsible 
for coordinating integration and installation activities at the \dword{surf}.
Detector integration and installation activities are supported by the
\dword{dune} consortia, which maintain responsibility for ensuring
proper installation and commissioning of their subsystems.  External
\dword{dune} interfaces with the on-site integration and installation
activities are managed through the \dword{jpo}.

\section{Global Project Organization}   
\label{sec:exec-tc-partners}

\subsection{Global Project Partners}

The \dword{lbnf} project is responsible for providing both the
\dword{cf} and supporting infrastructure (cryostats and
cryogenics systems) that house the \dword{dune} \dword{fd}
modules. 
The international \dword{dune}
collaboration under the direction of its management team is
responsible for the detector components.  The \dword{dune} \dword{fd}
construction project encompasses all activities required for designing
and fabricating the detector elements and incorporates contributions
from a number of international partners.  The organization of 
\dword{lbnf-dune}, which encompasses both project elements, is
shown in Figure~\ref{fig:DUNE_global}.

\begin{dunefigure}[Global project organization]{fig:DUNE_global}
  {The global \dword{lbnf-dune} organization.}
  \includegraphics[width=0.95\textwidth]{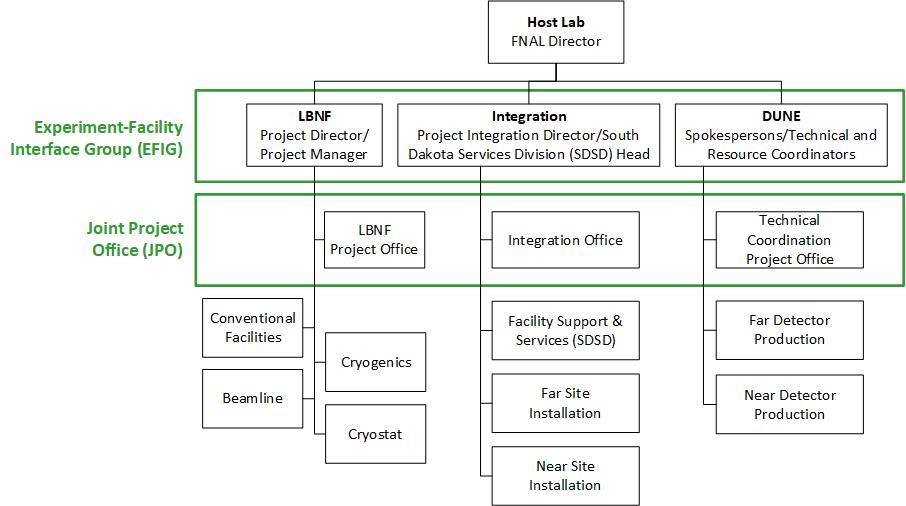}
\end{dunefigure}

The overall
coordination of installation activities in the underground caverns 
is managed as a separate element of \dword{lbnf-dune} under the
responsibility of the \dword{ipd}, who is appointed by and reports
to the \dword{fnal} director.  To ensure coordination across
all elements of \dword{lbnf-dune}, the \dword{ipd} connects to both
the facilities and detector construction projects through ex-officio
positions on the \dword{lbnf} project management board and
\dword{dune} \dword{exb}, respectively.  The \dword{ipd} receives support from the \dword{sdsd},
a \dword{fnal} division established to 
provide the necessary supporting infrastructure for installation, commissioning, and operation 
of the \dword{dune} far detector. 

The \dword{efig} is the body responsible for the required high-level
coordination between the \dword{lbnf} and \dword{dune} construction 
projects. 
The \dword{efig} is augmented by the \dword{jpo} that supports both 
the \dword{lbnf} and \dword{dune} projects as well as the integration
effort that connects the two together. The \dword{jpo} combines
project support functions that exist within the different elements 
of the global project to ensure proper coordination across the entire 
\dword{lbnf-dune} enterprise.  Project functions coordinated globally 
through the \dword{jpo} are shown in Figure~\ref{fig:DUNE_jpo} along 
with the team members 
currently supporting these functions  
within the \dword{jpo} framework. 
These team members are drawn from the \dword{lbnf} project office, \dword{dune} \dword{tc}, 
and \dword{lbnf-dune} \dword{integoff} personnel.  
\begin{dunefigure}[JPO functions]{fig:DUNE_jpo}
  {\dword{jpo} global support functions and teams}
  \includegraphics[width=0.85\textwidth]{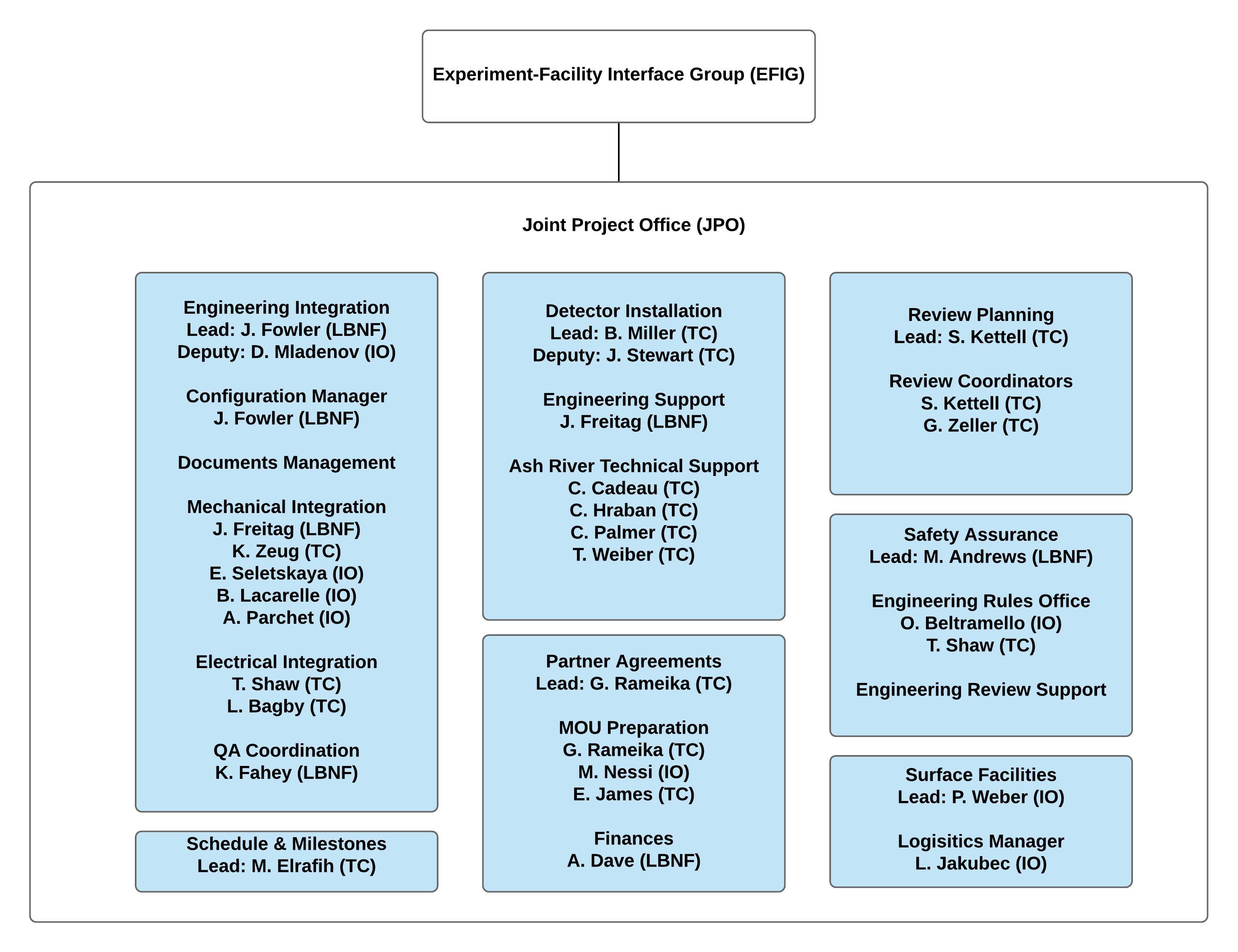}
\end{dunefigure}

\subsection{Coordinated Global Project Functions}

Project support functions requiring \dword{jpo} coordination include
safety, engineering integration, change control and document 
management, scheduling, review planning and oversight, and development 
of partner agreements.  

Planning activities related to detector installation and the provision 
of surface facilities are also currently embedded within the framework 
of the \dword{jpo} to ensure that all project elements are properly 
incorporated.  At the time when \dword{lbnf} \dword{fscf} delivers 
\dword{aup} of the underground detector caverns at \dword{surf}, the 
coordination of on-site activities associated with detector installation 
and the operation of surface facilities will be fully embedded within 
the \dword{lbnf}/\dword{dune} \dword{integoff} under the direction of the \dword{ipd}.  


\subsection{Coordinated Safety Program}    
\label{sec:dune_safety}

To ensure a consistent approach to safety across \dword{lbnf-dune},
a single \dword{lbnf-dune} \dword{esh} manager reports 
to the \dword{lbnf} project director, the \dword{ipd}, and \dword{dune}
management (via the \dword{dune} \dword{tcoord}).  This individual
directs separate safety teams responsible for implementing the
\dword{lbnf-dune} \dword{esh} program within 
the individual \dword{lbnf} 
and \dword{dune} projects as well as the coordinated \dword{lbnf}/\dword{dune}
installation activities at \dword{surf}. The safety organization 
is shown in Figure~\ref{fig:dune_esh}.

\begin{dunefigure}[\dshort{lbnf-dune} \dshort{esh}]{fig:dune_esh}
  {High level \dword{lbnf-dune} \dword{esh} organization.}
  \includegraphics[width=0.85\textwidth]{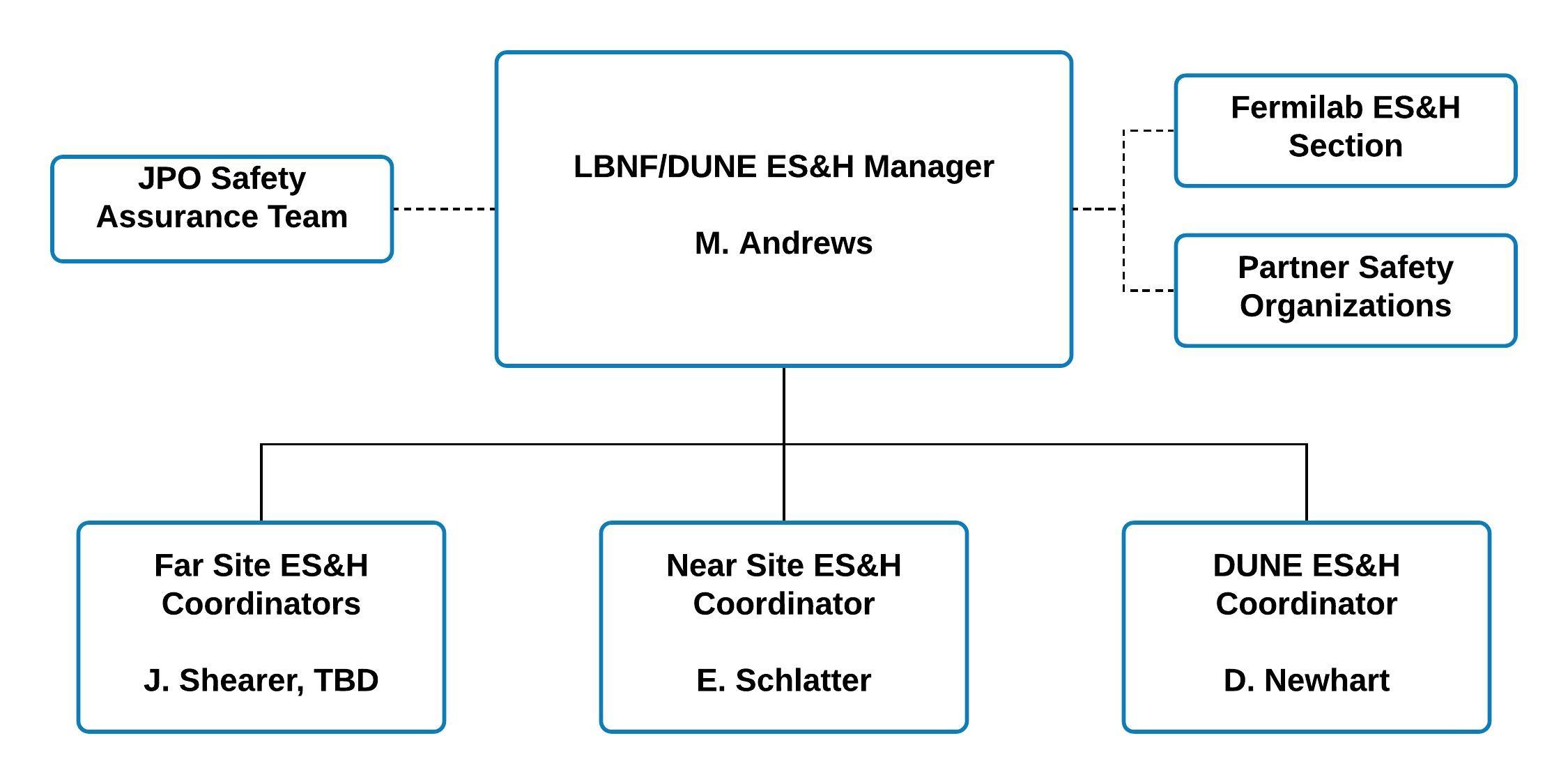}
\end{dunefigure}
The \dword{lbnf-dune} \dword{esh} manager works with the \dword{fnal} 
and \dword{surf} safety organizations to ensure that all project-related 
activities comply with the rules and regulations of the host 
organizations.  

The \dword{jpo} engineering safety assurance team defines a common 
set of design and construction rules (mechanical and electrical) to 
ensure consistent application of engineering standards and engineering 
documentation requirements across \dword{lbnf-dune}.  
Following lessons learned from the processes used for the 
\dword{protodune} detectors, an important mandate of the engineering 
safety assurance team is to ensure that safety issues related to 
component handling and installation are incorporated within the 
earliest stages of the design review process.  

\subsection{
Detector Integration}   
\label{sec:dune_engineering}

A central \dword{jpo} engineering team is responsible for building 
an integrated model of the detectors within their supporting
infrastructure and the \dword{fscf} that house them.  
This team incorporates approved changes as they 
are received and checks to ensure that no errors or space conflicts 
are introduced into the model.    After receiving the appropriate sign-offs from all 
parties, the 
team tags a new frozen release of the model 
and makes it available to the design teams as the current release 
against which the next set of design changes will be generated.

Electrical engineers are incorporated within 
this team to ensure proper integration of the detector electrical 
components. 

The \dword{jpo} engineering team is also responsible for documenting and
controlling the interfaces between the \dword{lbnf} and \dword{dune} 
projects as well as the interfaces between these projects and the 
\dword{lbnf}/\dword{dune}  installation activities at \dword{surf}.  

The \dword{lbnf-dune} project partners have agreed to adopt 
the formal change control process developed previously for the 
\dword{lbnf} project.  The change control process applies to 
proposed modifications of requirements, technical designs, 
schedule, overall project scope, and assigned responsibilities 
for individual scope items. 

\subsection{Schedule and Milestones}   
\label{sec:dune_schedule}

The \dword{jpo} team is responsible for creating a single project
schedule for \dword{lbnf-dune} that incorporates all \dword{lbnf} and
\dword{dune} activities together with the installation activities at
\dword{surf}, incorporating all interdependencies.  This schedule
will be used to track the status of the global enterprise.   
\dword{doe} activities will be 
tracked using the formal \dword{evms} procedures required for the 
\dword{doe} project activities; non-\dword{doe} activities will be tracked through regular assessments 
of progress towards completion by the management teams responsible 
for those activities.

\subsection{Partner Agreements and Financial Reporting}   
\label{sec:dune_agreements}

Partner contributions to all project elements will be detailed 
in a series of written agreements.  In the case of \dword{lbnf}, 
these contributions will be spelled out in bilateral agreements 
between the \dword{doe} and each of the contributing partners.  In 
the case of \dword{dune},  a \dword{mou} will 
detail the contributions of all participating partners.  
A series of more technical agreements describing the exact 
boundaries between partner contributions and the terms and 
conditions under which they will be delivered will 
accompany the primary agreements.

\section{DUNE Far Detector Organization}   
\subsection{Detector Design and Construction}   
\label{sec:es-tc-det-const}

The \dword{dune} \dword{fd} construction project refers collectively 
to the activities associated with the design and construction of 
necessary detector components.  \dword{dune} collaboration management 
is responsible for overseeing this portion of  \dword{lbnf-dune} and 
ensuring its successful execution.  The high-level \dword{dune} 
collaboration management team consisting of the co-spokespersons, 
\dword{tcoord}, and \dword{rcoord} is responsible for the
management of the construction project.  

Construction of the \dword{fd} modules is carried out by 
consortia of collaboration institutions who assume responsibility 
for detector subsystems.  Each consortium plans and executes the 
construction, installation, and commissioning of its subsystem.

Each consortium is managed by an overall consortium leader 
and a technical lead.  The consortium leader chairs an institutional 
board composed of one representative from each of the 
institutions contributing to the activities of the consortium.  Major 
consortium decisions such as technology selections and assignment of 
responsibilities 
among the institutions 
pass  through its institutional board.  These decisions are then passed 
as recommendations to the \dword{dune} \dword{exb} for formal collaboration approval.

Because the consortia operate as self-managed entities, a strong
\dfirst{tc} organization is required to ensure overall integration 
of the detector elements and successful execution of the detector
construction project.  \Dword{tc} areas of responsibility include 
general project oversight, systems engineering, \dword{qa}, and 
safety.  \Dword{tc} also supports the planning and execution 
of integration and installation activities at \dword{surf}.
The \dword{tcoord} manages the overall detector construction 
project through regular technical and project board meetings with 
the consortium leadership teams and members of the \dword{tc} 
organization.

The \dword{tc} organization, headed by the \dword{tcoord}, supports the work of 
the consortia and has responsibility for a number of major project 
support functions prior to the delivery of detector components to 
\dword{surf}, including
\begin{itemize}
\item ensuring that each consortium has a well defined and complete
  scope, that interactions between consortia are sufficiently 
  well defined, and that any missing scope outside of the 
  consortia is provided through other sources such as collaboration
  common funds;
\item defining and documenting scope boundaries and technical 
  interfaces both between consortia and with \dword{lbnf};  
\item developing an overall schedule with appropriate dependencies
  between activities covering all phases of the project; 
\item ensuring that appropriate engineering and safety standards 
  are developed, understood, and agreed to by all key stakeholders 
  and that these standards are conveyed to and understood by each
  consortium;
\item ensuring that all \dword{dune} requirements on \dword{lbnf} 
  for \dword{fscf}, cryostat, and cryogenics are clearly defined and 
  agreed to by each consortium;
\item ensuring that each consortium has well developed and reviewed
  component designs, construction plans, \dword{qc} processes, and 
  safety programs; and
\item monitoring the overall project schedule and the progress of 
  each consortium towards delivering its assigned scope. 
\end{itemize}
The \dword{dune} \dword{tc} organizational structure is shown 
in Figure~\ref{fig:DUNE_tc}. 
\begin{dunefigure}[DUNE technical coordination organization]{fig:DUNE_tc}
  {\dword{dune} \dfirst{tc} organizational chart.}
  \includegraphics[width=0.99\textwidth]{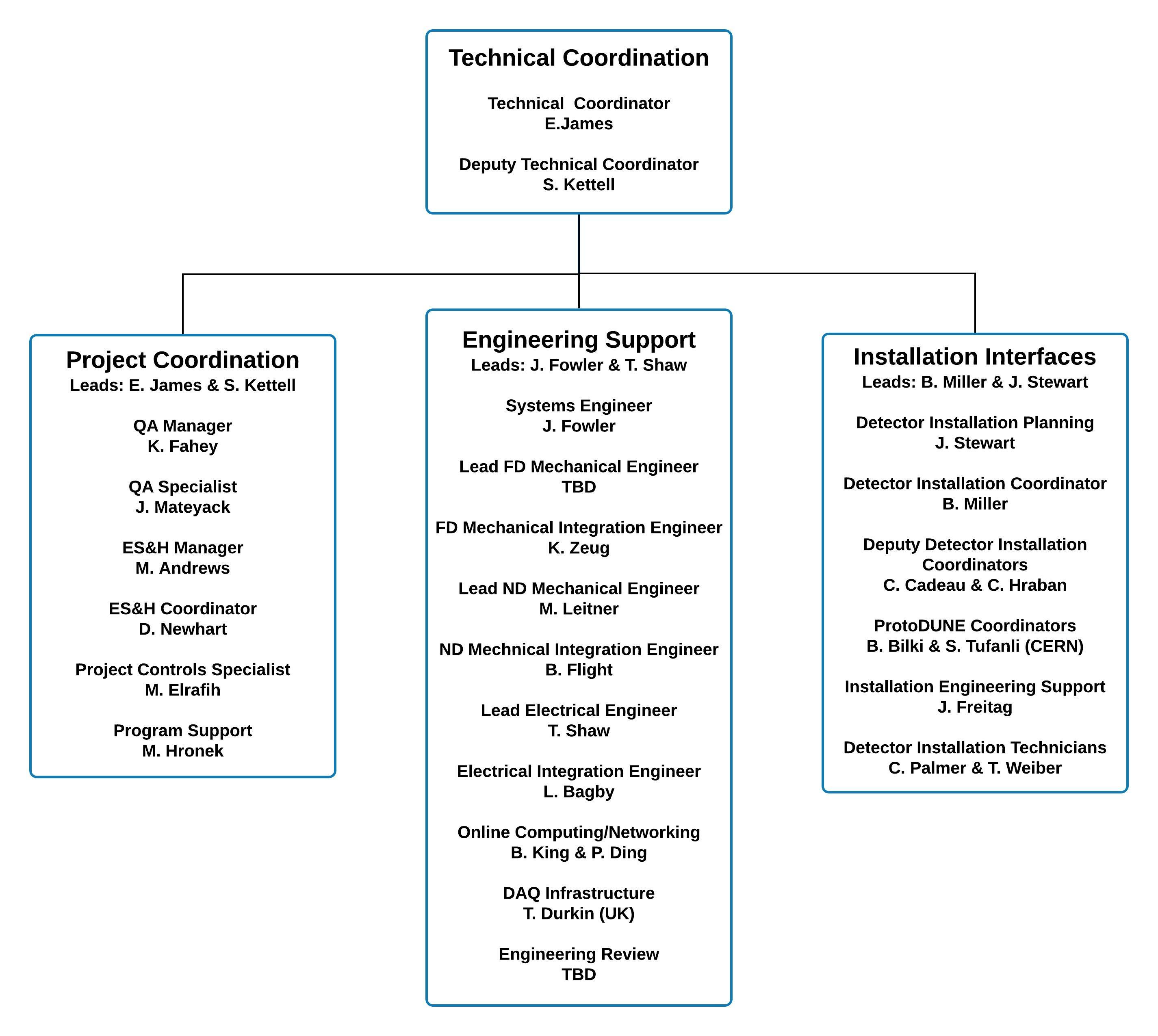}
\end{dunefigure}

The \dword{tc} project coordination team incorporates \dword{esh}, 
\dword{qa}, and project controls specialists.  Overall integration 
of the detector elements is coordinated through the \dword{tc} 
engineering support team headed by the \dword{lbnf-dune} systems 
engineer and lead \dword{dune} electrical engineer.  Planning 
coordinators for integration and installation activities at 
\dword{surf} (who sit within the \dword{lbnf-dune} \dword{integoff}) also head the \dword{tc} installation interfaces team.  
The dual placement of these individuals facilitates the required 
coordination of integration and installation planning efforts between 
the core team directing these activities and the \dword{dune} 
consortia. 

\subsection{Detector Installation and Commissioning}  
\label{sec:es-tc-det-instal}

The \dfirst{ipd} has
responsibility for coordinating the planning and execution of 
the \dword{lbnf-dune} installation activities, both 
in the underground detector caverns at \dword{surf} and in 
nearby surface facilities. 

The \dword{lbnf-dune} \dword{integoff} will evolve over 
time to incorporate the team in South Dakota responsible for the 
overall coordination of on-site installation activities.  In the 
meantime, the installation planning team within the \dword{integoff} works with 
the \dword{dune} consortia and \dword{lbnf} project team members 
to plan these activities.  This installation team is responsible for specification 
and procurement of common infrastructure items associated with 
installation of the detectors. 
The organization of the on-site team is 
shown in Figure~\ref{fig:io-org-chart}. 

\begin{dunefigure}[Integration office installation team org chart]{fig:io-org-chart}
  {\Dword{integoff} installation team organization chart}
  \includegraphics[width=0.95\textwidth]{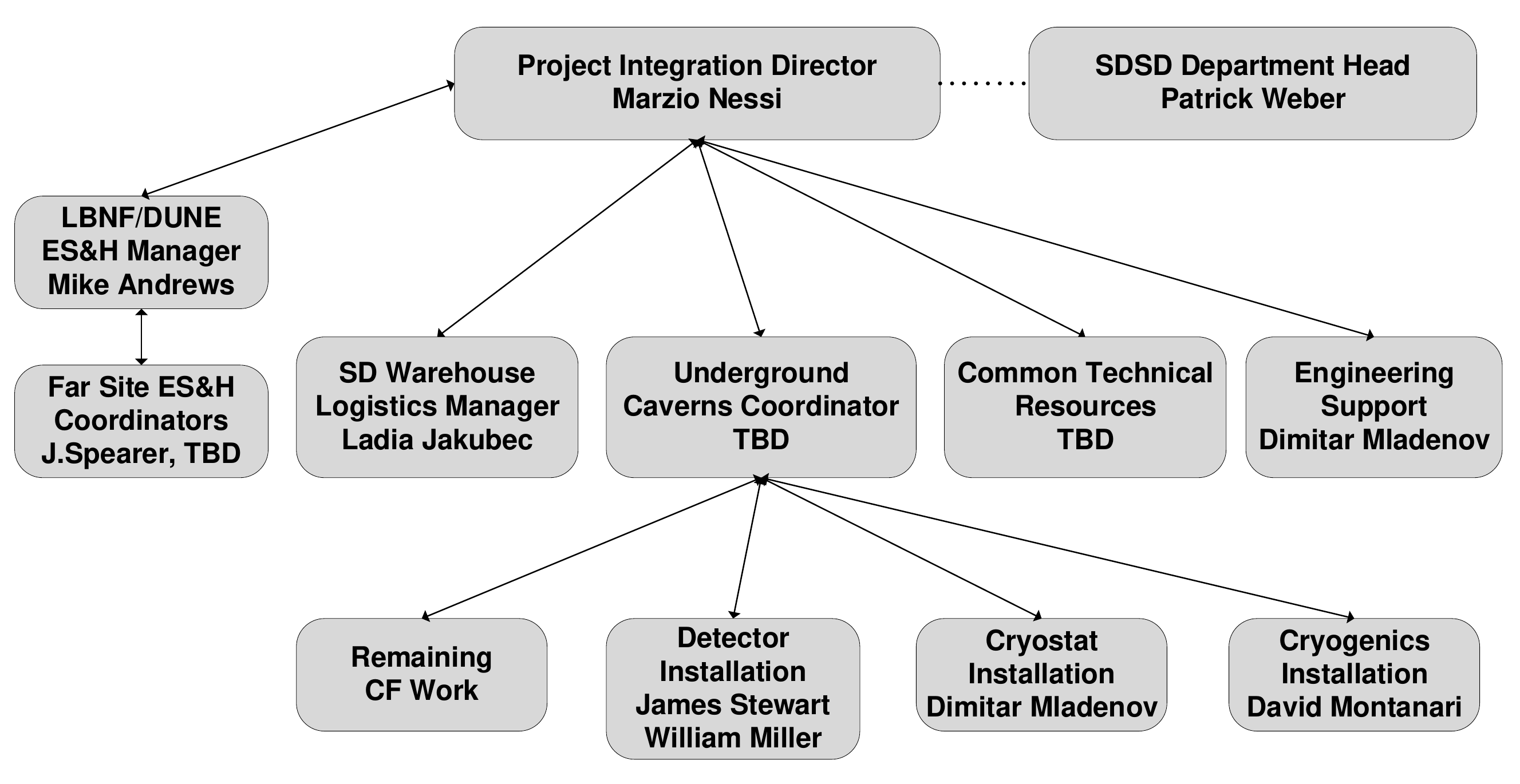}
\end{dunefigure}

The full on-site \dword{integoff} team includes rigging teams responsible for moving 
materials in and out of the shaft, through the underground drifts, 
and within the detector caverns, and personnel responsible 
for overseeing safety and logistics planning. 

The underground caverns coordinator is responsible for managing all 
activities in the two underground detector caverns and the
\dword{cuc}. The detector installation teams, distinct from the \dword{integoff}  installation team, 
incorporate a substantial number of scientific and
technical personnel from the \dword{dune} consortia.  \Dword{integoff} coordinators 
of the detector installation effort are jointly placed within 
\dword{dune} \dword{tc} to facilitate consortia involvement in  
detector installation activities.  Any modifications to the facilities 
occurring after \dword{aup} are managed by the underground cavern 
coordinator under the direction of the \dword{ipd}.

The \dword{lbnf-dune}
\dword{esh} manager heads the on-site safety organization and reports
to the \dword{ipd} to support the execution of this
responsibility. The far site \dword{esh} coordinators 
oversee the day-to-day execution
of the installation work.

\section{Facility Description}
\label{sec:es-tc-facility}

The \dword{dune} underground campus at the \dword{surf} \dword{4850l} is shown in
Figures~\ref{fig:caverns} and~\ref{fig:dune-underground}. The primary path for both personnel 
and material access to the underground excavations is through the Ross Shaft.
\begin{dunefigure}[Underground campus]{fig:dune-underground}
  {Underground campus at the \dword{4850l}.}
  \includegraphics[width=0.75\textwidth]{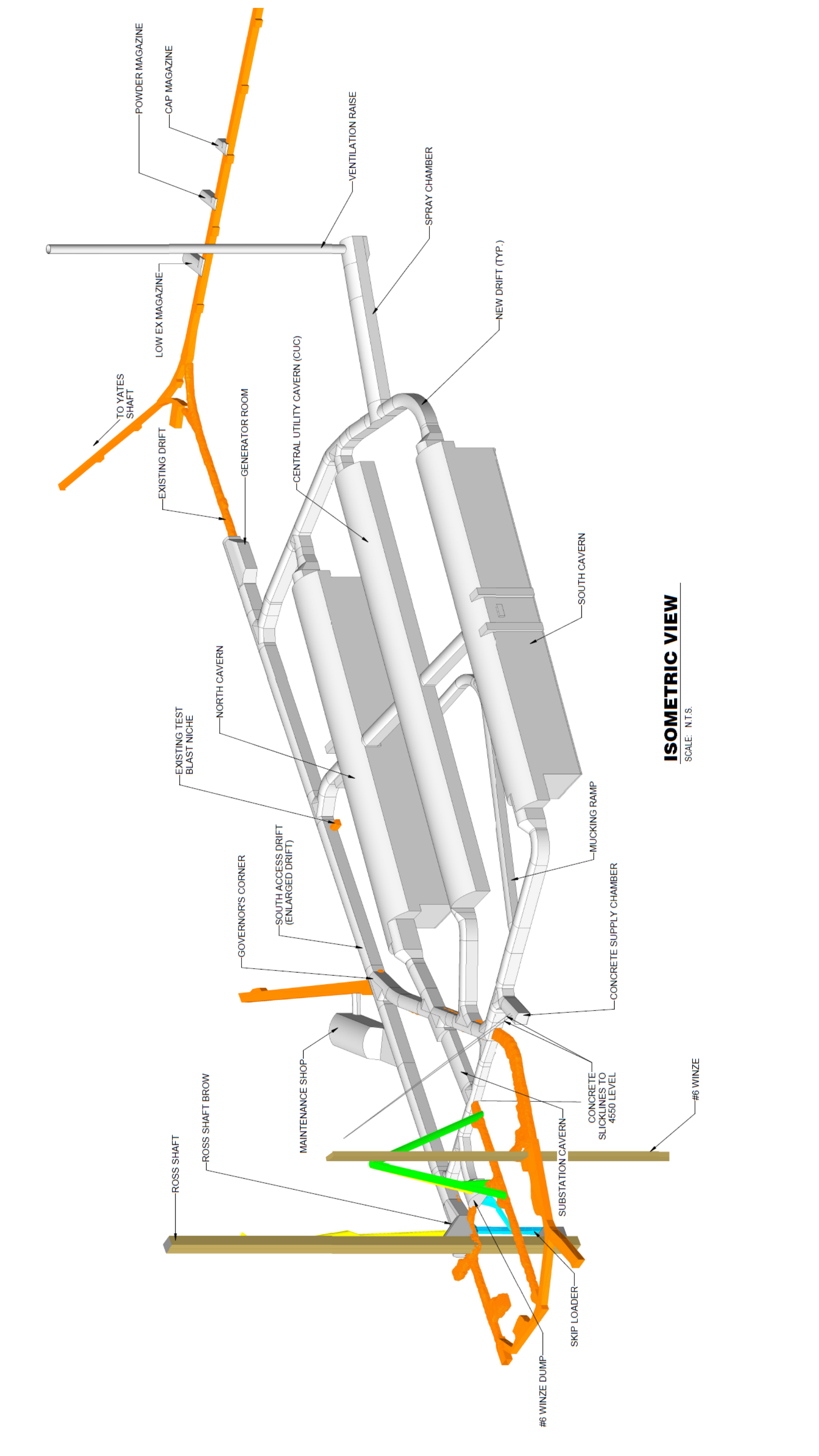}
\end{dunefigure}

\dword{lbnf} will provide facilities and services, on the surface and
underground, to support the \dword{dune} \dword{fd}.  This includes
logistical, cryogenics, electrical, mechanical, cyber, and environmental
facilities and services.  All of these facilities are provided for the
safe and productive operation of the \dwords{detmodule}.

On the surface, a new compressor building is being constructed
adjacent to the Ross headframe.  This building will house the cryogenics
systems for receiving cryogenic fluids and preparing them for delivery
down the Ross Shaft.  New piping is being installed down the 
shaft compartment to transport \dword{gar} and nitrogen underground
where they will be reliquefied.

Two large detector caverns
are being excavated.  Each of these caverns will support two
\larmass{}-capacity cryostats.  The caverns, labeled north and
south, are \SI{144.5}{\meter} long, \SI{19.8}{\meter} wide,  and 
\SI{28.0}{\meter} high. The tops of the cryostats are approximately
aligned with the \dword{4850l} of \dword{surf}, with the bottoms resting
at the 4910L.  A \SI{12}{\meter} space between the cryostats will
be used 
for the detector installation process, for placement of
cryogenic pumps and valves, and for access to the 4910L.  The
\dword{cuc}, between the north and south caverns, is \SI{190}{\meter}
long, \SI{19.3}{\meter} wide, and \SI{10.95}{\meter} high. 

The \dword{sdwf} is planned as a leased \SI{5000}{\square\meter} facility, hosted by 
\dword{sdsd}, to be located within a maximum one-day roundtrip of \dword{surf}.  
It must be in place for receiving cryostat and detector 
components approximately six months before \dword{aup}
of the underground detector caverns is received. 
 Laydown space near the Ross headframe is extremely 
limited.  For this reason, the transportation of materials from 
the \dword{sdwf} to the top of the Ross shaft requires careful 
coordination. The \dword{lbnf-dune} logistics manager works 
with the \dword{cmgc} through the end of excavation activities  
and with other members of the \dword{integoff} team to coordinate transport 
of materials into the underground areas.  Since no detector materials or 
equipment can be shipped directly to \dword{surf}, 
the \dword{sdwf} will be used for both short- and long-term storage, as 
well as for any re-packaging of items required prior to transport 
into the underground areas.

\section{Far Detector Construction Management}
\label{sec:es-tc-det-mgmt}

Eleven \dword{fd} consortia have been formed to cover 
the subsystems required for the \dword{sp} and \dword{dp} detector technologies  (Figure~\ref{fig:DUNE_consortia}). 
Three consortia (SP-APA, SP-TPC
Electronics and SP-Photon Detection) pursue subsystems specific to
the \dword{sp} design and another three consortia (DP-CRP, DP-TPC
Electronics, and DP-Photon Detection) pursue designs for \dword{dp}-specific 
subsystems.  
Five consortia (HV System, \dword{daq},
\dword{cisc}, Calibration, and Computing)
have responsibility for subsystems common to both detector
technologies.    

\begin{dunefigure}[DUNE far detector consortia]{fig:DUNE_consortia}
  {Consortia associated with the \dword{fd} construction effort along with their 
current leadership teams. CL refers to consortium leader
    and TL refers to technical lead.}
  \includegraphics[width=0.99\textwidth]{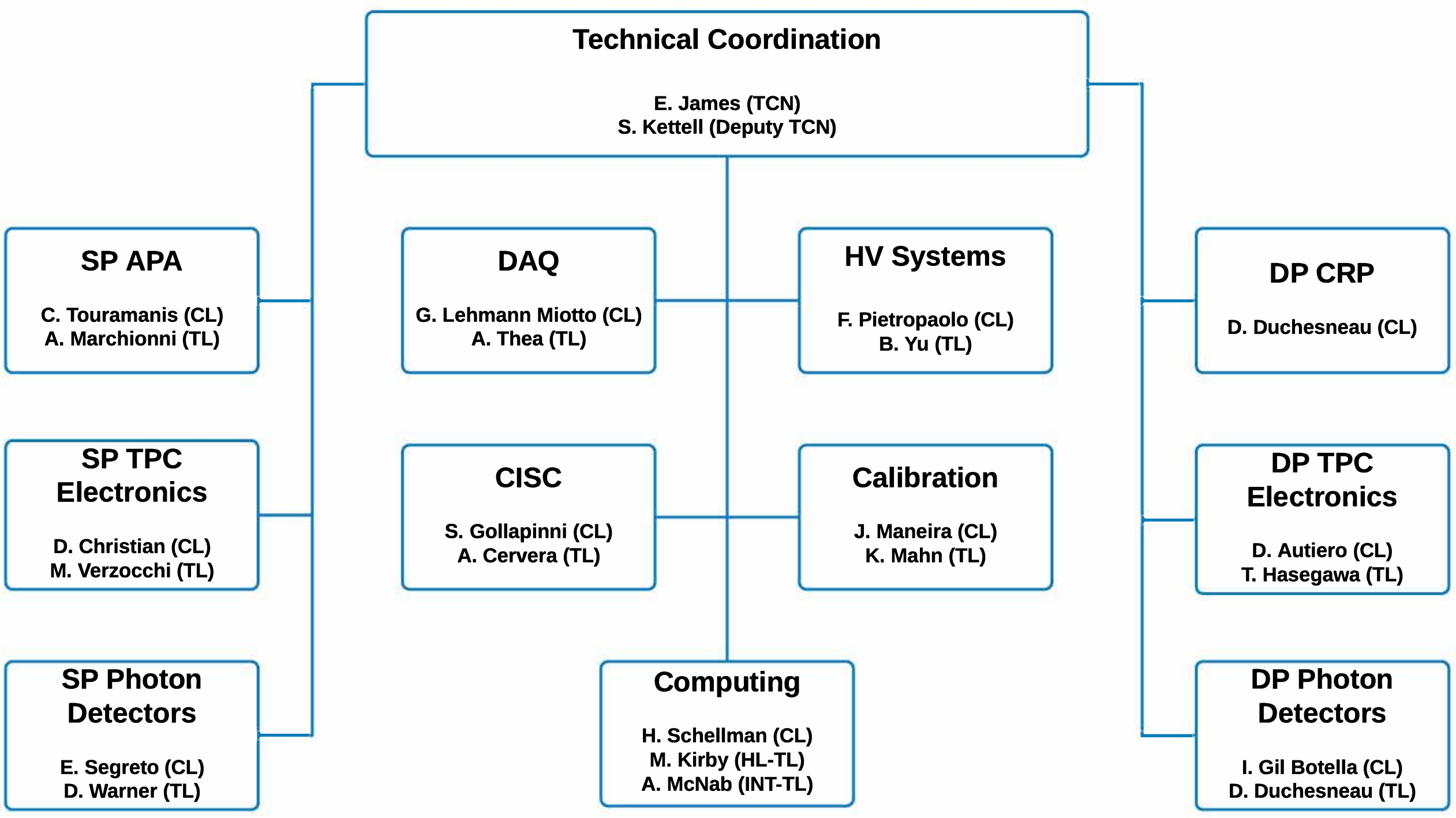}
\end{dunefigure}

The complete scope of the \dword{dune} construction project is captured in a 
\dword{wbs} to define and document the distribution of deliverables among 
the consortia.  In combination with interface documentation, the 
\dword{wbs} is used to validate that all necessary scope is covered.  The 
\dword{wbs} is also used as a framework for building \dword{dune} 
detector cost estimates.

The highest-level layers of the \dword{dune} \dword{wbs} are summarized 
in Figure~\ref{fig:WBS_level2}.  At level 1 the \dword{wbs} is broken down into 
six elements, which correspond to \dword{tc} (TC in the figure), 
four 
\dword{fd} modules, and a \dword{nd}.  The scope documented 
in this \dword{tdr} is fully contained within the elements of level-1 items 1 through 3, the \dword{tc}, a \dword{sp} \dword{fd} module, and a \dword{dp} \dword{fd} module.

\begin{dunefigure}[DUNE WBS at level 2]{fig:WBS_level2}
  {High level \dword{dune} \dword{wbs} to level 2.}
  \includegraphics[width=0.75\textwidth]{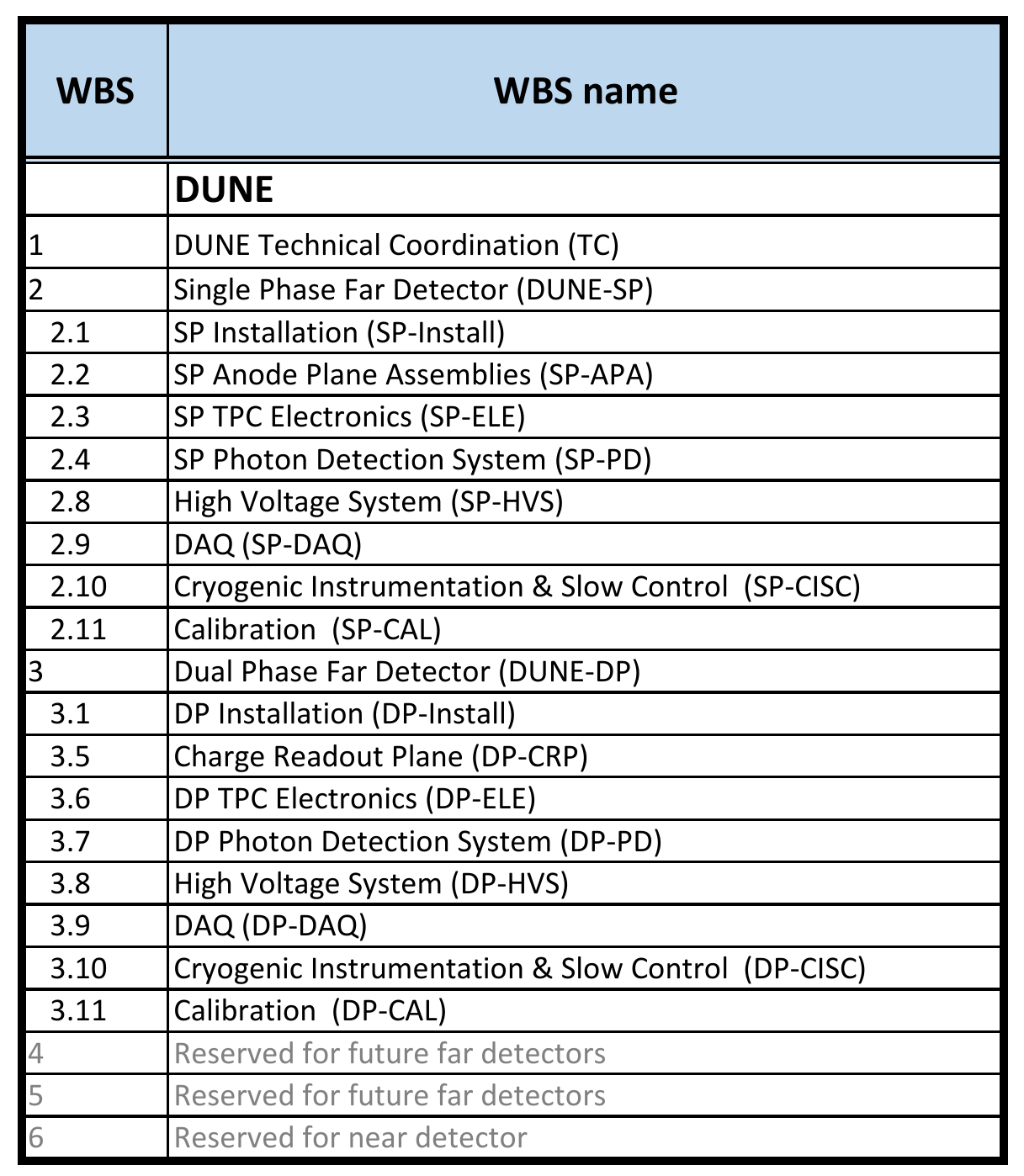}
\end{dunefigure}

\section{Integration Engineering}
\label{sec:es-coord-integ-sysengr}

Integration engineering for \dword{dune} encompasses three principal focus areas. First, it 
covers configuration of the mechanical and electrical systems of each \dword{detmodule} and management of 
the interfaces within them; this includes verifying that subassemblies
and their interfaces 
conform to the approved design of each detector element.
A second area is assurance that the \dwords{detmodule} can be integrated and
installed into their final configurations. Third, it covers 
integration of the necessary services provided by \dword{fscf} 
with the \dwords{detmodule}. The overall effort involves 
the \dword{jpo} engineering team, who maintains
subsystem component documentation for detector
configuration management, and 
the consortia, who provide engineering data for their detector subsystems to the \dword{jpo} team for incorporation into the global configuration files.

An integration mechanism has been developed to manage and create an
overall model of interfaces both within a \dword{detmodule} and
between a \dword{detmodule} and facilities. The mechanism defines
integration nodes, 
between which the \dword{jpo} engineering team carries out and
manages interfaces. 
Figure~\ref{fig:integration_nodes} shows the interfaces and nodes between a
\dword{detmodule} and the facilities it requires. The \dword{jpo} engineering
team also ensures that the interfaces are appropriately defined
and managed for the \dword{daq} room in the \dword{cuc} and the surface
control and network rooms. Interfaces with \dword{lbnf}
are managed at the boundaries of each integration node.  
Interface documents are developed and maintained to manage the
interfaces between consortia and between each consortium and 
\dword{lbnf}.

\begin{dunefigure}[Integration nodes and interfaces]{fig:integration_nodes}
  {Overall integration nodes and interfaces. The items
provided by \dword{lbnf} within the cavern are shown on the left and the items provided by
\dword{dune} are on the right.}
  \includegraphics[width=0.7\textwidth]{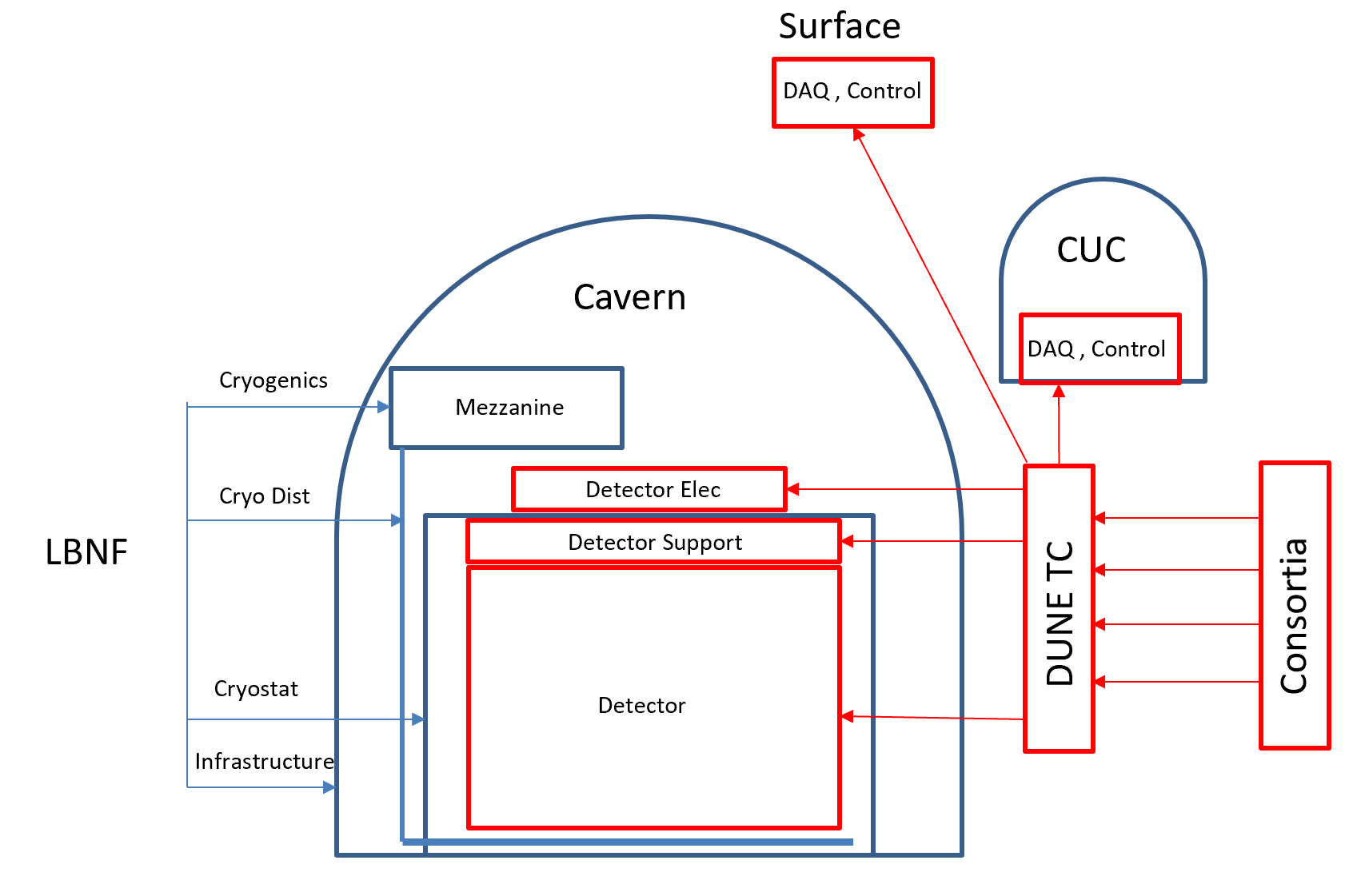}
\end{dunefigure}

\section{Reviews}
\label{sec:es-tc-reviews}

The \dword{integoff} and \dword{tc} review all stages of detector development
and work with each consortium to arrange reviews of the design
(\dword{cdrev}, \dword{pdr} and \dword{fdr}), production (\dword{prr}
and \dword{ppr}), installation (\dword{irr}), and operation
(\dword{orr}) of their system. The reviews are organized by the
\dword{jpo} \dword{ro}.  These
reviews provide information to the \dword{tb}, \dword{exb}, and \dword{efig}
in evaluating technical decisions. 

Review reports are tracked by the \dword{jpo} \dword{ro} and \dword{tc} and provide
guidance on key issues that require engineering oversight by the
\dword{jpo} engineering team. The \dword{ro}  maintains a
calendar of \dword{dune} reviews. 

\section{Quality Assurance}
\label{sec:es-tc-qa}

\dshort{dune} \dword{tc} monitors technical contributions from
collaborating institutions and provides centralized project
coordination functions. One part of this project coordination is
standardizing \dfirst{qa}/\dfirst{qc} practices, a facet
of which is to assist consortia in defining and implementing
\dword{qa}/\dword{qc} plans that maintain uniform, high
standards across the entire detector construction
effort. Figure~\ref{fig:fnal_qa} shows how \dword{dune} \dword{tc}
derives its \dword{qa} program from the principles of the \fnal \dword{qa} program:
requirements 
flow down through the \dword{lbnf-dune}
\dword{qa} program into the \dword{qc} plans developed for consortium fabrication of
detector components and integration and installation of the detector.
\begin{dunefigure}[Quality assurance flowdown]{fig:fnal_qa}
  {Flow-down of \fnal \dword{qa} to consortia}
  \includegraphics[width=0.85\textwidth]{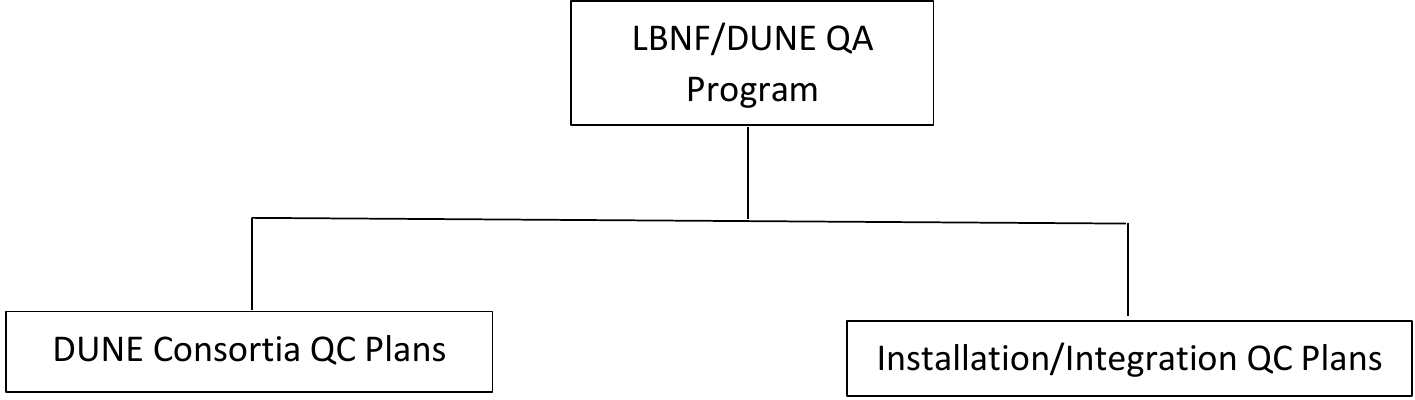}
\end{dunefigure}
The \dword{qa} effort includes design, production readiness, and
progress reviews as appropriate for the \dword{dune} detector
subsystems, as was done for \dword{pdsp} under \dword{tc}
oversight.

The primary objective of the \dword{lbnf-dune} \dword{qa} program is
to assure quality in the construction of the \dword{lbnf} facility and
\dword{dune} experiment while providing protection of
\dword{lbnf-dune} personnel, the public, and the environment. The
\dword{qa} plan aligns \dword{lbnf-dune} \dword{qa} activities, which
are spread around the world, with the principles of the \dword{fnal} Quality
Assurance Manual. The manual identifies the \dword{fnal} Integrated Quality
Assurance Program features that serve as the basis for the
\dword{lbnf-dune} \dword{qa} plan.

A key element of the \dword{lbnf-dune} \dword{qa} plan is the
concept of graded approach; that is, applying a level of analysis,
controls, and documentation commensurate with the potential for an
environmental, safety, health, or quality impact. To promote continuous improvement, \dword{dune} \dword{tc} will develop a
lessons learned program based on the \dword{fnal} Office of Project Support
Services lessons learned program.

The \dword{qa} plan
defines the \dword{qa} roles and responsibilities of the \dword{dune}
project. The \dword{dune} consortium leaders are responsible for identifying the
resources to ensure that their team members are adequately trained and
qualified to perform their assigned work.
All consortium members are responsible for the quality of the work that
they do and for using guidance and assistance that is available. All
have the authority to stop work and report adverse conditions that
affect quality of \dword{dune} products to their respective
\dword{dune} consortium leader and the \dword{lbnf-dune}
\dword{qa} manager.

\section{Environment, Safety, and Health}
\label{sec:es-tc-eshq}

\dword{lbnf-dune} is committed to protecting the health and safety of
staff, the community, and the environment, as stated in the
\dword{lbnf-dune} integrated \dword{esh} plan~\cite{bib:docdb291}.

The \dword{lbnf-dune} \dword{esh} program complies with applicable
standards and local, state, federal, and international legal
requirements through the \dword{fnal} Work Smart set of standards and the
contract between \dword{fra} and the \dword{doe}
Office of Science (FRA-DOE). \dword{fnal}, as the host laboratory,
established the \dword{sdsd} to provide facility support.
\dword{sdsd} is responsible for support of \dword{lbnf-dune}
operations at \dword{surf}.

The \dword{tcoord} and \dword{ipd} have responsibility for
implementation of the \dword{dune} \dword{esh} program for the construction and installation activities, respectively.  The
\dword{lbnf-dune} \dword{esh} manager reports to the
\dword{tcoord} and \dword{ipd} and is responsible for providing
\dword{esh} support and oversight for development and implementation of the 
\dword{lbnf-dune} \dword{esh} program. 

The \dword{dune} \dword{esh} coordinator reports to the
\dword{lbnf-dune} \dword{esh} manager and has primary responsibility
for \dword{esh} support and oversight of the \dword{dune} \dword{esh}
program for activities at collaborating institutions.  The far and near site
\dword{esh} coordinators are responsible for providing daily field support and
oversight for all installation activities at the \dword{surf}
and \dword{fnal} sites.

The \dword{lbnf-dune} \dword{esh} plan defines the \dword{esh}
requirements applicable to installation activities at the \dword{surf}
site. A key element of an effective \dword{esh} program is the hazard
identification process. Hazard identification allows production of a
list of hazards within a facility, so these hazards can be screened
and managed through a suitable set of controls. All work activities
are subject to work planning and \dword{ha}.  
All work planning documentation is reviewed and
approved by the \dword{dune} \dword{esh} coordinator and the \dword{dune}
\dword{irr} or \dword{orr} committees prior to the start of work activities.

A Safety Data Sheet (SDS) will be available for all chemicals and
hazardous materials that are used on-site. All chemicals and hazardous
materials brought to the \dword{surf} site must be reviewed and approved by the
\dword{dune} \dword{esh} coordinator and the \dword{surf} \dword{esh}
department before arriving at the site.

\dword{sdsta} will maintain an emergency response incident command
system and an emergency response team (\dword{ert}) on all shifts that can access the
underground sites with normal surface fire department response
times. This team provides multiple response capabilities for both
surface and underground emergencies.

Fire and life safety requirements for \dword{lbnf-dune} areas were
analyzed in the \dword{lbnf-dune} Far Site Fire and Life Safety
  Assessment. All caverns will be equipped with
fire detection and suppression systems, with both visual and audible
notification.  All fire alarms and system supervisory signals will be
monitored in the \dword{surf} Incident Command Center.  The
\dword{surf} \dword{ert} will respond with additional support from the
Lead and Deadwood Fire Departments and the county's emergency management
department. The caverns will be equipped with
an \dword{odh} monitoring and alarm system, with independent visual and
audible notification systems.

All workers on the \dword{dune} project have the
authority to stop work in any situation that presents an imminent
threat to safety, health, or the environment. Work may not resume
until the circumstances are investigated and the deficiencies corrected,
including the concurrence of the \dword{dune} \dword{ipd}
and \dword{lbnf-dune} \dword{esh} manager.

\cleardoublepage

\appendix

\chapter{The Near Detector Purpose and Conceptual Design}
\label{ch:appx-nd}

\section{Overview of the DUNE Near Detector}
\label{sec:appx-nd-overview}

\subsection{Motivation} 
\label{sec:appx-nd:BriefOverview-need}

A primary aim of the \dword{dune} experiment is to measure the oscillation probabilities for muon neutrino and muon antineutrinos to either remain the same flavor or oscillate to electron (anti)neutrinos. 
Measuring these probabilities as a function of the neutrino energy will allow definitive determination of the neutrino mass ordering, observation of leptonic \dword{cp} violation for a significant range of $\delta_{\rm{CP}}$ values, and precision measurement of \dword{pmns} parameters.

The role of the \dword{nd} is to serve as the experiment's control. The \dword{nd} establishes the null hypothesis (i.e., no oscillations) and constrains systematic errors. It measures the initial unoscillated \numu and \nue energy spectra, and that of the corresponding antineutrinos.  Of course, neutrino energy is not measured directly.  What is seen in the detector is a the convolution of flux, cross section, and detector response to the particles produced in the neutrino interactions, all of which have energy dependence. The neutrino energy is reconstructed from observed quantities.  \footnote{In experimental neutrino physics, it is common practice to refer to the neutrino energy (and spectra) when, in fact, it is the reconstructed neutrino energy (spectra) which is meant, along with all of the flux, cross section, and detector response complexities that implies.} 

To first order, a ``far/near'' ratio (or migration matrix), derived from the simulation, can predict the unoscillated energy spectra at the \dword{fd} based on the \dword{nd} measurements.  The energy spectra at the \dword{fd} are then sensitive to the oscillation parameters, which can be extracted via a fit.  The \dword{nd} plays a critical role in establishing what the oscillation signal spectrum should look like in the \dword{fd} because the expectations for the spectra in both the disappearance and appearance signals are based on the precisely measured spectra for $\nu_{\mu}$ and $\overline{\nu}_{\mu}$ interactions in the \dword{nd}.

To achieve the precision needed for \dword{dune}, the experiment will have to operate beyond the first-order paradigm. With finite energy resolution and nonzero biases, the reconstructed energy spectrum is an unresolved convolution of cross section, flux, and energy response. The \dword{nd} must independently constrain each of those components.  The \dword{nd} must provide information that can be used to model well each component. Models of the detector, beam, and interactions fill in holes and biases left by imperfect understanding and they are used to estimate the size of many systematic effects.  When imperfect models are not able to match observations, the \dword{nd} must provide the information needed to deal with that and estimate the impact of the imperfect modeling on final measurements. In general, this requires that the \dword{nd} significantly outperform the \dword{fd}, which is limited by the need for a large, underground mass. The \dword{nd} must have multiple methods for measuring neutrino fluxes as independently of cross section uncertainties as possible. With the necessity of relying on models, the \dword{nd} needs to measure neutrino interactions with much better detail than the \dword{fd}. This includes having a larger efficiency across the kinematically allowed phase space of all relevant reaction channels, superior identification of charged and neutral particles, better energy reconstruction, and better controls on experimental biases. The \dword{nd} must also have the ability to measure events in a similar way to the \dword{fd}, so that it can determine the ramifications of the more limited \dword{fd} performance, provide corrections, and take advantage of effects canceling to some extent in the near to far extrapolation.

The conceptual design of the \dword{nd} is based on the collective experience of the many \dword{dune} collaborators who have had significant roles in the current generation of neutrino experiments (\dword{minos}, MiniBooNE, \dword{t2k}, \dword{nova}, \dword{minerva}, and the \dword{sbn} program).  These programs have provided (and will provide) a wealth of useful data and experience that has led to improved neutrino interaction models, powerful new analyses and reconstruction techniques, a deep appreciation of analysis pitfalls, and a better understanding of the error budget. 
These experiments, while similar to \dword{dune}, were all 
done with a lower precision, in a different energy range, or with very different detector technologies. While the existing and projected experience and data from those experiments provide a strong base for \dword{dune}, it is not sufficient to enable \dword{dune} to accomplish its physics goals without a highly performing \dword{nd}.  

The \dword{dune} \dword{nd} will also have a physics program of its own measuring cross sections, non-standard interactions (\dshort{nsi}), searching for sterile neutrinos, dark photons, and other exotic particles. These are important aims that expand the physics impact of the \dword{nd} complex.  
Furthermore, the cross section program is coupled to the oscillation measurement insofar as the cross sections will be useful as input to theory and model development.   (Note that many of the \dword{nd} data samples are incorporated into the oscillation fits directly.) 

\subsection{Design} 
\label{sec:appx-nd:BriefOverview}

The \dword{dune} \dword{nd} is formed from three primary detector components and the capability for two of these components to move off the beam axis. The three detector components serve important individual and overlapping functions with regard to the mission of the \dword{nd}.  Because these components have standalone features, the \dword{dune} \dword{nd} is often discussed as a suite or complex of detectors and capabilities.  The movement off axis provides a valuable extra degree of freedom in the data which is discussed extensively in this report.  The power in the \dword{dune} \dword{nd} concept lies in the collective set of capabilities.  It is not unreasonable to think of the component detectors in the \dword{dune} \dword{nd} as being somewhat analogous to subsystems in a collider experiment, the difference being that, with one important exception (higher momentum muons), individual events are contained within the subsystems.  
The \dword{dune} \dword{nd} is shown in the \dword{dune} \dword{nd} hall in Figure~\ref{fig:NDHallconfigs}.  Table~\ref{tab:NDsummch} provides a high-level overview of the three components of the \dword{dune} \dword{nd} along with the off-axis capability that is sometimes described as a fourth component.  

The core part of the \dword{dune} \dword{nd} is a \dword{lartpc} called \dword{arcube}.  The particular implementation of the \dword{lartpc} technology in this detector is described in Section~\ref{sec:appx-nd:lartpc} below.  
This detector has the same target nucleus and shares some aspects of form and functionality with the \dword{fd}, while the differences are necessitated by the expected intensity of the beam at the \dword{nd}.  This similarity in target nucleus and, to some extent, technology, reduces sensitivity to nuclear effects and detector-driven systematic errors in the extraction of the oscillation signal at the  \dword{fd}.  The \dword{lartpc} is large enough to provide high statistics ($\num{1e8}{\numu \text{-CC events/year}}$ on axis) and a sufficient volume to provide good hadron containment.  The tracking and energy resolution, combined with the mass of the \dword{lartpc}, will allow for the measurement of the flux in the beam using several techniques, including the rare process of $\nu$-e$^{-}$ scattering.

The \dword{lartpc} begins to lose acceptance for muons above $\sim$\SI{0.7}{GeV/c} due to lack of containment. Because the muon momentum is a critical component of the neutrino energy determination, a magnetic spectrometer is needed downstream of the \dword{lartpc} to measure the charge sign and momentum of these muons. In the \dword{dune} \dword{nd} concept, this function is accomplished by the \dword{mpd}, which consists of a \dword{hpgtpc} surrounded by an \dword{ecal} in a \SI{0.5}{T} magnetic field. The \dword{hpgtpc} provides a lower density medium with excellent tracking resolution for the muons from the \dword{lartpc}. In addition, with this choice of technology for the tracker, neutrinos interacting on the argon in the \dword{hpgtpc} constitute a large (approximately \num{1e6}$\numu$-\dword{cc} events/year on axis) independent sample of $\nu$-Ar events that can be studied with a very low momentum threshold for tracking charged particles,  excellent resolution, and with systematic errors that differ from the liquid detector. These events will be valuable for studying the charged particle activity near the interaction vertex, since this detector can access lower-momentum protons than the \dword{lartpc} and has better particle identification of charged pions.  Misidentification of pions as knock-out protons (or vice versa) causes a mistake in the reconstructed neutrino energy, moving it away from its true value by the amount of a pion mass.  This mistake can become quite significant at the lower-energy second oscillation maximum. The gas detector will play an important role in mitigating this mistake, since pions are rarely misidentified as protons in the \dword{hpgtpc}.  In addition, the relatively low level of secondary interactions in the gas samples will be helpful for identifying the particles produced in the primary interaction and modeling secondary interactions in denser detectors, which are known to be important effects\cite{Friedland:2018vry}. The high pressure increases the statistics for these studies, improves the particle identification capabilities, and improves the momentum resolution. 
The \dword{mpd} is discussed further in Section~\ref{sec:appx-nd:mpd}.

The \dword{lartpc} and \dword{mpd} can move to take data in positions off the beam axis.  This capability is referred to as \dword{duneprism}. As the detectors move off-axis, the incident neutrino flux spectrum changes, with the mean energy dropping and the spectrum becoming somewhat monochromatic.  Though the neutrino interaction rate drops off-axis, the intensity of the beam and the size of the \dword{lartpc}  combine to yield ample statistics even in the off-axis positions. 
Data taken at different off-axis angles allow deconvolution of the neutrino flux and interaction cross section and the mapping of the reconstructed versus true energy response of the detector.  This latter mapping is applicable at the \dword{fd} up to the level to which the near and far \dword{lar} detectors are similar.  Stated a different way, it is possible to use information from a linear combination of the different fluxes to create a data sample at the \dword{nd} with an effective neutrino energy distribution that is close to that of the oscillated spectrum at the \dword{fd}.  This data-driven technique will reduce systematic effects coming from differences in the energy spectra of the oscillated signal events in the \dword{fd} and the \dword{nd} samples used to constrain the interaction model. Finally, the off-axis degree of freedom provides a sensitivity to some forms of mismodeling in the beam and/or interaction models. The \dword{duneprism} program is discussed further in Section~\ref{sec:appx-nd:DP}.

The final component of the \dword{dune} \dword{nd} suite is the beam monitor, called the \dword{sand}.  The core part of it, the \dword{3dst}, is a plastic scintillator detector made of \SI{1}{\cubic\centi\meter} cubes read out along each of three orthogonal dimensions.  The design eliminates the typical planar-strip geometry common to scintillator detectors, leading to improved acceptance at large angles relative to the beam direction. It is mounted  
inside an envelope of high-resolution, normal pressure \dwords{tpc} and an \dword{ecal}, all 
of which are surrounded by a magnet, as illustrated in Figure~\ref{fig:3dst-geometry}.  The reference design uses a repurposed magnet and \dword{ecal} from the \dword{kloe} experiment.
The \dword{3dst} serves as a dedicated neutrino spectrum monitor that never moves off-axis. 
It also provides an excellent on-axis, neutrino flux determination using many of the methods discussed in Section~\ref{sec:appx-nd:fluxappendix}. 
The neutrino flux determined using this detector, with  
technologies, targets, and interaction systematic errors that are different from \dword{arcube}, is an important point of comparison and a systematic cross-check for the flux as determined by \dword{arcube}.

\dword{sand} provides very fast timing and can isolate small energy depositions from neutrons in three dimensions.  This provides the capability to  incorporate neutrons in the event reconstruction using energy determination via time-of-flight with a high efficiency. 
This capability is expected to be useful for the low-$\nu$ flux determination since it allows for tagging of events with a significant neutron energy component\footnote{The low-$\nu$ technique involves measuring the flux for events with low energy transfer because the cross section is approximately constant with energy for this sample.  It provides a nice way to measure the shape of the spectrum.  This is discussed further in Section~\ref{sec:appx-nd:fluxappendix}.}.   
The inclusion of the neutron reconstruction also provides a handle for improving the neutrino energy reconstruction in $\overline{\nu}_\mu$ \dword{ccqe} events, which is helpful for the $\overline{\nu}_\mu$ flux determination.
The  
different mass number $A$ of the carbon target relative to argon may prove to be useful for developing models of nuclear effects and building confidence in the interaction model and the size of numerous systematic errors.  The addition of the neutron reconstruction capability extends the DUNE ND theme of including regions of phase space in neutrino interactions not seen in previous experiments.  This capability may provide insights that foster improvements in the neutrino interaction model on carbon.  Though extrapolating such improvements to argon is not straightforward, the development of current generators has benefited from data taken with different nuclear targets, including carbon. 
The \dword{sand} component of the \dword{nd} is discussed more in Section~\ref{sec:appx-nd:mpt-3dst}.

Table~\ref{tab:fluxrates} shows the statistics expected in the different \dword{nd} components for a few processes that are important for constraining the neutrino flux.  Some additional information on constraining the flux is provided in Section~\ref{sec:appx-nd:fluxappendix}.

\begin{dunetable}[Event rates for flux constraining processes]{llll}{tab:fluxrates}{Event rates for processes that can be used to constrain the neutrino flux. The rates are given per year for a \SI{1}{ton} (\dword{fv}) \dword{hpgtpc}, a \SI{25}{ton} (\dword{fv}) \dword{lartpc} \cite{bib:docdb12388}, and a \SI{9}{t} (\dword{fv}) \dword{3dst}. The flux for the \dword{hpgtpc} and \dword{lartpc} is from the simulated ``2017 engineered'' \dword{lbnf} beam with a primary momentum of \SI{120}{GeV/c} and \SI{1.1e21}{POT/year}. The flux for the \dword{3dst} is the \SI{80}{GeV}, three-horn, optimized beam with \SI{1.46e21}{POT/year}.  The detectors are assumed to be on-axis. Fiducial volumes are analysis dependent and in the case of the \dword{lartpc}, it is likely the volume could be made larger by a factor of two for many analyses.}
Event class & \dword{lartpc} & \dword{hpgtpc} & \dword{3dst} \\ \toprowrule
\numu + $e^-$ elastic ($E_e>\SI{500}{MeV}$) & \num{3.3e3} & \num{1.3e2} & \num{1.1e3} \\ \colhline
\numu low-$\nu$ ($\nu<\SI{250}{MeV})$ & \num{5.3e6} & \num{2.1e5} & \num{1.48e6} \\ \colhline
\numu \dword{cc} coherent & \num{2.2e5} & \num{8.8e3} &  \\ \colhline
\anumu \dword{cc} coherent & \num{2.1e4} & \num{8.4e2} &  \\ 
\end{dunetable}

\section{Role of the ND in the DUNE Oscillation Program}
\label{sec:appx-nd:exsum-nd-role}

Oscillation experiments need to accomplish three main tasks. First, they must identify the flavor of interacting neutrinos in \dword{cc} events, or identify the events as \dword{nc} interactions. Second, they need to measure the energy of the neutrinos since oscillations occur as a function of baseline length over neutrino energy, \dword{l/e}. Third, they need to compare the observed event spectrum in the \dword{fd} to  predictions based on differing sets of oscillation parameters, subject to constraints from the data observed in the \dword{nd}.  That comparison and how it varies with the oscillation parameters allows for the extraction of the  measured oscillation parameters and errors.

The connection between the observations in the \dword{nd} and the \dword{fd} is made using a simulation that convolves models of the neutrino flux, neutrino interactions, nuclear effects, and detector response.
This gives rise to a host of complicating effects that 
muddy the simple picture. They come from two main sources. First, the identification efficiency is not \SI{100}{\%} and there is 
some background (e.g., \dword{nc} events with a $\pi^0$ are a background to \nue \dword{cc} interactions). Both the efficiency and the background are imperfectly known. Generally, it is helpful to have a  \dword{nd} that is as similar as feasible to the  \dword{fd} because a bias in the efficiency as a function of energy will cancel between the two detectors. Since the background level tends to be similar between the two detectors, it is helpful if the \dword{nd} is more capable than the \dword{fd} at characterizing backgrounds, either due to its technology, or by leveraging the much larger statistics and freedom to take data in alternative beam configuration modes (e.g., different horn currents or movement off the beam axis). 

The second major source of complication occurs because the \dword{fd} (and the similar \dword{nd}) has to be made of heavy nuclei rather than hydrogen. Neutrino interactions can be idealized as a three stage process: (1) a neutrino impinges on a nucleus with nucleons in some initial state configuration, (2) scattering occurs with one of the nucleons, perhaps creating mesons, and (3) the hadrons reinteract with the remnant nucleus on their way out (so called \dword{fsi}). The presence of the nucleus impacts all three stages in ways that ultimately drive the design of the  \dword{nd} complex. To better understand this it is useful to consider what would happen if the detectors were made of hydrogen.


In a detector made of hydrogen, the initial state is a proton at rest and there are no \dword{fsi}. The scattering consists of a variety of processes. The simplest is \dword{qe} scattering: $\bar{\nu}_\ell p \to \ell^+ n$. The detector sees a lepton (which establishes the flavor of the neutrino), no mesons, and perhaps a neutron interaction away from the lepton's vertex. Because there are no mesons the kinematics is that of two body scattering and the neutrino energy can be reconstructed from the the lepton's angle (with respect to the $\nu$ beam) and energy. This is independent of whether the neutron is observed.

For $\nu_\ell$ interactions on hydrogen there is no \dword{qe} process. The simplest scattering channel is single pion production $\nu_\ell p \to \ell^- \pi^{(+,0)} (n,p)$. In that case the neutrino energy may be reconstructed from the energy of the muon and pion, and their angles with respect to the beam\footnote{The nucleon does not need to be observed. This is a consequence of having four energy-momentum conservation constraints, which allows $E_\nu$ and $\vec{p}_N$ to be computed.}. In both cases, the neutrino energy can be measured without bias so long as the detector itself measures lepton and meson momenta and angles without bias.  The neutrino energy in complicated scattering channels, such as ones with multiple pions or heavy baryons can be measured in a similar way (at least in principle).

A key simplifying feature offered by a hypothetical hydrogen detector is simply that there are enough constraints to measure the neutrino energy without needing to measure the single nucleon (especially a neutron escaping the detector). Additionally, the cross sections for different scattering channels (particularly the simpler ones) can be expressed in terms of leptonic and hadronic currents. The leptonic current is well understood. The structural elements of the hadronic current are known on general theoretical grounds. The current is often represented by form factors that are constrained by electron scattering experiments, beta decay, and neutrino scattering measurements that the detector can make itself (or take from other experiments).

The situation is significantly more complicated in a detector with heavy nuclei. The nucleons in the initial state of the nucleus are mutually interacting and exhibit Fermi motion. This motion ruins the key momentum conservation constraint available in hydrogen due to the target being at rest. Scattering at lower momentum transfer is suppressed because the nucleon in the final state would have a momentum that is excluded by the Pauli principle. 

The nucleon momentum distribution in heavy nuclei is commonly modeled as a Fermi gas with a cutoff momentum $k_F \approx \SI{250}{MeV/c}$ \cite{Smith:1972xh}.
This picture is overly simplistic.  For example,  there are nucleons with momenta larger than $k_F$ due to short-range correlated nucleon-nucleon interactions (\dword{src})\cite{Bodek:2014jxa}. Scattering on a nucleon with $p>k_F$ implies that there is a spectator nucleon recoiling against the target with a significant momentum. \dword{src} have been the subject of much investigation but are not fully understood or fully implemented in neutrino event generators.

Additionally, there is a second multi-body effect. For the few-GeV neutrinos of interest to \dword{dune}, the typical momentum transfer corresponds to a probe that has a wavelength on par with the size of a nucleon. In this case, the scattering can occur on two targets in the nucleus which may be closely correlated (\dword{2p2h} scattering). Experiments can easily confuse this process for \dword{qe} scattering since there are no mesons and one or both of the two nucleons may have low energy, evading detection. The presence of two nucleons in the initial and final state again ruins the kinematic constraints available in hydrogen. It is now known that \dword{2p2h} scattering is a significant part of the total scattering cross section at \dword{dune} energies \cite{Ruterbories:2018gub}. The \dword{2p2h} cross section is difficult to compute because it cannot be expressed as the sum over cross sections on individual nucleons. The dependence on atomic number and the fine details of the interaction (e.g., the final energies of the two particles) are also currently unknown. Finally, it is widely expected that there are components of \dword{2p2h} and \dword{src} scattering that result in meson production. Event generators do not currently include such processes.

Neutrino scattering on nuclei is also subject to \dword{fsi}. \dword{fsi} collectively refers to the process by which nucleons and mesons produced by the neutrino interaction traverse the remnant nucleus. The hadrons reinteract with a variety of consequences: additional nucleons can be liberated; ``thermal'' energy can be imparted to the nucleus; pions can be created and absorbed; and pions and nucleons can undergo charge exchange scattering (e.g., $\pi^- p \to \pi^0 n$).  Event generators include phenomenological models for \dword{fsi}, anchoring to hadron-nucleus scattering data.

The heavy nuclei in a detector also act as targets for the particles that have escaped the struck nucleus. Generally speaking, the denser the detector and the more crudely it samples deposited energy, the more difficult it is to observe low-energy particles. Negatively and positively charged pions leave different signatures in a detector since the former are readily absorbed while the latter are likely to decay.  Neutrons can be produced from the struck nucleus, but also from follow-on interactions of the neutrino's reaction products with other nuclei. The energy carried away by neutrons is challenging to detect and can bias the reconstructed neutrino energy. 

Finally, it is important to note that a significant fraction of the neutrino interactions in DUNE will come from deep inelastic scattering rather than the 
simpler \dword{qe} scattering discussed above.  This leads typically to a more complex morphology for events (beyond the heavy nucleus complications) and greater challenges for the detector and the modeling.  

\section{Lessons Learned} 
\label{sec:appx-nd:overview-lessons}

\subsection{Current Experiments}
Neutrino beams are notoriously difficult to model at the precision and accuracy required for modern accelerator-based experiments.  Recent \dword{lbl} experiments make use of a  \dword{nd} placed close to the beam source, where oscillations are not yet a significant effect.  The beam model, the neutrino interaction model, and perhaps the detector response model are tuned, or calibrated, by the data recorded in the  \dword{nd}. The tuned model is used in the extraction of the oscillation signal at the  \dword{fd}. Known effects that are not understood or modeled well must be propagated into the final results as part of the systematic error budget.  Unknown effects that manifest as disagreements between the model and observations in the  \dword{nd} also must be propagated into the final results as part of the systematic error budget.  These kinds of disagreements have happened historically to every precision accelerator oscillation experiment.  When such disagreements arise, some assumption or range of assumptions must be made about the source of the disagreement.  Without narrowing down the range of possibilities, this can become a leading systematic error.

Since the final results depend on the comparison of what is seen in the  \dword{fd} to that in the  \dword{nd}, having functionally identical detectors (i.e., the same target nucleus and similar detector response) is helpful.  In a similar vein, differences between the neutrino spectrum at the  \dword{nd} and the oscillated spectrum seen at the  \dword{fd} lead to increased sensitivity to systematic effects propagated from the  \dword{nd} to the  \dword{fd}.

The past experience of the neutrino community is a driving force in the design of the \dword{dune}  \dword{nd} complex. 
The performance of  current, state-of-the-art long baseline oscillation experiments  provides a practical guide to many of the errors and potential limitations \dword{dune} can expect to encounter, as well as case studies of issues that arose which were unanticipated at the design stage. 

The \dword{t2k} experiment uses an off-axis neutrino beam that has a narrow energy distribution peaked below \SI{1}{GeV}. This means, relative to \dword{dune}, interactions in \dword{t2k} are predominantly \dword{ccqe} and have relatively simple morphologies.  The data sample has little feed-down from higher energy interactions.  The \dword{t2k}  \dword{nd} (plastic scintillator and
 \dword{tpc}) technology is very different from its   \dword{fd} (water Cerenkov), though the \dword{nd} contains embedded water targets that provide samples of interactions on the same target used in the \dword{fd}.
The experiment relies on the flux and neutrino interaction models, as well as the \dword{nd} and  \dword{fd} response models to extrapolate the constraint from the  \dword{nd} to the  \dword{fd}.   In the most recent oscillation results released by \dword{t2k}, the  \dword{nd} data constraint reduces the flux and interaction model uncertainties at the  \dword{fd} from 11---14\% down to 2.5---4\%~\cite{Abe:2018wpn}. Inclusion of the water target data was responsible for a factor of two reduction in the systematic uncertainties, highlighting the importance of measuring interactions on the same target nucleus as the \dword{fd}.\footnote{These numbers are not used directly in the analysis but were extracted to provide an indication of the power of the \dword{nd} constraint.}

The  \dword{nova}  experiment uses an off-axis neutrino beam from \dword{numi} that has a narrow energy distribution peaked around \SI{2}{GeV}.  The  \dword{nova}   \dword{nd} is functionally identical to its  \dword{fd}.  Still, it is significantly smaller than the  \dword{fd} and it sees a different neutrino spectrum due to geometry and oscillations.  Even with the functionally identical near and  far detectors,  \dword{nova}  uses a model to subtract \dword{nc} background and relies on a model-dependent response matrix to translate what is seen in the  \dword{nd} to the ``true'' spectrum, which is then extrapolated to the  \dword{fd} where it is put through a model again to predict what is seen in the  \dword{fd}~\cite{NOvA:2018gge, WolcottNUINT2018}.  Within the extrapolation, the functional similarity of the near and  far detectors reduces but does not eliminate many systematic effects.  Uncertainties arising from the neutrino cross section model dominate the  \dword{nova}  $\nu_{e}$ appearance systematic error budget and are among the larger errors in the $\nu_{\mu}$ disappearance results.  The  \dword{nd} constraint is significant.  For the $\nu_{e}$ appearance signal sample in the latest  \dword{nova}  results, for example, a measure of the systematic error arising from cross section uncertainties without using the  \dword{nd} extrapolation is 12\,\% and this drops to 5\,\% if the  \dword{nd} extrapolation is used \cite{WolcottNUINT2018}.

The process of implementing the  \dword{nd} constraint in both \dword{t2k} and  \dword{nova}  is less straightforward than the typical description implies.  It will not be any more straightforward for \dword{dune}.  One issue is that there are unavoidable near and far differences. Even in the case of functionally identical detectors, the beam spectrum and intensity are very different near to far.  For \dword{dune}, in particular, 
\dword{arcube} is smaller than the \dword{fd} and is divided into modular, optically isolated regions that have a pixelated readout rather than the wire readout of the \dword{fd}.  Space charge effects will differ near to far.  All of this imposes model dependence on the extrapolation from near to far.  This is mitigated by collecting data at differing off-axis angles with \dword{duneprism}, where an analysis can be done with an \dword{nd} flux that is similar to the oscillated \dword{fd} flux (see Section~\ref{sec:appx-nd:DP}). (Data from \dword{protodune} will also be useful to understand the energy-dependent detector response for the \dword{fd}.)  Regardless, near to far differences will persist and must be accounted for through the beam, detector, and neutrino interaction models.  

Although long baseline oscillation experiments use the correlation of fluxes at the \dword{nd} and the \dword{fd} to reduce sensitivity to flux modeling, the beam model is a critical component in understanding this correlation.  Recently, the \dword{minerva} experiment used spectral information in the data to diagnose a discrepancy between the expected and observed neutrino event energy distribution in the \dword{numi} medium energy beam \cite{JenaNUINT2018}. In investigating this issue, \dword{minerva} compared the observed and simulated neutrino event energy distribution for low-$\nu$ events, as shown in Figure~\ref{fig:minervameflux}.  Since the cross section is known to be relatively flat as a function of neutrino energy for this sample, the observed disagreement as a function of energy indicated a clear problem in the flux model or reconstruction.      
\dword{minerva} believes the observed discrepancy between the data and simulation is best described by what is a mismodeling in horn focusing combined with an error in the muon energy reconstruction (using range traversed in the downstream spectrometer).  This is notable, in part, because the two identified culprits in this saga would manifest differently in the extrapolation to the far detector in an oscillation experiment. The spectral analysis provided critical information in arriving at the final conclusion.  This experience illustrates the importance of good monitoring/measurements of the neutrino beam spectrum.  

\begin{dunefigure}[MINERvA medium energy \dshort{numi} flux for low-$\nu$ events]{fig:minervameflux}
{Reconstructed \dword{minerva} medium energy \dword{numi} neutrino event spectrum for low-energy transfer events compared to simulation (left) and same comparison shown as a ratio (right). From \cite{JenaNUINT2018}.}
\includegraphics[width=0.49\textwidth]{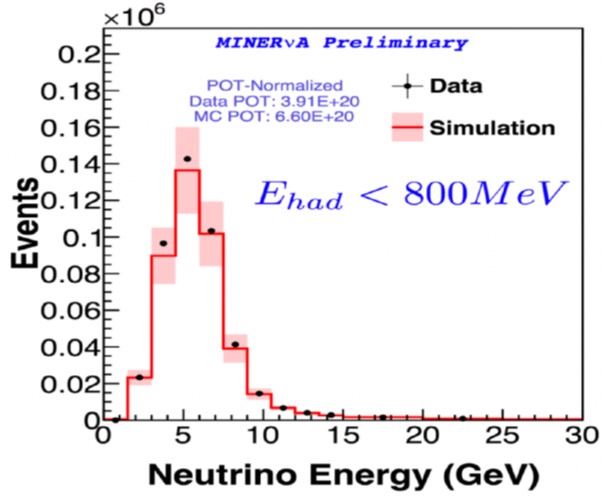}
\includegraphics[width=0.49\textwidth]{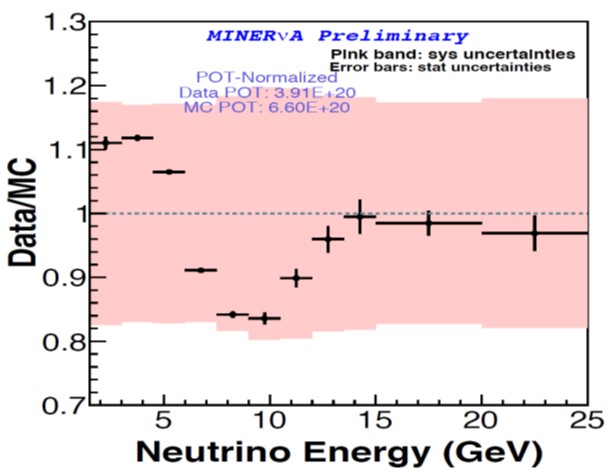}
\end{dunefigure}

Another important issue is that the neutrino interaction model is not perfect, regardless of the experiment and implementation.  With an underlying model that does not describe reality, even a model tuned to  \dword{nd} data will have residual disagreements with that data.  These disagreements must be accounted for in the systematic error budget of the ultimate oscillation measurements.  Although the model(s) may improve before \dword{dune} operation, the degree of that improvement cannot be predicted and the \dword{dune}  \dword{nd} complex should have the capability to gather as much information as possible to help improve and tune the model(s) during the lifetime of the experiment.  In other words, the  \dword{nd} needs to be capable of narrowing the range of plausible possibilities giving rise to data-model differences at the  \dword{nd} in order to limit the systematic error incurred in the results extracted from the  \dword{fd}.   

Recent history provides illustrations of progress and continuing struggles to improve neutrino interaction models.  The MiniBooNE collaboration published results in 2010 showing a disagreement between the data and the expected distribution of \dword{ccqe} events as a function of Q$^{2}$ \cite{AguilarArevalo:2010cx,Gran:2006jn}.   They brought the model into agreement with the data by increasing the axial mass form factor used in the model.  K2K \cite{Gran:2006jn} and \dword{minos} \cite{Adamson:2014pgc} made similar measurements.  It has since been shown that the observed disagreement is due to the need to include multi-nucleon processes and that the use of the large effective axial mass form factor used by these experiments to fit the data leads to a misreconstruction of the neutrino energy.  

The importance of modeling multi-nucleon (\dword{2p2h}) processes for oscillation experiments is underscored by the fact that such interactions when reconstructed as a \dword{ccqe} (1p1h) process lead to a significant low-side tail in the reconstructed neutrino energy \cite{Martini:2012uc}.  Multi-nucleon processes also change the hadronic calorimetric response.  The first  \dword{nova}  $\nu_{\mu}$ disappearance oscillation results had a dominant systematic error driven by the disagreement of their model to the data in their hadronic energy distribution \cite{Adamson:2016xxw}.  In more recent work, the inclusion of multi-nucleon processes in the interaction model contributed to a substantial reduction of this disagreement \cite{NOvA:2018gge}.

The \dword{minerva} experiment has compiled a significant catalog of neutrino and antineutrino results and recently developed a model tuned to their \dword{qe}-like  (\dword{numi} low energy) data \cite{Ruterbories:2018gub}.  The tune is based on a modern neutrino interaction generator (\dword{genie} 2.8.4 \cite{Andreopoulos:2009rq}, using a global Fermi gas model \cite{Smith:1972xh}  with a Bodek-Ritchie tail \cite{Bodek:1981wr} and the INTRANUKE-hA \dword{fsi} model \cite{Dytman:2007zz}).  Even so, \dword{minerva} scales down non-resonance pion production \cite{Rodrigues:2016xjj}, includes a random phase approximation model (RPA) \cite{Nieves:2004wx,Gran:2017psn}, and incorporates a multi-nucleon model \cite{Nieves:2011pp, Gran:2013kda, Schwehr:2016pvn} with an empirical enhancement in the dip region between the \dword{qe} and $\Delta$  region that is determined by a fit to the neutrino data \cite{Ruterbories:2018gub}.  The same tune as developed on the neutrino data also fits well the \dword{minerva} antineutrino \dword{qe}-like data (with no additional tuning or ingredient).  The required enhancement of the multi-nucleon contribution to the model implies shortcomings in the interaction model, but the decent fit to data for both neutrinos and antineutrinos implies that the tune is effectively making up for some imperfections in the model. 

More recent versions of \dword{genie} include some of the modifications incorporated by \dword{minerva} in the tune discussed above \cite{Alam:2015nkk}.  This illustrates the dynamic nature of neutrino interaction modeling and the interplay between the experiments and generator developers.  The evolution of the field continues as illustrated with a snapshot of some of the current questions and areas of focus:
\begin{itemize}
    \item There is a pronounced deficit of pions produced at low Q$^{2}$ in \dword{cc}1$\pi^{\circ}$ events as compared to expectations \cite{BercellieNUINT2018,Altinok:2017xua,Aliaga:2015wva,McGivern:2016bwh,novaminosPC}.  Current models take this into account by tuning to data without any underlying physical explanation for how or why this happens.
    \item The \dword{minerva} tune that fits both neutrino and antineutrino \dword{ccqe} data involves a significant enhancement and distortion of the \dword{2p2h} contribution to the cross section.  The real physical origin of this cross section strength is unknown.  Models of multi-nucleon processes disagree significantly in predicted rates.
    \item Multi-nucleon processes likely contribute to resonance production.  This is neither modeled nor well constrained.
    \item Cross section measurements used for comparison to models are a convolution of what the models view as initial state, hard scattering, and final state physics.   Measurements able to deconvolve these contributions are expected to be very useful for model refinements.  
    \item Most neutrino generators make assumptions about the form of form factors and factorize nuclear effects in neutrino interactions into initial and final state effects via the impulse approximation.  These are likely oversimplifications.  The models will evolve and the systematic errors will need to be evaluated in light of that evolution. 
    \item  Neutrino detectors are largely blind to neutrons and low-momentum protons and pions (though some $\pi^{+}$ are visible via Michel decay).  This leads to smearing in the  reconstructed energy and transverse momentum, as well as a reduced ability to accurately identify specific interaction morphologies.  The closure of these holes in the reconstructed particle phase space is expected to provide improved handles for model refinement.
    \item There may be small but significant differences in the $\nu_{\mu}$ and $\nu_{e}$ \dword{ccqe} cross sections which are poorly constrained \cite{Day-McFarland:2012}.
\end{itemize}
Given the critical importance of neutrino interaction models and the likelihood that the process of refining these models will continue through the lifetime of \dword{dune}, it is important the \dword{dune}  \dword{nd} suite be highly capable.   

\subsection{Past Experience}
\label{sec:appx-nd:overview-experience}

The philosophy driving the \dword{dune}  \dword{nd} concept is to provide sufficient redundancy to address areas of known weaknesses in previous experiments and known issues in the interaction modeling insofar as possible, while providing a powerful suite of measurements that is likely to be sensitive to unanticipated issues and useful for continued model improvements.  Anything less reduces \dword{dune}'s potential to achieve significantly improved systematic errors over previous experiments in the \dword{lbl} analyses. 

The \dword{dune}  \dword{nd} incorporates many elements in response to lessons learned from previous experiments. 
The massive  \dword{nd} \dword{lartpc} has the same target nucleus and a similar technology to the  \dword{fd}. These characteristics reduce the detector and target systematic sensitivity in the  extrapolation of flux constraints from this detector to the  \dword{fd}.  This detector is capable of providing the primary  sample of \dword{cc} $\nu_{\mu}$ interactions to constrain the flux at the  \dword{fd}, along with other important measurements of the flux from processes like $\nu$-e$^{-}$ scattering and low-$\nu$.  Samples taken with this detector at off-axis angles (\dword{duneprism}) will allow the deconvolution of the flux and cross section errors and provide potential sensitivity to mismodeling.  The off-axis data can, in addition, be used to map out the detector response function and construct effective  \dword{nd} samples that mimic the energy distribution of the oscillated sample at the  \dword{fd}. 

The \dword{dune}  \dword{nd} provides access to particles produced in neutrino interactions that have been largely invisible in previous experiments, such as low-momentum protons and charged pions measured in the \dword{hpgtpc} and neutrons in the \dword{3dst} and \dword{ecal}. The \dword{hpgtpc} provides data on interactions that minimize the effect of secondary interactions on the produced particles.  These capabilities improve the experiment's ability to identify specific interaction morphologies, study samples with improved energy resolution, and extract samples potentially useful for improved tuning of model(s) of multi-nucleon processes. The neutron content in neutrino and antineutrino interactions is different and this will lead to differences in the detector response. For an experiment that is measuring \dword{cpv}, data on neutron production in neutrino interactions is likely to be an important handle in the tuning of the interaction model and the flavor-dependent detector response function model.

The \dword{3dst} provides dedicated beam spectrum monitoring on axis, as well as high statistics samples  useful for the on-axis flux determination as a crosscheck on the primary flux determination (which has different detector and target systematic errors). The beam spectrum monitoring is useful for identifying and diagnosing unexpected changes in the beam.  This proved useful for \dword{numi} and is likely to be more important for DUNE given the need to associate data taken at different times and off-axis angles.

The large data sets that will be accumulated by the three main detectors in the  \dword{nd} suite will allow for differential studies and the use of transverse kinematic imbalance variables, where each detector brings its unique strengths to the study: the \dword{lartpc} has good tracking resolution and containment and massive statistics; the \dword{hpgtpc} has excellent tracking resolution, very low charged particle tracking thresholds, and unambiguous track charge sign determination; and the \dword{3dst} has good containment and can include neutrons on an event-by-event basis. The neutrino interaction samples acquired by this array of detectors will constitute a powerful laboratory for deconvoluting the initial state, hard scattering, and final state physics, which, in turn, will lead to improved modeling and confidence in the final results extracted from the  \dword{fd}.

\section{Constraining the Flux in the ND}
\label{sec:appx-nd:fluxappendix}

The \dword{dune}  \dword{fd} will not measure the neutrino oscillation probability directly. Instead, it will measure the neutrino interaction rate for different neutrino flavors as a function of the reconstructed neutrino energy. It is useful to formalize the measurements that are performed in the near and  far \dwords{detmodule} in the following equations:

\begin{align}
\label{eq:fdrate}
\frac{dN^{FD}_{x}}{dE_{rec}}(E_{rec}) & = 
\int \Phi^{FD}_{\numu}(E_\nu)P_{\numu\rightarrow x}(E_\nu)\sigma^{Ar}_x(E_\nu)T^{FD,Ar}_x(E_\nu,E_{rec})dE_\nu\\
\frac{dN^{ND}_{x}}{dE_{rec}}(E_{rec}) & = 
\int \Phi^{ND}_{x}(E_\nu)\sigma^m_x(E_\nu)T^{d,m}_x(E_\nu,E_{rec})dE_\nu\
\end{align}
with
\begin{itemize}
    \item $x$  = \nue , \numu 
   \item $d$  = \mbox{detector index}(\dword{nd},\dword{fd}) 
   \item $m$  = \mbox{interaction target/material, (e.g., H, C, or Ar)}
   \item $E_\nu$  = \mbox{true neutrino energy}
   \item $E_{rec}$  = \mbox{reconstructed neutrino energy} 
   \item $T^{d,m}_x(E_\nu,E_{rec})$  = \mbox{true-to-reconstruction transfer function} 
   \item $\sigma^m_x(E_\nu)$  = \mbox{neutrino interaction cross section} 
   \item $\Phi^{d}_x(E_\mu)$  = \mbox{un-oscillated neutrino flux} 
   \item $\frac{dN^{d}_{x}}{dE_{rec}}(E_{rec})$  = \mbox{measured differential event rate per target (nucleus/electron)} 
\end{itemize}

There are equivalent formulae for antineutrinos. For simplicity, the instrumental backgrounds (wrongly selected events) and the intrinsic beam contaminations (\nue interactions in case of the appearance measurement) have been ignored. But an important function of the  \dword{nd} is also to quantify and characterize those backgrounds.

It is not possible to constrain the \dword{fd} neutrino flux directly, but the near-to-far flux ratio is believed to be tightly constrained by existing hadron production data and the beamline optics. As such Equation~\ref{eq:fdrate} can be rewritten as

\begin{align}
\frac{dN^{FD}_{x}}{dE_{rec}}(E_{rec}) & = 
\int \Phi^{ND}_{\numu}(E_\nu)R(E_\nu)P_{\numu\rightarrow x}(E_\nu)\sigma^{Ar}_x(E_\nu)T^{d,Ar}_x(E_\nu,E_{rec})dE_\nu\\
\end{align}
with
\begin{align}R(E_\nu) = \frac{\Phi^{FD}_{\numu}(E_\nu)}{\Phi^{ND}_{\numu}(E_\nu)}
\end{align}
taken from the beam simulation. 
It is not possible to measure only a near-to-far event ratio and extract the oscillation probability since many effects do not cancel trivially.  This is due to the non-diagonal true-to-reconstruction matrix, which not only depends on the underlying differential cross section, but also on the detector used to measure a specific reaction.
\begin{align}
\frac{dN^{FD}_{x}}{dE_{rec}}(E_{rec})/{\frac{dN^{ND}_{\numu}}{dE_{rec}}(E_{rec})} & \neq  R(E_\nu)P_{\numu\rightarrow x}(E_\nu)\frac{\sigma^{Ar}_x(E_\nu)}{\sigma^{m}_{\numu}(E_\nu)}
\end{align}
It is therefore important that the \dword{dune} \dword{nd} suite constrain as many components as possible.

While the near-to-far flux ratio is tightly constrained to the level of \SIrange{1}{2}{\%}, the same is not true for the absolute flux itself. \dword{t2k}, using hadron production data obtained from a replica target, can constrain the absolute flux at the  \dword{nd} to \SIrange{5}{6}{\%} in the peak region and to around 10\% in most of its energy range. The \dword{numi} beam has been constrained to 8\% using a suite of thin target hadron production data. The better the \dword{nd} flux is known, the easier it is to constrain modeling uncertainties by measuring flux-integrated cross sections. Predicting the event rate at the  \dword{fd} to a few percent will require additional constraints to be placed with the  \dword{nd} or substantial improvements in our understanding of the hadron production and focusing uncertainties. 

Several handles to constrain the flux 
are addressed below. Briefly they offer the following constraints:

\begin{itemize}
    \item The overall flux normalization and spectrum can be constrained by measuring neutrino scattering off of atomic electrons.
    \item The energy dependence (``shape'') of the \numu and \anumu 
     flux can be constrained using the ``low-$\nu$'' scattering process.
    \item The flux ratio $\anumu/\numu$ can be constrained using \dword{cc} coherent neutrino scattering.
    \item The $\nue/\numu$ flux ratio in the energy region where standard oscillations occur is well-constrained by the beam simulation. The experiment can also measure the $\nue/\numu$ interaction ratio and constrain the flux ratio using cross section universality.
\end{itemize}

\subsection{Neutrino-Electron Elastic Scattering}
\label{sec:appx-nd:fluxintro-e-nu-scatt}

Neutrino-electron scattering ($\nu \ e \rightarrow \nu \ e$) is a pure electroweak process with calculable cross section at tree level. The final state consists of a single electron, subject to the kinematic constraint 

\begin{equation}
1 - \cos \theta = \frac{m_{e}(1-y)}{E_{e}},
\end{equation}

where $\theta$ is the angle between the electron and incoming neutrino, $E_{e}$ and $m_{e}$ are the electron mass and total energy, respectively, and $y = T_{e}/E_{\nu}$ is the fraction of the neutrino energy transferred to the electron. For \dword{dune} energies, $E_{e} \gg m_{e}$, and the angle $\theta$ is very small, such that $E_{e}\theta^{2} < 2m_{e}$. 

The overall flux normalization can be determined by counting $\nu \ e \rightarrow \nu \ e$ events. Events can be identified by searching for a single electromagnetic shower with no other visible particles. Backgrounds from $\nu_{e}$ \dword{cc} scattering can be rejected by looking for large energy deposits near the interaction vertex, which are evidence of nuclear breakup. Photon-induced showers from \dword{nc} $\pi^{0}$ events can be distinguished from electrons by the energy profile at the start of the track. The dominant background is expected to be $\nu_{e}$ \dword{cc} scattering at very low $Q^{2}$, where final-state hadrons are below threshold, and $E_{e}\theta^{2}$ happens to be small. The background rate can be constrained with a control sample at higher $E_{e}\theta^{2}$, but the shape extrapolation to $E_{e}\theta^{2} \rightarrow 0$ is uncertain at the \SIrange{10}{20}{\%} level.

For the \dword{dune} flux, approximately \num{100} events per year per ton of fiducial mass are expected with electron energy above \SI{0.5}{GeV}. For a \dword{lartpc} mass of 25 tons, this corresponds to \num{3300} events per year. The statistical uncertainty on the flux normalization from this technique is expected to be $\sim$1\%. \dword{minerva} has achieved a systematic uncertainty just under 2\% 
and it seems plausible that \dune could do at least as well\cite{Valencia:2019mkf}. 
The \dword{3dst} can also do this measurement with significant statistics and with detector and reconstruction systematics largely uncorrelated with \dword{arcube}.  The signal is independent of the atomic number $A$ and the background is small; so, it seems plausible the samples can be combined to good effect.

\subsection{The Low-$\nu$ Method}
\label{ssec:intro-low-nu}
The inclusive cross section for \dword{cc} scattering $(\nu_l+N\rightarrow l^-+X)$ does not depend on the neutrino energy in the limit where the energy transferred to the nucleus $\nu = E_\nu-E_{l} $ is zero~\cite{bib:original_lownu}.  In that limit, the event rate is proportional to the flux, and by measuring the rate as a function of energy, one can get the flux ``shape.'' This measurement has been used in previous experiments and has the potential to provide a constraint in \dune with a statistical uncertainty $<1\%$.

In practice, one cannot measure the rate at $\nu=0$. Instead it is necessary to restrict $\nu$ to be less than a few \SI{100}{MeV}.  This introduces a relatively small $E_\nu$ dependence into the cross section that must be accounted for to obtain the flux shape. Thus the  measurement technique depends on the cross section model but the uncertainty is manageable~\cite{bib:bodek_lownu}. This is particularly true if low-energy protons and neutrons produced in the neutrino interaction can be detected. 

\subsection{Coherent Neutrino-Nucleus Scattering}

The interactions $\nu_\ell + A \rightarrow \ell^- + \pi^+ + A$ and 
$\overline{\nu}_\ell + N    \rightarrow \ell^+ + \pi^- + N$  
occur with very low three momentum transfer to the target nucleus (A).  As such, the interactions proceed coherently with the entire nucleus, and do not suffer from nuclear effects (though background channels certainly do). These coherent interactions are most useful as a constraint on the $\anumu/\numu$ flux ratio. Identifying with high efficiency and purity requires a detector with excellent momentum and angular resolution.

\subsection{Beam \nue Content}
\label{ssec:beam-nue}
Electron neutrinos in a wide-band beam come from two primary sources: kaon decays and muon decays. These ``beam'' \nue are an irreducible background in $\numu \to \nue$ oscillation searches. As such, the \dword{lbnf} beam was optimized to make the \nue flux as small as possible while maximizing the \numu flux. In the energy range relevant for oscillations (\SI{0.5}{GeV} - \SI{4.0}{GeV}) the predicted $\nue/\numu$ ratio varies between 0.5\% and 1.2\% as a function of energy. The beam \nue flux in the same energy range is strongly correlated with the \numu flux due to the decay chain $\pi^+\to\mu^+\numu$ followed by $\mu^+ \to \anumu{} e^+ \nue $ (and likewise for \anue). As a result, the \dword{lbnf}  beam simulation predicts that the uncertainty on the $\nue/\numu$ ratio varies from \SIrange{2.0}{4.5}{\%}. At the  \dword{fd}, in a 3.5 year run, the statistical uncertainty on the beam \nue component is expected to be 7\% for the $\nu$ mode beam and 10\% for the $\bar{\nu}$ mode beam. The systematic uncertainty on the beam \nue flux is therefore subdominant, but not negligible.

\section{Movable components of the ND and the DUNE-PRISM program}
\label{sec:appx-nd:nd-movable}

\subsection{Introduction to DUNE-PRISM}

One of the primary challenges for \dword{dune} will be controlling systematic uncertainties from the modeling of neutrino-argon interactions. The relationship between the observable final state particles from a neutrino interaction and the incident neutrino energy is currently not understood with sufficient precision to achieve \dword{dune} physics goals.  This is due in part to mismodeling of the outgoing particle composition and kinematics and due to missing energy from undetected particles, such a neutrons and low energy charged pions, and misidentified particles. The latter effects tend to cause  a ``feed-down" in reconstructed neutrino energy relative to the true energy. Since neutrino energy spectra at the \dword{fd} and \dword{nd} have substantially different features due to the presence of oscillations at the  \dword{fd}, these mismodeling and neutrino energy feed-down effects do not cancel in a far-to-near ratio as a function of neutrino energy, and lead to biases in the measured oscillation parameters.

Understanding  \dword{nd} constraints on neutrino-nucleus interaction uncertainties is challenging, since no complete model of neutrino-argon interactions is available. If it were possible to construct a model that was known to be correct, even with a large number of unknown parameters, then the task of a  \dword{nd} would much simpler: to build a detector that can constrain the unknown parameters of the model. However, in the absence of such a model, this procedure will be subject to unknown biases due to the interaction model itself, which are difficult to quantify or constrain.

The \dword{duneprism}  \dword{nd} program consists of a mobile  \dword{nd} 
that can perform measurements over a range of angles off-axis from the neutrino beam direction in order to sample many different neutrino energy distributions, as shown in Figure~\ref{fig:offaxisfluxes}. By measuring the neutrino-interaction final state observables over these continuously varying incident neutrino energy spectra, it is possible to experimentally determine the relationship between neutrino energy and what is measured in the detector (i.e., some observable such as reconstructed energy).

\begin{dunefigure}[Variation of neutrino energy spectrum as a function of off-axis angle]{fig:offaxisfluxes}
{The variation in the neutrino energy spectrum is shown as a function of detector off-axis position, assuming the nominal \dword{nd} location \SI{574}{m} downstream from the production target.}
\includegraphics[width=0.8\textwidth]{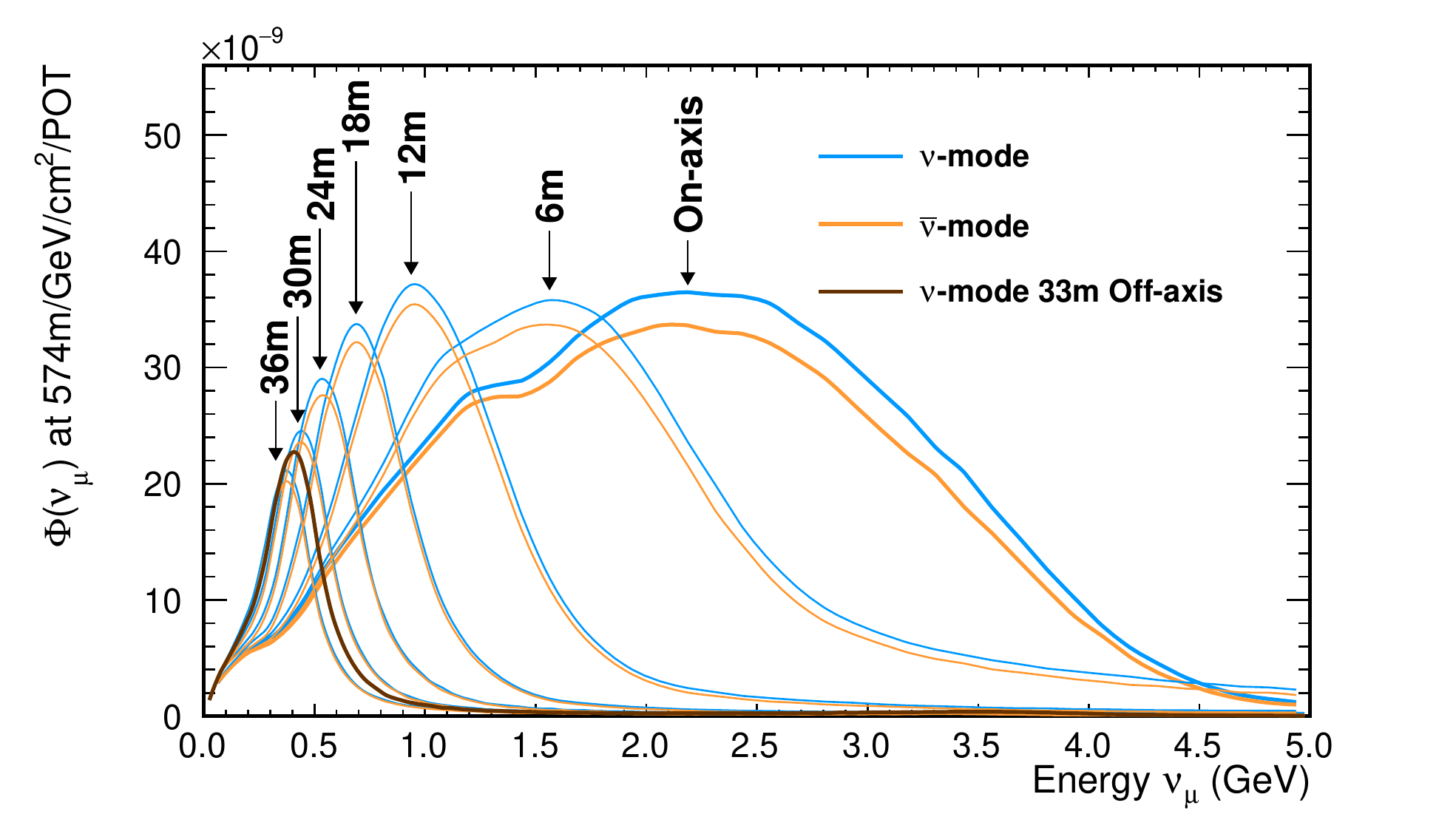}
\end{dunefigure}

In the DUNE \dword{nd}, the movable components of the detector that are used in the \dword{duneprism} program are \dword{arcube} and the \dword{mpd}.  These components of the \dword{nd} will take data both on the beam axis and off-axis.  In the following sections, \dword{arcube} and the \dword{mpd} will be described in some detail and then the \dword{duneprism} program will be described in more detail.

\subsection{LArTPC Component in the DUNE ND: ArgonCube}
\label{sec:appx-nd:lartpc}


As the \dword{dune} \dwords{fd} have \dword{lar} targets, there needs to be a major \dword{lar} component in the \dword{dune}  \dword{nd} complex in order to reduce cross section and detector systematic uncertainties for oscillation analyses~\cite{Acciarri:2016crz, Acciarri:2015uup}. However, the intense neutrino flux and high event rate at the  \dword{nd} makes traditional, monolithic, projective wire readout \dwords{tpc} unsuitable.  This has motivated a program of R\&D into a new \dword{lartpc} design, suitable for such a high-rate environment, known as \dword{arcube}~\cite{argoncube_loi}. \dword{arcube} utilizes detector modularization to improve drift field stability, reducing \dword{hv} and the \dword{lar} purity requirements; pixelized charge readout~\cite{Asaadi:2018oxk, larpix}, which provides unambiguous \threed imaging of particle interactions, drastically simplifying the reconstruction; and new dielectric light detection techniques with \dword{arclt}~\cite{Auger:2017flc}, which can be placed inside the \dword{fc} to increase light yield, and improve the localization of light signals. Additionally, \dword{arcube} uses a resistive field shell, instead of traditional field shaping rings, to maximize the active volume, and to minimize the power release in the event of a breakdown~\cite{bib:docdb10419}. 

The program of \dword{arcube} R\&D has been very successful to date, working on small component prototypes and is summarized in references~\cite{ Ereditato:2013xaa, Zeller:2013sva, art_cold_ero, Asaadi:2018oxk, Cavanna:2014iqa, larpix, bib:docdb10419, Auger:2017flc}. 
With the various technological developments demonstrated with small-scale \dwords{tpc}, the next step in the \dword{arcube} program is to demonstrate the scalability of the pixelized charge readout and light detection systems, and to show that information from separate modules can be combined to produce high-quality event reconstruction for particle interactions. To that end, a mid-scale (\SI[product-units=repeat]{1.4x1.4x1.2}{\metre}) modular \dword{tpc}, dubbed the \dword{arcube} 2$\times$2 demonstrator, with four independent \dword{lartpc} modules arranged in a 2$\times$2 grid has been designed, and is currently under construction. 

After a period of testing at the University of Bern, the \dword{arcube} 2$\times$2 demonstrator will be placed in the \dword{minos}  \dword{nd} hall at \dword{fnal} where it will form the core of a prototype \dword{dune}  \dword{nd}, \dword{pdnd}~\cite{bib:docdb12571}.   As part of \dword{protodune} \dword{nd}, the \dword{arcube} concept can be studied and operated in an intense, few-GeV neutrino beam.  This program aims to demonstrate stable operation and the ability to handle backgrounds, relate energy associated with a single event across \dword{arcube} modules, and connect tracks to detector elements outside of \dword{arcube}.  The \dword{arcube} 2$\times$2 demonstrator is described below in some detail since the \dword{dune}  \dword{nd} modules are anticipated to be very similar.

\subsubsection{ArgonCube in ProtoDUNE-ND}
\label{sec:appx-nd:2x2-design}

The  \dword{arcube} concept is a detector made of self-contained \dword{tpc} modules sharing a common cryostat. Each module is made of a rectangular box with a square footprint and a height optimized to meet the physics goals and/or sensitivity constraints. The \dword{arcube} 2$\times$2 demonstrator module will be housed within an existing \lntwo-cooled and vacuum-insulated cryostat, 
which is $\sim$\SI{2.2}{\metre} in diameter and $\sim$\SI{2.8}{\metre} deep, for a total volume of $\sim$\SI{10.6}{\metre\cubed}. The size of the cryostat sets the dimensions of the modules for the demonstrator. The square base of each module will be \SI{0.67 x 0.67}{\metre}, and the height will be \SI{1.81}{\metre}. This makes the modules comparable in size to, but slightly smaller than, the proposed \dword{arcube} \dword{dune}  \dword{nd} modules, which will have a base of \SI{1 x 1}{\metre} and a \SI{3.5}{\metre} height.

\begin{dunefigure}[\dshort{arcube} 2$\times$2 demonstrator] 
{fig:2x2_extraction}
{Illustration of the \dword{arcube} 2$\times$2 demonstrator module. The four modules are visible, with one of them partly extracted, on the right. This figure has been reproduced from Ref.~\cite{argoncube_loi}.}
\includegraphics[width=\textwidth]{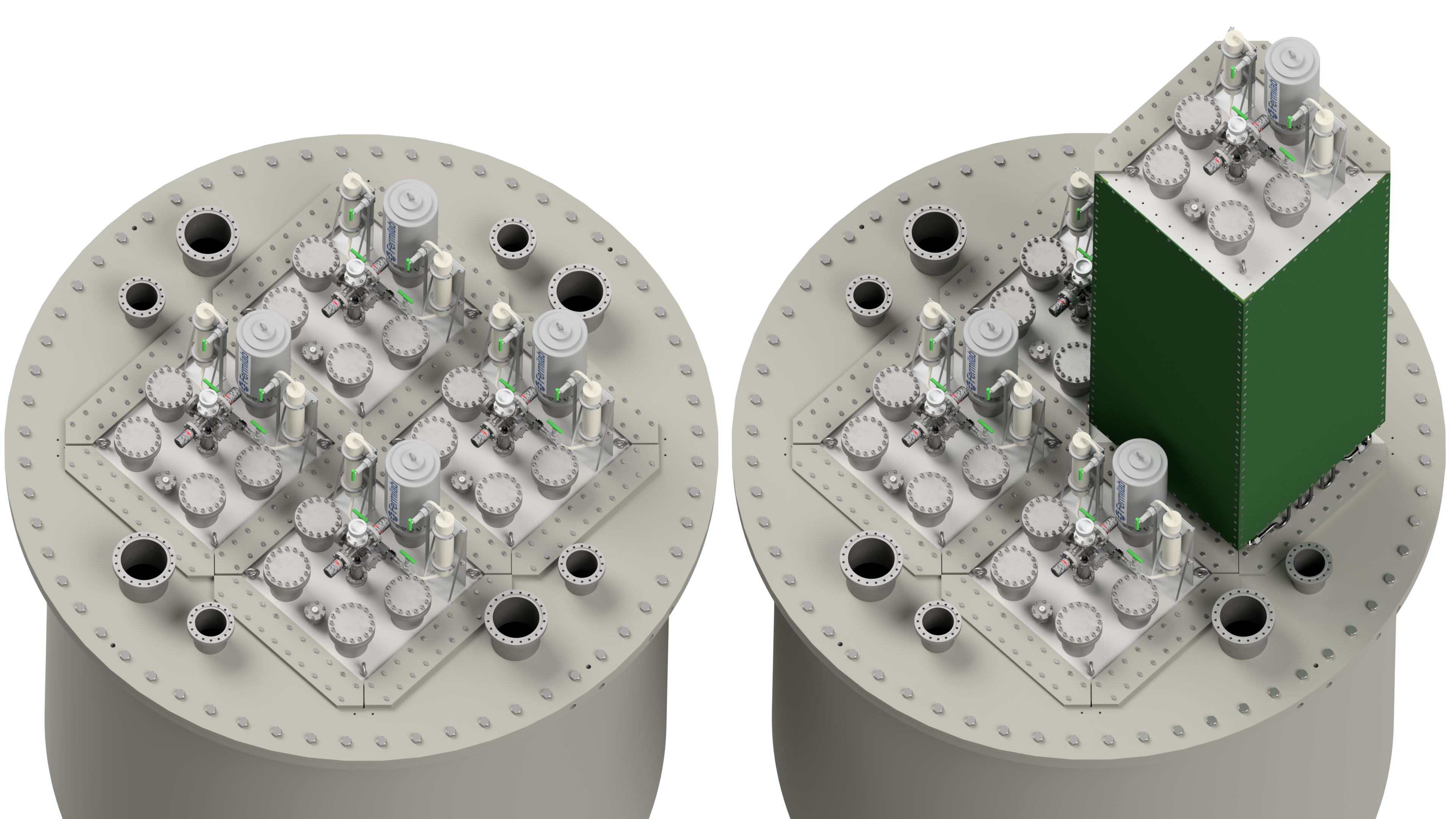}
\end{dunefigure}

Individual modules can be extracted or reinserted into a common \dword{lar} bath as needed, as is illustrated in Figure~\ref{fig:2x2_extraction}. This feature will be demonstrated during a commissioning run at the University of Bern, but is not intended to be part of the detector engineering studies in the \dword{minos}-\dword{nd} hall. The pressure inside the modules is kept close to the bath pressure, putting almost no hydrostatic force on the module walls.  This allows the walls to be thin, minimizing the quantity of inactive material in the walls. The purity of the \dword{lar} is maintained within the modules, independent of the bath, as will be described below. The argon surrounding the modules need not meet as stringent purity requirements as the argon inside. Under normal operating conditions, all modules are inserted with  clearance distances of only \SI{1.5}{\milli\metre} between modules. Cooling power to the bath is supplied by liquid nitrogen circulated through lines on the outer surface of the inner cryostat vessel.

\begin{dunefigure}[Cutaway drawing of an ArgonCube 2$\times$2 demonstrator module]{fig:ac_module}
{Cutaway drawing of a \SI{0.67 x 0.67 x 1.81}{\metre} \dword{arcube} module for the 2$\times$2 demonstrator module. For illustrative purposes the drawing shows traditional field-shaping rings instead of a resistive field shell. The G10 walls will completely seal the module, isolating it from the neighboring modules and the outer \dword{lar} bath. 
The 2$\times$2 modules will not have individual pumps and filters.}
\includegraphics[width=0.8\textwidth]{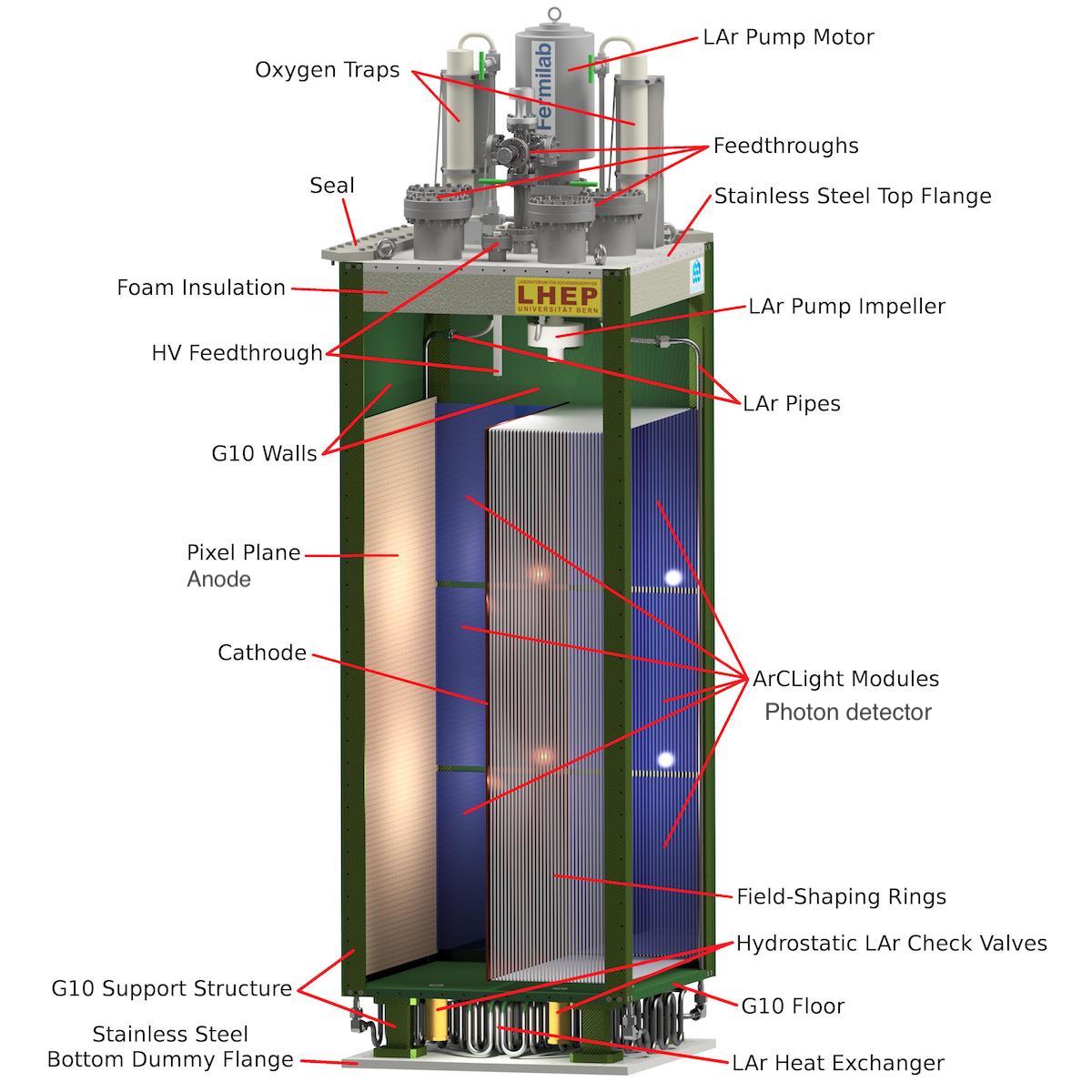}
\end{dunefigure}

A cutaway drawing of an individual 2$\times$2 module is shown in Figure~\ref{fig:ac_module}. The side walls of each module are made from \SI{1}{\centi\metre} G10 sheets, to which the resistive field shell is laminated. The G10 radiation length ($X_{\mathrm{0}} = \SI{19.4}{\centi\metre}$) and hadronic interaction length ($\lambda_{\mathrm{int}} = \SI{53.1}{\centi\metre}$)~\cite{Tanabashi:2018oca} 
are both comparable to \dword{lar} (14.0~cm and 83.7~cm respectively). 
G10 provides a strong dielectric, capable of \SI{200}{\kilo\volt\per\centi\metre} when \SI{1}{\centi\metre} thick~\cite{G10Breakdown}. This dielectric shielding eliminates the need for a clearance volume between the \dwords{tpc} and the cryostat, while also shielding the \dword{tpc} from field breakdowns in a neighboring module. 

Each module is split into two \dwords{tpc} by a central cathode made of an additional resistive layer on a G10 substrate. The segmented drift length does not require a high cathode voltage, and minimizes stored energy. For the 2$\times$2 module footprint of \SI{0.67 x 0.67}{\metre}, and an \efield of \SI{1}{\kilo\volt\per\centi\metre}, a cathode potential of only \SI{33}{\kilo\volt} is required. Operating a \dword{lartpc} at this voltage is feasible without a prohibitive loss of active volume~\cite{Zeller:2013sva}.  The high field is helpful for reducing drift time and the potential for pileup, minimizing the slow component of the scintillation light, reducing space charge effects, and providing robustness against loss of \dword{lar} purity.  

The detector is oriented such that the cathodes are parallel to the beam. This minimizes the load on the readout electronics by spreading the event over more channels and reducing the required digitization rate for hit channels. In turn, this reduces the heat load generated at the charge readout and prevents localized boiling.

During filling and emptying of the cryostat, the argon flow is controlled by hydrostatic check valves located at the lower flange of the module, which require a minimal differential pressure of \SI{15}{\milli\bar} to open. The purity inside each module is maintained by means of continuous \dword{lar} recirculation through oxygen traps. Dirty argon is extracted from the base of the module, and is then pushed through oxygen traps outside the cryostat, clean argon then re-enters the module above the active volume. For optimal heat transport, the argon flow is directed along the cold electronics. To prevent dirty argon from the bath entering the modules, their interior is held at a slight over-pressure. For the 2$\times$2, the dirty argon from all four modules is extracted by a single pump at the base of the cryostat with a four-to-one line, and after being filtered and cooled, the clean argon is pumped back in the module via a one-to-four line.
A more extensive version of the same scheme is envisaged for the \dword{dune} \dword{nd}.

\dword{arcube} offers true \threed tracking information using the \dword{larpix} cryogenic \dword{asic}~\cite{Dwyer:2018phu} pixelated charge readout. \dword{larpix} \dwords{asic} amplify and digitize the charge collected at single-pixels in the cold to mitigate the need for analogue signal multiplexing, and thus produce unambiguous \threed information. Sixty-four pixels can be connected to a single \dword{larpix} \dword{asic}. The baseline design for the 2$\times$2 is a \SI{4}{\milli\metre} pixel pitch, corresponding to 62.5k pixels m$^{-2}$. Pixelated anode planes are located on the two module walls parallel to the cathode; each plane is \SI[product-units=repeat]{1.28x0.64}{\metre\squared}. The total area across all four modules is \SI{6.6}{\metre\squared}, which corresponds to 410k pixels. The readout electronics utilize two \dword{fpga} boards per module, connected to a single Ethernet switch. It should be noted that the pixel pitch may be reduced as prototypes develop, but this can be accommodated in the readout design. 

\begin{figure}[!ht]
	\centering
	\subfloat[\dword{arclt} paddle] {\includegraphics[width=0.454\textwidth]{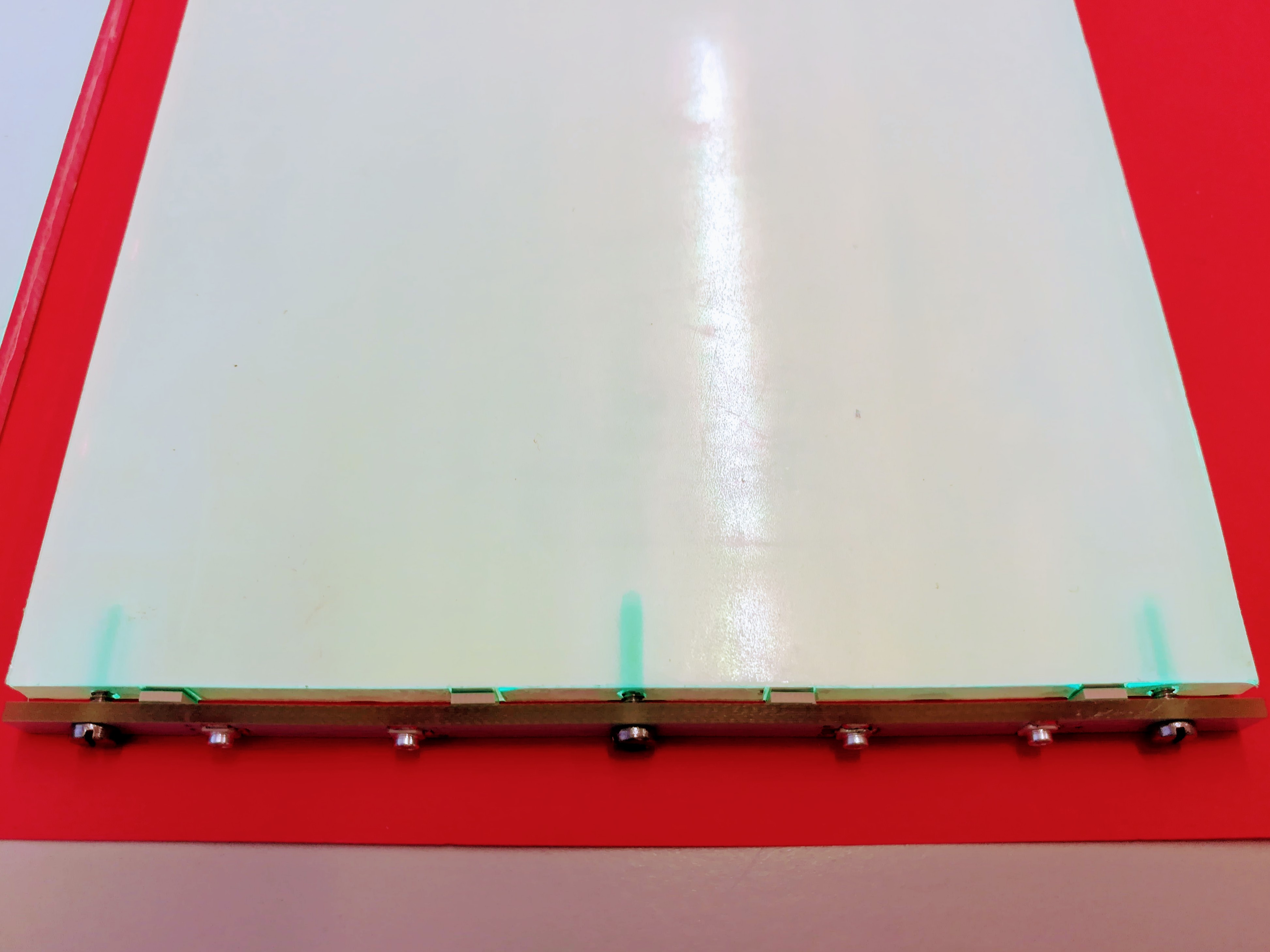}}
	\subfloat[\dword{arclt} mounted on a pixel readout PCB]  {\includegraphics[width=0.51\textwidth]{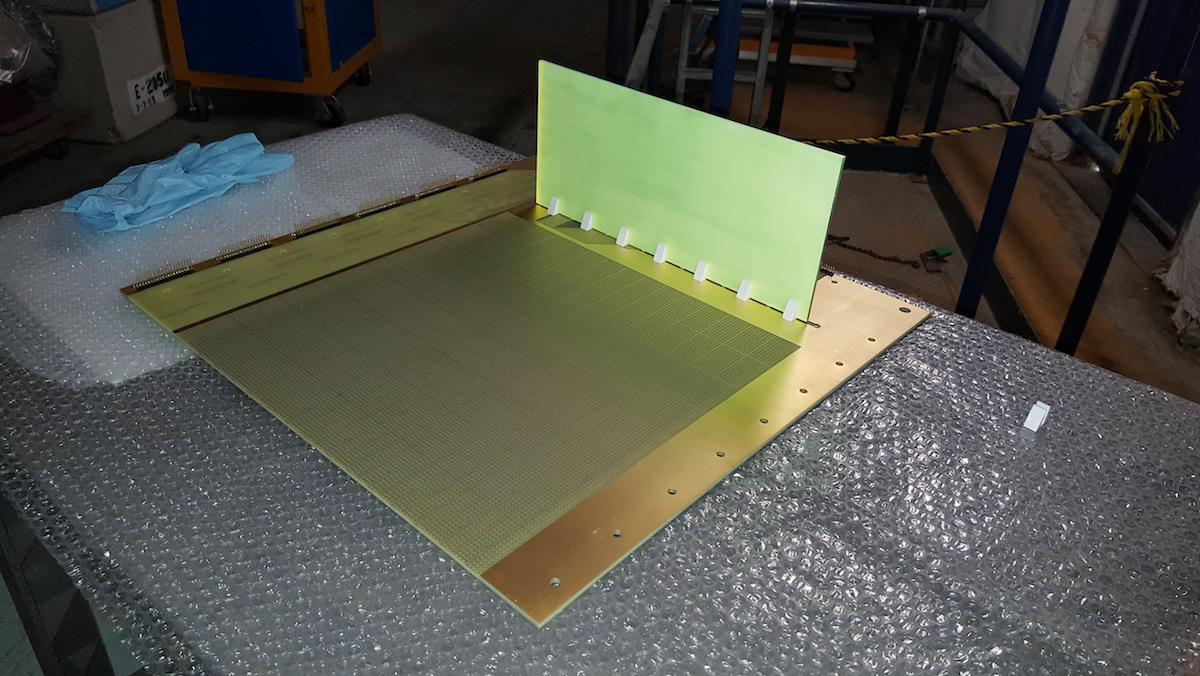}}
	\caption[A prototype ArgonCube light readout paddle and a mounted ArCLight paddle]{(a) A prototype \dword{arcube} light readout paddle. The paddle is \SI{50}{cm} long and \SI{10}{cm}, with four \dwords{sipm} coupled to one end. Reproduced from Ref.~\cite{argoncube_loi}. (b) \dword{arclt} paddle mounted on the PixLAr pixelated charge readout plane, as used in test beam studies at \dword{fnal}.}
	\label{fig:arclight}
\end{figure}

The charge readout window (drift time) of \SI{137}{\micro\second} is long compared to the \SI{10}{\micro\second}~\cite{Adamson:2015dkw} beam spill length in the \dword{numi} and \dword{lbnf} beams.
For a \SI{1}{MW} beam intensity, the expected rate of neutrino interactions at the \dword{dune} \dword{nd} is roughly 0.5 per spill per \dword{arcube} module.  
With \dword{larpix}, reconstruction issues are greatly simplified compared to a projective readout \dword{tpc}.
Tracks and connected energy deposits will frequently overlap in any \twod projection, but can be resolved easily with the full \threed readout.
However, disconnected energy deposits, such as those from photon conversions or neutron interactions in the detector, cannot be associated easily to a specific neutrino interaction.
This problem can be solved by incorporating fast timing information from the prompt scintillation light emitted in \dword{lar}.
The module's opaque cathode and walls contain scintillation light within each \dword{tpc} (half module), improving the detection efficiency of the prompt component of the scintillation light. 
Furthermore, attenuation due to Rayleigh scattering, characterized by an attenuation length of \SI{0.66}{\metre} in \dword{lar}~\cite{Grace:2015yta}, is mitigated by the maximum photon propagation length of \SI{0.3}{\metre}. 
It is desirable to have a large area \dword{pds} to maximize the utility of scintillation light signals in the detector. 
To minimize any dead material within the active volume, it is also desirable that the light detection be as compact as possible. 
The solution pursued for the \dword{arcube} effort is \dword{arclt}~\cite{Auger:2017flc}, which is a very compact dielectric light trap that allows for light collection from a large area, inside high \efield{}s. 
An example \dword{arclt} sheet is shown in Figure~\ref{fig:arclight}. These sheets are mounted on the walls of the module, inside the field shell, aligned with the drift direction, between the anode and the cathode. 
The additional \SI{5}{\milli\metre} deep dead volume is similar to the one caused by the charge readout in the perpendicular direction.

\subsubsection{Dimensions of the ArgonCube Component of the DUNE  ND}\label{sec:appx-nd:had_containment}

Since it is unrealistic to build a \SI{25}{\metre} long \dword{lartpc} in order to contain a \SI{5}{\giga\electronvolt} muon, the \dword{lartpc} dimensions have instead been optimized for hadronic shower containment~\cite{lartpcSizeChris}, relying on a downstream spectrometer to analyze crossing muons.
Hadronic showers are defined as contained if a reasonable efficiency across a wide range of kinematics is maintained, and there is no phase space with zero acceptance. 
The specific metric used is that \textgreater95\% of hadronic energy has to be contained for interactions in the \dword{fv}, excluding neutrons and their descendants.

To assess the efficiency, detector volumes of varying sizes were simulated in a neutrino beam.
This provides a good measure of the efficiency of a given volume to contain different events, but it is not necessarily a good quantity to assess the required detector size.
Many events are not contained because of their specific location and/or orientation.
Cross section coverage remedies this deficiency by looking at the actual extent of the event, instead of its containment, at a random position inside a realistic detector volume.
However, events extending through the full detector will very likely never be contained in a real detector due to the low probability of such an event happening in exactly the right location (e.g., at the upstream edge of the detector).
Therefore, the maximum event size needs to be smaller than the full detector size.
For the \dword{nd} simulation this buffer was chosen to be \SI{0.5}{\metre} in all directions.
In this way, this measure of cross section coverage allows us to look for phase-space regions which are inaccessible to particular detector volume configurations.

To find the optimal detector size in each dimension, two are held constant at their nominal values, while the third dimension is varied and the cross section coverage is plotted as a function of neutrino energy. 
This is shown for the dimension along the beam direction in Figure~\ref{fig:dune-nd_lartpc-size}. In this case, Figure~\ref{fig:dune-nd_lartpc-size} shows us that
\SI{4.5}{\metre} would be sufficient, but to avoid model dependencies, \SI{5}{\metre} has been selected.
Increasing the length beyond \SI{5}{\metre} does little to improve cross section coverage, but reducing to \SI{4}{\metre} begins to limit coverage at higher energies.
Note that 1 minus the cross section coverage gives the fraction of events that cannot be well reconstructed no matter where their vertex is, or how they are rotated within the \dword{fv}. The optimized dimensions found using this technique were \SI{3}{\metre} tall, \SI{4}{\metre} wide, and \SI{5}{\metre} along the beam direction. There is also a need to measure large angle muons that do not go into the \dword{hpgtpc}. Widening the detector to \SI{7}{\metre} accomplishes that goal without the added complication of a side muon detector.

\begin{dunefigure}[Influence of the \dshort{lartpc} size on hadron containment]{fig:dune-nd_lartpc-size}
{Influence of the \dword{lartpc} size on hadron containment, expressed in terms of cross section coverage as a function of neutrino energy.
		Two dimensions are held constant at their nominal values, while the third is varied, in this case the height is held at \SI{2.5}{\metre} and the width at \SI{4}{\metre}.
		The optimal length is found to be \SI{5}{\metre}.
		See text for explanation of cross section coverage~\cite{lartpcSizeChris}.}
	\includegraphics[width=0.5\textwidth]{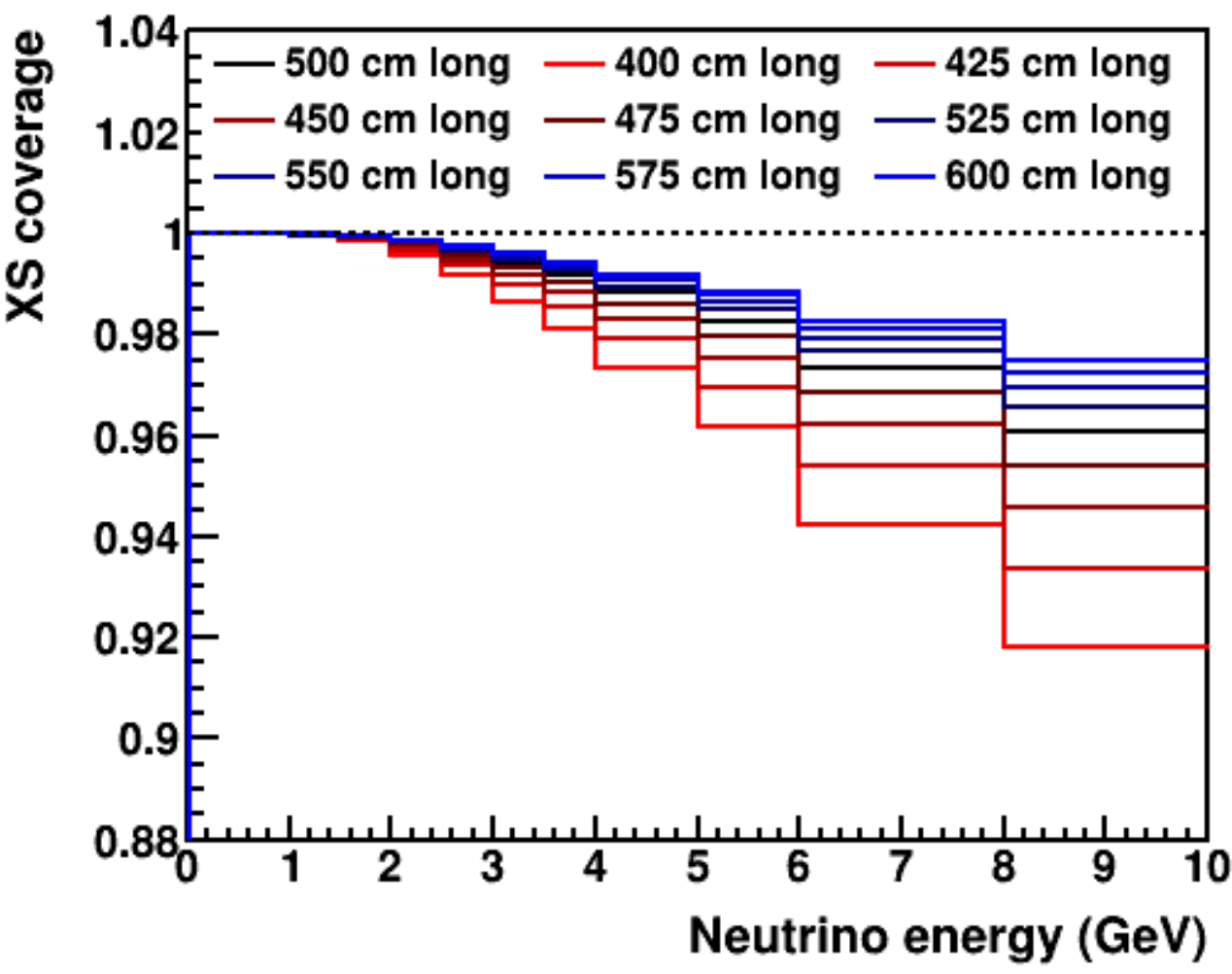}
\end{dunefigure}

\subsubsection{ArgonCube Module Dimensions}

The \dword{dune}  \dword{nd} \dword{arcube} module dimensions are set to maintain a high drift field, \SI{1}{\kilo\volt\per\centi\metre}, with minimal bias voltage, and to allow for the detection of prompt scintillation light while mitigating the effects of diffusion on drifting electrons.
The prompt scintillation light, $\tau<$\SI{6.2}{\nano\second}~\cite{Heindl:2015yaa}, can be efficiently measured with a dielectric light readout with $\mathcal{O}\left(1\right)\,\mathrm{ns}$ timing resolution, such as \dword{arclt}~\cite{Auger:2017flc}.
To reduce attenuation and smearing due to Rayleigh scattering, the optical path must be kept below the \SI{0.66}{\metre}~\cite{Grace:2015yta} scattering length.   Additionally, the slow scintillation component can be further suppressed by operating at higher \efield{}s~\cite{PhysRevB.20.3486}, effectively reducing the ionization density~\cite{PhysRevB.27.5279} required to produce excited states. 

A module with a \SI{1x1}{\metre} footprint split into two \dwords{tpc} with drift lengths of \SI{50}{\centi\metre} requires only a \SI{50}{\kilo\volt} bias.
With \dword{arclt} mounted either side of the \SI{1}{\metre} wide \dword{tpc}, the maximal optical path is only \SI{50}{\centi\metre}.
For a nonzero drift field, diffusion needs to be split into longitudinal and transverse components. Gushchin~\cite{gushchin} report a transverse diffusion of \SI{13}{\centi\metre\squared\per\second} at \SI{1}{\kilo\volt\per\centi\metre}.
This results~\cite{Chepel:2012sj} in a transverse spread of \SI{0.8}{\milli\metre} for the drift time of \SI{250}{\micro\second}, well below the proposed pixel pitch of \SI{3}{\milli\metre}.
The longitudinal component is smaller than the transverse ~\cite{Chepel:2012sj},  and is therefore negligible.

\subsubsection{ND Dimensions}
\label{sec:appx-nd:det_dimensions}

Though the acceptance study discussed in Section~\ref{sec:appx-nd:had_containment} indicated a width of \SI{4}{\metre} is sufficient to contain the hadronic component of most events of interest, the width has been increased to 
\SI{7}{\metre} in order to mitigate the need for a side-going muon spectrometer.
Figure~\ref{fig:actual-size} shows the overall dimensions of the planned \dword{arcube} deployment in the \dword{dune}  \dword{nd}. 
With an active volume of \SI{1x1x3}{\metre} per module, the full \dword{arcube} detector corresponds to seven modules transverse to the beam direction, and five modules along it. 
It should be noted that the cryostat design is currently based on \dword{protodune}~\cite{Abi:2017aow}, and will be optimized for the  \dword{nd} pending a full engineering study.

\begin{dunefigure}[The current \dshort{arcube} dimensions for the \dshort{dune}  \dshort{nd}]{fig:actual-size}
{The current \dword{arcube} Dimensions for the \dword{dune}  \dword{nd}. The cryostat is based on \dword{protodune}~\cite{Abi:2017aow}, and yet to be optimized for the \dword{dune}  \dword{nd}.}
	\includegraphics[width=0.7\textwidth]{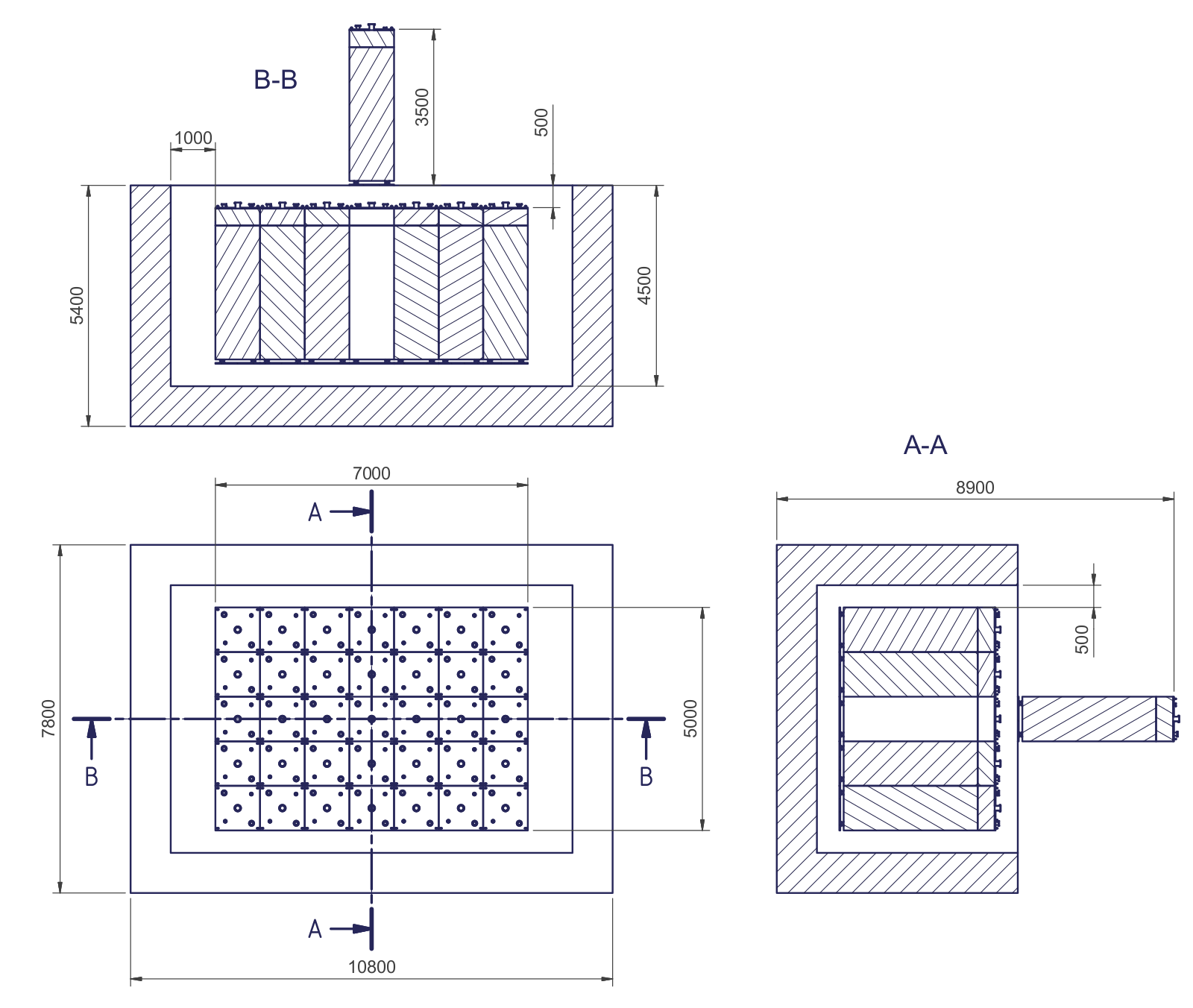}
\end{dunefigure}

\subsubsubsection{Statistics in Fiducial Volume}\label{sec:appx-nd::rates}

Figure~\ref{fig:all_ey} shows 37 million total \dword{cc} $\numu$ neutrino events per year within a \SI{25}{\tonne} \dword{fv} in \dword{fhc} mode at \SI{1.07}{\mega\watt} (on-axis). Figure~\ref{fig:hadContNorm_ey} shows only the event rate for events where the visible hadronic system is fully contained, for the same \dword{fv} and beam configuration. Note that for the visible hadronic system to be contained, all energy not associated with the outgoing lepton, or outgoing neutrons, was required to be contained.

For hadronic containment, there is a \SI{30}{\centi\metre} veto region upstream and on all sides of the active volume, and \SI{50}{\centi\metre} veto region downstream. The \dword{fv} is then defined as \SI{50}{\centi\metre} from all edges, with \SI{150}{\centi\metre} downstream.  Within the \SI{25}{\tonne} \dword{fv} in \dword{fhc} mode at \SI{1.07}{\mega\watt} the number of fully reconstructed (contained or matched muon, discussed below, plus contained hadrons) \dword{cc} $\numu$ events per year is 14 million.

\begin{dunefigure}[All neutrino events in the nominal \SI{25}{\tonne} \dshort{arcube} fiducial volume]{fig:all_ey}
{All neutrino events in the nominal \SI{25}{\tonne} \dword{fv}, in \dword{fhc} at \SI{1.07}{\mega\watt}, per year, rates are per bin. The elasticity is the fraction of the original neutrino energy carried by the outgoing lepton.}
	\includegraphics[width=0.6\textwidth]{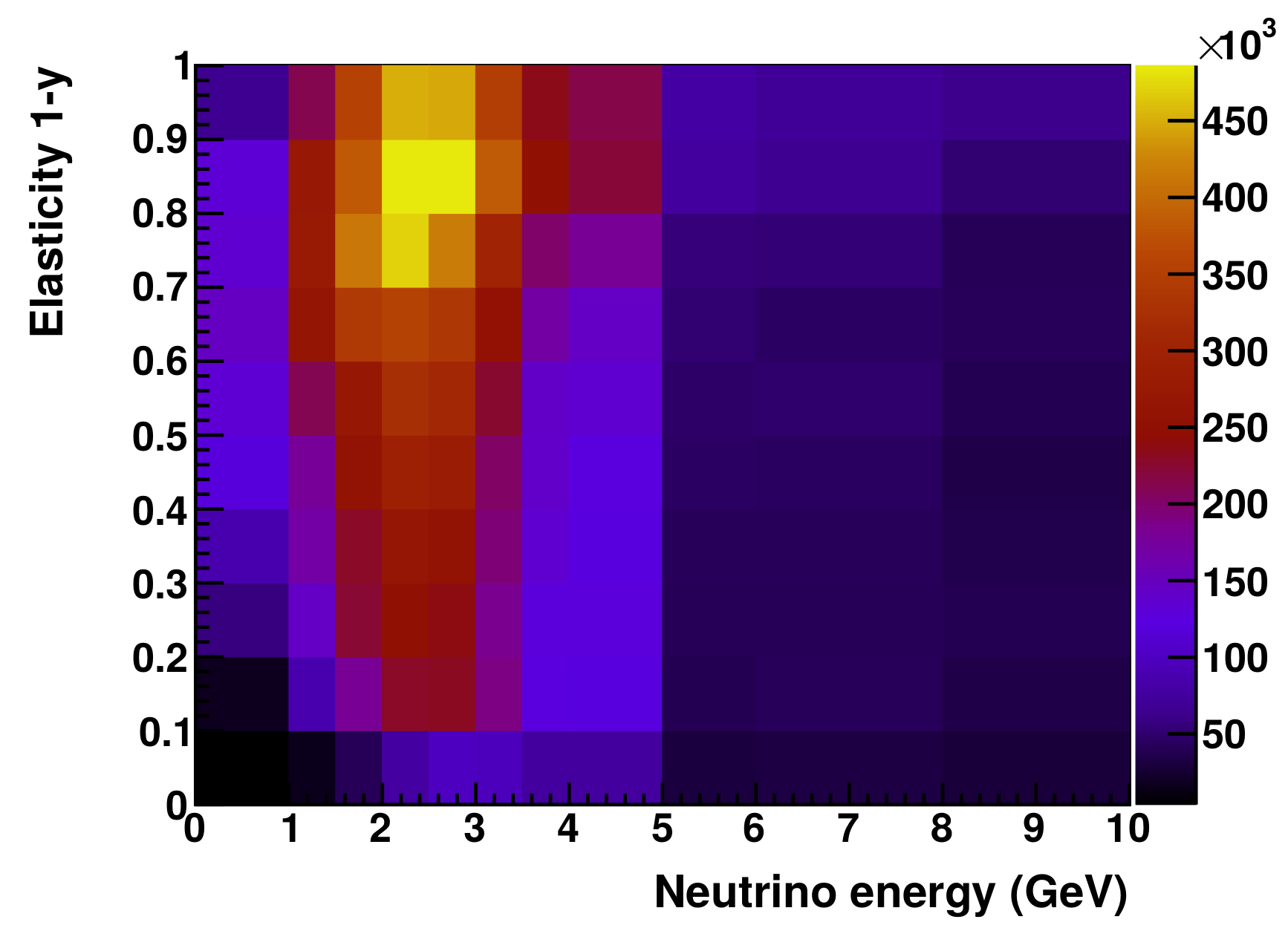}
\end{dunefigure}

\begin{dunefigure}[Events where the visible hadronic system is contained in \dshort{arcube} 
fiducial volume]{fig:hadContNorm_ey}
{Events where the visible hadronic system is contained within the nominal \SI{25}{\tonne} \dword{fv}, in \dword{fhc} at \SI{1.07}{\mega\watt}, per year, rates are per bin.  The elasticity is the fraction of the original neutrino energy that is carried by the outgoing lepton.}
	\includegraphics[width=0.6\textwidth]{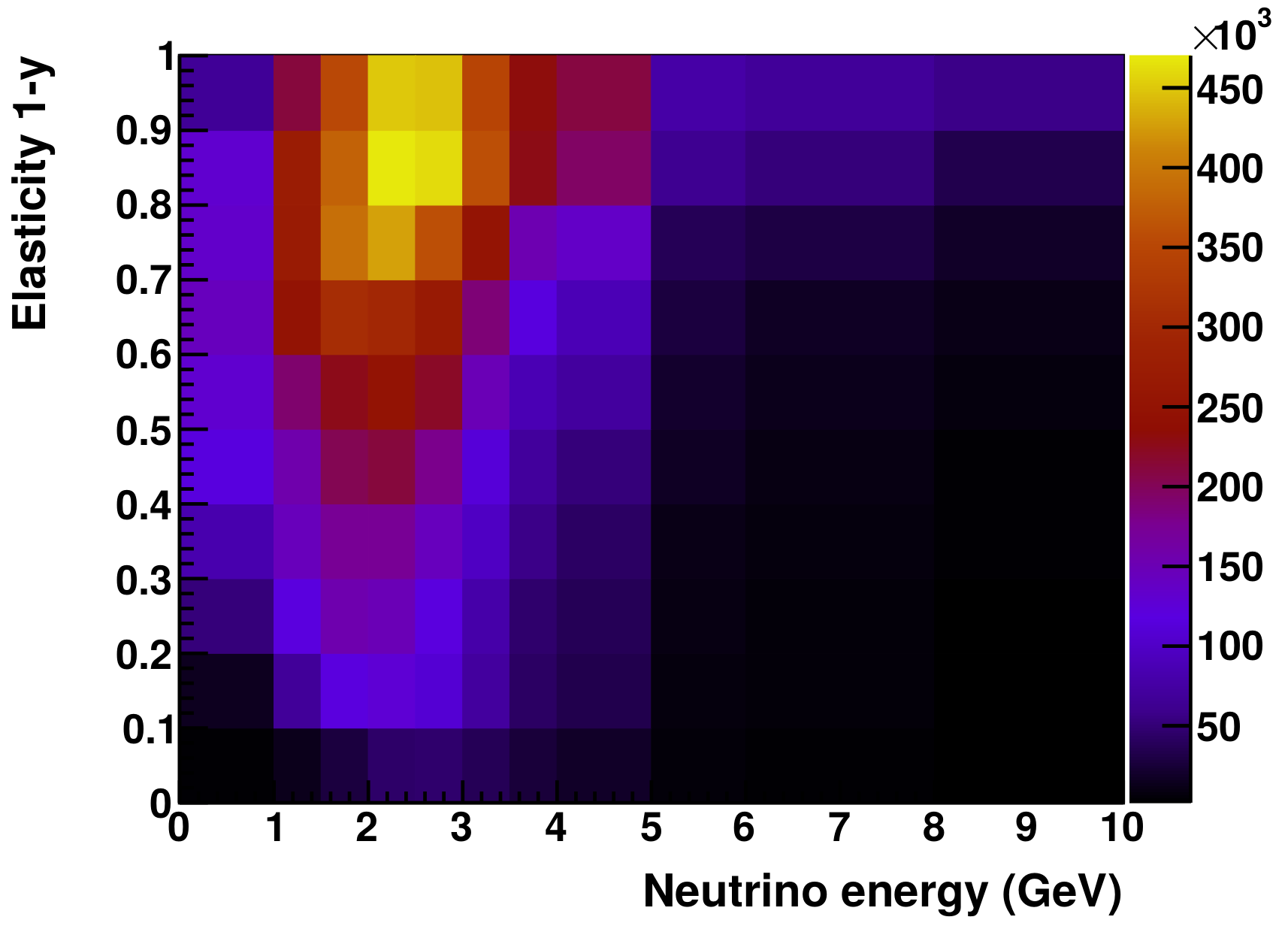}
\end{dunefigure}

\subsubsubsection{Muon Acceptance}\label{sec:appx-nd:muacc}

Muons are considered as useful for physics if they stop in the active region of \dword{arcube} or if they leave the \dword{lar} detector and are reconstructed in a magnetic spectrometer downstream.  Under the assumption that the downstream magnetic spectrometer is the multipurpose detector described in Section~\ref{sec:appx-nd:mpd}, Figure~\ref{fig:muonacc}  shows the muon acceptance as a function of true neutrino energy (on the left) and muon energy (on the right). The acceptance dip at \SI{1}{GeV} in muon energy is from muons that exit \dword{arcube} and are not reconstructed in the \dword{mpd} downstream. This dip can be reduced by minimizing the passive material between the liquid argon and high pressure gaseous argon detectors.

\dword{icarus} and \dword{microboone} have used multiple Coulomb scattering to determine muon momentum \cite{Abratenko:2017nki}.  
This technique may prove to be useful for muons in \dword{arcube} and could  mitigate somewhat the size of the dip in Figure~\ref{fig:muonacc}.

\begin{dunefigure}[Muon acceptance as a function of true neutrino energy and true muon energy]{fig:muonacc}
{Muon acceptance shown as a function of true neutrino energy (left) and true muon energy (right).  The acceptance for muons that stop in \dword{arcube} is shown in red and that for muons reconstructed in the downstream magnetic spectrometer is shown in blue.}
      \includegraphics[width=0.45\textwidth]{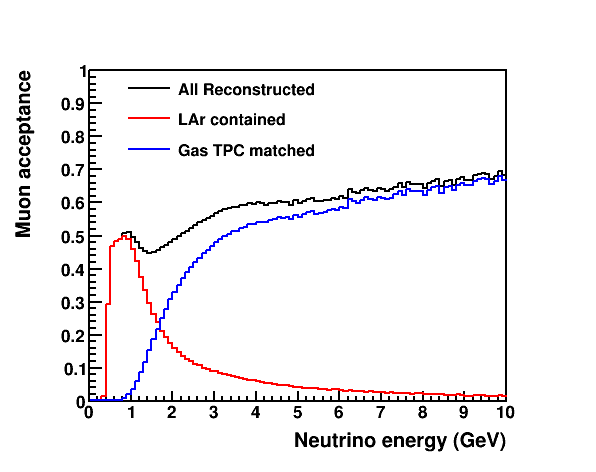}
      \includegraphics[width=0.45\textwidth]{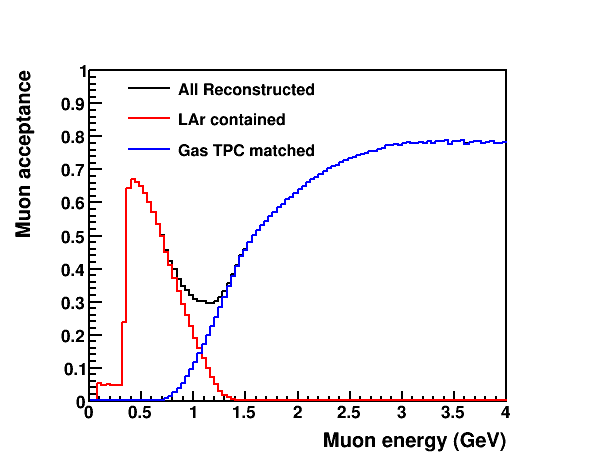}
\end{dunefigure}

\subsubsection{Acceptance vs.\ energy and momentum transfer}

The acceptance of \dword{arcube} with the \dword{mpd} acting as a downstream spectrometer can be studied in a more nuanced way by looking at it as a function of the energy $q_0$ and three-momentum $q_3$  transferred to the target nucleus. The energy transfer is simply $q_0=E_\nu - E_\mu$. The three-momentum transfer is related to the four-momentum transfer $Q$ and $q_0$ by $q_3 = \sqrt{Q^2 + q_0^2}$. These variables have long been used to study nuclear structure in electron scattering experiments. 

\begin{dunefigure}[Neutrino acceptance as a function of energy and momentum transfer]{fig:q0q3acc}
{Neutrino acceptance shown as a function of energy transfer and momentum transfer ($q_0$ and $q_3$) to the target nucleus. The units for $q_0$ and $q_3$ are GeV and GeV/c, respectively. The figures show the event rate (left) and the acceptance (right) for reconstructing the muon and containing the hadronic system. The top row was made for neutrinos with true neutrino energy between \num{1.0} and \SI{2.0}{GeV} and the bottom was made for neutrinos between \num{4.0} and \SI{5.0}{GeV}.}
      \includegraphics[width=0.45\textwidth]{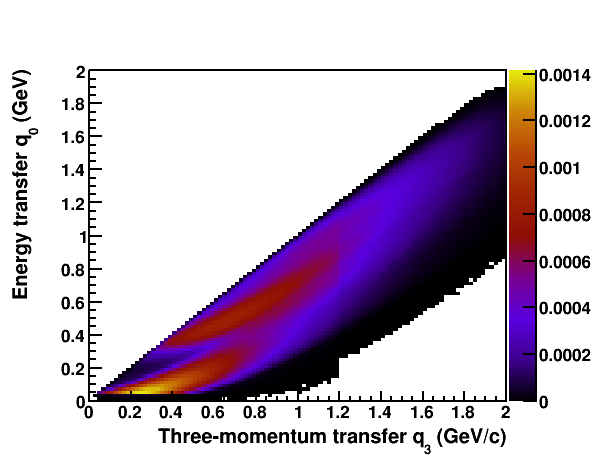}
      \includegraphics[width=0.45\textwidth]{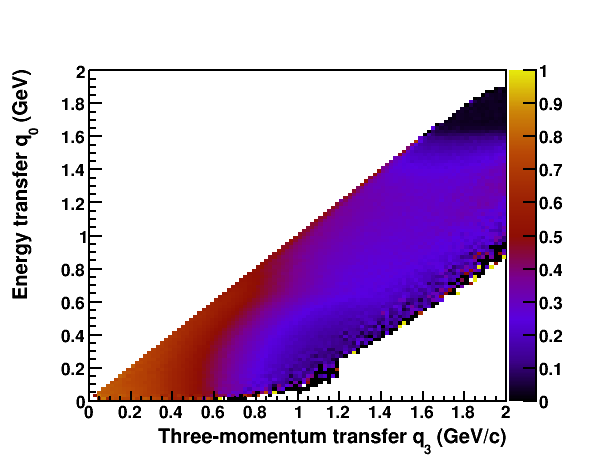}
      \includegraphics[width=0.45\textwidth]{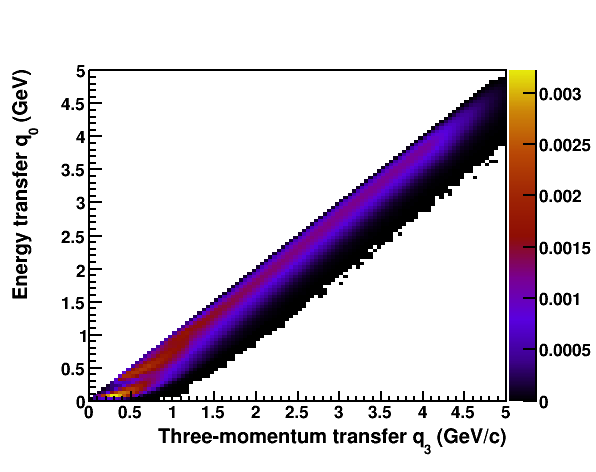}
      \includegraphics[width=0.45\textwidth]{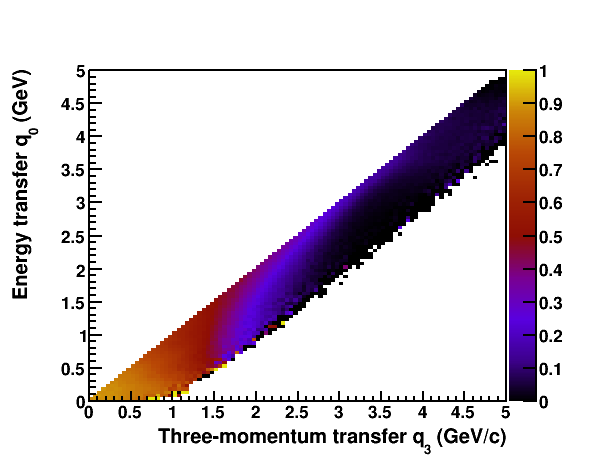}
\end{dunefigure}

Figure~\ref{fig:q0q3acc} shows the event rate (left figures) and acceptance (right figures) in bins of $(q_3,q_0)$. The rows correspond to two neutrino energy bins. The top row is for $E_\nu$ between \num{1.0}-\SI{2.0}{GeV} and it covers the first oscillation maximum. The second bin is for $E_\nu$ between \num{4.0}-\SI{5.0}{GeV}.  The rate histograms have ``islands''  corresponding to hadronic systems with fixed invariant mass. The islands are smeared by Fermi motion and decay width. The lower island in $(q_3,q_0)$ corresponds to the quasi-elastic peak while the upper corresponds to the $\Delta$ resonance. One should note that the axes in the lower row cover a larger range of kinematic space than those in the upper row. 

The acceptance is generally very good in the kinematic region where the vast majority of the events occur but is nowhere perfect. This is not necessarily a problem because the loss is chiefly geometrical. Losses typically occur in events with a vertex near one boundary of the detector where the muon, or hadronic system exits out that boundary.  However for each lost event there is generally a set of symmetric events that are accepted because the final state is rotated by some angle about the neutrino beam axis ($\phi$ symmetry) or is closer to the center of the fiducial volume (x,y symmetry).

Regions where the acceptance is zero are problematic because they will introduce model dependence into the prediction of the rate at the far detector (which has a nearly $4\pi$ acceptance). Acceptances of even a few \% in some kinematic regions are not necessarily a problem as long as the event rate is large enough to accumulate a statistically significant number of events. 
There is a potential danger if the acceptance varies quickly as a function of the kinematic variables because a small mismodeling of the detector boundaries or neutrino cross-sections could translate into a large mismodeling in the number of accepted events. 

The size of the accepted event set decreases as a function of both $q_0$ and $q_3$ (and therefore $E_\nu$) due to more energetic hadronic systems and larger angle muons. This can clearly be seen in the transition from the colored region to the black region in the $\num{4.0} < E_\nu < \SI{5.0}{GeV}$ acceptance histogram shown in the lower right-hand corner of Figure~\ref{fig:q0q3acc}. The transition is smooth and gradual. 

The acceptance for $\num{1.0} < E_\nu < \SI{2.0}{GeV}$ (shown in the upper right-hand corner of Figure~\ref{fig:q0q3acc}) is larger than 10\% except in a small region at high $q_0$ and $q_3$. Events in that region have a low-energy muon and are misidentified as neutral-current according to the simple event selection applied in the study. The fraction of events in that region is quite small, as can be seen in the upper left-hand plot of Figure~\ref{fig:q0q3acc}. 

\begin{dunefigure}[Neutrino acceptance in the $(q_3,q_0)$ plane as a function of neutrino energy]{fig:q0q3acc_vs_enu}
{This figure summarizes the neutrino acceptance in the $(q_3,q_0)$ plane, as shown in Figure~\ref{fig:q0q3acc}, for all bins of neutrino energy (plotted in GeV). Here the quantity on the vertical axis is the fraction of events that come from bins in $(q_3,q_0)$ with an acceptance greater than $A_{cc}$. As an example we consider the \num{4.0}-\SI{5.0}{GeV} neutrino energy bin. The $A_{cc}>0.03$ curve in that neutrino energy bin indicates that 96\% of events come from $(q_3,q_0)$ bins that have an acceptance greater than 3\%. }
      \includegraphics[width=0.6\textwidth]{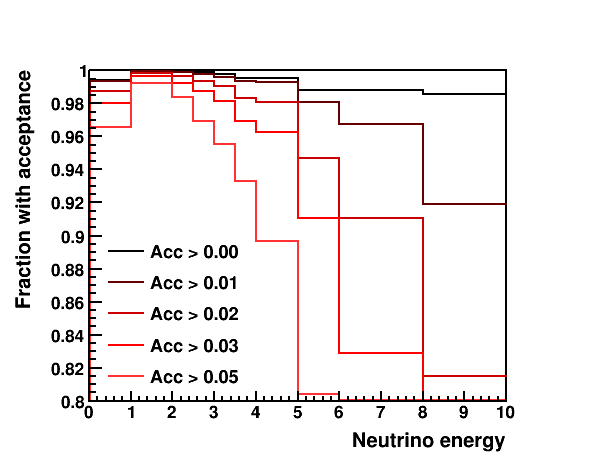}
\end{dunefigure}

Figure~\ref{fig:q0q3acc_vs_enu} summarizes the neutrino acceptance in the $(q_3,q_0)$ plane as function of neutrino energy. The $y$ axis shows the fraction of events coming from $(q_3,q_0)$ bins with an acceptance greater than $A_{cc}$. The $A_{cc}>0.00$ curve shows the fraction of events for which there is nonzero acceptance. For $E_\nu < \SI{5.0}{GeV}$ (the oscillation region) that fraction is greater than 99\%. So, there are no significant acceptance holes.  In the same energy region, more than 96\% of events come from regions where the acceptance is greater than 3\%. 

\subsubsection{Muon and Electron Momentum Resolution and Scale Error}


For muons stopping in the \dword{lar} and for those with momentum measured in the downstream tracker (\dword{mpd}), the energy scale uncertainty from \dword{arcube} is driven by the material model of the \dword{lar} and passive materials.  This is expected to be known to \textless 1\%.  Note that the B field in the \dword{mpd} is expected to be known to about 0.5\% from simulation and field maps made with Hall and NMR probes.

For electrons, the energy will be measured calorimetrically, rather than by range.  The \dword{mip} energy scale (charge/MeV) will be set by rock muons.  The scaling to more dense deposits from EM showers can give rise to uncertainties, i.e., recombination could be different.  Such uncertainties can be reduced by taking data with \dword{arcube} modules in a test beam.  Outside of this, a useful calibration sample
of electrons up to \SI{50}{MeV} comes from Michel electrons from stopping rock muons. The $\pi^0$ invariant mass peak is another good standard candle.

\subsubsection{Tagging Fast Neutrons}

Studies have shown that contained prompt scintillation light provides an important handle for neutron tagging, allowing for the association of detached energy deposits to the correct neutrino interaction using timing information. Such neutron tagging is important for minimizing the uncertainty on neutrino energy reconstruction, both for neutrons generated at a neutrino vertex and for hadronic showers that fluctuate to neutrons. 

Figure~\ref{fig:NDSpill} shows a simulated beam spill in the \SI[product-units=repeat]{5x4x3}{\metre} \dword{lar} component of the \dword{dune}  \dword{nd}\footnote{Note that this study was performed before the detector width was increased to \SI{7}{m}, as described in Section~\ref{sec:appx-nd:det_dimensions}.}. 
It highlights the problem of associating fast-neutron induced energy deposits to a neutrino vertex using only collected charge.

\begin{dunefigure}[A beam spill in the \dshort{lar} component of the \dshort{dune} ND]{fig:NDSpill}
{A beam spill in the \dword{lar} component of the \dword{dune} \dword{nd}. 
		The detector volume is \SI[product-units=repeat]{5x4x3}{\metre}.
		Fast-neutron induced recoiling proton tracks, with an energy threshold greater than $\sim\,$\SI{10}{\mega\electronvolt}, are shown in white.
		The black tracks are all other energy deposits sufficient to cause charge collected at the pixel planes.}
\includegraphics[width=.7\textwidth]{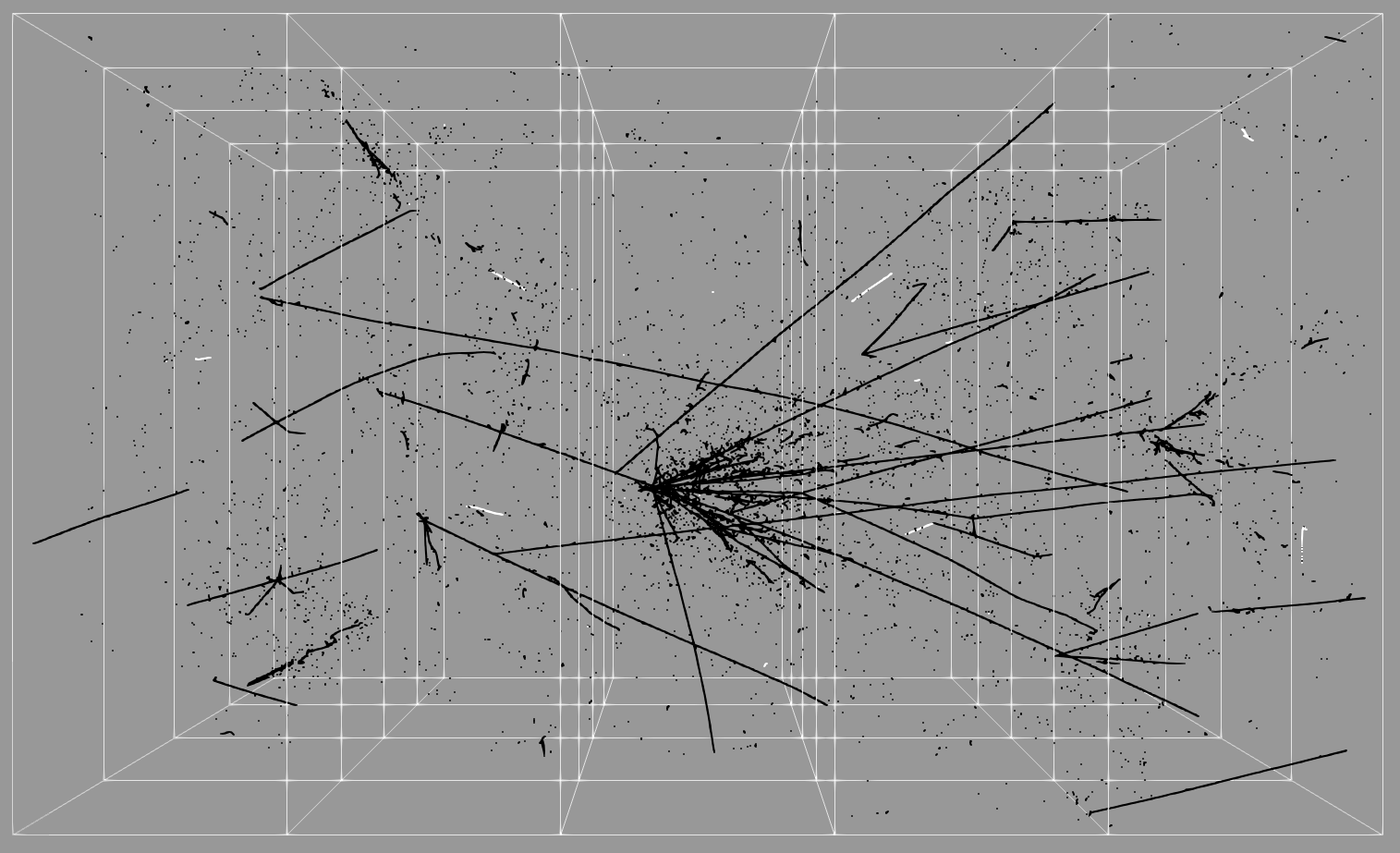}
\end{dunefigure}

By containing scintillation light, prompt light signals can be used to associate fast-neutron induced deposits back to a neutrino vertex anywhere within the detector.
Figure~\ref{fig:Timing} shows the temporal distribution of neutrino vertices within a representative, randomly selected, beam spill.
The mean separation of neutrino vertices is \SI{279}{\nano\second}, with all fast-neutron induced energy deposits occurring $<$\SI{10}{\nano\second} after each neutrino interaction.

\begin{dunefigure}[Temporal distribution of $\nu$ vertices within a beam spill in the ND LAr component] 
{fig:Timing}
{The temporal distribution of neutrino vertices (red lines) within a beam spill in the \dword{lar} component of \dword{dune} \dword{nd}.
		The mean separation of neutrino vertices is \SI{279}{\nano\second}. The filled bins show the number of hits due to recoiling protons, crosses indicate a hit due to a recoiling $^{2}$H, $^3$H, $^2$He or $^3$He nucleus.
		All fast-neutron induced energy deposits occur $<$\SI{10}{\nano\second} after each neutrino interaction.}
\includegraphics[width=0.7\textwidth]{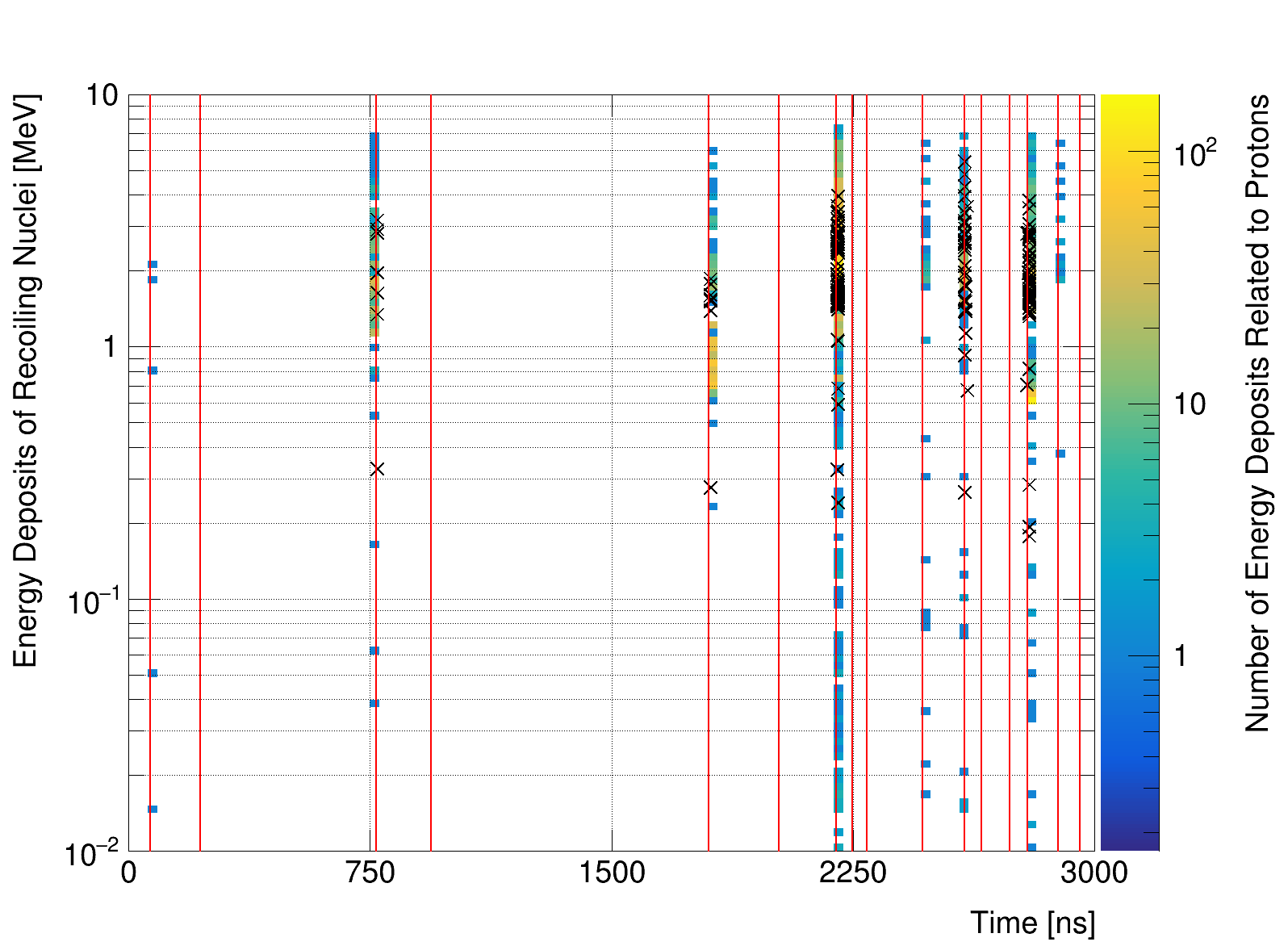}
\end{dunefigure}
	
\subsubsection{Neutrino-Electron Elastic Scattering}
\label{sec:appx-nd:lartpc-nu-electron-scatt}

Neutrino scattering on atomic shell electrons, $\nu_{l}(\overline{\nu}_{l}) + e^{-} \rightarrow \nu_{l}(\overline{\nu}_{l}) + e^{-}$,
is a purely electroweak process with a known cross section as function of neutrino energy, $E_{\nu}$, in which all neutrino flavors participate, albeit with different cross sections. This process is not affected by nuclear interactions and has a clean signal of a single very forward-going electron. \dword{minerva}~\cite{Park:2015eqa} has used this technique to characterize the \dword{numi} beam flux normalization (running in the \dword{numi} low-energy mode), although the rate and detector resolution were insufficient to make a shape constraint. It has been investigated as a cross section model-independent way to constrain the neutrino flux at the \dword{dune}  \dword{nd}.

For a neutrino-electron  sample, $E_{\nu}$ could, in principle, be reconstructed event-by-event in an ideal detector using the formula
\begin{equation}
  E_{\nu} = \frac{E_{e}}{1 - \frac{E_{e}(1-\cos\theta_{e})}{m_{e}}},
\label{eq:nue}
\end{equation}
\noindent where $m_e$ and $E_e$ are the electron mass and outgoing energy, and $\theta_e$ is the angle between the outgoing electron and the incoming neutrino direction. The initial energy of the electrons are low enough to be safely neglected ($\sim$\SI{10}{keV}). It is clear from Equation~\ref{eq:nue} that the ability to constrain the shape of the flux is critically dependent on the energy- and, in particular, angular-resolution of electrons. For a realistic detector, the granularity of the $E_{\nu}$ shape constraint (the binning) depends on its performance. Additionally, the divergence of the beam (few \si{mrad}) at the \dword{dune}  \dword{nd} site is a limiting factor to how well the incoming neutrino direction is known.

In work described in Ref.~\cite{PhysRevD.101.032002}, the ability for various proposed \dword{dune} \dword{nd} components to constrain the \dword{dune} flux is shown using the latest three-horn optimized flux and including full flavor and correlation information.  This was used to determine what is achievable relative to the best performance expected from hadron production target models. When producing the input flux covariance matrix, it was assumed that an \dword{na61}~\cite{Laszlo:2009vg} style replica-target experiment was already used to provide a strong prior shape constraint. Detector reconstruction effects and potential background processes are included, and a constrained flux-covariance is produced following the method used in Ref.~\cite{Park:2015eqa}.

\begin{figure}[htbp]
  \centering
  \subfloat[FHC pre-fit]  {\includegraphics[width=0.45\textwidth]{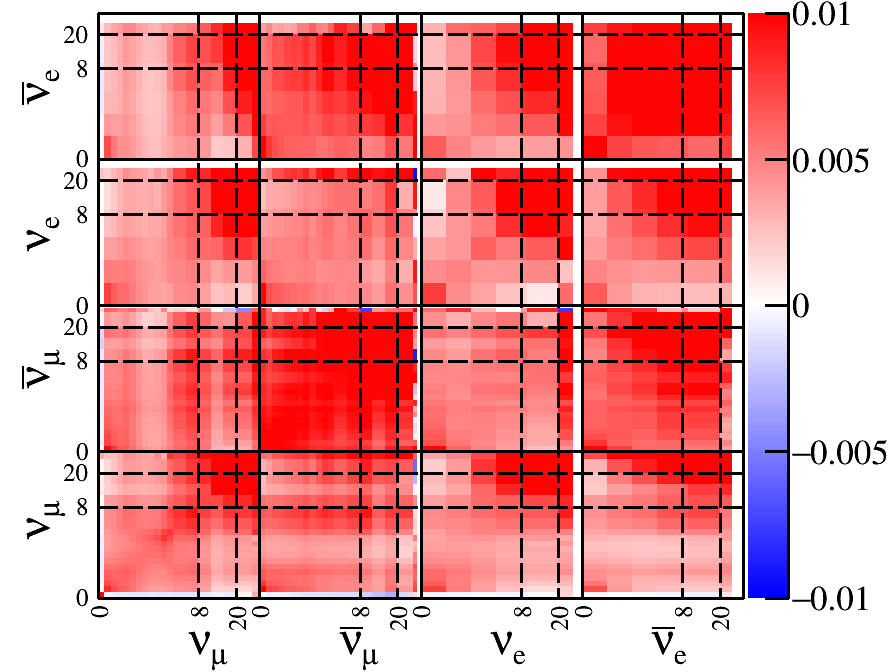}}
  \subfloat[FHC post-fit] {\includegraphics[width=0.45\textwidth]{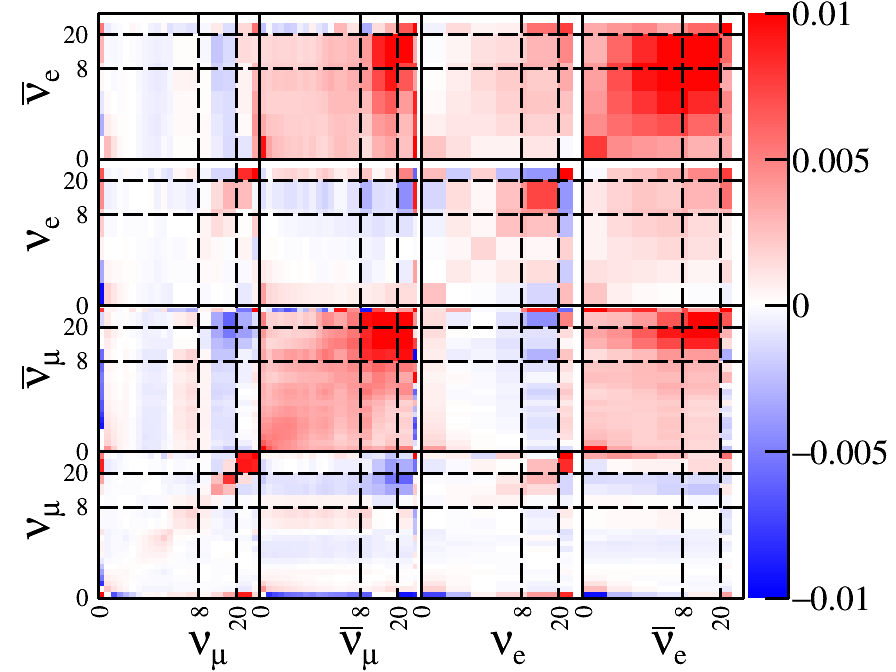}}
  \caption[\dshort{fhc} flux covariance matrices for nominal \SI{35}{t} \dshort{arcube}]{Pre- and post-fit \dword{fhc} flux covariance matrices for the nominal \SI{35}{t} \dword{arcube} \dword{lar} detector using a five-year exposure.}
  \label{fig:LAR_nominal_covariances}
\end{figure}
The impact of the neutrino-electron scattering constraint on the flux covariance is shown in Figure~\ref{fig:LAR_nominal_covariances} for \dword{fhc} and a five year exposure of the nominal \SI{35}{t} \dword{arcube} \dword{lar} detector (corresponding to $\sim$22k neutrino-electron events). It is clear that the overall uncertainty on the flux has decreased dramatically, although, as expected, an anticorrelated component has been introduced between flavors (as it is not possible to tell what flavor contributed to the signal on an event-by-event basis). Similar constraints are obtained for \dword{rhc} running.

\begin{figure}[htbp]
  \centering
  \subfloat[Rate+shape]  {\includegraphics[width=0.45\textwidth]{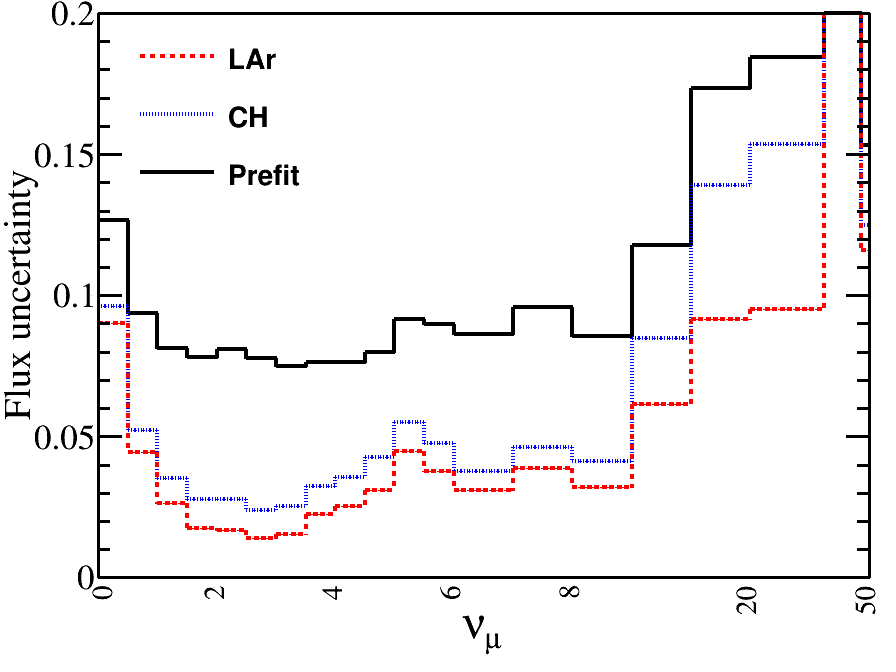}}
  \subfloat[Shape-only]  {\includegraphics[width=0.45\textwidth]{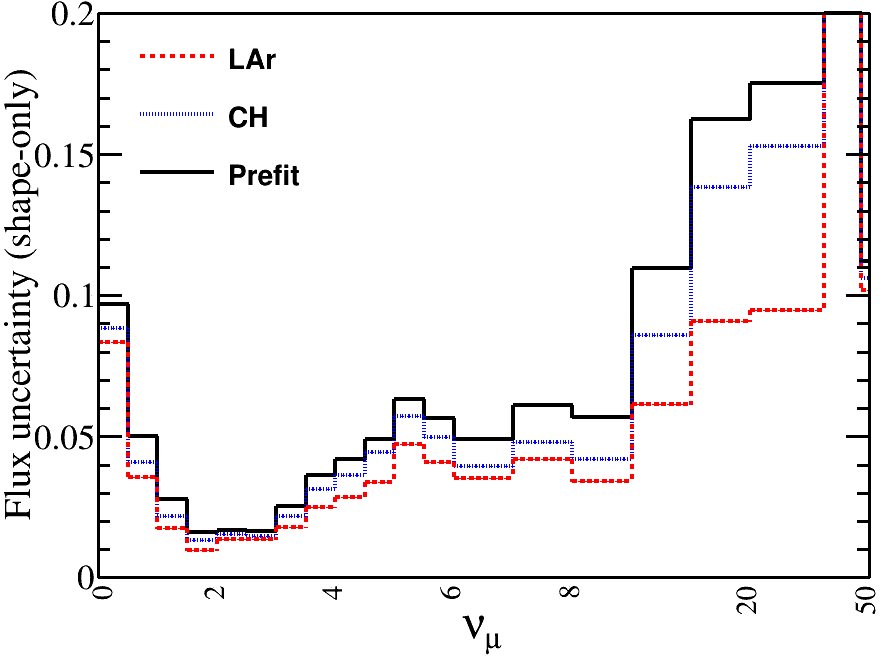}}
  \caption[Rate+shape and shape-only bin-by-bin flux uncertainties]{Rate+shape and shape-only bin-by-bin flux uncertainties as a function of neutrino energy for a five year exposure with various detector options, compared with the input flux covariance matrix before constraint.}
  \label{fig:nominal_det_constraint}
\end{figure}
Figure~\ref{fig:nominal_det_constraint} shows the flux uncertainty as a function of $E_{\nu}$ for the $\nu_{\mu}$-\dword{fhc} flux, for a variety of \dword{nd} options. In each case, the constraint on the full covariance matrix is calculated (as in Figure~\ref{fig:LAR_nominal_covariances}), but only the diagonal of the $\nu_{\mu}$ portion is shown. In the flux peak of $\sim$2.5 GeV, the total flux uncertainty can be constrained to $\sim$2\% for the nominal \dword{lar} scenario, and the constraint from other detector types is largely dictated by the detector mass. Clearly the neutrino-electron scattering sample at the \dword{dune} \dword{nd} will be a powerful flux constraint. However, it is also clear that the ability to constrain the shape of the flux is not a drastic improvement on the existing flux covariance matrix, and none of the possible detectors investigated added a significantly stronger constraint. That said, the neutrino-electron sample is able to make in situ measurements of the flux prediction, and is able to diagnose problems with the flux prediction in a unique way.


\subsection{Multipurpose Detector}
\label{sec:appx-nd:mpd}
The multipurpose detector (\dword{mpd}) extends and enhances the capabilities of the \dword{lartpc}.  It does this by providing a system that will measure the momentum and sign of charged particles exiting the \dword{lartpc} and, for neutrino interactions taking place in the \dword{mpd}, it will extend charged particle measurement capabilities to lower energies than achievable in the far or near \dwords{lartpc}. This capability enables further constraints of systematic uncertainties for the \dword{lbl} oscillation analysis.  
The \dword{mpd} is a magnetized system consisting of a high-pressure gaseous argon time projection chamber (\dword{hpgtpc}) and a surrounding \dword{ecal}. The detector design will be discussed in more detail in a later section.

\paragraph{\dword{mpd} goals}
\begin{itemize}

\item {Measure particles that leave the LAr \dword{nd} component and enter the \dword{mpd} 
    
The LAr component of the DUNE \dword{nd} will not fully contain high-energy muons or measure lepton charge.  A downstream \dword{mpd} will be able to determine the charge sign and measure the momenta of the muons that enter its acceptance, using the curvature of the associated track in the magnetic field. }

    \item {Constrain neutrino-nucleus interaction systematic uncertainties

In its 1-ton gaseous argon \dword{fv}, the \dword{mpd} will collect \num{1.5e6} \dword{cc} muon neutrino interactions per year (plus \num{5e5} \dword{nc} muon neutrino interactions). The very low energy threshold for reconstruction of tracks in the \dword{hpgtpc} gives it a view of interactions that is more detailed than what is seen in the \dword{lar}, and on the same target nucleus. The associated \dword{ecal} provides excellent ability to detect neutral pions, enabling the \dword{mpd} to measure this important component of the total event energy while also tagging the presence of these pions for interaction model studies.

The ability to constrain ``known unknowns'' is a high priority of the \dword{mpd}. One example is nucleon-nucleon correlation effects and meson exchange currents in neutrino-nucleus scattering.  Although a few theoretical models that account for these effects are available in neutrino event generators, no model reproduces well the observed data in \dword{nova}, \dword{minerva}, or \dword{t2k}.  These experiments therefore use empirical models tuned to the limited observables in their detector data.  Tuning results in better agreement between simulation and data, although still not perfect. In addition, this type of empirical tuning leaves some large uncertainties, such as the four-momentum transfer response, the neutrino energy dependence of the cross sections (where models disagree, and a ``model spread'' is typically used for the uncertainty), and the relative fractions of final state nucleon pairs ($pp$ vs. $np$).

Another example of a ``known unknown'' for which the \dword{mpd} will provide a more stringent cross section constraint than the \dword{lartpc} is the case of single and multiple pion production in \dword{cc} neutrino interactions. An \dword{mpd}-based measurement of these processes will be implemented in the DUNE \dword{lbl} oscillation analysis in the near future, making use of the high-purity samples of \dword{cc}-$0\pi$, \dword{cc}-$1\pi$, and \dword{cc}-multi-$\pi$ events in the gaseous argon, separated into $\pi^+$ and $\pi^-$ subsamples and binned in neutrino energy and other variables of interest. Figure~\ref{fig:chgpi_diffs} illustrates two simple differences among the \dword{hpgtpc} \dword{cc}-$1\pi$ subsamples; it is still to be determined which variables will be the most useful in the \dword{lbl} oscillation analysis.

The relative lack of secondary interactions for particles formed in neutrino interactions in the gaseous argon \dword{fv} will provide samples with a less model dependent connection to the particles produced in the primary interaction.  These secondary interactions are a significant effect in denser detectors \cite{Friedland:2018vry} and this crosscheck/validation of the reinteraction model is likely to be useful in understanding the full energy response of the liquid detectors.

The \dword{mpd} will  measure ratios of inclusive, semi-exclusive, and exclusive cross sections as functions of neutrino energy, where the flux cancels in the ratio. These ratios will be measured separately for \dword{nc} and \dword{cc} events, and the \dword{nc} to \dword{cc} ratio itself will be measured precisely with the \dword{mpd}.  The ratios of cross sections for different pion, proton, and kaon multiplicity will help constrain interaction models used in the near and far liquid detectors.

The \dword{hpgtpc} will have better capability than the \dword{lartpc} to distinguish among particle species at low momentum using $dE/dx$ measurements.  Some muon/pion separation is possible via $dE/dx$ for very low momenta, and protons are very easily distinguished from pions, muons, and kaons for momenta below 2~GeV/$c$, as shown in Figure~\ref{fig:ALICE_dEdx}.  At higher momenta, the magnet makes it possible to easily distinguish $\pi^+$ from $\mu^-$ (or $\pi^-$ from $\mu^+$), as has been done in T2K near-detector fits for oscillation analyses. The fact that pions will interact hadronically far less often in the gas than in the liquid will give another important handle for constraining uncertainties in the \dword{lartpc}.  These aspects give the \dword{mpd} a significant complementarity to the \dword{lartpc}, which is not magnetized.
Since the target nucleus in the \dword{mpd} is the same as that in the near and far \dwords{lartpc} this information feeds directly into the interaction model constraints without complicating nuclear physics concerns.

Finally, having a \dword{nd} that can see one level deeper than the far detector keeps open the possibility to investigate ``unknown unknowns'' as well. Since the \dword{mpd} will identify and measure interactions more accurately than can be done in the \dword{lartpc}, it will provide the ability to investigate more deeply our observations in the liquid argon, and the flexibility to address other unexpected things we may encounter. 

}

\item {{\bf Precisely and accurately measure all components of the neutrino flux}  

The magnetic field of the \dword{mpd} enables the precise determination of momenta of charged particles escaping the upstream \dword{lartpc}.  
Because the \dword{nd} is necessarily smaller than the \dword{fd}, near-far differences arising from the different containment fractions are compensated by the fact that the \dword{nd} has a magnetic spectrometer.  Also, higher-energy particles from the neutrino interaction will be measured better in the \dword{mpd} than in the liquid \dword{nd} or \dword{fd} (for example, non-contained muons), constraining the effects of energy feed-down in the liquid detectors.

The ability to separate charge signs will allow the \dword{mpd} to measure the relative contributions of $\nu_\mu$ and $\bar{\nu}_\mu$ in both the neutrino beam and the antineutrino beam, as well as distinguishing $\nu_e$ from $\bar{\nu}_e$ components.  These components are important to anchor the oscillation fit. Otherwise, reliance on the beam modeling is needed to predict the small but uncertain fractions of wrong-sign neutrinos in the beams.  Stopping muons' Michel signatures can be used on a statistical basis in the far detector, as the decay rates differ for $\mu^+$ and $\mu^-$, but that is after oscillation has distorted the spectrum.  No corresponding test is present for $\nu_e$.
}

\item {{\bf Constrain $\pi^0$ backgrounds to $\nu_e$ events}

An accurate measurement of backgrounds to the $\nu_e$ appearance measurement is a critical input for far detector oscillation analyses. In the \dword{lartpc}, the largest background to $\nu_e$'s is \dword{nc}-$\pi^0$ interactions in which one photon is not detected and the other is mistakenly identified as an electron. The \dword{hpgtpc} and \dword{ecal} together provide a unique capability to constrain \dword{nc}-$\pi^0$ backgrounds that are misidentified as $\nu_e$~\dword{cc} in the \dword{lartpc}. The \dword{hpgtpc} will collect a reduced background sample of $\sim20$k $\nu_e$~\dword{cc} events per year. The \dword{lartpc} detector measures $\nu_e$+ mis-ID'ed~$\pi^0$ events, while the \dword{mpd} measures $\nu_e$~\dword{cc} events alone (by rejecting all $\pi^0$ events using the \dword{ecal}). The \dword{mpd} sample will reduce backgrounds from \dword{nc}-$\pi^0$ events because the photon conversion length in gas is much greater than that in the liquid, and photons from $\pi^0$ decays will not often convert in the gas volume of the \dword{hpgtpc} in such a way as to fake $e^\pm$ from $\nu_e$ interactions. The \dword{ecal}, however, will have excellent ability to detect the $\pi^0$ decays, enabling the \dword{mpd} to reject $\pi^0$ events as background to $\nu_e$'s.

The \dword{mpd} measurement of $\nu_e$~\dword{cc} events can be scaled to the \dword{lartpc} density and volume and corrected to the same acceptance as the \dword{lartpc} in order to provide a constraint on the $\pi^0$-misID. The difference of the two: $(\nu_e^{{\textrm{LAr}}}$ + mis-ID'ed~$\pi^0)-(\nu_e^{\textrm{GAr-scaled-to-LAr}})$ yields the $\pi^0$-misID rate in \dword{lartpc}. This measurement of the backgrounds to $\nu_e$'s would not be possible at all if the \dword{mpd} were replaced by a simple muon range detector. It would also not be easy to extrapolate to the \dword{lartpc} if the target material of the \dword{mpd} were not argon.

}

\item{{\bf Measure energetic neutrons from $\nu$-Ar interactions via time-of-flight with the \dword{ecal}}

Neutron production in neutrino and antineutrino interactions is highly uncertain, and is a large source of neutrino energy misreconstruction. In the \dword{hpgtpc}+\dword{ecal} system, a preliminary study of the time-of-flight from the \dword{hpgtpc} neutrino interaction point to hits in the \dword{ecal} is encouraging, indicating that ToF can be used to detect and correctly identify neutrons.  With the current \dword{ecal} design, an average neutron detection efficiency of 60\% is achieved by selecting events in which an \dword{ecal} cell has one hit with more than \SI{3}{MeV}. This is still very preliminary work, and further studies to understand the impact of backgrounds and \dword{ecal} optimization (strip vs. tile, overall thickness) are underway.

}

\item{{\bf Reconstruct neutrino energy via spectrometry and calorimetry}
    
Although all neutral particles from an event must be measured with the \dword{ecal} in the \dword{mpd}, the \dword{hpgtpc} will be able to make very precise momentum measurements of charged particle tracks with a larger acceptance than the upstream \dword{lartpc}, including tracks created by high-momentum exiting particles, which allows the measurement of the entire neutrino spectrum.  In addition, short and/or stopping tracks will be measured via $dE/dx$.  The sum of this capability provides a complementary event sample to that obtained in the  \dword{lar}, whose much higher density leads to many secondary interactions for charged particles. The two methods of measurement in the \dword{mpd} will help in understanding the  \dword{lar} energy resolution.
}

\item{{\bf Constrain LArTPC detector response and selection efficiency} 
    
The \dword{mpd} will collect large amounts of data in each of the exclusive neutrino interaction channels, with the exception of $\nu-e$ elastic scattering, where the \dword{hpgtpc} sample will be too small to be useful. The high statistics $\nu$-Ar interaction samples will make it possible to directly crosscheck every kinematic distribution that will be used to constrain the fluxes and cross sections.  Typically these checks will be over an extended range of that variable.  The high purity of the \dword{mpd} samples and low detection threshold for final state particles in the \dword{hpgtpc} will give a benchmark or constraint on \dword{lartpc} detector response and selection efficiencies for each of the interaction channels.

Using the events collected in the \dword{hpgtpc} (where selection and energy reconstruction are easy), the performance of  \dword{lar} event selection and energy reconstruction metrics can be tested by simulating the well-measured \dword{hpgtpc} four-vectors in the \dword{lartpc}.  This allows the validation of the \dword{lartpc} reconstruction performance on these events. This process is expected to reduce the  errors in the \dword{lartpc} detector energy response model.

}

\end{itemize}   
\paragraph{\dword{mpd} strengths}

The strengths of the \dword{mpd} enable it to reach the goals above and to augment the capabilities of the \dword{lartpc}. Below are a few examples of its strengths relative to the \dword{lartpc}:
\begin{itemize}
     \item High-fidelity particle charge determination via magnetic curvature. This is the only detector that can measure electron and positron charge.
     
     \item Precise and independent measurement of particle momentum, via magnetic curvature, will allow the measurement of the momentum of higher-energy charged particles without requiring containment. This extends the utility of the  \dword{nd}, especially for the high-energy beam tune. The absolute momentum scale is easily calibrated in the magnetic spectrometer and provides a cross-check on energy loss through ionization measurements.  Calibration strategies for the magnetic tracking include pre-assembly field maps, {\it in situ} NMR probes, and $K^0_s$ and $\Lambda\!^0$ reconstruction.
     
      \item Particle identification through $dE/dx$. The gaseous argon TPC has better separation power of particle species by $dE/dx$ than the liquid because the momentum can be measured along with energy loss. 
      
      \item High-resolution imaging of particles emerging from the $\nu$-Ar vertex (including nucleons).  The reconstruction threshold in the gas phase is much lower than the threshold in liquid because particles travel further in the low density medium, e.g., a proton requires only \SI{3.7}{MeV} kinetic energy to make a \SI{2}{cm} track in 10~atmospheres of gaseous argon, while a \SI{3.7}{MeV} proton in liquid can only travel \SI{0.02}{cm}. Figure~\ref{fig:LArGArThresholds} demonstrates the difference in the thresholds for reconstructing protons in the \dword{hpgtpc} and the \dword{lartpc} in light of the energy spectra of final state protons from a selection of types of neutrino interactions at the DUNE \dword{nd}. The \dword{lartpc} threshold is what has been achieved in \dword{microboone} up to now, and the \dword{hpgtpc} threshold is what has been achieved with the tools discussed in Section~\ref{sec:TPC_ML}.

    \item Separation of tracks and showers for less-ambiguous reconstruction.  Particle tracks are locally helical and tend to bend away from each other in the magnetic field as they travel from a dense vertex.  Electromagnetic showers do not occur in the gas, but in the physically separate \dword{ecal}.  By contrast, in a \dword{lartpc} tracks and showers frequently overlap.  The measurement resolution scales are comparable between the liquid and the gas, but the distance scales on which interactions occur are much longer in the gas, allowing particles to be identified and measured separately more easily.
   
    \item The measurement of energetic neutrons through time-of-flight with \dword{ecal} is a potential game-changer for validating energy reconstruction. Preliminary studies of the \dword{hpgtpc}+\dword{ecal} system indicate that an average neutron detection efficiency of 60\% can be achieved via a time-of-flight analysis. A study of the impact of backgrounds is underway, but initial studies are encouraging.

    \item{The \dword{hpgtpc} is able to measure the momentum of particles over almost the full solid angle.  Particles that are emitted at a large angle with respect to the beam have a high probability of exiting the \dword{lar} without leaving a matching track in the \dword{mpd}. However, events collected in the \dword{hpgtpc}, with its $\simeq 4\pi$ coverage, can be used in the regions of phase space where the exiting fraction is high in the liquid argon to ensure that the events are accurately modeled in all directions in the \dword{fd}. }
    
    \item{The \dword{mpd} neutrino event sample, while smaller than the \dword{lartpc} sample, is a statistically independent sample. Moreover, the systematic uncertainties of the \dword{mpd} will be very different than the \dword{lartpc} and likely smaller in many cases. This will allow the \dword{mpd} to act as a systematics constraint for the \dword{lartpc}.
    }
\end{itemize}

\begin{dunefigure}[Reconstructed $\nu$ energy spectra for \dshort{cc} $\nu_{\mu}$ interactions with charged pions]{fig:chgpi_diffs}
{Representative differences among subsamples of \dword{cc} $\nu_{\mu}$ interactions with one $\pi^+$ (solid lines) and \dword{cc} $\bar{\nu}_{\mu}$ interactions with one $\pi^-$ (dashed lines). The forward- and reverse- horn current samples are shown in black and red, respectively. Left: Reconstructed neutrino energy spectra, normalized to the same number of protons on target. Right: Angle of outgoing muon relative to neutrino direction, normalized to unit area for shape comparison.}
    \includegraphics[width=0.49\textwidth]{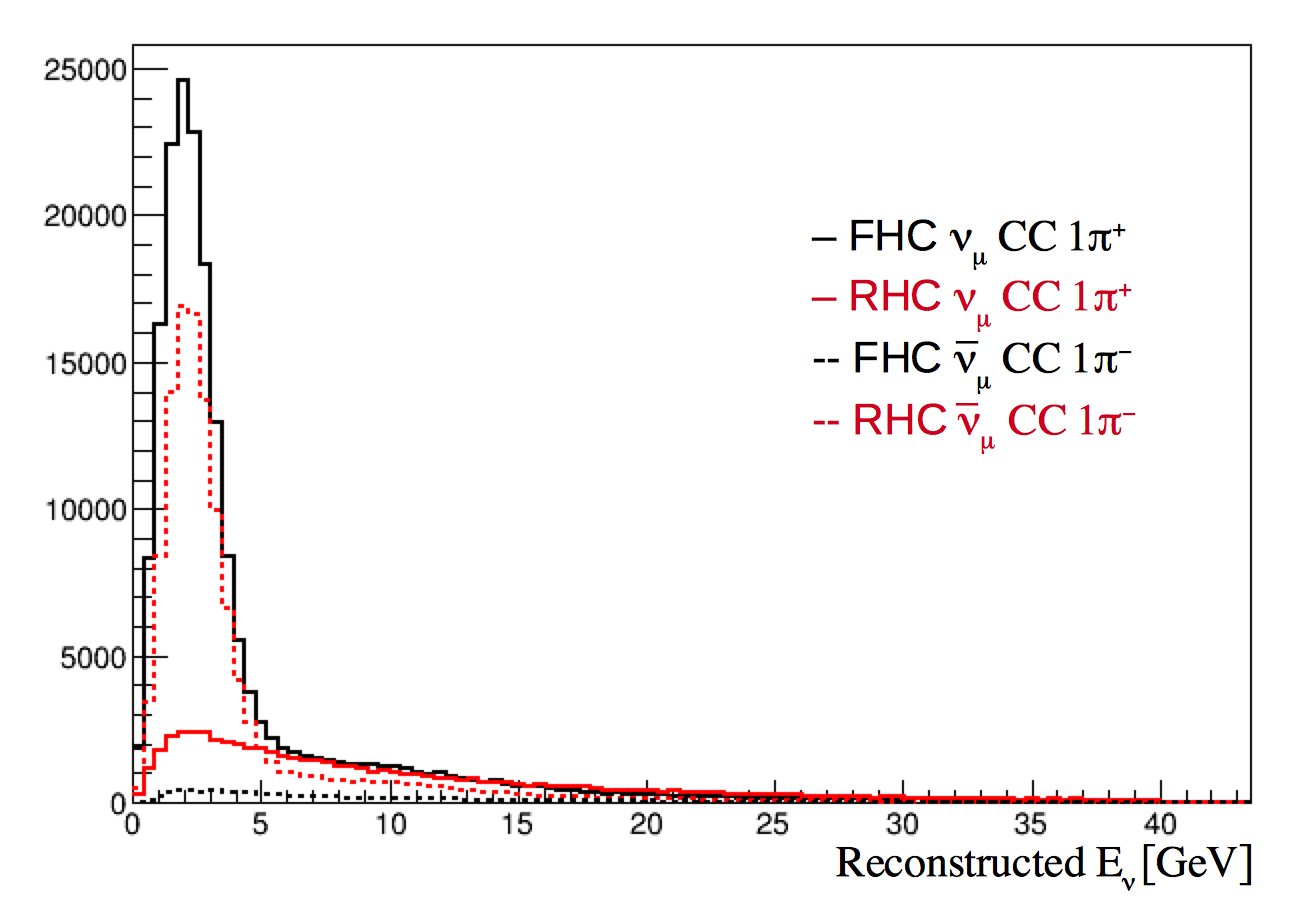}
    \includegraphics[width=0.49\textwidth]{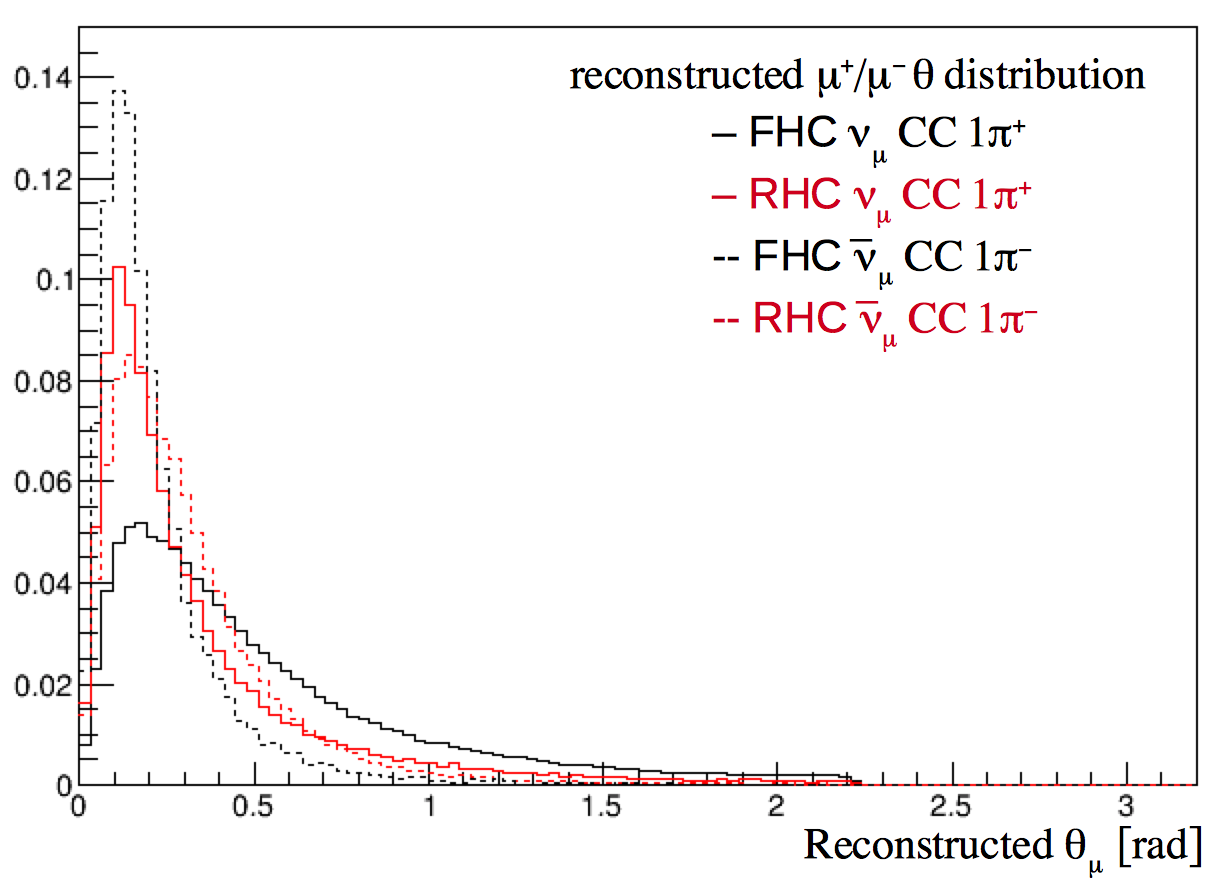}
\end{dunefigure}

\begin{dunefigure}[Momentum spectra of protons ejected from $\nu$ interactions in Ar] 
{fig:LArGArThresholds}
{The momentum spectra of protons ejected from neutrino interactions in argon, for several categories of interaction types. The vertical lines indicate the lowest momentum protons that have been reconstructed using existing automated reconstruction tools, where the dotted line is the \dword{hpgtpc} threshold, and the solid line is the \dword{lartpc} threshold (from \dword{microboone}).}
    \includegraphics[width=0.6\textwidth]{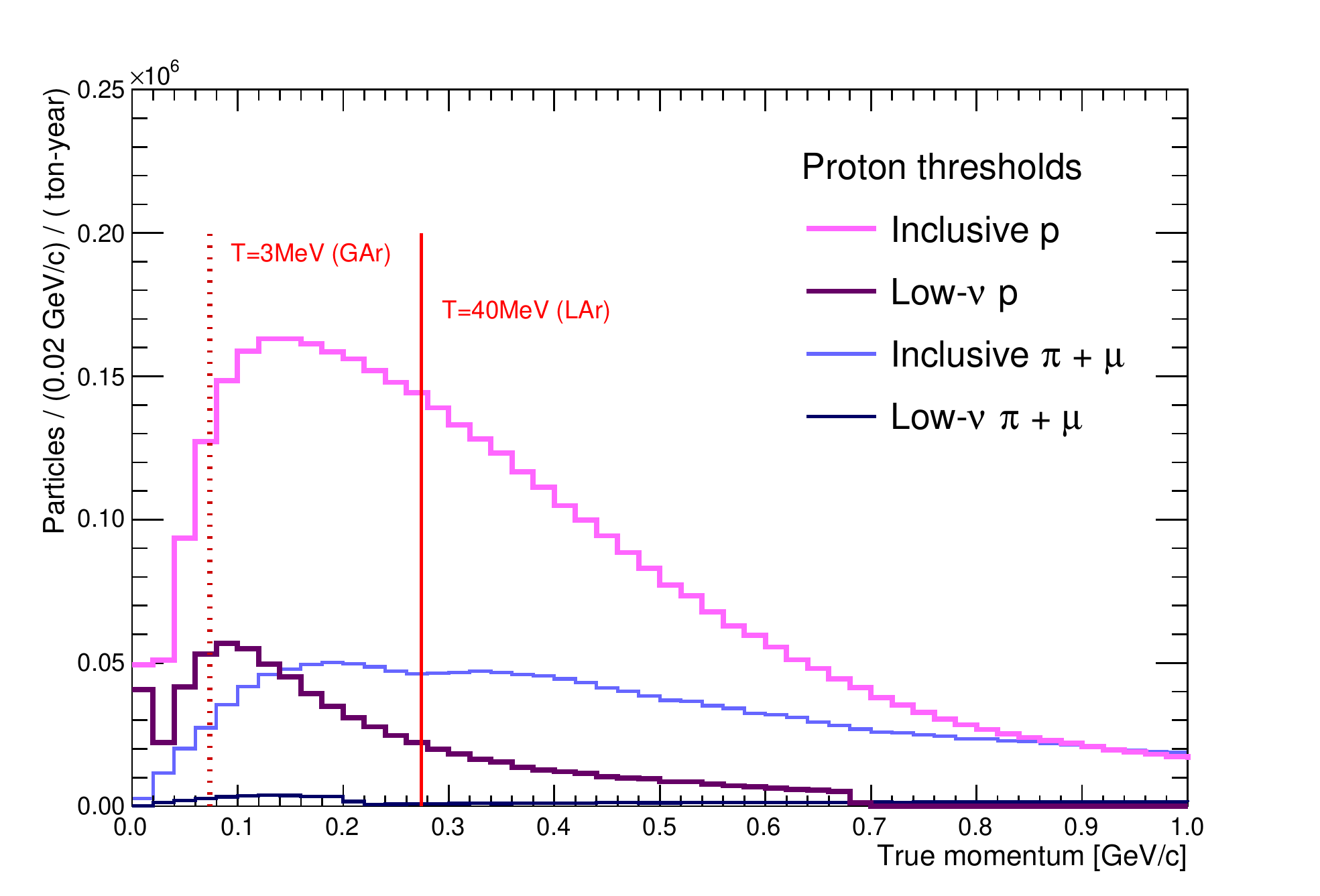}
\end{dunefigure}


\subsubsection{\dshort{mpd} Technical Details}
\subsubsubsection{High-Pressure Gaseous Argon TPC}
The basic geometry of the \dword{hpgtpc} is a gas-filled cylinder with a \dword{hv} electrode at its mid-plane, providing the drift field for ionization electrons. It is oriented inside the magnet such that the magnetic and \efield{}s are parallel, reducing transverse diffusion to give better point resolution. Primary ionization electrons drift to the end plates of the cylinder, which are instrumented with multi-wire proportional chambers to initiate avalanches (gas gain) at the anode wires.  Signals proportional to the avalanches are induced on cathode pads situated behind the wires; readout of the induced pad signals provides the hit coordinates in two dimensions.  The drift time provides the third coordinate of the hit.

The details of the \dword{hpgtpc} design will be based closely on the design of the ALICE detector~\cite{Dellacasa:2000bm} shown in Figure~\ref{fig:ALICETPC}. Two readout planes sandwich a central \dword{hv} electrode (25$\,\mu$m of aluminized mylar) at \dword{hv} that generates the drift field, which is parallel to a \SI{0.5}{T} magnetic field. On each side of the electrode, primary ionization electrons drift up to \SI{2.5}{m} to reach the endplates, which are segmented azimuthally into 18 trapezoidal regions instrumented with \dwords{roc} that consist of \dword{mwpc} amplification regions and cathode pad planes to read out the signals. A cross sectional view of an ALICE MWPC-based \dword{roc} is shown in Figure~\ref{fig:ALICE_ROC_MWPC}. The \dwords{roc} are built in two sizes: a smaller \dword{iroc} and a larger \dword{oroc}. The trapezoidal segments of the endplates are divided radially into inner and outer sections, and the \dwords{iroc} and \dwords{oroc} are installed in those sections. The existing \dwords{iroc} and \dwords{oroc} will become available in 2019, when they are scheduled to be replaced by new GEM-based \dwords{roc} for upgraded pileup capability in the high rate environment of the \dword{lhc}. For the DUNE \dword{hpgtpc}, the existing \dwords{roc} are more than capable of providing the necessary performance in a neutrino beam.

\begin{dunefigure}[The ALICE TPC]{fig:ALICETPC}
{Diagram of the ALICE \dword{tpc}, from Ref.~\cite{Alme:2010ke}. The drift \dword{hv} cathode is located at the center of the \dword{tpc}, defining two drift volumes, each with \SI{2.5}{m} of drift along the axis of the cylinder toward the endplate. The endplates are divided into 18 sectors, and each endplate holds 36 readout chambers.}
    \includegraphics[width=0.7\textwidth]{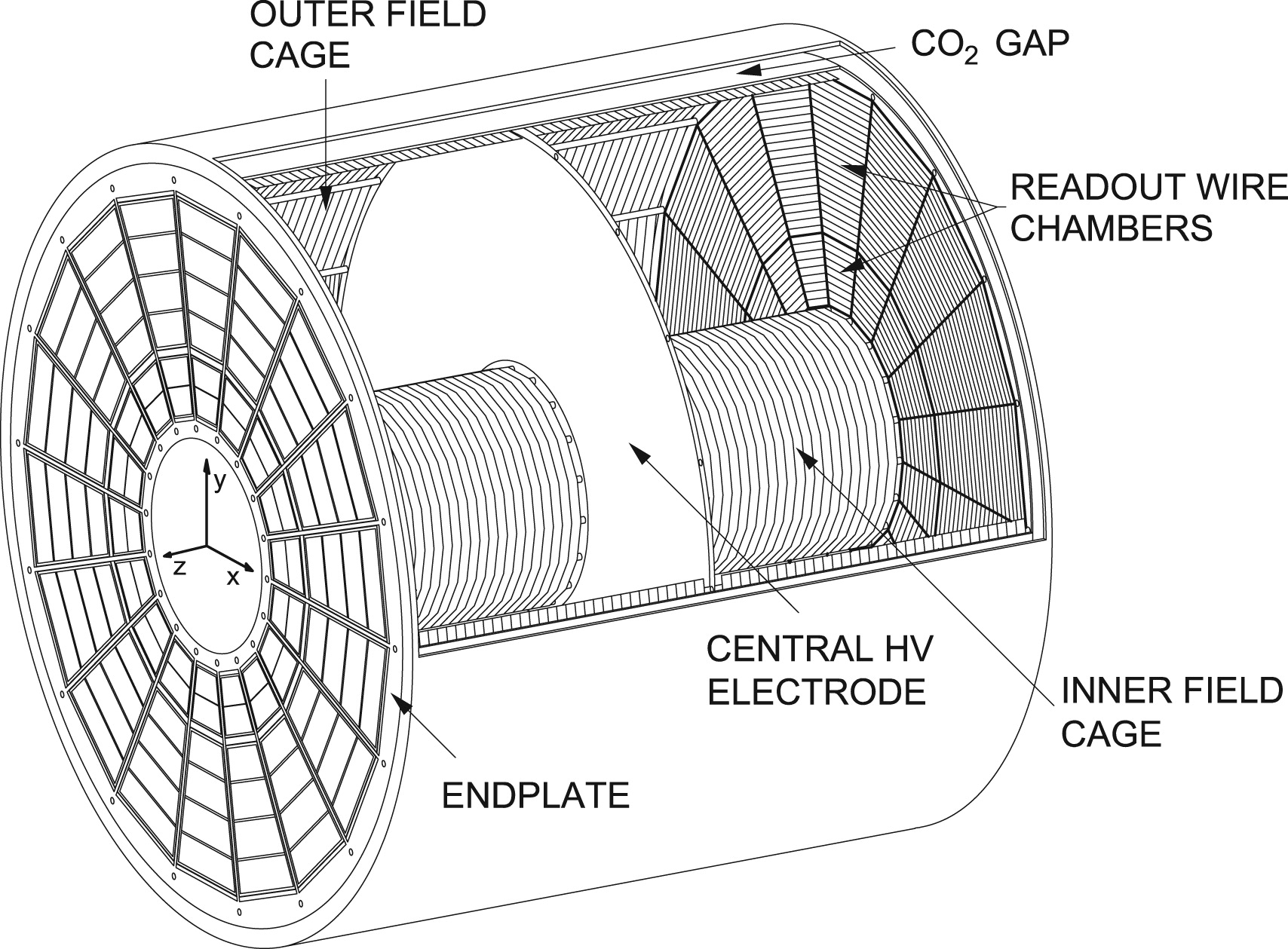}
\end{dunefigure}

\begin{dunefigure}[The ALICE MWPC-based ROC with pad plane readout]{fig:ALICE_ROC_MWPC}
{Schematic diagram of the ALICE MWPC-based \dword{roc} with pad plane readout, from Ref.~\cite{Alme:2010ke}.}
    \includegraphics{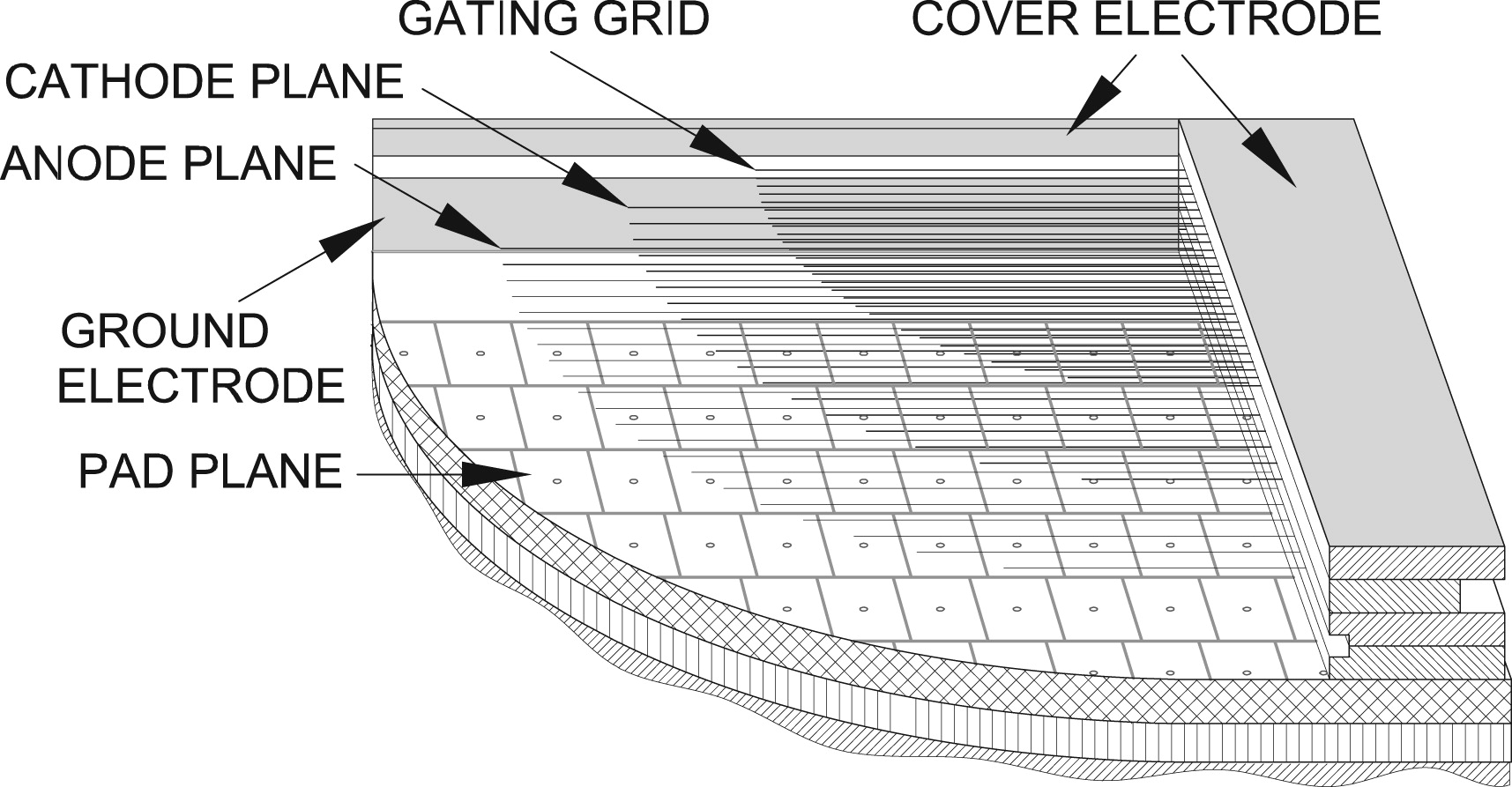}
\end{dunefigure}

In the ALICE design, the innermost barrel region was isolated from the \dword{tpc} and instrumented with a silicon-based inner tracker; for the DUNE \dword{hpgtpc}, the inner field cage labeled in Figure~\ref{fig:ALICETPC} will be removed, and the entire inner region will be combined to make a single gas volume for the \dword{tpc}. New \dwords{roc} will be built to fill in the central uninstrumented region, which is \SI{1.6}{m} in diameter, left by reusing the existing ALICE chambers.  The active dimensions of the \dword{hpgtpc} will be \SI{5.2}{m} in diameter and \SI{5}{m} long, which yields an active mass of $\simeq$ \SI{1.8}{t}.

\paragraph{\dword{mpd} Pressure Vessel}\label{sec:TPC_PV}

The preliminary design of the pressure vessel, presented in Figure~\ref{fig:TPC_PV}, accounts for the additional volume needed to accommodate the TPC field cage, the \dword{roc} support structure, \dword{fe} electronics, and possibly part of the \dword{ecal}.

The pressure vessel can be fabricated from aluminum or stainless steel, has a cylindrical section that is 6~m in diameter and \SI{6}{m} long and utilizes two semi-elliptical end pieces with flanges. The walls of the cylinder barrel section are $\simeq$~1.6X$_0$ in thickness in the case of stainless steel or $\simeq$ 0.3X$_0$ in the case of Al 5083.  Further reduction of the thickness in radiation lengths can be accomplished  with the addition of stiffening rings.   This preliminary design includes two flanged endcaps.  However, these large flanges are the cost driver for the pressure vessel and, therefore, vessel designs with a single flange will also be considered.  As an example, DOE/NETL-2002/1169 (Process Equipment Cost Estimation
Final Report) indicates that a horizontal pressure vessel of the size indicated here and rated at \SI{1034}{kPag} (\SI{150}{psig}) (approximately 10~atmospheres) is costed at \$150k ($\simeq$ \$210k in 2019 dollars).  The budgetary estimate for a vessel with two flanges was \$1.2M with the flanges driving the cost.  DOE/NETL-2002/1169 also indicates that pressure is not a significant cost driver.  Reducing the pressure from \SI{1034}{kPag} to \SI{103}{kPag} (\SI{15}{psig}) only reduces the basic (\$150k) vessel cost by a factor of two.


\begin{dunefigure}[Pressure vessel preliminary design]{fig:TPC_PV}
{Pressure vessel preliminary design.}
\includegraphics[width=0.6\textwidth]{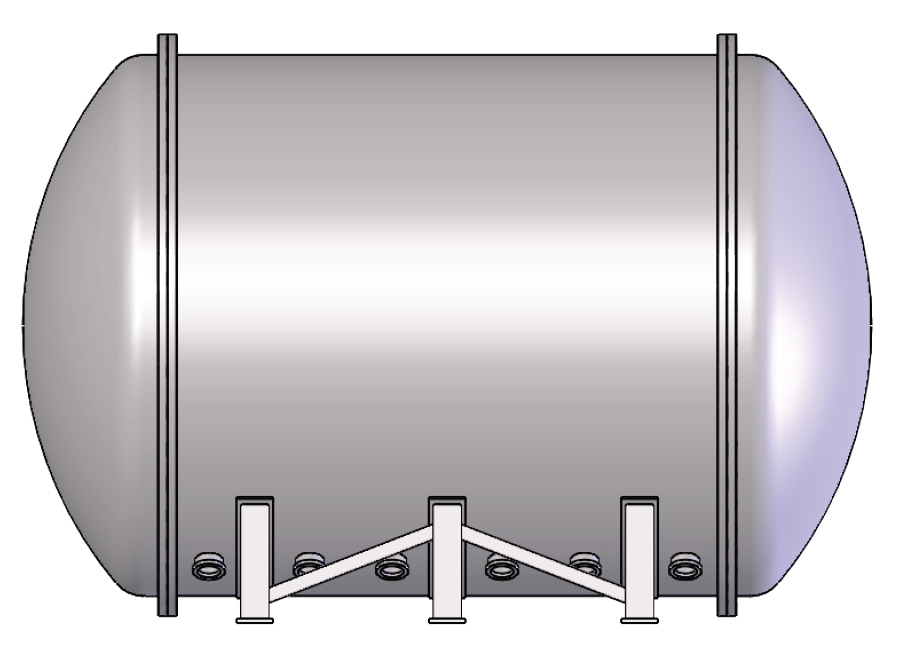}
\end{dunefigure}

\subsubsubsection{Electromagnetic Calorimeter (\dshort{ecal})}

The \dword{mpd} \dword{ecal} concept is based on a high granularity calorimeter to provide direction information in addition to the energy measurement of electromagnetic showers and an efficient rejection of background. The principal role of the \dword{ecal} is to reconstruct photons directly produced in neutrino interactions and originating from $\pi^0$ decays, providing a measurement of the photon's energy and direction to enable the association of photons to interactions observed in the \dword{hpgtpc} and the determination of the decay vertex of the $\pi^0$s. In the case of $\nu_e$ measurements in the \dword{hpgtpc}, the \dword{ecal} will play an important role in rejecting events with $\pi^0$ decays, which represent a background to $\nu_e$ interactions in the \dword{lartpc}. The \dword{ecal} can also be used to reject external backgrounds, such as rock neutrons and muons, providing a sub-nanosecond timestamp \cite{Simon:2013zya} for each hit in the detector. As the \dword{ecal} uses hydrogen-rich scintillator, it is assumed to have capabilities to provide neutron detection, and studies are underway to determine the performance of neutron detection.

\paragraph{\dword{ecal} Design}

The \dword{ecal} design is inspired by the design of the CALICE analog hadron calorimeter (AHCAL) \cite{collaboration:2010hb}. 

\begin{dunefigure}[\dshort{mpd} \dshort{ecal} conceptual design]{fig:ConceptDesign_NDECAL}
{On the left, the conceptual design of the \dword{mpd} system for the \dword{nd}. The \dword{ecal} (orange) is located outside the \dword{hpgtpc} pressure vessel. On the right, a conceptual design of the \dword{ecal} endcap system.}
\includegraphics[width=0.45\textwidth]{graphics/MPDdrawing.jpg}
\includegraphics[width=0.42\textwidth]{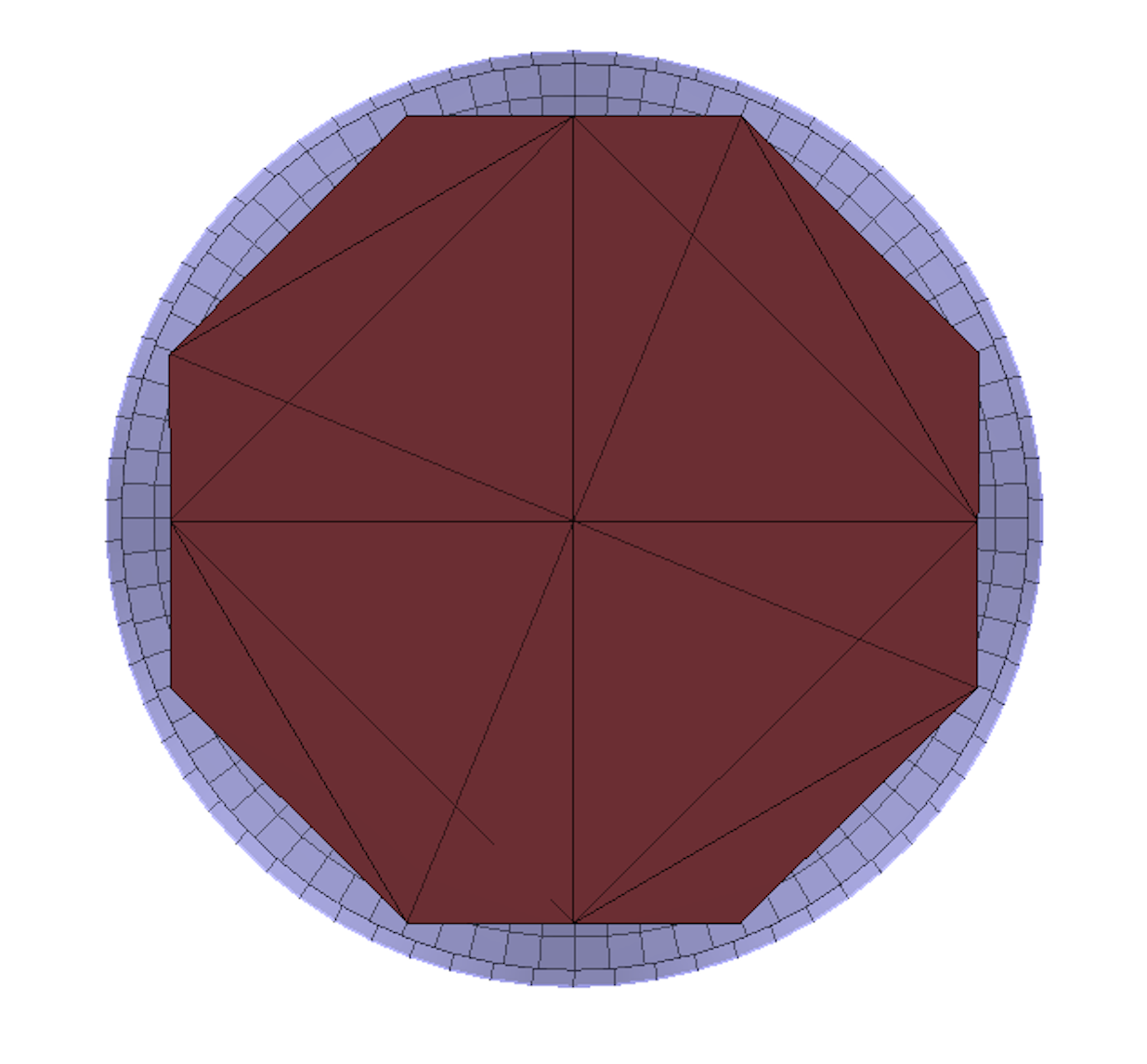}
\end{dunefigure}
The \dword{ecal} is shown in Figure~\ref{fig:ConceptDesign_NDECAL}.  The barrel has an octagonal shape with each quadrant composed of several trapezoidal modules. The \dword{ecal} endcap has a similar design providing hermeticity and a large solid-angle coverage. Each module consists of scintillating layers of polystyrene as active material read out by \dwords{sipm}, sandwiched between absorber sheets. The scintillating layers consist of a mix of tiles with dimensions between $2\times2$ cm$^2$ to $3\times3$ cm$^2$ (see Figure~\ref{fig:ConceptTile_NDECAL}) and cross-strips with embedded wavelength shifting fibers to achieve a comparable effective granularity. The high-granularity layers are concentrated in the front part of the detector, since that has been shown to be the most relevant factor for the angular resolution \cite{Emberger:2018pgr}. With the current design, the number of channels is of the order 2.5 to 3 million. A first design of the \dword{ecal} and the simulated performance has already been studied in \cite{Emberger:2018pgr}.

\begin{dunefigure}[Conceptual layout of the \dshort{mpd} \dshort{ecal}] 
{fig:ConceptTile_NDECAL}
{Conceptual layout of the \dword{ecal} showing the absorber structure, scintillator tiles, \dwords{sipm} and \dword{pcb}.}
\includegraphics[width=0.8\textwidth]{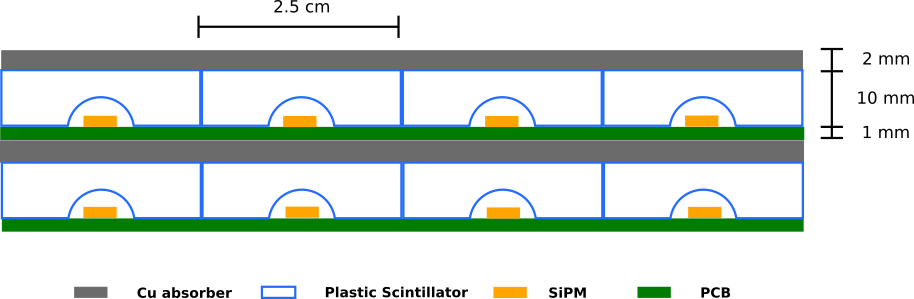}
\end{dunefigure}

In the preliminary design, it was assumed that the full \dword{ecal} barrel is outside the pressure vessel.  The thickness of the pressure vessel has an impact on the calorimeter energy resolution \cite{Emberger:2018pgr}, and more recent designs of the pressure vessel have reduced its thickness.
Currently, the \dword{ecal} design is undergoing a detailed design study in order to further optimize the detector design, cost, and performance.  
\subsubsubsection{Magnet}
\label{sssec:nd:appx:mpd-magnet}
Two magnet designs are under consideration to house the \dword{hpgtpc} and the \dword{ecal}. One is a \dword{ua1}-type conventional electromagnet, the other is based on a superconducting Helmholtz-coil-like design. The common requirement is a central magnetic field of 0.5\,T with $\pm$20\% uniformity over the TPC volume (5\,m long and 5\,m in diameter). With the current design of the access shaft (11.8\,m diameter), the clear diameter is about 7.8\,m. Recent studies for the construction of an electromagnet similar to the \dword{ua1} magnet predict that the cost of the design, procurement, infrastructure (power and cooling) and assembly will be in excess of \$20 million, with operation costs of approximately \$1.6M per year of running.  Because of this, the main focus has been on the superconducting design.
\paragraph{Superconducting Magnet}
The SC magnet design is a Helmholtz-coil-like configuration, air core,  five coil magnet system. Three central coils produce the analyzing field and two outer shielding coils help contain stray field. The advantage of this design is that little or no iron is used for field containment or shaping. This eliminates background coming from neutrino interactions in the iron, which for the normal-conducting magnet case is the largest component of the background. Figure~\ref{fig:dune_nd_magnet_sc_layout} shows the magnet concept indicating the five-coil arrangement and support structure. 

\begin{dunefigure}[Helmholz coil arrangement]{fig:dune_nd_magnet_sc_layout}
{Helmholz coil arrangement for \dword{mpd} superconducting magnet.}
\includegraphics[width=0.95\textwidth]{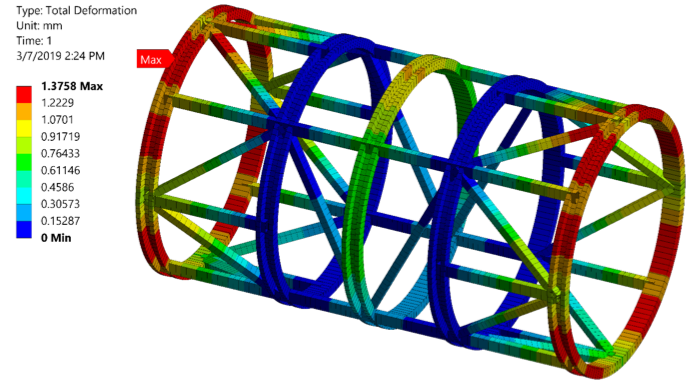}
\end{dunefigure}
All five coils have the same inner radius of \SI{3.5}{m}. The center and shielding coils are identical with the same number of ampere-turns. The side coils are placed at \SI{2.5}{m}, the shielding coils at \SI{6}{m} from the magnet center along $z$.  The case where the shielding coils are at \SI{5}{m} from the magnet center so that the magnet system would be the same width as the  \dword{lar} detector is also being examined.  The magnet system will have a stored energy of about \SI{110}{MJ}, using a conventional NbTi superconducting cable design, a \dword{ssc}-type Rutherford cable soldered in a copper channel with a 50\% margin. All coils should be wired in series to reduce imbalanced forces during a possible quench. Small transverse centering force components are possible due to coil de-centering from mechanical errors. 
Shown in Figure~\ref{fig:dune_nd_magnet_sc_fieldmap} is the field along the $z$-axis at different radii. The peak field in the coils is \SI{2.14}{T} (center), \SI{5}{T} (side) and \SI{2.03}{T} (shield). The resulting forces are only along the $z$-axis, $F_{z}$ is \SI{0.0}{MN} (center), \SI{-6.81}{MN} (side) and \SI{2.2}{MN} (shield). The fringe field at the shielding coil is rather large but can be reduced further; more studies will be needed. There is a preliminary mechanical support design. A first glimpse at the radiative heat load assumes a coil and support surface of 180\,m$^{2}$, resulting in a load of \SI{5.4}{W} from \SI{77}{K} to \SI{4.5}{K}. The coil support and leads will likely have a much larger contribution (power leads usually have \SI{15}{W} for \SI{10}{kA}). With a mass of \SI{42}{t} the magnets are in some aspects similar to the \dword{mu2e} solenoids.

\begin{dunefigure}[Field map of the superconducting magnet along the $z$ axis]{fig:dune_nd_magnet_sc_fieldmap}
{Field map of the superconducting magnet along the $z$ axis. The colors represent different radii from the center line.}
\includegraphics[width=0.70\textwidth]{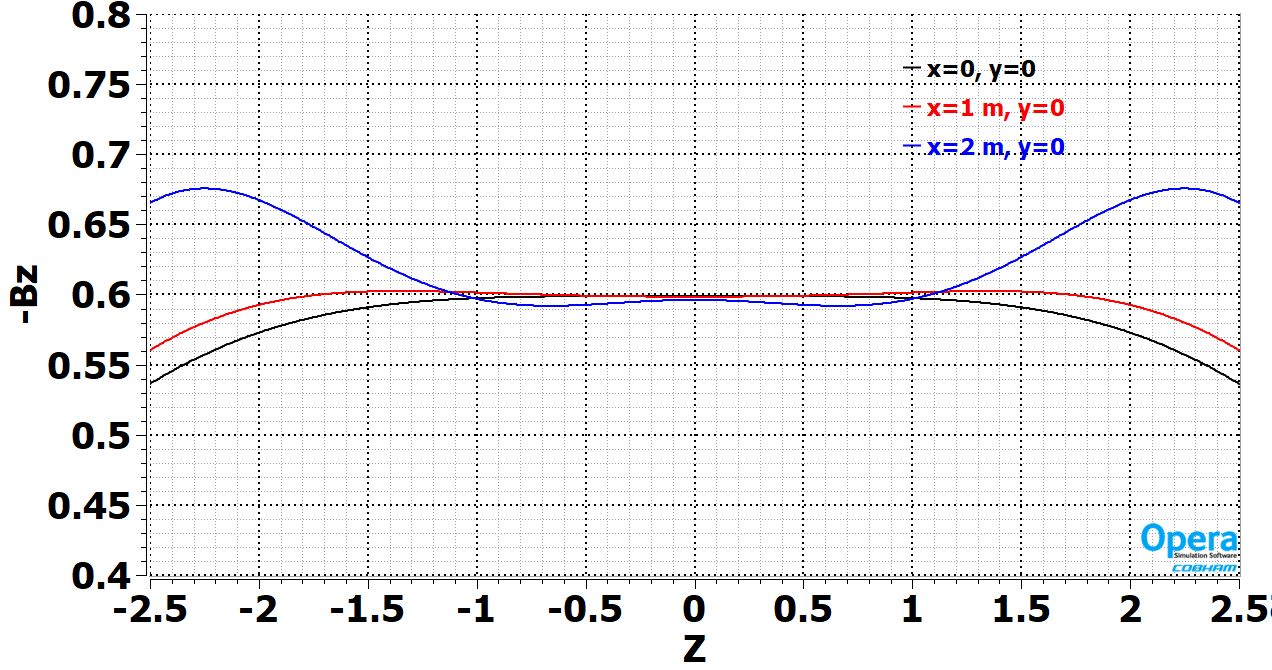} 
\end{dunefigure}
%
%
\paragraph{Normal Conducting Magnet}

Although the SC magnet design is the favored option, the normal conducting magnet design produced for the LBNE CDR is also being revised and studied.  Due to the cylindrical geometry imposed by the \dword{hpgtpc}, a cylindrical coil design for the normal conducting magnet is the baseline. The cooling requirement of the coil is approximately \SI{3.5}{MW} and involves a substantial cooling water flow. A thermal shield between the coils and the detector volume is required in order to minimize heat flow to the \dword{hpgtpc} and the \dword{ecal}. The coil thickness becomes excessive (in order to maintain a maximum 5$^\circ$ C temperature in the coil) if the thermal shield is not used.  The shield does take up space in the magnet volume, however.  
Note: the iron end-walls will most likely not be needed. The estimated magnet weight is well over 1\,kt, and this mass provides a significant source of background for the high pressure gas TPC and, perhaps, the  \dword{lar}.  There is a significant amount of material between the \dword{lartpc} and the \dword{hpgtpc} in the \dword{mpd} in this configuration, which will affect the acceptance for muons emanating from events in the  \dword{lar}. This option will continue to be studied as part of the optimization process.

\subsubsubsection{Size optimization}
    
 The process of optimizing the design of the \dword{mpd} is in progress. One of the more critical issues is the size of the \dword{mpd}.  This is an important factor in the angular acceptance of particles exiting the upstream \dword{lartpc}. A preliminary study of geometries shows that reducing the \dword{hpgtpc} diameter by more than 1 meter, or reducing the length by more than 1.5 meters would have significant consequences on the acceptance. Reducing the \dword{hpgtpc} diameter from its nominal 5~meters to a slightly smaller 4.5~meters while increasing its length in the direction transverse to the neutrino beam improves acceptance, since the \dword{hpgtpc} would better match the 7-meter width of the \dword{lartpc} in the transverse direction. It should be noted, however, that reducing the diameter may actually result in a higher-cost \dword{mpd}, since the ALICE TPC readout chambers would not be used in the configuration for which they were designed. Increasing the length of the \dword{hpgtpc} is feasible, but will require additional studies of high voltage stability in the gas, since HV breakdown in gas is proportional to the pressure (in the absence of field enhancements).  The \dword{hpgtpc} operating pressure will be nominally 10 times that of ALICE, so extending the drift distance from 2.5 meters to 3 meters while keeping the same drift velocity will require raising the drift HV by approximately 20~kV.

\subsubsection{\dshort{mpd} performance}
The expected performance of the \dword{mpd} is summarized in Table~\ref{tab:TPCperformance}. Details of the \dword{hpgtpc} performance are based upon experience from operation of the PEP-4~\cite{PEP4_results_Layter,PEP4_Stork,Madaras:1982cj} and ALICE~\cite{Alessandro:2006yt} time projection chambers. Performance of the \dword{ecal} is based on experience from operation of similar \dwords{ecal} and on simulations. 

\begin{dunetable}[\dshort{mpd} performance parameters]{l|c|c}{tab:TPCperformance}{Expected \dword{mpd} performance, extrapolated from \dword{alice}}
Parameter	               & Value	                      & units \\ \toprowrule
$\sigma_x$ 		           & 250	                      & $\mu$m\\ \colhline
$\sigma_y$ 		           & 250	                      & $\mu$m\\ \colhline
$\sigma_z$ 		           & 1500	                      & $\mu$m\\ \colhline
$\sigma_{r\phi}$ 	       & <1000	                      & $\mu$m\\ \colhline
Two-track separation       & 1		                      & cm \\ \colhline
Angular resolution	       & 2-4	                      & mrad \\ \colhline
$\sigma$($dE/dx$)		       & 5		                      & \% \\ \colhline
$\sigma_{p_T}/p_T$	       & 0.7	                      & \% (10-1 GeV/c)\\ \colhline
$\sigma_{p_T}/p_T$	       & 1-2	                      & \% (\SIrange{1}{0.1}{GeV/c})\\ \colhline
Energy scale uncertainty    & $\lesssim$ 1              & \% (dominated by $\delta_p/p$) \\ \colhline
Charged particle detection thresh. & 5                    & MeV (K.E.)\\ \colhline
ECAL energy resolution	           & 5-7/$\sqrt{E/{\rm{GeV}}}$	  & \% \\ \colhline
ECAL pointing resolution &  $\simeq 6$ at 500 MeV         & degrees\\
\end{dunetable}

\subsubsubsection{Track Reconstruction and Particle Identification}

The combination of very high resolution magnetic analysis and superb particle identification from the \dword{hpgtpc}, coupled with a high-performance \dword{ecal} will lead to excellent event reconstruction capabilities and potent tools to use in neutrino event analysis.  
As an example of this capability, the top panel of Figure~\ref{fig:GAr} shows a $\nu_e + {}^{(N)}Ar \xrightarrow{} e^- + \pi^+ +n + {}^{(N-1)}Ar$ in the \dword{hpgtpc} with automatically-reconstructed tracks.  The same event was simulated in a \dword{fd} \dword{spmod}, 
and is shown in the bottom panel of Figure~\ref{fig:GAr}.

\begin{dunefigure}[Track-reconstructed $\nu_e$ \dshort{cc} event in the \dshort{hpgtpc}]{fig:GAr}
{(Top) Track-reconstructed $\nu_e$ \dword{cc} event in the \dword{hpgtpc}, simulated and reconstructed with GArSoft.  The annotations are from \dword{mc} truth. (Bottom) The same $\nu_e$ \dword{cc} event, but simulated in a \dword{spmod} using \dword{larsoft}.  The topmost blue panel shows the collection-plane view, the middle blue panel shows the $V$ view, and the bottom blue panel shows the $U$ view.  Wire number increases on the horizontal axes and sample time along the vertical axes.   The wire number in the collection view is labeled on the top of the panel, while the $V$ and $U$ wire numbers are below their respective panels. Simulated \dword{adc} values are indicated by the colors.  The curve in the bottom-most panel is a simulated waveform from a collection-plane wire.  The annotations are from \dword{mc} truth.}
    \includegraphics[width=0.99\textwidth]{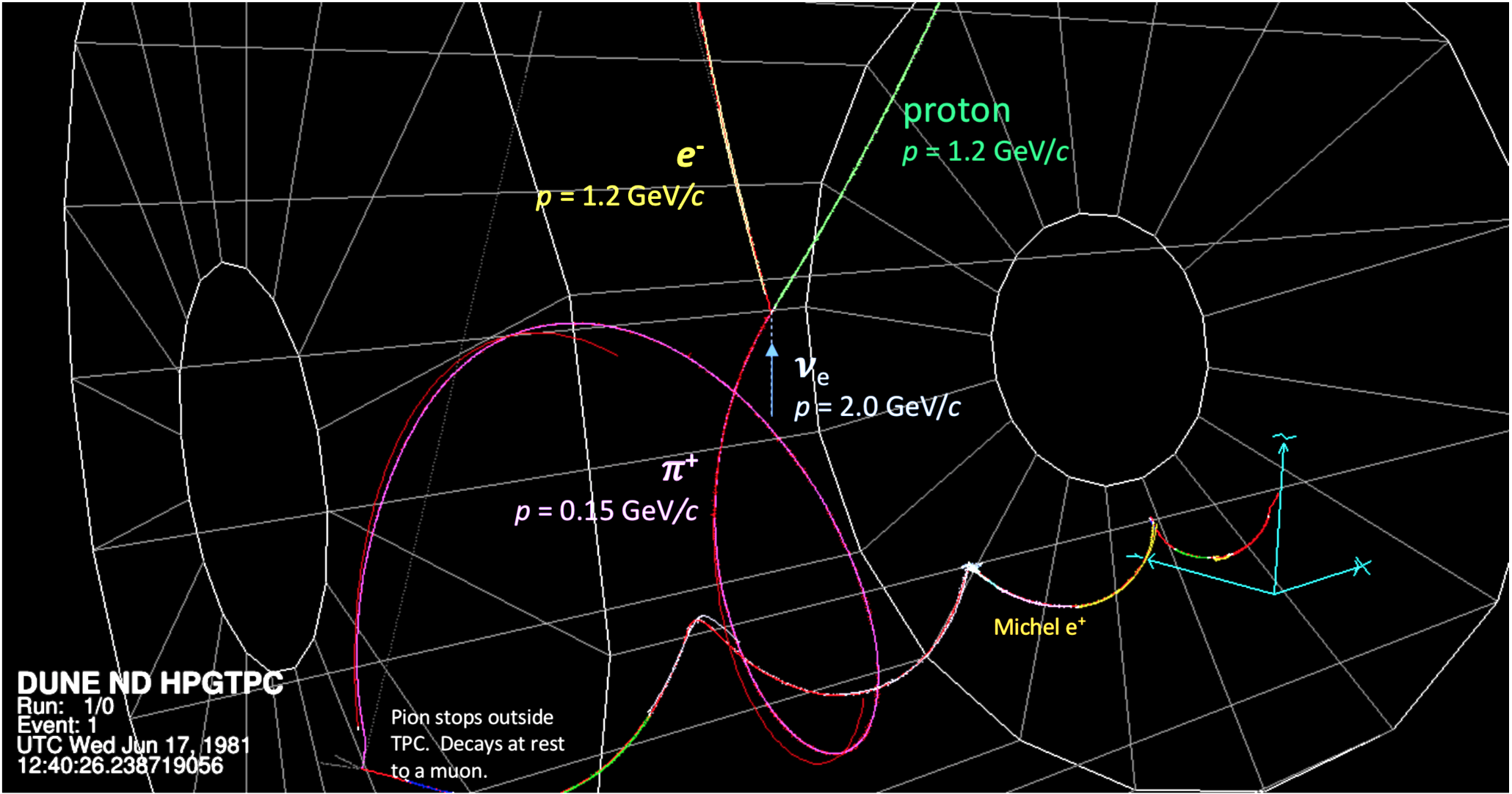}
    \includegraphics[width=0.99\textwidth]{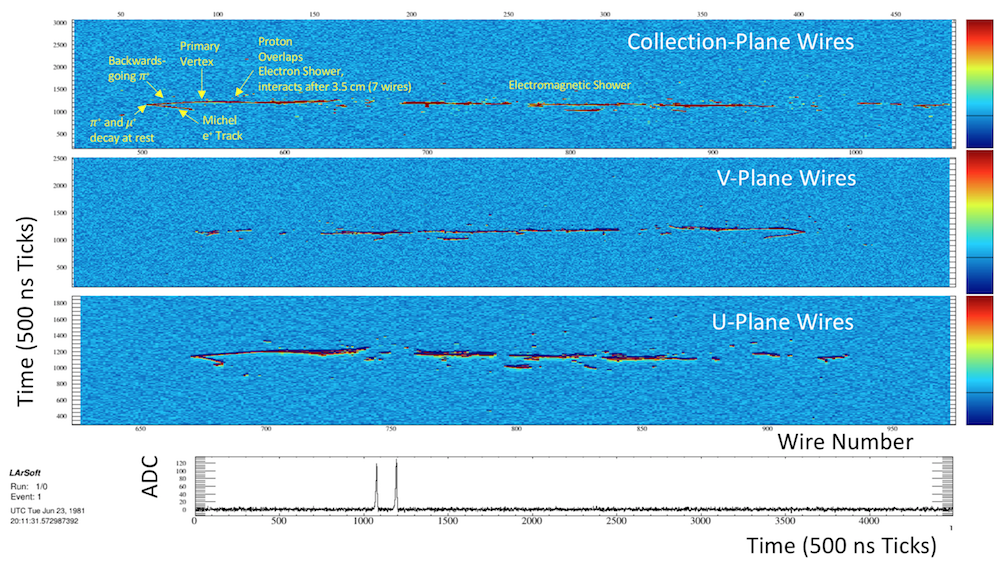}
\end{dunefigure}

Since important components of the hardware and design for the \dword{hpgtpc} are taken from or duplicated from the ALICE detector, the ALICE reconstruction is a useful point of reference in this discussion.
Track reconstruction in ALICE is achieved by combining hits recorded on the ROC pads into tracks following a trajectory that a charged particle traveled through the TPC drift volume.  The \dword{hpgtpc} is oriented so that the neutrino beam is perpendicular to the magnetic field, which is the most favorable orientation for measuring charged particles traveling along the neutrino beam direction.   

The GArSoft simulation and reconstruction package borrows heavily from  \dword{lartpc}, and is based on the {\it art} event processing framework and {\tt GEANT4}.  It is designed to be able to reconstruct tracks with a full $4\pi$ acceptance.   GArSoft simulates a 10~atmosphere gaseous argon detector with readout chambers filling in the central holes in the ALICE geometry.  GArSoft's tracking efficiency has been evaluated in a large sample of \dword{genie} $\nu_\mu$ events interacting in the TPC gas at least 40 cm from the edges, generated using the optimized \dword{lbnf} forward horn current beam spectra. The efficiency
for reconstructing tracks associated with pions and muons as a function of track momentum $p$ is shown in  Figure~\ref{fig:garsoft_efficiency}.  The efficiency is above 90\% for tracks with $p>40$~MeV/$c$, and it steadily rises with increasing momentum.  

Also shown is the efficiency for reconstructing all charged particles with $p>\SI{200}{MeV/c}$ as a function of $\lambda$,  the track angle with respect to the center plane.  The tracking efficiency for protons is shown in Figure~\ref{fig:TEpr} as a function of kinetic energy, $T_p$.  Currently, the tracking works well down to $T_p \simeq \SI{20}{MeV}$. For $T_p < \SI{20}{MeV}$, a machine-learning algorithm is in development, targeting short tracks near the primary vertex. This algorithm, although currently in a very early stage of development, is already showing good performance, and efficiency improvements are expected with more development. The machine learning algorithm is described in Section~\ref{sec:TPC_ML}.

The ALICE detector, as it runs at the LHC, typically operates with particle densities ranging from 2000 to 8000 charged particles per unit rapidity ($dN/dy$) for central Pb-Pb interactions~\cite{Cheshkov:2006ym}. The expected particle densities in the DUNE neutrino beam will be much lower and less of a challenge for the reconstruction.

\begin{dunefigure}[Efficiency of track finding in the HPgTPC]{fig:garsoft_efficiency}
{(Left) The efficiency to find tracks in the \dword{hpgtpc} as a function of momentum, $p$, for tracks in a sample of \dword{genie} events simulating \SI{2}{GeV} and $\nu_\mu$ interactions in the gas, using GArSoft. (Right) The efficiency to find tracks as a function of $\lambda$, the angle with respect to the center plane, for tracks with $p>200\,$MeV/$c$.}
    \includegraphics[width=0.49\textwidth]{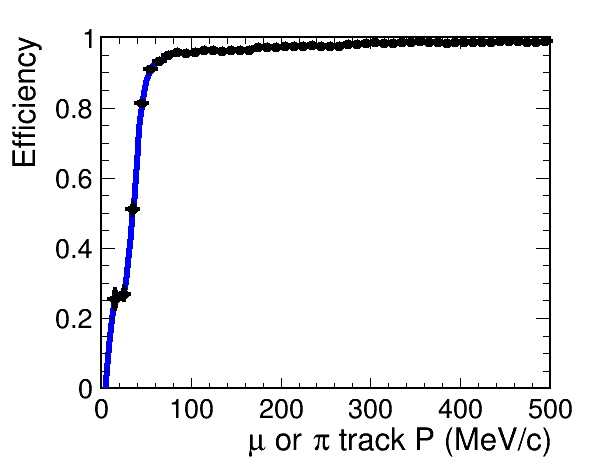}\includegraphics[width=0.49\textwidth]{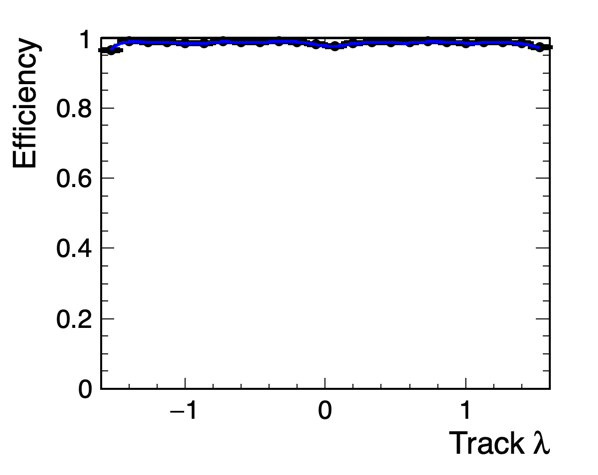}
\end{dunefigure}

\begin{dunefigure}[Tracking efficiency for protons in the HPgTPC]{fig:TEpr}{Tracking efficiency for protons in the \dword{hpgtpc} as a function of kinetic energy.} 
\includegraphics[width=0.65\columnwidth]{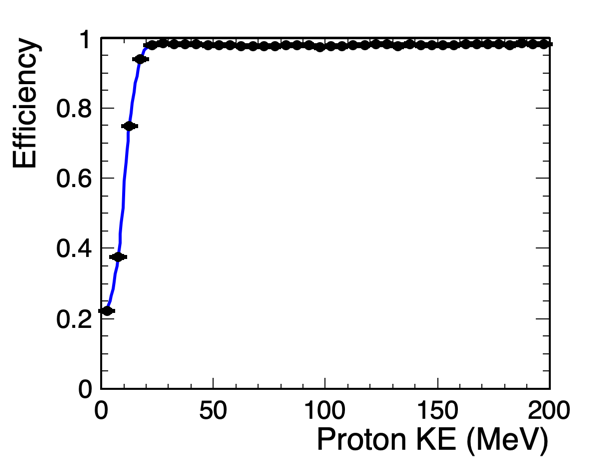} 
\end{dunefigure}

ALICE chose to use neon, rather than argon, for the primary gas in their first run; the decision was driven by a number of factors, but two-track separation capability was one of the primary motivations due to the extremely high track multiplicities in the experiment.  Neon performs better than argon in this regard.  A better comparison for the \dword{hpgtpc}'s operation in DUNE is the two-track separation that was obtained in PEP4~\cite{PEP4_Stork}.  PEP4 ran an 80-20 mixture of Ar-CH$_4$ at 8.5~atmospheres, yielding a two-track separation performance of \SI{1}{cm}.

In ALICE, the ionization produced by charged particle tracks is sampled by the TPC pad rows (there are 159 pad rows in the TPC) and a truncated mean is used for the calculation of the PID signal. Figure~\ref{fig:ALICE_dEdx} (left) shows the ionization signals of charged particle tracks in ALICE for pp collisions at $\sqrt{s} = 7$~TeV. The different characteristic bands for various particles are clearly visible and distinct at momenta below a few GeV.  When repurposing ALICE as the \dword{hpgtpc} component of the \dword{mpd},  better performance is expected for particles leaving the active volume, since the detector will be operating at higher pressure (10~atmospheres vs. the nominal ALICE 1~atmosphere operation), resulting in ten times more ionization per unit track length available for collection. Figure~\ref{fig:ALICE_dEdx} (right) shows the charged particle identification for PEP-4/9~\cite{Grupen:1999by}, a higher pressure gas TPC that operated at 8.5~atmospheres, which is very close to the baseline argon gas mixture and pressure of the DUNE \dword{hpgtpc}.

\begin{dunefigure}[ALICE and PEP-4 $dE/dx$-based particle identification as a function of momentum]{fig:ALICE_dEdx}
{Left: ALICE TPC $dE/dx$-based particle identification as a function of momentum (from~\cite{ALICE_Lippmann}). Right: PEP-4/9 TPC (80:20 Ar-CH4, operated at 8.5~Atm, from~\cite{Grupen:1999by}) $dE/dx$-based particle identification.}
\includegraphics[width=0.49\textwidth]{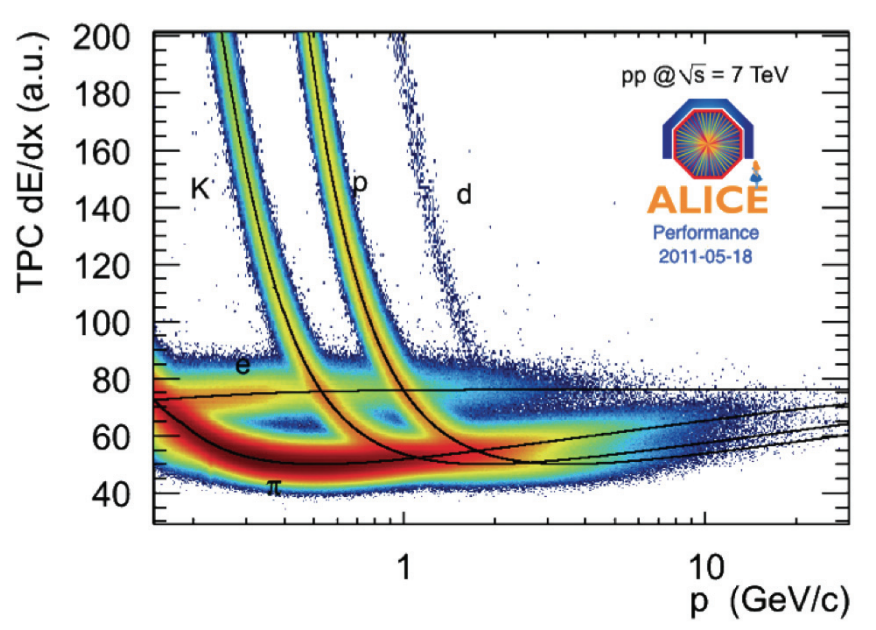}
\includegraphics[width=0.49\textwidth]{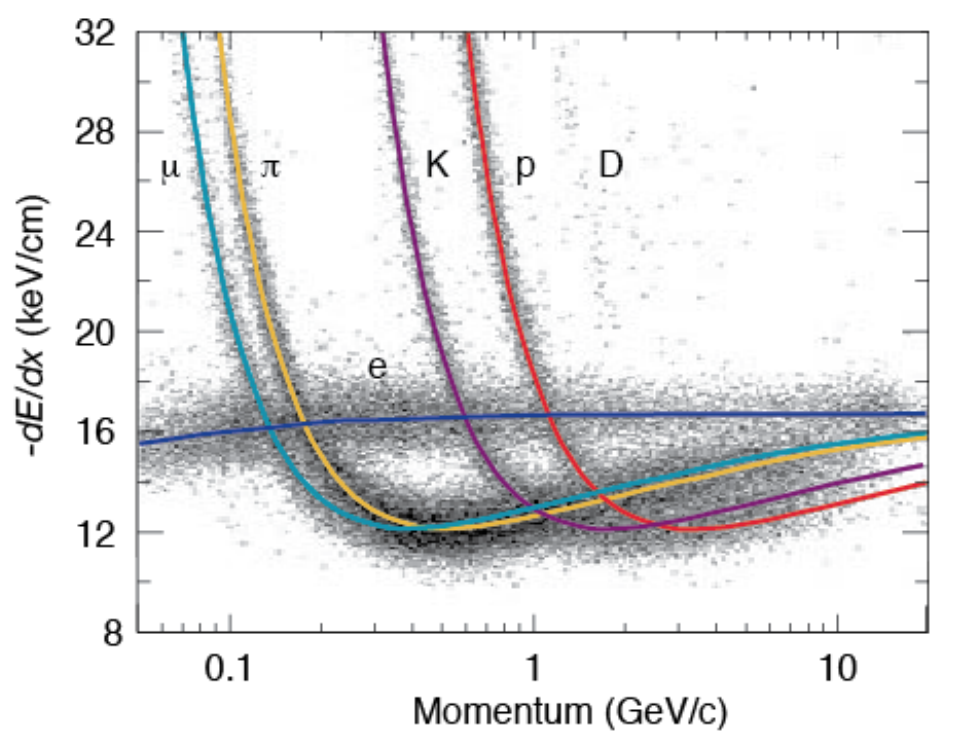} 
\end{dunefigure}

\subsubsubsection{Momentum and Angular Resolution for Charged Particles}
The ability to determine the sign of the charge of a particle in the \dword{hpgtpc} tracking volume is limited by the spatial resolution of the measured drift points in the plane perpendicular to the magnetic field, as well as multiple Coulomb scattering (MCS) in the gas. For a fixed detector configuration, the visibility of the curvature depends on the particle's $p_{\rm{T}}$, its track length in the plane perpendicular to the field, and the number and proximity of nearby tracks.  Because primary vertices are distributed throughout the tracking volume, the distribution of the lengths of charged-particle tracks is expected to start at very short tracks, unless sufficient \dword{fv} cuts are made to ensure enough active volume remains to determine particle's track sign.  The kinetic energies of particles that leave short tracks and stop in the detector will be better measured from their tracks' lengths than from their curvatures.  Protons generally stop before their tracks curl around, but low-energy electrons loop many times before coming to rest in the gas.

Within the \dword{fv} of the \dword{hpgtpc}, charged particles can be tracked over the full 4$\pi$ solid angle.  Even near the central electrode, tracking performance will not be degraded due to the very thin (25 $\mu$m of mylar) nature of the central electrode.   Indeed, tracks crossing the cathode provide an independent measurement of the event time, since the portions of the track on either side of the cathode will only line up with a correct event time assumed when computing drift distances. The 4$\pi$ coverage is true for all charged particles.  ALICE ran with a central field of 0.5~T and their momentum resolution from $p$--Pb data~\cite{Abelev:2014ffa} is shown in Figure~\ref{fig:ALICE_MOMres}.

\begin{dunefigure}
[The TPC stand-alone p$_T$ resolution in ALICE for $p$--Pb collisions]
{fig:ALICE_MOMres}
{The black squares show the TPC stand-alone p$_T$ resolution in ALICE for $p$--Pb collisions. From Ref.~\cite{Abelev:2014ffa}.}
\includegraphics[width=0.65\columnwidth]{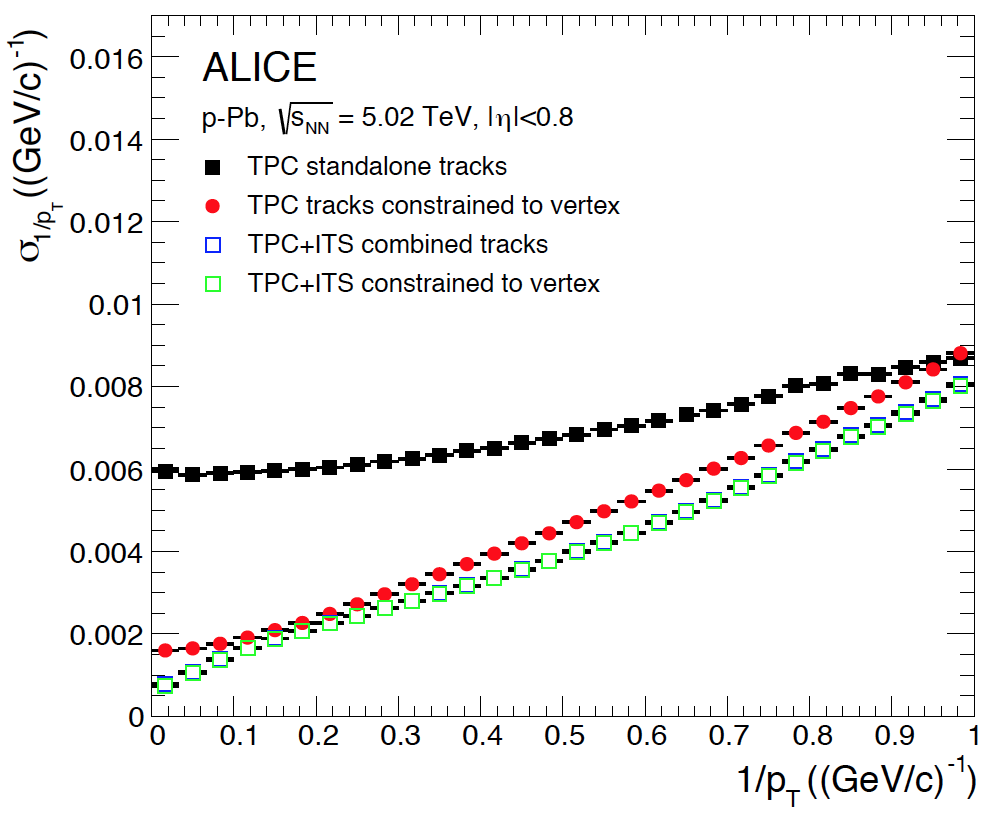}
\end{dunefigure}
The momentum resolution of muons in neutrino scatters using the GArSoft simulation and reconstruction package is shown in Figure~\ref{fig:garsoftpamuonres1}, using a sample of 2~GeV $\nu_\mu$~\dword{cc} events.  This resolution differs from ALICE's achieved resolution due to the higher pressure, the heavier argon nucleus compared with neon, the non-centrality of muons produced throughout the detector, and the fact that the GArSoft simulation and reconstruction tools have yet to be fully optimized.  The momentum resolution achieved for muons is $\Delta p/p = 4.2$\%, and is expected to improve with optimization of the simulation and reconstruction tools.  The 3D angular resolution of muons is approximately 0.8~degrees, as shown in Figure~\ref{fig:garsoftpamuonres1}.

\begin{dunefigure}[Momentum and angular resolutions for muons in GArSoft]{fig:garsoftpamuonres1}
{Left: the momentum resolution for reconstructed muons in GArSoft, in a sample of \SI{2}{GeV} $\nu_\mu$~\dword{cc} events simulated with \dword{genie}.  The Gaussian fit to the $\Delta p/p$ distribution has a width of 4.2\%. Right:  the \threed angular resolution for the same sample of muons in GArSoft.}
\includegraphics[width=0.49\columnwidth]{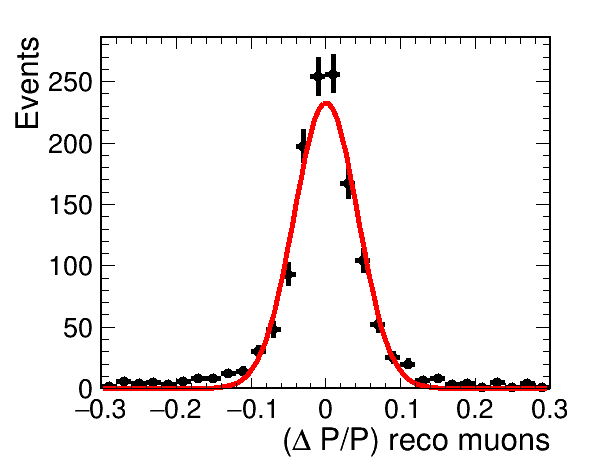}\includegraphics[width=0.49\columnwidth]{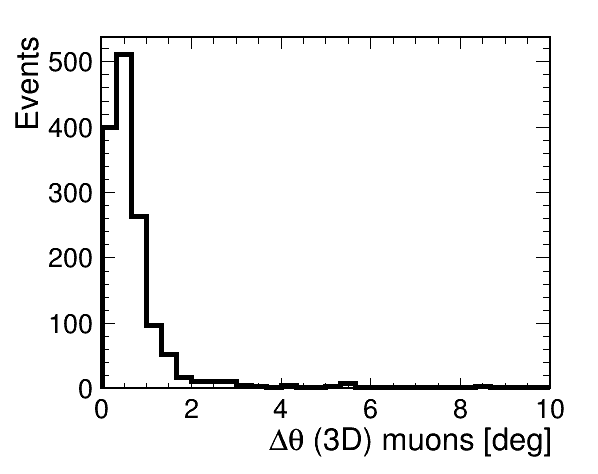} 
\end{dunefigure}

\subsubsubsection{Machine Learning for Low Energy Protons}\label{sec:TPC_ML}                                                     

As a complement to the existing reconstruction, an initial exploration of several machine learning methods has been performed.
The main goal of this effort has been to attempt to reconstruct very low energy protons and pions where traditional           
tracking methods might struggle.   
While this study is still in very early stages, there has been success so far in using a fully connected multi-layer perceptron (MLP) to both regress
the kinetic energy of and classify between protons and pions.  Additionally a Random Sample Consensus (RANSAC) based          
clustering algorithm has been developed to group hits into short tracks for events where there are multiple particles.        
Together, these two algorithms can be used to measure the kinetic energy of multiple particles in a single event.             

As a demonstration, a test sample of multiple proton events was generated where each event has:
\begin{itemize}
\item 0-4 protons, number determined randomly with equal probabilities
\item all protons share a common starting point (vertex) whose position in the TPC is randomly determined
\item each proton is assigned independently and randomly:
\begin{itemize}
\item a direction in space (isotropically distributed)
\item a scalar momentum between 0 and 200 MeV/$c$ (flat distributed)
\end{itemize}
\end{itemize}

The RANSAC-based clustering algorithm assigns individual hits to proton candidate sets of hits which are
passed to a MLP that was trained on a set of individual proton events in the TPC to predict kinetic energy.  Figure~\ref{fig:ML_residuals} shows the kinetic energy residuals, the reconstruction efficiency,
and a 2D scatter plot of the measured kinetic energy versus the true kinetic energy
for each individual proton with kinetic energy between 3 and 15 MeV in the test sample.  Additionally, the residual for the total kinetic energy in each multi-proton event is given. 

\begin{dunefigure}[Machine learning residuals for protons in the \dshort{mpd}]{fig:ML_residuals}
{(Top left) Kinetic energy residual, (Top right) measured KE vs. true KE, and (Bottom right) reconstruction efficiency for individual protons with \SIrange{3}{15}{MeV} KE in the test set.  (Bottom left) Residual of the total kinetic energy of all protons in each event in the test sample.}
    \includegraphics[width=0.49\textwidth]{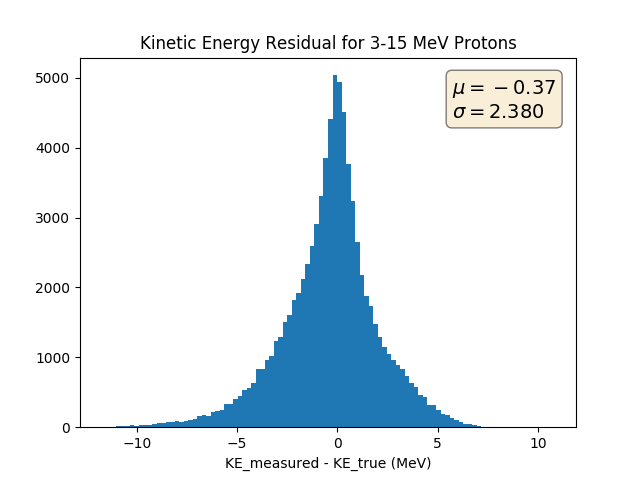}                                                    
    \includegraphics[width=0.49\textwidth]{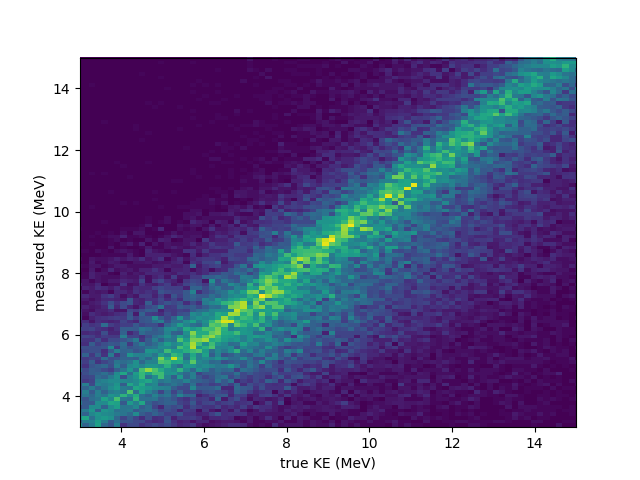}
    \vspace{1mm}
    \includegraphics[width=0.49\textwidth]{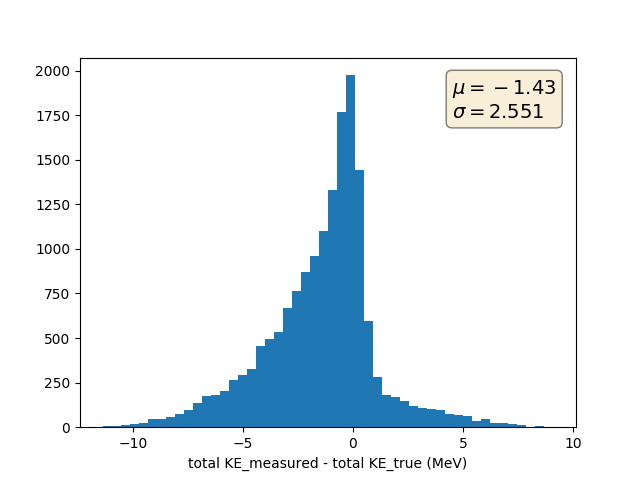}
    \includegraphics[width=0.49\textwidth]{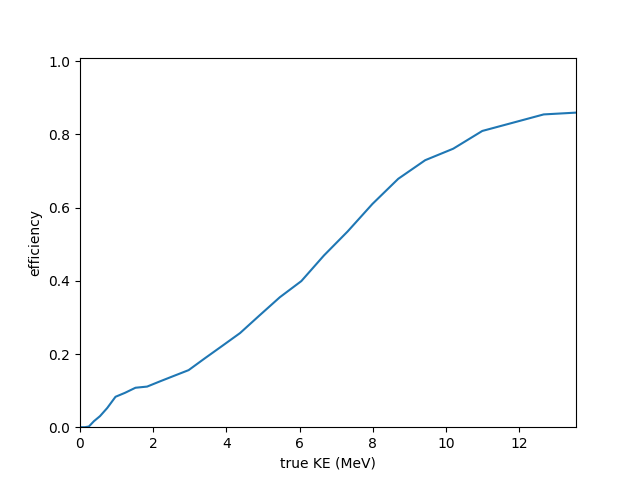}
\end{dunefigure}

\subsubsubsection{\dword{ecal} Performance}

The expected performance of the calorimeter was studied with Geant4-based \cite{Agostinelli:2002hh} simulations and GArSoft \cite{GArSoftwebsite}. In the following, a first scenario referred to as scenario A (shown by the red curve in the figures below) in which the \dword{ecal} is located inside the pressure vessel is considered. The barrel geometry consists of 55 layers with the following layout:
\begin{itemize}
  \item 8 layers of \SI{2}{\mm} copper + \SI{10}{\mm} of $2.5\times2.5$ cm$^2$ tiles + \SI{1}{\mm} FR4
  \item 47 layers of \SI{4}{\mm} copper + \SI{10}{\mm} of cross-strips \SI{4}{\cm} wide
\end{itemize}
For the present studies, copper has been chosen as absorber material as initial studies have shown that this material provides a good compromise between calorimeter compactness, energy resolution, and angular resolution. However, the choice of absorber material is still under study. The choice of granularity, scintillator thickness, and the arrangement of tiles and strips is still under optimization in order to reduce the number of readout channels while keeping the calorimeter performance. Two alternative scenarios are shown below: scenario B (black curve) has a different arrangement of the tile and strip layers, and scenario C (blue curve) has thinner absorbers in the front layers.
Digitization effects are accounted for by introducing an energy threshold of 0.25~MIPs ($\sim$\SI{200}{\keV}) for each detector cell/strip, a Gaussian smearing of \SI{0.1}{\MeV} for the electronic noise, SiPM saturation effects, and single photon statistics.

\paragraph{Energy Resolution} The energy resolution is determined by fitting the visible energy with a Gaussian. Converted photons are rejected based on Monte-Carlo information. A fit function of the form $\frac{\sigma_{E}}{E} = \frac{A}{\sqrt{E}} \oplus \frac{B}{E} \oplus C$ is used, where $A$ denotes the stochastic term, $B$ the noise term, $C$ the constant term, and $E$ is in GeV. Figure~\ref{fig:EResARes_NDECAL} shows the energy resolution as a function of the photon energy. For scenario A, shown in red, the energy resolution is around $\frac{6.7\%}{\sqrt{E}}$. With further optimization, it is believed that an energy resolution of (or below) $\frac{6\%}{\sqrt{E}}$ is achievable. It should be noted that due to the lack of non-uniformities, dead cells, and other effects in the simulation, the energy resolution is slightly optimistic.

\begin{dunefigure}[Energy and angular resolutions for photons in the \dshort{mpd} ECAL]{fig:EResARes_NDECAL}
{Left: energy resolution in the barrel as a function of the photon energy for three \dword{ecal} scenarios. The energy resolution is determined by a Gaussian fit to the visible energy. Right: the angular resolution in the barrel as a function of the photon energy for the three \dword{ecal} scenarios. The angular resolution is determined by a Gaussian fit to the 68\% quantile distribution. For both figures, the scenario A is shown by the red curve, scenario B by the black curve and scenario C by the blue curve. The fit function is of the form $\frac{\sigma_{E}}{E} = \frac{A}{\sqrt{E}} \oplus \frac{B}{E} \oplus C$.}
\includegraphics[width=0.45\textwidth]{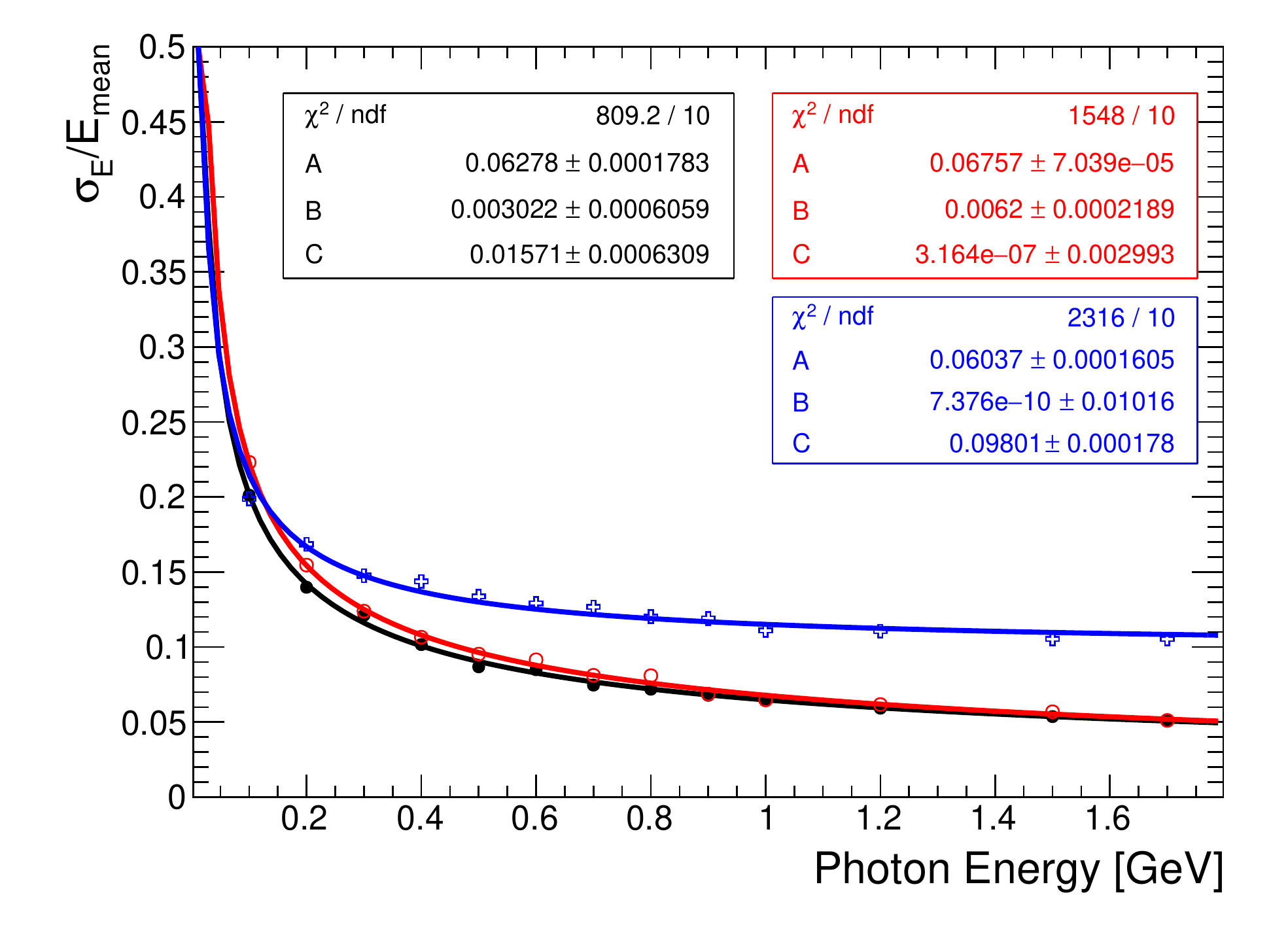}
\includegraphics[width=0.45\textwidth]{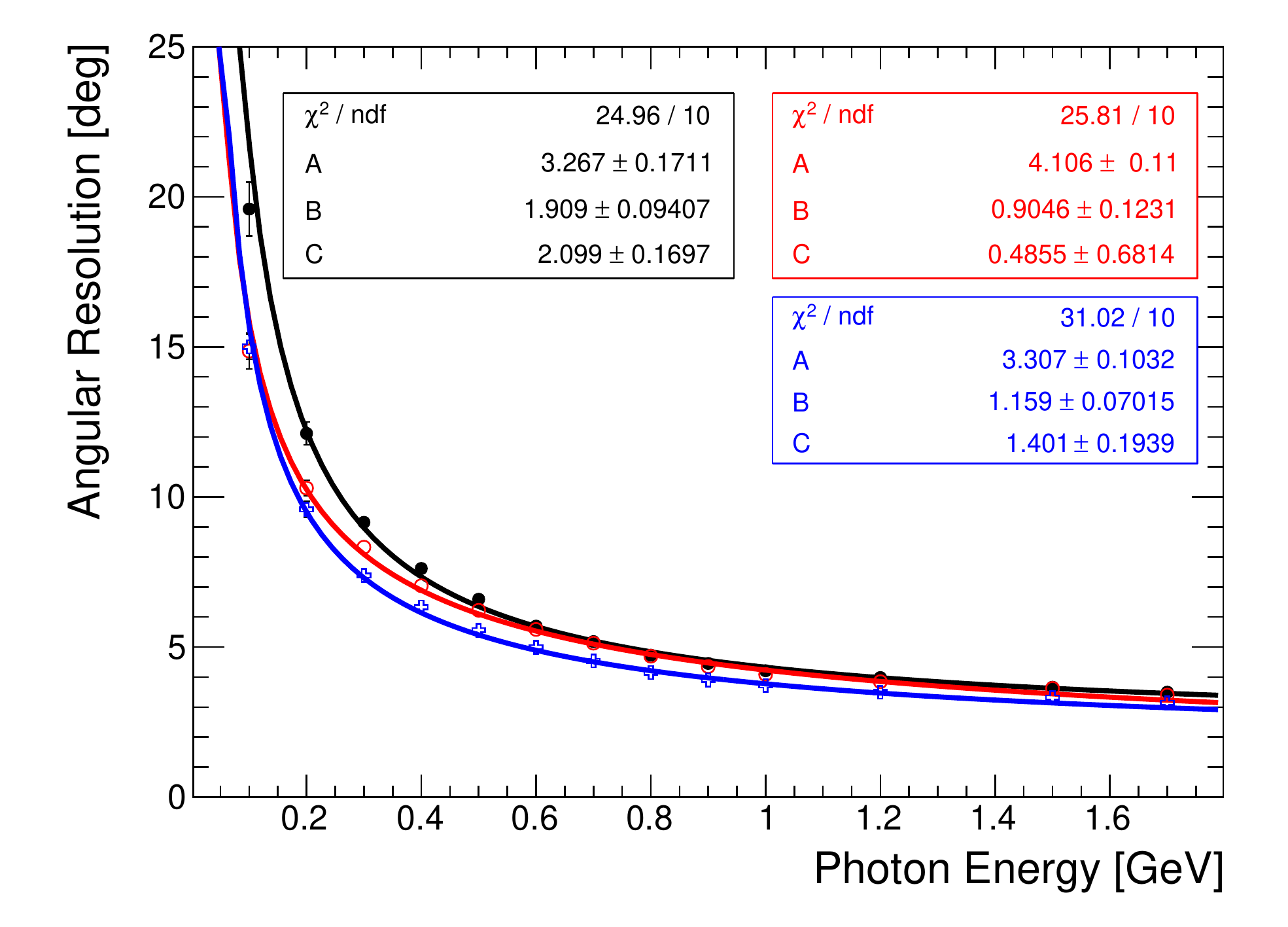}
\end{dunefigure}

\paragraph{Angular Resolution} The angular resolution of the calorimeter has been determined using a principal component analysis (PCA) of all reconstructed calorimeter hits. The direction is taken as the first eigenvector (main axis) of all the reconstructed hits. The angular resolution is determined by taking the 68\% quantile of the reconstructed angle distribution and fitting a Gaussian distribution. The mean of the Gaussian is taken as the angular resolution and the error as its variance. Figure~\ref{fig:EResARes_NDECAL} shows the angular resolution as a function of the photon energy. In scenario A, shown in red, an angular resolution of $\frac{\SI{3.85}{\degree}}{\sqrt{E}} \oplus \SI{2.12}{\degree}$ can be achieved. This can potentially be further improved with a different arrangement of the tile and strip layers, an optimization of the absorber thickness, and an improved reconstruction method. However, the requirements will be further refined and will impact the detector optimization. The angular resolution is mainly driven by the energy deposits in the first layers of the \dword{ecal}. Using an absorber with a large $X_{0}$ creates an elongated shower that helps in determining the direction of the shower. In general, high granularity leads to a better angular resolution, however, studies have shown that there is no additional benefit to having cell sizes below $2\times2$ cm$^2$ \cite{Emberger:2018pgr}.

\paragraph{Neutron detection} The \dword{ecal} is sensitive to neutrons due to the scintillator containing hydrogen. Previous simulation studies showed that a detection efficiency above 60\% can be achieved for neutron energies greater than \SI{50}{\MeV}. However, the energy measurement is not very accurate (around 50-60\% below \SI{600}{\MeV}) \cite{Emberger:2018pgr}. Other methods of detection such as time of flight (ToF) could be used to improve the neutron energy measurement by measuring precisely the hit time of the neutron and its travel distance in the calorimeter.  This is currently under study.

\paragraph{$\pi^0$ reconstruction} For identification of neutral pions, both the energy and angular resolution are relevant. In an initial study, the position of the neutral pion is determined by using a $\chi^2$-minimization procedure taking into account the reconstructed energy of the two photons and the reconstructed direction of the photon showers \cite{Emberger:2018pgr}. The location of the decay vertex of the neutral pion can be determined with an accuracy between \SIrange{10}{40}{\cm}, depending on the distance from the downstream calorimeter and the $\pi^0$ kinetic energy. This is sufficient to associate the $\pi^0$ to an interaction in the \dword{hpgtpc}, since the gas will have less than one neutrino interaction per beam spill.
The pointing accuracy to the pion decay vertex may be further improved by a more sophisticated analysis technique and by using precision timing information, and is a subject of current study.


\subsection{The DUNE-PRISM Program}
\label{sec:appx-nd:DP}

The goals of the off-axis measurements are twofold:
\begin{itemize}
\item {\bf To identify problems in the cross section modeling.} By comparing  \dword{nd} data to \dword{mc} at many off-axis locations with different energy spectra, the neutrino interaction model will be more tightly constrained than it would be with only on-axis measurements, and the potential for biases in the measured oscillation parameters can be identified, i.e., the off-axis data might be sensitive to mismodelings that are degenerate or indeterminate with only on-axis measurements.
\item {\bf To overcome problems in the cross section modeling.} The most important novel feature of a \dword{duneprism} detector is that measurements at different off-axis positions can be linearly combined to determine any set of observables for any user-defined neutrino energy spectrum. In particular, it is possible to predict the expected distribution of an observable, such as the reconstructed neutrino energy, for a neutrino flux with oscillations  using linear combinations of  \dword{nd} off-axis spectra.  This will greatly reduce the dependence on neutrino interaction modeling within the oscillation analysis.
\end{itemize}

\subsubsection{Impact of Cross Section Modeling on Neutrino Oscillations}

One strategy to understand the potential impact of using imperfect neutrino interaction models is to extract oscillation parameters from a ``fake'' data set that is different from the model used in the analysis.  This fake data set represents a reality that includes effects unknown to or not accounted for properly by the model used in the analysis to fit the data. In this way, it is possible to understand potential biases in the measured oscillation parameter values extracted from a full near+far detector fit due to the use of an incorrect cross section model in the fit.

The fake data set considered here assumes that 20\% of the kinetic energy that the interaction model originally assigned to protons was instead carried away by neutrons. The resulting model is then further modified by adjusting the differential cross section in proton energy as a function of true neutrino energy until the measured kinematic distributions in the on-axis  \dword{nd} match the prediction from the default model. This procedure is similar to actions that are routinely taken in actual neutrino oscillation experiments to resolved discrepancies between  \dword{nd} data and the Monte Carlo simulation. There are many potential modifications to the cross section model that can be chosen to resolve such disagreements. Incorrect choices can lead to biased oscillation parameter measurements because the associated incorrect particle composition and cross section model can lead to an incorrect relation between reconstructed and true energy.

The resulting fake data is analyzed as though it were the actual data taken by the experiment. The \dword{nd} and  \dword{fd} data are fit simultaneously  to constrain nuisance parameters in the flux and cross section models, and to extract the measured value of the neutrino oscillation parameters. The results of this fit are shown in Figure~\ref{fig:duneprismfit}. The fit to the fake data shows a clear bias in the measured oscillation parameter values that lie outside the 95\% confidence limit contours.

\begin{dunefigure}[Oscillation fits to nominal and fake data sets for \dshort{duneprism} fake data study]{fig:duneprismfit}
{The results of a full two-flavor, antineutrino+neutrino, near+far oscillation fit are shown for a fit to the nominal \dword{mc} (dashed) and a fit to the fake data set (solid). The true values of the oscillation parameters in each of the data sets are indicated by the dashed yellow lines. Clear biases can be see in all oscillation parameters that are well outside the 1$\sigma$ (black), 2$\sigma$ (red), and 3$\sigma$ (blue) contours.}
      \includegraphics[width=0.85\textwidth]{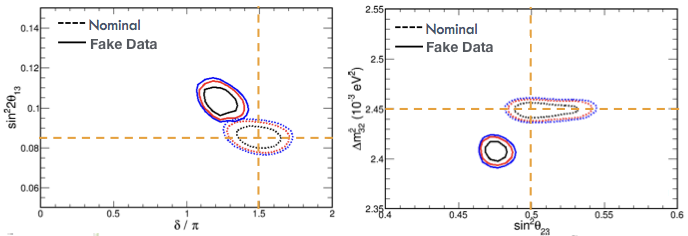}
\end{dunefigure}

A comparison of the fake data and the nominal Monte Carlo reconstruction energy distributions is shown in Figure~\ref{fig:duneprismerec}. In the on-axis location, good agreement is seen, as was intended in the construction of the fake data samples. This good agreement is assured since the model is tuned to the on-axis data.  Conversely, clear disagreement is seen between these samples when moving off-axis. As the off-axis location is varied, this comparison can be made across a wide range of neutrino energy distributions.

\begin{dunefigure}[Reconstructed energy distributions for  nominal and fake data sets; on- and off-axis]{fig:duneprismerec}
{A comparison between the fake data (green) and nominal Monte Carlo (red) reconstructed neutrino energy distributions are shown for the on-axis  \dword{nd} location (left) and a position \SI{18}{m} off-axis (right).}
      \includegraphics[width=0.45\textwidth]{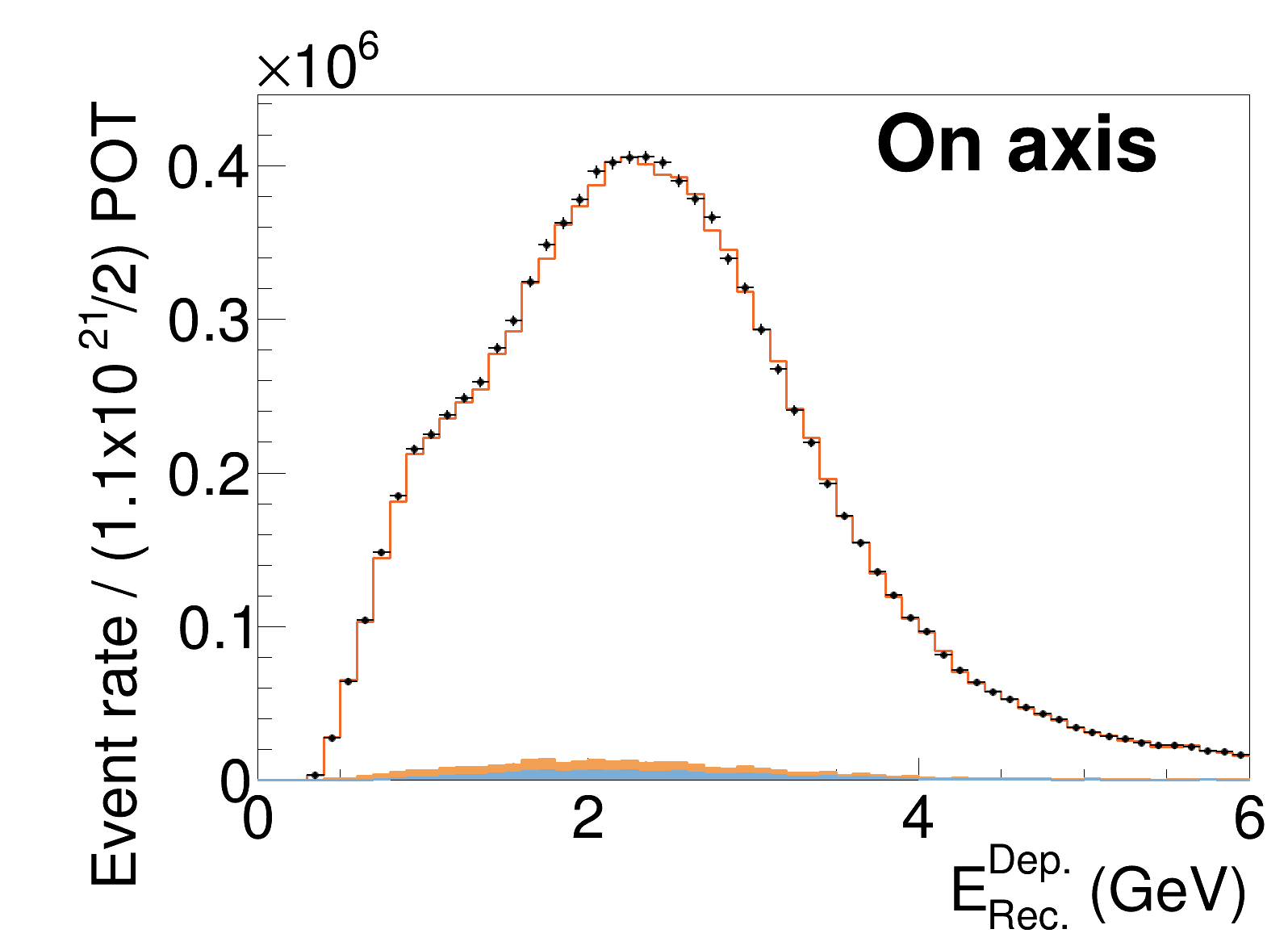}
      \includegraphics[width=0.45\textwidth]{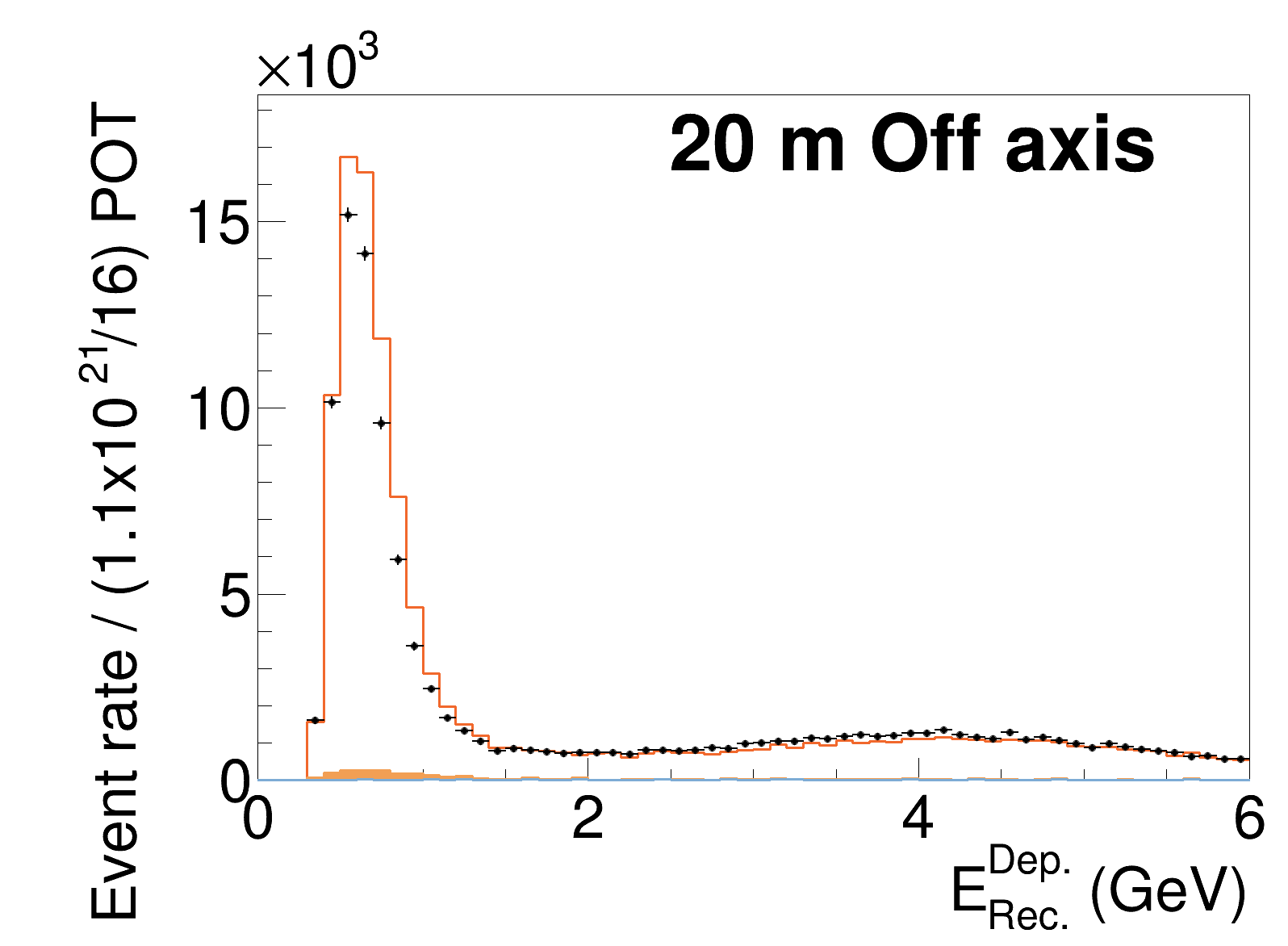}
\end{dunefigure}

\subsubsection{DUNE-PRISM Linear Combination Analysis}

In addition to identifying problems in cross section modeling, \dword{duneprism} measurements provide a mechanism for creating  \dword{fd} predictions directly from the  \dword{nd} data that is largely independent of neutrino interaction modeling. By constructing linear combinations of measurements taken under exposure to different neutrino fluxes, it is possible to determine the distribution of any observable (e.g. reconstructed neutrino energy) for a different neutrino flux of interest. This means, for example, from the \dword{nd} data alone it is possible to create a distribution of the expected reconstructed neutrino energy distribution at the \dword{fd} in the event of oscillations with a specific set of parameters.  This distribution, created using this data-driven technique, can then be compared to that seen in the \dword{fd} with a reduced dependence on the flux and neutrino interaction models and their associated uncertainties.

A few example fits of the off-axis \dword{nd} muon neutrino spectra to an oscillated  \dword{fd} muon neutrino energy spectrum are shown in Figure~\ref{fig:duneprismfluxfits}. Good agreement is seen near the first and second oscillation maxima at 2.5~GeV and 0.7~GeV, respectively. The ability to fit the \dword{fd} spectrum breaks down outside the central range of energy because the constituent off-axis spectra used in the fit extend only slightly outside this range and cannot duplicate the spectrum in any combination.  This does not pose a significant problem for the oscillation analysis because the fit is good for the bulk of the events and in the region that drives the CP sensitivity.  This technique can also be applied to match the off-axis muon neutrino spectra to the  \dword{nd} intrinsic electron neutrino spectrum, in order to make a precise measurement of $\sigma(\nu_e)/\sigma(\nu_\mu)$ with a common flux, or to the  \dword{fd} oscillated electron neutrino energy spectra for the measurement of $\delta_{CP}$.

\begin{dunefigure}[Linear combinations of off-axis fluxes giving \dshort{fd} oscillated spectra; range of 
parameters]{fig:duneprismfluxfits}
{Linear combinations of off-axis fluxes giving far-detector oscillated spectra for a range of oscillation parameters. The  \dword{fd} oscillated flux is shown in black, the target flux is shown in green, and the linearly combined flux obtained with the nominal beam MC is shown in red. Systematic effects due to 1 $\sigma$ variations of the decay pipe radius (green), horn current (magenta) and horn cooling water layer thickness (teal) are shown.}
	\includegraphics[width=0.7\textwidth]{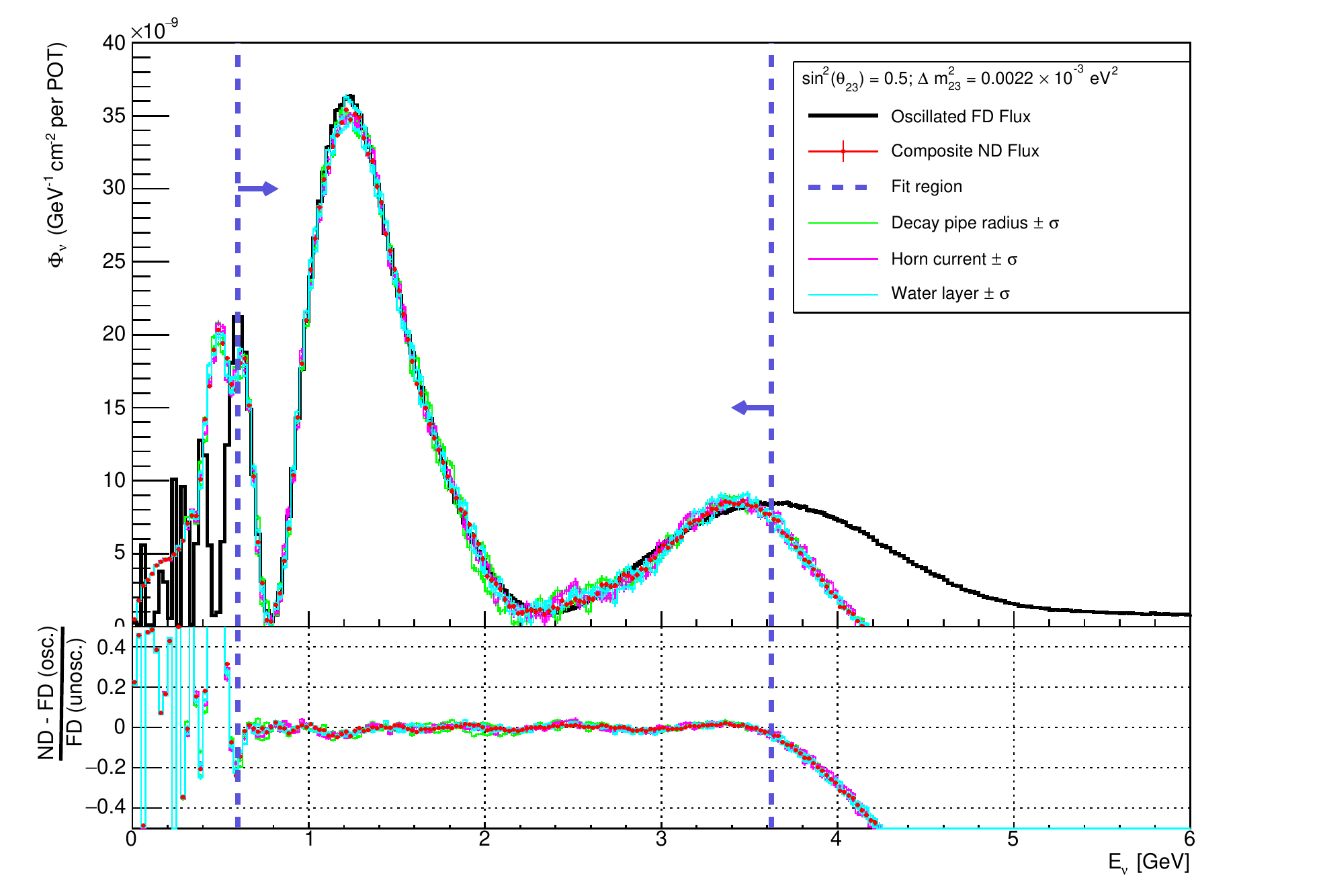}
	\includegraphics[width=0.7\textwidth]{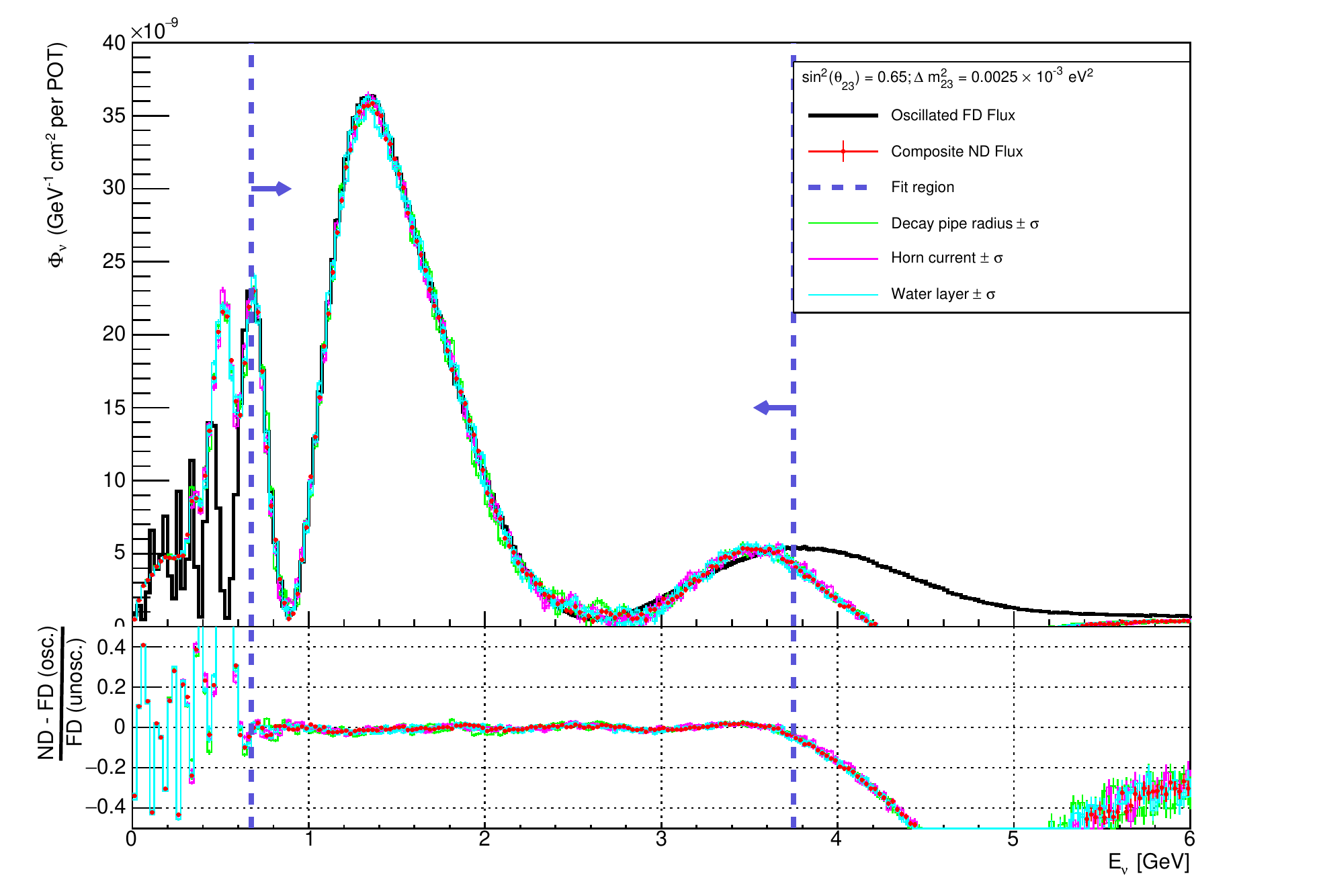}
\end{dunefigure}

\section{Fixed On-axis Component of the \dshort{dune} \dshort{nd}}

\subsection{Motivation and Introduction}
\label{sec:appx-nd:exsum-nd-onaxis-mission}

In spite of tremendous efforts to ensure stable operation, neutrino beams are dynamic in nature.  Experiments must track changes in the beam as a function of time in order to  understand and model the neutrino flux spectrum well enough to achieve their physics goals. In addition, neutrino interactions observed by experiments in the beam, if taken at a sufficient rate with good energy and spatial resolutions, can provide unique and invaluable information as a beam diagnostic. The issue of beam stability is particularly important for DUNE because the wide-band beam is sensitive to a large range of beamline changes and the \dword{duneprism} program makes use of spectral shifts induced by off-axis translation which must be distinguished from time-dependent neutrino spectral changes intrinsic to the beam.  On-axis beam monitoring, in particular, is critical because the \dword{dune} far detector sees that beam.  The on-axis spectrum must be modeled and used for the extraction of the oscillation parameters at the far detector.  

There are many valuable lessons for \dword{dune}/\dword{lbnf} in the \dword{numi} experience. One of those lessons is that unexpected things happen to neutrino beamline elements that can take them well outside the typical error tolerance.  One example of this at \dword{numi} was degradation in the target due to broken upstream target fins.  This was observed and diagnosed prior to target autopsy by the change in the observed near detector neutrino event spectrum \cite{Holin2017}.  Figure~\ref{fig:minosspectrum} shows MINOS near detector data in bins of reconstructed neutrino event energy.  Within each bin are points representing successive periods of the data taking run.  A time-dependent shift in the three peak bins, i.e., a change in the spectral shape, is obvious in the plot. Another example of a significant and noticeable change in the beam spectra at \dword{numi} was caused by magnetic horn tilt due to degradation of a supporting washer \cite{Hylen2016}.  Yet another significant wiggle in the beam spectral shape has been observed in the \dword{minerva} and \dword{nova} medium energy run \cite{JenaNUINT2018}.  Notably, this wiggle is not as significant in the off-axis data taken by the \dword{nova} near detector as it is for on-axis \dword{minerva}.  \dword{minerva} studies indicate the observed spectral shift is best modeled by a shifted horn position or a slight change in the inner horn radius relative to the expected value.

\begin{dunefigure}[The \dshort{minos} near detector event spectrum shown in run periods]{fig:minosspectrum}
{The low energy run \dword{minos} event rate as a function of neutrino energy broken down in time. From~\cite{Holin2017}.}
  \includegraphics[width=5.in]{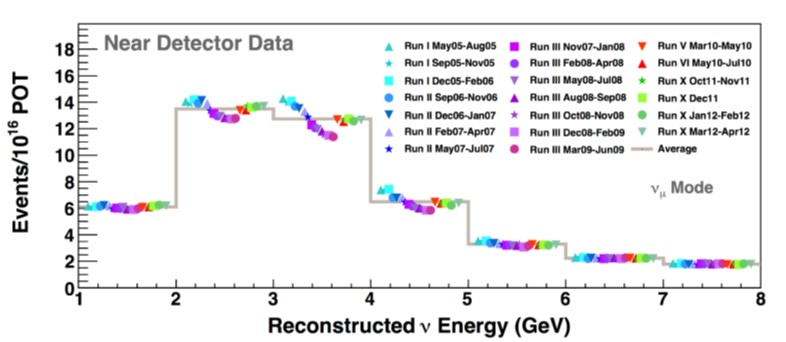}
\end{dunefigure}

Given the past experience from \dword{numi}, it is thought to be critically important that the \dword{dune} \dword{nd} monitor and track the beam spectrum over time.    \dword{arcube} and the \dword{mpd} will make on-axis neutrino beam spectrum measurements when they are located on-axis.  There are two broadly different concepts under discussion for how to monitor the on-axis beam when \dword{arcube} and the \dword{mpd} move off-axis as required by the DUNE-PRISM program.  One concept is to have a dedicated on-axis beam monitor that is capable of measuring the neutrino beam spectrum as a function of time.  Such a capability would require target mass and a magnet or range stack with tracking to measure the momentum of muons arising from \dword{cc} \numu interactions.  The other concept assumes a dedicated on-axis neutrino interaction rate monitor.  Both the on-axis rate monitor and the off-axis spectrometer would each track beam stability. In the case of an observed instability of either the rate or the off-axis spectrum, \dword{arcube} and the \dword{mpd} would move back on-axis to make a spectrum measurement. It is likely the DUNE-PRISM run plan would include intermittent on-axis measurements as well regardless of observed instabilities.  Since the first concept involves a magnet or large range stack, it is likely to be more expensive to implement than the option that makes use of the dedicated rate monitor on-axis.  On the other hand, the latter option with intermittent on-axis spectral measurements involves additional movements of the large, DUNE-PRISM detectors, precision comparison of spectral measurements separated in time with detectors that have moved in the interim, and accepting the risk that rate and off-axis spectral monitoring are both less sensitive to some changes in the beam than on-axis spectral monitoring.  Figure~\ref{fig:BeamMoncompare} provides an illustration, for a small subset of relevant beam parameters, of the differing sensitivities of integrated rate monitoring as compared to spectral monitoring of the beam.  Additional studies are in progress.

\begin{dunefigure}[Rate vs spectral monitoring of $\nu$ beam for 1~$\sigma$ shifts of the horn positions]{fig:BeamMoncompare}
{Comparison of rate monitoring and spectral monitoring of the neutrino beam for  1-sigma shifts of the horn positions.  On the left is shown the significance of the variation in the observed rate for one week of running with a seven ton fiducial mass detector.  It is shown as a function of off-axis angle (including zero, i.e., on-axis). On the right is shown the significance of the shape change as a function of energy for one week of running with an 8.7 ton fiducial mass on-axis spectrometer.}
  \includegraphics[width=3.4in]{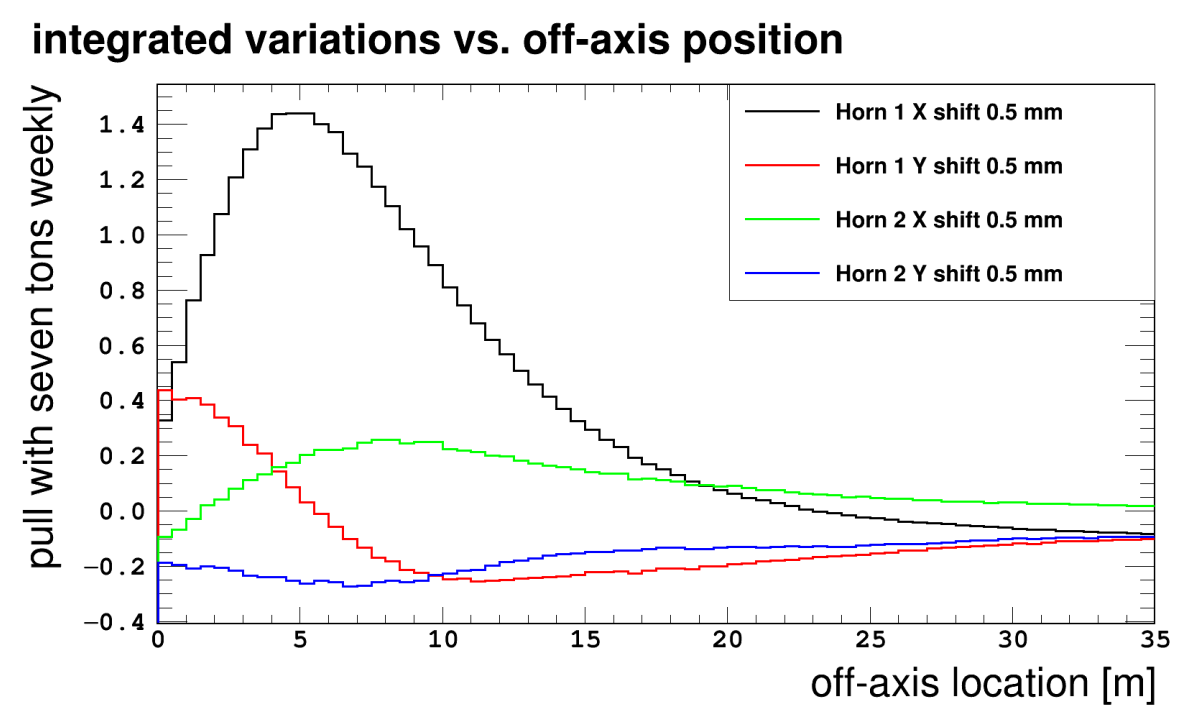}
    \includegraphics[width=3.4in]{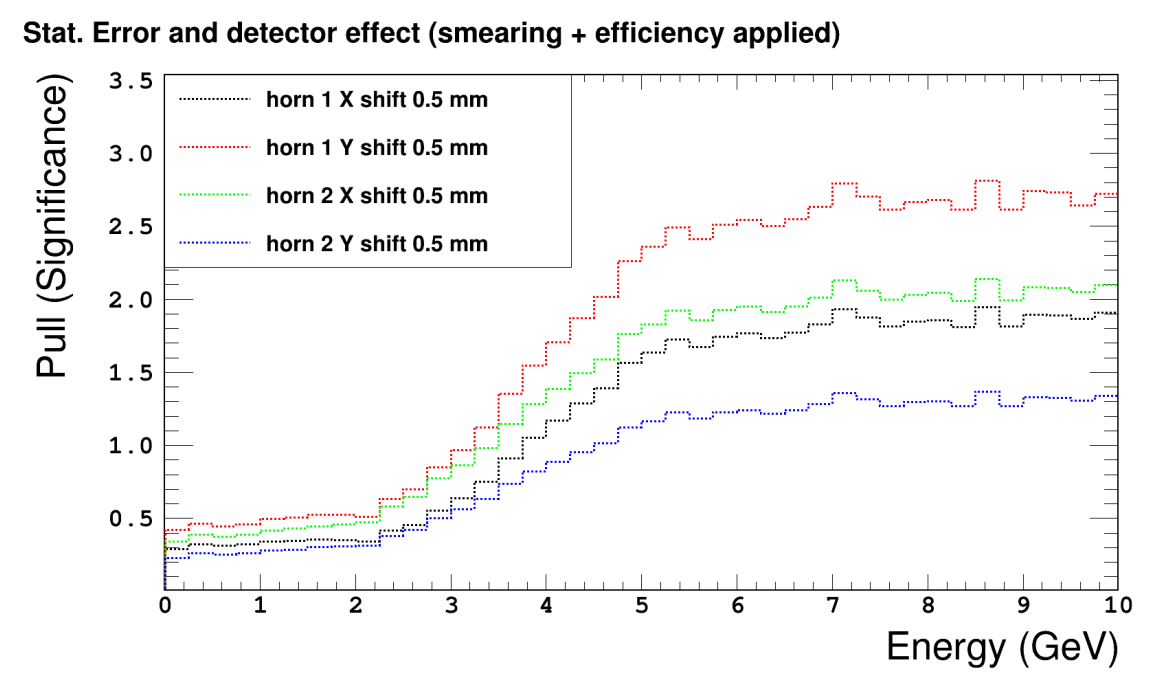}
\end{dunefigure}

The reference design described below in Section~\ref{sec:appx-nd:mpt-3dst} assumes the beam monitoring is done with a dedicated on-axis magnetic spectrometer.  The device described is a capable beam monitor in that it has the required mass and muon momentum resolution as will be shown in Section~\ref{sssec:appx:nd:3dst-bm}.  In addition, the on-axis spectrometer described below has capabilities that go beyond beam monitoring, which are useful for building confidence in the flux model and providing information that is potentially useful for the evolution of the neutrino interaction model.  Other concepts, including one that utilizes an on-axis rate monitor, are under consideration.

\subsection{Three-Dimensional Projection Scintillator Tracker Spectrometer}
\label{sec:appx-nd:mpt-3dst}

\dword{sand} consists of an active target core of scintillator called the \dword{3dst} surrounded by \dwords{tpc} and an \dword{ecal} in a \SI{0.5}{T} magnetic field.  This system has two main goals in the context of the larger \dword{nd} complex.  First, \dword{sand} functions as an on-axis, high mass target and muon spectrometer that is capable of producing a statistically significant neutrino beam spectrum measurement in a short period.  This dedicated, on-axis beam monitoring will be important in light of the movement of \dword{arcube} and the \dword{mpd} within the context of the \dword{duneprism} program.  Second, the \dword{3dst} can measure neutrons on an event-by-event basis, including those at a lower neutron kinetic energy than those seen by the other components of the \dword{nd}.  The inclusion of neutrons in the event reconstruction on an event-by-event basis is a powerful new tool that can be used for flux measurements and to probe neutrino interaction physics and modeling.

The \dword{3dst} is a fully active plastic scintillator detector made up of optically isolated 1 cm$^{3}$ cubes \cite{Sgalaberna:2017khy}.  The cubes are read out by wavelength shifting (WLS) fibers along 3 orthogonal axes providing three two-dimensional projections that yield effective three-dimensional reconstruction.

The \dword{3dst} is dense enough to provide a large statistics sample with reasonable containment of hadrons and photons from neutrino interactions. The high statistics and granularity  of the \dword{3dst} will allow for timely beam monitoring, flux determination via different methods (with charge separation), and the study of many different neutrino interaction morphologies.  The sub-ns timing resolution provides the  capability to include neutrons in the event reconstruction via Time-of-Flight (ToF) with a reasonably high efficiency.  

To date, neutrino experiments have been largely blind to neutrons on an event-by-event basis.  This is not ideal, in part because the neutron content of neutrino and antineutrino interactions differ, and in part because neutrons can carry a significant part of the outgoing energy and momentum of a neutrino interaction which compromises the reconstruction of events with missing neutrons.  Preliminary studies show the \dword{3dst} is likely to be able to measure neutrons to a lower neutron KE (KE$_{n}$) than the other component detectors of the ND and pursue event-by-event analysis with fully reconstructed final state particles, including neutrons. The addition of neutrons in event reconstruction will open the avenue to improved single transverse variable analyses that are expected to yield an improved neutrino energy resolution for flux (particularly antineutrinos) determination and for studies that may aid the evolution of the neutrino interaction model.

The \dword{3dst} uses the same technology as the SuperFGD detector that is being constructed for the \dword{t2k} \dword{nd} upgrade \cite{Abe:2019whr}.  The two detectors are functionally identical, though somewhat different in size.  The SuperFGD will be installed 2021 and will act effectively as a prototype for the larger \dword{3dst} in the \dword{dune}  \dword{nd}.

\subsubsection{Detector Configuration}

The \dword{3dst} detector concept is shown in Figure~\ref{fig:cube}.
The scintillator composition is polystyrene doped with 1.5\% of paraterphenyl (PTP) and 0.01\% of POPOP. After fabrication, the scintillator surface of the cubes is etched with a chemical agent that results in the formation of a white, reflecting polystyrene micro-pore deposit over the scintillator. Three orthogonal through holes of 1.5 mm diameter are drilled in the cubes to accommodate WLS fibers. 
This novel geometry provides full angular coverage for particle produced in neutrino interactions.  The momentum threshold for protons is about 300 MeV/c (if at least three hits are requested).

\begin{dunefigure}[A few plastic scintillator cubes assembled with WLS fibers.]{fig:cube}
{A few plastic scintillator cubes assembled with \dword{wls} fibers.}
  \includegraphics[width=3.in]{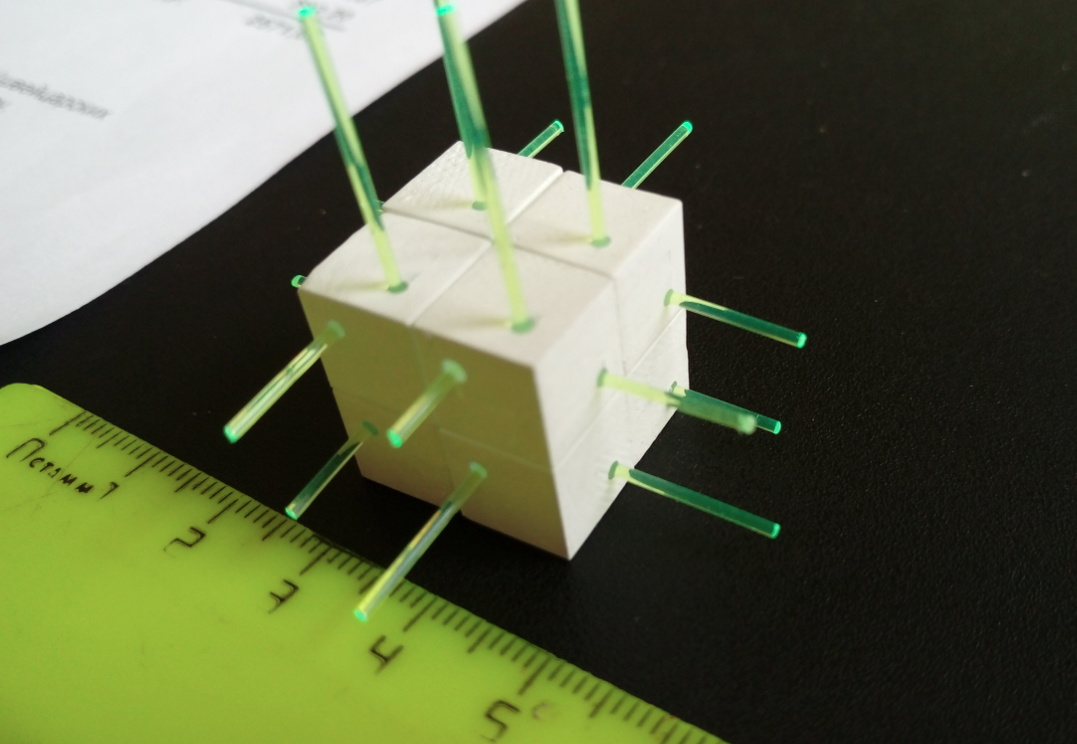}
\end{dunefigure}

The \dword{3dst} and surrounding elements are shown in Figure~\ref{fig:3dst-geometry}.  The size of the \dword{3dst} detector is under discussion.  Detectors of size 2.4$\times$2.4$\times$2.0~m$^{3}$, 3.0$\times$2.0$\times$2.0~m$^{3}$, and 2.0$\times$2.0$\times$2.0~m$^{3}$ have been used in different studies.  The primary considerations that drive the size are space, statistics, and neutron containment.

The \dword{3dst} is surrounded by low mass trackers to measure the kinematics of charged particles produced but not stopping in \dword{3dst}, and an \dword{ecal} to identify and reconstruct photons and electrons exiting the \dword{3dst}. The trackers are \dwords{tpc} in the reference design.  Straw tube trackers are also under consideration.  All the detectors will be immersed in a 0.5~T magnetic field provided by the \dword{kloe} magnet. The TPCs are envisioned to be similar to those being constructed for the T2K ND280 detector upgrade, described in \cite{Abe:2019whr}.  They are approximately 80~cm thick and use a gas mixture of Ar-CF4-iC4H10 (95\% - 3\% - 2\%).  

The \dword{kloe} magnet and \dword{ecal} already exist and have been used successfully for many years in a running experiment. The work and infrastructure necessary to transport the devices to Fermilab, as well as to install and operate them in the \dword{nd} hall are under study.

\begin{dunefigure}[The \dshort{sand} detector configuration]{fig:3dst-geometry}
{The \dword{3dst} inside the KLOE magnet. The drawing shows the \dword{3dst} in the center (white), \dword{tpc}s (orange), \dword{ecal} (green), magnet coil (yellow), and the return yoke (gray).}
  \includegraphics[width=7.in]{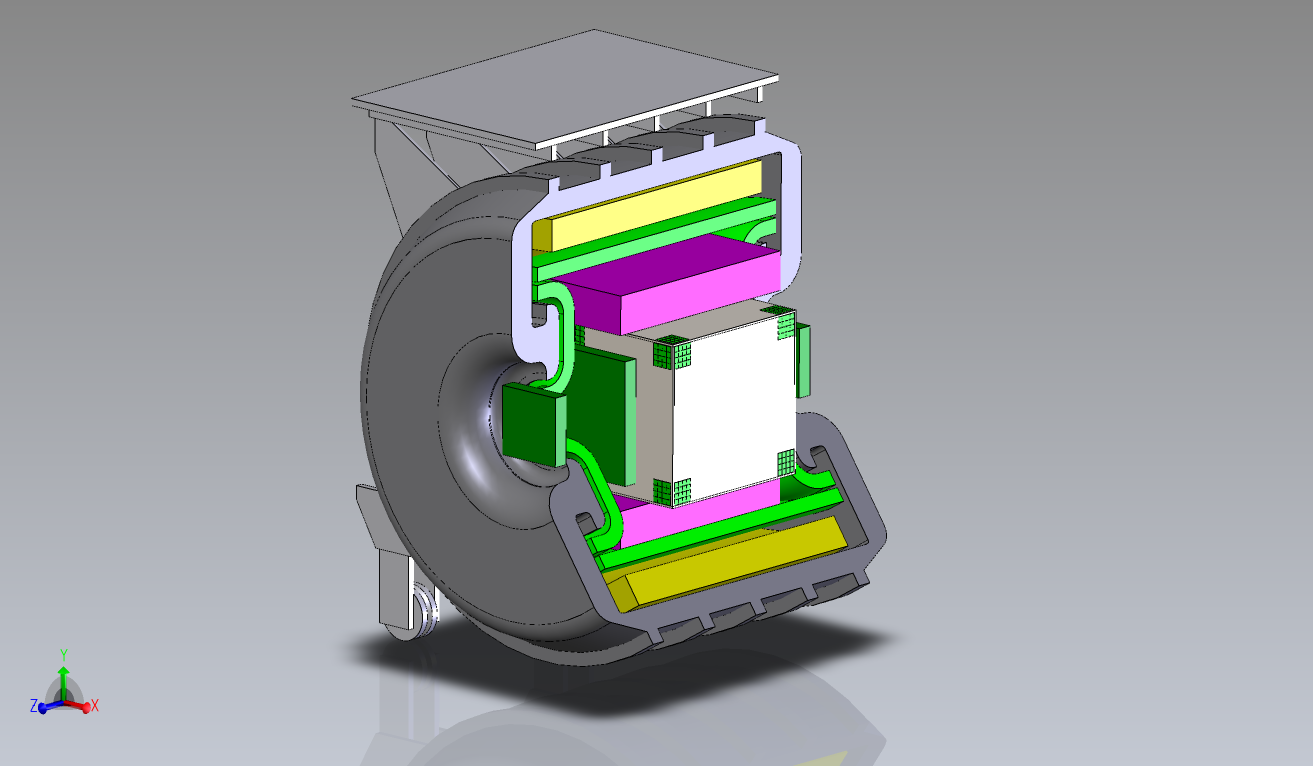}
\end{dunefigure}

\subsubsection{3DST Detector Performance}

The performance of devices built on the \dword{3dst} concept have been tested in several test beams at CERN \cite{Mineev:2018ekk}.
A small prototype of $5\times5\times5$ cubes collected data in the T10 test-beam area at CERN in 2017, with the goal of characterizing the response of the plastic scintillator cubes.
The detector was instrumented with 75 WLS fibers (1 mm diameter Y11(200) Kuraray S-type of 1.3 m length). One end of the fiber was attached to a photosensor while the other end was covered by a reflective Al-based paint (Silvershine). The photosensors in the beam test were Hamamatsu MPPCs 12571-025C with a $1\times1~\text{mm}^2$ active area and 1600 pixels. The data were collected with a 16-channel CAEN digitizer DT5742 with 5 GHz sampling rate and 12-bit resolution.

The average light yield was about 40 p.e./MIP in a single fiber, and the total light yield from two fibers in the same cube was measured on an event-by-event basis to be about 80 p.e., as expected.
The light cross-talk probability between a cube fired by a charged particle and a neighboring cube was studied. The light measured in the neighboring cube was about 3.4\% of the light collected from the fired cube. 
The timing resolution for a single fiber was measured to be $\sim$0.95~ns. If the light of a cube is read out by two WLS fibers, the timing resolution becomes better than 0.7~ns and would improve further if the light collected by all the three WLS fibers is taken into account.
In Figure~\ref{fig:testbeam2017} the light yield and the time spectra obtained from two fibers reading out the light in the same cube are shown.

\begin{dunefigure}[Charge and time spectra for a single 3DST cube]{fig:testbeam2017}
{Charge and time spectra for a single \dword{3dst} cube. Charge signal is a sum from two fibers, the time is an average time between two fibers.}
  \includegraphics[width=3.5in]{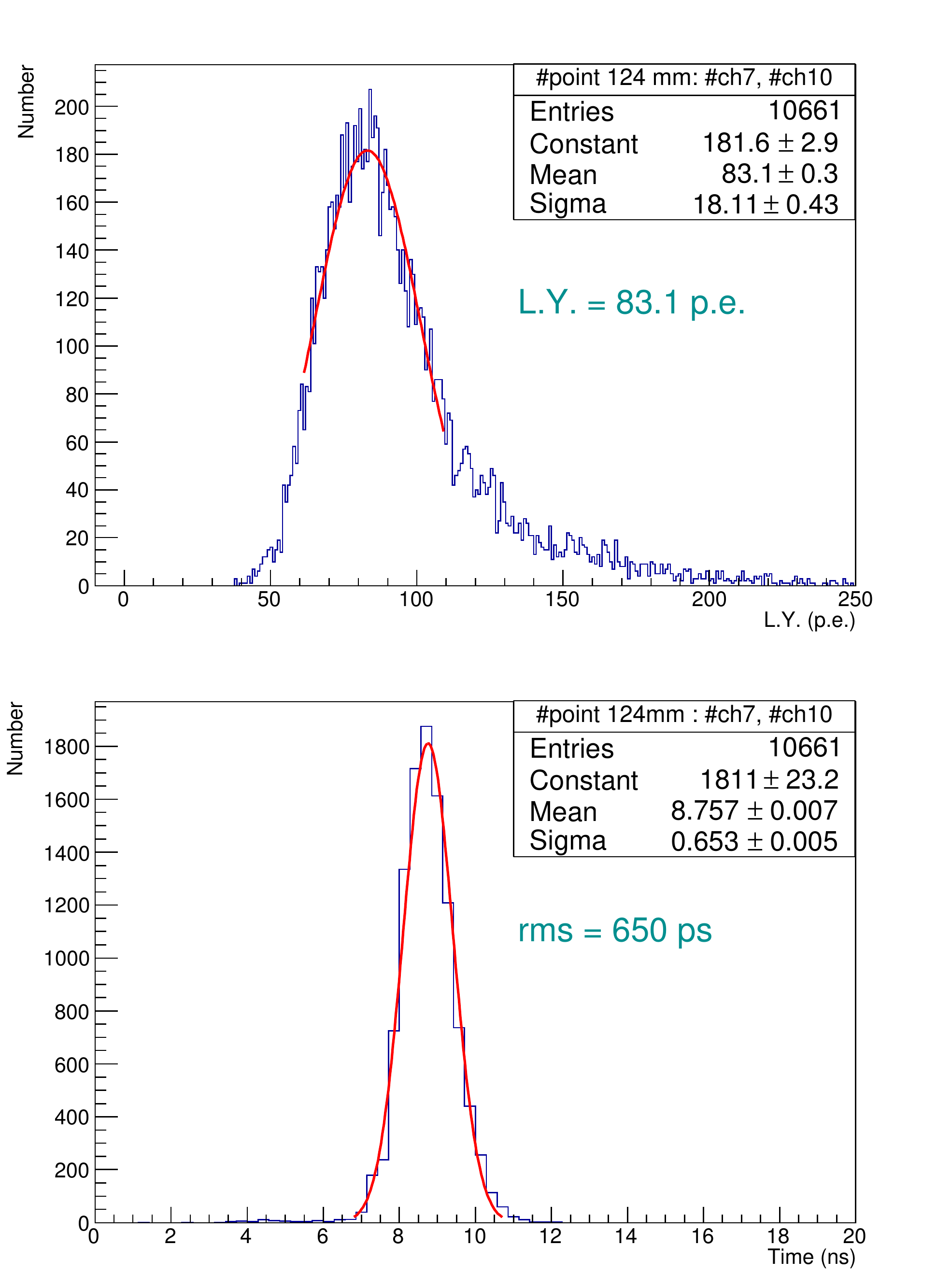}
\end{dunefigure}

In the summer of 2018, a prototype made of 9,216 cubes with a size of 8$\times$24$\times$48~cm$^{3}$  collected data in the CERN T9 test-beam line.
A different electronic readout was used, which was based on the CITIROC chip used in the Baby MIND experiment.
Preliminary results confirmed the light yield performances of the 2017 test-beam data. A more detailed analysis of the data is currently ongoing.
Some event displays are shown in Figure~\ref{fig:testbeam2018}.

\begin{dunefigure}[Event displays from the 2018 test beam]{fig:testbeam2018}
{Event displays showing the three two-dimensional projections of energy from a photon conversion (top) and a stopping proton (bottom). From data collected at the 2018 test beams at the CERN T9 area.}
\includegraphics[width=7.5in]{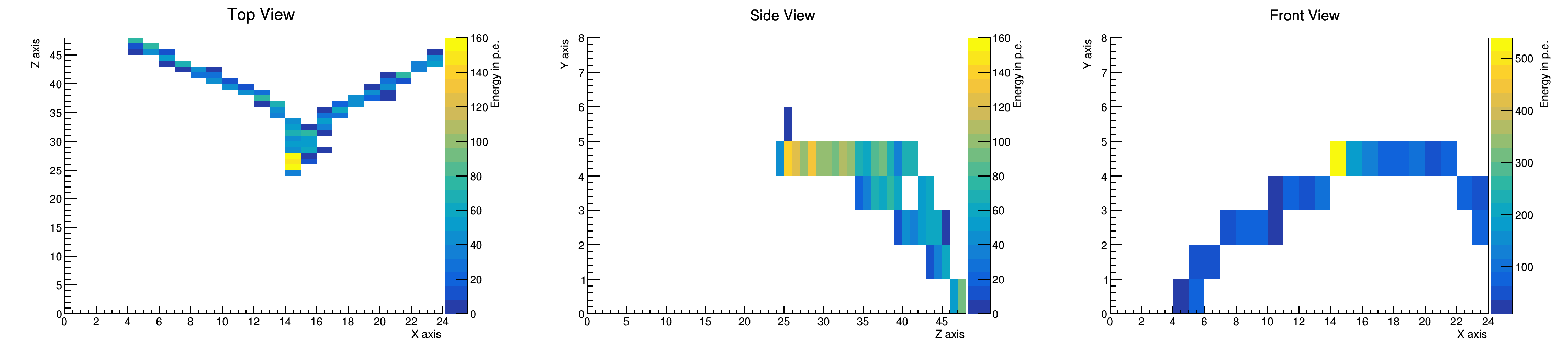}
\includegraphics[width=7.5in]{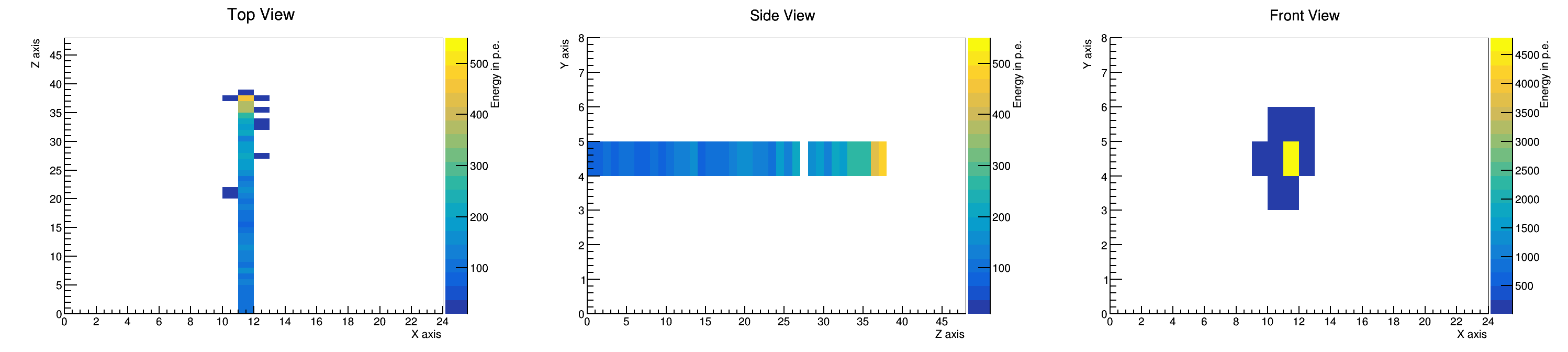}
\end{dunefigure}

\subsubsection{Expected Statistics}

The default size of the \dword{3dst}, 2.4$\times$2.4$\times$2.0~m$^{3}$, gives a total target mass of 12 metric tons.  Implementing a generic veto region around each side of the detector of 10 cm, gives a fiducial mass of 8.7~tons.
Table~\ref{tab:3dststats} gives the number of events expected per year in the \dword{fv} of such a \dword{3dst} detector.  The numbers given in the table are assuming the 80 GeV, 3 horn, optimized \dword{lbnf} beam flux and 1.46$\times10^{21}$~POT/year.

\begin{dunetable}[\dshort{sand} event rates]{c|c|c}{tab:3dststats}{This table summarizes the projected event rates per year for a \SI{2.4x2.4x2.0}{m} \dword{3dst} detector, assuming the 80 GeV, three horn, optimized \dword{lbnf} beam. A 10~cm veto region at each side was required.}
Channel & $\nu$ mode & $\bar{\nu}$ mode \\ \toprowrule
$\nu_{\mu}$ \dword{cc} inclusive & 13.6$\times$10$^{6}$ & 5.1$\times$10$^{6}$ \\ \colhline
\dword{ccqe} & 2.9$\times$10$^{6}$ & 1.6$\times$10$^{6}$ \\ \colhline
\dword{cc} $\pi^{\circ}$ inclusive & 3.8$\times$10$^{6}$ & 0.97$\times$10$^{6}$ \\ \colhline
\dword{nc} total & 4.9$\times$10$^{6}$ & 2.1$\times$10$^{6}$ \\ \colhline
$\nu_{\mu}$-e$^{-}$ scattering & 1067 & 1008 \\ \colhline
$\nu_{e}$ \dword{cc} inclusive & 2.5$\times$10$^{5}$ & 0.56$\times$10$^{5}$ \\ 
\end{dunetable}

\subsubsection{Beam monitoring}
\label{sssec:appx:nd:3dst-bm}

\dword{sand} is a capable on-axis beam monitor. The plot in Figure~\ref{fig:BeamMonitoring3DST} shows the number of events seen in 2.5 days of nominal running across the face of the 3DST.  The beam center can be measured to 11~cm in that time.  
The right plot of Figure~\ref{fig:BeamMoncompare}, previously shown, shows the significance of the change in the reconstructed neutrino energy spectra as a function of reconstructed energy for one week of nominal running for 1$\sigma$ shifts in the transverse position of horns 1 and 2.  
The leptonic and hadronic energies were smeared with a parameterization appropriate for  \dword{sand}.

\begin{dunefigure}[\dshort{sand} beam monitoring capability.]{fig:BeamMonitoring3DST}
{Beam center determination after 2.5 days of running with \dword{sand}.}
\includegraphics[width=5.in]{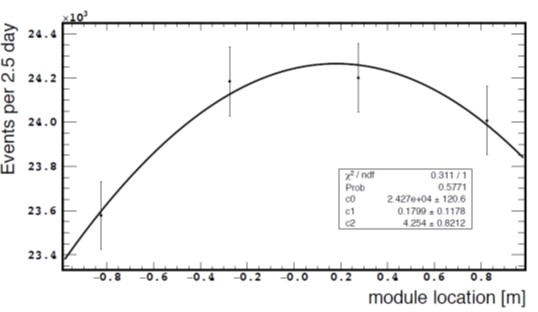}
\end{dunefigure}

\subsubsection{Neutron Detection Performance}

The \dword{minerva} experiment demonstrated the ability of measuring neutrons produced in neutrino interactions with a plastic scintillator detector~\cite{Elkins:2019vmy}. 
The \dword{3dst} should be able to do this far better than \dword{minerva} because of its high granularity and exquisite timing resolution (both much better than \dword{minerva}). 

Neutron scattering can  be seen clearly in \dword{3dst} simulations. 
Figure~\ref{fig:NeutronDisplay} shows an example of an
$\overline{\nu}_{\mu}$ \dword{cc}  single charged pion interaction. The neutron-induced energy deposition due to proton recoil can be seen apart from the vertex region. 
Inspired by \dword{minerva}, recent studies (described below) have shown that the \dword{3dst} can tag the presence of neutrons as well as determine the neutron energy via time-of-flight. 
This capability is likely to be helpful for 
 improving both neutrino and antineutrino interaction models, and of potential use when faced with "unknown unknown" sources of systematic uncertainties. The argon-based detectors in the \dword{nd} complex are expected to have some ability to detect neutrons, but studies indicate the sensitivity will be limited to regions of relatively high neutron kinetic energy (due to backgrounds and event confusion that arise at lower neutron kinetic energy where the considered event time window must be larger).  The \dword{3dst} will be sensitive to neutrons down to significantly lower kinetic energy. 

The neutron measurements in the \dword{3dst} are on carbon and likely will have limited direct bearing on tuning the neutron model for argon.  However, the analysis of events including neutrons on an event-by-event basis may lead to improvements in the neutrino interaction model for carbon.  Insights and neutrino interaction model improvements on carbon may inform the model used for argon.  A notable example of this from the recent past is the 2p2h process, which is included in many current interaction models for argon even though the evidence for multi-nucleon processes was 
 extracted from data taken on hydrocarbon targets.

Since work to date has focused on establishing the ability and quality of the neutron detection in the 3DST (as shown below), detailed studies making use of the neutron reconstruction in simulated analyses are in an early stage.  Simulations show the selection of $\overline{\nu}_\mu$ \dword{ccqe} events with small missing transverse momentum, using a technique described in \cite{Abe:2018pwo}, yields a sample with a substantially improved energy resolution.  This sample consists of events with relatively small nuclear effects which is useful for flux determination and studies of nuclear effects in neutrino interactions.  It is also expected that neutron multiplicity can  be used as an indication that multi-nucleon interactions or large \dword{fsi} effects are present, which may be helpful for selecting  events particularly useful for exploration of the interaction model.

\begin{dunefigure}[Sample antineutrino
interaction in the \dshort{3dst}]{fig:NeutronDisplay}
{An example of the antineutrino
interaction in a 2.4$\times$2.4$\times$2.0~m$^{3}$ \dword{3dst}. 
The number of \phel{}s (PE) is plotted.
An isolated cluster of hits 
corresponds to a neutron indirect signature produced by the antineutrino interaction.}
  \includegraphics[width=5in]{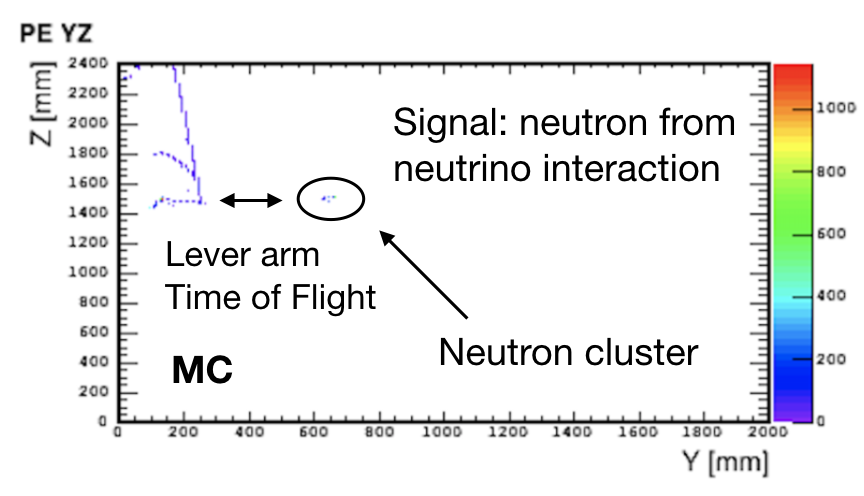}
\end{dunefigure}

With a 2.4$\times$2.4$\times$2.0~m$^{3}$ \dword{3dst} detector, Figure~\ref{fig:nProposal_4} shows 
the reconstructed neutron energy residual for 100 MeV kinetic energy neutron using time-of-flight with a lever arm (distance between neutron hit and neutrino vertex) larger than 0.5~m and smaller than 1~m.
This study was conducted with a neutron particle gun simulation.
The tail is due to both the timing resolution as well as the mis-reconstructed neutron flight distance due to non-visible interactions like elastic scattering with carbon.
The neutron energy resolution is about 18\%. \\
\begin{dunefigure}[Reconstructed neutron energy residual in the \dshort{3dst}]{fig:nProposal_4}
{Reconstructed neutron energy residual with lever arm larger than 0.5~m and smaller than 1~m for 100 MeV for a 2.4$\times$2.4$\times$2.0~m$^{3}$ \dword{3dst} detector.}
  \includegraphics[width=5in]{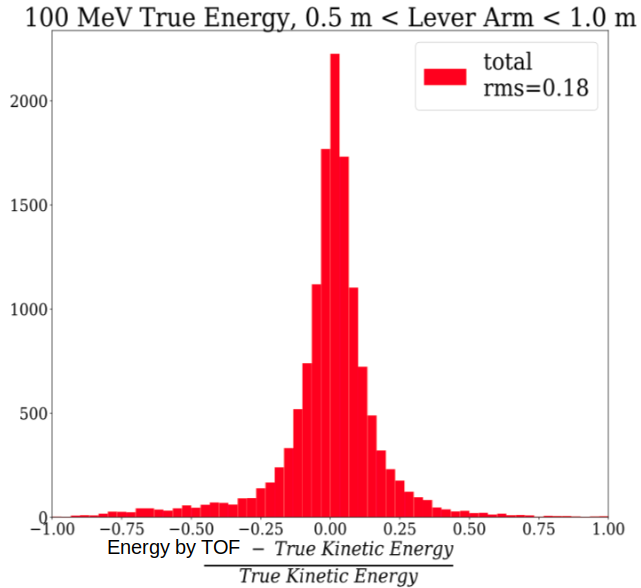}
\end{dunefigure}

 Neutrons produced by neutrino interactions happening outside the \dword{3dst} \dword{fv} (out-\dword{fv}), such as in the \dword{ecal}, Magnet, front detector, and rock can act as a background to the neutron signal from neutrino interactions. 
A simulation study was performed to understand the significance of background. In this study, the \dword{sand} detector was place in an underground alcove and significant dead material was placed upstream.  The \dword{fv} was taken to be an inner core of 1.0$\times$1.0$\times$1.0~m$^{3}$ of scintillator inside a \dword{3dst} of size 2.0$\times$2.0$\times$2.0~m$^{3}$.  Neutrino beam spills of separation 1.3~s were used.  Within each spill, the neutrinos were distributed uniformly in time.
For each neutrino interaction occurring inside the \dword{fv},  
 the earliest neutron-induced hit leaving an energy greater than 0.5 MeV in one cube was recorded. This threshold is thought to be conservative for the \dword{3dst} system because of the large light yield expected.  If that hit was from the neutrino interaction vertex, it was considered a signal neutron-induced hit. On the other hand, if that hit was created by a particle from outside the \dword{fv}, it was considered a background neutron-induced hit.
Figure~\ref{fig:neutronBG1} shows the time difference between the neutrino interaction vertex time ($t_{vtx}$) and the following earliest neutron-induced hit time ($t_{neutron}$). 
Note that a pure signal neutron sample can be obtained by cutting on ($t_{neutron} - t_{vtx}$). 

\begin{dunefigure}[Time between $\nu$ interaction vertex in \dshort{3dst} and earliest neutron-induced hit]{fig:neutronBG1}
{Time difference between the neutrino interaction vertex time inside the 1.0$\times$1.0$\times$1.0~m$^{3}$ \dword{fv} core of the \dword{3dst} and the earliest neutron-induced hit time. The neutron-induced hit leaves at least 0.5 MeV in a single cube. The neutron-induced background hits arise from neutrons produced in neutrino interactions outside the \dword{fv}.}
  \includegraphics[width=5.2in]{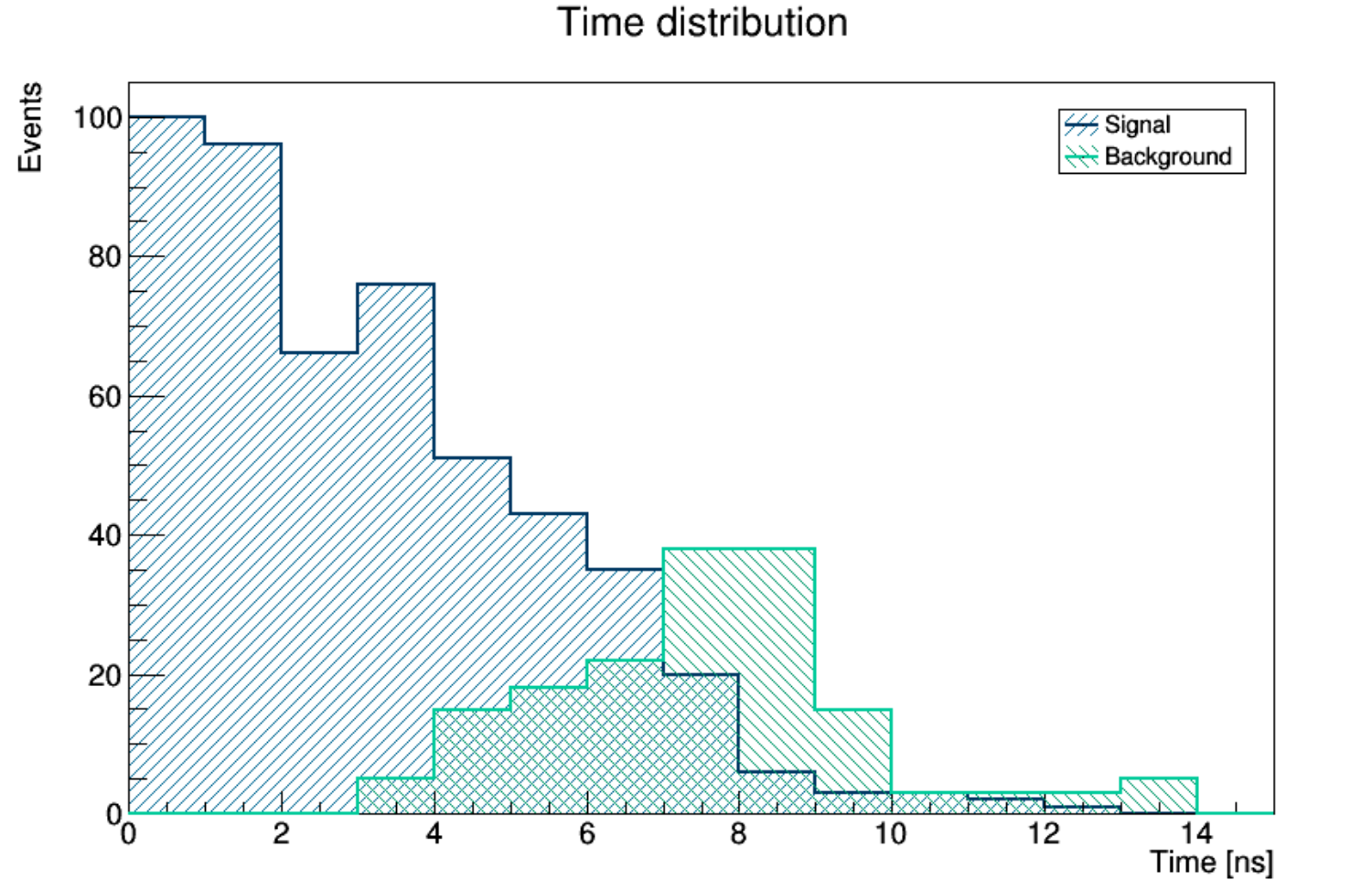}
\end{dunefigure}

It is likely to be possible to veto  \dword{cc} and \dword{nc} interactions with pions in the materials surrounding the \dword{3dst}.  Such a veto would reduce backgrounds from neutrons generated by these out-\dword{fv} events.  In this study, such a veto was not used. This will be investigated in the future.

To quantify the background, the purity is defined as the ratio of events where the first neutron-induced hit by time is from the signal vertex to all events which have a neutron-induced hit in the \dword{fv}. 
Figure~\ref{fig:nominal_purity} shows the purity in time - lever arm space. Lines indicate regions populated by neutrons with different kinetic energies.

\begin{dunefigure}[Purity of neutron-induced hit in the (time, lever arm) space for the 3DST]{fig:nominal_purity}
{Purity of the neutron-induced hit in the (time, lever arm) space.
The dashed line corresponds to the cut required to select an almost 100\% pure sample of signal neutrons. The solid lines are theoretical curves for neutrons with different kinetic energies.
Note that this study was performed with a total volume of 2$\times$2$\times$2 m$^3$.
See text for details.}
  \includegraphics[width=5in]{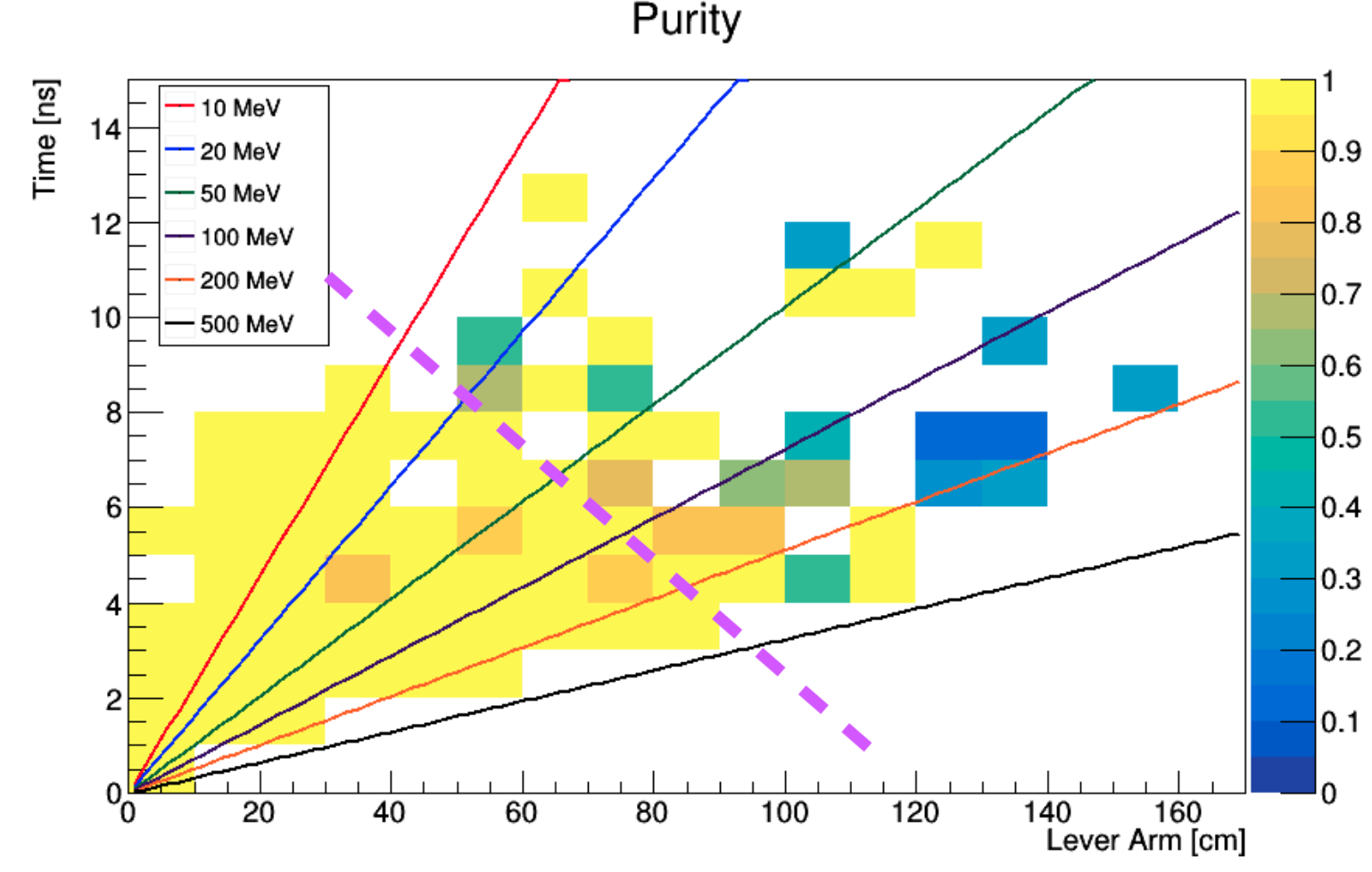}
\end{dunefigure}

The reconstructed energy resolution in the same (time, lever arm) 2D space was studied. For this work, the time was smeared by 0.58~ns, corresponding to a per fiber time resolution of 1~ns (the documented performance in the CERN test beam is 0.9~ns).
Though higher light yield can help improve the time resolution, this effect has not been taken into account. Figure~\ref{fig:nominal_resolution} shows the reconstructed-by-ToF neutron energy resolution. In general, $\sim 20 \%$ energy resolution can be reached with most of the lever arm and time windows, 
in the region selected by the background cut. \\

\begin{dunefigure}[Energy resolution for neutron candidates in the (time, lever arm) space for the 3DST]{fig:nominal_resolution}
{Energy resolution of the neutron candidates in the (time, lever arm) space. 
The dashed line corresponds to the cut required to select an almost 100\% pure sample of signal neutrons. The solid lines are theoretical curves for neutrons with different kinetic energies.
Note that this study was performed with a total volume of 2.0$\times$2.0$\times$2.0~m$^3$.
See text for details.}
  \includegraphics[width=5in]{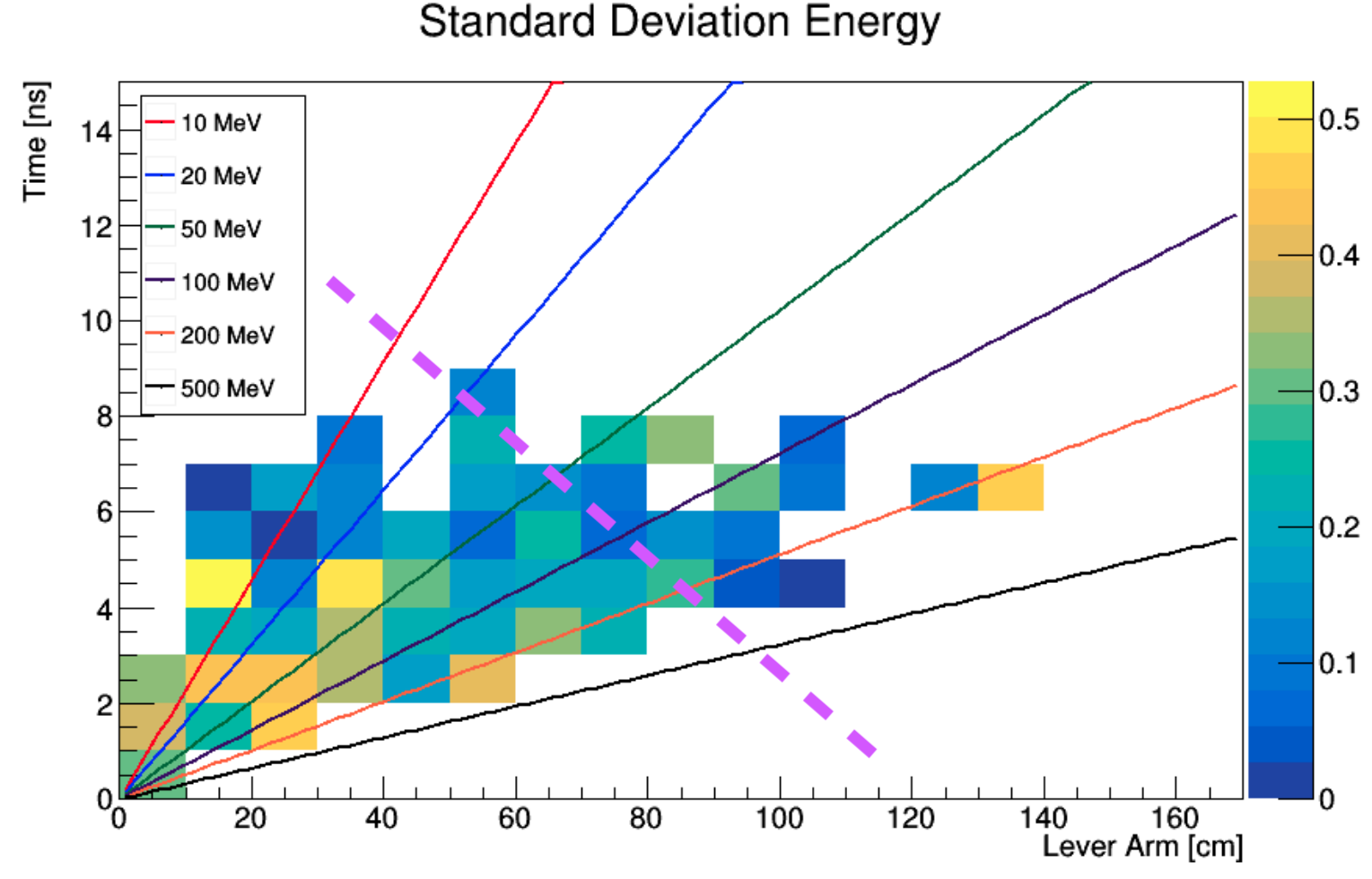}
\end{dunefigure}

\section{Meeting the Near Detector Requirements}
\label{sec:appx-nd:requirements}

As discussed in Sections~\ref{sec:appx-nd:exsum-nd-role} and \ref{sec:appx-nd-overview}, the \dword{dune}  \dword{nd} complex has many missions, and the components of the  \dword{nd}  address these missions in a complementary fashion. In this section we first discuss the key overarching requirements driving the \dword{nd} complex. We then discuss some thought experiments and case studies that illustrate how different parts of the complex work together. These case studies naturally suggest more detailed capabilities, performance statistics, and technical requirements that we are in the process of tabulating.  Most of these require additional studies before numerical values can be tabulated.

\subsection{Overarching Requirements}

\begin{itemize}
    \item {{\bf Predict the neutrino spectrum at the \dword{fd}:}} The \dword{nd} must provide a prediction for the energy spectrum of \numu, \anumu, \nue and \anue at the \dword{fd}. The prediction must be provided as a function of the oscillation parameters and systematic uncertainties must be small enough to achieve the required \dword{cp} coverage. This is the primary requirement of the \dword{dune} \dword{nd}.
    
    \item{{\bf Measure interactions on argon:}} The \dword{nd} must measure neutrino interactions on argon to reduce uncertainties due to nuclear modeling. The \dword{nd} must be able to determine the neutrino flavor and measure the full kinematic range of the interactions that will be seen at the FD.
    
    \item{{\bf Measure the neutrino energy:}} The \dword{nd} must be able to reconstruct the neutrino energy in \dword{cc} events and control for any biases in energy scale or resolution, keeping them small enough to achieve the required \dword{cp} coverage. These measurements must also be transferable to the \dword{fd}. 
    
    \item{{\bf Constrain the cross-section model:}} The \dword{nd} must measure neutrino cross-sections in order to constrain the cross-section model used in the oscillation analysis. In particular, cross-section mismodeling that causes incorrect \dword{fd} predictions as a function of neutrino flavor and true or reconstructed energy must be constrained well enough to achieve the required \dword{cp} coverage. 
    
    \item{{\bf Measure neutrino fluxes:}} The \dword{nd} must measure neutrino fluxes as a function of flavor and neutrino energy. This allows for neutrino cross section measurements to be made and constrains the beam model and the extrapolation of neutrino energy spectra from the \dword{nd} to the \dword{fd}.
    
    \item{{\bf Obtain data with different fluxes:}} The \dword{nd} must measure neutrino interactions in different beam fluxes (especially ones with different mean energies) to deconvolve flux and cross section, verify the beam model, and guard against systematic uncertainties on the neutrino energy reconstruction.
    
    \item{{\bf Monitor the neutrino beam:}} The \dword{nd} must monitor the neutrino beam energy spectrum with sufficient statistics to be sensitive to intentional or accidental changes in the beam that could affect the oscillation measurement.
    
    
\end{itemize}

\subsection{Event Rate and Flux Measurements}

The three most straightforward flux measurements are described here. Other powerful but more complex techniques are described in Section~\ref{sec:appx-nd:fluxappendix}.

\subsubsection{CC $\nu_{\mu}$ and $\overline{\nu}_{\mu}$ Interactions}
Each core component of the \dword{nd} complex will have large data samples with which to constrain the flux model:  \dword{arcube} will accumulate \num{3.7e7} \dword{cc} $\nu_{\mu}$ events per year (on axis, less when off axis);  The \dword{3dst} will see \num{1.4e7} \dword{cc} $\nu_{\mu}$ per year on axis; and the \dword{hpgtpc} will see \num{1.6e6} \dword{cc} $\nu_{\mu}$ events per year (on axis, less when off axis).

\subsubsection{Intrinsic Electron Neutrino and Antineutrino Fluxes}

The intrinsic $\nu_{e}$ and $\overline{\nu}_{e}$ component of the beam is discussed in Section~\ref{ssec:beam-nue}.  This is an important component to quantify in the  \dword{nd} since it represents an irreducible background for the appearance oscillation analysis at the FD.  The number of \dword{cc} $\nu_{e}$ events expected in the \dword{nd} per year (on axis) are \num{6.7e5}, \num{2.5e5}, and \num{2.5e4} for \dword{arcube}, the \dword{3dst}, and the \dword{hpgtpc}, respectively.  The primary background comes from \dword{nc}$\pi^{\circ}$ production.  The systematics are dominated by the flux model and the interaction model (which enters in the background subtraction).  In the past, statistics has been a limitation.  That will not be the case for \dword{dune} \dword{nd}.  With large samples, \dword{arcube} and the \dword{3dst} each will measure this component of the beam fairly quickly with somewhat different systematic errors.  Although accumulating statistics more slowly, the \dword{hpgtpc} will provide the best overall measurements the $\nu_{e}$ and $\overline{\nu}_{e}$ components of the beam.  Photons mostly do not convert in the gas.  This eliminates the primary background to electron (anti)neutrino identification and the accompanying interaction model uncertainty in the background subtraction.  In addition, the \dword{hpgtpc} has a magnetic field that allows for the sign separation of $\nu_{e}$ and $\overline{\nu}_{e}$.

\subsubsection{Neutrino-Electron Scattering}
This process and estimates of the  \dword{nd} performance measuring the flux using this technique is discussed in Sections~\ref{sec:appx-nd:fluxintro-e-nu-scatt} and~\ref{sec:appx-nd:lartpc-nu-electron-scatt}.  Measuring the flux using this process is a critical \dword{nd} mission because it is independent of nuclear effects.  This is a rare process that can be used by both \dword{arcube} and \dword{sand} components of the \dword{nd} to measure the neutrino flux.  The target nucleus is irrelevant for neutrino-electron scattering.  The measurement places a premium on the overall target mass (for statistics) as well as electron energy and angular resolutions.  The primary backgrounds are \dword{cc} interactions of intrinsic beam $\nu_{e}$ and \dword{nc}$\pi^{\circ}$ interactions.  \dword{arcube} will do this measurement well as indicated by the results of a study shown in Figure~\ref{fig:nominal_det_constraint}.  Also, that study shows a \dword{minerva}-like scintillator detector can do the measurement fairly well.  \dword{sand} will have better angular resolution than the detector used in the study. Note that the detector and reconstruction systematic errors will be different for the two very different detectors.  For such an important measurement, the duplication is good, and with many uncorrelated errors it may be possible to combine the \dword{arcube} data set with that from  \dword{sand} for a somewhat improved constraint. 

\subsection{Control of Systematic Errors}

\subsubsection{PRISM program}

The PRISM program of on- and off-axis measurements is an essential component of the  \dword{nd} complex and plays a key role in reducing systematic uncertainties on neutrino cross-sections and reconstructed energy. PRISM is described in Section~\ref{sec:appx-nd:DP}. 

\subsubsection{Absolute muon energy scale}

The  \dword{nd} complex must provide knowledge of the absolute muon energy scale in the \larnd with sufficient accuracy to meet the oscillation physics requirement 
and the ancillary low-nu capability. 
The complex will utilize \dword{mpd} magnetic field survey information, 
as well as K$_{s}$ and $\Lambda$ decays to charged hadrons 
within the \dword{mpd}, to establish the charged particle momentum scale in the \dword{mpd}. 
The measurement will be translated to the \larnd by tracking and momentum analyzing, in the \dword{mpd}, cosmic ray muons going through the \dword{mpd} into the \larnd{}. 
The \larnd will then measure the momentum of the muons (particularly stopping muons) by range and multiple coulomb scattering (MCS) to establish its muon momentum scale and verify the reconstruction and detector simulation. 

The momentum scale will be translated to the FD by measuring and comparing the range and MCS of stopping tracks in the data and the simulation.


\subsubsection{Hadronic Energy Scale; Hadronic Response of the LArTPC}

The  \dword{nd} complex must calibrate the response (energy measured vs true energy) of the \larnd and FD to the hadronic system in neutrino interactions with an accuracy 
to meet the oscillation physics goals. The complex will utilize a simulation benchmarked against the single particle response measured in ProtoDUNE as a baseline. It is expected that the response will differ for different particle species and will not be constant as a function of energy. Additional calibration is needed due to the imperfectly known particle spectra in neutrino interactions, and the confusion that the \dword{lar}  \dword{nd} and \dword{fd} will experience in identifying the composition of particles in the final state. The  \dword{nd} complex will accomplish the calibration by first observing $\numu$-\dword{cc} interactions in the \dword{mpd}. The \dword{mpd} will identify the protons, charged pions and photons in the final state, 
and precisely reconstruct their energy via curvature in the magnetic field and, for photons, energy deposition in the calorimeter. 
The \dword{mpd} will also observe, statistically, the energy going into neutrons using time of flight. 

The muon kinematics will then be used to select one or more semi-inclusive set of events occurring in the \larnd and \dword{mpd} that have identical hadronic final states. A comparison of the reconstructed hadronic energy in the \larnd with the precisely measured \dword{mpd} calibrates the response and establishes the hadronic energy scale in LAr. It will also be possible to take the reconstructed data events from the \dword{mpd} and simulate them in the LAr to compare to the actual events in the LAr. This will serve as a cross-check and as a way of studying the response. For example, one could drop any primary neutron information from the \dword{mpd} reconstructed events and simulate the rest in \larnd to compare to the \larnd data as a way of understanding the amount of neutron energy seen by the \larnd.

\subsubsection{Beam Monitoring}
Previous experience with the NUMI, JPARC, and Booster neutrino beams has shown that DUNE must prepare for changes in the beam that occur on timescales as short as a few days but that are not readily detected by primary and secondary beam monitors. Sudden changes in the beam have the potential of polluting the PRISM program if they occur when the \larnd and \dword{mpd} are taking data at off-axis locations. The  \dword{nd} complex addresses this problem with the \dword{sand} detector which will continuously measure the \numu and \anumu energy spectrum on-axis.

\subsubsection{LArTPC energy resolution}

The PRISM measurement program demands that the energy resolution of the \larnd and FD be as similar as possible, and that any differences can be understood and corrected for\footnote{In fact this is true if  the experiment only makes on-axis measurements.}. The resolution can be constrained using a similar approach as followed for the hadronic response.

\subsubsection{\larnd acceptance}

Neutrino flavor change due to oscillation occurs over a broad energy range which demands that the  \dword{nd} complex is able to achieve a broad energy coverage. The \larnd has acceptance limits (relative to the FD) at large muon angle ($\theta > 25\deg$) and high hadronic energy due to its limited size. Moreover, there is a strongly energy dependent and deep acceptance dip for $1.0 < E_\mu < 1.5~GeV/c$ due to dead material between the  active portions of the \larnd and \dword{mpd} TPC. The experiment must understand these features of the \larnd event sample in order to use it to make predictions for the FD\rrt{LArNDAcceptance}. 

To address this shortcoming the \dword{mpd} must have a nearly $4\pi$ acceptance for charged tracks and photons, a high tracking efficiency, and the ability to reconstruct events with high $E_\nu$\rrt{MPDAcceptance}. A comparison of \dword{mpd} and \larnd event rates as a function of kinematic variables will verify the \larnd acceptance model.

\subsubsection{LAr detection thresholds}

The energy threshold for detecting charged particles in LAr will be verified using the \dword{mpd} in a way that is similar to the more general \larnd acceptance study described above. 

\subsubsection{Managing pile-up}
At the location of the  \dword{nd}, the \dword{lbnf} beam is expected to generate approximately 1 neutrino interaction per 10 tons per 10~$\mu$s spill.   
Each of the core detectors in the \dword{nd} complex can eliminate most of the pile-up background with timing from optical elements. For \dword{arcube}, it is estimated that there will be approximately 0.5 neutrino interactions per spill per \dword{arcube} module. Prompt scintillation light from the argon detected in \dword{arclt} detectors or something similar is used to provide t$_{o}$ and separate events (both connected and disconnected parts of the event).  For the \dword{hpgtpc}, an estimate based on a significantly (15$\times$) more massive magnet than the superconducting option presented in section~\ref{sssec:nd:appx:mpd-magnet} suggests there will be approximately 75 tracks per 10~$\mu$s spill from interactions in surrounding materials passing through the \dword{tpc}.  The excellent <10~ns timing resolution of the \dword{ecal} surrounding the \dword{tpc} will be used to provide a t$_{o}$ and to define a time window for pileup rejection.  Similarly the exquisite (sub-ns) time resolution of the cubes in the \dword{3dst} can be used to generate a narrow window in time around neutrino interactions and limit the potential for overlapping events.

\subsubsection{Energy Carried by Neutrons}

The  \dword{nd} complex must measure or otherwise account for the neutrino energy that goes into neutrons because much of that energy ends up being undetectable by a \dword{lartpc}. The fraction of hadronic energy carried by neutrons is sizable and it also differs between \numu and \anumu: 20\% vs 40\% for the flux peak. To meet the oscillation physics goals these fractions must be known sufficiently well. 

The complex will approach this challenge in multiple ways. First, the PRISM measurement program is required to map the relationship between true and reconstructed energy using inclusive \dword{cc} scattering. 
These measurements are sensitive to cumulative mismodelings but may have trouble pinning down their origin. To augment the PRISM program, the \dword{mpd} and \dword{sand} are required to measure the energy carried by neutrons using calorimetry and time of flight. 
The \dword{mpd} has the advantage of measuring neutron production off of an Ar target, but the disadvantage of doing so with lower statistics than  \dword{sand}. The \dword{mpd} measurement is also challenging due to the interaction rate in its calorimeter and the composition of the calorimeter driven by its multi-role nature. The advantage of  \dword{sand} is in finer granularity and a better ability to reconstruct neutron energy on an event by event basis.

\subsubsection{\nue-CC rate and $\pi^0$/$\gamma$ background}

Neutral current events with a final state $\pi^0$, or a single $\gamma$, are a potentially problematic background to the \nue and \anue appearance measurements. The  \dword{nd} complex will address this background by using the \dword{mpd} to precisely measure the rate of \nue (and \anue) \dword{cc} interactions as a function of energy and other kinematic variables. 
The measurement will have relatively small $\pi^0$ and $\gamma$ backgrounds because photons have a low conversion probability in the \dword{mpd} TPC. The \larnd will make a similar measurement. The efficiency and background of that measurement will benchmarked using the data from the \dword{mpd}.

\chapter{ND Hall and Construction}
\label{sec:appx-nd:exsum-nd-hall}

Figure~\ref{fig:NDhall} shows the current design of the underground hall as required for the  \dword{nd} construction concept. The underground hall must house the detector components and allow for the required movement. The layout shows the space required for the detector, safety, and egress.  This is work in progress. 

\begin{dunefigure}[DUNE ND hall from above and side (transverse to beam)]{fig:NDhall}
{\dword{dune}  \dword{nd} hall shown from above (top) and from the side transverse to the beam (bottom). The \dword{lartpc}, \dword{mpd}, and \dword{3dst} detectors are shown in position on the beam axis in both drawings. }
\includegraphics[width=0.8\textwidth]{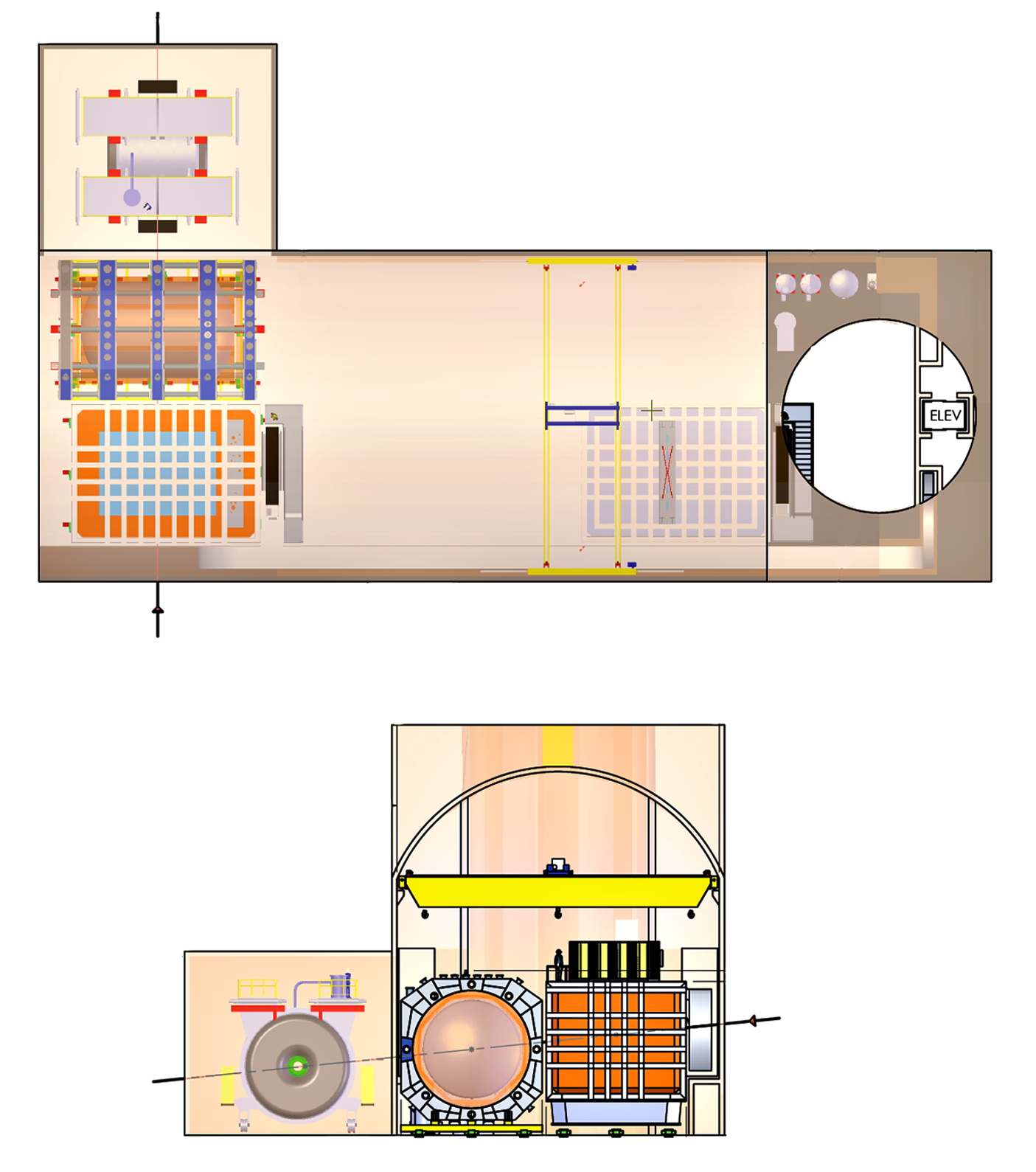}
\end{dunefigure}

The overall construction method places requirements on the conventional facilities. 
The primary access shaft is large enough for lowering the pressure vessel and the magnet coils. The \dword{lar} cryostat is shown in its construction position near the main shaft. The multipurpose detector and the \dword{lar} detector are also shown in the on-axis position. Since the \dword{3dst} detector does not need to move for \dword{duneprism}, it is shown in a dedicated alcove downstream of the \dword{lar} and multipurpose detectors.

The overall method of detector construction must be consistent with the construction concepts of each of the elements as outlined in previous sections. The construction method must also allow for parallel activities on major components and reduce demand on individual facilities. The underground hall will be the last facility to be completed. Therefore, insofar as possible, components must be constructed elsewhere and lowered as large assemblies.

The current assumptions for the overall construction involving the major components are listed below.  Only the major components are considered, as they place the main constraints on the conventional facilities.
\begin{itemize}
    \item The primary access shaft diameter is sufficient to  accommodate the lowering of the pressure vessel and magnet coils separately. The coils and pressure vessel are constructed on the surface and lowered. This allows for remote and parallel construction.
    \item Two transport frames and moving systems are built in the cavern.  These frames will be used to support and move the multipurpose detector and the \dword{lar} detector.
    \item Articulated carriers are built to carry services supporting the moving detectors.
    \item The components of  \dword{sand} are constructed remotely and assembled in the underground alcove cavern.  
    \item The five magnet coils are constructed and integrated with cryostats remotely.  The magnet coils and cryostats are lowered into the  \dword{nd} hall and assembled together to form the magnet system on one of the transport frames.
    \item The pressure vessel is fabricated remotely and transported to the  \dword{nd} hall surface building.  The TPC and some of the \dword{ecal} components are constructed in the surface building and installed in the pressure vessel.  The pressure vessel must be fabricated and certified by a qualified fabricator.
    \item The \dword{lar} cryostat is constructed in the cavern on the second transport frame near the main shaft.  The construction of the cryostat starts by erecting and assembling the warm exoskeleton from pre-fabricated structural steel members. The warm membrane is then installed and welded in situ. Insulation is then installed inside the warm vessel. The cold membrane is the last component and is welded in situ to form the final containment vessel for the \dword{lar}. A thin window is installed on the side facing the multipurpose detector. 
    \item The \dword{lar} modules are constructed remotely and lowered down the shaft and installed in the cryostat. The modules are inserted/extracted from the  top of the cryostat using an overhead lifting device. 
    \item The \dword{lar} services are installed.
    \item The \dword{hpgtpc} inside the pressure vessel is lowered in the hall and then mounted inside the magnet system. 
    \item The \dword{ecal} segments are lowered into the hall and mounted around the pressure vessel. 
    \item  Services are installed.
\end{itemize}
The basic requirement for \dword{duneprism} is that both the \dword{mpd} and \dword{arcube} can move horizontally to a position off the beam axis. The direction of the motion is to one side of the beam and the total motion is approximately \SI{30.5}{\meter}. 

Though the \dword{mpd} and \dword{arcube} will be moved together to different positions for operations, they will be able to move independently for engineering, construction, and maintenance reasons.  The specific method of movement is not yet determined. However, it is anticipated that tracks and rollers will be used in a fashion similar to what has been done for other large particle physics detectors. The driving mechanism may be a rack and pinion drive, or a similar system, which also allows for continuous motion. 
It is planned that the speed of movement will allow for the entire motion to be completed in one 8-hour shift. This requires a speed of approximately \SI{6}{cm/min}. If it is desired that data can be taken during the movement, a slower speed may be used. If moved at a speed of about \SI{0.6}{cm/min}, an entire round trip would take about one week. 

Services for both the \dword{mpd} and \dword{arcube} will need to remain connected while moving, or be disconnected and reconnected at intermediate positions. Ideally, the former will be possible, and articulated service carriers will maintain the connections during movement. In the case of \dword{arcube} this presents particular challenges and will require flexible conduits. 
\cleardoublepage

\chapter{Computing Roles and Collaborative  Projects} 
\label{appx:es-comp}

\section{Roles}
\label{appx:comp-roles}

This appendix lists computing roles for \dshort{dune}  derived from a comparison with existing similar roles on the \dword{lhcb} experiment at \dshort{cern}.  \dword{lhcb} is similar in size and data volumes to \dword{dune}. 

\begin{description}

\item {Distributed Computing Development and Maintenance - 5.0 FTE}

This role includes oversight of all software engineering and development activities for packages needed to operate on distributed computing resources. The role requires a good understanding of the distributed computing infrastructure used by \dword{dune} as well as the \dword{dune} computing model.

\item {Software and Computing Infrastructure Development and Maintenance - 6.0 FTE}

This 
role includes software engineering, development, and maintenance for central services operated by \dword{dune} to support software and computing activities of the project.   

\item {Database Design and Maintenance - 0.5 FTE}

This role includes designing, maintaining, and scaling databases for  
tasks within \dword{dune}. 

\item {Data Preservation Development - 0.5 FTE}

This role includes activities related to reproducibility of analysis as well as data preservation, which requires expert knowledge of analysis and the computing model.

\item {Application Managers and Librarians - 2.0 FTE}

Application managers handle software applications for data processing, simulation, and analysis, and also coordinate activites in the areas of development, release preparation, and 
deployment of software package releases needed by \dword{dune}. Librarians organize the overall setup of software packages needed for releases. 

\item {Central Services Manager and Operators - 1.5 FTE}

The site manager and operators are responsible for the central infrastructure and services of the \dword{dune} distributed computing infrastructure. This includes coordination with the host laboratory for services provided to \dword{dune}. 

\item {Distributed Production Manager - 0.5 FTE}

Distributed production managers are responsible for the setup, launch, monitoring, and 
completion of processing campaigns executed on distributed computing resources for the experiment. 
These include data processing, \dword{mc} simulation, and working group productions.

\item {Distributed Data Manager - 0.5 FTE}

The distributed data manager is responsible for operational interactions with distributed computing disk and tape resources. The role includes but is not limited to helping to establish new storage areas and data replication, deletion, and movement. 

\item {Distributed Workload Manager - 0.5 FTE}

The distributed workload manager is responsible for operational interactions with distributed computing resources. The role includes activities such as helping to establish grid and cloud sites.

\item {Computing Shift Leaders - 1.4 FTE}

The shift leader is 
responsible for the experiment's distributed computing operations for a week-long period  
starting on a Monday to the following Sunday.  Shift leaders chair regular operations meetings during their week and attend general \dword{dune} operations meetings as appropriate. 

\item {Distributed Computing Resource Contacts - 0.5 FTE}

Distributed computing resource contacts are the primary contacts for the \dword{dune} distributed computing operations team and for the operators of large (Tier-1) sites and regional federations. They interact directly with the computing shift leaders at operations meetings. 

\item {User Support - 1.0 FTE}

User support (software infrastructure, applications, and distributed computing) underpins all user activities of the \dword{dune} computing project. 
User support 
personnel 
respond to questions from users on mailing lists, Slack-style chat systems, and/or ticketing systems, 
and are responsible for documenting solutions in knowledge bases and wikis.

\item {Resource Board Chair - 0.1 FTE}

This role is responsible for chairing quarterly meetings of the Computing Resource Board, which includes representatives from the 
various national funding agencies that support \dword{dune}, to discuss 
funding for and delivery of the computing resources required for successful processing and exploitation of \dword{dune} data. 

\item {Computing Coordination - 2.0 FTE}

Coordinators oversee management of the computing project. 
\end{description}

\section{Specific Collaborative Computing Projects}
\label{ch:exec-comp-gov-coop}

The \dword{hep} computing community has come together to form a \dword{hsf}\cite{Alves:2017she} that, through working groups, workshops, and white papers, guides the next generation of shared \dword{hep} software.  
The \dword{dune} experiment's time scale, particularly the planning and evaluation phase, is almost ideal for allowing the \dword{hsf} to develop effective contributions. Our overall strategy for computing infrastructure is to carefully evaluate existing and proposed field-wide solutions, to participate in useful designs, and to build our own solutions only where common solutions do not fit and additional joint development is not feasible.   This section describes some of these common activities.

\subsection{LArSoft for Event Reconstruction}

Several neutrino experiments using the \dword{lartpc} technology share the \dword{larsoft}\cite{Snider:2017wjd} reconstruction package.  \dword{microboone}, \dword{sbnd}, \dword{dune}, and others share in developing a common core software framework that can be customized for each experiment. This software suite and earlier efforts in other experiments made the rapid reconstruction of the \dword{pdsp} data possible.  \dword{dune} will contribute heavily to  the future evolution of this package, in particular, by introducing full multi-threading to allow parallel reconstruction of parts of large events, thus anticipating the very large events expected from the full detector. 

\subsection{WLCG/OSG and the HEP Software Foundation}

The  \dword{wlcg} organization~\cite{Bird:2014ctt}, which combines the resource and infrastructure missions of the \dword{lhc} experiments, has proposed a governance structure called \dword{sci} that splits out dedicated resources for \dword{lhc} experiments from the general middleware infrastructure used to access those resources.  In a white paper submitted to the European Strategy Group in December 2018~\cite{bib:BirdEUStrategy}, a formal \dword{sci} organization is proposed. Many other experiments worldwide are already using this structure.  As part of the formal transition to \dword{sci}, the \dword{dune} collaboration was provisionally invited to join the \dword{wlcg} management board as observers and to participate in the Grid Deployment Board and task forces. Our participation allows us to contribute to the  technical decisions on global computing infrastructure while also contributing to that infrastructure. Many of these contributions involve the broader \dword{hep} Software Foundation efforts. 

Areas of collaboration are described in the following sections. 

\subsubsection{Rucio Development and Extension}

 \dword{rucio}\cite{Barisits:2019fyl}
is a data management system originally developed by the \dword{atlas} collaboration and is now an open-source project.  \dword{dune} has chosen to use \dword{rucio} for large-scale data movement.  Over the short term, it is combined with the \dword{sam} data catalog used by \dword{fnal} experiments.  \dword{dune} collaborators at \dword{fnal} and in the UK are actively collaborating on the \dword{rucio} project, adding value for both \dword{dune} and the wider community. 

Besides \dword{dune}, the global \dword{rucio} team now includes \dword{fnal} and \dword{bnl} staff, \dword{cms} collaborators, and the core developers on \dword{atlas} who initially wrote \dword{rucio}.  \dword{dune} \dword{csc} members have begun collaborating on several projects:  (1) making object stores (such as Amazon S3 and compatible utilities) work with \dword{rucio} (a large object store in the UK exists for which \dword{dune} has a sizable allocation);  (2) monitoring  and administering the \dword{rucio} system, and leveraging the landscape system at \dword{fnal}; and  (3) designing a  data description engine that can be used to replace the \dword{sam} system we now use.

\dword{rucio} has already proved to be a powerful and useful tool for moving defined datasets from point A to point B.  
\dword{rucio} appears to offer a good solution for file localization but it lacks 
the detailed tools for data description and granular dataset definition available in the 
\dword{sam} system.  The rapidly varying conditions in the test beam have shown that we need a sophisticated data description database interfaced to \dword{rucio}'s location functions. 

Efficient integration of caching into the \dword{rucio} model will be an important component for \dword{dune} unless we can afford to have most data on disk to avoid staging.  The dCache model, a caching front end for a tape store, is used in most \dword{fnal} experiments. In contrast, \dword{lhc} experiments such as \dword{atlas} and \dword{cms} work with disk storage and tape storage that are independent of each other.


\subsubsection{Testing New Storage Technologies and Interfaces}

The larger \dword{hep} community\cite{Berzano:2018xaa} currently has a \dword{doma} task force
 in which several \dword{dune} collaborators participate. 
 It includes groups working on authorization, caching, third party copy, hierarchical storage, and quality of service. All are of interest to \dword{dune} because they will determine the long-term standards for common computing infrastructure in the field. 
Authorization is of particular interest; they are covered in Section~\ref{ch-comp-auth}.

\subsubsection{Data Management and Retention Policy Development}

A data life cycle is built into the \dword{dune} data model.  Obsolete samples (old simulations and histograms and old commissioning data) need not be maintained indefinitely.  
We are organizing the structure of lower storage to store the various retention types separately for easy deletion when necessary.  

\subsubsection{Authentication and Authorization Security and Interoperability}\label{ch-comp-auth}

Within the next few years, we expect the global \dword{hep} community to change significantly the methods of authentication and authorization of computing and storage. 
Over that period, \dword{dune} must collaborate with the USA and European \dword{hep} computing communities on improved authentication methods  that will allow secure but transparent access to storage and other resources such as logbooks and code repositories.  The current model requires individuals to 
be authenticated through different mechanisms for access to USA and European resources. 
Current efforts to expand the trust realm between \dword{cern} and \dword{fnal} should allow a single sign-on for access to the two laboratories. 

\subsection{Evaluations of Other Important Infrastructure}

The \dword{dune} \dword{csc} is still evaluating some major infrastructure components, notably databases, and workflow management systems.

The \dword{fnal} \textit{conditions} database is being used for the first run of \dword{protodune}, but the Belle II~\cite{Ritter:2018jxh} system supported by \dword{bnl} is being considered for subsequent runs~\cite{Laycock:2019ynk}. 

We are evaluating \dword{dirac}~\cite{Falabella:2016waj} as a workflow management tool and plan to investigate PANDA~\cite{Megino:2017ywl}, as well, to compare against the current GlideInWMS, HT Condor, and POMS solution that was successfully used for the 2018 \dword{protodune} campaigns.
Both \dword{dirac} and PANDA are being integrated with \dword{rucio} and several \dword{lhc} and non-\dword{lhc} experiments use them.

\cleardoublepage


\cleardoublepage
\printglossaries

\cleardoublepage
\cleardoublepage
\renewcommand{\bibname}{References}
\bibliographystyle{utphys} 
\bibliography{common/tdr-citedb}

\end{document}